\documentclass[final,3p,times,nopreprintline]{elsarticle}

\usepackage{amsmath}
\usepackage{amssymb}
\usepackage{booktabs}
\usepackage{braket}
\usepackage[pdftex]{color}
\usepackage{dcolumn}
\usepackage{enumitem}
\usepackage{epsfig}
\usepackage{etoolbox}
\usepackage{graphicx}
\usepackage{hyperref}
\hypersetup{
    colorlinks=true,       
    linkcolor=blue,          
    citecolor=blue,        
    filecolor=blue,      
    urlcolor=blue           
}
\usepackage{latexsym}
\usepackage{longtable}
\usepackage{mathrsfs}
\usepackage{mathtools}
\usepackage{multirow}
\usepackage{rotating}
\usepackage{rotfloat}
\usepackage[squaren,mediumqspace]{SIunits}
\usepackage{slashed}
\usepackage{stackrel}
\usepackage{wrapfig}

\usepackage[utf8]{inputenc}
\usepackage{afterpage}
\usepackage{xspace}
\usepackage{placeins}
\usepackage{natbib}

\usepackage[symbol]{footmisc}

\usepackage[capitalise]{cleveref}

\crefname{section}{Sec.}{Secs.}

\Crefname{section}{Section}{Sections}

\biboptions{sort&compress}

\DeclareGraphicsExtensions{.pdf,.png,.jpg,.eps}

\usepackage{orcidlink}

\definecolor{red}{rgb}{0.8,0.0,0.0}
\definecolor{green}{rgb}{0.0,0.6,0.0}
\definecolor{darkblue}{rgb}{0.0,0.1,0.7}
\definecolor{brown}{rgb}{0.6,0.1,0.0}
\definecolor{grey}{rgb}{0.6,0.6,0.6}
\definecolor{darkgreen}{rgb}{0.0, 0.545098, 0.0}
\definecolor{applegreen}{rgb}{0.55, 0.71, 0.0}
\definecolor{purple}{rgb}{0.5,0.0,0.5}
\definecolor{babypink} {rgb}{0.64, 0.44, 0.44}
\definecolor{orange}{rgb}{1.0,0.5,0.0}

\definecolor{DARKBLUE}{rgb}{0.0,0.1,0.7}
\allowdisplaybreaks

\newcommand{\toright}[1]{\hspace*{\fill}{\footnotesize{#1}}}

\newcommand{\amuexpresult}{116\, 592\, 071.5(14.5)}

\newcommand{\amuHVPLOresultsection}{713.2(6.1)}
\newcommand{\amuHVPLOresult}{7132(61)}
\newcommand{\amuHVPNLOresultsection}{-9.96(13)}
\newcommand{\amuHVPNLOresult}{-99.6(1.3)}
\newcommand{\amuHVPNNLOresultsection}{1.24(1)}
\newcommand{\amuHVPNNLOresult}{12.4(1)}
\newcommand{\amuHVPtotalresult}{7045(61)}

\newcommand{\amuHLbLdataresult}{103.3(8.8)}
\newcommand{\amuHLbLlatticeresult}{122.5(9.0)}
\newcommand{\amuHLbLNLOdataresult}{2.6(6)}
\newcommand{\amuHLbLaverageresult}{112.6(9.6)}
\newcommand{\amuHLbLtotalresult}{115.5(9.9)}

\newcommand{\amuQEDresult}{116\,584\,718.8(2)}
\newcommand{\amuEWresult}{154.4(4)}

\newcommand{\amuSMresult}{116\,592\,033(62)}
\newcommand{\amudiffresult}{38(63)}

\newcommand{\F}{\mathcal{F}}
\newcommand{\dd}{d}

\newcommand{\GeV}{\,\text{GeV}}
\newcommand{\MeV}{\,\text{MeV}}
\newcommand{\fm}{\,\text{fm}}
\newcommand{\ppb}{\,\text{ppb}}
\newcommand{\ppt}{\,\text{ppt}}

\newcommand{\FFP}{F_{P\gamma^*\gamma^*}}
\newcommand{\etap}{\eta^{\prime}}

\newcommand{\chisq}{\ensuremath{\chi^2/\text{dof}}}

\newcommand{\amumacro}[1]{\smash{#1}}

\newcommand{\amuSM}{\ensuremath{a_\mu^\text{SM}}}

\newcommand{\amuexp}{\ensuremath{a_\mu^\text{exp}}}

\newcommand{\amuHVP}{\ensuremath{a_\mu^\text{HVP}}}

\newcommand{\amuHVPLO}{\ensuremath{a_\mu^\text{HVP, LO}}}

\newcommand{\amuHVPNLO}{\ensuremath{a_\mu^\text{HVP, NLO}}}
\newcommand{\amuHVPNNLO}{\ensuremath{a_\mu^\text{HVP, NNLO}}}

\newcommand{\amuHLbL}{\ensuremath{a_\mu^\text{HLbL}}}
\newcommand{\amuHLbLNLO}{\ensuremath{a_\mu^\text{HLbL, NLO}}}

\newcommand{\amuQED}{\ensuremath{a_\mu^\text{QED}}}
\newcommand{\amuEW}{\ensuremath{a_\mu^\text{EW}}}

\newcommand{\amuHVPLOud}{\ensuremath{a_\mu^\text{HVP, LO}(ud)}}
\newcommand{\amuHVPLOs}{\ensuremath{a_\mu^\text{HVP, LO}(s)}}
\newcommand{\amuHVPLOsdisc}{\ensuremath{a_\mu^\text{HVP, LO}(s+\text{disc})}}
\newcommand{\amuHVPLOc}{\ensuremath{a_\mu^\text{HVP, LO}(c)}}
\newcommand{\amuHVPLOb}{\ensuremath{a_\mu^\text{HVP, LO}(b)}}
\newcommand{\amuHVPLOdisc}{\ensuremath{a_{\mu}^\text{HVP, LO}(\text{disc})}}

\newcommand{\amuW}{\amumacro{a_\mu^\mathrm{W}}}
\newcommand{\deltaW}{\amumacro{\delta a_\mu^\mathrm{W}}}
\newcommand{\amuWiso}{\amumacro{a_\mu^\mathrm{W}(\mathrm{iso})}}
\newcommand{\amuWud}{\amumacro{a_\mu^\mathrm{W}(ud)}}
\newcommand{\amuWs}{\amumacro{a_\mu^\mathrm{W}(s)}}
\newcommand{\amuWsdisc}{\ensuremath{a_\mu^\text{W}(s+\text{disc})}}
\newcommand{\amuWc}{\amumacro{a_\mu^\mathrm{W}(c)}}
\newcommand{\amuWdisc}{\amumacro{a_\mu^\mathrm{W}(\mathrm{disc})}}
\newcommand{\amuSD}{\amumacro{a_\mu^\mathrm{SD}}}
\newcommand{\deltaSD}{\amumacro{\delta a_\mu^\mathrm{SD}}}
\newcommand{\amuSDiso}{\amumacro{a_\mu^\mathrm{SD}(\mathrm{iso})}}
\newcommand{\amuSDud}{\amumacro{a_\mu^\mathrm{SD}(ud)}}
\newcommand{\amuSDs}{\amumacro{a_\mu^\mathrm{SD}(s)}}
\newcommand{\amuSDsdisc}{\ensuremath{a_\mu^\text{SD}(s+\text{disc})}}
\newcommand{\amuSDc}{\amumacro{a_\mu^\mathrm{SD}(c)}}
\newcommand{\amuSDb}{\amumacro{a_\mu^\mathrm{SD}(b)}}
\newcommand{\amuSDdisc}{\amumacro{a_\mu^\mathrm{SD}(\mathrm{disc})}}
\newcommand{\amuLD}{\amumacro{a_\mu^\mathrm{LD}}}
\newcommand{\amuLDiso}{\amumacro{a_\mu^\mathrm{LD}(\mathrm{iso})}}
\newcommand{\amuLDud}{\amumacro{a_\mu^\mathrm{LD}(ud)}}
\newcommand{\amuLDs}{\amumacro{a_\mu^\mathrm{LD}(s)}}
\newcommand{\amuLDdisc}{\amumacro{a_\mu^\mathrm{LD}(\mathrm{disc})}}
\newcommand{\amuLDsdisc}{\ensuremath{a_\mu^\text{LD}(s+\text{disc})}}
\newcommand{\amuLDc}{\amumacro{a_\mu^\mathrm{LD}(c)}}

\newcommand{\amuHVPLOiso}{\ensuremath{a_\mu^\text{HVP, LO}(\text{iso})}}
\newcommand{\deltaHVPLO}{\ensuremath{\delta a_\mu^\text{HVP, LO}}}

\newcommand{\aem}{\alpha}

\newcommand{\phokhara}{{\sc Phokhara}}
\newcommand{\mcmule}{{\sc McMule}}
\newcommand{\babayaga}{{\sc BabaYaga@NLO}}
\newcommand{\afkqed}{{\sc AfkQed}}
\newcommand{\kkmc}{{\sc KKMC}}
\newcommand{\mcgpj}{{\sc MCGPJ}}
\newcommand{\sherpa}{{\sc Sherpa}}
\newcommand{\mesmer}{{\sc MESMER}}

\def\babar{\textsc{BaBar}}

\def\fb{~${\rm fb}^{-1}$}

\newcommand{\pipig}{\ensuremath{\pi^+\pi^-(\gamma)}\xspace}
\newcommand{\mumug}{\ensuremath{\mu^+\mu^-(\gamma)}\xspace}
\newcommand{\KKg}{\ensuremath{K^+K^-(\gamma)}\xspace}

\newcommand{\eeg}{\ensuremath{e^+e^-(\gamma)}\xspace}

\newcommand{\epemtopipigamma}{\ensuremath{e^+e^-\to\pi^+\pi^-(\gamma)}\xspace}
\newcommand{\epemtomumugamma}{\ensuremath{e^+e^-\to\mu^+\mu^-(\gamma)}\xspace}

\newcommand{\eff}{\ensuremath{\varepsilon}}

\newcommand{\eetoxg}{\ensuremath{e^+e^-\to X\gamma}\xspace}

\newcommand{\eetopipipiz}{\ensuremath{e^+e^-\to\pi^+\pi^-\pi^0}\xspace}

\newcommand{\invfb}{\ensuremath{\mbox{\,fb}^{-1}}\xspace}
\newcommand{\pipi}{\ensuremath{\pi^+\pi^-}\xspace}
\newcommand{\mumu}{\ensuremath{\mu^+\mu^-}\xspace}

\newcommand{\mee}{e^+e^-}

\newcommand{\amuEWl}{{a_\mu ^{\text{EW(1)}}}}
\newcommand{\amub}{a_{\mu;\text{bos}}^{\rm EW(2)}}
\newcommand{\amuf}{a_{\mu;\text{ferm}}^{\rm EW(2)}}

\newcommand{\amufrestH}{a_{\mu;\text{f-rest,H}}^{\rm EW(2)}}

\newcommand{\amufrestnoH}{a_{\mu;\text{f-rest,no H}}^{\rm EW(2)}}

\newcommand{\mpi}{M_\pi}
\newcommand{\Order}{\mathcal{O}}
\renewcommand{\Im}{\text{Im}\,}
\renewcommand{\Re}{\text{Re}\,}

\newcommand{\expref}{Muong-2:2025xyk,Muong-2:2023cdq,Muong-2:2024hpx,Muong-2:2021ojo,Muong-2:2021vma,Muong-2:2021ovs,Muong-2:2021xzz,Muong-2:2006rrc}

\newcommand{\QEDref}{Aoyama:2012wk,Volkov:2019phy,Volkov:2024yzc,Aoyama:2024aly,Parker:2018vye,Morel:2020dww,Fan:2022eto}

\newcommand{\EWref}{Czarnecki:2002nt,Gnendiger:2013pva,Ludtke:2024ase,Hoferichter:2025yih}

\newcommand{\latticeHVPref}{RBC:2018dos,Giusti:2019xct,Borsanyi:2020mff,Lehner:2020crt,Wang:2022lkq,Aubin:2022hgm,Ce:2022kxy,ExtendedTwistedMass:2022jpw,RBC:2023pvn,Kuberski:2024bcj,Boccaletti:2024guq,Spiegel:2024dec,RBC:2024fic,Djukanovic:2024cmq,ExtendedTwistedMass:2024nyi,MILC:2024ryz,FermilabLatticeHPQCD:2024ppc}

\newcommand{\latticeHLbLref}{Blum:2019ugy,Chao:2021tvp,Chao:2022xzg,Blum:2023vlm,Fodor:2024jyn}

\newcommand{\dataHLbLref}{Colangelo:2015ama,Masjuan:2017tvw,Colangelo:2017fiz,Hoferichter:2018kwz,Eichmann:2019tjk,Bijnens:2019ghy,Leutgeb:2019gbz,Cappiello:2019hwh,Masjuan:2020jsf,Bijnens:2020xnl,Bijnens:2021jqo,Danilkin:2021icn,Stamen:2022uqh,Leutgeb:2022lqw,Hoferichter:2023tgp,Hoferichter:2024fsj,Estrada:2024cfy,Ludtke:2024ase,Deineka:2024mzt,Eichmann:2024glq,Bijnens:2024jgh,Hoferichter:2024bae,Holz:2024diw,Cappiello:2025fyf}

\newcommand{\HVPref}{RBC:2018dos,Giusti:2019xct,Borsanyi:2020mff,Lehner:2020crt,Wang:2022lkq,Aubin:2022hgm,Ce:2022kxy,ExtendedTwistedMass:2022jpw,RBC:2023pvn,Kuberski:2024bcj,Boccaletti:2024guq,Spiegel:2024dec,RBC:2024fic,Djukanovic:2024cmq,ExtendedTwistedMass:2024nyi,MILC:2024ryz,FermilabLatticeHPQCD:2024ppc,Keshavarzi:2019abf,DiLuzio:2024sps,Kurz:2014wya}

\newcommand{\HLbLref}{Colangelo:2015ama,Masjuan:2017tvw,Colangelo:2017fiz,Hoferichter:2018kwz,Eichmann:2019tjk,Bijnens:2019ghy,Leutgeb:2019gbz,Cappiello:2019hwh,Masjuan:2020jsf,Bijnens:2020xnl,Bijnens:2021jqo,Danilkin:2021icn,Stamen:2022uqh,Leutgeb:2022lqw,Hoferichter:2023tgp,Hoferichter:2024fsj,Estrada:2024cfy,Ludtke:2024ase,Deineka:2024mzt,Eichmann:2024glq,Bijnens:2024jgh,Hoferichter:2024bae,Holz:2024diw,Cappiello:2025fyf,Colangelo:2014qya,Blum:2019ugy,Chao:2021tvp,Chao:2022xzg,Blum:2023vlm,Fodor:2024jyn}

\newcommand{\SMref}{Aoyama:2012wk,Volkov:2019phy,Volkov:2024yzc,Aoyama:2024aly,Parker:2018vye,Morel:2020dww,Fan:2022eto,Czarnecki:2002nt,Gnendiger:2013pva,Ludtke:2024ase,Hoferichter:2025yih,RBC:2018dos,Giusti:2019xct,Borsanyi:2020mff,Lehner:2020crt,Wang:2022lkq,Aubin:2022hgm,Ce:2022kxy,ExtendedTwistedMass:2022jpw,RBC:2023pvn,Kuberski:2024bcj,Boccaletti:2024guq,Spiegel:2024dec,RBC:2024fic,Djukanovic:2024cmq,ExtendedTwistedMass:2024nyi,MILC:2024ryz,FermilabLatticeHPQCD:2024ppc,Keshavarzi:2019abf,DiLuzio:2024sps,Kurz:2014wya,Colangelo:2015ama,Masjuan:2017tvw,Colangelo:2017fiz,Hoferichter:2018kwz,Eichmann:2019tjk,Bijnens:2019ghy,Leutgeb:2019gbz,Cappiello:2019hwh,Masjuan:2020jsf,Bijnens:2020xnl,Bijnens:2021jqo,Danilkin:2021icn,Stamen:2022uqh,Leutgeb:2022lqw,Hoferichter:2023tgp,Hoferichter:2024fsj,Estrada:2024cfy,Deineka:2024mzt,Eichmann:2024glq,Bijnens:2024jgh,Hoferichter:2024bae,Holz:2024diw,Cappiello:2025fyf,Colangelo:2014qya,Blum:2019ugy,Chao:2021tvp,Chao:2022xzg,Blum:2023vlm,Fodor:2024jyn}

\begin{document}

\title{
{\footnotesize{CERN-TH-2025-101}}
\toright{FERMILAB-PUB-25-0344-T}\\[-0.2cm]
{\footnotesize{INT-PUB-25-015}}
\toright{IPARCOS-UCM-25-029}\\[-0.2cm]
{\footnotesize{KEK Preprint 2025-22}}
\toright{LTH 1403}\\[-0.2cm]
{\footnotesize{MITP-25-037}}
\toright{UWThPh 2025-15 }\\[-0.2cm]
\toright{ZU-TH 37/25}\\[0.9cm]
The anomalous magnetic moment of the muon in the Standard Model: an update}

\renewcommand{\theaffn}{\arabic{affn}}

\renewcommand{\thefootnote}{\fnsymbol{footnote}}

\author[Mainz,MainzP]{R.~Aliberti\,\orcidlink{0000-0003-3500-4012}}
\author[ISSP]{T.~Aoyama\,\orcidlink{0009-0009-0491-1024}}
\author[PadovaUniversity,Padova]{E.~Balzani\,\orcidlink{0009-0005-9971-9473}}
\author[Hidalgo,Huelva]{A.~Bashir\,\orcidlink{0000-0003-3183-7316}}
\author[Illinois,ICASU]{G.~Benton\,\orcidlink{0009-0005-1576-3654}}
\author[Lund]{J.~Bijnens\,\orcidlink{0000-0002-1618-2844}}
\author[Mainz,MainzP]{V.~Biloshytskyi\,\orcidlink{0000-0001-5274-9812}}
\author[Uconn,RIKENBNL]{T.~Blum\,\orcidlink{0000-0002-2866-7689}}
\author[SaoPaulo]{D.~Boito\,\orcidlink{0000-0002-4426-7984}}
\author[Milano,MilanoINFN]{M.~Bruno\,\orcidlink{0000-0002-5127-4461}}
% ...

\address[Mainz]{Institute for Nuclear Physics, Johannes Gutenberg University of Mainz, 55128 Mainz, Germany}
\address[MainzP]{PRISMA$^+$ Cluster of Excellence, Johannes Gutenberg University of Mainz, 55099 Mainz, Germany }
\address[ISSP]{Institute for Solid State Physics, University of Tokyo, Kashiwa, 277-8581, Japan}
\address[PadovaUniversity]{Dipartimento di Fisica e Astronomia ``G. Galilei'', Universit\`a di Padova, Via F.\ Marzolo 8, 35131 Padova, Italy}
\address[Padova]{Istituto Nazionale di Fisica Nucleare (INFN), Sezione di Padova, Via F.\ Marzolo 8, 35131 Padova, Italy}
\address[Hidalgo]{Instituto de F\'isica y Matem\'aticas, Universidad Michoacana de San Nicol\'as de Hidalgo, Morelia, Michoac\'an 58040, M\'exico}
\address[Huelva]{Department of Integrated Sciences and Center for Advanced Studies in Physics, Mathematics and Computation, University of Huelva, 21071~Huelva, Spain}
\address[Illinois]{Department of Physics, University of Illinois Urbana-Champaign, Urbana, IL 61801, USA}
\address[ICASU]{Illinois Center for Advanced Studies of the Universe, University of Illinois Urbana-Champaign, Urbana, IL 61801, USA}
\address[Lund]{Division of Particle and Nuclear Physics, Department of Physics, Lund University, Box 118, 221 00 Lund, Sweden}
\address[Uconn]{Department of Physics, 196 Auditorium Road, Unit 3046, University of Connecticut, Storrs, CT 06269-3046, USA}
\address[RIKENBNL]{RIKEN BNL Research Center,  Brookhaven National Laboratory, Upton, NY 11973, USA}
\address[SaoPaulo]{Instituto de F\'isica de S\~ao Carlos, Universidade de S\~ao Paulo, IFSC --- USP, 13560-970, S\~ao Carlos, SP, Brazil}
\address[Milano]{Dipartimento di Fisica ``Giuseppe Occhialini'', Universit\`a degli Studi di Milano-Bicocca, Piazza della Scienza 3, 20126 Milan, Italy}
\address[MilanoINFN]{Istituto Nazionale di Fisica Nucleare (INFN), Sezione di Milano-Bicocca, Piazza della Scienza 3, 20126 Milan, Italy}
% ...

\begin{abstract}
We present the current Standard Model (SM) prediction for the muon anomalous magnetic moment, $a_\mu$, updating  the first White Paper (WP20)~\cite{Aoyama:2020ynm}. 
The pure QED and electroweak contributions have been further
consolidated, while hadronic contributions continue to be responsible for the bulk of the uncertainty of the SM prediction.
Significant progress has been achieved in the hadronic light-by-light scattering contribution 
using both the data-driven dispersive approach as well as
lattice-QCD calculations, leading to a reduction of 
the uncertainty by almost a factor of two.
The most important development since WP20 is the change in the estimate
of the leading-order hadronic-vacuum-polarization (LO HVP) contribution.
A new measurement of the $e^+e^-\to\pi^+\pi^-$
cross section by CMD-3 has increased the tensions among data-driven dispersive
evaluations of the LO HVP contribution 
to a level that makes it impossible to combine the results in a meaningful way. At the same time, the attainable precision of lattice-QCD calculations
has increased substantially and allows for a consolidated lattice-QCD
average of the LO HVP contribution with a precision of about 0.9\%.
Adopting the latter in this update
has resulted in a major upward shift 
of the total SM prediction, which now reads
$\amuSM = \amuSMresult\times 10^{-11}$ (530\,ppb).
When compared against the current
experimental average based on the E821 experiment and runs~1--6 of E989
at Fermilab, one finds $\amuexp - \amuSM =\amudiffresult\times 10^{-11}$, 
which implies that there is no tension between the SM and experiment at the current level of precision. 
The final precision of E989 (127\,ppb) is the target of
future efforts by the Theory Initiative.
The resolution of the tensions among data-driven dispersive evaluations
of the LO HVP contribution will be a key element in this endeavor.
\end{abstract}

\numberwithin{equation}{section}

\maketitle

\emph{We dedicate this paper to the memory of Simon Eidelman.}

\newpage
\tableofcontents
\newpage

\setcounter{footnote}{0}

\setcounter{section}{-1}
\section{Executive summary} 
\label{sec:execsumm}

\nocite{Charpak:1962zz,Bailey:1968rxd,Bailey:1978mn,Muong-2:2006rrc,Muong-2:2021ojo,Muong-2:2023cdq,Muong-2:2025xyk,Abe:2019thb}

\begin{table}[t]
\renewcommand{\arraystretch}{1.1}
\begin{centering}
\small
	\begin{tabular}{l  l l r l }
	\toprule
	   Contribution & Section & Equation & Value   $\times 10^{11}$ & References\\ \midrule
	   Experiment (E989, E821)     &                   & \cref{eq:amuexpresult} & $\amuexpresult$ & Refs.~\cite{\expref}  \\\midrule
HVP LO (lattice) & \cref{sec:lattice_HVP_world_average}  &\cref{eq:amuLOHVWP25}  & $\amuHVPLOresult$ & Refs.~\cite{\latticeHVPref}\\
\textcolor{gray}{HVP LO ($e^+e^-,\tau$)} & \textcolor{gray}{\cref{sec:dataHVP}} & \textcolor{gray}{\cref{{tab:summary_ee}}} & \multicolumn{2}{l}{\textcolor{gray}{Estimates not provided at this point}}\\
HVP NLO ($e^+e^-$) & \cref{sec:NLO} & \cref{HVPNLO} & $\amuHVPNLOresult$ & Refs.~\cite{Keshavarzi:2019abf,DiLuzio:2024sps}\\
HVP NNLO ($e^+e^-$) & \cref{sec:NLO} & \cref{HVPNNLO} & $\amuHVPNNLOresult$ & Ref.~\cite{Kurz:2014wya} \\
HLbL (phenomenology) & \cref{sec:finalnumber} & \cref{eq:amuHLbLdata} & $\amuHLbLdataresult$ & Refs.~\cite{\dataHLbLref}\\
HLbL NLO (phenomenology) & \cref{sec:finalnumber} & \cref{eq:amuHLbLNLOdata} & $\amuHLbLNLOdataresult$ & Ref.~\cite{Colangelo:2014qya}\\
HLbL (lattice) & \cref{sec:amuHLbLlatticeaverage} & \cref{eq:amuHLbL_aver} & $\amuHLbLlatticeresult$ & Refs.~\cite{\latticeHLbLref}\\
HLbL (phenomenology + lattice) & \cref{sec:conclusionsWP} & \cref{eq:HLbL_comb} & $\amuHLbLaverageresult$ & Refs.~\cite{\dataHLbLref,\latticeHLbLref}\\
\midrule
QED             &  \cref{sec:amuQEDfinal}  &  \cref{amuQEDfinal}   &  $\amuQEDresult$  & Refs.~\cite{\QEDref}\\
EW     &  \cref{sec:EW} &   \cref{amuEWNew}  & $\amuEWresult$   & Refs.~\cite{\EWref}\\
HVP LO (lattice) + HVP N(N)LO ($e^+e^-$) & \cref{sec:conclusionsWP} & \cref{HVPtotal} & $\amuHVPtotalresult$ &  Refs.~\cite{\HVPref}\\
HLbL (phenomenology + lattice + NLO) & \cref{sec:conclusionsWP}  & \cref{{eq:amuHLbLtotal}} & $\amuHLbLtotalresult$ & Refs.~\cite{\HLbLref}\\ 
      Total SM Value  & \cref{sec:conclusionsWP}            & \cref{eq:amuSM} & $\amuSMresult$  & Refs.~\cite{\SMref}\\
      Difference:    $\Delta a_\mu\equiv\amuexp - \amuSM$ & \cref{sec:conclusionsWP}  & \cref{amudiff} & $\amudiffresult$ & \\
        \bottomrule
        \renewcommand{\arraystretch}{1.0}
	\end{tabular}
	\caption{Summary of the contributions to $\amuSM$. The experimental number refers to the world average completely dominated by E989. The subsequent block summarizes the pertinent hadronic contributions from \cref{sec:dataHVP,sec:latticeHVP,sec:dataHLbL,sec:latticeHLbL} as well as the combination of data-driven and lattice-QCD evaluations of HLbL scattering from  
    \cref{sec:conclusionsWP}. 
	The second block summarizes the quantities entering our recommended SM value, in particular, the total HVP contribution (using lattice QCD for LO and $e^+e^-$ for higher-order iterations) and the total HLbL number. The construction of the total HLbL contribution takes into account correlations among the terms at different orders, and the final rounding includes subleading digits at intermediate stages. The experimental world average has been updated including the final results from the Fermilab experiment.}     
\label{tab:summary}
\end{centering}
\end{table}

The experimental program to study the muon's anomalous magnetic moment, $a_\mu \equiv (g-2)_\mu/2$, which started over sixty years ago at CERN, continues to inspire theoretical efforts to obtain a Standard-Model (SM) prediction with matching precision. 
In the SM, the dominant contributions to $a_\mu$ are due to QED corrections, followed by hadronic contributions (governed by QCD) and electroweak (EW) corrections. The CERN experiments provided tests of the $(\alpha/\pi)^2$ \cite{Charpak:1962zz} and $(\alpha/\pi)^3$ \cite{Bailey:1968rxd} QED contributions and sensitivity to the leading-order (LO) hadronic corrections~\cite{Bailey:1978mn}. 
The BNL E821 experiment \cite{Muong-2:2006rrc} reached a precision of 540 parts-per-billion (ppb), yielding sensitivity also to EW and higher-order hadronic contributions and intriguing tensions with the SM predictions available at that time.
This provided motivation to launch the Fermilab E989 experiment, which started taking data in 2017, and has obtained measurements with increasing precision, from 460 ppb in 2021~\cite{Muong-2:2021ojo} to 200 ppb in 2023~\cite{Muong-2:2023cdq}, all consistent with each other as well as with E821. 
The Fermilab experiment has just announced its final measurement result, with a precision of 127 ppb, and hence will enable, in principle, an exquisitely precise test of the SM.\footnote{This paper was posted on arXiv on May 28, 2025.  \Cref{sec:dataHVP,sec:latticeHVP,sec:comparison,sec:dataHLbL,sec:latticeHLbL,sec:QED,sec:EW} and all numbers pertaining to the SM prediction have remained unchanged, but the experimental world average has been updated according to the E989 announcement on June 3, 2025~\cite{Muong-2:2025xyk}, and the description in \cref{sec:execsumm,sec:introWP,sec:conclusionsWP} has been adapted accordingly. In particular, the experimental results in abstract, \cref{tab:summary}, \cref{fig:summary_plot_data_hvp,fig:finalHVPsummary,fig:summary_plot_final}, and \cref{sec:conclusionsWP} have been updated.}   Meanwhile, a new experiment that will employ a largely different experimental method is currently under preparation at J-PARC~\cite{Abe:2019thb}, allowing an independent cross-check of these results.

On the SM side, the QED and EW corrections are at less than 5 ppb already well quantified, while the notoriously difficult-to-calculate hadronic contributions are the dominant sources of theory error, rendering SM predictions of $a_\mu$ still considerably less precise than experiment. They, therefore, continue to be the main focus of the Muon $g-2$ Theory Initiative. A lot has happened regarding these corrections since the publication of the first White Paper (WP20)~\cite{Aoyama:2020ynm}. 

The hadronic-vacuum-polarization (HVP) contribution was estimated in WP20 from $e^+e^-$ hadronic cross-section data in the dispersive, data-driven method, while lattice-QCD methods were not mature enough to yield reliable results at the required precision. In the present review, we find that this situation is reversed. In the data-driven method, some tensions in the dominant $e^+e^- \to \pi^+\pi^-$ channel were present in WP20 and were accounted for by inflating uncertainties. New cross-section measurements of the same channel that became available after WP20 have increased these tensions to a level that prevents forming any meaningful average for use in obtaining a precise and reliable evaluation of the HVP contribution. Moreover, at present no scientific grounds have been identified that would allow one to disregard any of the data sets relevant for the HVP evaluation. 

Resolving the tensions in the $e^+e^-$ hadronic cross-section data will require new analyses and measurements as well as an improved understanding of higher-order radiative corrections. Both are currently being pursued by several experiments and theorists. Despite the present difficult situation, a data-driven estimate of HVP remains an important goal of the Theory Initiative and an essential cornerstone on which to build future precise SM predictions. In this regard, also the  $\tau$-based HVP determination, which relies on consistent data sets, is being reconsidered in view of recent  progress towards a model-independent evaluation of the required isospin-breaking corrections. While promising, this is still work in progress: the current status of a $\tau$-based HVP estimate is reviewed here, but is not yet included in the overall evaluation.

On the other hand, thanks to dedicated efforts by the world-wide lattice field theory community, lattice-QCD calculations of HVP have matured considerably. The present review is informed by more than a dozen new papers (since WP20), which enable consolidated averages of (almost) all the components that are computed separately to construct the total LO HVP contribution in lattice QCD. The result is a reliable determination of the LO HVP contribution at $0.9\%$ precision that enters the SM prediction of $a_\mu$. The change from a data-driven to a lattice-based estimate for the HVP contribution has resulted in a significant shift in the central value.

For the hadronic light-by-light (HLbL) contribution, significant progress since WP20 has been made in both the dispersive method and lattice QCD. 
On the data-driven side, improved calculations of short-distance constraints and a number of subleading contributions have become available, leading to a reduction of the uncertainty by about a factor of two compared to WP20. At the same time, new lattice-QCD calculations have reached a similar level of precision. The two averages, from phenomenology and lattice QCD, are combined into a final average with a precision below $10\%$.

\cref{tab:summary} gives a summary of the contributions to the SM prediction, along with the locations in this White Paper (WP25) where the results are discussed.

\clearpage

\section{Introduction}
\label{sec:introWP}

The anomalous magnetic moment of the muon $a_\mu$ has been an enduring testbed of the SM and of its possible extensions (generically defined as beyond the SM, or BSM, physics). Both experiment and SM theory provide superb precision beyond parts-per-million (ppm), allowing one to put constraints on BSM physics through the comparison between the two. 
The {\em Muon $g-2$ Theory Initiative} (TI) was started in 2017 with the goal of bringing together the community of theorists and experimentalists working on this topic and of coordinating their efforts to provide a single, consensus number for the theoretical estimate of $a_\mu$ within the SM. The basis and the methodology behind such a number would be detailed in an extensive review, a so-called White Paper. The first edition was published in 2020 (WP20)~\cite{Aoyama:2020ynm}. Since then, new measurements have been announced by the Fermilab $g-2$ experiment, the first result in 2021~\cite{Muong-2:2021ojo} with a precision of 460~ppb, and the second result in 2023~\cite{Muong-2:2023cdq} with 200~ppb. These new measurements are consistent with, but more precise than, the ones of the BNL experiment, and provide the basis of the current world average. The agreement of the two most precise experiments lends a high degree of reliability to this world average. The Fermilab E989 experiment has just announced its final results with the full set of all experimental data collected. Another experiment with a largely different experimental method is currently under preparation at J-PARC~\cite{Abe:2019thb}, allowing an independent cross-check of these results.

For the physical interpretation of the muon $g-2$ experimental measurement, it is crucial to have a reliable SM prediction. Ideally, uncertainties on the SM prediction would be reduced to the same level as the experimental ones to maximize the BSM sensitivity. In the SM, $a_{\mu}$ is calculated from a perturbative expansion in the fine-structure constant $\alpha$. The most important contributions to this quantity arise from pure QED diagrams, but starting from order $\alpha^2$ also hadronic contributions arise, in particular HVP and HLbL. Corrections due to EW bosons are small because of their large mass. They have been calculated at next-to-leading order (NLO) and are relevant at the present level of precision. The uncertainties are currently dominated by hadronic contributions due to the nonperturbative nature thereof, which make the corresponding theoretical calculations highly nontrivial. The goal of this second edition of the White Paper (WP25), as for the previous one, is to provide a detailed snapshot of the present situation and a new consensus SM number, before the final Fermilab result is announced.\footnote{This paper was posted on arXiv on May 28, 2025.  \Cref{sec:dataHVP,sec:latticeHVP,sec:comparison,sec:dataHLbL,sec:latticeHLbL,sec:QED,sec:EW} and all numbers pertaining to the SM prediction have remained unchanged, but the experimental world average has been updated according to the E989 announcement on June 3, 2025~\cite{Muong-2:2025xyk}, and the description in \cref{sec:execsumm,sec:introWP,sec:conclusionsWP} has been adapted accordingly. In particular, the experimental results in abstract, \cref{tab:summary}, \cref{fig:summary_plot_data_hvp,fig:finalHVPsummary,fig:summary_plot_final}, and \cref{sec:conclusionsWP} have been updated.}

In recent years, the theoretical and experimental work aimed at reducing the uncertainties of the SM prediction has been particularly intense and relied on higher-order calculations, dispersive methods, lattice QCD, effective field theories, as well as new data inputs from experiments. For HLbL these efforts have indeed been successful and the change from WP20 to the present review is that the uncertainty has been reduced by about a factor of two. Concerning HVP the situation is more complicated as reflected by the fact that the present review does not present a number for a data-driven estimate of this contribution, since not yet understood discrepancies among different experiments have emerged. On the other hand, there has been significant progress in lattice-QCD calculations of this quantity, and the consensus number for the HVP contribution presented in this review is based on them. The change from a data-driven to a lattice estimate of this contribution also means that the central value has shifted, unfortunately, by more than the uncertainty quoted in WP20.

Activities of the initiative are coordinated by a Steering Committee that consists of theorists, experimentalists, and representatives from the Fermilab and J-PARC experiments. This committee also functions as the Advisory Committee for the workshops organized by the TI. Since the release of WP20, we organized four plenary workshops as well as focused mini-workshops. 
The format of the recent plenary workshops follows the previous editions with plenary sessions consisting of updates from $g-2$ experiments, updates from HVP and HLbL evaluations from dispersive approaches as well as lattice QCD. The plenary workshop in 2021 was held virtually during the period of the pandemic and was hosted by KEK~\cite{PlenaryWS-2021}. After 2022, the three plenary workshops were held annually as in-person events in Edinburgh~\cite{PlenaryWS-2022}, Bern~\cite{PlenaryWS-2023}, and Tsukuba~\cite{PlenaryWS-2024}, respectively. The focused mini-workshops, held virtually, were devoted to high-precision calculations of HVP in lattice QCD~\cite{MiniWS-Lattice}, the CMD-3 result for $e^+e^-\to\pi^+\pi^-$~\cite{MiniWS-CMD3-1,MiniWS-CMD3-2}, WP25 preparations prior to the Tsukuba workshop~\cite{MiniWS-Sprin2024}, and the HVP evaluation from hadronic $\tau$ decays~\cite{MiniWS-tau}.
At the next plenary workshop, planned to take place in Paris~\cite{PlenaryWS-2025}, the TI will discuss the next steps after  WP25 toward reducing the theory errors to match the experimental precision. A summary of the prospects for such future improvements was compiled as contribution to the US Community Study on the Future of Particle Physics (Snowmass 2021)~\cite{Colangelo:2022jxc}.

The Steering Committee is chaired by {\it Aida El-Khadra}, supported by {\it Christoph Lehner} and {\it Michel Davier} as co-chairs. Other committee members are\footnote{Simon Eidelman, a founding member of the Theory Initiative, served on the Steering Committee until he passed away in June 2021.} {\it Gilberto Colangelo, Martin Hoferichter, Laurent Lellouch, Tsutomu Mibe, Lee Roberts, Thomas Teubner}, and {\it Hartmut Wittig}. The Steering Committee's roles are the long-term planning of the Theory Initiative, the organization of the workshops, and the coordination of writing the White Paper and its updates. This review (WP25) has kept the same structure as the previous one and is organized in four main chapters, namely data-driven HVP, lattice HVP, analytic HLbL, and lattice HLbL, the writing of which was delegated to four corresponding working groups. The coordinators of the working groups are:
\begin{itemize}
    \item Data-driven HVP: {\it Vincenzo Cirigliano, Achim Denig, Fedor Ignatov, Bogdan Malaescu} 
    \item Lattice HVP: {\it Steven Gottlieb, Antonin Portelli}
    \item Analytic HLbL: {\it Johan (Hans) Bijnens, Anton Rebhan} 
    \item Lattice HLbL: {\it Luchang Jin, Harvey Meyer}
\end{itemize}
In addition, sections on the QED and EW contributions are coordinated by {\it Makiko Nio} and {\it Dominik St\"ockinger}, respectively.

This paper was written under the following agreements within the initiative.
All participants of the past workshops, members of working groups, and their collaborators were invited to become co-authors of WP25. 
Essential input papers that were accepted for publication by May 22, 2025 were fully considered and included in the averages, while brief descriptions of unpublished papers available on arXiv before Apr 22, 2025 are also included.

The rest of the review is organized as follows: we first discuss the evaluations of data-driven HVP calculations in \cref{sec:dataHVP} and lattice-QCD calculations in \cref{sec:latticeHVP}, followed by comparisons between them in \cref{sec:comparison}. Then, we describe evaluations of HLbL by using data-driven and analytic approaches in \cref{sec:dataHLbL}, and lattice QCD in \cref{sec:latticeHLbL}. Lastly, QED and EW contributions are given in \cref{sec:QED,sec:EW}, respectively. We summarize the conclusions and outlook for the current SM prediction in \cref{sec:conclusionsWP}.

\clearpage

\section{Data-driven calculations of HVP}
\label{sec:dataHVP}

\noindent
\begin{flushleft}
\emph{R.~Aliberti, G.~Benton, D.~Boito, M.~Bruno, C.~M.~Carloni Calame, V.~Cirigliano, G.~Colangelo, L.~Cotrozzi, M.~Cottini, M.~Davier, A.~Denig, V.~Druzhinin, M.~Golterman, A.~Hoecker, M.~Hoferichter, B.-L.~Hoid, S.~Holz, F.~Ignatov, A.~Keshavarzi, B.~Kubis, A.~Kupich, S.~Laporta, T.~Leplumey, Q.~Liu, I.~Logashenko, G.~L{\'o}pez Castro, A.-M.~Lutz, B.~Malaescu, K.~Maltman, A.~Miranda, S.~E.~M{\"u}ller, A.~Nesterenko, D.~Nomura, M.~Passera, S.~Peris, F.~Piccinini, R.~Pilato, L.~Polat, P.~Roig, J.~Ruiz de Elvira, A.~Signer, P.~Stoffer, T.~Teubner, G.~Toledo, Y.~Ulrich, G.~Venanzoni, A.~Wright, E.~Zaid, Z.~Zhang}
\end{flushleft}

\subsection{Introduction}
\label{sec:intro_data_HVP}

The calculation of the LO HVP contribution in terms of hadronic cross sections proceeds via the master formula~\cite{Bouchiat:1961lbg,Brodsky:1967sr,Lautrup:1969fr,Gourdin:1969dm}
\begin{equation}
\label{amu_HVP_master}
    \amuHVPLO=\bigg(\frac{\alpha m_\mu}{3\pi}\bigg)^2\int_{s_\text{thr}}^\infty ds \frac{\hat K(s)}{s^2}R_\text{had}(s)\,,
\end{equation}
with muon mass $m_\mu$, fine-structure constant $\alpha=e^2/(4\pi)$, and
kernel function 
\begin{align}
\label{eq:K_def}
\hat K(s)&=\frac{3s}{m_\mu^2}K(s)\,,\qquad 
x=\frac{1-\beta_\mu}{1+\beta_\mu}\,,\qquad 
\beta_\mu=\sqrt{1-\frac{4m_\mu^2}{s}}\,,\notag\\ 
K(s) &= \frac{x^2}{2}(2-x^2) +
\frac{(1+x^2)(1+x)^2}{x^2}\left( \log(1+x)-x+\frac{x^2}{2}\right) +
\frac{1+x}{1-x}x^2\log x\,.
\end{align}
Here, the hadronic $R$-ratio, 
\begin{equation}
\label{R_ratio}
R_\text{had}(s)=\frac{3s}{4\pi\alpha^2}\sigma\big[e^+e^-\to\text{hadrons}(+\gamma)\big] \,, 
\end{equation}
is defined photon-inclusively, so that the integration starts at the threshold $s_\text{thr}=M_{\pi^0}^2$ due to the $e^+e^-\to\pi^0\gamma$ channel. The challenge in evaluating $\amuHVPLO$ at sub-percent level thus lies in the extraordinary precision requirements for the measurement of $e^+e^-\to\text{hadrons}$ cross sections, especially, for the crucial $e^+e^-\to\pi^+\pi^-$ channel. In this section, various aspects of this program are discussed, including reports from the $e^+e^-$ experiments (\cref{sec:e+e-}), indirect cross-section measurements via $\tau$ decays (\cref{sec:tau}), Monte-Carlo (MC) tools (\cref{sec:MC}), global data combinations (\cref{sec:data_combinations}), theory developments
(\cref{sec:disp,sec:disp_app,sec:lattice_comparison}), higher orders (\cref{sec:NLO}), and the MUonE project (\cref{sec:MUonE}). A summary of the current situation together with an outlook to future prospects is provided in \cref{sec:HVP_disp_summary,sec:HVP_disp_summary_outlook}.

Throughout, in addition to the total LO HVP contribution as defined in \cref{amu_HVP_master}, also so-called Euclidean-window observables will be considered. First introduced by RBC/UKQCD~\cite{RBC:2018dos}, weight functions in Euclidean time are introduced that separate the entire integral into a short-distance (SD), intermediate (W), and long-distance (LD) component, see \cref{sec:windows} for the precise definitions. Importantly, as detailed in \cref{sec:windows_data_driven}, these weight functions can be translated into center-of-mass (CM) energy $\sqrt{s}$, to be inserted into \cref{amu_HVP_master}, so that data-driven evaluations of the same quantities become possible, a connection that is addressed in more detail in \cref{sec:comparison}.

\subsection{Status and perspectives of \texorpdfstring{$e^+e^-$}{} experiments}
\label{sec:e+e-}

\subsubsection{CMD-3}
\label{TI-CMD3}

\nocite{Aulchenko:2001je,Khazin:2008zz,Achasov:2009zza,Shatunov:2016bdv,Shwartz:2016fhe,Ignatov:2019omb}

\begin{figure}[t]
\centering
\includegraphics[width=0.49\linewidth]{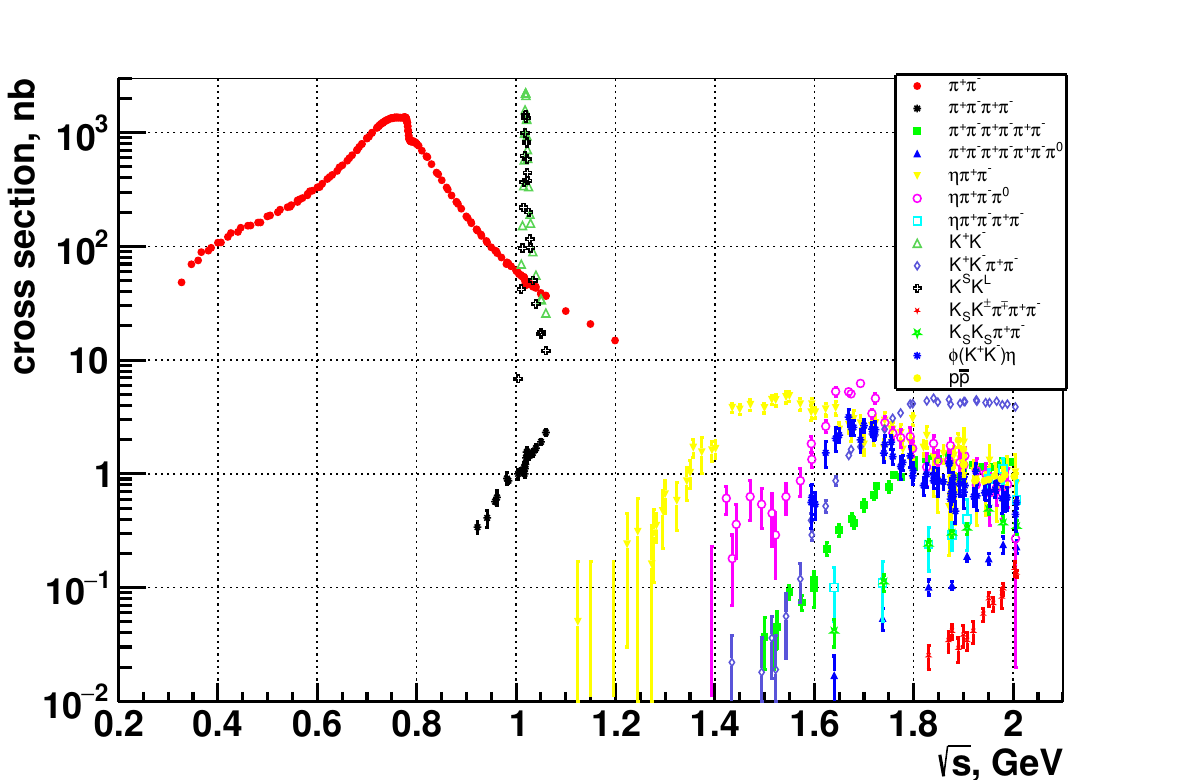}
\includegraphics[width=0.49\linewidth]{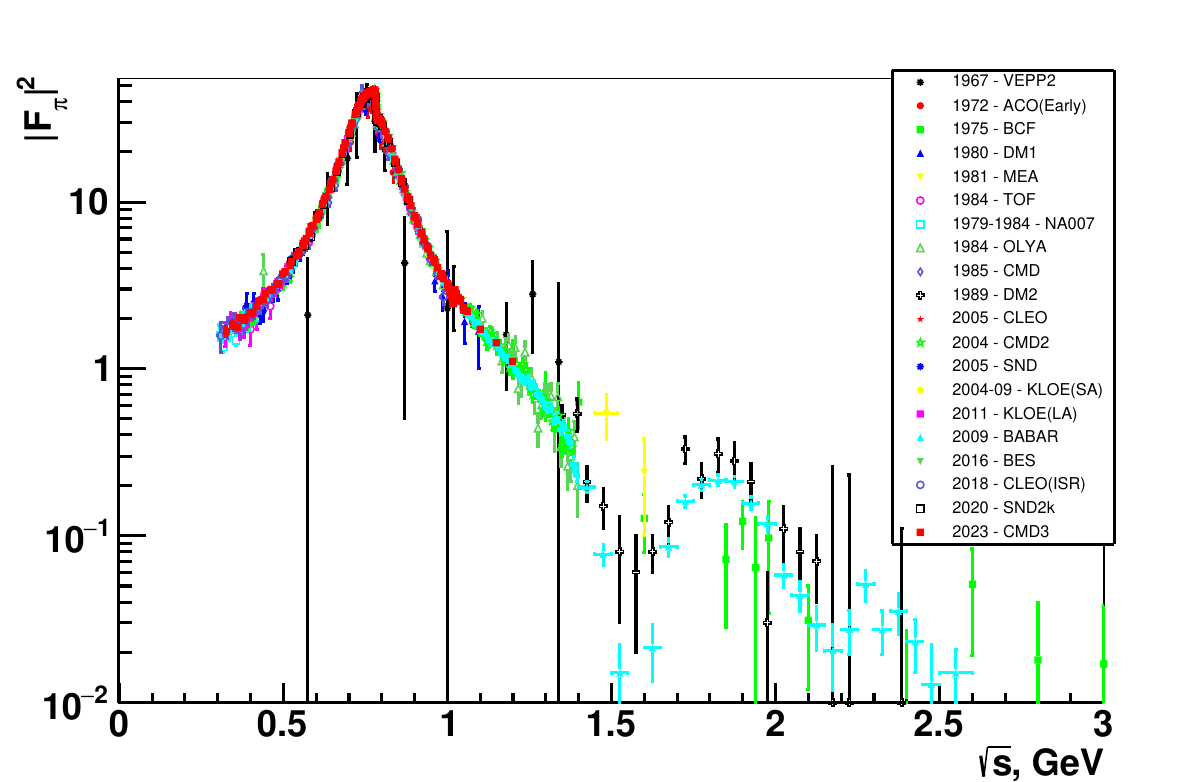}
\caption{Left: Published measurements of $e^+e^-\to \text{hadrons}$ cross sections by the CMD-3 experiment~\cite{CMD-3:2023rfe,CMD-3:2023alj,Akhmetshin:2016dtr,CMD-3:2013nph,CMD-3:2019ufp,Gribanov:2019qgw,CMD-3:2017tgb,Kozyrev:2017agm,Shemyakin:2015cba,CMD-3:2016nhy,CMD-3:2022qhm,CMD-3:2019cvx,Ivanov:2019crp,CMD-3:2015fvi}.
Right: Status of the $e^+e^-\to\pi^+\pi^-$ cross-section measurement with different experiments contributing over the years~\cite{CMD-3:2023rfe,CMD-3:2023alj,SND:2020nwa,Auslander:1967xma,Auslander:1969ak,Balakin:1971zh,Balakin:1972vg,Benaksas:1972ps,Cosme:1976ft,Bollini:1975pn,Quenzer:1978qt,Esposito:1977xg,Esposito:1980bz,Vasserman:1981xq,Amendolia:1983di,DM2:1988xqd,Barkov:1985ac,CMD-2:2001ski,CMD-2:2003gqi,CMD-2:2005mvb,CMD-2:2006gxt,Aulchenko:2006dxz,Achasov:2005rg,Achasov:2006vp,KLOE:2004lnj,KLOE:2008fmq,KLOE:2010qei,KLOE:2012anl,KLOE-2:2017fda,CLEO:2005tiu,Seth:2012nn,Xiao:2017dqv,BaBar:2012bdw,BESIII:2015equ}.}
\label{fig:CMD3crosssections}
\end{figure}

The CMD-3~\cite{Aulchenko:2001je,Khazin:2008zz} and SND~\cite{Achasov:2009zza} experiments have been in operation
at the electron--positron collider VEPP-2000~\cite{Shatunov:2016bdv,Shwartz:2016fhe} since 2010.
The collider provides an instantaneous luminosity of up to $10^{32} \rm cm^{-2}\rm s^{-1}$ at the maximum CM energy 
$\sqrt{s} = 2\GeV$, a world record for single-bunch luminosity. 
Overall, more than 1 fb$^{-1}$ of data has already been collected by each experiment across the entire available CM energy range from 0.32 to 2.007\,GeV. Today, VEPP-2000 is the only collider operating at these energies.

The new generation CMD-3 detector was designed and constructed with a major upgrade of all subsystems compared to its predecessor CMD-2 experiment. In particular, a new drift chamber provided higher efficiency and more than twice better momentum resolution, and a new liquid-xenon
(LXe) calorimeter added multi-layer tracking capabilities and shower profile measurement. The detector also has completely new up-to-date
electronics and an elaborate trigger system. The main goals of experiments at VEPP-2000 include the
high-precision measurement of cross sections of various modes of $e^+e^- \to \text{hadrons}$ at low energies. All major channels are
under analysis with final states of up to 7 pions, or two kaons and three pions~\cite{Ignatov:2019omb}. Many results have already been
published by CMD-3 as shown in \cref{fig:CMD3crosssections}(left), where $e^+e^-\to K_SK_S\pi^+\pi^-$~\cite{CMD-3:2019cvx}, $K_SK^{\pm}\pi^{\mp}\pi^+\pi^-$~\cite{CMD-3:2022qhm}, and $\pi^+\pi^-$~\cite{CMD-3:2023rfe,CMD-3:2023alj} have appeared since WP20, and many more channels are still being analyzed.

The most crucial channel for evaluation of $\amuHVPLO$ is the simplest $e^+e^-\to \pi^+\pi^-$ production. The study of this process has a long history since the first $e^+e^-$ colliders, with numerous experiments that have contributed over the years, as shown in \cref{fig:CMD3crosssections}(right). This enormous effort is still not sufficient; to match the expected precision of the $g-2$ experiments, this channel should be known with a precision better than 0.2\%. The latest, and one of the most precise measurements, was provided by the CMD-3 experiment~\cite{CMD-3:2023rfe,CMD-3:2023alj}. This new measurement was intensely scrutinized and reviewed by the community within a specially organized series of seminars~\cite{MiniWS-CMD3-1,MiniWS-CMD3-2}, where a comprehensive list of questions covering all aspects of the analysis was asked by a panel of experts nominated by the $g-2$ TI Steering Committee. No pitfalls were found.

The $e^+e^-\to \pi^+\pi^-$ CMD-3 analysis is based on the largest ever data set of $34\times10^{6}$ selected $\pi^+\pi^-$ events at $\sqrt{s}<1\GeV$, which is an order of magnitude larger statistics compared to most of the previous measurements.
The large statistics were crucial to study various systematic effects in detail and to obtain a sharper view of possible detector effects. It would be impossible to apply the analysis procedure, developed for the CMD-3 measurement, to the data collected by the predecessor CMD-2 experiment, as many of the systematic effects would be hidden under statistical fluctuations. 
The estimation of the possible systematic effects in the CMD-3 analysis is done in the most conservative way, with the goal to achieve a high confidence in the final declared precision even at the expense of possible increase in error.

One of the key features of the analysis, in contrast to other measurements, is the use of three independent sources of information for measuring the number of detected $\pi^+\pi^-$ events: using momentum distributions of two particles measured in the tracking system, using detected energy depositions in the LXe calorimeter, or using the polar angular distribution. All three methods are consistent within 0.2\%.

\begin{figure}[t]
\centering
\includegraphics[width=0.49\linewidth]{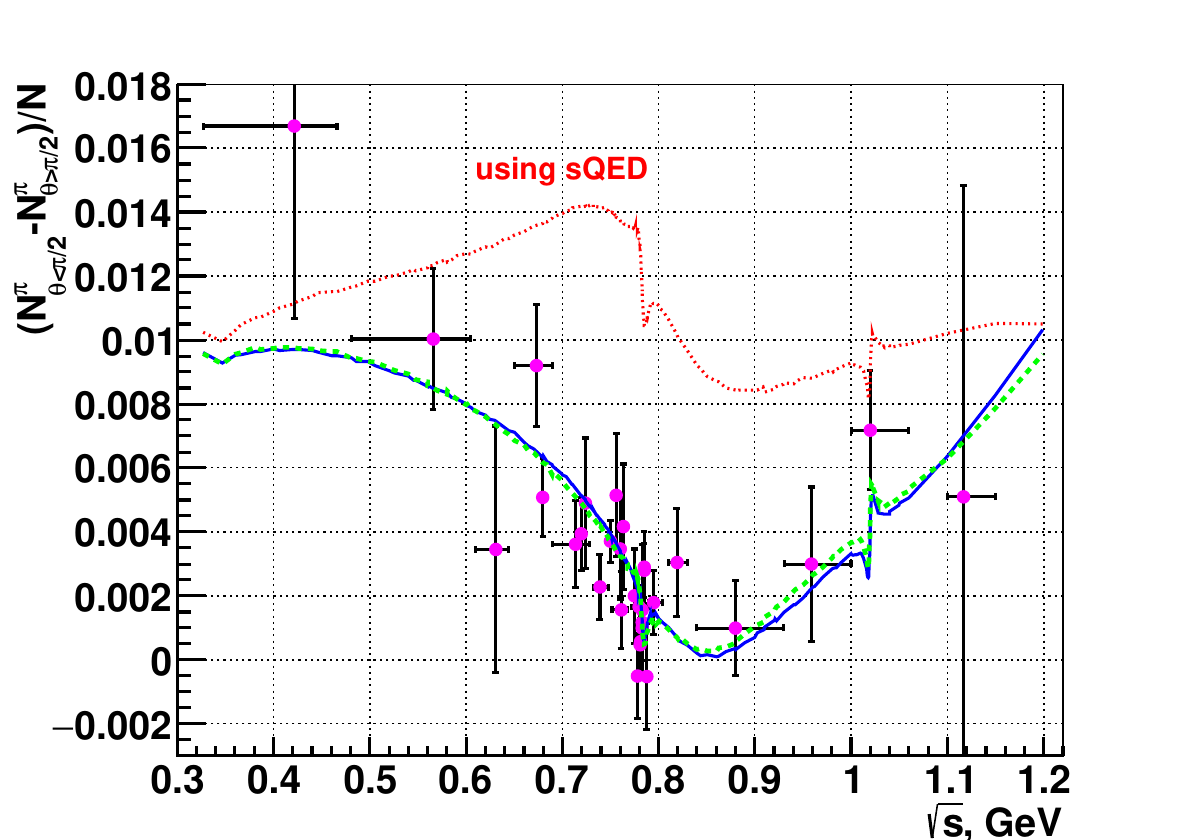}
\includegraphics[width=0.49\linewidth]{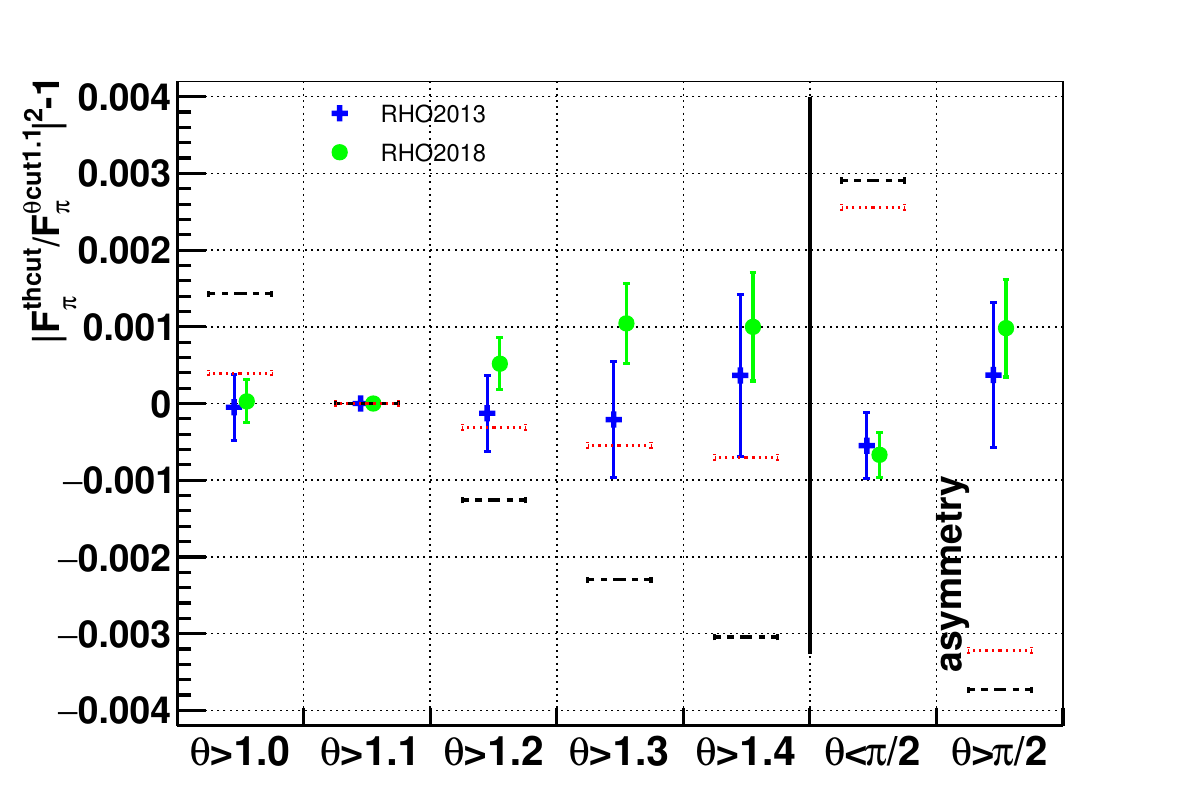}
\caption{
Left: The measured asymmetry in the $\pi^+\pi^-$ process at CMD-3 in comparison with
        the prediction based on the conventional sQED approach (red dotted line),
        the GVMD model~\cite{Ignatov:2022iou} (blue line), and the
        dispersive calculation~\cite{Colangelo:2022lzg} (green dashed line). 
Right: The relative differences between the pion form factors in the cases of the various
         $\theta^{cut}$ cut values. Different data taking seasons are shown by crosses (RHO2013)
         and circles (RHO2018). Also shown is the expected behavior in the case of the presence of a 0.5\% systematic uncertainty in $|F_\pi^V|^2$ from the
         polar angle determination, either due to a $Z$-scale detector miscalibration (black dashed lines)
         or due to a constant shift in the reconstructed angle (red dotted lines). Figures adapted from Ref.~\cite{CMD-3:2023alj}.
}
\label{fig:CMD3asym} 
\end{figure}

Another important aspect of the performed analysis by CMD-3 was a comprehensive study of the detector acceptance systematic uncertainty
due to the determination of the polar angle. As a by-product of this investigation, the forward--backward charge asymmetry of $\pi^+\pi^-$
was measured with an integrated statistical precision of about 0.025\%. A strong 1\% deviation was observed from the prediction based
on the conventional scalar QED (sQED) approach used for the calculation of radiative corrections, as shown in
\cref{fig:CMD3asym}(left), which was never noted before. The improved generalized-vector-meson-dominance (GVMD) model was proposed in Ref.~\cite{Ignatov:2022iou}, which gives
remarkable agreement with the experimental data. This was later confirmed by a calculation in the dispersive formalism~\cite{Colangelo:2022lzg}, see \cref{sec:radiative_corrections}.
At first the dispersive approach prediction still displayed some visible deviation from the data, but a subsequent identification of an incorrect treatment of endpoint singularities in the dispersive calculation integrals led to an astonishing consistency with the experimental $\pi^+\pi^-$ asymmetry and the GVMD prediction, see Refs.~\cite{Budassi:2024whw,Colangelo:2022lzg} and \cref{fig:CMD3asym,fig:asymmetry}.
Independent beyond-sQED calculations within the GVMD and the dispersive
formalism were also performed in Ref.~\cite{Budassi:2024whw}, and some of the cross-checks were facilitated by the {\it RadioMonteCarLow~2} effort, see \cref{sec:MC}. 
The observed limitation of the sQED approach also suggests to revisit the radiative corrections used in the initial-state-radiation (ISR) measurements, where this can contribute to the $C$-even two-photon-exchange corrections and consequently affect the experimental extraction of the pion vector form factor (VFF) $F_\pi^V$, see Refs.~\cite{Abbiendi:2022liz,Aliberti:2024fpq}. 
The dispersive approach offers a more comprehensive description of the pion VFF, as it incorporates the correct analytic properties by construction. 
However, the GVMD model can provide a much simpler framework for more complicated calculations involving additional hard photons, and it also allows for iterative refinement of the VFF through the inclusion of additional terms. Ideally, having both calculations available would be beneficial for cross-checks. 

The dominant systematic contribution in the CMD-3 $\pi^+\pi^-$ measurement comes from the fiducial-volume determination and was
conservatively estimated as 0.5\%/0.8\% (RHO2018/2013) as shown in the
summary systematic Table~2 of Ref.~\cite{CMD-3:2023alj}.
The achieved consistency at the permil level in the measured
asymmetries of $e^+e^-$, $\pi^+\pi^-$ processes in CMD-3 together
with overall stability of the result over different detector regions gives a quite confident level of control of this source of systematics. 
Such a cross-check with the extracted pion VFF within the different polar angle selections is shown in \cref{fig:CMD3asym}(right).
For comparison, dashed lines show how the experimental points should look like in case
of the presence of a 0.5\% systematic effect from the fiducial volume in the measured cross section due to the most natural scenarios,
either by a miscalibrated $Z$-scale of the detector or by a constant
bias of the reconstructed polar angle for tracks of one of the
charges. It is seen that the data is about 5 times more consistent than
the declared assigned systematics. 
It should also be noted that some other previous measurements, which
showed the dependence of results versus the detection polar angle, demonstrated variations an order of magnitude larger than those achieved by CMD-3.

Another important test was the measurement of $e^+e^-\to \mu^+\mu^-$ cross section, predicted by QED. It was done for the momentum-based analysis for $\sqrt{s}<0.7\GeV$ only, where the momentum resolution of the tracking system allowed to separate muons from other particles. The observed average ratio of the measured cross section to the QED prediction $1.0017(16)(61)$ proves the consistency of most parts of the analysis procedure, including the separation procedure, detector effects, evaluation of radiative corrections, etc.

Reaching improved precision requires support by MC tools. One of the drawbacks of the CMD-3 analysis was that only one generator,
\mcgpj, with the required precision was available for the $e^+e^-\to \pi^+\pi^-$ process. Since the CMD-3 publication, the
Pavia group presented a calculation of the $\pi^+\pi^-(\gamma)$ hadronic channel at NLO matched to a parton shower
algorithm, with implementation in the new version of the \babayaga{}
MC event generator~\cite{Budassi:2024whw}.
The values of integrated radiative corrections
and the forward--backward charge asymmetry in $\pi^+\pi^-$ were confirmed. The strong {\it RadioMonteCarLow~2} community effort, going
towards next-to-next-to-leading-order (NNLO) precision and above, will help further clarify and strengthen the
theoretical side of the experimental $\pi^+\pi^-(\gamma)$ measurements,
as discussed in \cref{sec:MC}. This effort will be mandatory if
there is a wish to reach permil-level experimental precision in this channel.
This aspect is even more critical for the ISR measurements right now, as the LO of the ISR processes starts at NLO precision of MC tools for the energy-scan experiments.

The total systematic precision of the CMD-3  $e^+e^-\to\pi^+\pi^-$ measurement was conservatively estimated as 0.7\% in the central
region near the $\rho$ resonance. It is important to give
a dedicated clarification regarding the treatment of the provided  experimental systematic uncertainties. Being dominated mainly by a
single source, the CMD-3 publication provides only the total  systematic error for each point and suggests using a single-source
full systematic covariance matrix accordingly, for simplicity of further external use.
This yields an even more conservative evaluation of the uncertainty of the
$\amuHVPLO$ integral compared to splitting over different statistically independent systematic contributions. Such a
prescription in the case of direct point-by-point integration
conventionally assumes using 100\% correlated systematic uncertainty.
However, strictly speaking, the exact energy dependence of the required systematic corrections are unknown. Therefore, it is  essential to
consider the uncertainties of the uncertainty, as argued in Sec.~2.3.6 of WP20. This is especially important given
the reached advancements in statistical precision and the entirely dominant systematic uncertainties of the latest CMD-3 measurement. 
A strict interpretation of the provided systematic error values is that they accommodate an underlying systematic effect with possible internal correlations up to 100\%. Furthermore, these values allow for any energy-dependent functional form of the required systematic correction, provided it remains within the specified systematic band.
This prescription was commonly assumed by experimentalists in most of the $e^+e^-\to \text{hadrons}$ measurements and for most of the
separate systematics sources (even when a correlation matrix was provided).
Applications that exploit correlations between data and consider only 100\% strict correlation from the first to the last energy point of different split systematic contributions 
can produce a significant underestimate of the uncertainty on the quantity being evaluated.
This, for example, applies to the all analyticity-constrained fits described in \cref{sec:disp}, where using
only ``100\% correlated systematic'' of each separate source strongly
reduces the degrees of freedom of the fit and consequently shrinks the systematic part of errors from that. Such an
underestimation of the systematic error of the evaluated  $\amuHVPLO$ integrals may in some cases be greater than a factor 1.5,
as shown in Ref.~\cite{Leplumey:2025kvv} and noted in~\cref{sec:disp_2pi} for the CMD-3 data treatment case.

The published CMD-3 pion VFF result was based on the full data sample collected at energies below 1~GeV before 2024 (62 pb$^{-1}$) and on a small subset of data, about 24~pb$^{-1}$, collected at energies above 1~GeV.
The full data sample above 1 GeV is under analysis. The preliminary results on the pion VFF were obtained for data collected in 2019--2021 (about 170 pb$^{-1}$). The major data sample of 580 pb$^{-1}$ was collected in 2022 and 2023; these data are yet to be incorporated into the ongoing analysis. Data analysis at these CM energies differs from the one implemented in Ref.~\cite{CMD-3:2023alj}:
\begin{itemize}
\item The momentum resolution of the CMD-3 tracking system does not allow one to distinguish momenta of $e$, $\mu$, and
$\pi$ in $e^+e^- \to e^+e^-$, $\mu^+\mu^-$, and $\pi^+\pi^-$, correspondingly. Thus, the final-state identification is based on the analysis of energy deposition in the calorimeter. In order to cross-check the identification procedures, two independent methods are being developed---one is based on data from the LXe calorimeter only (similar to the method described in Ref.~\cite{CMD-3:2023alj}) and another is based on the data from the full calorimeter and machine learning.
Muons and pions cannot be robustly distinguished by the CMD-3 calorimeter response, thus the number of $\mu^+\mu^-$ pairs is evaluated from the number of detected $e^+e^-$. The identification procedure based on the analysis of angular distributions is statistically limited and is applied only as an overall cross-check, not point-by-point.
\item At energies above 1 GeV, the cross section for the 2$\pi$ final state drops, but it rises for many other hadronic final states, most notably for $K^+K^-$. In order to keep background low, strict selection cuts are used, which introduces larger efficiency corrections, which differ for $e^+e^-$, $\mu^+\mu^-$, and $\pi^+\pi^-$.
\end{itemize}

New data taking at energies below 1 GeV started in 2024 and is expected to continue through 2025 with the goal to scan the whole CM energy range from 0.32 to 1 GeV and to collect 2--3 times more data compared to previous scans. The CMD-3 result~\cite{CMD-3:2023alj} is systematically limited. Nevertheless, the gain in statistics will be beneficial for the pion VFF measurement as well, as it will open additional possibilities for the systematics studies. It should be noted that a significant bulk of the data collected (and planned to be collected) in 2024/2025 is taken with the directions of $e^+$ and $e^-$ beams reversed; that allows one to cross-check various detector-specific efficiency corrections.

The luminosity provided by VEPP-2000 at $\rho(770)$ energies allows for statistical accuracy of (0.1--0.2)\% for the pion VFF. Such precision is required in order to match the accuracy of the $\amuHVPLO$ evaluation in the dispersive approach to the precision of the final FNAL result. Reduction of the systematic uncertainty to the same level requires a significant upgrade of the CMD-3 detector. The key directions of the upgrade are: a new tracking system with improved momentum resolution; the possibility for the precise and robust measurement of the polar angles of the tracks and the position of the vertex along the beam axis; a dedicated system to study nuclear interactions of pions with the material of the beam pipe and the tracking system. The upgrades are being designed. It is estimated that the data taking with the upgraded detector can start around 2030. It should be mentioned that 0.1\% precision for the pion VFF also sets very high requirements for the theoretical models and corresponding MC codes used to evaluate the radiative corrections.

The same data sets discussed here are used for the measurement of cross sections of many exclusive channels of $e^+e^-\to\text{hadrons}$ beyond $e^+e^-\to \pi^+\pi^-$. The second most important channel after $2\pi$ for the
$\amuHVPLO$ evaluation is $e^+e^-\to \pi^+\pi^-\pi^0$. The first result for this channel was published as a by-product of the $2\pi$ measurement \cite{CMD-3:2023alj}, based on the small subset of $\pi^+\pi^-\pi^0$ identified as a background to $\pi^+\pi^-$. A dedicated analysis of the full $\pi^+\pi^-\pi^0$ data set is in progress and is expected to be released in 2025.

\subsubsection{SND}

\begin{figure}[t]
\includegraphics[width=0.49\linewidth]{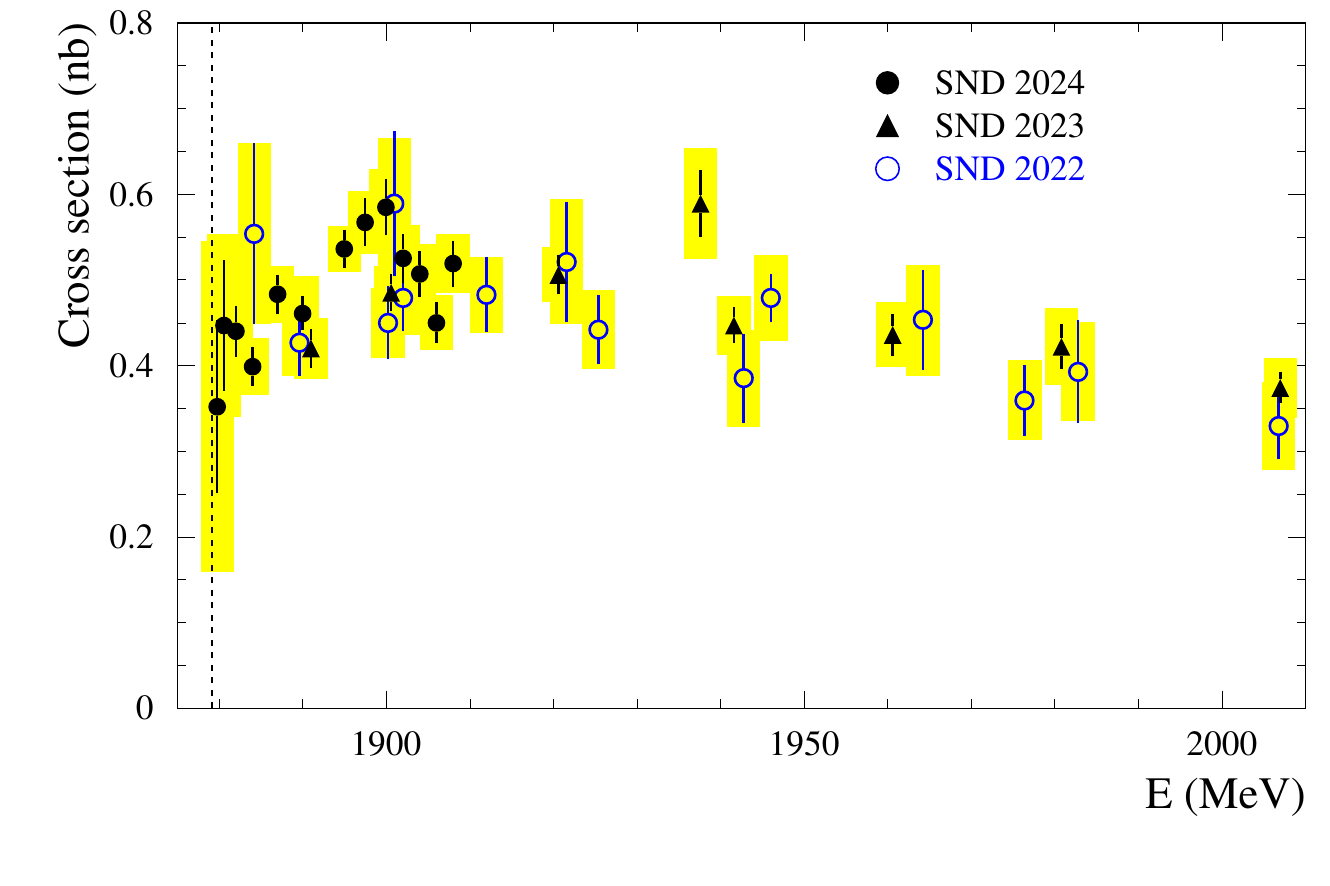}
\includegraphics[width=0.49\linewidth]{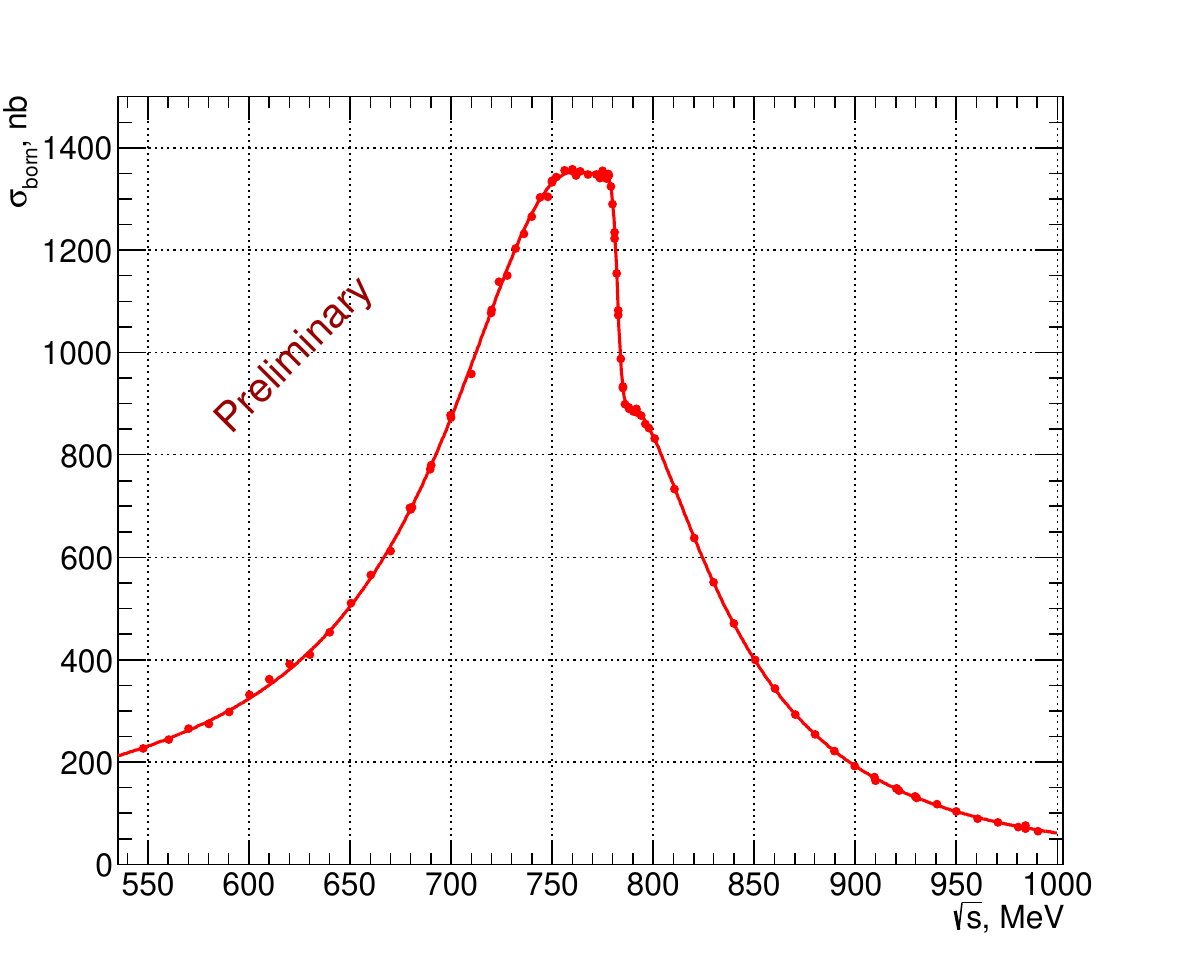}
\caption{Left: The $e^+e^-\to n\bar{n}$ cross section measured by SND~\cite{SND:2022wdb,SND:2023fos,Achasov:2024pbk}. The error bars and shaded boxes represent the statistical and systematic
uncertainties, respectively. Right: The $e^+e^-\to \pi^+\pi^-$ born cross section measured by SND. The red line is a fit curve.}
\label{fig:SND_fig}
\end{figure}

Since 2020 several hadronic cross sections were measured at the SND experiment
at the VEPP-2000 $e^+e^-$ collider. A precise measurement of the
$e^+e^-\to K_SK_L$ cross section was carried out in the vicinity of the $\phi$
meson resonance~\cite{SND:2024kbi}. The systematic uncertainty in the measured cross section
at the maximum of the $\phi$ resonance is 0.9\%. 
The $e^+e^-\to \pi^+\pi^-\pi^0$ cross section was measured in the energy range
$(1.075\text{--}1.975)\GeV$~\cite{Achasov:2024tfh}. The SND result is consistent with the previous \babar{} 
measurement~\cite{BABAR:2021cde}, but is more accurate. The SND data on the 
$e^+e^-\to K^+K^-\pi^0$ cross section below 2 GeV \cite{SND:2020qmb} are also in reasonable
agreement with the \babar{} measurement~\cite{BaBar:2007ceh}. The $e^+e^-\to n\bar{n}$ process was
studied in the energy range from threshold up to 2 GeV. The measured
cross section~\cite{SND:2022wdb,SND:2023fos,Achasov:2024pbk} is presented in \cref{fig:SND_fig}(left). New data on the 
processes with multiphoton final state $e^+e^-\to \eta\pi^0\gamma$ \cite{SND:2020kgt}, 
$e^+e^-\to \eta\eta\gamma$ \cite{SND:2021amz}, $e^+e^-\to \eta\gamma$ \cite{SND:2023sur}, 
$e^+e^-\to \eta^\prime\gamma$ \cite{SND:2024qaq} were obtained.

Currently, the SND collaboration works with $\pi^+\pi^-$ and $\pi^+\pi^-\pi^0$ channels in the $0.548 \GeV <\sqrt{s}<1 \GeV$ and $0.548 \GeV <\sqrt{s}<1.1 \GeV$ energy regions, respectively, using 69 pb$^{-1}$ data. They have faced serious challenges due to the fact that detector performance was very inconsistent during the 2017--2018 experiment. Despite these problems, it is possible to perform high-precision measurements of the $e^+e^- \to \pi^+\pi^-$ cross section, using a modified version of the reconstruction software. A new algorithm of $e$/$\pi$ separation allows one to suppress $e^+e^- \to e^+e^-$ background by a factor of $10^3$ in the whole energy region, thus limiting the growth of the systematic error with decrease of $\sqrt{s}$. An improved reconstruction algorithm allows one to reduce the number of poorly reconstructed tracks. Simulated $e^+e^- \to \pi^+\pi^-$ events now better describe experimental inefficiencies of the cosmic veto and energy-deposition cut, after introduction of the light quenching effects in the plastic scintillator of the muon system and NaI(Tl) of the electromagnetic (EM) calorimeter. The current number of the selected $e^+e^- \to \pi^+\pi^-$ events makes the statistical uncertainty negligible compared to the systematic one. Due to the increased tensions between the most precise measurements of the $e^+e^- \to \pi^+\pi^-$ cross section, a blinding procedure was introduced for the earlier stages of the SND analysis. The efficiency of $e^+e^- \to \pi^+\pi^-$ was modified via multiplying it by $1+a_0+a_1\times(\sqrt{s}-775 \MeV)$, where $a_0$ and $a_1$ are randomly generated parameters, unknown to the researchers.
The preliminary results of the unblinded $e^+e^-\to \pi^+\pi^-$ 
cross-section measurement are shown in \cref{fig:SND_fig}(right). Current estimates of the systematic uncertainty are (0.6--0.8)\% in the whole energy range, due to much higher statistics and improvements in the simulation and reconstruction algorithms, compared to the 2020 SND analysis.  
Progress with the $\pi^+\pi^-\pi^0$ channel was delayed, since this analysis relies on the high-precision measurements of the integrated luminosity from the $\pi^+\pi^-$ analysis. Preliminary estimates of the systematic uncertainty of $e^+e^- \to \pi^+\pi^-\pi^0$ cross-section measurements show it to be no greater than 1.4\%.

\subsubsection{KLOE}
\label{Sec:KLOE}

The KLOE experiment took data during 2001--2006, with a total integrated luminosity of 2.5\,$\mathrm{fb^{-1}}$ at $\sqrt{s} = 1.0194\,\mathrm{GeV}$ (the $\phi$ peak)
and 250\,$\mathrm{pb^{-1}}$  off-peak at $\sqrt{s} = 1\,\mathrm{GeV}$ (in 2006). Peak luminosity was reached during the 2005 run with a luminosity of 8.5\,$\mathrm{pb^{-1}}$/day.
KLOE has published three precise cross-section measurements of $\sigma(e^{+}e^{-} \rightarrow \pi^{+} \pi^{-}\gamma(\gamma))$: KLOE08~\cite{KLOE:2008fmq}, KLOE10~\cite{KLOE:2010qei}, and KLOE12~\cite{KLOE:2012anl}, which have been used for the evaluation of the 2$\pi$ contribution to $\amuHVPLO$~\cite{Aoyama:2020ynm}.
The KLOE08 and KLOE12 analyses were performed with a small-angle (SA) selection where the photon is required to be within $\theta_{\gamma}<15\degree$ or $\theta_{\gamma}>165\degree$ (i.e., missing momentum) resulting in 240\,$\mathrm{pb^{-1}}$ ($\simeq$ 3.5 million $\pi \pi \gamma $ events) of data taken in 2002. The KLOE10 analysis was performed on 232\,$\mathrm{pb^{-1}}$ from 2006 (corresponding to 0.6 million events), after the photon polar angle is selected at large angle (LA), i.e., $50\degree<\theta_{\gamma}<130\degree$. Further details of the analyses can be found in Refs.~\cite{Aoyama:2020ynm,Aliberti:2024fpq,Venanzoni:2017ggn}, and details on the agreement and combination of the three results are found in Ref.~\cite{KLOE-2:2017fda}.

In the KLOE08, KLOE10, and KLOE12 analyses, the MC generator \phokhara{} (version 5) was used, including NLO ISR, final-state-radiation (FSR) corrections, and simultaneous emission of one ISR and one FSR photon~\cite{phokhara, mueller_phokhara}. \phokhara{} was interfaced with the standard KLOE MC GEANFI to compute MC efficiencies as a function of the reconstructed $M^2_{\pi\pi}$, and to compute the radiator function $H$ used to relate the measured differential cross section for $e^+e^-\to\pi^+\pi^-\gamma$ to the cross section $\sigma_{\pi\pi}$. In addition, \phokhara{} was used to correct for the shift between the measured $M^2_{\pi\pi}$ and the invariant mass of the intermediate photons for FSR events. The FSR treatment was done assuming point-like pions (F$\times$sQED, see Ref.~\cite{Aliberti:2024fpq} for definitions).

The KLOE12 analysis normalized the cross section using a $\mu^+\mu^-\gamma$ cross-section measurement with the photon at SA. 
The error quoted by the authors of \phokhara{} on the ISR part of the generator is 0.5\%, this error cancels out in the extraction of the pion VFF $|F_\pi^V|^2$ from the $\pi^+\pi^-\gamma/\mu^+\mu^-\gamma$ ratio. In the KLOE12 analysis, a global uncertainty of 0.2\% was estimated for the factorization approximation when subtracting FSR for muons, and for the correction due to the FSR shift for pions. The effects of additional diagrams not present in \phokhara{} (v5) for muons was estimated to be 0.1\% at most. The SA selection greatly suppresses events with only one hard photon from FSR and no ISR, as well as a large part of events with one photon from ISR and one from FSR. For KLOE08, the uncertainty due to the treatment of FSR events is estimated to be 0.3\%. For KLOE10, the uncertainty on the treatment of FSR has been estimated to be 0.5\% due to the uncertainty of the pion VFF at $\sqrt{s}= 1\GeV$ and 0.6\% due to a comparison of FSR modeled with SU(3) chiral perturbation theory (ChPT) \cite{Aliberti:2024fpq,Gasser:1983yg} with respect to the treatment used in \phokhara. 

Within the {\it RadioMonteCarLow~2} efforts \cite{Aliberti:2024fpq}, ``tuned comparisons'' (see Ref.~\cite{WorkingGrouponRadiativeCorrections:2010bjp} for the definition) between \phokhara{} 10.0\footnote{Differences between \phokhara{}5 (which was used by KLOE) and \phokhara{}10 have been investigated in Ref.~\cite{Campanario:2019mjh}.} (which has a full NLO matrix element for  $\pi^+\pi^-\gamma$ and $\mu^+\mu^-\gamma$ final state) and other MC generators---namely, \afkqed{}, \babayaga{}, \kkmc{}, and \mcmule{}, all described in Ref.~\cite{Aliberti:2024fpq}---were performed~\cite{gv_ti24,Aliberti:2024fpq}, in order to investigate radiative correction effects both due to missing NNLO contributions in \phokhara{} and also to test the validity of the generator at NLO, as questioned in Ref.~\cite{Davier:2023fpl}. 
The following configurations have been studied: 
\begin{itemize}
    \item $e^{+}e^{-} \xrightarrow{} \mu^{+} \mu^{-}\gamma(\gamma)$ before and after SA angular acceptance, considering both ISR and FSR emission;
    \item $e^{+}e^{-} \xrightarrow{} \pi^{+} \pi^{-}\gamma(\gamma)$ before and after SA and LA angular acceptance for only ISR emission---due to limitations of other MC generators used in this comparison~\cite{gv_ti24}. In the case of LA acceptance, it was also possible to compare \afkqed{} vs.\ \phokhara{} after the trackmass cut;\footnote{The trackmass is a kinematic variable defined in Ref.~\cite{KLOE:2008fmq}. The tails in its distribution are mainly due to radiative corrections, therefore, it is sensitive to (N)NLO effects.} this was not possible in the case of SA due to the \afkqed{} limitations for the KLOE configuration.
\end{itemize}

\begin{figure}[t]
\centering
\includegraphics[width=.8\linewidth]{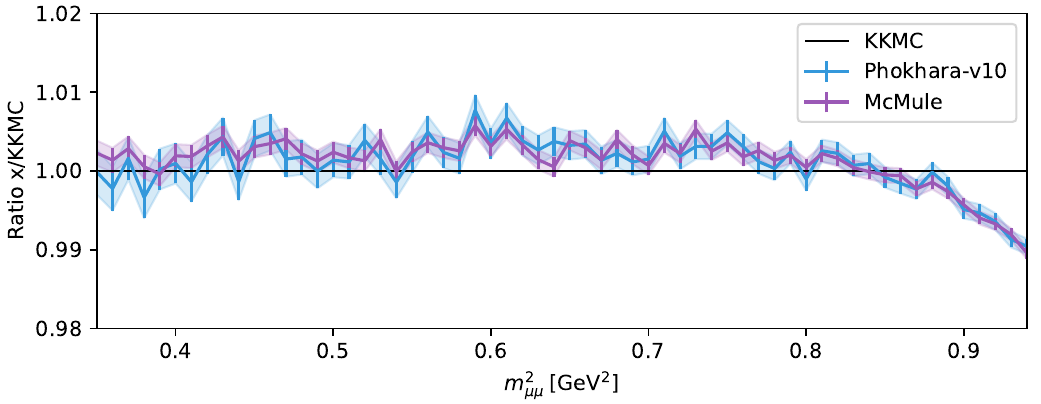}
\caption{Cross section of $e^+e^-\to\mu^+\mu^-\gamma$ vs.\ the invariant mass of two muons, in the KLOE SA acceptance selection scenario, with ISR and FSR. The results from \phokhara{} and \mcmule{} were compared to \kkmc{} (black line), showing an agreement better than $0.5\%$ for most of the spectrum.}
\label{SA_isrfsr_mmg}
\end{figure}

In this study, the $e^+e^-\xrightarrow{} \pi^+\pi^-\gamma$ and $e^+e^-\xrightarrow{} \mu^+\mu^-\gamma$ cross sections as a function of the invariant mass of the two pions and muons from \phokhara{} were compared with the predictions from other generators~\cite{Venanzoni:2017ggn,Aliberti:2024fpq}. Vacuum polarization (VP) was switched off and detector effects (involving smearing of momenta) were not included. The selection cuts used for the SA and LA comparisons are described in Ref.~\cite{Aliberti:2024fpq}, in the sections ``KLOE-like small-angle scenario'' and ``KLOE-like large-angle scenario,'' respectively.

The studies on muons showed good agreement with the SA acceptance between \phokhara{}, \mcmule{}, and \kkmc{}, with a difference below $0.5\%$ between \phokhara{} and \kkmc{} for most of the $M^2_{\mu\mu}$ spectrum, as shown in \cref{SA_isrfsr_mmg}. We note, see Fig.~32 of Ref.~\cite{Aliberti:2024fpq}, that this is not the case for the so-called ``$B$-scenario'' where differences between generators (particularly \kkmc{} and \afkqed{}) reach a few percent at $M_{\mu\mu}<1\GeV$. 
In the inclusive selection (i.e., without angular acceptance cuts), there are known differences for $M^2_{\mu\mu}>0.85\GeV^2$ between \phokhara{} and \kkmc{} \cite{Jadach:2005gx}, up to $\simeq1\%$, due to missing exponentiation in \phokhara, but with negligible effects on $\amuHVPLO$ (in the pion case)---which is dominated by the low-energy region.

\begin{figure}[t]
\centering
\includegraphics[width=0.49\linewidth]{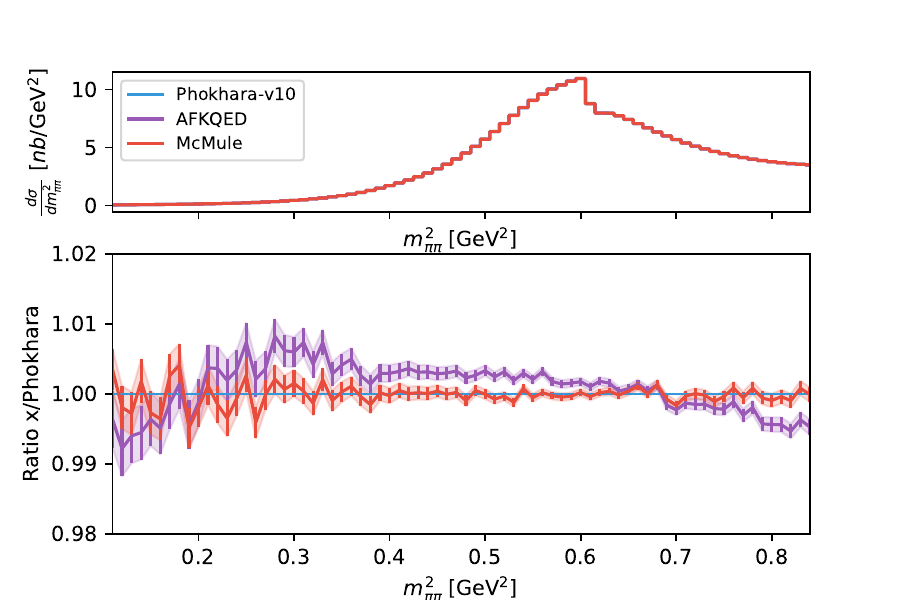}
\includegraphics[width=0.49\linewidth]{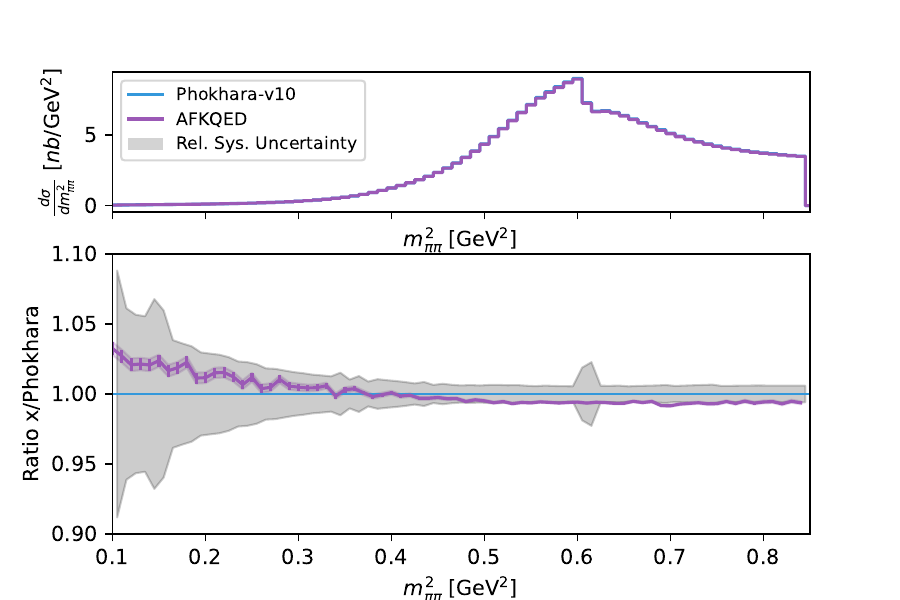}
\caption{Left: Cross section of $e^+e^-\to\pi^+\pi^-\gamma$ vs. the invariant mass of two pions, in the KLOE LA acceptance selection scenario, with ISR only, no trackmass cut is applied. The results from \afkqed{} and \mcmule{} are compared to \phokhara{} (blue line), showing an agreement better than $0.5\%$ at the $\rho$ peak. Right: Cross section of $e^+e^-\to\pi^+\pi^-\gamma$ vs. the invariant mass of two pions, in the KLOE LA acceptance selection scenario, with ISR only, with the trackmass cut applied. The results from \afkqed{} are compared to \phokhara{} (blue line), showing an agreement better than $1\%$ around the $\rho$ peak ($0.6\GeV^2$). The systematic uncertainty for KLOE10 LA analysis is shown as a function of the invariant mass of the two pions (gray band).}
\label{LA_isr_ppg}
\end{figure}

The studies on pions showed excellent agreement between \phokhara{} and \mcmule{} in the inclusive and SA~\cite{gv_ti24} and LA acceptance selections, see \cref{LA_isr_ppg}(left). This is good confirmation of the correctness of the implementation of full NLO matrix elements in \phokhara. The LA acceptance selection (KLOE10) was studied also after the trackmass cut, see \cref{LA_isr_ppg}(right), which is sensitive to (N)NLO effects. With acceptance cuts only, an agreement below 0.5\% was found around the $\rho$ peak ($M^2_{\pi\pi}\simeq 0.6\GeV^2$) between \phokhara{} and \afkqed. With an additional cut on the trackmass variable on top of acceptance cuts, the agreement at the $\rho$ peak was found to be below 1\%, 
as shown in \cref{LA_isr_ppg}(right). The larger differences observed below $0.4\GeV^2$ are expected to originate from the different treatment of radiative corrections. While the size of the differences is still within the systematic uncertainty of KLOE10~\cite{KLOE:2010qei}, see \cref{LA_isr_ppg}(right), for the future analyses this difference is a further indication of the need for a NNLO MC generator, on which work is in progress~\cite{Aliberti:2024fpq}.

Preliminary results, which do not yet account for realistic detector effects, indicate the following:

\begin{itemize}
    \item  \phokhara{} shows excellent agreement with \mcmule{} for $\mu\mu\gamma$ and $\pi\pi\gamma$. These results confirm the correct implementation of NLO matrix elements in \phokhara{} and exclude possible issues associated with \phokhara{} as questioned in scenario 1 of Ref.~\cite{Davier:2023fpl}.
    
    \item For $\mu\mu\gamma$, the agreement between \phokhara{} and \kkmc{} is mostly within $\leq 0.5\%$ for ISR, ISR+FSR in both inclusive and SA selections, before the trackmass selection is applied. Differences above $0.85$ GeV$^2$ (in $m^2_{\mu\mu}$), reaching up to approximately $1\%$ are due to missing exponentiation in \phokhara{} (and have negligible effects on $\amuHVPLO$ in the pion case). Work is currently in progress to estimate the effects of radiative corrections due to the trackmass cut on muons, which can be sensitive to the presence of additional photon emission.
    
    \item For $\pi\pi\gamma$ a comparison with \afkqed{} in the LA selection reveals differences below 0.5\% when considering acceptance cuts, and below $1\%$ when including also the trackmass cut, at the $\rho$ peak. These findings limit the impact of NNLO effects on the LA analysis (KLOE10) as possible explanation for the discrepancy with \babar{} and CMD-3 experiments.
\end{itemize}

More studies are in progress, specifically: enlarging the statistics; including realistic detector effects (GEANFI); including more refined hadron--photon models for FSR (beyond F$\times$sQED)~\cite{Aliberti:2024fpq} and ISR NNLO radiative corrections; and applying the trackmass selection for SA $\pi\pi\gamma$ (KLOE08) and $\mu\mu\gamma$  (KLOE12) analyses. 

A new SA KLOE 2$\pi$ analysis on 2004--2005 data, normalized to $\mu \mu \gamma $, has started with the aim of clarifying the current tension with respect to other HVP evaluations. It is based on $\simeq$ 25 million $\pi \pi \gamma $ events, corresponding to $7$ times the statistics of the KLOE published analyses, which have not yet been analyzed.
The 2006 off-peak data will also be used for additional cross-checks and systematic studies. This new analysis, called KLOE-nxt, will employ new software tools and additional validation tests aiming to a precision of 0.4\% on the $2\pi$ contribution to $\amuHVPLO$, representing a two-fold improvement in accuracy compared to the previous KLOE12 result~\cite{estifaa}. It will also benefit from the ongoing work towards a NNLO implementation of \phokhara{}~\cite{Aliberti:2024fpq}. KLOE-nxt will provide an important cross-check of the published  results
and will be conducted in a blinded manner~\cite{PunziTesi}. 

In addition, KLOE is analyzing the three-pion cross section using the radiative-return method with 1.7\,fb$^{-1}$ of data collected at the $\phi$ meson mass~\cite{Cao:2020jus}.

\subsubsection{\babar}
\label{Sec:BABAR}

A number of hadronic processes have been studied by \babar\ since WP20, including $e^+e^- \rightarrow 2(\pi^+\pi^-)\pi^0\pi^0 + \pi^0/\eta$~\cite{BaBar:2021rki} and $\pi^+\pi^-\pi^0\pi^0\pi^0 + \pi^0/\eta$~\cite{BaBar:2021gyu} in 2021, as well as $K^+K^-\pi^0\pi^0\pi^0,\ K^0_SK^\pm\pi^\mp\pi^0\pi^0,\ K^0_SK^\pm\pi^\mp\pi^+\pi^-$~\cite{BaBar:2022ahi} in 2022. In particular, the $\pi^+\pi^-\pi^0$ cross section, which represents the second largest input of the HVP contribution to $a_\mu$, was measured in 2021~\cite{BABAR:2021cde} using 469\fb\ of data collected by the experiment near the $\Upsilon(4S)$ resonance, with CM energies ranging from 0.62 to 3.5 GeV via the ISR method. The analysis requires all final-state particles to be detected, namely two good-quality opposite-sign charged tracks and at least three photons, one being the ISR candidate with a CM energy larger than 3 GeV while the remaining photons must form one or more $\pi^0$ candidates with invariant masses $0.1<M_{\pi^0}<0.17$\GeV. A kinematic fit is performed on selected events and provides a $\chi^2$ quality value used to reject background processes, estimated with MC simulation. The cross section is studied separately below and above 1.1\GeV\ because of the detector resolution which significantly distorts the $3\pi$ mass spectrum in the lower region. From threshold to 2.0 GeV, the measured HVP contribution of this channel is $45.86(14)(58)\times 10^{-10}$, with a systematic uncertainty of 1.3\% near the $\omega$ and $\phi$ resonances, dominated by the detection efficiency, radiative correction, and luminosity. This result improves the precision by a factor of about 2 compared to calculations based on previous $\pi^+\pi^-\pi^0$ analyses.

The last \babar\ analysis to measure the \pipig\ cross section was published in 2009 and 2012~\cite{BaBar:2009wpw,BaBar:2012bdw} with around half the \babar\ statistics. To cancel out the ISR photon efficiency and the VP, the measured $\pi^+\pi^-(\gamma)$ mass spectrum is divided by the $\mu^+\mu^-(\gamma)$ spectrum, equivalent to the ratio of each final state's bare cross section. The separation between pions and muons was based on particle identification (PID) which required both charged tracks to have momenta larger than 1\GeV\ in order to make the muon identification more reliable. PID was found to be the dominant source of systematic uncertainties on the $a_\mu$ prediction, with a total relative systematic error of 0.5\% for energies from 0.5 to 1 GeV.

An upcoming \babar\ analysis, intended to be published in 2025, aims at improving the precision on the HVP contribution to $a_\mu$ from \pipig. For that purpose, the entirety of \babar\ data will be studied, while PID requirements on the tracks will be removed. An angular fit is considered as a new method~\cite{Davier_2017} to distinguish the main signal (\pipig, \mumug) and remaining background (\KKg, \eeg) processes, based on the cosine of the angle between the negative charge track and the ISR photon in the 2-track CM frame. To further differentiate the shapes of the dipion and dimuon distributions, the track momentum selection required in the previous analysis is released, increasing the statistics at the same time. Overall, this new study will result in an independent measurement of the \pipig\ cross section.

A first step towards this objective was taken with a study of additional radiation in ISR processes, published in 2023~\cite{BaBar:2023xiy} and relying on all the data collected by \babar. The intent was to evaluate how accurate the \phokhara\ and \afkqed\ generators are in simulating data and to measure the relative proportions of additional radiations in \mumug\ and \pipig\ processes with ISR, at ``LO'' (no additional photon), ``NLO'' ($+1$ real photon), and ``NNLO'' ($+2$ real photons, simulated only by \afkqed). Tracks are assumed to have pion masses, while dipion and dimuon events are identified according to tight PID selections. Kinematic fits are performed on events according to the different configurations in which additional photons are emitted. In the case of ``NLO'' events, two fits are applied depending on whether the single additional photon is detected in the EM calorimeter at a large angle (LA) from the beams, between 0.35 and 2.4 radians, or it is emitted at a small angle (SA), assumed to be collinear with the beams. For ``NNLO'' events, three fits are devised considering the two additional photons can both make large angles (2LA), small angles (2SA), or one of each configuration (LA+SA).

Inputs to the fits are the measured energy and direction of the ISR photon, the momenta and angles of both charged tracks, as well as the measured energy and angles of detected additional photons in the LA fit. Returned quantities are $\chi^2$ values and fit kinematic variables. Events are classified depending on the smallest $\chi^2$, with the additional requirement that fit energies are larger than 200 MeV either in the laboratory frame for LA photons or in the $e^+e^-$ CM frame for SA photons. Events below these thresholds are classified as ``LO.''

The distinction between ``NLO'' LA radiations from ISR and FSR is performed by fitting the distribution of the minimum angle between the additional photon and the charged tracks, using templates from \afkqed\ (FSR, no LA ISR simulated) and \phokhara\ (ISR after subtraction of FSR). The separation is fixed at 20 degrees in both \pipig\ and \mumug\ samples, with FSR peaking below 10 degrees and LA ISR forming a wide bump centered around 60 degrees.
Distribution shapes of LA photon energy simulated by both generators are observed to be in good agreement with data.

\begin{figure}[t]
\centering
{\includegraphics[width=0.45\columnwidth]{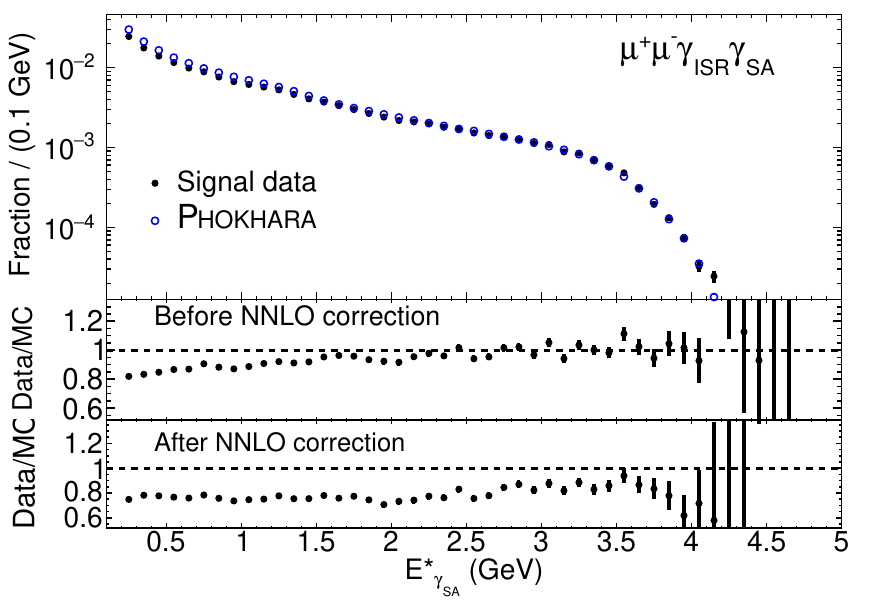}}
\hspace{0.4cm}
{\includegraphics[width=0.45\columnwidth]{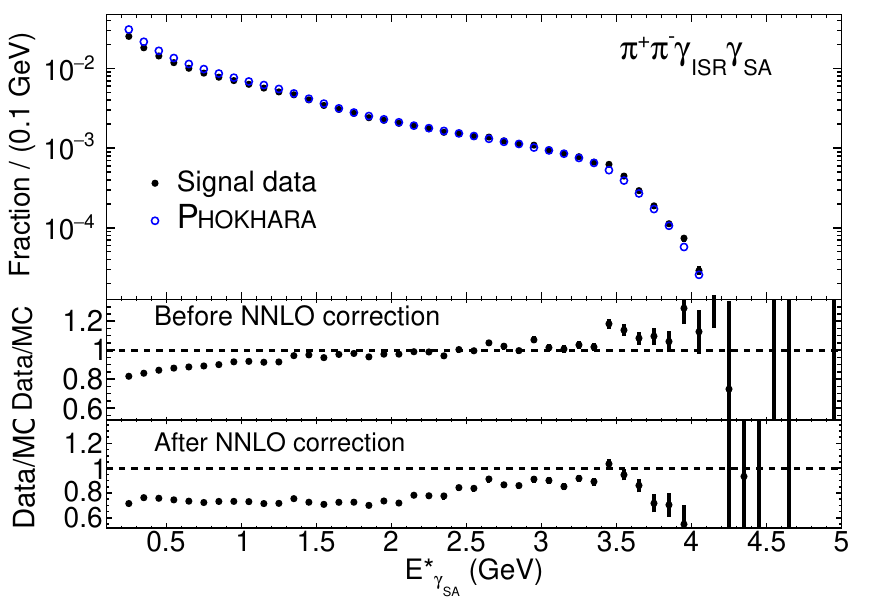}}
\vspace{-0.2cm}
\caption{
Comparison of the fit energy $E^*_{\gamma_{\mathrm{SA}}}$ in the CM frame of the additional ``NLO'' SA photon, before and after ``NNLO'' correction, in the \mumug\ (left) and \pipig\ (right) data and \phokhara\ samples. Figures taken from Ref.~\cite{BaBar:2023xiy}.}
\label{fig:NLO_SA} 
\end{figure}

In contrast, an excess of ``NLO'' SA photons, all ISR, is observed in \phokhara\ samples compared to data, evolving with a positive slope as a function of the additional photon's energy in the CM frame as seen in \cref{fig:NLO_SA}.
This excess has been shown not to be due to the collinear assumption in the ``NLO'' SA fits with the help of an alternative zero-constraint calculation that does not assume any collinearity. It is however affected by feed-through of ``NNLO'' 2SA photons, originating from a same beam, into the ``NLO'' SA category. After correction, the slope disappears and an almost constant excess of simulated events is observed with an overall data/MC ratio of $0.750(8)$ in \mumug\ and $0.763(19)$ in \pipig.

Significant ``NNLO'' contributions are found at the above-mentioned photon energy thresholds,  $3.47(38)\%$ and $3.36(39)\%$ of events in the muon and pion channels, respectively, more than 3\% being due to two additional ISR photons. \afkqed\ shows overall good performance in describing the rates and energy distributions of data at ``NNLO.'' A slightly high data/MC ratio of $1.061(15)$ for muons and $1.043(10)$ for pions is found up to the maximum generated energy.

The consequences for past $\pi^+\pi^-(\gamma)$ cross-section measurements vary depending on the experiment. The previous \babar\ result is unaffected as ``NLO'' and higher orders were already included in the analysis. The event acceptance determined with \phokhara\ must be corrected by a factor $0.3(1) \times 10^{-3}$, negligible compared to the total 0.5\% systematic uncertainty.
This independence from the treatment of additional radiation in \phokhara\ confirms the robustness of the \babar\ analysis with respect to radiative corrections. On the contrary, unaccounted shortcomings from this generator could potentially affect the results of other experiments which apply more stringent ``LO'' selections and rely on \phokhara\ for additional radiations.

\subsubsection{BESIII}
\label{Sec:BES}

The BESIII Collaboration is committed to providing experimental inputs to the dispersive evaluation of HVP by measuring the most relevant hadronic channels in electron--positron annihilation.
The timelike pion VFF, and related cross section, $\sigma(e^+e^-\to\pi^+\pi^-$), 
which represents the most important contribution to $\amuHVPLO$, was measured in 2015 in the $\rho(770)$ region with a systematic accuracy of 0.9\%~\cite{BESIII:2015equ}.
The analysis makes use of a data sample of 2.9\,fb$^{-1}$ collected at a CM energy of 3.77\,GeV and the ISR technique is employed to access the relevant energy range.
The event selection requires the detection of the $\pi^+\pi^-$ pair as well as the ISR photon in the fiducial volume of the apparatus. The four-momenta of these particles are subject to a four-constraint kinematic fit (4C), which enforces energy--momentum conservation under the assumption of the $e^+e^- \to \pi^+\pi^-\gamma$ final state. 
Conventional particle-identification methods are applied to reject background contributions from Bhabha scattering events, while for the separation of signal events from the QED process $e^+e^-\to \mu^+\mu^-\gamma$ a neural network method has been worked out. The latter is retained as control sample for various cross-checks and for a QED test.
The \phokhara~\cite{Campanario:2019mjh} event generator (version 7) is employed to determine selection efficiencies and corrections (e.g., VP, FSR), as well as to extract the radiator function at NLO.
A normalization is obtained by means of the integrated luminosity, which is determined by measuring large-angle Bhabha events using the \babayaga~\cite{CarloniCalame:2003yt} event generator.

\begin{figure}[t]
\begin{minipage}{0.49\linewidth}
    \centering
    \includegraphics[width=\linewidth]{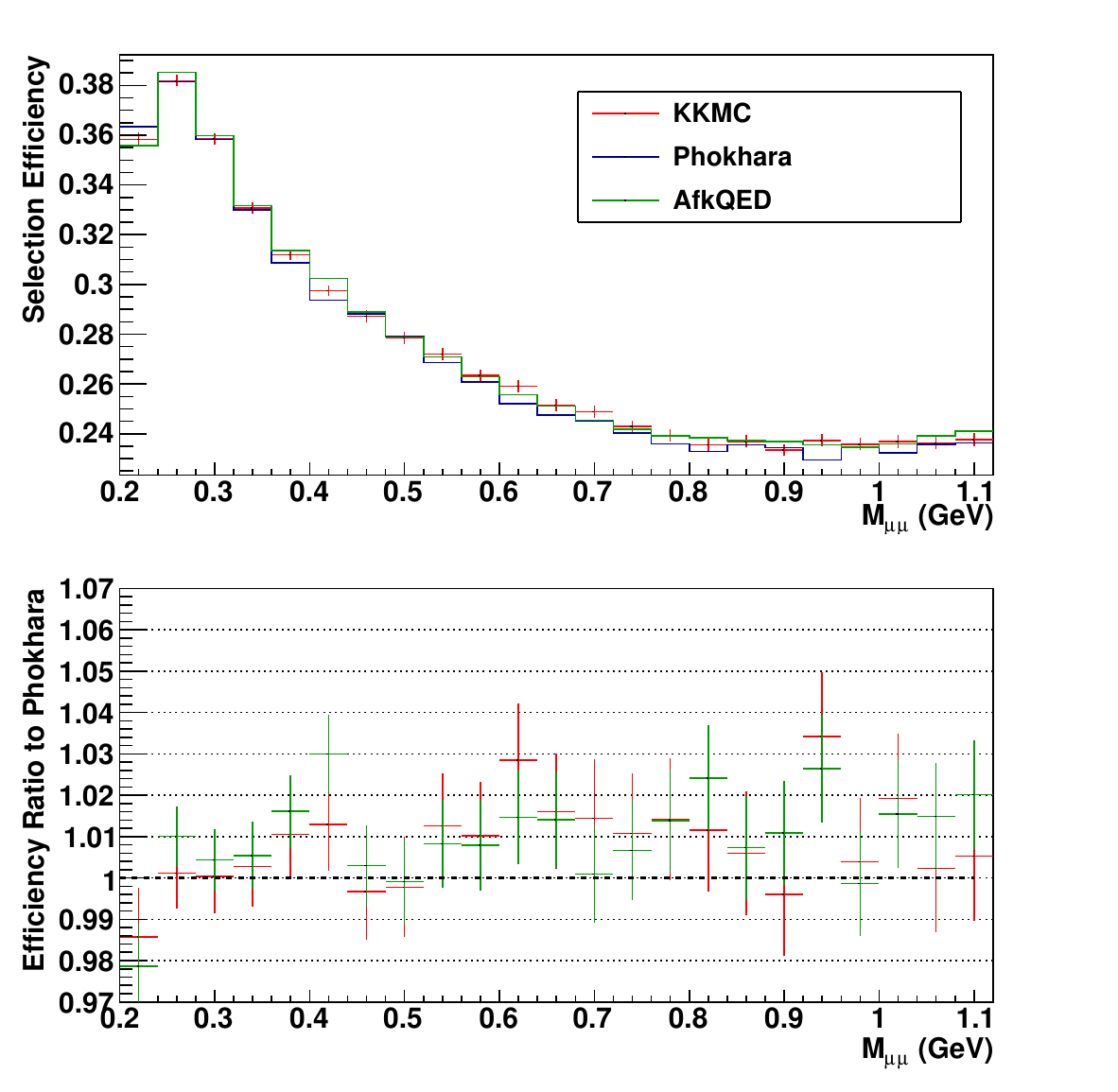}
\end{minipage}
\begin{minipage}{0.49\linewidth}
    \centering
    \includegraphics[width=\linewidth]{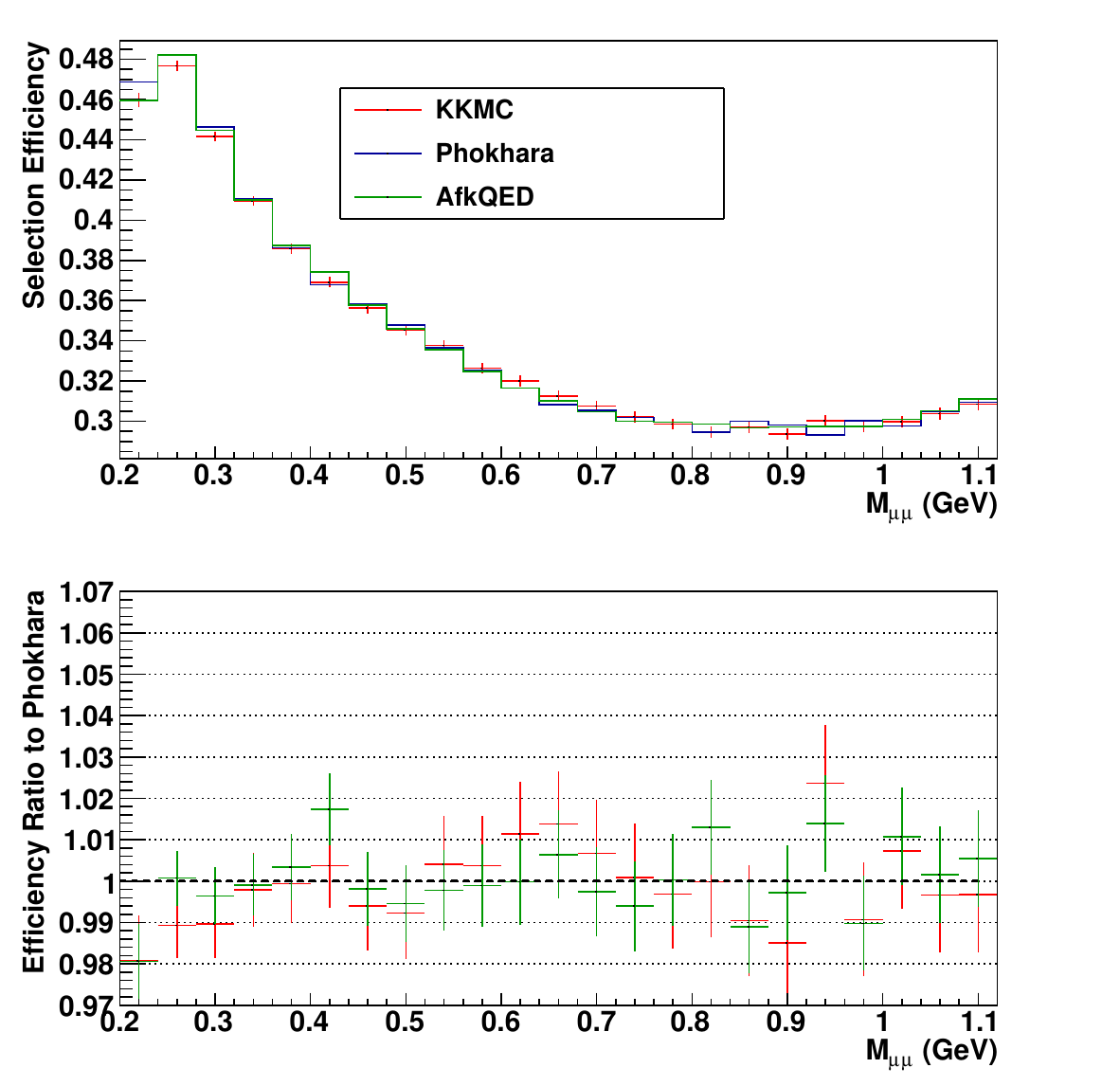}
\end{minipage}
    \caption{Comparison of the pion VFF selection efficiency, after a kinematic fit with four (left) and one constraints (right), as evaluated with different MC generators.}
\label{fig:BESIII-Eff-Comp}
\end{figure}

Recent observations of the \babar\ collaboration~\cite{BaBar:2023xiy} have identified a limitation of the \phokhara{} event generator in the description of those NLO events in which two ISR photons are present (one at small angles, one at large angles). Such a limitation can introduce a mismatch in the $\chi^2$ distribution of the 4C kinematic fit between the \phokhara{} simulation and real data, thus leading to an incorrect evaluation of the event-selection efficiency. As claimed in Ref.~\cite{Davier:2023fpl}, the impact of this \phokhara{} limitation on the BESIII analysis might be as large as 3.2\% in the case of the first of the two scenarios discussed in the paper, while the published BESIII result will be unaffected in the second scenario. The estimate of scenario 1, however, does not consider the effects of the photon efficiency calibration, which is based on a 1C fit. It has indeed been demonstrated that selections based on a 1C kinematic fit are much less affected by an imprecise description of NLO photons as these selections are more inclusive in NLO photon radiation. Effectively, the photon efficiency calibration procedure therefore mitigates the possible \phokhara{} issues found by the \babar\ collaboration. A quantitative investigation has demonstrated that the impact on the BESIII pion VFF measurement can be constrained to $\leq 1\%$. Such a result is also confirmed independently by detailed comparisons of the selection efficiencies calculated using the \phokhara, \afkqed, and \kkmc~\cite{Jadach:2000ir} event generators in the dimuon control sample. 
\Cref{fig:BESIII-Eff-Comp}(left) shows the selection efficiencies for the three generators (upper plot) and the relative difference with respect to \phokhara{} (lower plot) for realistic selection cuts including the 4C kinematic fit and the photon efficiency calibration. Also in this case an agreement at the 1\% level is observed among the three generators. \Cref{fig:BESIII-Eff-Comp}(right) shows the same comparison for a selection in which the 1C kinematic fit is employed. Here, an even better agreement between the three event generators can be observed, which is due to the much more inclusive nature of the 1C kinematic fit. From these studies the scenario 1 of Ref.~\cite{Davier:2023fpl} with a big impact on the BESIII measurement cannot be confirmed. The development of additional event generators suited for ISR measurements, as well as the further development of those presently available, will be crucial for pinning down the actual uncertainty due to the employed version of \phokhara. 

An improved measurement of the pion VFF at BESIII is foreseen in the upcoming years and aims to achieve $\mathcal{O}($0.5\%$)$ accuracy, taking advantage of the additional 17\,fb$^{-1}$ collected in the past years at the CM energy of 3.77\,GeV, increasing thereby the statistics by a factor of seven with respect to the 2015 result. The BESIII collaboration intends to improve the accuracy in a staged approach with the first result to be expected in 2025.
New analysis techniques are presently being worked out, most importantly a selection without the explicit detection of the ISR photon. This is possible by constraining the event kinematics to the production of a $\pi^+\pi^-$ pair and one (undetected) photon in a 1C kinematic fit. 
As mentioned above, this method is largely insensitive to the possible limitations of the \phokhara{} event generator thanks to an increased acceptance of events with additional (soft) photons emitted.
The updated analysis strategy is applied to the 2.9\,fb$^{-1}$ at 3.77\,GeV considered in the 2015 analysis and to a data set of 3.1\,fb$^{-1}$ at 4.18\,GeV, to produce an intermediate result with normalization to luminosity and a target precision of 0.7\%. This measurement represents an important cross-check of the published measurement in terms of both analysis strategy and background contributions, thanks to the different CM energies of the analyzed data sets.

The ultimate precision of $\mathcal{O}($0.5\%$)$ will be achieved by analyzing the full 20\,fb$^{-1}$ data sample, which opens the avenue to a normalization to $e^+e^-\to \mu^+\mu^-\gamma$ events with the additional advantage that the systematic uncertainties related to the radiator function and the luminosity drop in such an approach. In the intermediate future, corresponding high-precision analyses for the $3\pi$, $4\pi$, and $K^+K^-$ channels are planned.

\begin{figure}[t]
\centering
\includegraphics[height=4.5cm]{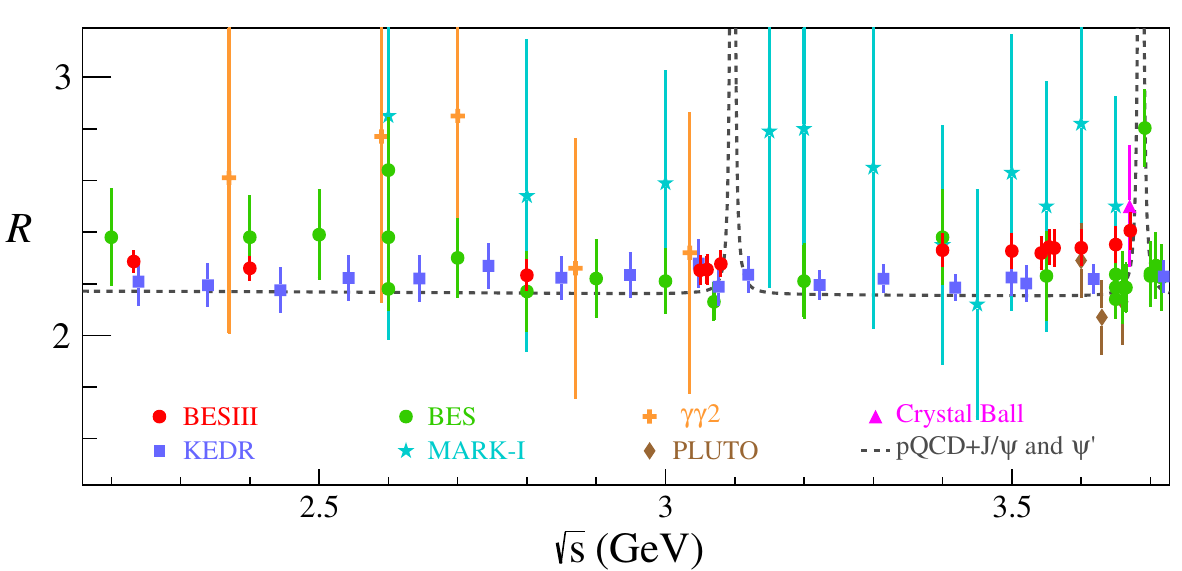}\qquad
\includegraphics[height=4.5cm]{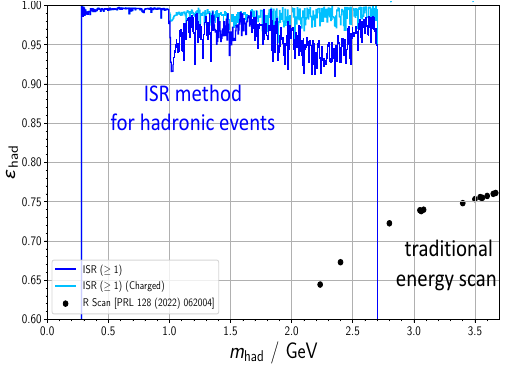}
\caption{Left: Result from the BESIII $R$-value measurement (figure adapted from Ref.~\cite{BESIII:2021wib}). Right: Preliminary selection efficiency for an inclusive measurement of $R$ below 2\,GeV with the ISR method.}
\label{BES3:FigEff}
\end{figure}

Recently, the BESIII Collaboration has published an inclusive $R$-value measurement in the energy region between 2.2 and 3.7\,GeV, based on the energy-scan technique, with unprecedented precision at most of the considered energy points~\cite{BESIII:2021wib}.
The analysis employs, as first in this kind of measurements, two independent event generators for simulating the hadron production: a fully theoretical model based on the Lund model~\cite{Andersson:1997xwk}, called LUARLW~\cite{Andersson:1999ui}, and the hybrid generator~\cite{Ping:2016pms}, which makes use of the well-established \phokhara~\cite{Campanario:2019mjh} and the data-driven ConExc~\cite{Ping:2013jka} event generators in combination with LUARLW used only for the remaining channels. The measured $R$-value shows some tensions with perturbative-QCD (pQCD) predictions, reaching a significance of 2$\sigma$ above 3\,GeV, see \cref{BES3:FigEff}(left).
The data sets analyzed in the published measurement represent a small fraction of the scan-data energy points collected by BESIII between 1.8 and 4\,GeV, thus, further results are to be expected in the near future.

The energy region below approximately 2\,GeV is inaccessible for inclusive $R$-value measurements using the traditional energy-scan method at $e^+e^-$ colliders. This is mostly due to the dominant production of two-prong final states, e.g., $\pi^+\pi^-(\pi^0)$ or $ K^+K^-(\pi^0)$, which can hardly be distinguished from the overwhelming QED background (mainly Bhabha events).
Currently, an alternative approach based on the ISR technique is under development at BESIII for an inclusive $R$ measurement below 2 GeV. It is based on the detection of the ISR photon in the central region of the detector and the reconstruction of (at least) one charged particle.
The emission of a hard ISR photon
boosts the hadronic system within the detector acceptance, such that the selection efficiency ($>$95\%) is found to be significantly higher than in traditional energy-scan measurements, see \cref{BES3:FigEff}(right). As a consequence, the final result has a minimal dependence on fully inclusive event generators (e.g., LUARLW), which is currently the largest contribution to the systematic uncertainty in traditional measurements. Moreover, in large-angle radiative events, the background contribution from Bhabha scattering significantly reduces and can be effectively suppressed by simple particle-identification conditions.
Preliminary studies suggest an accuracy of $\mathcal{O}(1\%)$ to be in reach, which will imply a valuable input to the dispersive evaluation of $\amuHVPLO$ as this inclusive method is a novel and independent approach for determining the HVP contribution.

\subsubsection{Belle II}
\label{sec:BelleII}

The \epemtopipigamma measurement at Belle II has adopted the same approach as the original \babar{} analysis~\cite{BaBar:2012bdw} and will use a data set of 427\invfb. ISR events are used to measure the cross section for \eetoxg at the reduced collision energy
\begin{equation}
\sqrt{s'}=m_X\,(X=\pipi, \mumu)\,,
\end{equation}
where $s'$ is given by
\begin{equation}
s' = s\left(1-\frac{2E_{\gamma}^*}{\sqrt{s}}\right)\,,
\end{equation}
and $E_{\gamma}^*$ is the ISR photon energy in the CM frame.
Muons and pions will be identified using the Belle-II particle-identification (PID) detectors and  the ratio of cross sections of the pion pair to the muon pair will be measured.
NLO QED effects and other backgrounds are validated or rejected using kinematic fits with multiple-photon hypotheses.
Control-sample-based efficiency corrections $\left(\epsilon^{\rm data}/\epsilon^{\rm MC}\right)_{X}$ related to different sources $X$ are determined: trigger, PID, tracking, and fit $\chi^2$. Thus, the data efficiency $\epsilon^{\rm data}$ is estimated from the MC efficiency $\epsilon^{\rm MC}$ as
\begin{equation}
    \eff^{\rm{data}} =  \eff^{\rm{MC}}\left(\frac{\eff^{\rm{data}}}{\eff^{\rm{MC}}}\right)_{\rm{trigger}}\left(\frac{\eff^{\rm{data}}}{\eff^{\rm{MC}}}\right)_{\rm{PID}}\left(\frac{\eff^{\rm{data}}}{\eff^{\rm{MC}}}\right)_{\rm{tracking}}\left(\frac{\eff^{\rm{data}}}{\eff^{\rm{MC}}}\right)_{\chi^2}\,.
\end{equation}
To validate the analysis, the \epemtomumugamma cross section will be measured based on the luminosity derived from other QED processes~\cite{Belle-II:2024vuc}.

\begin{figure}[t]
    \centering
    \includegraphics[width=0.49\textwidth]{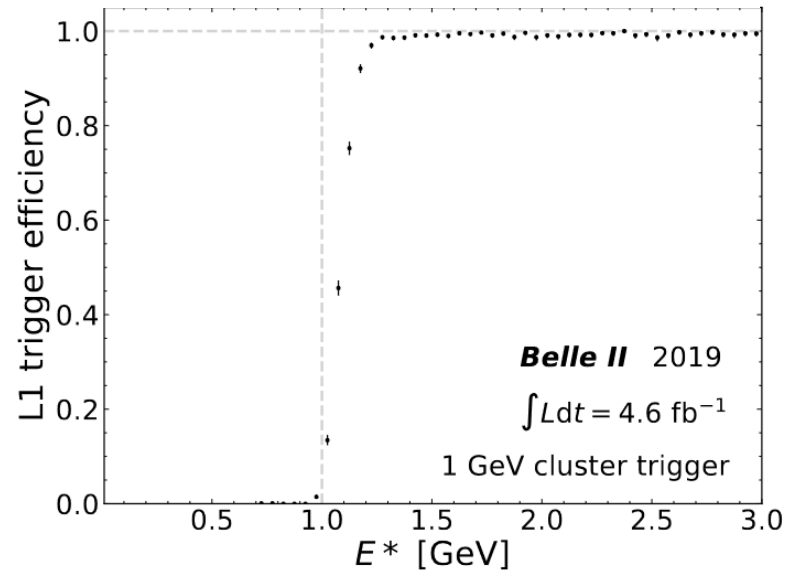}
    \includegraphics[width=0.49\textwidth]{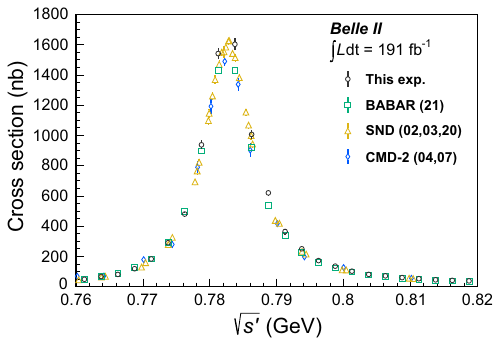}
    \caption{Left: L1 trigger efficiency for high-energy photons as a function of the ISR photon energy in the CM frame. Right: Observed \eetopipipiz cross section as a function of energy compared with previous results at the $\omega$ resonance (figure taken from Ref.~\cite{Belle-II:2024msd}).}
    \label{fig:BelleII} 
\end{figure}

The ISR-trigger efficiency is determined using a sample of \epemtomumugamma events selected by an independent track trigger. \Cref{fig:BelleII}(left) shows the trigger efficiency as a function of $E^{\ast}_{\gamma}$ in the barrel region of the EM calorimeter; the efficiency is 99.9\% for $E^{\ast}_{\gamma}>2~\mathrm{GeV}$.
The kinematic fits and background simulations were validated with a $1.8~\mathrm{fb}^{-1}$ subsample of data. Regarding tracking efficiency, a data-driven method tested with signal MC was developed. Studies of background, PID, and trigger are ongoing.

An \eetopipipiz measurement using 191\invfb Belle II data has been published recently~\cite{Belle-II:2024msd}, with the main result shown in \cref{fig:BelleII}(right). The LO HVP contribution of this analysis to the muon magnetic anomaly is $a_\mu^{3\pi}=48.91(0.23)(1.07)\times 10^{-10}$ in the $(0.62\text{--}1.8)\GeV$ energy range. This is 2.5$\sigma$ larger than the current most precise measurement from \babar~\cite{BABAR:2021cde} and the global fit~\cite{Hoferichter:2023bjm}, see \cref{sec:disp_3pi}.

\subsection{Tau data and isospin-breaking corrections}
\label{sec:tau}

Since the   hadronic weak charged current and the isovector component of the EM current  belong to the 
same  triplet of the isospin group,   semi-leptonic  $\tau$ decays can provide  independent input for the 
computation of  HVP, as first pointed out in Ref.~\cite{Alemany:1997tn}.  
In the limit of exact isospin symmetry, 
a simple kinematic factor relates  
the hadronic invariant mass distributions in  $e^+ e^- \to h$ 
and in $\tau \to \nu_\tau h^\prime$, where $h$ and $h^\prime$ denote 
hadronic states related by an isospin rotation.  
However,  an analysis of  HVP  to sub-percent level requires 
an appropriate set of isospin-breaking (IB) corrections. 
In what follows, we first 
 briefly  review the experimental data sets  (\cref{Sec:DHMZtau}),  
then discuss
the theoretical status and prospects 
for the  IB corrections (\cref{Sect:tauIBcorrections,sec:tau-ChPT,sect:tau-dispersive,sect:tau-lattice,Sect:th-summary}),  
and finally   present  the current HVP evaluation based on $\tau$ data (\cref{Sect:th-summary}).

\subsubsection{Use of hadronic data from \texorpdfstring{$\tau$}{} decays}
\label{Sec:DHMZtau}

\begin{figure}[t] \centering
 \includegraphics[width=0.55\columnwidth]{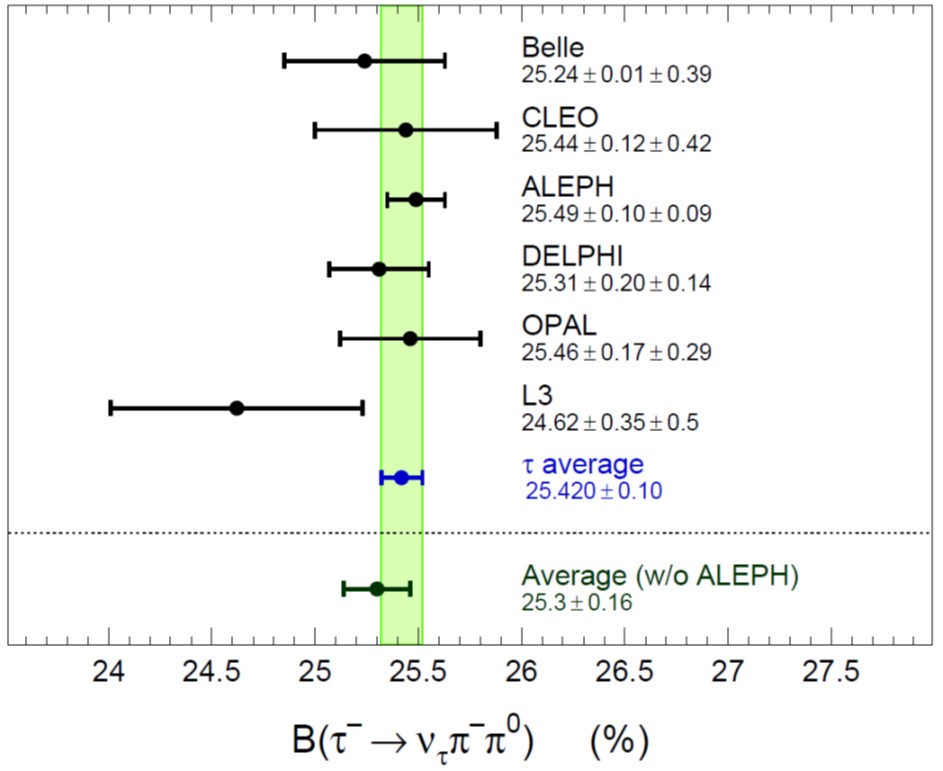}
 \caption{
Measured values of the branching ratio for $\tau\rightarrow \pi \pi^0 (\gamma) \nu_\tau$. Good consistency is observed among the different experiments. Figure adapted from Ref.~\cite{Davier:2010fmf}.
  } 
  \label{fig:BRtau-pipi0} 
\end{figure}

Several high-quality data sets 
for semi-leptonic $\tau$ decays into vector states 
were collected more than 20 years ago at LEP and also at $B$ factories, first by CLEOc, later by Belle and \babar. The available experimental information was very recently reconsidered~\cite{Davier:2023fpl}, in view of the poor consistency among $e^+e^-$ results. 
The  spectral function in the dominant $\tau$ decay mode $\tau^-\to \pi^-\pi^0\nu_\tau$, defined as
\begin{equation}
v_{\pi\pi^0}(s)=\frac{m^2_\tau}{6|V_{ud}|^2}\frac{\mathcal{B}_{\pi\pi^0}}{\mathcal{B}_e}\frac{dN_{\pi\pi^0}}{N_{\pi\pi^0}ds}\times\left[\left(1-\frac{s}{m^2_\tau}\right)^2\left(1+\frac{2s}{m^2_\tau}\right)\right]^{-1}\,,
\end{equation}
has been precisely measured by several experiments~\cite{ALEPH:2005qgp,Davier:2013sfa,OPAL:1998rrm,CLEO:1999dln,Belle:2008xpe} under very different conditions at LEP and the $B$ factories. Here $m_\tau$ is the $\tau$ lepton mass, $|V_{ud}|$ the CKM matrix element, $\mathcal{B}_{\pi\pi^0}$ and $\mathcal{B}_e$ are the branching fractions of $\tau^-\to \pi^-\pi^0\nu_\tau(\gamma)$ (FSR is implied) and of $\tau^-\to e^-\bar{\nu}_e\nu_\tau$, and $dN_{\pi\pi^0}/N_{\pi\pi^0}ds$ is the normalized invariant mass spectrum of the hadronic final state. 
The precision achieved in the experiments for the branching fractions (0.4\%) and the agreement between the different results, as seen in \cref{fig:BRtau-pipi0}, provide a highly precise normalization of the spectral functions, even superior to that obtained in $e^+e^-$ data. There is also good agreement between the spectral function results as shown in Ref.~\cite{Davier:2013sfa}.
These measured spectral functions have been widely used (see, e.g., Ref.~\cite{Davier:2005xq}) for a number of applications including in particular the evaluation of $\amuHVPLO$ and 
$\Delta\alpha^{(5)}_\mathrm{had}$ as originally proposed in Ref.~\cite{Alemany:1997tn}. 
The evaluation of $\amuHVPLO$ 
using the $\tau$ hadronic decay has been valuable in earlier years when the $e^+e^-$ data were not yet precise enough and in recent years given the large discrepancy among the most precise measurements from \babar~\cite{BaBar:2009wpw,BaBar:2012bdw}, CMD-3~\cite{CMD-3:2023rfe,CMD-3:2023alj}, and KLOE~\cite{KLOE:2008fmq,KLOE:2010qei,KLOE:2012anl,KLOE-2:2017fda}.
In order to achieve the required  precision in the $\tau$-based evaluation of $\amuHVPLO$,   
IB corrections have to be understood and  applied---a topic that   we discuss in \cref{Sect:tauIBcorrections,sec:tau-ChPT,sect:tau-dispersive,sect:tau-lattice,Sect:th-summary}.

\subsubsection{Theoretical input  for the HVP analysis based on \texorpdfstring{$\tau$}{} data: generalities}
\label{Sect:tauIBcorrections}

We  focus on the dominant   $\tau\to\pi\pi\nu_\tau (\gamma)$ 
 channel and  denote with $s$ the $\pi \pi$ invariant mass squared. 
 The   photon-inclusive  differential decay 
spectrum   $d\Gamma_{\pi\pi(\gamma)}/ ds$ 
 can be used to evaluate $\amuHVPLO[\pi\pi]$ according to the following dispersive
  formula~\cite{Cirigliano:2001er,Cirigliano:2002pv,Davier:2010fmf} (with threshold $s_\text{thr}=4M_{\pi^\pm}^2$)
 \begin{equation}\label{eq.amu_dispersive}
    \amuHVPLO[\pi\pi, \tau]=\frac{1}{4\pi^3}\int_{s_{\text{thr}}}^{\infty}ds \,K(s)\,\left[\frac{K_\sigma(s)}{K_\Gamma(s)}\frac{d\Gamma_{\pi\pi[\gamma]}}{ds}\right]\times\frac{R_{\text{IB}}(s)}{S_{\text{EW}}^{\pi\pi}}\,,
\end{equation}
where $K(s)$ is the QED kernel \cite{Bouchiat:1961lbg,Brodsky:1967sr,Lautrup:1969fr,Gourdin:1969dm}, see \cref{eq:K_def} for the explicit expression,   
\begin{equation}
     K_{\sigma}(s)=\frac{\pi \alpha^2}{3s}\,,  \qquad 
    K_\Gamma(s)= \frac{\Gamma_e \vert V_{ud}\vert^2}{2 m_\tau^2} \,\left(1-\frac{s}{m_\tau^2}\right)^2\left(1+\frac{2s}{m_\tau^2}\right)\,, 
\label{eq:kintau}
\end{equation}
and the IB  corrections are encoded in the product of several $s$-dependent factors
\begin{equation}\label{RIB}
    R_{\text{IB}}(s)=\frac{\text{FSR}(s)}{G_{\text{EM}}(s)}\frac{\beta^3_{\pi^{+}\pi^{-}}(s)}{\beta^3_{\pi^\pm \pi^0}(s)}\left\vert\frac{F_\pi^V(s)}{f_{+}(s)}\right\vert^2\,.     
\end{equation}
The physical origin  and meaning of the various IB corrections can be summarized as follows: 
\begin{itemize}

\item  The factor $S_\text{EW}^{\pi\pi} = S_\text{EW} / S_\text{EW}^{\rm sub, lept}$~\cite{Davier:2002dy}  is  the $s$-independent  short-distance EM correction to the weak semi-leptonic  decay  $\tau \to \pi \pi \nu_\tau$. 
The numerator 
\begin{equation}
S_\text{EW}  = 1 + \frac{2 \alpha}{\pi}\log \frac{M_Z}{m_\tau}    + \cdots
\end{equation}
encodes the universal  leading logarithmic correction that affects all semi-leptonic  decays~\cite{Sirlin:1981ie,Marciano:1985pd,Marciano:1988vm,Marciano:1993sh,Braaten:1990ef,Erler:2002mv,Cirigliano:2023fnz} 
and is now known to next-to-leading logarithmic (NLL) accuracy~\cite{Erler:2002mv,Cirigliano:2023fnz}. 
The denominator    
\begin{equation}
S_\text{EW}^\text{sub, lept} =  1+\frac{\alpha(m_\tau)}{2\pi}\bigg(\frac{25}{4}-\pi^2\bigg)
\end{equation}
encodes the radiative corrections to the  purely leptonic decay width  
\begin{equation}
 \Gamma_e \equiv \Gamma (\tau \to e \nu_\tau \bar \nu_e) = \frac{G_F^2 m_\tau^5}{192 \pi^3} \times \bigg[1 + {\mathcal O}\bigg(\frac{m_e^2}{m_\tau^2}\bigg)\bigg]    \times S_\text{EW}^\text{sub, lept}\,,   
\end{equation}
and needs to be included because   the measurements provide  $\tau \to \pi \pi \nu_\tau$ 
normalized to the leptonic mode.  This is made explicit by expressing $K_\Gamma (s)$ in terms of $\Gamma_e$ in \cref{eq:kintau}.

 \item The $G_{\text{EM}}(s)$ factor \cite{Cirigliano:2001er,Cirigliano:2002pv,Flores-Baez:2006yiq,Miranda:2020wdg} includes the long-distance QED corrections to the $\tau^-\to\pi^-\pi^0\nu_\tau$ decay with virtual- plus real-photon radiation. 
 The ${\rm FSR}(s)$  factor, first  introduced in $R_{\text{IB}}$ in Ref.~\cite{Davier:2010fmf}, 
 originates from   FSR corrections to the $\pi^+\pi^-$ channel~\cite{Schwinger:1989ix,Drees:1990te}.  
 Both FSR$(s)$ and $G_\text{EM}(s)$ are so far computed at ${\mathcal O}(\alpha)$ and with the help of hadronic models as specified below. 
However, the scheme-dependent  nonlogarithmic term in $G_\text{EM}(s)$  is not currently known.  In both cases, \cref{RIB} implies a factorization assumption, separating long-range radiative effects subsumed in FSR$(s)$ and $G_\text{EM}(s)$ from radiative corrections to the matrix elements, see \cref{sec:radiative_corrections} for the case of $F_\pi^V(s)$.

 \item The $\pi^\pm$--$\pi^0$ mass difference generates a mismatch in the phase-space factors resulting in 
 the correction term  $\beta^{3}_{\pi^-\pi^+}(s)/\beta^{3}_{\pi^-\pi^0}(s)$,  
 with $\beta_{PP'}(s) = 2 p_{\rm cms} (s)/\sqrt{s}$, where  $p_{\rm CM}$  is the magnitude of the pion momentum  in the dipion CM frame. 
 This kinematic correction is large for $s$ near threshold and carries negligible uncertainty. 
 
 \item The last factor in \cref{RIB} amounts to the ratio between the EM  and the weak  pion form factors $F_\pi^V(s)$ and $f_{+}(s)$, respectively,  with IB only due to quark mass difference and to radiatively induced hadronic mass differences. 
Currently, this is the correction that carries the largest uncertainty, in particular because it is difficult to characterize in a model-independent way. 

\end{itemize} 

The shift in $a_\mu$ caused by the IB corrections, denoted by $\Delta \amuHVPLO[\pi\pi, \tau]$, is obtained by replacing  
$R_{\text{IB}}(s)/S_{\text{EW}} \to R_{\text{IB}}(s)/S_{\text{EW}} -1$ in \cref{eq.amu_dispersive}.

We begin by discussing  the impact of the $s$-independent short-distance corrections,  which are not sensitive to  hadronic structure. 
We use $S_\text{EW}^{\pi \pi}=1.0233(3)$~\cite{Castro:2024prg},  obtained by  including the NLL correction  from Ref.~\cite{Cirigliano:2023fnz} and estimating the uncertainty to be of the order of missing nonlogarithmic terms of ${\mathcal O}(\alpha \alpha_s/\pi^2)$. 
The corresponding shift to  HVP induced by  $S^{\rm \pi \pi}_\text{EW}$  is 
$\Delta \amuHVPLO[\pi\pi, \tau]=  -12.16(15) \times 10^{-10}$~\cite{Castro:2024prg}, consistent with the 
result of Ref.~\cite{Davier:2023fpl}.
The above results do not account for the uncertainty in $S_\text{EW}^{\pi \pi}$ associated to the scheme dependence of the short-distance Wilson coefficient at NLL.   We discuss this in \cref{Sect:th-summary}. 
Next,  the phase-space correction  generates the shift $\Delta \amuHVPLO[\pi\pi, \tau] =  -7.52 \times 10^{-10}$, with negligible intrinsic uncertainty, but the result does depend on the input for the spectral function, thus the small difference to Ref.~\cite{Davier:2023fpl}.

In the remainder of this section,  we discuss the status and prospects for computing the remaining  hadron-structure-dependent  corrections in various approaches: 
first, we summarize the current results based on a combination of ChPT and resonance hadronic models. 
This approach  produces the corrections used to arrive at the current $\tau$-based results for the HVP contribution in \cref{Sect:th-summary}.
We then discuss ongoing efforts based on dispersive methods and lattice QCD.

\subsubsection{Chiral perturbation theory and phenomenological  models}
\label{sec:tau-ChPT}

We summarize here the  status of  IB  corrections 
obtained through a combination of  ChPT,  resonance chiral theory (R$\chi$T), and resonance models for radiative corrections 
and for  the pion VFF. 
This approach has been refined over the years, starting from the early work found in Refs.~\cite{Alemany:1997tn,Cirigliano:2001er,Cirigliano:2002pv,Davier:2002dy,Flores-Baez:2006yiq,Flores-Tlalpa:2006snz,Flores-Baez:2007vnd,Davier:2010fmf}. 
The  state-of-the-art results of this approach have been recently presented in Refs.~\cite{Davier:2023fpl,Castro:2024prg}. 
While differing in some details, these two references reach mutually consistent results, as shown below in Table~\ref{tab:summary_tau_IB}.
For the sake of simplicity, in this section we closely follow the  analysis of  Ref.~\cite{Castro:2024prg}, pointing out differences 
from Ref.~\cite{Davier:2023fpl} as needed 
(a comparison with lattice-QCD results in Euclidean-time windows, see \cref{sec:windows}, was reported in Ref.~\cite{Masjuan:2023qsp}).

The correction induced by  the function  $G_\mathrm{EM}(s)$  
 is    $\Delta \amuHVPLO[\pi\pi, \tau] =   (-1.67)^{+0.60}_{-1.39} \times 10^{-10}$~\cite{Castro:2024prg}, which corresponds to the reference value reported in Ref.~\cite{Miranda:2020wdg} and is in good agreement with results obtained in Ref.~\cite{Davier:2010fmf} using Refs.~\cite{Cirigliano:2002pv,Flores-Baez:2006yiq}. 
This result updates the computation in Ref.~\cite{Cirigliano:2002pv}, which used  R$\chi$T~\cite{Ecker:1988te, Ecker:1989yg} and included  
only those resonance operators that---upon integrating out the resonances---contribute to the $\mathcal{O}(p^4)$ chiral low-energy constants (LECs). 
The uncertainty of the updated result is estimated by assessing the spread of results obtained  by 
 including a subset of  resonance couplings that contribute to the 
chiral LECs  at $\mathcal{O}(p^6)$ \cite{Cirigliano:2006hb, Kampf:2011ty},  precisely those couplings that are 
determined  by short-distance QCD constraints (ensuring  an appropriate asymptotic behavior of suitable two- and three-point Green functions). 
This  analysis concerns entirely the effects of structure-dependent real-photon emission ($\tau \to \nu_\tau \pi \pi \gamma$) 
and  lacks the structure-dependent virtual-photon contribution. This missing piece, estimated as in Ref.~\cite{Escribano:2023seb} based on the results for the one-meson modes \cite{Arroyo-Urena:2021nil, Arroyo-Urena:2021dfe}, is covered by the quoted uncertainty. Measurements of the $\tau\to\pi\pi\nu_{\tau}\gamma$ spectrum would greatly  help   reduce the model dependence of the corrections induced by  $G_\text{EM} (s)$. 
The FSR$(s)$ correction is based on sQED, producing a shift of $\Delta a_{\mu}^{\text{HVP, LO}}[\pi\pi, \tau] =  4.62(46) \times 10^{-10}$, where missing structure-dependent virtual- and real-photon corrections are assigned a 10\% uncertainty~\cite{Davier:2010fmf,Castro:2024prg}.

The remaining IB corrections enter the ratio of form factors in \cref{RIB}.
The modifications affecting $F_\pi^V(s)/f_{+}(s)$ can be split into the following sources:
\begin{itemize}

\item The $\rho$--$\omega$ and---to a much smaller extent---the $\rho$--$\phi$ mixing 
(leading to $\Delta \amuHVPLO[\pi\pi,\tau]= 2.87(8) \times 10^{-10}$~\cite{Castro:2024prg}). 

\item The $M_{\pi^\pm}$--$M_{\pi^0}$ difference, affecting the $\rho^{\pm,0}$ widths. An analogous mass difference for the kaons is considered in the Guerrero--Pich (GP) form factor \cite{Guerrero:1997ku} (and other chiral-based descriptions  \cite{GomezDumm:2013sib,Gonzalez-Solis:2019iod}), but not in the Gounaris--Sakurai (GS) \cite{Gounaris:1968mw} or K\"uhn--Santamar\'ia (KS) \cite{Kuhn:1990ad} models 
(leading to $\Delta \amuHVPLO[\pi\pi,\tau]=  3.37  \times 10^{-10}$~\cite{Castro:2024prg}).

\item The $M_{\rho^\pm}$--$M_{\rho^0}$ difference. For given $\rho^{\pm,0}$ mass values, the dominant on-shell $\rho^{\pm,0}\to \pi\pi$ widths are predictions in GP and related parameterizations, once short-distance QCD constraints \cite{Ecker:1988te, Ecker:1989yg} have been accounted for 
(leading to $\Delta \amuHVPLO[\pi\pi,\tau]= 1.95^{+ 1.56}_{-1.55}  \times 10^{-10}$~\cite{Castro:2024prg}).

\item The  contributions to the $\rho^{\pm}$--$\rho^ 0$ width difference dominated by their $\pi\pi(\gamma)$ decay channels~\cite{Flores-Baez:2007vnd}, where a $10\%$ error is assigned to structure-dependent uncertainties~\cite{Castro:2024prg, Davier:2010fmf, Flores-Baez:2007vnd} 
(leading to $\Delta \amuHVPLO[\pi\pi,\tau]= -6.66(73) \times 10^{-10}$~\cite{Castro:2024prg}). 
We note that the quoted references did not include IB in the $\rho\pi\pi$ coupling.
The origin and possible size of this correction is discussed in \cref{Sect:th-summary}.
\end{itemize}

These corrections were evaluated using a number of different parameterizations in Ref.~\cite{Castro:2024prg}, taking inputs for IB in the $\rho$ masses and widths consistent with Ref.~\cite{Davier:2010fmf} and the PDG \cite{ParticleDataGroup:2024cfk}. Based on analyticity tests, the dispersive form factor based on Ref.~\cite{Colangelo:2018mtw} is  taken as the reference result. The central value is obtained from the dispersive result in the case where it has a conformal polynomial of fourth degree (constrained to comply with $P$-wave behavior) accounting for inelasticities. This yields $\Delta \amuHVPLO[\pi\pi,\tau]= 1.53^{+1.74}_{-1.73} \times 10^{-10}$ for the form factor correction~\cite{Castro:2024prg}. The sum of all IB  corrections entering \eqref{eq.amu_dispersive} gives $\Delta \amuHVPLO[\pi\pi, \tau]=(-15.20)^{+1.90}_{-2.27}\times 10^{-10}$. The associated systematic uncertainties are estimated from the differences among the various dispersive results, and adding linearly a $2\%$ uncertainty to account for the small difference with respect to the GS value, leading to $\Delta \amuHVPLO[\pi\pi,\tau]=(-15.20)^{+2.26}_{-2.63}\times10^{-10}$~\cite{Castro:2024prg}. 

Finally, we comment on $\rho$--$\gamma$ mixing.  
First, we note that for a proper treatment of the $\rho$ one just needs to correctly describe the resonant behavior of pion form factors, either of the weak or the EM currents. Gauge invariance, leading to the constraint $F_\pi^V(0)=1$, as is respected in any theoretically sound approach, ultimately settles the issue of a possible $\rho$--$\gamma$ mixing, since the $\rho$ is never needed as an asymptotic state. This is explicit in a dispersive approach, since the $\rho$ does not even enter as a degree of freedom, but also holds in any Lagrangian framework  
that couples hadrons to external sources representing the currents (this is the case in ChPT,  R$\chi$T, and any resonance model made consistent with chiral symmetry).
The $\rho$--$\gamma$ mixing issue may arise in intermediate steps of calculations within a  given Lagrangian model. 
For example, in Ref.~\cite{Jegerlehner:2011ti}   the effect of such a mixing was studied within the hidden-local-symmetry model. 
In QFT the mixing of two fields can be addressed through a field redefinition. 
Since physical results should not depend on the field parameterization and given that 
the $\rho$ meson is not an asymptotic state, performing the field redefinition that diagonalizes the $\rho$--$\gamma$ quadratic terms is a possible, but far from necessary choice. In most approaches it turned out to be more convenient not to perform this diagonalization. However, if one performs the redefinition, the re-defined ``$\rho$'' field couples to all charged particles, 
and the resulting $\rho ee$ and $\rho \mu \mu$ couplings should be taken into account, 
given that they affect the leptonic QED cross section used to normalize $\sigma (e^+ e^- \to \pi^+ \pi^-)$.  
This effect and its interplay with higher-order corrections  was not discussed in Ref.~\cite{Jegerlehner:2011ti}. 
Moreover,  the large mixing corrections at  $\sqrt{s} \gg M_\rho$  and the large slope at $\sqrt{s} \ll M_\rho$,  
leading to sizable numerical  effects  in Ref.~\cite{Jegerlehner:2011ti}, 
seem to arise from the fact that the pion VFF adopted does not incorporate the basic  short-distance asymptotic conditions imposed by QCD (i.e., vanishing at high momentum transfer), a constraint whose relevance even at low energy has been thoroughly discussed in Ref.~\cite{Ecker:1989yg}.

\subsubsection{The dispersive approach}
\label{sect:tau-dispersive}

\begin{figure}[tb]
    \centering
    \begin{minipage}{0.3\linewidth}
       \includegraphics[width=\linewidth]{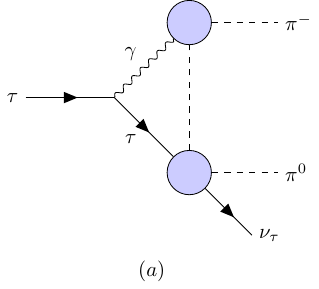} 
    \end{minipage}\hspace{1cm}
    \begin{minipage}{0.3\linewidth}
       \includegraphics[width=\linewidth]{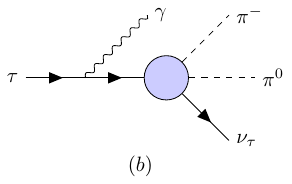} 
    \end{minipage}\\
    \begin{minipage}{0.3\linewidth}
       \includegraphics[width=\linewidth]{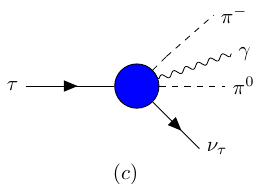} 
    \end{minipage}\hspace{1cm}
    \begin{minipage}{0.3\linewidth}
       \includegraphics[width=\linewidth]{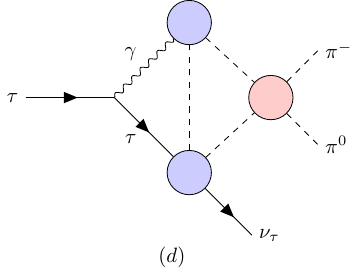} 
    \end{minipage}
    \caption{Virtual $(a)$ and real $(b)+(c)$ contributions to the radiative corrections to $\tau^-\to\pi^-\pi^0\nu_\tau$ in the dispersive framework at first order in IB. Light-blue blobs indicate the pion VFF. Diagram $(c)$ includes bremsstrahlung off the charged pion, but the blue blob represents the general matrix element~\cite{Bijnens:1992en}. A rescattering correction is shown in $(d)$, where the light-red blob refers to the $\pi\pi$ scattering amplitude.}
    \label{fig:tau_dispcont_diags}
\end{figure}

To first order in IB the photon-inclusive differential decay rate for $\tau^-(l_1) \to \pi^-(q_1) \pi^0(q_2) \nu_\tau(l_2) [\gamma(k)]$ is given by~\cite{Cirigliano:2001er,Cirigliano:2002pv}
\begin{equation}
    \frac{d\Gamma_{\pi\pi[\gamma]}}{d s} =  S_{\rm EW}^{\pi \pi} \ K_\Gamma (s) \ \beta^3_{\pi^\pm \pi^0} (s) \ \big|f_+(s) \big|^2 \ G_\text{EM}(s) \, ,
\end{equation}
where  $s=(q_1+q_2)^2$ and  $f_+(s)$ is the $\pi^-\pi^0$ form factor. 
$G_\text{EM}(s)$ the EM correction factor defined as:
\begin{equation}
\label{Eq:GEM_def}
    G_\text{EM}(s) = \frac{\int_{t_\text{min}(s)}^{t_\text{max}(s)} dt\, D(s,t) \big[1+2f_\text{elm}^\text{loop}(s,t)+g_\text{rad}(s,t)\big]}{\int_{t_\text{min}(s)}^{t_\text{max}(s)} dt\, D(s,t)} \, ,
\end{equation}
with Mandelstam variable $t=(l_1-q_1)^2$, $t_\text{min,max}(s)$ its lower, upper phase-space borders, and 
\begin{equation}
    D(s,t)=\frac{m_\tau^2}{2}(m_\tau^2-s) + 2 M_\pi^2 M_{\pi^0}^2 - 2 t (m_\tau^2 - s + M_\pi^2 + M_{\pi^0}^2) + 2 t^2 \, .
\end{equation}
The functions $f_\text{elm}^\text{loop}(s,t)$ and $g_\text{rad}(s,t)$ in \cref{Eq:GEM_def} account for virtual and real corrections, respectively. While previously these functions were calculated in ChPT or R$\chi$T~\cite{Cirigliano:2001er,Cirigliano:2002pv,Miranda:2020wdg,Castro:2024prg}, in this section we outline progress in a dispersive approach~\cite{Colangelo2025}. 
First, the approximation of only taking into account intermediate hadronic states up to two pions is justified phenomenologically: in the energy region and in the channel of interest inelastic effects are known to be small. Radiative corrections to them are therefore considered to be negligible. In this approximation the contribution to $f_\text{elm}^\text{loop}$ in the dispersive framework is depicted in \cref{fig:tau_dispcont_diags}$(a)$, where the light-blue blobs denote the pion VFF.

In analogous fashion to how radiative corrections to $e^+e^- \to \pi^+\pi^-$ are handled in a dispersive approach in Ref.~\cite{Colangelo:2022lzg}, a dispersive representation is used for the pion VFF in the unitarity diagram of \cref{fig:tau_dispcont_diags}$(a)$,
\begin{equation}
\label{Eq:tau_pvff_rep}
    F_\pi^V(s) =1+ \frac{s}{\pi}\int_{4M_\pi^2}^{\infty} ds'\, \frac{\operatorname{Im} F_{\pi}^V(s')}{s'(s'-s-i\epsilon)} = \frac{1}{\pi}\int_{4M_\pi^2}^{\infty} ds'\, \frac{\operatorname{Im} F_{\pi}^V(s')}{s'-s-i\epsilon} \, .
\end{equation}
In contrast to the approach in Ref.~\cite{Colangelo:2022lzg}, an unsubtracted dispersion relation for the VFF (with an implicit sum rule for the value at $s=0$) is chosen to ensure that the resulting representation of the triangle diagram is manifestly UV finite. However, IR singularities remain: these are handled through dimensional regularization and cancel with the IR-divergent interference of real contributions in \cref{fig:tau_dispcont_diags}$(b)$~and~$(c)$. In this approach the effects of the VFF inside the loop integral are rigorously accounted for, an effect that was shown to be essential to describe the charge asymmetry in $e^+e^-\to\pi^+\pi^-$ due to resonance enhancement of the IR-finite remainder~\cite{Ignatov:2022iou,Colangelo:2022lzg}. In previous model-based approaches such effects have only been crudely approximated.

Since the use of an unsubtracted dispersion relation for the pion VFF introduces sensitivity to its high-energy behavior, a matching procedure to the low-energy behavior calculable in ChPT is essential to mitigate this effect. Taking the virtual corrections in ChPT including virtual photons and leptons~\cite{Weinberg:1968de,Gasser:1983yg,Gasser:1984gg,Urech:1994hd,Knecht:1999ag} at $\mathcal{O}(e^2p^2)$~\cite{Cirigliano:2001er,Cirigliano:2002pv}, denoted by $f_\text{elm}^{\text{loop, ChPT}}$, the matched virtual contribution can be expressed as
\begin{equation}
\label{matching}
    f_\text{elm}^{\text{loop, match}}(s,t) = f_\text{elm}^{\text{loop, VFF}}(s,t) - f_\text{elm}^{\text{loop, VFF}}(0,0) + f_\text{elm}^{\text{loop, ChPT}}(0,0)\, ,
\end{equation}
where $f_\text{elm}^{\text{loop, VFF}}(s,t)$ refers to the dispersive result based on the unsubtracted dispersion relation. While $f_\text{elm}^{\text{loop, VFF}}(s,t)$ carries an IR divergence canceled by the interference of ISR and FSR, the term $f_\text{elm}^{\text{loop, ChPT}}(0,0)$ covers the IR divergences of the remaining contributions, e.g., self-energy corrections, which are canceled by real-photon emission of either the initial or final state.
The matching to ChPT in \cref{matching} entails several further advantages: first, the chiral logarithms only partially captured by the low-energy limit of the dispersive calculation of \cref{fig:tau_dispcont_diags}$(a)$ are fully restored. Second, the chiral LECs define a convenient bridge to the short-distance amplitude, as their scale dependence needs to cancel in physical observables. While previously only resonance estimates of their finite parts were available, the matching to lattice-QCD calculation~\cite{Feng:2020zdc,Ma:2021azh,Yoo:2023gln} and incorporation of RG corrections~\cite{Braaten:1990ef,Cirigliano:2023fnz} now presents the opportunity to define a consistent set of correction factors $S_\text{EW}$ and $G_\text{EM}(s)$, such that  the scale dependence disappears up to a given order in the perturbative expansion.

Work along these lines is ongoing and should provide improved results for the combination of $S_\text{EW}$ and $G_\text{EM}(s)$ with controlled uncertainties.  Corrections include higher intermediate states and $\pi\pi$ rescattering corrections as shown in \cref{fig:tau_dispcont_diags}$(d)$, which will also be addressed in a dispersive approach. For a complete estimate of IB corrections two main classes of additional corrections need to be considered:
\begin{enumerate}
    \item FSR in $e^+e^-\to\pi^+\pi^-$ and $\rho$--$\omega$ mixing: these corrections can be determined from dispersive analyses of $e^+e^-\to\pi^+\pi^-$ and are therefore already under reasonable control~\cite{Colangelo:2018mtw,Colangelo:2022prz};
    \item $2\pi$ matrix element in charged and neutral current: these corrections can be expressed as a ratio of form factors 
    $|f_+(s)/F_\pi^V(s)|$,
     where $f_+(s)$ refers to $\pi^-\pi^0$ system and 
     $F_\pi^V(s)$
      to the $\pi^+\pi^-$ one. 
\end{enumerate}
While frequently expressed in terms of $\rho$ Breit--Wigner parameters and couplings, leading to model errors that are difficult to control,
from a dispersive perspective the second class of corrections should be interpreted as IB in the respective QCD matrix elements. Accordingly, a better understanding 
of IB in the pion VFF is ultimately required, including pion-mass-difference effects and radiative corrections. Work along these lines was started in Ref.~\cite{Monnard:2021pvm} and is in progress, see also \cref{sec:radiative_corrections}.

\subsubsection{Lattice-QCD approach} 
\label{sect:tau-lattice}

As discussed above, addressing IB effects in the $\pi^-\pi^0$ decay mode of the $\tau$ lepton from first principles is highly desirable. In this context lattice QCD+QED simulations can provide useful information, as we briefly review below, see also \cref{sec:latticeHVP}, in particular \cref{sec:tau_IB_lattice}.
The first difficulty that one encounters is in the exclusivity of the mode. In fact, even in the isosymmetric limit of QCD, providing the two-pion contribution to the spectral density is a challenging problem for a lattice calculation, due to the rotation to the Euclidean metric. The latter is by-passed in the limited elastic region (below the four-pion threshold) via the finite-volume formalism, while it requires inverse Laplace methods in the inelastic regime, methods that are currently being intensively developed in the community (see, e.g., Refs.~\cite{Hansen:2017mnd,Hansen:2019idp,Bailas:2020qmv,Lupo:2023qna,Bergamaschi:2023xzx,Bruno:2024fqc}). IB effects constitute another layer of difficulty, so for these reasons in Ref.~\cite{Bruno:2018ono} it was first proposed to study the problem in a fully inclusive manner.

\begin{figure}[t]
    \centering
\begin{minipage}{0.3\linewidth}
        \includegraphics[height=3.7em]{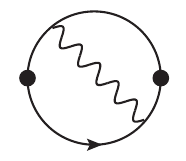}
        \includegraphics[height=3.7em]{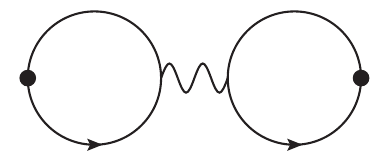}
\end{minipage}\hspace{.75cm}
\begin{minipage}{0.3\linewidth}
        \includegraphics[height=4em]{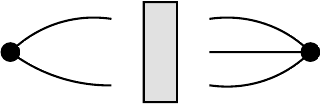}
\end{minipage}\hspace{.6cm}
\begin{minipage}{0.3\linewidth}    
        \includegraphics[height=5.em]{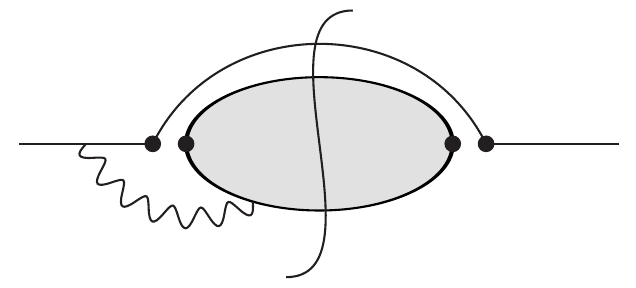}
\end{minipage}        
    \caption{Left: Diagrams contributing to the $\rho$ mass shift (figure adapted from Ref.~\cite{RBC:2018dos}). The black dots represent the external isovector currents, and only quark lines are drawn. The exchange of gluons between the two disconnected bubbles in implicitly assumed for the second diagram. Center: Phenomenological interpretation of the two-point isoscalar--isovector correlator, namely $G_{01}^\gamma$, with 2- and 3-pion states coupled by an IB operator. 
    Right: An example of a mixed contribution where the photon is exchanged between the leptonic and hadronic sectors. The two horizontal external lines represent the $\tau$ lepton, the upper line the neutrino, and the gray blob the hadronic system. The vertical curvy line denotes a final-state cut, while the black dots represent the four-Fermi interaction vertices. \label{fig:mixed}}
\end{figure}

More specifically, the difference of the inclusive isovector spectral densities, charged and neutral, was examined. Since their relation to Euclidean correlators proceeds via a Laplace transform, e.g., for the HVP contribution to $a_\mu$ one starts from
\begin{equation}
    C(x_0) = \int d\omega \, e^{-\omega x_0} \, \rho(\omega^2) \, \omega^2\,,
\end{equation}
with $\rho(\omega^2)$ trivially related to the experimental $R$-ratio, the corresponding difference of two-point Euclidean correlators, denoted by $G_{11}^W-G_{11}^\gamma$, is examined in 
\cref{sec:tau_IB_lattice}.  Thanks to large cancellations it was shown that it is of purely EM nature and only the diagrams in \cref{fig:mixed}(left) contribute. As pointed out in Ref.~\cite{Bruno:2024tau}, this difference could be used as an intermediate quantity to address some open questions on the model dependence of some effects in $R_\text{IB}(s)$, as defined in \cref{RIB}. In simple terms, starting from a model of the charged and neutral pion form factor, after a Laplace transform, one could fit their (squared) difference to the lattice correlators to fix a few of their free parameters.
More specifically, despite receiving EM corrections originating also from four-pion states (due to the inclusive nature of the Euclidean correlators considered), in analogy to the charged-neutral pion mass correction, the aforementioned difference may be used to fix the shift of the $\rho$ mass parameter induced by EM interactions. An attractive feature is that being a property of the spectrum, this would not require the renormalization of the charged current.

While the latter does not constitute per se a conceptual obstacle  for a lattice calculation, its matching from a scheme suitable for lattice calculations to the typical schemes used to match the EFT without the weak bosons to the SM, e.g., $\mathrm{\overline{MS}}$, is still missing. As a consequence this still prevents the usage of lattice data to fix other model-dependent quantities such as the EM shift of the $\rho$ decay width,  a rather important effect as outlined in \cref{tab:summary_tau_IB}, which should be prioritized.

Another interesting object is the two-point correlator involving the isoscalar and isovector neutral currents, denoted as $G_{01}^\gamma$ in  \cref{sec:tau_IB_lattice}.
From a phenomenological point of view, it contains the mixing between two- and three-pion states and therefore information on the $\rho$--$\omega$ mixing parameters, as sketched in \cref{fig:mixed}(center). While on  the one hand this could be used again to fix a few model-dependent parameters, its inclusive nature shows the difficulty in using it for a direct estimate of the $2\pi$ IB effects alone, since in practice one would have to account and remove the IB effects of the three-pion channel, and neglect the effects of even higher multiplicity channels.

The remaining term that could be ideally addressed in a lattice calculation is the $G_\text{EM}$ function. The current strategy so far pursued in Ref.~\cite{Bruno:2018ono} consists of borrowing the available knowledge on $G_\text{EM}$ from ChPT~\cite{Cirigliano:2001er, Cirigliano:2002pv}, since a complete lattice calculation would require a study of the necessary triangle diagrams, see \cref{fig:mixed}(right), from Euclidean space--time, where problems with analytic continuation are present. Developments in this direction are currently being pursued for simpler quantities, but progress is certainly to be expected as these methods become more and more mature. From the short-distance perspective performing the entire calculation using the lattice regulator would presumably simplify the renormalization pattern, leaving only the matching with $S_\text{EW}$ as the open question.

In summary, while an exclusive study of IB effects for the $2\pi$ channel remains a challenging problem for a lattice calculation, an inclusive approach is currently being developed and intermediate quantities, such as the difference of isovector charged and neutral correlators may turn out to be useful, in the short term, to constrain model-dependent parameters.

\subsubsection{Summary on isospin-breaking corrections and \texorpdfstring{$\tau$}{}-based HVP result}
\label{Sect:th-summary}

In this section we summarize  the theoretical IB corrections  needed for a $\tau$-based analysis of HVP, 
discuss the robustness of  the associated uncertainties, and provide recommended values.
\cref{tab:summary_tau_IB}   compiles  results for various IB effects  from recent state-of-the-art analyses,
as discussed in previous subsections, and includes in the last column  ``our estimate,''
based on our assessment  of the  uncertainties.

Before discussing  each line in \cref{tab:summary_tau_IB}, we observe  that estimating uncertainties in this mostly  nonperturbative regime
is not a simple exercise, due to the ensuing model-dependence  of most  results. 
The current main analyses~\cite{Davier:2023fpl,Castro:2024prg}  address this issue  by  
assigning systematic uncertainties associated with the  spread of results obtained with different models. 
We will not repeat   this exercise.  
Rather,  we adopt this approach:  (i)  for most IB corrections, we combine the results  from Refs.~\cite{Davier:2023fpl,Davier:2010fmf,Castro:2024prg,Colangelo:2022prz,Hoferichter:2023sli};  
(ii)  where appropriate,  we identify  
uncertainties  not included in current analyses 
and provide recommended numerical values for them.

\begin{table}[t]
\small
	\centering	\renewcommand{\arraystretch}{1.1}
	\begin{tabular}{lrrrrr}
	\toprule
	& &  Refs.~\cite{Davier:2023fpl,Davier:2010fmf} & Ref.~\cite{Castro:2024prg} & Refs.~\cite{Colangelo:2022prz,Hoferichter:2023sli} & Our estimate\\\midrule
    Phase space  & & $-7.88$ & $-7.52$ & -- & $-7.7(2)$ \\
    $S_{\text{EW}}$ & & $-12.21(15)$  & $-12.16(15)$ & -- &   $-12.2(1.3)$ \\
    $G_\text{EM}$ & & $-1.92(90)$ & $(-1.67)^{+0.60}_{-1.39}$ & -- & $-2.0(1.4)$ \\
    FSR   & &  $4.67(47)$ &  $4.62(46)$ & $4.42(4)$ & $4.5 (3)$ \\
    $\rho$--$\omega$ mixing & &  $4.0(4)$ & $2.87(8)$ & $3.79(19)$ &  $3.9(3)$
   \\\midrule 
     &  $ \Delta M_\rho  $  &     $0.20(^{+27} _{-19})(9)$    & $1.95^{+1.56}_{-1.55}$ & -- &    \\
      &  $\Delta \Gamma_\rho ( \Delta M_\pi )  $  &      $4.09(0)(7)$      & $3.37$ & -- &    \\
            $\frac{F_\pi^V}{f_+}$ (w/o $\rho$--$\omega$)     &  $ \Delta \Gamma_\rho (\pi \pi \gamma)$  &      $ -5.91(59)(48)$      & $-6.66(73)$ & -- &  \\
                              &  $ \Delta \Gamma_\rho ( g_{\rho \pi \pi} )$  &     --     & -- &     --      &   \\
    &   Total     &      $-1.62(65)(63)$      & $(-1.34)^{+1.72}_{-1.71}$ & -- &    $-1.5(4.7)$   
    \\\midrule
    Sum & & $-14.9(1.9)$ &  
       $(-15.20)^{+2.26}_{-2.63}$ &
    -- & 
     $-15.0 (5.1)$
    \\
	\bottomrule
	\renewcommand{\arraystretch}{1.0}
	\end{tabular}
	\caption{Summary of the different classes of IB corrections contributing to 
    $\Delta \amuHVPLO[\pi\pi,\tau]$
    (in units of $10^{-10}$).
    For Refs.~\cite{Davier:2023fpl,Davier:2010fmf}, the second errors due to the difference between the GS and KS models are added linearly to the quadratic sum of all other uncertainties.
    For Ref.~\cite{Castro:2024prg}, the total uncertainty includes an estimate of the systematic uncertainty 
    arising from using different dispersive parameterizations. 
    An additional $2\%$ uncertainty is added linearly to account for the  difference between the result based on the dispersive and the GS parameterizations.
    The entries in the last column  are discussed in the main text.     }
 \label{tab:summary_tau_IB}
\end{table}

\begin{itemize}

\item \emph{Phase space:} The small differences reflect the use of different form-factor  parameterizations. 
We  adopt the midpoint of the results shown in the second and third column of \cref{tab:summary_tau_IB}, 
assigning an uncertainty to cover the full range.   Also note  that higher-order IB corrections could play a role here,  depending on whether  one uses
the  spectra from $\tau$ decay or $e^+ e^-$ data.

\item \emph{Scheme dependence in $S_\text{EW}$:}
As anticipated earlier, current analyses do not account for the ${\mathcal O}(\alpha/\pi)$ scheme dependence of the
Wilson coefficient whose square gives the factor $S_{\rm EW}$.
As is well known, to NLL, in general the Wilson coefficients 
acquire a scheme dependence of ${\mathcal O}(\alpha/\pi)$, to be absorbed by corresponding long-distance matrix elements
($G_\text{EM} (s)$ in this case).
As discussed in   Ref.~\cite{Cirigliano:2023fnz} in the context of $\beta$ decays, 
the current result  $S_\text{EW}^{\pi \pi}=1.0233(3)$ 
is based on  a particular definition of  ``scheme-independent'' NLL Wilson coefficient~\cite{Buras:1989xd} that
effectively factors out the scheme dependence of $\mathcal{O}(\alpha)$ and {\it assumes} it to be reabsorbed in the
LECs entering the long-distance 
correction $G_\text{EM}(s)$. Since currently we have no control over the scheme dependence in $G_\text{EM}(s)$,
we should add this perturbative scheme-dependence uncertainty to $S_\text{EW}$.
Taking the ambiguity in the decay rate to be  $\alpha (m_\tau)/\pi \simeq 0.24\%$  leads to 
$S_\text{EW}^{\pi \pi}=1.0233(3)(24)$, which entails  an additional uncertainty  of $1.3 \times 10^{-10}$ in 
$\Delta \amuHVPLO[\pi\pi,\tau]$, adopted  in \cref{tab:summary_tau_IB}.
Note that if we had assumed an ${\mathcal O}(\alpha/\pi)$ uncertainty at the level of  the Wilson coefficient,
the impact on $S_{\rm EW}$   would double, leading to an ambiguity in  $\Delta \amuHVPLO[\pi\pi,\tau]$ of $2.5 \times 10^{-10}$.

\item  \emph{Structure-dependent virtual corrections in $G_\text{EM}$:}
Combining the  results from Refs.~\cite{Davier:2023fpl,Davier:2010fmf} and  Ref.~\cite{Castro:2024prg} 
leads to the $G_\text{EM}$-induced  correction   $\Delta \amuHVPLO[\pi\pi,\tau] = -2.0(1.0) \times 10^{-10}$.
As discussed in  Ref.~\cite{Castro:2024prg},  
existing results for $G_\text{EM}(s)$  do not include the effect of hadronic structure dependence in the 
virtual-photon corrections to $\tau \to \pi \pi \nu_\tau$---pions are treated as point-like objects.
In Ref.~\cite{Escribano:2023seb} it was argued that  the effect of the missing structure-dependent loops can be  estimated  
by analogy with  the single-meson decay modes  $\tau \to P \nu_\tau$ ($P = \pi, K$),  
for which structure effects have been studied~\cite{Arroyo-Urena:2021nil, Arroyo-Urena:2021dfe}. 
There is, however, one class of diagrams  that may lead to a substantial difference between the corrections to  $\tau \to \pi  \nu_\tau$  and $\tau \to \pi \pi \nu_\tau$. 
Before integrating out the $W$, these take the form of box diagrams, in which two gauge bosons ($\gamma$ and $W$) are exchanged between the lepton and pion lines (after integrating out the $W$ boson this reduces to the diagram in \cref{fig:tau_dispcont_diags}(a)).
In the case of $\tau \to \pi \nu_\tau$, the weak current involves just the pion decay constant, while in the case of $\tau \to \pi \pi \nu_\tau$ 
the weak $\pi \pi$ form factor appears and the box diagram involves the interplay of weak and EM form factors.  
In Refs.~\cite{Ignatov:2022iou,Colangelo:2022lzg} it was shown that such diagrams are sensitive to structure-dependent corrections, as the remainder after the cancellation of IR divergences can be strongly enhanced by the $\rho$ resonance. In fact, the measurement of the forward--backward asymmetry by CMD-3~\cite{CMD-3:2023alj,CMD-3:2023rfe} showed that these effects were critical to explain the data, substantially increasing the size of the asymmetry in the $\rho$ peak, see \cref{TI-CMD3,sec:radiative_corrections}.  
A similar effect could also be present in the virtual corrections to the $\tau$ decay, whose size depends on the strength of the IR enhancement and the interplay with the kernel function in the $a_\mu$ integral, as currently under investigation
(see \cref{sect:tau-dispersive}).  Given the large numerical impact observed in the $e^+e^-$ case, it is prudent to assign  to $G_\text{EM}$ an additional uncertainty,
which is currently hard to estimate.  
As a reference point, we  assign to the structure-dependent virtual-photon corrections a similar uncertainty as the one
emerging from the structure-dependent real-photon corrections, leading to 
the $G_\text{EM}$-induced  correction   $\Delta \amuHVPLO[\pi\pi,\tau] = -2.0(1.0)_\text{real} (1.0)_\text{virtual} \times 10^{-10} =  - 2.0(1.4) \times 10^{-10}$.

\item \emph{FSR effects in $e^+ e^- \to \pi^+ \pi^-$:}
The results reported   in \cref{tab:summary_tau_IB}   are quite consistent and the small differences are  likely due to the form-factor input.
All calculations use sQED as theoretical input. 
The result of Ref.~\cite{Hoferichter:2023sli} includes the estimate from Ref.~\cite{Moussallam:2013una} for the non-IR-enhanced contributions, 
which amounts to  an effect of $\simeq 0.2 \times 10^{-10}$ in  $\Delta \amuHVPLO[\pi\pi,\tau]$  
and shows that in this case sQED indeed captures the dominant effect.
We take as our estimate the average of the sQED results  and assign an uncertainty 
to cover the spread of central values, which is  large enough to cover possible effects beyond sQED~\cite{Moussallam:2013una}.
This leads to  $\Delta \amuHVPLO[\pi\pi,\tau] = 4.5 (3) \times 10^{-10}$.

\item \emph{$\rho$--$\omega$ mixing:}
For this contribution the result of Ref.~\cite{Castro:2024prg}  differs considerably from the results of 
Ref.~\cite{Davier:2023fpl}  and Refs.~\cite{Colangelo:2022prz,Hoferichter:2023sli}.
The source of this difference can be traced back to the phase of the $\rho$--$\omega$ mixing parameter,
which  is $\simeq 10\degree$ in the reference dispersive analysis of Ref.~\cite{Castro:2024prg},
while it is $\simeq 4\degree$ in the other analyses.  In Ref.~\cite{Colangelo:2022prz}  this phase has been bound
fairly robustly and therefore in our combination we give more weight to the results from 
Refs.~\cite{Davier:2023fpl,Colangelo:2022prz,Hoferichter:2023sli}, leading to
$\Delta \amuHVPLO[\pi\pi,\tau] = 3.9 (3) \times 10^{-10}$.\footnote{Reference~\cite{Castro:2024prg} obtains $3.87(8)$ for the $\rho$--$\omega$ mixing correction using the GS form factor parameterization.
Again, the difference can be traced back to the  phase of the $\rho$--$\omega$ mixing parameter.
Other (partially compensating) differences are found between dispersive and GS parameterizations in Ref.~\cite{Castro:2024prg}, 
and they  are accounted for by increasing  the overall uncertainty, as discussed in the text.}

\item \emph{$F_\pi^V(s)/ f_+(s)$ and impact of IB in the  $\rho \pi \pi$ coupling:}
The ratio of form factors leads to a number of corrections that largely compensate in the total,
as shown in \cref{tab:summary_tau_IB}. The results of Refs.~\cite{Davier:2023fpl,Davier:2010fmf}  and  Ref.~\cite{Castro:2024prg}
are fairly consistent.  However, they both omit a potentially large source of IB in the width of the $\rho$ resonance,  as discussed below.  
IB effects in the width of the $\rho$ resonance are usually discussed in the context of
resonance models, assuming the  form 
$\Gamma_\rho (s)  \propto  \, g_{\rho \pi \pi}^2  \, \sqrt{s} \, \beta_{\pi \pi}^3(s) \, (1 + \delta_{\rho \pi\pi} )$
for the  off-shell $\rho \to \pi \pi$  width~\cite{Flores-Baez:2007vnd,Davier:2010fmf}.   Here, $\delta_{\rho \pi \pi}$ encodes long-distance radiative corrections, including
the effect of the $\pi \pi \gamma$ decay channels.  
In this context there are four effects that can generate the splitting $\Delta \Gamma_\rho \equiv \Gamma_{\rho^0} - \Gamma_{\rho^+}$:  
the coupling $g_{\rho \pi \pi}$, 
the $\rho$ mass, 
the kinematic factor $\beta_{\pi \pi}^3 (s)$, and the radiative corrections $\delta_{\rho \pi \pi}$. 
In the literature only the last three effects are discussed and it is found~\cite{Flores-Baez:2007vnd,Davier:2010fmf} that 
they partially cancel.  Here we point out that the effect of $g_{\rho^0 \pi^+ \pi^-} \neq g_{\rho^\pm \pi^\mp \pi^0}$
is of similar size to the ones included, and therefore, absent a detailed analysis, its impact should be reflected in the 
total uncertainty. 
In a purely phenomenological parameterization, one should adopt $g_{\rho^0 \pi^+ \pi^-} \neq g_{\rho^\pm \pi^\mp \pi^0}$.
In the context of Lagrangian models such as R$\chi$T, for example, 
this kind of IB can be induced
by operators involving 
two EM spurions, following common practice in ChPT~\cite{Urech:1994hd} 
and resonance chiral Lagrangians~\cite{Bijnens:1996nq}.
The impact of $\delta_g \equiv ( g_{\rho^0 \pi^+ \pi^-} - g_{\rho^+ \pi^- \pi^0} )/g_{\rho^0 \pi^+ \pi^-}$ on $\Delta \Gamma_\rho$
can be estimated by observing that 
\begin{equation}
\Delta \Gamma_\rho  \simeq 2 \Gamma_\rho \, \delta_g  \simeq  300 \, \delta_g \MeV\,.
\label{eq:DeltaGammaRho}
\end{equation}
$\delta_g$ could  range anywhere from    $\delta_g \simeq \alpha/\pi \simeq 0.23\%$ to  $\delta_g \simeq 1\%$ (the typical size of form-factor-related IB effects, e.g., FSR and $\rho$--$\omega$ mixing),
corresponding to   $\Delta \Gamma_\rho \simeq 0.7\MeV$ and  $\Delta \Gamma_\rho \simeq 3\MeV$, respectively.
The choice  $\delta_g \simeq \alpha/\pi$ is likely too aggressive,  because a similar estimate would not capture the actual size of the long-distance radiative
corrections to $\Gamma (\rho \to \pi \pi [\gamma])$ (these  induce a shift  $\Delta \Gamma_\rho = 1.8(2)\MeV$~\cite{Flores-Baez:2007vnd,Davier:2010fmf}).
On the other hand, the choice  $\delta_g \simeq 1\%$   would lead to $\Delta \Gamma_\rho \simeq 3\MeV$,
which exceeds the current PDG average $\Delta \Gamma_\rho = 0.3(1.3)\MeV$~\cite{ParticleDataGroup:2024cfk}.\footnote{Alternatively, Ref.~\cite{ParticleDataGroup:2024cfk} also quotes separate values $\Gamma_{\rho^0}=147.4(8)\MeV$, $\Gamma_{\rho^+}=149.1(8)\MeV$, and thus $\Delta\Gamma_\rho=-1.7(1.1)\MeV$, suggesting a somewhat larger uncertainty.}
Taking this into account,  we  adopt as a compromise value  $|\delta_g| \leq 2 \alpha/\pi \simeq 0.46 \%$,
which leads to $|\Delta \Gamma_\rho| \leq 1.4\MeV$, reflecting the current PDG uncertainty. Scaling the effect from the pion mass difference $\Delta\Gamma_\rho(\Delta M_\pi)=1.1\MeV$ from Refs.~\cite{Castro:2024prg,Davier:2023fpl,Davier:2010fmf}, this corresponds to an  additional  uncertainty of  $\simeq 4.7 \times 10^{-10}$ in
$\Delta \amuHVPLO[\pi\pi,\tau]$, motivating the corresponding entry  in \cref{tab:summary_tau_IB} for $F_\pi^V/f_+$ .

As an alternative to using theoretical predictions for IB in the $\rho$ parameters, a data-driven approach  
using the two-pion $\tau$ and $e^+e^-$ spectral functions
has very recently been proposed in Ref.~\cite{Davier:2025jiq}. 
Given the direct relevance of this work for the topic discussed here, we briefly summarize it, also noting that there has not been enough time for detailed scrutiny. 
This approach  proposes to 
decouple the normalization of the measured spectral function from its shape, 
the latter characterized mainly by the mass and width of the $\rho$  resonance, 
with some small residual contribution from higher resonances.  
For the correction to the dispersive integral from the pion form factors Ref.~\cite{Davier:2025jiq} finds 
$\Delta a_\mu [\pi\pi,\tau] (F_\pi^V/f_+)  =  + 1.68 (2.92)  (1.39)  \times 10^{-10}$, 
where the first uncertainty is dominated by the $\tau$ data and the second is due to the line-shape dependence. 
This data-based estimate agrees with the previous results from two different groups as quoted in 
\cref{tab:summary_tau_IB},    $-1.62(65)(63) \times 10^{-10}$~\cite{Davier:2023fpl,Davier:2010fmf},
 and $-1.34(1.71) \times 10^{-10}$~\cite{Castro:2024prg}, respectively at the 1.0 and $0.8\sigma$ level. 
 It also agrees with the new theoretical estimate in \cref{tab:summary_tau_IB}, $-1.5(4.7) \times 10^{-10}$. 
These figures    emphasize  the need for model-independent input for the ratio  
 of matrix elements   $F_\pi^V(s)/f_+(s)$.

\end{itemize}

Putting all the corrections together, the individual analyses arrive at 
$\Delta \amuHVPLO[\pi\pi,\tau] =  - 14.9 (1.9) \times 10^{-10}$ ~\cite{Davier:2023fpl,Davier:2010fmf},
$\Delta \amuHVPLO[\pi\pi,\tau] =  (-15.20)^{+2.26}_{-2.63}  \times 10^{-10}$ ~\cite{Castro:2024prg},
and $\Delta \amuHVPLO[\pi\pi,\tau] =  - 12.2 (3.4) \times 10^{-10}$ ~\cite{Davier:2025jiq}.  
The latter data-based determination is not used in the following  because it appeared late  in the WP25 schedule  and it is not published (at the time of writing). 
Altogether,  based on the above discussion and \cref{tab:summary_tau_IB}, 
our current best estimate for the  IB corrections to be used in a $\tau$-based evaluation of HVP is 
\begin{equation}
 \Delta \amuHVPLO[\pi\pi,\tau] =  - 15.0(5.1)  \times 10^{-10} \quad \text{(our estimate)}\,. 
\label{eq:best-estimate}
 \end{equation}
 While the central value fully  reflects the state-of-the-art  published  analyses,   
 the uncertainty  attempts to  account for  sources of IB that are  not yet fully addressed in the literature. 
 Work to tackle the issues summarized above is  ongoing 
 and   there are good prospects of integrating  in the near future new information from the 
  dispersive approach   (see~\cref{sect:tau-dispersive}), the lattice-QCD 
 approach (see~\cref{sect:tau-lattice}), and  data-driven constraints  
(see Ref.~\cite{Davier:2025jiq}).

\begin{figure}[t]
 \centering
 \includegraphics[width=0.6\linewidth]{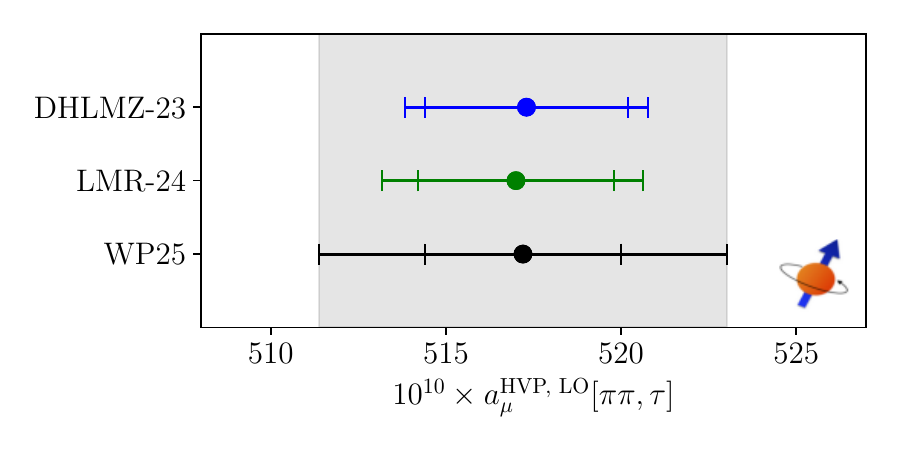}
 \caption{Summary of evaluation of $\amuHVPLO[\pi\pi]$ from $\tau$ decays. The points refer to DHLMZ-23~\cite{Davier:2023fpl}, LMR-24~\cite{Castro:2024prg}, and \cref{eq:atau-ourestimate}, respectively. The inner error bar reflects the current experimental uncertainties, while the outer error represents the combination of experimental and theoretical uncertainties.}
 \label{fig:summary_plot_tau}
 \end{figure} 

In summary, there are two  consistent evaluations of HVP based on $\tau$ data~\cite{Davier:2023fpl,Castro:2024prg}, corresponding to the first two columns in 
\cref{tab:summary_tau_IB}. 
They lead to 
$\amuHVPLO[\pi\pi,\tau]=517.3(1.9)_{\rm  stat}(2.2)_{\rm  syst}(1.9)_{\rm th}\times 10^{-10}$~\cite{Davier:2023fpl} and 
$\amuHVPLO[\pi\pi,\tau]=517.0(2.8)_{\rm exp}  \big(^{+2.3}_{-2.6}\big)_{\rm th}\times 10^{-10}$ ~\cite{Castro:2024prg}. 
Based on our conservative estimate of the IB uncertainties, larger than the ones quoted in the individual analyses,  
we arrive at 
\begin{equation}
 \amuHVPLO[\pi\pi,\tau]=517.2(2.8)_{\rm exp}  (5.1)_{\rm th}\times 10^{-10} \quad \text{(our estimate)}\,,
 \label{eq:atau-ourestimate}
\end{equation}
see \cref{fig:summary_plot_tau} for a summary. Adding the offset $187.3(2.2)\times 10^{-10}$ from WP20 for the remainder of the LO HVP contribution, one arrives at a total LO HVP value of
\begin{equation}
\label{LOHVP_tau}
   \amuHVPLO[(\pi\pi,\tau)+\text{WP20}]=704.5(6.2)\times 10^{-10}\,. 
\end{equation}
The above offset from WP20 is not updated in this work, we instead focus on the major tensions in the $2\pi$ channel. However, we emphasize that relevant new information on the subleading channels has become available, e.g., new measurements for $e^+e^-\to 3\pi$~\cite{BABAR:2021cde,Belle-II:2024msd}, $e^+e^-\to K_SK_L$~\cite{SND:2024kbi}, and the inclusive $R$-ratio~\cite{BESIII:2021wib}. As described in \cref{sec:BelleII,sec:disp_3pi}, tensions between the Belle-II $3\pi$ data and previous measurements are now visible, other tensions are present in the $K^+K^-$ channel and in the comparison of the BESIII inclusive $R$-ratio measurement with pQCD. While the overall effect is minor compared to the tension seen in the $2\pi$ channel, these effects will need to be carefully studied for a new, complete data-driven evaluation of $\amuHVPLO$, for which continued progress in the overall $R_\text{had}(s)$ measurement program is essential.

\subsection{Monte-Carlo generators and radiative corrections}
\label{sec:MC}

The {\it RadioMonteCarLow~2} community effort, which started in 2022~\cite{Abbiendi:2022liz,Durham22,Zurich23,Mainz24}, is dedicated to improving the
theoretical predictions for hadron and lepton production at low-energy
$e^+e^-$ colliders.  In Phase~I, whose results are documented in Ref.~\cite{Aliberti:2024fpq}, the focus was on $2\to{2}$ and
$2\to{3}$ processes $e^+\,e^-\to X^+\,X^-(+\gamma)$ for $\sqrt{s} =
(1\text{--}10)\GeV$ with $X\in\{e,\mu,\pi\}$. This concerns
scan-based measurements (no hard photon required) and radiative return
measurements (hard photon required).

The long-term goal of the project is to preserve the well-established codes; to provide canonical versions of codes; and to further develop and improve codes.
An important aspect is the inclusion of new ideas and approaches.
This includes exploiting the technical progress since the last
systematic review of low-energy MC tools in 2010~\cite{WorkingGrouponRadiativeCorrections:2010bjp}, which is leading to new codes getting developed or extended to cover the processes of interest.

{\it RadioMonteCarLow~2} is committed to open science and has released all codes, input data, and output together with a report as part of a living review at\\
\begin{indent}
    \url{https://radiomontecarlow2.gitlab.io}
\end{indent}\\
Phase~I was mostly a theoretical exercise in which seven  codes (\afkqed, \babayaga, \kkmc, \mcgpj, \mcmule, \phokhara, and \sherpa) were used in five
scenarios inspired by former and current experiments.
It should be stressed that the goal was \emph{not} to prove any one
code wrong or obtain a theory error for the prediction. The main
purpose was to ascertain the importance of various effects,
exploiting the wide variation of approaches used in
the codes. As documented in the review, the differences among the
results of the codes can be understood, giving valuable information
for further progress.
Beyond Phase~I,  this community effort is expected to continue for many years yet during  which the living review will be continuously updated as new results become available.

\subsubsection{Terminology}
When discussing the features of the different codes
and the effects they include it is important to use a systematic terminology.
An often used one is the fixed-order counting in $\alpha$ where LO, NLO, and NNLO refer (almost) universally to the full contribution at $\alpha^2$, $\alpha^3$, and $\alpha^4$ for $ee\to XX$ (or $\alpha^3$, $\alpha^4$, $\alpha^5$ for $ee\to XX\gamma$).
This includes \emph{per definition} both real and virtual corrections as either would not be finite or physical on their own.
Beyond fixed order, a MC might utilize some form of
approximations of soft and/or collinear effects---potentially
resummed---to capture dominant effects that are related to
additional photon emission.
However, it is important to stress that even if a given code can generate
configurations with two or more extra photons this does not mean it is
NNLO exact. The proper use of the term (N)NLO, indicating
that \emph{all} terms of a given order in $\alpha$ are taken into
account, is encouraged.

Resummation of logarithms in the electron mass involves modeling the splitting $\ell\to\ell\gamma$, which can be done (semi-) analytically using structure and fragmentation functions or fully numerically using a parton shower.
This leads to a good description of observables involving the lepton but does not model the transverse momentum $p_\perp$ of electrons and photons unless special care is taken.
Soft logarithms can be resummed using the Yennie--Frautschi--Suura (YFS) formalism~\cite{Yennie:1961ad}, leaving an IR finite perturbative expansion.
The calculation of these finite residuals allows for the matching of the YFS resummation
with higher-order corrections. This can be achieved at the amplitude
squared level using the exclusive exponentiation (EEX)~\cite{Jadach:1988gb,Krauss:2022ajk}
or directly at the amplitude level using coherent exponentiation (CEEX)~\cite{Jadach:2000ir}.

Ultimately the goal should always be to combine a fixed-order calculation with some form of approximate higher-order corrections as both effects contribute at the percent level.
The current state of the art is NNLO (NLO) or NLO (LO) matched to
resummation/additional approximate emission for $2\to 2$ ($2\to3$) scattering.

To further classify the QED aspects,
initial-state corrections (ISC), final-state corrections (FSC), mixed
corrections (interferences between ISC and FSC), and VP corrections (VPC) are distinguished. 
This is done by assigning formally different charges to the initial
electron and outgoing muon or pion and all parts are gauge invariant
and IR finite. The reason for such a distinction
is of a purely technical nature. It allows one to use adapted techniques
for the various parts. 
It is quantum mechanically invalid to interpret any photon as being emitted from either the initial or final state.
Hence, Ref.~\cite{Aliberti:2024fpq} refrains from using the term ``ISR process'' for $e^+\,e^-\to
X^+\,X^-\,\gamma$. Furthermore, it is stressed that a factorized treatment
of the corrections into ISC and FSC is always an approximation, the validity of which depends strongly on process and observable.

Beyond the purely perturbative QED aspects, the
nonperturbative nature of the hadronic states needs to be considered. First, this concerns
the hadronic VPC. While there are different compilations available, in
the {\it RadioMonteCarLow~2} report all results were produced with the
same HVP, based on Refs.~\cite{Ignatov:2008bfz,Ignatov:hvp,WorkingGrouponRadiativeCorrections:2010bjp}. However, there were different
implementations in that some codes resum the HVP corrections, others
follow an approach closer to the fixed-order approach. Second,
final-state pions require a nonperturbative treatment.  In principle
the hadronic matrix elements of $\pi\pi\to n\gamma^{(*)}$ is
well-defined.  However, especially if some or all of the photons
involved are hard, there is not always sufficient information
available for practical calculations.
\begin{description}
    \item[$\boldsymbol{n=1}$]
        The matrix element reduces to the pion VFF
        $F_\pi^V(Q^2)$ which is known sufficiently well. All results in
        the report were produced with the same version of the VFF.

    \item[$\boldsymbol{n=2}$]
        The two-pion pole contribution (with two insertions of $F_\pi^V$, either using GVMD or dispersively) can be an acceptable approximation, which is called FsQED.
        On the other hand, a simple multiplication of the sQED result with a form factor evaluated at some scale to ensure finiteness---a method referred to as F$\times$sQED---is known to fail in producing the correct asymmetries, see \cref{TI-CMD3,sec:radiative_corrections}.

    \item[$\boldsymbol{n=3}$]
        This contribution is not yet systematically studied beyond F$\times$sQED and will require an improved understanding of the hadronic matrix element.
        Unfortunately, this contribution is especially important for the full NLO corrections to $ee\to\pi\pi\gamma$.

    \item[beyond $\boldsymbol{n=3}$]
        These matrix elements, or approximations of them, will be required for a full NNLO description of $ee\to\pi\pi\gamma$.
        Needless to say, this is still some time away.
\end{description}

It should once again be stressed that despite these shortcomings it is possible to generate approximate higher-order effects using resummation tools.
However, these results are not NLO or even NNLO for $ee\to\pi\pi\gamma$.
It is vital that investigations beyond F$\times$sQED be carried out to estimate the impact of this approximation.

\subsubsection{Monte-Carlo comparison}

A major part of the {\it RadioMonteCarLow~2} report for Phase I was the comparison of seven MC codes in five scenarios to loosely approximate the experiments on which they are based: CMD, KLOE small-angle, KLOE large-angle, BES, and a generic $B$ factory.
Designing these was a trade-off between complexity, computing requirements, and realism.
A major goal of Phase II is streamlining this process and improving the realism of these scenarios.

In the following, some broad conclusions from the comparison of different codes that were used as proxies for the different effects and approximations will be discussed.
However, the reader is strongly recommended to refer to the review for a more detailed and nuanced discussion.
Specifically, the spread of results for the different codes (``code spread'') cannot be used as an estimate for the theory error.

Starting with $2\to{2}$ processes, for simple, well-behaved
observables the code spread is as low as $0.2\%$. The notion of
``well-behaved'' refers to the fact that the result is nonvanishing at
LO for the $2\to2$ kinematics. Outside this region, the spread gets significantly larger. This is
not surprising, since the technical accuracy is reduced by one
order. In this region multiple emission becomes the main effect.  In
most cases that were studied, the resummation methods were dominated by
the emission of one extra photon.  This is not to say that these
effects are small; indeed they can become substantial (several
percent) depending on the specific observable under consideration.

The importance of the interference effects between ISC and FSC can be most plainly seen in asymmetries.
This is especially important since the ISC are conceptually the simplest and here improvements are expected first.
Further, unless special care is taken to match a resummation
(completely or approximately) to an NLO-or-better fixed-order result,
these effects will not be captured. The treatment of the
$\pi^+\,\pi^-$ final state for the $2\to{2}$ process is not 
too severe an issue. FsQED and GVMD appear to be reasonable
approximations. 

Turning to $2\to{3}$ processes, the situation is more delicate.
Most codes are either NLO without resummation or LO with resummation, allowing one to estimate both the importance of fixed-order and multi-photon topologies.
In the case of $ee\to\mu\mu\gamma$, the fixed-order result is unambiguous and extensions to NNLO are underway, making it a valuable testing ground for various approximations.
The importance of supplementing a parton shower with $p_\perp$-effects as is done, e.g., in \babayaga, both for capturing the correct shape and the correct cross section, is noted.
Similarly, the interference terms that are included in CEEX have proved to be very important and are capable of capturing most of the NLO dynamics though the effect is less pronounced when only considering the shape of the distribution rather than its absolute value.
To achieve a precision below the percent level effects beyond NLO need to be incorporated, though the exact details of course depend on scenario and cuts.

For $ee\to\pi\pi\gamma$, often small FSC at LO were witnessed, $(0.1\text{--}0.5)\%$, due to suppression from the VFF since this leads to a larger $Q^2$ in the form factor.
However, this trend does not continue to NLO and large FSC at NLO around $1\%$ were observed, even if FSC were negligible at LO.
It is, therefore, important to realize that extrapolating to the next uncomputed order is fraught with difficulty.
Even at LO, there are notable exceptions such as the KLOE-like large-angle scenario where the cuts prefer the photon closer to pion, leading to FSC effects of more than $10\%$.

The importance of FSC in $ee\to\pi\pi\gamma$ is a major issue, especially at NLO.
This $n=2$ case was not treated really satisfactorily in any of the codes at the time of the comparison.
If it is considered at all, it is done using F$\times$sQED, the validity of which has been questioned, at least in some cases.
Especially worrying are topologies that involve real or virtual corrections to the pion VFF.
These are $C$-even, i.e., contributing to the total cross section rather than merely an asymmetry, and at least partially FSC.
Without a calculation beyond F$\times$sQED for $ee\to\pi\pi\gamma$, it is difficult to estimate how large this effect really is.

When it comes to resummation for $ee\to\pi\pi\gamma$ it is possible to use similar technology as for $ee\to\mu\mu\gamma$.
For soft logarithms nothing changes because the soft limit is independent of the spin of the emitter, so YFS can be directly used at LO,  and can be extended to higher orders by the inclusion of the corresponding amplitudes (even if changes to the code may be required).
For collinear logarithms the $\pi\to\pi\gamma$ splitting function needs to be used instead.
Both methods inherently assume that the emitting pion is nearly on-shell as part of their constructions.

Finally, many scenarios were seen in which VPC were small at NLO but not at NNLO, where their interplay with the non-VP corrections can be complicated.
The size of the VPC depends strongly on the scenario but their careful treatment is very important, especially where narrow resonances such as the $J/\psi$ become relevant.
This again shows the danger of extrapolating to the next uncomputed order.

\subsubsection{Next steps}

So far, only the first phase of {\it RadioMonteCarLow~2} has concluded, most of which was a review of the existing state-of-the-art.
Actual improvements to the codes are expected to take place during Phase II 
(see, e.g., \babayaga~\cite{Budassi:2024whw}).
The {\it RadioMonteCarLow~2} community strives to update their living review to reflect these developments as they get released.

A major improvement will be the development of NNLO results for $2\to3$ where possible.
Even for $ee\to\mu\mu\gamma$, a full calculation without any approximation would be challenging.
However, there are well-established procedures to obtain sufficiently precise results in the near future.
These tools could also be used for $ee\to\pi\pi\gamma$ in F$\times$sQED but it is unclear how good an approximation this would be.

Another point that is of high priority for all the developers of MC tools is the inclusion of structure-dependent corrections for pion production.
In the case of $ee\to\pi\pi$, this was already done by a few codes (most recently in \babayaga~\cite{Budassi:2024whw} using GVMD or FsQED).
For $ee\to\pi\pi\gamma$, these effects were identified to be problematic as they could enter the cross section rather than just asymmetries (cf.\ Figs.~14 and 15 of Ref.~\cite{Aliberti:2024fpq}).
First investigatory steps are being taken in cases where GVMD or FsQED are assumed to be reliable approximations.
The development of a full model of the $n=2$ case is also underway.

The existing NLO calculations and future NNLO calculations will have to be matched with resummation techniques, either based on parton shower or YFS.
This will improve the reliability of the prediction and estimate of the truncation error.
Further, a strict NLO (NNLO) calculation for $2\to3$ processes will never generate events that have more than four (five) photons.
If such events turn out to be experimentally important, the exclusive generation of the resummation techniques will be able to at least approximate these events.

Finally,  only a few fairly simple scenarios have been considered  and only for $ee\to ee,\mu\mu,\pi\pi(+\gamma)$.
As a first step it is planned to consider more complicated experimental scenarios by streamlining the production pipelines.
It is also planned to consider more types of hadronic final states with $ee\to\pi^+\pi^-\pi^0$ as a first likely candidate.
Phase II will also hopefully see estimates of theory errors for some of the processes and scenarios under consideration.

\subsection{New developments for the evaluation of HVP}
\label{sec:data_combinations}

\subsubsection{The DHLMZ studies}
\label{Sec:DHLMZ}

\paragraph{Data combination and comparisons of mass spectra in the \texorpdfstring{$\pipi$}{} channel} 

The DHMZ studies employ since 2009 a procedure for combining cross-section data with arbitrary point spacing or binning, redistributed in a fine common binning using spline-based interpolations, as implemented in the HVPTools software~\cite{Davier:2010fmf,Davier:2009zi,Davier:2010nc,Davier:2017zfy,Davier:2019can,Davier:2023fpl}.
For each narrow final bin, a $\chi^2$ is minimized to get the average weights and locally test the level of agreement among the input measurements.
The average weights also account for the different bin sizes and point-spacing of measurements, in order to compare their precisions on the same footing.
This procedure has been validated through a closure test. 
It features full and realistic~(i.e., not too optimistic) treatment of uncertainties and correlations, between the measurements~(data points or bins) of a given experiment, between experiments and between different channels.
It also fully accounts for systematic tensions between experiments.

\begin{figure}[t]
  \centering
  \includegraphics[width=7.9cm]{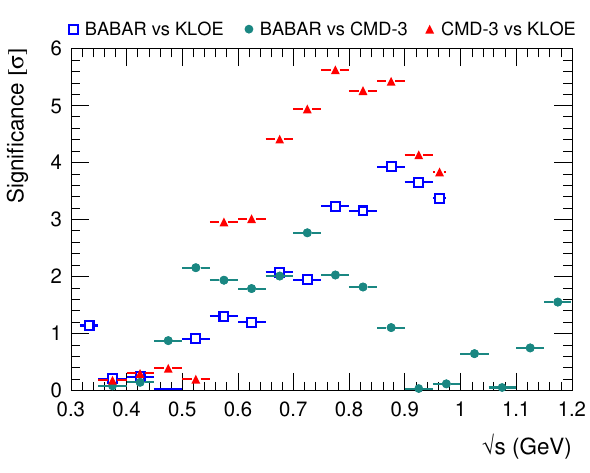}
  \hspace{0.3 cm}
  \includegraphics[width=7.9cm]{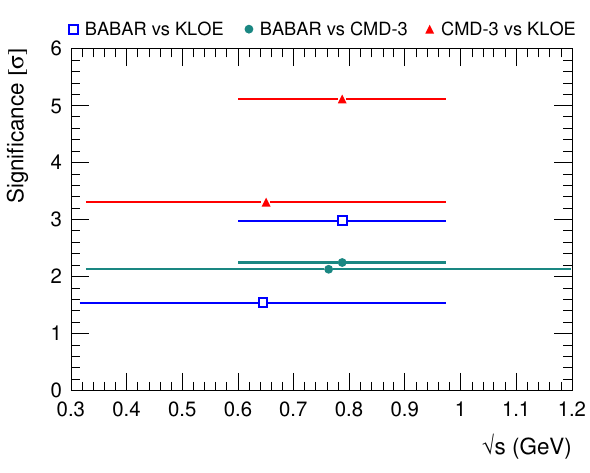}
  \caption{\small 
Significance of the difference between pairs of the three most precise $\mee \to \pipi$ cross section measurements~(\babar~\cite{BaBar:2009wpw,BaBar:2012bdw}, CMD-3~\cite{CMD-3:2023alj}, KLOE~\cite{KLOE:2008fmq,KLOE:2010qei,KLOE:2012anl,KLOE-2:2017fda}) for narrow energy intervals of $50~{\rm MeV}$ or less~(left) and larger energy intervals~(right) indicated by the horizontal lines. 
Figures taken from Ref.~\cite{Davier:2023fpl}.
}
\label{Fig:amuDataComparisons}
\end{figure}

\begin{figure}[tbp]
  \centering
  \includegraphics[width=7.9cm]{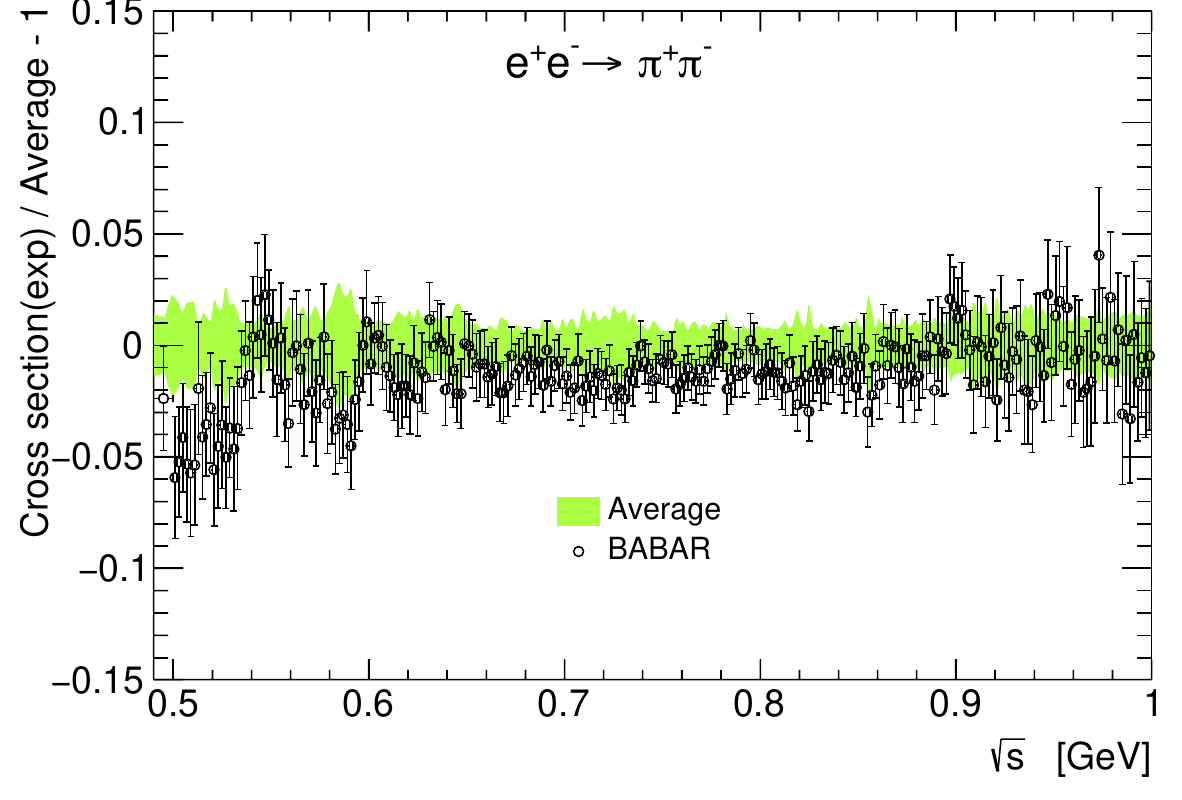}
  \hspace{0.3 cm}
  \includegraphics[width=7.9cm]{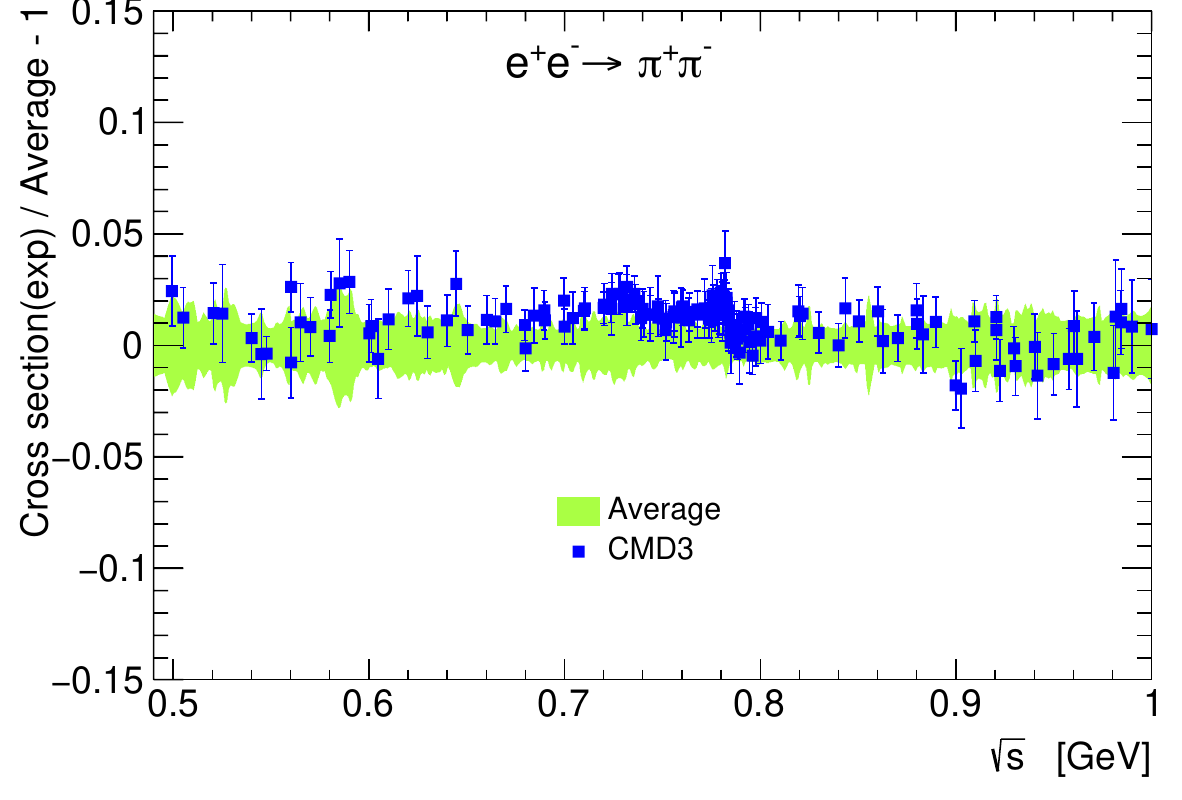} \\
  \includegraphics[width=7.9cm]{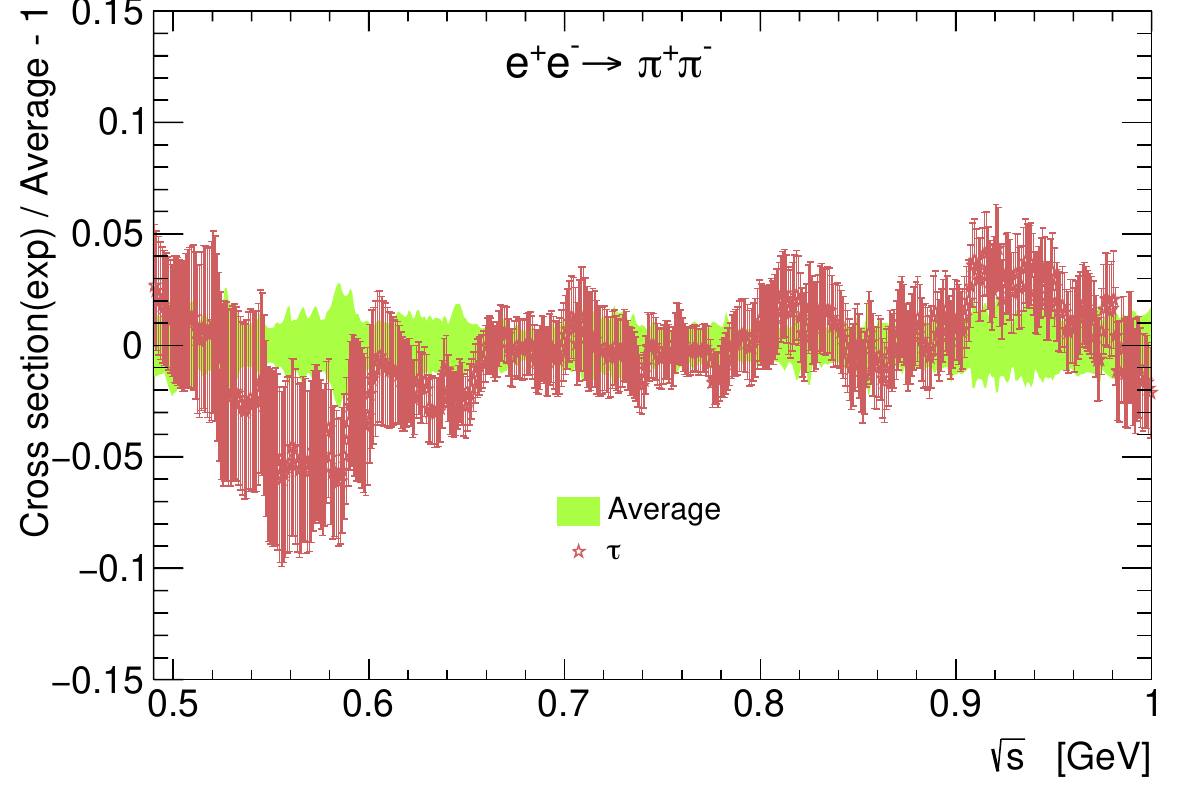}
  \hspace{0.3 cm}
  \includegraphics[width=7.9cm]{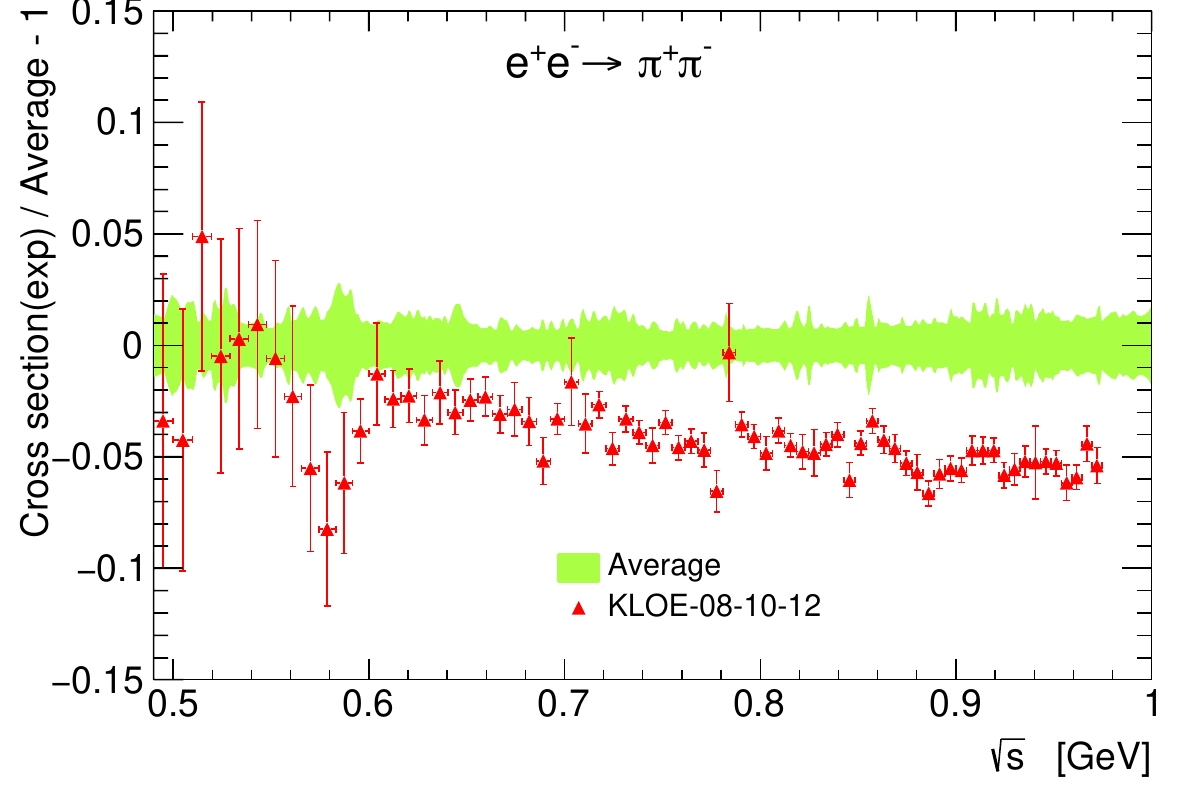} \\
  \includegraphics[width=7.9cm]{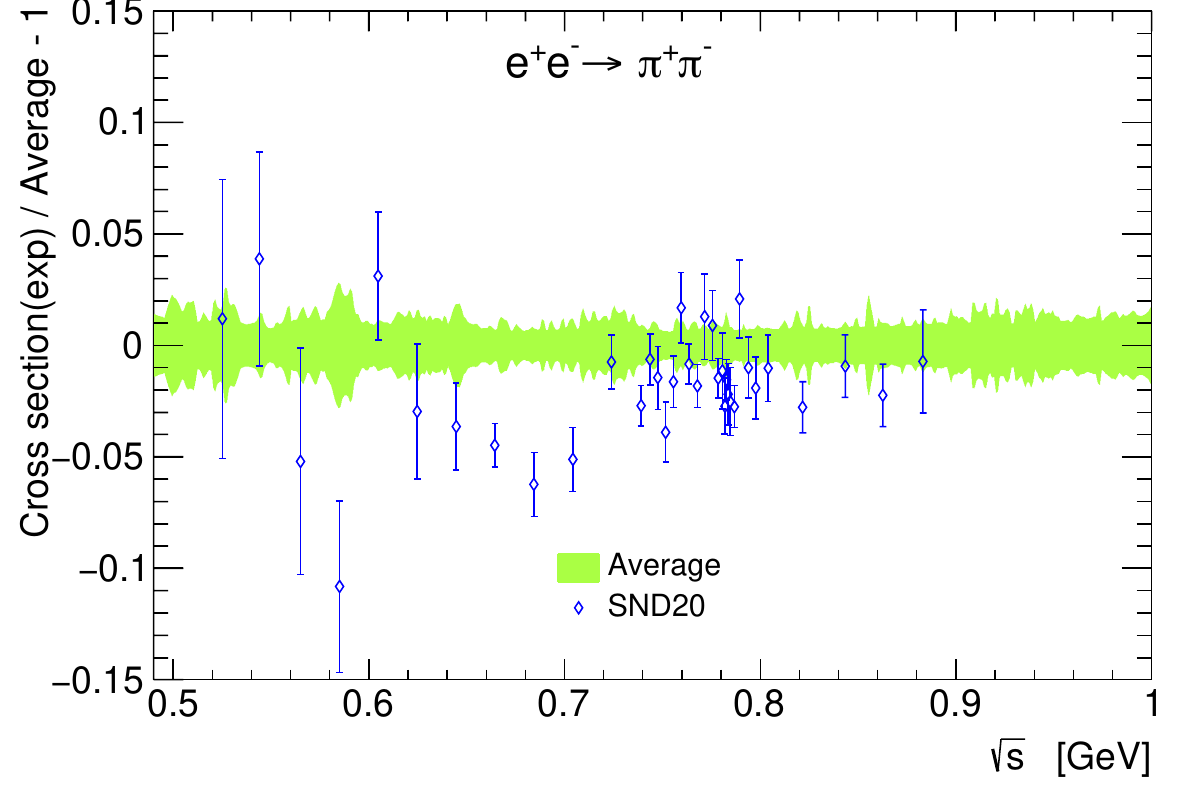}
  \hspace{0.3 cm}
  \includegraphics[width=7.9cm]{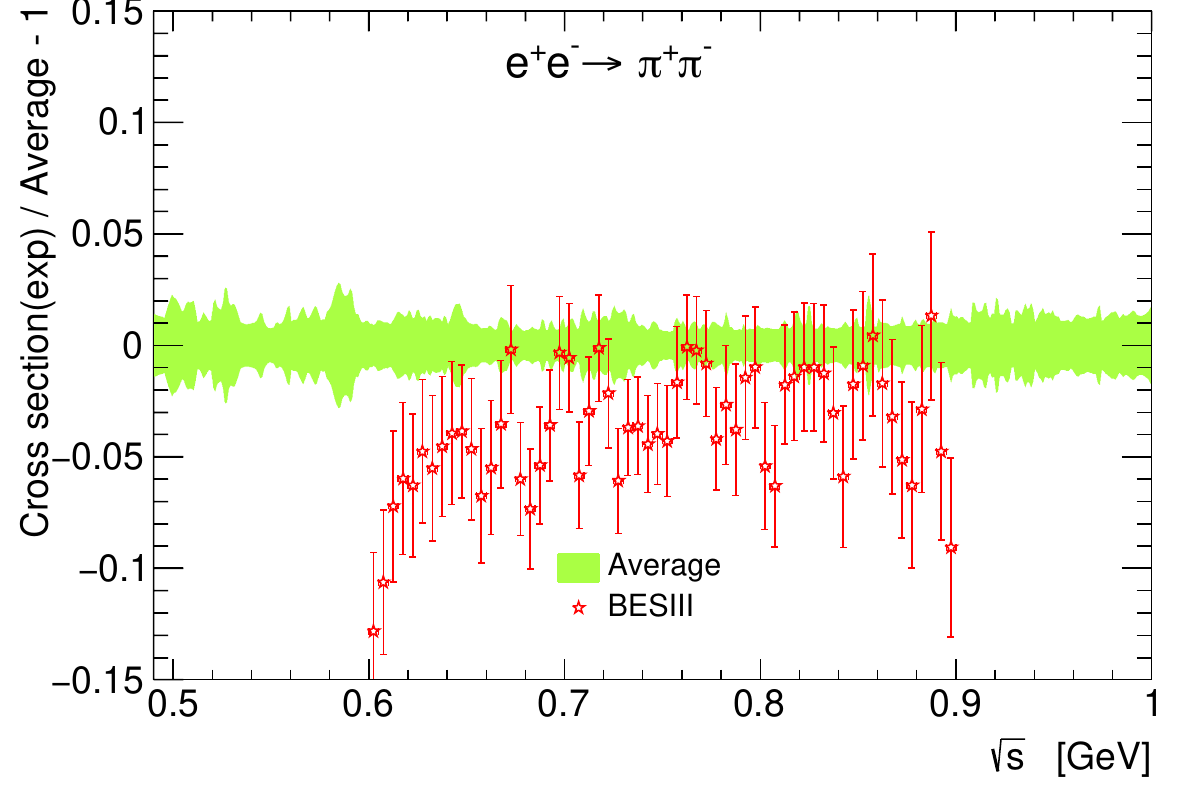}
  \caption{\small 
Relative difference between the $\pi \pi$ cross sections from \babar~(top left)~\cite{BaBar:2009wpw,BaBar:2012bdw}, CMD-3~(top right)~\cite{CMD-3:2023alj}, the IB-corrected $\tau$ data~(middle left)~\cite{Davier:2013sfa}, and their average. The IB corrections and uncertainties employed here correspond to the ones from Ref.~\cite{Davier:2023fpl}.
Relative difference with respect to the same average for the measurements from KLOE~\cite{KLOE-2:2017fda}~(middle right), SND20~\cite{SND:2020nwa}~(bottom left), and BESIII~\cite{BESIII:2015equ}~(bottom right).
Figures taken from Ref.~\cite{Bogdan-KEK-2024-dhmz}.
}
\label{Fig:DHMZspectraRatios}
\end{figure}

The DHMZ average is dominated by the most precise experiments (\babar~\cite{BaBar:2009wpw,BaBar:2012bdw}, CMD-3~\cite{CMD-3:2023alj}, KLOE~\cite{KLOE:2008fmq,KLOE:2010qei,KLOE:2012anl}, SND20~\cite{SND:2020nwa}, followed by CMD-2~\cite{CMD-2:2001ski,CMD-2:2003gqi,CMD-2:2005mvb,CMD-2:2006gxt,Aulchenko:2006dxz}, BESIII~\cite{BESIII:2015equ}, and SND06~\cite{Achasov:2005rg}), \babar\ covering the full energy range of interest.
Taking the ratio between various measurements and this average, it is found that in the $[0.5,1]\GeV$ range the \babar\ and SND20 data overlap rather well with the average, KLOE is systematically below it, while CMD-3 is above~\cite{Davier:2023fpl}.
These tensions, especially between KLOE and CMD-3, which provide the smallest and respectively largest cross sections in the $\rho$ region, are also reflected by the enhanced values of $\chi^2/{\rm dof}$.
For KLOE slopes are observed when comparing the 2010 and 2012 data with the 2008 ones.
These tensions were quantified through fits and found to be at the $(2.5\text{--}3)\sigma$ level. No significant slope is present between 2010 and 2012 data~(see Ref.~\cite{Aoyama:2020ynm} and references therein).

In order to further quantify these tensions, the integrals for individual experiments were computed and compared in various restricted energy ranges.
The significance of the difference between pairs of experiments is determined, taking into account the correlations of the uncertainties, see \cref{Fig:amuDataComparisons}.
The largest observed tensions are between CMD-3 and KLOE, going beyond $5\sigma$ on the $\rho$ peak~\cite{Davier:2023fpl}.
The impact of these tensions on the comparison of the experimental result to the SM expectation for $a_\mu$ is displayed in the Fig.~5 from Ref.~\cite{Davier:2023fpl}~(see more detailed discussion below).

The presence of these tensions among experimental measurements represents a clear indication of underestimated uncertainties.
This calls for a conservative uncertainty treatment in combination fits and in the determination of the averaging weights, as implemented in the DHMZ approach~\cite{Davier:2019can,Aoyama:2020ynm}.
These systematic tensions go well beyond the effects accounted through the local $\chi^2/{\rm dof}$ rescaling.
This had already motivated the inclusion of the dominant \babar--KLOE systematic by DHMZ, since the studies reported in Ref.~\cite{Davier:2019can}. 
However, the tensions are larger now and therefore require one to understand their actual source.

\paragraph{Impact of higher-order photon emissions: the implications of a unique ``(N)NLO'' \babar\ study} 

As discussed in \cref{Sec:BABAR}, the higher-order photon emissions~(i.e., in addition to the hard ISR photon), were studied in-situ with \babar\ data~\cite{BaBar:2023xiy}, for the first time at NLO and NNLO, in the $\mee \to \mumu \gamma$ and $\mee \to \pipi \gamma$ channels.
This allows one to test the most frequently used MC generators, \phokhara{} and \afkqed.
The ``(N)NLO'' order counting in data and simulations is performed based on the number of additional photons in the final state, having the energy above some given threshold.

It is found that the rate of ``NLO'' small-angle ISR in \phokhara{} is higher than in data, while the data/MC ratios for large-angle photon emissions are consistent with unity~\cite{BaBar:2023xiy}.
An independent confirmation of the \phokhara{} problem has been provided by the measurement of the $\pi^+\pi^-\pi^0$ channel performed by the Belle-II Collaboration~\cite{Belle-II:2024msd}.
The ``NNLO'' contributions are also clearly observed in data, while they are missing in \phokhara.
At the same time, \afkqed{} provides a reasonable description of the rate and energy distributions for ``(N)NLO'' data.

The discrepancies observed between BABAR and \phokhara{} in the energy distributions of additional ISR photons reveal  \phokhara{} shortcomings that cannot be fully interpreted due to the absence of complete NNLO MC generators.  A full range of scenarios with extremes labeled 1 and 2, has been considered according to presently unknown NNLO calculations~\cite{Davier:2023fpl}. Scenario 1 questions the validity of \phokhara{} at the NLO level in addition to missing NNLO, while in scenario 2 the discrepancy observed by BABAR would originate solely from missing NNLO. The realistic situation is expected to lie between these two extremes. These findings have triggered further studies performed by BESIII and KLOE. Tests of MC generators regarding the $\mu\mu/\pi\pi$ mass distributions are now available (see \cref{Sec:KLOE,Sec:BES,sec:MC}) which confirm the validity of NLO \phokhara{} at the $1\%$ level~(where the uncertainties of the other generators employed in the comparisons are relevant too). This is to be compared with a systematic uncertainty of $0.5\%$ for \phokhara{} as quoted in the KLOE/BESIII publications. However, similar comparisons should be done for the energy distribution of additional ISR photons, playing an important role in the event selection. For the moment, the question of re-evaluating \phokhara{}-related systematic uncertainties for the published KLOE/BESIII results has not yet been considered, particularly for KLOE, which has competitive precision. It is worth noting that the sensitivity to these effects does depend on analysis strategy choices that are somewhat less critical for BESIII.
It is important to note that the \babar\ measurements~\cite{BaBar:2009wpw,BaBar:2012bdw} employ a loose selection, incorporating ``NLO'' and higher-order radiation, minimizing hence the dependence on MC simulations.

\paragraph{A new perspective on \texorpdfstring{$\amuHVPLO$}{}} 

The $a_\mu$ and central $\amuW$ calculations~(see \cref{sec:windows,sec:windows_data_driven}) were performed employing the dispersive approach, based on the most precise measurements available in the $\pipi(\gamma)$ channel~(see pages 7, 8, 18, and 19 in Ref.~\cite{Bogdan-KEK-2024-dhmz}), from \babar~\cite{BaBar:2009wpw,BaBar:2012bdw}, CMD-3~\cite{CMD-3:2023alj}, and KLOE~\cite{KLOE-2:2017fda}, as well as from hadronic $\tau$ decays~\cite{Davier:2013sfa} as discussed in \cref{Sec:DHMZtau}, see the Fig.~5 from Ref.~\cite{Davier:2023fpl}.
For KLOE, both the full available range~(${\rm KLOE}_{\rm wide}$) and a restricted range of $(0.6\text{--}0.975)\GeV$~(${\rm KLOE}_{\rm peak}$) were considered.
For the latter, the data are most precise and KLOE's weight in the combination is largest.
These various $\amuHVPLO$ integrals are completed with the combination of all the available measurements in the $\pi\pi$ channel~\cite{Davier:2023fpl}, in order to cover the full mass range of interest, as well as with contributions from other hadronic channels~(as evaluated in Ref.~\cite{Aoyama:2020ynm}), in view of the comparisons with the BMW lattice QCD result~\cite{Borsanyi:2020mff} and with the experimental measurement~\cite{Muong-2:2023cdq}.

The $\tau$-based HVP contribution is close to the values provided by \babar\ and CMD-3, see the Fig.~5 from Ref.~\cite{Davier:2023fpl}.
Their combination is compatible with BMW for $a_\mu$, but a $2.9\sigma$ tension persists for $\amuW$.
Combining \babar, CMD-3, $\tau$ (and BMW), a difference of $2.5\sigma$ ($2.8\sigma$) is found with respect to the experiment.
When including KLOE in the dispersive calculation, the difference to the experiment becomes larger than $5\sigma$.
This situation concerning the most precise experiments can be illustrated at the level of the cross sections by comparison with the average of \babar~\cite{BaBar:2009wpw,BaBar:2012bdw}, CMD-3~\cite{CMD-3:2023alj}, and the IB-corrected $\tau$ data~\cite{Davier:2013sfa}, showing their consistency, while KLOE~\cite{KLOE-2:2017fda} disagrees, both in magnitude and shape, see \cref{Fig:DHMZspectraRatios}.\footnote{The average of \babar, CMD-3, and the IB-corrected $\tau$ data is employed here just as a reference in the comparisons. It enables a clear visualization of agreement/differences in terms of shape and normalization among various measurements, which would not be the case if, e.g., a full combination were used instead.}
The SND20~\cite{SND:2020nwa} and BESIII~\cite{BESIII:2015equ} measurements show also some tensions with this same average, although within somewhat larger uncertainties and hence less statistically significant.
To further investigate these problems, tests of MC generators must be performed for experiments which rely on them to correct for missing higher-order contributions. 
These studies should focus on the rate and angular distributions of additional photons, which impact the selection efficiency~\cite{Davier:2023fpl}.

These findings provide additional insight into the longstanding deviation among the muon $g-2$ measurement, the SM prediction using the data-driven dispersive approach for calculation of HVP, and the comparison with lattice-QCD calculations.

\subsubsection{The KNTW studies}
\label{Sec:KNTW}

\begin{figure}[t]
  \centering
  \includegraphics[width=15cm]{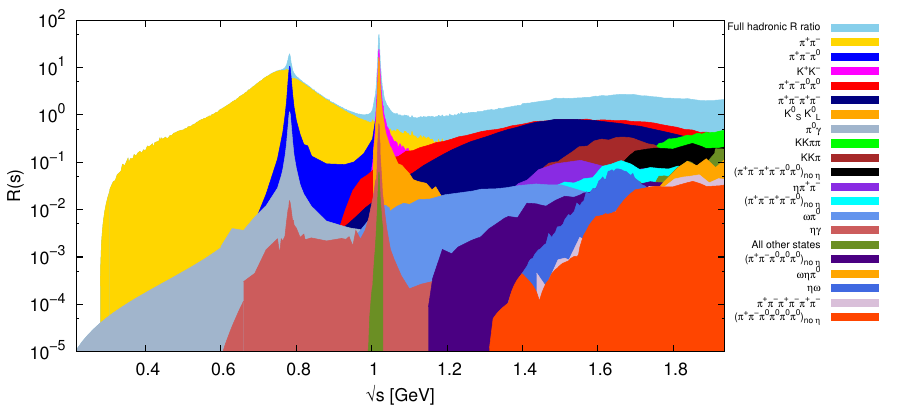}
  \caption{\small 
Contributions to the KNT data compilation of the total hadronic $R$-ratio from the different hadronic final states below 1.937 GeV~\cite{Keshavarzi:2018mgv,Keshavarzi:2019abf}. The full $R$-ratio is shown in light blue. Each final state is included as a new layer on top in decreasing order of the size of its contribution to $\amuHVPLO$. Figure adapted from Ref.~\cite{Keshavarzi:2018mgv}.
} \label{fig:KNT-Rratio}
\end{figure}

For the first edition of WP20~\cite{Aoyama:2020ynm}, KNT (now KNTW) provided their data-driven HVP
compilation \cite{Keshavarzi:2018mgv, Keshavarzi:2019abf}. Their data
for the full hadronic cross section, see \cref{fig:KNT-Rratio}, result in the most precise evaluation of $\amuHVPLO$ to date of $\amuHVPLO[{\rm KNT19}] = 692.8(2.4) \times 10^{-10}$,
with overall good agreement observed when compared to other
groups. The results were used in the merging procedure between data-driven analyses, which determined the adopted consistent value $\amuHVPLO[{\rm WP20}] = 693.1(4.0) \times 10^{-10}$ in Ref.~\cite{Aoyama:2020ynm}. KNT have also provided the prediction for the NLO HVP contribution, $\amuHVPNLO[{\rm WP20}] = -9.83(7) \times 10^{-10}$, the data input for the NNLO HVP contribution, $\amuHVPNNLO[{\rm WP20}] = 1.24(1) \times 10^{-10}$ \cite{Kurz:2014wya}, predictions for the HVP contributions to the electron and tau $g-2$ (both at LO and NLO), 
for the running of the EM coupling,  $\alpha(Q^2)$, and for the hyperfine structure of muonium. Their publicly available, open-access 
data compilation includes: (1) the full hadronic cross section 
ranging from the production threshold to 1 TeV, complete with a full 
covariance matrix for all data, (2) cross-section data and covariance 
matrices for all individual hadronic modes, and (3) a software package
(\texttt{knt\_vp}) that produces $\alpha(Q^2)$ for both spacelike and
timelike $Q^2$.\footnote{All data are available upon request by 
contacting the authors directly.} These data have been used by 
the wider particle physics community including in many studies 
related to the muon $g-2$, notably for predictions of 
window-based quantities.

The philosophy of the KNTW compilation is summarized by the main
features: (i) Prioritize, wherever possible, direct input from data
without modeling, parameterization, or other constraints. (ii)
Include all published data unless they are known to be
defective. (iii) Understand and minimize/avoid all possible bias
that could be present in the analysis. (iv) Incorporate full
correlation information between different data points due to
systematic uncertainties in the combination to
constrain the fits by using all available experimental data. 
(v) Evaluate and incorporate additional
theoretical systematic uncertainties arising from the analysis. 
(vi) Be fully open-access with regard to the sharing of studies, results, and resulting combination data.

At the time of Ref.~\cite{Aoyama:2020ynm}, the KNT data combination could accommodate tensions between the data sets, still achieving an acceptable fit quality. This was reflected in a global $\chi^2/{\rm dof} = 1.26$ for the most important two-pion channel. For the evaluation of the $a_\mu$ integral, an energy-dependent chi-square inflation was used, which amounted to a $14\%$ error inflation for $a_\mu^{\pi^+\pi^-}$ in the range $0.305 < \sqrt{s} < 1.937$ GeV. This situation changed dramatically with the publication of the CMD-3 two-pion data~\cite{CMD-3:2023rfe, CMD-3:2023alj}. Despite sustained efforts by the community, coordinated by the Muon $g-2$ Theory 
and the {\it RadioMonteCarLow~2} 
initiatives, see \cref{TI-CMD3,sec:MC}, no explanations for the discrepancy have been found so far. With further dedicated efforts on the calculation of higher-order radiative corrections and their implementation in MC generators, 
and new data analyses in the two-pion channel underway, the picture is
not yet settled. Therefore, the persistent strong tension with the previous data across the entire energy range in the dominant $\pi^+\pi^-$ channel means that the most precise data sets and their combination are highly inconsistent, and extra caution should be taken before trying to proceed with the direct combination as before. KNTW have consequently refrained from providing an updated compilation.

\begin{figure}[t]
    \centering
   \includegraphics[width=0.5\textwidth]{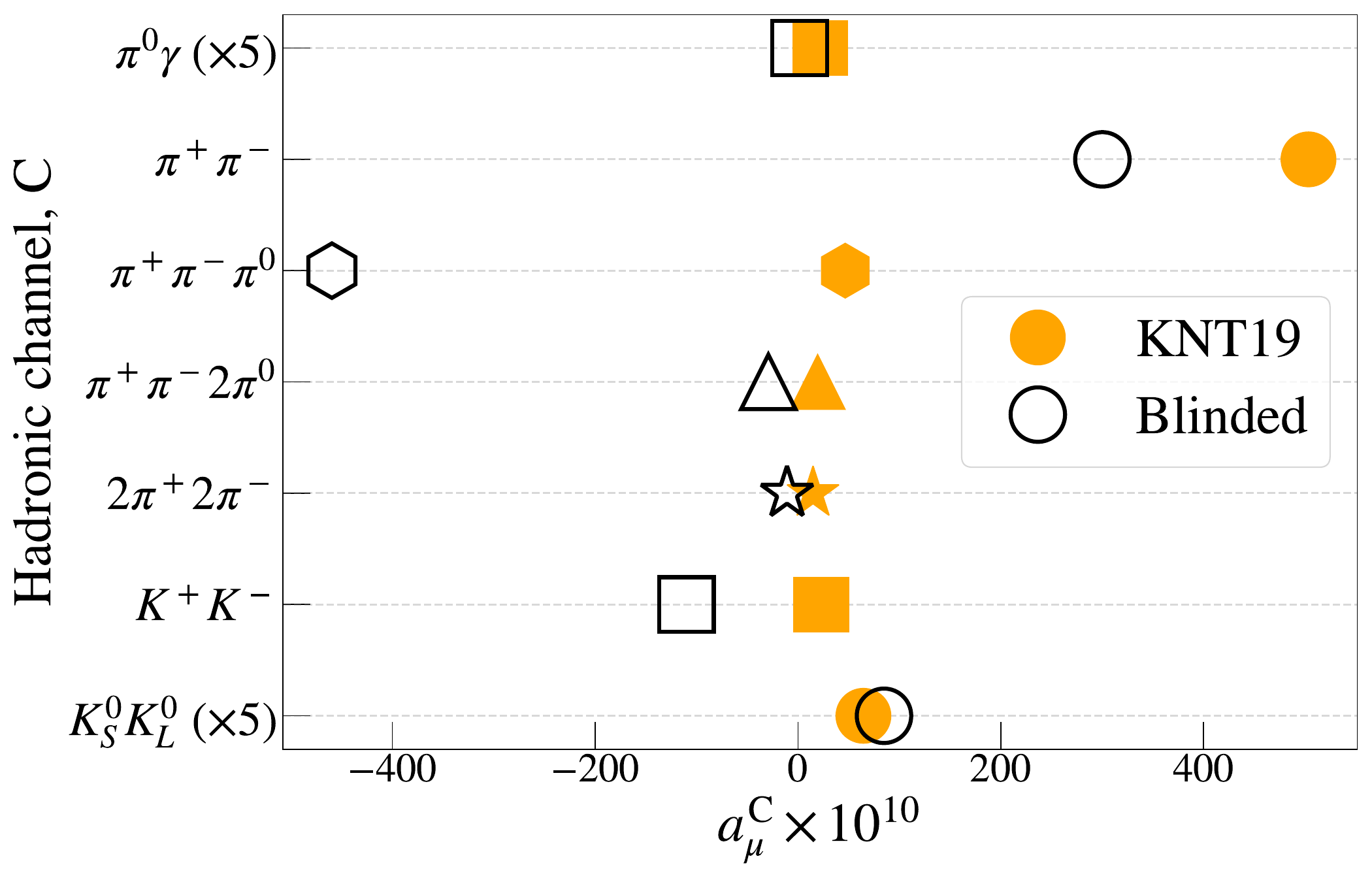}
   \caption{Values for $\amuHVPLO$ for different hadronic channels from the original unblinded KNT19 analysis~\cite{Keshavarzi:2019abf} (filled), and what the blinded results of that analysis would have looked like prior to unblinding had the new KNTW blinding procedure~\cite{Keshavarzi:2024wow} been employed (unfilled). For the offsets of this specific example, the blinded results for $\pi^0 \gamma$ and $K_S^0 K_L^0$ are randomly similar to their unblinded counterparts so both have been scaled by $\times5$ to make the differences visible. Figure taken from Ref.~\cite{Keshavarzi:2024wow}.}
   \label{fig:KNTW-blinding}
\end{figure}

Instead, KNTW are taking the opportunity to fully scrutinize, overhaul, and modernize
their analysis and data combination procedure in preparation for crucial, new hadronic 
cross-section data expected in the future, particularly for the 
$\pi^+\pi^-$ channel, see \cref{sec:e+e-}. The aim is to 
improve all aspects of the data treatment, data combination, and error estimation, 
and to provide a versatile and modern database and software tool for wider use. 
This is particularly important given that current tensions in different evaluations
of $\amuHVPLO$ and in the $e^+e^-\to\pi^+\pi^-$ cross-section data indicate either a discovery of new physics or a multi-method confirmation of the SM. 
Crucially, different analysis choices in $e^+e^-\to\text{hadrons}$ data combinations 
by different groups can lead to different results and, in Ref.~\cite{Aoyama:2020ynm}, have been shown 
to differ at the level of the uncertainty on the combined cross 
section. It follows that future data-driven 
determinations of $\amuHVPLO$ must attempt to avoid analysis bias wherever possible,
including any past or future analysis choices on how to combine the available data. 
Consequently, implementing analysis blinding in data-driven determinations of 
$\amuHVPLO$ is paramount and critical before including new data whose impact 
on the resulting $\amuHVPLO$ will be influenced by such analysis choices. The 
first blinding scheme for data-driven evaluations of HVP has been developed by
KNTW and is in place as part of their new, ongoing KNTW analysis, see \cref{fig:KNTW-blinding}. The full blinding procedure is described in Ref.~\cite{Keshavarzi:2024wow}.
Unblinding will occur when the analysis is complete and at an appropriate time with 
respect to the release of new or updated hadronic cross-section data.

\subsection{Dispersive representations for exclusive channels}
\label{sec:disp}

For the dominant exclusive channels in a data-based evaluation of HVP, the form of the required hadronic matrix elements is sufficiently simple that strong constraints from general principles of QCD can be derived, including analyticity, unitarity, crossing symmetry, and chiral low-energy theorems. In Ref.~\cite{Aoyama:2020ynm} such constraints were already described for the $2\pi$~\cite{Colangelo:2018mtw,Ananthanarayan:2018nyx,Davier:2019can} and $3\pi$~\cite{Hoferichter:2019mqg} channels, here, we review the new developments in dispersive representations in these cases, as well as new applications to $\pi^0\gamma$ and $\bar K K$.

\subsubsection{\texorpdfstring{$2\pi$}{}}
\label{sec:disp_2pi}

\nocite{Omnes:1958hv}

\begin{figure}[!t]
\begin{center}
\includegraphics[width=0.7\textwidth]{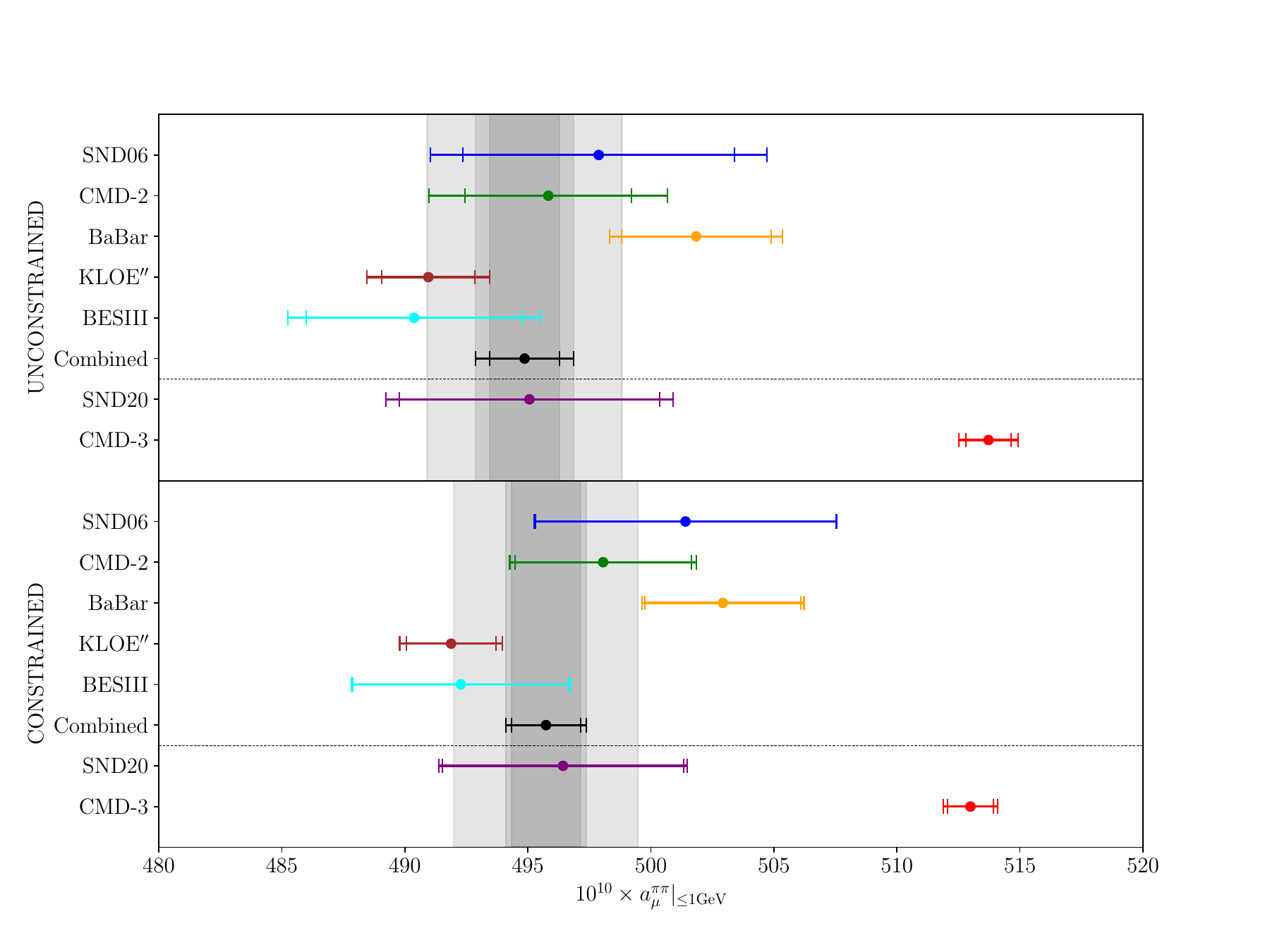}
\caption{$2\pi$ contribution to $a_\mu$ for $\sqrt{s}\leq 1\GeV$, in the unconstrained dispersive representation from Refs.~\cite{Colangelo:2022prz,Stoffer:2023gba} (upper panel) and imposing the absence of zeros in the VFF~\cite{Leplumey:2025kvv} (lower panel). The data sets are SND06~\cite{Achasov:2005rg,Achasov:2006vp}, CMD-2~\cite{CMD-2:2001ski,CMD-2:2003gqi,Aulchenko:2006dxz}, \babar~\cite{BaBar:2009wpw,BaBar:2012bdw}, KLOE~\cite{KLOE:2008fmq,KLOE:2010qei,KLOE:2012anl,KLOE-2:2017fda}, BESIII~\cite{BESIII:2015equ}, SND20~\cite{SND:2020nwa}, and CMD-3~\cite{CMD-3:2023rfe,CMD-3:2023alj}. The combined fit includes all data sets apart from SND20 and CMD-3. Inner error bars are derived from the experimental uncertainties, the outer ones include theory uncertainties from the dispersive representation. In the case of the combined fit, the outmost gray error band includes, in addition, an estimate of the \babar--KLOE tension as in Ref.~\cite{Aoyama:2020ynm}.
Figure taken from Ref.~\cite{Leplumey:2025kvv}.}
\label{fig:2pidisp}
\end{center}
\end{figure}

The central object that defines the $2\pi$ contribution to $a_\mu$ is the pion VFF $F_\pi^V(s)$, related to the cross section via
\begin{equation}
    \sigma(e^+e^-\to\pi^+\pi^-)(s)=\frac{\pi \alpha^2}{3s} \sigma_\pi^3(s) \big| F_\pi^V(s) \big|^2 \, , \quad \sigma_\pi(s) = \sqrt{1-\frac{4\mpi^2}{s}}\,.
\end{equation}
To derive dispersive constraints, it is most useful to separate the contributions from different cuts according to
\begin{equation}
\label{VFF_decomp}
	F_\pi^V(s) = \Omega_1^1(s) G_\omega(s) G_\text{in}^N(s)\,,
\end{equation}
where the Omn\`es factor $\Omega_1^1(s)$~\cite{Omnes:1958hv} accounts for $2\pi$ intermediate states, $G_\omega(s)$ includes the IB $3\pi$ cut, and further inelastic intermediate states such as $4\pi$ are expanded into a conformal polynomial $G_\text{in}^N(s)$ (with $N-1$ degrees of freedom). In the formalism from Ref.~\cite{Colangelo:2018mtw}, the strength of $G_\omega(s)$ is parameterized by the $\rho$--$\omega$ mixing parameter $\epsilon_\omega$, which can be interpreted as the residue at the $\omega$ pole, and apart from a tiny phase from the analytic continuation into the complex plane has to be a real quantity. In Ref.~\cite{Colangelo:2022prz}, this formalism was extended to account for radiative intermediate states mediating the $\rho$--$\omega$ transition, such as $\rho\to\pi^0\gamma\to\omega$, which can produce a small but relevant phase in $\epsilon_\omega$. The improved parameterization of $G_\omega(s)$ takes the form~\cite{Colangelo:2022prz}
\begin{align}
\label{eq:GomegaMod}
	G_\omega(s) = 1 &+ \frac{s}{\pi} \int_{9\mpi^2}^\infty ds^\prime \frac{\Re\epsilon_\omega}{s^\prime(s^\prime-s)} \Im\left[ \frac{s'}{(M_\omega - \frac{i}{2} \Gamma_\omega)^2 - s'} \right]  \left( \frac{1 - \frac{9\mpi^2}{s^\prime}}{1 - \frac{9\mpi^2}{M_\omega^2}} \right)^4 \nonumber\\
		&+ \frac{s}{\pi} \int_{M_{\pi^0}^2}^\infty ds^\prime \frac{\Im\epsilon_\omega}{s^\prime(s^\prime-s)} \Re\left[ \frac{s'}{(M_\omega - \frac{i}{2} \Gamma_\omega)^2 - s'} \right]  \left( \frac{1 - \frac{M_{\pi^0}^2}{s^\prime}}{1 - \frac{M_{\pi^0}^2}{M_\omega^2}} \right)^3\,,
\end{align}
ensuring the absence of imaginary parts below the respective thresholds and their correct threshold behavior above. Moreover, the size of the expected phase of $\epsilon_\omega$ could be estimated to $\delta_\epsilon=3.5(1.0)\degree$, based on a narrow-width  approximation for the intermediate-state vector mesons in $\pi^0\gamma$, $\pi\pi\gamma$, and $\eta\gamma$. A fit to the different $e^+e^-\to\pi^+\pi^-$ data sets was performed in Refs.~\cite{Colangelo:2022prz,Stoffer:2023gba}, leading to the picture in the upper panel of \cref{fig:2pidisp}.\footnote{The figure shows the implementation from Ref.~\cite{Leplumey:2025kvv}, which also involves improved error estimates for the $P$-wave elasticity parameter in the solution of $\pi\pi$ Roy equations~\cite{Roy:1971tc,Ananthanarayan:2000ht,Garcia-Martin:2011iqs,Caprini:2011ky} and the truncation of the conformal expansion.} As key result, it was found that all but the SND20 data set could be described in a statistically satisfactory manner, which was therefore excluded from the combined fit of experiments prior to CMD-3. Already in this unconstrained fit, the tension between CMD-3 and the combination evaluates to $7.3\sigma$ (excluding the additional \babar--KLOE tension represented by the outmost gray band in \cref{fig:2pidisp}).

\begin{figure}[!t]
\begin{center}
\raisebox{0.15cm}{\includegraphics[width=0.45\textwidth]{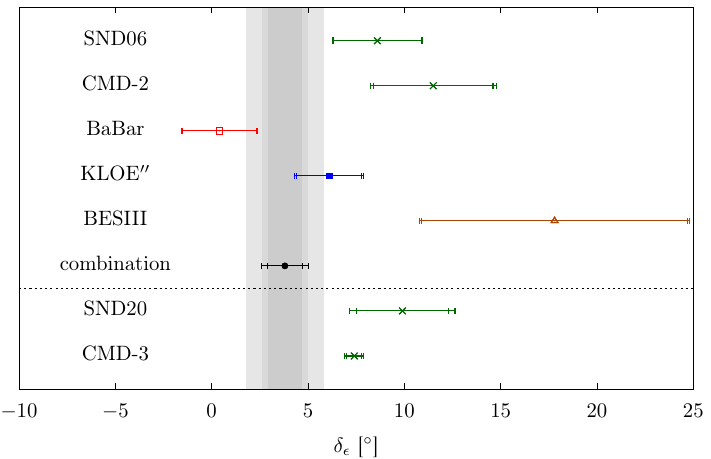}}
\includegraphics[width=0.52\textwidth]{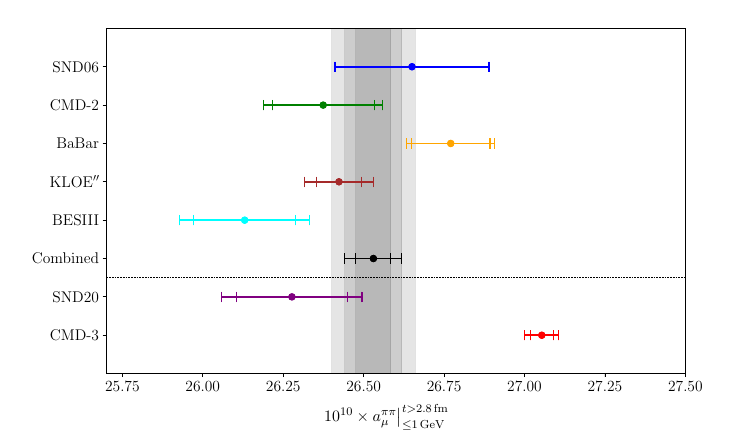}
\caption{Left: Phase $\delta_\epsilon$ of the $\rho$--$\omega$ mixing parameter, according to the dispersive representation~\cref{eq:GomegaMod}.
 Right: Two-pion contribution to the $t>2.8\,\text{fm}$ Euclidean time window (\cref{eq:win}, with $t_1=2.8\,\text{fm}$), obtained from low-energy fits of the VFF constrained to be free of zeros. Legends as in \cref{{fig:2pidisp}}, figures taken from Refs.~\cite{Stoffer:2023gba,Leplumey:2025kvv}.}
\label{fig:2piphase+VLDwindow2pi}
\end{center}
\end{figure}

The largest uncertainty in the dispersive representation of the VFF from Refs.~\cite{Colangelo:2018mtw,Colangelo:2022prz,Stoffer:2023gba} is a systematic (theory) uncertainty related to variations with the order $N$ of the conformal polynomial $G_\text{in}^N(s)$ describing the inelastic effects. In Ref.~\cite{Leplumey:2025kvv}, it was observed that these large variations coincide with the appearance of zeros in the VFF fits for $N\ge5$. Although the assumption that the VFF is free of complex zeros is not proved, there are theoretical arguments against the presence of such zeros~\cite{Leutwyler:2002hm}. In Refs.~\cite{Ananthanarayan:2011xt,RuizArriola:2024gwb}, zeros have been excluded in a large region of the complex $s$-plane. In Ref.~\cite{Leplumey:2025kvv} it was shown that imposing the absence of zeros as a constraint in the fit removes the largest systematic uncertainty without a significant impact on the fit quality. The results of these constrained fits are shown in the lower panel in~\cref{fig:2pidisp}. While qualitatively leading to similar results, the additional constraint exacerbates the discrepancies between dispersive fits to different experiments. Notably, the discrepancy between the fits to KLOE and CMD-3 reaches $8.9\sigma$~\cite{Leplumey:2025kvv}.

In case that the VFF has no zeros (or the position of zeros is known), its phase can be reconstructed dispersively from the modulus~\cite{PhysRev.172.1645,Geshkenbein:1998gu,Leutwyler:2002hm}. In Ref.~\cite{RuizArriola:2024gwb}, such a modulus dispersion relation was used to determine the phase of the VFF up to $s=(2.5\GeV)^2$ from \babar{} data, by parameterizing the data on the modulus with a GS~\cite{Gounaris:1968mw} function. This approach can also be combined with an Omn\`es representation for the elastic $\pi\pi$ rescattering contribution, resulting in a hybrid representation that requires input for the modulus only above the inelastic threshold~\cite{Ananthanarayan:2011xt,Ananthanarayan:2013zua}. In Ref.~\cite{Leplumey:2025kvv}, such a hybrid representation was fit to the different experiments at low energies and to \babar{} data~\cite{BaBar:2009wpw,BaBar:2012bdw} above $1.4\GeV$. The results are compatible with the unconstrained or constrained low-energy Omn\`es fits, but lead to even smaller uncertainties, at the price of introducing a parameterization dependence for the modulus above the inelastic threshold. While the representation of inelastic contributions in terms of a conformal polynomial $G_\text{in}^N(s)$ in \cref{VFF_decomp} is restricted to energies below $\simeq 1\GeV$, the hybrid representation has the advantage that it can be used also above $1\GeV$. An assessment of the parameterization dependence and a further reduction of the uncertainties would be possible with more high-statistics data on the modulus of the VFF above the inelastic threshold.

The dispersive representation of the VFF has the feature that it provides a global fit function: it thus allows one to evaluate each experiment in the entire region below $1\GeV$, not only in the range in which data were taken, and to sharpen the tensions among the experiments by taking into account that the allowed form of the cross section is highly constrained by analyticity and unitarity. Moreover, the fit parameters can be compared as well, implying interesting correlations with other low-energy observables, in particular the pion charge radius, see \cref{sec:corr_obs}. An example is provided in \cref{fig:2piphase+VLDwindow2pi}(left), where $\delta_\epsilon$ is shown for the fits to the different experiments. Again, considerable spread is observed among the experiments, but in a different order than for the cross section, comparing to \cref{fig:2pidisp}, with some experiments preferring phases much bigger than the narrow-width estimate $\delta_\epsilon=3.5(1.0)\degree$. Another example is given by Euclidean window quantities: as shown in Ref.~\cite{Leplumey:2025kvv}, the dispersive constraints correlate the discrepancies in the $\rho$-peak region with tensions even at very long Euclidean distance (corresponding to very low energies), see \cref{fig:2piphase+VLDwindow2pi}(right) for the spread of results for the two-pion contribution below $1\GeV$ to the Euclidean window $t>2.8\fm$.

In contrast to some other experiments, CMD-3~\cite{CMD-3:2023rfe,CMD-3:2023alj} does not provide detailed information on the bin-to-bin correlations of the systematic uncertainties, see \cref{TI-CMD3}. The fits of Refs.~\cite{Stoffer:2023gba,Leplumey:2025kvv} were performed assuming fully correlated systematic uncertainties. Since this assumption might not apply to all systematic effects, Ref.~\cite{Leplumey:2025kvv} also investigated a toy scenario of a de-correlation of the systematic uncertainties, ranging from fully correlated uncertainties to only diagonal errors, which shows some shifts in the central fit values: in the considered de-correlation scenario for CMD-3, the maximal shift of the central value for $a_\mu^{\pi\pi}|_{\le1\GeV}$ amounts to $1.8$ times the fit uncertainty obtained with fully correlated systematics, whereas for all other experiments, the shifts in the central values induced by neglecting correlations remain within the fit uncertainties. Since no detailed information is available on the systematic correlations, for a most conservative treatment of systematic errors the CMD-3 fit uncertainties should be somewhat enlarged. Since the observed shifts for CMD-3 are towards larger values of $a_\mu^{\pi\pi}|_{\le1\GeV}$, these studies indicate that reduced correlations would lead to even larger differences between CMD-3 and the other experiments than the treatment with fully correlated systematic uncertainties. In either case, these studies emphasize the need for improved information on the correlation of systematic uncertainties, especially in cases such as CMD-3 in which both statistical and truncation uncertainties in the conformal expansion are small.

\subsubsection{\texorpdfstring{$3\pi$}{} and \texorpdfstring{$\pi^0\gamma$}{}}
\label{sec:disp_3pi}

The $e^+e^-\to3\pi$ cross section can be described in a dispersive approach by its underlying $\gamma^*\to3\pi$ decay amplitude $\F(s,t,u;q^2)$ defined as
\begin{equation}
    \langle0|j_\mu(0)|\pi^+(p_+)\pi^-(p_-)\pi^0(p_0)\rangle = -\epsilon_{\mu\nu\alpha\beta}p_+^\nu p_-^\alpha p_0^\beta \F(s,t,u;q^2)\,,
\end{equation}
where $s$, $t$, and $u$ are  Mandelstam variables and $q^2$ is the photon virtuality. A dispersive representation of this amplitude was constructed in Refs.~\cite{Niecknig:2012sj,Schneider:2012ez,Hoferichter:2012pm,Hoferichter:2014vra,Hoferichter:2019mqg,Niehus:2021iin} in terms of a Khuri--Treiman equation~\cite{Khuri:1960zz}, and was used to evaluate the $3\pi$ contribution to $a_\mu$. The new developments since then mainly concern IB effects, with $\rho$--$\omega$ mixing and radiative corrections being incorporated into the dispersive representation in Ref.~\cite{Hoferichter:2023bjm}. The first effect, $\rho$--$\omega$ mixing, can be predicted by the coupled-channel formalism from Ref.~\cite{Holz:2022hwz}, from which the correction factor
\begin{equation}
   g_\pi(q^2)=1-\frac{g_{\omega\gamma}^2\epsilon_\omega}{e^2}\Pi_\pi(q^2)\,,
\end{equation}
depending on the $2\pi$ VP function $\Pi_\pi(q^2)$,
emerges. In particular, the $\rho$--$\omega$-mixing parameter $\epsilon_\omega$ is introduced in a way consistent with that of $e^+e^-\to2\pi$, see \cref{eq:GomegaMod}. The value of the mixing parameter obtained from the $3\pi$ global fit, including the data sets of SND~\cite{Achasov:2000am,Achasov:2002ud,Achasov:2003ir,SND:2020ajg}, CMD-2~\cite{Akhmetshin:1995vz,Akhmetshin:1998se,CMD-2:2003gqi,Akhmetshin:2006sc}, and \babar~\cite{BABAR:2021cde}, is in agreement with $\epsilon_\omega$ extracted from the $2\pi$ channel at the level of 1.9$\sigma$. An estimate of FSR was derived by focusing on IR-enhanced corrections, and a correction factor $\eta_{3\pi}$ analogous to  $\eta_{2\pi}$ in $2\pi$ channel was obtained for the first time, see Ref.~\cite{Hoferichter:2023bjm} for details. The total HVP contribution amounts to $a_\mu^{3\pi}|_{\leq 1.8\GeV}=45.91(53)\times 10^{-10}$, of which $a_\mu^{\rho\text{--}\omega}[3\pi]=-2.68(70)\times 10^{-10}$ and $a_\mu^\text{FSR}[3\pi]=0.51(1)\times 10^{-10}$ arise from IB.

The dispersive representation for $\F(s,t,u;q^2)$ is constrained by the Wess--Zumino--Witten (WZW) anomaly for $3\pi\gamma$~\cite{Wess:1971yu,Witten:1983tw}, which, in the chiral limit, predicts $\F(0,0,0;0)=F_{3\pi}^\text{WZW}=1/(4\pi^2 F_\pi^3)$ in terms of the pion decay constant $F_\pi$. Including quark-mass corrections~\cite{Bijnens:1989ff,Hoferichter:2012pm,Hoferichter:2017ftn}, this constraint anchors the dispersive representation directly at low energies and indirectly via a sum rule. In Ref.~\cite{Hoferichter:2025lcz} an extended data fit was performed to test this input, obtaining $F_{3\pi}/F_{3\pi}^\text{WZW}=1.028(53)$ and thus validating the anomaly constraint at the $5\%$ level. In addition to the global fit, also the recent Belle II data~\cite{Belle-II:2024msd} were considered, see \cref{sec:BelleII}, observing tensions with the dispersive constraints, the width parameters of $\omega$ and $\phi$, and the chiral anomaly; the latter reflecting the tension in the $3\pi$ contribution to $a_\mu$.

\nocite{Hoid:2020xjs,Achasov:2000zd,Achasov:2003ed,CMD-2:2004ahv,Achasov:2016bfr,Achasov:2018ujw}

\begin{table}[t]
\small
	\centering	\renewcommand{\arraystretch}{1.1}
	\begin{tabular}{lccccc}
	\toprule
	& $e^+e^-\to \pi^0\gamma$ & $e^+e^-\to\bar K K$ & \multicolumn{2}{c}{$e^+e^-\to 3\pi$} &\\
	& Ref.~\cite{Hoid:2020xjs} & Ref.~\cite{Stamen:2022uqh} & Ref.~\cite{Hoferichter:2019mqg} & Ref.~\cite{Hoferichter:2023bjm}  & PDG~\cite{ParticleDataGroup:2022pth}\\\midrule
	$M_\omega \ [\text{MeV}]$ & $782.584(28)$ & -- & $782.631(28)$ & $782.697(32)$ & $782.53(13)$\\
	$\Gamma_\omega \ [\text{MeV}]$ & $8.65(6)$ & -- & $8.71(6)$ & $8.711(26)$ & $8.74(13)$\\
	$M_\phi \ [\text{MeV}]$ & $1019.205(55)$ & $1019.219(4)$ & $1019.196(21)$ & $1019.211(17)$ & $1019.201(16)$\\
	$\Gamma_\phi \ [\text{MeV}]$ & $4.07(13)$ & $4.207(8)$ & $4.23(4)$ & $4.270(13)$ & $4.249(13)$\\
	\bottomrule
	\renewcommand{\arraystretch}{1.0}
	\end{tabular}
	\caption{VP-subtracted resonance parameters of $\omega$ and $\phi$ from $e^+e^-\to 3\pi$~\cite{Hoferichter:2019mqg,Hoferichter:2023bjm}, $e^+e^-\to\pi^0\gamma$~\cite{Hoid:2020xjs}, and $e^+e^-\to\bar K K$~\cite{Stamen:2022uqh}. The last column gives the PDG values~\cite{ParticleDataGroup:2022pth}, with VP removed using the corrections from Ref.~\cite{Holz:2022hwz}. Table from Ref.~\cite{Hoferichter:2023bjm}. The values from Ref.~\cite{Hoferichter:2019mqg} are superseded by Ref.~\cite{Hoferichter:2023bjm}, illustrating the impact of the \babar{} data~\cite{BABAR:2021cde}. At this level of precision, tensions among the different channels start to emerge, most notably in $\Gamma_\phi$ between $e^+e^-\to\bar K K$ and $e^+e^-\to 3\pi$.}
	\label{tab:vector_mesons}
\end{table}

Similarly, the $\pi^0\gamma$ channel is based on a dispersive representation of the singly-virtual $\pi^0\to \gamma \gamma^*$ transition form factor. A once-subtracted representation was implemented as~\cite{Hoid:2020xjs}
\begin{equation}
F_{\pi^0\gamma^*\gamma^*}(q^2,0)=F_{\pi\gamma\gamma}
+\frac{1}{12\pi^2}\int_{4M_\pi^2}^\infty d s' \frac{q_\pi^3(s')(F_\pi^{V}(s'))^* }{s'^{3/2}}
\times \bigg\{f_1(s',q^2)-f_1(s',0)
+\frac{q^2}{s'-q^2}f_1(s',0)\bigg\}\,.
\end{equation}
This process is closely related to the $3\pi$ channel since the $\gamma^*\to3\pi$ $P$-wave amplitude $f_1(s,q^2)$ turns out to be a building block of its unitarity relation. The resonance parameters were then fit to the existing data sets~\cite{Achasov:2000zd,Achasov:2003ed,CMD-2:2004ahv,Achasov:2016bfr,Achasov:2018ujw} below 1.4\GeV. The final combined fit produced the HVP contribution to $a_\mu$ from this lowest radiative channel, $a_\mu^{\pi^0\gamma}\big|_{\leq 1.35\GeV}=43.8(6)\times 10^{-11}$. As a by-product, the resonance parameters of $\omega$ and $\phi$ can be compared to the determinations from other channels, as shown in~\cref{tab:vector_mesons}.\footnote{The resonance parameters also play a crucial role in R$\chi$T and related approaches, see Refs.~\cite{Qin:2020udp,Benayoun:2021ody,Wang:2023njt,Qin:2024ulb} for recent applications to evaluations of the HVP contribution.} In particular, good agreement of the $\omega$ mass was found between $3\pi$ and $\pi^0\gamma$ in the dispersive approach, evading unphysical phases that afflict previous conflicting determinations in the same channel~\cite{CMD-2:2004ahv}.

\subsubsection{\texorpdfstring{$\bar K K$}{}}
\label{sec:disp_KKbar}

In Ref.~\cite{Stamen:2022uqh}, a dispersive analysis of the charged- and neutral-kaon EM form factors was performed, based on their lowest-lying singularities, and thus allowing one to correlate both time- and spacelike data.  For this purpose, the physical form factors are decomposed into isovector~($v$) and isoscalar~($s$) components according to
\begin{equation}
    F_{K^{\pm}}(s) = F_K^s(s) + F_K^v(s)\,, \qquad
    F_{K^{0}}(s) = F_K^s(s) - F_K^v(s)\,.
\end{equation}
The unitarity relation for the isovector part is, at low energies, entirely dominated by two-pion intermediate states, resulting in~\cite{Blatnik:1978wj}
\begin{equation}
\Im F^v_K(s)=\frac{s}{4\sqrt{2}}\sigma^3_\pi(s)\big(g_1^{1}(s)\big)^*F_\pi^V(s)\,. \label{eq:ImFvK}
\end{equation}
Here, $g_1^1(s)$ denotes the $\pi\pi\to\bar KK$ $P$-wave amplitude known from corresponding Roy--Steiner analyses~\cite{Buettiker:2003pp,Pelaez:2018qny,Pelaez:2020gnd}.  The isovector kaon form factor itself is calculated from \cref{eq:ImFvK} using an unsubtracted dispersion relation.  Higher intermediate states are modeled effectively by a $\rho'$-pole contribution, whose strength is adjusted to fulfill the normalization sum rule $F_K^v(0)=1/2$.  The isoscalar part, on the other hand, is dominated by the narrow $\omega(782)$ and $\phi(1020)$ resonances.  While the $\phi$ dominates the $e^+e^-\to\bar KK$ cross sections above threshold, and its spectral form needs to be described carefully using energy-dependent widths~\cite{Hoferichter:2014vra}, the $\omega$ lies in the unphysical region and can be constrained only to some extent as a background effect, from spacelike data~\cite{Dally:1980dj,Amendolia:1986ui}, or using SU(3) symmetry.  Once more, a heavier, effective, $\omega'$ pole is added to ensure the correct isoscalar normalization.

\begin{figure}[t]
\centering
\includegraphics[width=0.48\linewidth]{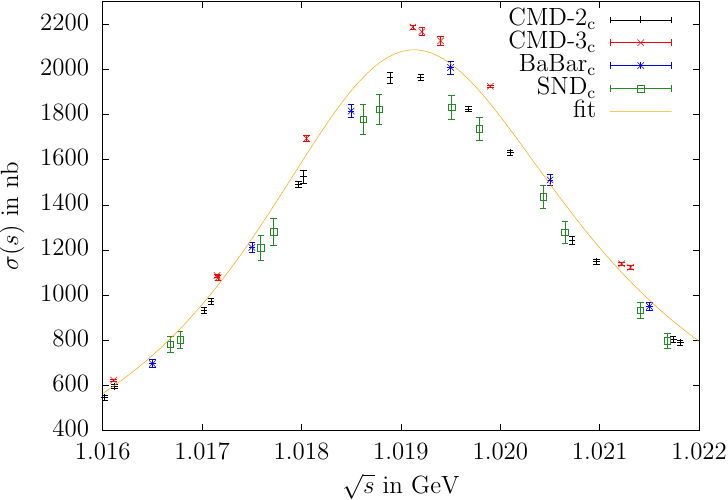}
\includegraphics[width=0.48\linewidth]{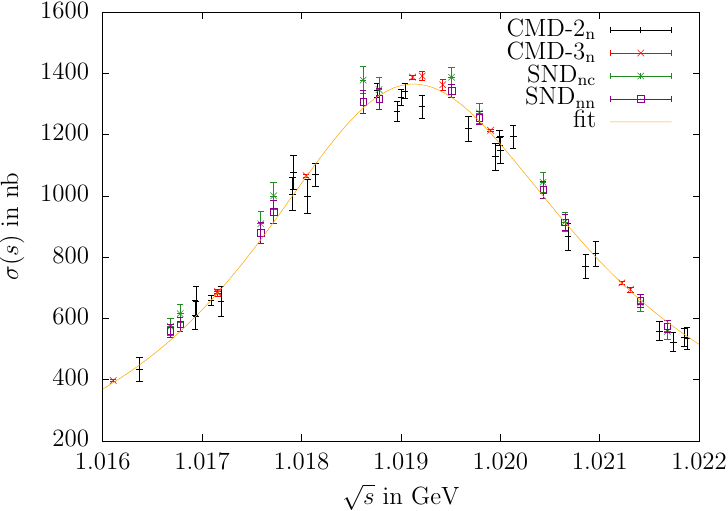}
\caption{Cross-section data and combined fits to all data sets for $e^+e^-\to K^+K^-$ (data from Refs.~\cite{CMD-2:2008fsu,Kozyrev:2017agm,BaBar:2013jqz,Achasov:2000am}, left) and $e^+e^-\to K_SK_L$ (data from Refs.~\cite{CMD-2:2003gqi,CMD-3:2016nhy,Achasov:2000am,Achasov:2000am}, right), respectively. Both are shown focused on the $\phi$ peak region. Figure adapted from Ref.~\cite{Stamen:2022uqh}. \label{fig:KK}}
\label{fig:fits}
\end{figure}

An interesting observation is that combined fits to the $e^+e^-\to K^+K^-$~\cite{CMD-2:2008fsu,Kozyrev:2017agm,Achasov:2000am,BaBar:2013jqz} and $e^+e^-\to K_SK_L$~\cite{CMD-2:2003gqi,CMD-3:2016nhy,Achasov:2000am} data sets suggest an IB difference in the $\phi$ residues at the level of $2.6(9)\%$.  This is found despite the kaon mass difference, universal FSR for the charged kaons, and the isovector background in the form factors being taken into account.  Mass and width of the $\phi$ come out consistently with determinations from the $3\pi$ and $\pi^0\gamma$ channels, cf.\ \cref{tab:vector_mesons}.
The combined HVP contributions in the $\phi$-resonance region are found to be
\begin{equation}
    \label{eq:KK-HVP}
 a_\mu^\text{HVP}[K^+K^-, \leq 1.05\GeV]=184.5(2.0)\times 10^{-11} \,, \qquad a_\mu^\text{HVP}[K_SK_L, \leq 1.05\GeV]=118.3(1.5)\times 10^{-11} \,,
\end{equation}
while results based on individual data sets reflect the tensions between the data, in particular \babar~\cite{BaBar:2013jqz} and \mbox{CMD-3}~\cite{Kozyrev:2017agm} for the charged kaons; see \cref{fig:KK} and Ref.~\cite{Stamen:2022uqh}.

The resulting spacelike form factors have also been used to determine the kaon-box contributions to HLbL, see \cref{sec:TwoMesonContributions}.
Besides, radii for both neutral and charged kaons can be extracted, yielding
\begin{equation}
    \langle r^2 \rangle_\text{n} =-0.060(4)\fm^2 \,,\qquad
    \langle r^2 \rangle_\text{c} =0.359(3)\fm^2 \,,
\end{equation}
which both present a significant gain in precision compared to the values cited by the PDG~\cite{ParticleDataGroup:2024cfk}.
Finally, knowledge of the form factors allows one to evaluate the elastic contributions to the EM mass shifts from the Cottingham formula~\cite{Cottingham:1963zz},
\begin{equation}
\label{Cottingham}
    \big(M_K^2\big)_\text{EM}
=\frac{\alpha}{8\pi}\int_0^\infty d s \,\big[F_{K}(-s)\big]^2 \bigg(4W+\frac{s}{M_{K}^2}\left(W-1\right)\bigg) \,,  \qquad W=\sqrt{1+\frac{4M_K^2}{s}} \,,
\end{equation}
which results in an estimate of the EM charged-to-neutral kaon mass difference $(\Delta M_K^2)_\text{EM}=2.12(18)\times 10^{-3}\GeV^2$ and allows one to disentangle EM from quark-mass-induced effects
in $\Delta M_K^2 = M_{K^\pm}^2-M_{K^0}^2$.  This separation serves as important input to the discussion of IB in \cref{sec:isospin_breaking}. In particular, the scheme implicitly defined by \cref{Cottingham}, and the resulting mass decomposition in \cref{mass_decomposition}, agree well with the FLAG-recommended convention in lattice QCD, see \cref{sec:Breakdown}. 

\subsection{Applications of dispersive representations}
\label{sec:disp_app}

\subsubsection{Correlations with other observables}
\label{sec:corr_obs}

Having a dispersive representation for the $2\pi$ cross section allows one to study interesting correlations with other low-energy observables, see, e.g., Ref.~\cite{Colangelo:2020lcg}. First, the resulting values of the $P$-wave $\pi\pi$ phase shift at $s_0=(0.8\GeV)^2$ and $s_1=(1.15\GeV)^2$, which enter as fit parameters in $\Omega_1^1(s)$ via the Roy-equation solution, can be compared to partial-wave analyses~\cite{Hyams:1973zf,Protopopescu:1973sh}. The result of the global VFF fit proves consistent with these partial-wave solutions, but appreciably more precise, to the extent that the corresponding phase shift enters as input for global analyses of $\pi\pi$ scattering~\cite{Pelaez:2024uav}. Surprisingly, even for the fits to CMD-3 the change in these phase-shift values is small~\cite{Stoffer:2023gba,Leplumey:2025kvv}, essentially realizing scenario (2) from Ref.~\cite{Colangelo:2020lcg} in which all changes occur in the conformal polynomial $G_\text{in}^N(s)$, see \cref{sec:disp_2pi}. This observation could potentially allow one to discriminate among the $2\pi$ data sets using additional input from $e^+e^-\to 4\pi, \pi\omega$ data, and work in this direction is ongoing~\cite{Chanturia:2022rcz,Heuser:2024biq}.

Next, in addition to studying $\delta_\epsilon$, the results for the $\omega$ pole parameters in VFF fits can be contrasted to determinations from $e^+e^-\to 3\pi$ and $e^+e^-\to\pi^0\gamma$, see \cref{sec:disp_3pi}. While the sensitivity cannot compete with the $3\pi$ channel, some deficit in $M_\omega$ tends to remain, correlated with $\delta_\epsilon$ (in line with a similar observation in Ref.~\cite{BaBar:2012bdw} in the context of GS fits).

\begin{figure}[t!]
\begin{center}
\includegraphics[width=0.65\textwidth]{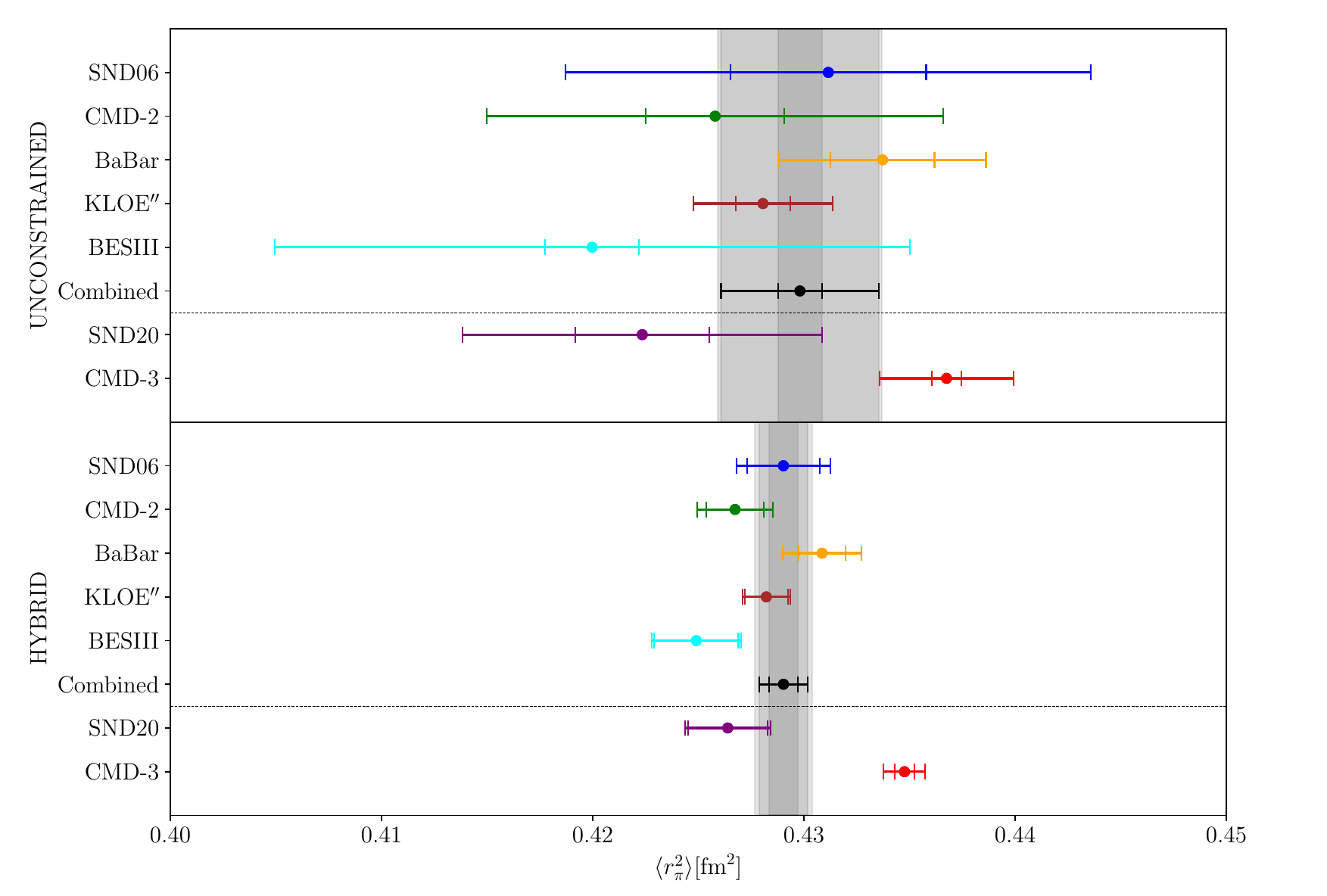}
\caption{Values of the pion charge radius $\langle r_\pi^2 \rangle$ (in fm$^2$) from the (unconstrained) Omn\`es representation \cref{VFF_decomp} and from the hybrid representation of Ref.~\cite{Leplumey:2025kvv}, fit to single experiments. Figure taken from Ref.~\cite{Leplumey:2025kvv}, legend as in \cref{{fig:2pidisp}}.}
\label{fig:pion_charge_radius}
\end{center}
\end{figure}

Finally, once the pion VFF is determined, the pion charge radius follows via the sum rule
\begin{equation}
	\label{eq:rpiSumRule}
	\langle r_\pi^2\rangle=6\frac{d F_\pi^V(s)}{d s}\bigg|_{s=0}=\frac{6}{\pi}\int_{4\mpi^2}^\infty d s\frac{\Im F_\pi^V(s)}{s^2}\,,
\end{equation}
defining another low-energy observable that could help discriminate among the $2\pi$ data sets in case lattice-QCD calculations~\cite{Koponen:2015tkr,Feng:2019geu,Wang:2020nbf,Gao:2021xsm} at the required level of precision became available. For such a comparison, the improvements in the evaluation of the inelastic contributions in Ref.~\cite{Leplumey:2025kvv} become critical, since, with the imaginary part entering in \cref{eq:rpiSumRule} in principle up to arbitrarily high energies, $\langle r_\pi^2\rangle$ displays an increased sensitivity  to $G_\text{in}^N$. In particular, the hybrid representation of Ref.~\cite{Leplumey:2025kvv} leads to significantly smaller uncertainties in $\langle r_\pi^2\rangle$ than the unconstrained Omn\`es representation \cref{VFF_decomp}, see \cref{fig:pion_charge_radius}, but in all fits the hybrid representation relies on \babar{} data above $1.4\GeV$. Therefore, the main results of Ref.~\cite{Leplumey:2025kvv} for the pion charge radius are based on the hybrid representation, fit to a combination of all data sets apart from SND20 and CMD-3, as well as the constrained Omn\`es representation, fit to CMD-3, leading to
\begin{equation}
	\langle r_\pi^2\rangle |^\text{comb} = 0.4290(17)\,\text{fm}^2 \, , \qquad
	\langle r_\pi^2\rangle |^\text{CMD-3} = 0.4367(24)\,\text{fm}^2 \, .
\end{equation}
Apart from these low-energy correlations, connections also exist with $Z$-pole observables via the hadronic running of the fine-structure constant, see Refs.~\cite{Passera:2008jk,Crivellin:2020zul,Keshavarzi:2020bfy,Malaescu:2020zuc} and \cref{sec:running_alpha}.

\subsubsection{Chiral extrapolation}
\label{sec:chiral_extrapolation}

A dispersive representation of the $2\pi$ cross section further allows one to improve chiral extrapolations. That is, the pion-mass dependence of \cref{VFF_decomp}, and thus the two-pion contribution to $a_\mu$, can be evaluated once chiral extrapolations of the phase shift $\delta_1^1(s)$ and the conformal polynomial are available. For the former, one may use the inverse-amplitude method (IAM)~\cite{Guo:2008nc,Colangelo:2021moe}, relying on input for the chiral amplitudes up to two loops~\cite{Bijnens:1995yn,Niehus:2020gmf} to be able to assess the convergence by comparing NLO and NNLO results. For the conformal polynomial, the pion-mass dependence can be estimated by matching to ChPT for the pion VFF, e.g., the charge radius, whose $\mpi$ dependence is also known up to two loops~\cite{Bijnens:1998fm}, allows one to include the first nontrivial term in the conformal expansion. In particular, resonance saturation for the required two-loop low-energy constant $r_{V1}^r$ is validated by lattice-QCD calculations of $\langle r^2_\pi\rangle$ at heavier-than-physical pion masses~\cite{Feng:2019geu,Colangelo:2021moe}.

\begin{figure}[t]
\begin{center}
\includegraphics[width=0.5\textwidth]{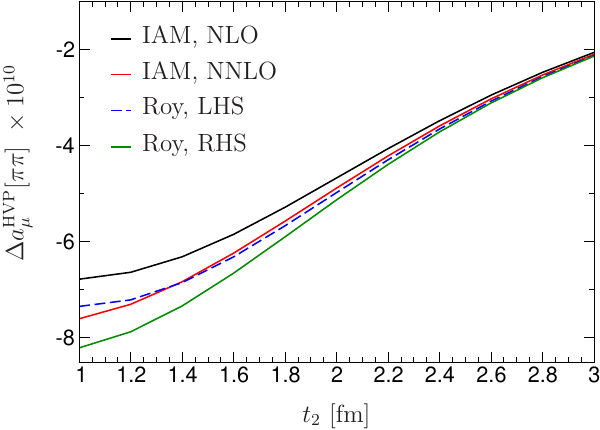}
\caption{Projection of the pion-mass-difference correction to the two-pion contribution $\Delta \amuHVPLO[\pi\pi]$ onto the Euclidean window $[t_2,\infty)$. The curves refer to the IAM at NLO and NNLO as well as the LHS and RHS of the Roy-equation solution, respectively. Based on Refs.~\cite{Colangelo:2021moe,Hoferichter:2023sli}.}
\label{fig:pion_mass_difference}
\end{center}
\end{figure}

A detailed analysis of the pion-mass dependence of the two-pion contribution along these lines was performed in Ref.~\cite{Colangelo:2021moe}. The results inform the chiral extrapolation (or interpolation) of lattice-QCD calculations, e.g., suggesting the presence of an $1/\mpi^2$ term in empirical fits in the range $\mpi\in[0.14,0.25]\GeV$. Moreover, also the important IB effect when performing iso-symmetric calculations at the mass of the neutral pion can be estimated, see \cref{fig:pion_mass_difference} for the projection of this correction onto the Euclidean window $[t_2,\infty)$. As a cross-check, the same effect was evaluated in Ref.~\cite{Hoferichter:2023sli} from a Roy-equation analysis, determining the $\pi\pi$ scattering lengths at the mass of the neutral pion from two-loop ChPT~\cite{Bijnens:1995yn,Niehus:2020gmf,Colangelo:2001df}, where the difference between left-hand side (LHS) and right-hand side (RHS) of the Roy equations provides an estimate of the systematic uncertainty. \Cref{fig:pion_mass_difference} shows that the different variants agree very well especially in the long-distance tail, while uncertainties increase towards $t_2=1\,\text{fm}$. This result enters the phenomenological estimate of IB corrections discussed in \cref{sec:isospin_breaking}.

\subsubsection{Isospin breaking}
\label{sec:isospin_breaking}

\nocite{Hoferichter:2022iqe}

\begin{table}[t]
\small
	\renewcommand{\arraystretch}{1.1}
	\centering
\begin{tabular}{crrrrrrrr}
\toprule
& \multicolumn{2}{c}{SD} & \multicolumn{2}{c}{W} & \multicolumn{2}{c}{LD} & \multicolumn{2}{c}{Full}\\
& $\Order(e^2)$ & $\Order(\delta)$& $\Order(e^2)$ & $\Order(\delta)$& $\Order(e^2)$ & $\Order(\delta)$& $\Order(e^2)$ & $\Order(\delta)$\\\midrule
$\pi^0\gamma$ & $0.16(0)$ & -- & $1.52(2)$  & -- & $2.70(4)$ & -- & $4.38(6)$ & --\\
$\eta\gamma$ & $0.05(0)$ & -- & $0.34(1)$ & -- & $0.31(1)$ &-- & $0.70(2)$ & --\\
$\omega(\to\pi^0\gamma)\pi^0$ & $0.15(0)$ & -- & $0.54(1)$ & -- & $0.19(0)$ &-- & $0.88(2)$ & --\\\midrule
FSR ($2\pi$) & $0.12(0)$ & -- & $1.17(1)$ & -- & $3.13(3)$ &-- & $4.42(4)$ & --\\
FSR ($3\pi$) & $0.03(0)$ & -- & $0.20(0)$ & -- & $0.28(1)$ &-- & $0.51(1)$ & --\\
FSR ($K^+ K^-$)  & $0.07(0)$ & -- & $0.39(2)$ & -- & $0.29(2)$ &-- & $0.75(4)$ & --\\\midrule
$\rho$--$\omega$ mixing ($2\pi$) & -- & $0.06(1)$ & -- & $0.86(6)$ & -- & $2.87(12)$ & -- & $3.79(19)$\\
$\rho$--$\omega$ mixing ($3\pi$) & -- & $-0.13(3)$ & -- & $-1.03(27)$ & -- & $-1.52(40)$ & -- & $-2.68(70)$\\\midrule
Pion mass ($2\pi$) & $0.04(8)$ & -- & $-0.09(56)$ & -- & $-7.62(63)$ & --& $-7.67(94)$ &--\\
Kaon mass ($K^+ K^-$)  & $-0.29(1)$ & $0.44(2)$ & $-1.71(9)$ & $2.63(14)$ & $-1.24(6)$ & $1.91(10)$ & $-3.24(17)$ & $4.98(26)$\\
Kaon mass ($\bar K^0 K^0$)  & $0.00(0)$ & $-0.41(2)$  & $-0.01(0)$ & $-2.44(12)$ & $-0.01(0)$ & $-1.78(9)$ & $-0.02(0)$ & $-4.62(23)$\\\midrule
Sum of channels &  $0.33(8)$ & $-0.04(4)$ & $2.34(57)$ & $0.02(33)$ & $-1.97(63)$ & $1.48(44)$ & $0.71(95)$ & $1.47(80)$\\
$\phi$ residue & $0.00(8)$ & $0.00(8)$ & $0.00(47)$ & $0.00(47)$ & $0.00(36)$ & $0.00(36)$ & $0.00(90)$ & $0.00(90)$\\
Missing channels & $0.00(49)$ & $0.00(49)$ & $0.00(55)$ & $0.00(55)$ & $0.00(12)$ & $0.00(12)$ & $0.00(1.16)$ & $0.00(1.16)$\\\midrule
Sum  & $0.33(51)$ & $-0.04(50)$ & $2.34(92)$ & $0.02(79)$ & $-1.97(74)$ & $1.48(58)$ & $0.71(1.75)$ & $1.47(1.67)$\\
\bottomrule
\renewcommand{\arraystretch}{1.0}
\end{tabular}
\caption{IB contributions to $\amuHVPLO$ from the various radiative channels, FSR, $\rho$--$\omega$ mixing, and threshold effects, separated into $\Order(e^2)$ and $\Order(\delta)$ contributions as well as the RBC/UKQCD SD, intermediate, and LD windows. $\rho$--$\omega$ mixing is booked as $\Order(\delta)$ following the LO R$\chi$T argument from Ref.~\cite{Urech:1995ry}, see Ref.~\cite{Colangelo:2022prz} for details.
The penultimate panel reflects the uncertainties from the quadratic sum of the individual-channel errors, the ambiguity in the $\phi$ residues, and a generic $1\%$ error for the channels not explicitly included, while the last line gives the total quadratic sum.  Table adapted from Ref.~\cite{Hoferichter:2023sli}, all entries in units of $10^{-10}$.}
	\label{tab:IB_channels}
\end{table}

The detailed understanding of the dominant exclusive channels summarized in \cref{sec:disp} presents an opportunity to estimate a number of IB effects from phenomenology. First, the radiative channels $\pi^0\gamma$, $\eta\gamma$, and $\pi^0\pi^0\gamma$ are naturally booked as QED contributions $\Order(e^2)$. They are large compared to the naive expectation $\frac{\alpha}{\pi}\amuHVPLO\simeq 1\times 10^{-10}$, especially considering the energy suppression of the latter two channels. Their size can be explained by resonance enhancement, due to $V=\omega,\rho,\phi$ for the $P\gamma$ channels and a double enhancement due to $\rho(1450)\to\omega\pi^0\to \pi^0\pi^0\gamma$. Next, FSR effects contribute to the $2\pi$, $3\pi$, and $K^+K^-$ channels, and can again be extracted from dispersive fits to data, as can the contribution due to $\rho$--$\omega$ mixing thanks to the improved representations discussed in \cref{sec:disp_2pi} and \cref{sec:disp_3pi}. For the pion-mass correction when defining the isospin limit by the mass of the neutral pion the treatment according to \cref{sec:chiral_extrapolation} applies. Finally, for the kaon channels one needs to further specify the isospin conventions, for which Refs.~\cite{Hoferichter:2023sli,Hoferichter:2022iqe} suggest the decomposition
\begin{equation}
\label{mass_decomposition}
M_{K^\pm} = \big(494.58-3.05_\delta + 2.14_{e^2}\big)\MeV\,,\qquad
M_{K^0} = \big(494.58+3.03_\delta\big)\MeV\,,
\end{equation}
into $\Order(e^2)$ and $\Order(\delta)$, $\delta=m_u-m_d$, contributions, as
motivated by the kaon self energies determined via the Cottingham approach~\cite{Stamen:2022uqh}, see \cref{sec:disp_KKbar}. \Cref{mass_decomposition} agrees well with the FLAG prescription in lattice QCD, see \cref{sec:Breakdown}.

The complete summary of all IB effects collected in Ref.~\cite{Hoferichter:2023sli} is reproduced in \cref{tab:IB_channels}, separated into the benchmark RBC/UKQCD  SD, intermediate, and LD Euclidean windows as defined in \cref{sec:windows}, as well as $\Order(e^2)$ and $\Order(\delta)$ contributions. For each channel and effect, the uncertainties are well defined and reasonably small, so that the final uncertainties are dominated by potential IB corrections that cannot be fully quantified. The first of these concerns the residue of the $\phi$ in the $\bar K K$ decay, since it is not clear with which channel the isospin limit should be identified. Accordingly, the difference between the possible choices is added as an additional source of uncertainty.\footnote{A similar effect also occurs for the $\omega$ residue in the $3\pi$ channel, which could subsume IB effects, but in this case no phenomenological way to estimate the potential impact is available. Assigning an additional $1\%$ uncertainty for the $3\pi$ channel would amount to $\{0.03, 0.19,0.25,0.46\}$, for SD, intermediate, LD, and total.} Finally, higher multiplicity channels such as $4\pi$ may also involve IB contributions, but in those cases no strong enhancements from resonances, threshold effects, or IR cancellations are expected, so that a generic $1\%$ error is assigned in \cref{tab:IB_channels}.

From the sum of exclusive channels one observes that individually large contributions tend to cancel in the sum, leaving remarkably small overall corrections. Moreover, for the SD window a null effect is predicted, with a large uncertainty dominated, as expected, by the $1\%$ estimate of missing channels. In the intermediate window, such effects still play an important role, while evidence for a nonvanishing effect is found largely driven by the radiative channels. The most robust prediction concerns the LD window, in which case the contribution from higher-multiplicity channels is very small and the dominant effects well under control.  As illustrated in \cref{fig:pion_mass_difference} for the case of the pion-mass correction, the uncertainties will reduce further if the lower boundary $t_2$ is increased.
The complementarity to lattice-QCD results is discussed in \cref{sec:totalHVP}, especially for the LD window. Within uncertainties, the full $\Order(\delta)$ result given in \cref{tab:IB_channels} agrees with the ChPT analysis of Ref.~\cite{James:2021sor}.

\subsubsection{Radiative corrections}
\label{sec:radiative_corrections}

At the required level of precision, radiative corrections for the main
$\pi^+\pi^-$ channel are important and need to be evaluated carefully.
In particular, since the $2 \pi$ final state is produced
in an $e^+e^-$ collision, one needs to remove all effects produced by photons
emitted by the electron--positron pair before the collision, i.e., all
ISR effects, see Ref.~\cite{Aliberti:2024fpq} and \cref{sec:MC} for the required MC framework. In this section, we address the application of dispersive techniques for the calculation of radiative corrections, starting from FSR and interference effects.
For both contributions, it is
necessary to perform calculations in QCD+QED and to rely on a perturbative
expansion only for the latter. At leading order in $\alpha$, one needs
hadronic matrix elements involving one, two, or three EM
currents in the nonperturbative regime, making model-independent calculations particularly challenging.

In principle, all experiments measuring $e^+e^- \to \pi^+\pi^-$ aim to determine the inclusive $2 \pi (\gamma)$ final state, so that a calculation of FSR effects would not be strictly necessary if a measurement with complete angular coverage were possible. In practice, corrections are so far evaluated using sQED multiplied by the pion VFF (F$\times$sQED in the classification of Ref.~\cite{Aliberti:2024fpq}), leading to a universal correction factor $\eta_{2\pi}(s)$ that captures the leading, IR-enhanced effects~\cite{Hoefer:2001mx,Czyz:2004rj,Gluza:2002ui,Bystritskiy:2005ib}, see also \cref{sec:disp_3pi}. Removing this effect is not only important for the application of dispersive techniques to the QCD matrix elements, see~\cref{sec:disp_2pi}, but also allows one to estimate one class of isospin corrections for $\tau$ decays, see \cref{sect:tau-dispersive}.  A first step towards a dispersive calculation of FSR corrections was performed in Ref.~\cite{Moussallam:2013una}, supporting the assumption that the IR-enhanced contributions are numerically dominant. Here, we discuss the recent approach from Refs.~\cite{Monnard:2021pvm,CCRdE-WiP1,CCRdE-WiP2}.
Since only the two-pion final state is considered, the quantity of interest becomes the VFF of the pion in QCD+QED, for which an expansion in
$\alpha$ is employed:
\begin{equation}
F_{\pi,\,\mathrm{QED}}^V(s)=F_{\pi}^V(s)+F_{\pi}^{V,\,\alpha}(s)+
F_{\pi}^{V,\,\alpha^2}(s) + \ldots \,.
\end{equation}
The first term is the VFF in pure QCD, while the goal of the dispersive analysis is a complete calculation of the second term, the
correction at $\mathcal{O}(\alpha)$, by reconstructing
this matrix element  from its unitarity cuts. For
$F_{\pi}^{V,\,\alpha}(s)$ there are three possible cuts that determine its discontinuity:
\begin{equation}
\text{disc}\raisebox{-1.1cm}{
\includegraphics[width=3cm]{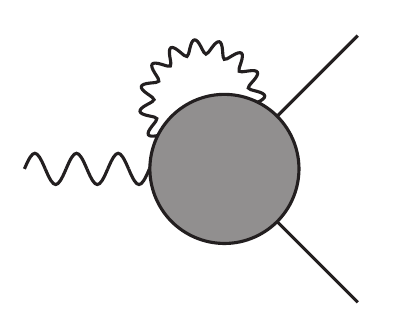}}
=
\raisebox{-1.1cm}{
\includegraphics[width=3cm]{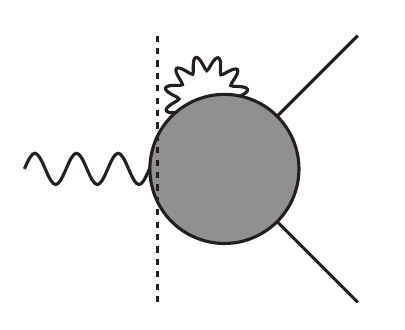}}
+
\raisebox{-1.1cm}{
\includegraphics[width=3cm]{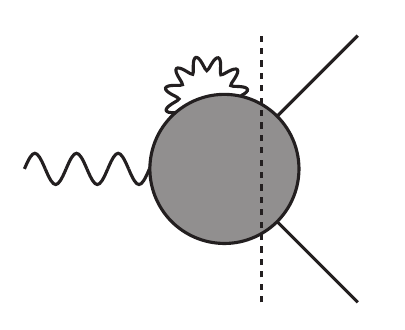}}
+
\raisebox{-1.1cm}{
\includegraphics[width=3cm]{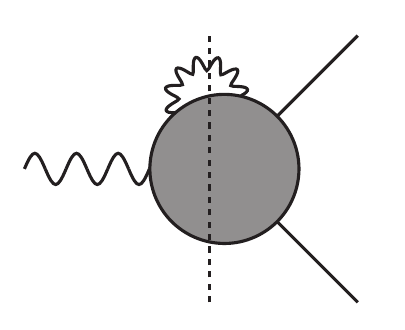}}\,,
\end{equation}
where the vertical dashed lines represent the unitarity cuts, the gray blob the hadronic matrix element, and when cutting through it, in principle, all possible hadronic intermediate
states have to be considered. However, since we are interested only in the energy region around 1 GeV
and below---for which inelastic contributions to $F_{\pi}^V(s)$
are known to be small---the analysis can be restricted to two-pion intermediate states in a first step.
This allows one to translate this schematic representation into a sum
of products of well-defined sub-amplitudes by replacing the cut hadronic blob with a two-pion state. These contributions are integrated over a two- or three-body phase space, so that the discontinuity becomes
\begin{align}
\frac{\text{disc}\,F_{\pi}^{V,\,\alpha}(s)}{2i} &=\frac{(2\pi)^4}{2} \left[\int
d\Phi_2 F_{\pi}^{V}(s) T_{\pi\pi}^{\alpha *}(s,t) +
\int d\Phi_2 F_{\pi}^{V,\,\alpha}(s) T_{\pi\pi}^{
  *}(s,t) + \int d\Phi_3
F_{\pi}^{V,\,\gamma}(s,t)T_{\pi\pi}^{\gamma *}(s,\{t_i\}) \right]\,,
\label{ImFvpi}
\end{align}
where each term corresponds to a cut in \cref{ImFvpi}. The first term
is the product of the purely hadronic VFF with the
$\mathcal{O}(\alpha)$ correction to the $\pi\pi$ scattering amplitude
$T_{\pi\pi}^{\alpha}(s,t)$. The second term is the product of
$F^{V,\,\alpha}_{\pi}(s)$ with the purely hadronic $\pi\pi$ scattering
amplitude. In the last term, $F_{\pi}^{V,\,\gamma}(s,t)$ and
$T_{\pi\pi}^{\gamma}(s,\{t_i\})$ are the transition amplitudes for the processes $\gamma^* \to \pi^+ \pi^- \gamma$ and $\pi^+\pi^-
\to \pi^+\pi^- \gamma$, respectively, where the latter involves five external particles, introducing a dependence on five Mandelstam variables ($s$ and $t_i$, $i \in \{1,2,3,4\}$).
The three-body phase-space integral of the last term is due to the presence of an additional photon in the intermediate state, contrary to the two first terms.

If one is able to determine the discontinuity of $F_{\pi}^{V,\,\alpha}(s)$
from this expression, a dispersive integral will allow one to reconstruct
the whole function. Among all the sub-amplitudes appearing on the RHS of \cref{ImFvpi},
only the pion VFF in pure QCD can be considered to
be known, while all other amplitudes need to be determined. A striking
feature of this equation is that the initial amplitude
$F_{\pi}^{V,\,\alpha}(s)$ also appears as a sub-amplitude in the two-body phase-space integral of the second term. Therefore, determining $F_{\pi}^{V,\,\alpha}(s)$
not only requires additional input, such as $T_{\pi\pi}^{\alpha}(s,t)$,
but also involves solving an implicit integral equation once all sub-amplitudes are known.
The building blocks $T_{\pi\pi}^{\alpha}(s,t)$,
$F_{\pi}^{V,\,\gamma}(s,t)$, and $T_{\pi\pi}^{\gamma}(s,\{t_i\})$ can also be determined
using a dispersive approach, following similar lines to those discussed here.
A preliminary account of this determination, along with a numerical estimate of these effects, is provided in Ref.~\cite{Monnard:2021pvm}.
Based on these findings, no significant shifts with respect to the F$\times$sQED treatment are expected. Ongoing studies of these effects are addressing, in particular, the following important aspects:
\begin{enumerate}
\item
Self-energy photonic corrections to the pions are responsible for the mass
difference between the charged and the neutral pions. These effects can be
analyzed by solving Roy equations in the presence of a nonzero pion mass
difference, see Ref.~\cite{CCRdE-WiP1}.
\item
An important step in the determination of $F_{\pi}^{V,\,\alpha}(s)$ within a
dispersive approach is represented by the matching to the chiral
representation of the same quantity (which is available in the
literature~\cite{Kubis:1999db,Descotes-Genon:2012fvr}). This matching is a necessary step also for the sub-amplitudes
occurring in the discontinuity of $F_{\pi}^{V,\,\alpha}(s)$. While a global
matching was performed in Ref.~\cite{Monnard:2021pvm},
an improved treatment should consider the matching
for the individual sub-amplitudes, which also rigorously takes into account the chiral counting, see Ref.~\cite{CCRdE-WiP2}.
\end{enumerate}
Finally, an important application of the study of FSR effects concerns the calculation of differences between the VFF in the $\pi^+\pi^-$ and $\pi^-\pi^0$ channels, presenting an avenue for a model-independent determination of this class of IB corrections in the analysis of $\tau$ decays, see \cref{{Sect:th-summary}}.

\begin{figure}[t!]
\begin{center}
\includegraphics[width=0.45\textwidth]{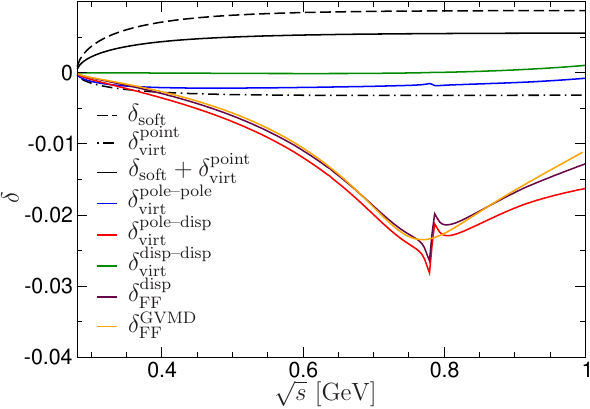}\quad
	\includegraphics[width=0.45\textwidth]{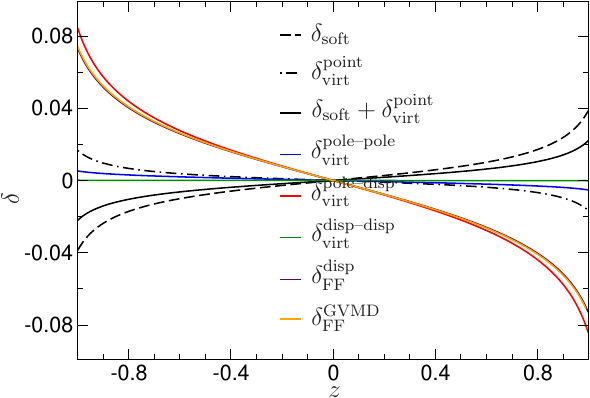}
	\caption{Correction factors $\delta$ for the charge asymmetry in $e^+e^-\to\pi^+\pi^-$ as a function of $\sqrt{s}$ for fixed $z=\cos(1)$ (left) and as a function of $z$ for fixed $\sqrt{s}=0.75\GeV$ (right), evaluated using a dispersive representation of the pion VFF~\cite{Colangelo:2022lzg}. The different lines refer to the point-like result (black dashed: real, dot-dashed: virtual, solid: sum of real and virtual), the pole--pole (blue), pole--dispersive (red), and dispersive--dispersive (green) contributions in the dispersive approach, and the GVMD model of Ref.~\cite{Ignatov:2022iou} for comparison (orange). In all cases, the logarithmic term of the IR divergence is subtracted in the conventions of Ref.~\cite{Ignatov:2022iou}. Figure taken from Ref.~\cite{Colangelo:2022lzg}.}
\label{fig:asymmetry}
\end{center}
\end{figure}

Beyond FSR corrections, dispersive techniques have been applied to the charge asymmetry in $e^+e^-\to\pi^+\pi^-$. Based on a GVMD model, it was observed in Ref.~\cite{Ignatov:2022iou} that the F$\times$sQED description is inadequate for this quantity, see \cref{TI-CMD3}. In a dispersive approach, these findings can be put onto more solid grounds, by avoiding
the systematic uncertainties due to unphysical imaginary parts below the $2\pi$ threshold. Keeping the dominant $2\pi$ cuts (the FsQED approximation in the classification of Ref.~\cite{Aliberti:2024fpq}), the results shown in \cref{fig:asymmetry} were obtained in Ref.~\cite{Colangelo:2022lzg}, demonstrating that in this case the difference between GVMD and the dispersive calculation is astonishingly small. Both calculations have already been implemented in MC generators~\cite{Budassi:2024whw}. In particular, the full dispersive calculation involves subtle effects related to the treatment of an end-point divergence in the imaginary part, and collaboration within the {\it RadioMonteCarLow~2} project, see \cref{sec:MC,TI-CMD3}, was critical to establish the correct finite terms.

For energy-scan experiments, the charge asymmetry is $C$ odd and therefore does not directly contribute to the $2\pi$ cross section that enters the HVP integral. Accordingly, in this case the above studies mainly serve as a consistency check for the CMD-3 data~\cite{CMD-3:2023rfe,CMD-3:2023alj}. However, as first pointed out in Ref.~\cite{Abbiendi:2022liz},  for the ISR process a similar effect is possible for the $C$-even cross section: this implies that resonance-enhanced corrections that are not captured by the current F$\times$sQED treatment~\cite{Campanario:2019mjh} could affect such measurements, see Ref.~\cite{Aliberti:2024fpq} for a more detailed discussion of these classes of diagrams. The calculation of these effects for ISR processes becomes significantly more challenging due to the appearance of $\pi\pi\to\gamma\gamma^*\gamma^*$ matrix elements (in addition to the well-studied pion Compton scattering $\pi\pi\to\gamma^*\gamma^*$~\cite{Garcia-Martin:2010kyn,Hoferichter:2011wk,Moussallam:2013una,Danilkin:2018qfn,Hoferichter:2019nlq,Danilkin:2019opj}), but work in this direction is ongoing and could be critical to help resolve the current situation in the data-driven evaluation of the HVP contribution.

\subsection{Data-driven results for comparisons to lattice QCD}
\label{sec:lattice_comparison}

With the advent of precise calculations of $\amuHVPLO$ from lattice
QCD, the quantitative comparison of data-driven results with lattice-QCD-based counterparts is, obviously,
necessary. A detailed comparison should go beyond simply comparing
results for $\amuHVPLO$ itself since, in both the
dispersive and lattice approaches, the full result is the sum of components
that are physical quantities in their own right. However, the very
nature of the lattice approach, which is quark based and inclusive,
makes a comparison of intermediate physical quantities obtained from the
lattice with data-driven ones built up by summing experimental data
for exclusive-mode hadronic cross sections not straightforward. The
lattice computation of $\amuHVPLO$, as reviewed in
\cref{sec:latticeHVP}, is commonly split into a number of
separately computed components. The dominant contributions arise
from the isospin-limit light- and strange-quark connected and disconnected
diagrams, with the light-quark connected accounting for about $90\%$ of
$\amuHVPLO$ and the sum of all strange- and light-quark
disconnected diagrams for about $6\%$. Additional, smaller,
contributions from charm and bottom quarks have also been computed.
IB is accounted for by including corrections from
EM (QED) and strong (SIB) origin, to first order in an
expansion in $\alpha$  and the up-down quark-mass difference $\delta=m_u-m_d$.
In order to scrutinize potential conflicts between the data-driven
and lattice approaches to $\amuHVPLO$, it is highly
desirable to obtain reliable data-driven estimates for different
components of the lattice-QCD result.

In a series of recent
papers~\cite{Boito:2022rkw,Boito:2022dry,Benton:2023dci,Benton:2023fcv,Benton:2024kwp},
it has been shown that it is possible to reliably determine, in an almost
purely data-driven manner, the light-quark connected ($ud$) and the sum
of the full strange- and light-quark disconnected ($s+\text{disc}$) contributions
to $\amuHVPLO$. Subtracting from the latter an averaged lattice
result for the strange-quark connected contribution also gives access
to an estimate of the total light- and strange-quark disconnected
components. The strategy starts from the isospin (SU(3)$_F$) decomposition
of the EM current into its $I=1$ (flavor $3$) and $I=0$ (flavor $8$)
parts, and, from this, the decomposition of the EM VP
into pure $I=1$ (flavor $33$), pure $I=0$ (flavor $88$), and
mixed-isospin (MI) (flavor $38$) parts.  It is well known that, in the isospin limit, the MI part vanishes,
and the $I=1$ part is purely light-quark connected (see, for example,
Refs.~\cite{DellaMorte:2010aq,Blum:2015you}). One has then
\begin{equation}\label{eq:lqc}
  \amuHVPLOud =\frac{10}{9}\,a_\mu^{I=1}\, ,\qquad
  \amuHVPLOsdisc = a_\mu^{I=0}-\frac{1}{9}\,a_\mu^{I=1}\, .
\end{equation}
The main task to be accomplished is then the identification, with
sufficient precision, of the $I=1$ and $I=0$ components of the EM spectral
function $\rho_{\rm EM}$. Once a procedure is established to accomplish
this goal, it is straightforward to obtain the $ud$ and $s+\text{disc}$ parts of the
total HVP contribution, or of any windowed quantity, including, in
particular, the three RBC/UKQCD windows \cite{RBC:2018dos}, introduced in \cref{sec:windows}. The decomposition into $I=1/0$ parts can
be achieved on a mode-by-mode basis in the exclusive-mode region of
the $R_\text{had}(s)$ data, with this region giving the largest contribution to
$\amuHVPLO$. The smaller contributions from the inclusive region are
obtained from pQCD supplemented with a conservative error
estimate for potential residual quark--hadron duality violations.
Finally, EM IB contributions have to be subtracted from the
data-based pure $I=1$ and $I=0$ components to obtain the
isospin-limit results to be compared with lattice-QCD determinations
(as shown in Ref.~\cite{Hoferichter:2023sli} they are not large enough
to explain discrepancies between data-driven and lattice results).
Below we outline how this is done in practice; see the original
publications~\cite{Boito:2022rkw,Boito:2022dry,Benton:2023dci,Benton:2023fcv,Benton:2024kwp}
for further details. Data-driven estimates for the RBC/UKQCD
window quantities were considered before in Ref.~\cite{Colangelo:2022vok},
but not the breakdown in $ud$ and $s+\text{disc}$ parts.

In the exclusive-mode region of the $R_\text{had}(s)$ data, there are two classes
of contributions. The first, which gives the dominant contribution to the
$ud$ results, arises from modes with well-defined $G$-parity: modes
with positive/negative $G$-parity have isospin $I=1/0$. We refer
to these as ``unambiguous modes.'' The large contributions from
$n\pi$ modes, for example, fall into this category.
The numerically most important $\pi^+\pi^-$ mode, in particular,
is $I=1$, and therefore linked directly with the $ud$ contribution.
The second class consists of contributions from modes without
well-defined isospin, referred to as ``ambiguous modes.'' These
are of two types: those from modes for which external information
can be used to separate the isospin components and those for which no
such information is available.

For ambiguous modes for which no external information is available,
which turn out to give small contributions, we rely on a maximally
conservative separation, based on the observation that, in the isospin
limit, due to the positivity of the $I=0,1$ spectral functions,
the $I=0$ and $I=1$ parts of a given mode contribution must
lie between zero and the full experimental $I=1+0$ total. The $I=1$
and $I=0$ components thus lie in the range ($50\pm 50$)\% of
that total, with $I=1$ and $0$ errors 100\% anticorrelated. Fortunately,
for the ambiguous modes that give large contributions, leading to
unacceptably large errors with this maximally conservative split,
external information can be used to achieve a sufficiently precise
isospin separation. This is the case for the $K\bar K$, $K\bar K \pi$
and $\pi^0/\eta+\gamma$ modes. In the case of the $K\bar K$
mode, \babar's results for the differential decay distribution
of $\tau \to K^-K^0\nu_\tau$~\cite{BaBar:2018qry} provide a
measurement of the charged-current $I=1$ vector spectral function. Using
the conservation of the vector current, one then obtains an estimate of the $I=1$ $K\bar K$ contribution to
$\rho_{\rm EM}$, and hence an estimate of the $I=1$ contribution from
this channel to integrated quantities up to near the end-point
of \babar's spectrum ($s=2.7556\GeV^2$). The decomposition
of the remaining small contribution from higher $s$ is obtained
via the maximally conservative separation treatment of the integrated
$K\bar{K}$-mode $R_\text{had}(s)$ contribution from this region. This approach is
crucial to achieving good control of the $K\bar K$ contributions,
dramatically reducing the final uncertainty for the $I=1/0$
components. A similar treatment of the $K\bar K\pi$ contribution is
possible thanks to \babar's Dalitz plot separation of $I=1/0$ parts of the
$e^+e^-\to K\bar K\pi$ cross sections~\cite{BaBar:2007ceh}. Finally,
for the radiative channels $\pi/\eta +\gamma$ a reliable decomposition
into pure $I=1/0$ and MI components is possible due to the strong dominance
of the observed cross sections by intermediate vector-meson
contributions. A detailed description of this separation is given in
App.~B of Ref.~\cite{Benton:2023fcv}.

The remaining ingredients for the data-driven determination of the
$ud$ and $s+\text{disc}$ contributions to $\amuHVPLO$ are: (i) the treatment
of the inclusive $R_\text{had}(s)$ data region, and (ii) the corrections for
IB (both EM and SIB effects). In the inclusive $R_\text{had}(s)$ region the
type of separation that was outlined above cannot be performed and one
has to turn to pQCD, in which the $I=0/1$ split is
straightforward. The value of $s$ at which the exclusive-mode region ends
and the inclusive region begins depends on the data compilation
used~\cite{Davier:2017zfy,Davier:2019can,Keshavarzi:2018mgv,Keshavarzi:2019abf},
but is always close to $s= (2\GeV)^2$. The pQCD
spectral function for massless quarks is exactly known to
$\alpha_s^4$~\cite{Baikov:2008jh} and
Refs.~\cite{Boito:2022rkw,Boito:2022dry,Benton:2023dci,Benton:2023fcv,Benton:2024kwp}
use these results supplemented with an estimate for the $\alpha_s^5$
coefficient. Given, however, that recent experimental results from
BESIII~\cite{BESIII:2021wib} show some tension with pQCD
below the charm threshold,
Refs.~\cite{Boito:2022rkw,Boito:2022dry,Benton:2023dci,Benton:2023fcv,Benton:2024kwp}
include a conservative error estimate based on a model for potential
residual duality violations, relying on previous work~\cite{Boito:2018yvl,Boito:2020xli}.
The pQCD contribution turns out to be small in all quantities
of interest, except, as expected, the SD RBC/UKQCD window.

To correct for EM and SIB effects, it is sufficient to define the
isospin-symmetric world as that in which all pions have mass
$M_{\pi^0}$. SIB effects, which to order $m_d-m_u$ are MI only, are
then split into those associated with the $2\pi$ and $3\pi$ exclusive
modes, where SIB effects are expected to be dominated by $\rho$--$\omega$
mixing, and those associated with the remaining exclusive modes
and perturbation theory above the inclusive threshold. For
the $2\pi$ and $3\pi$ modes, rather precise SIB results are available
from Refs.~\cite{Hoferichter:2023sli,Colangelo:2022prz,Hoferichter:2023bjm}, see also \cref{sec:isospin_breaking} and \cref{tab:IB_channels}, based on
dispersive representations of the  $2\pi$ and  $3\pi$ channels,
both for $\amuHVPLO$ and the RBC/UKQCD window quantities. For the
remaining modes, Refs.~\cite{Boito:2022rkw,Boito:2022dry,Benton:2023dci,Benton:2023fcv,Benton:2024kwp}
allow for an SIB-induced uncertainty equal to $1\%$ of the total
contribution. For the SD window, the only quantity for which the
perturbative contribution is not very small, an additional $1\%$
SIB uncertainty is assigned to that contribution.

EM corrections are the only building blocks for which no complete
data-driven estimates are available (see Ref.~\cite{Boito:2022dry} for a
discussion). Therefore, inclusive lattice results for the EM
contributions to $\amuHVPLO$ and the intermediate window from
Ref.~\cite{Borsanyi:2020mff} (with updates from Ref.~\cite{Boccaletti:2024guq})
were used to estimate these corrections, with conservative estimates for the
SD and LD windows for which such lattice results are not available.
For all quantities considered, these corrections are very small (the
largest being the $0.4\%$ for the LD window~\cite{Benton:2024kwp}), and
therefore, although relying on this small lattice-QCD input, the results of
Refs.~\cite{Boito:2022rkw,Boito:2022dry,Benton:2023dci,Benton:2023fcv,Benton:2024kwp}
are almost entirely data driven.

\subsubsection{Light-quark connected contributions}
\label{sec:lqc}

The isospin-limit data-driven $ud$ contribution $\amuHVPLOud$ to
$\amuHVPLO$ was first discussed in Ref.~\cite{Boito:2022dry} and
slightly updated in Refs.~\cite{Benton:2023fcv,Benton:2024kwp}. In this
case, the required exclusive-mode $\amuHVPLO$ contributions
can be found in tables in the publications of the two main $R_\text{had}(s)$ data
compilations by DHMZ~\cite{Davier:2017zfy,Davier:2019can} and
KNT~\cite{Keshavarzi:2018mgv,Keshavarzi:2019abf}. Supplementing this
information with the external results required for the separation
of ambiguous-mode contributions, the pQCD contribution
in the inclusive region, and estimates of IB corrections that
partially rely on BMW results~\cite{Borsanyi:2020mff,Boccaletti:2024guq},
as explained above, one obtains the following pre-CMD-3 results for
$\amuHVPLOud$, based on either the DHMZ or the KNT data
compilation~\cite{Benton:2024kwp,Boito:2022dry}
\begin{align}
\amuHVPLOud&=638.9(4.1) \times 10^{-10}\qquad \text{(DHMZ based)}\,,\notag\\
\amuHVPLOud&=635.8(2.6) \times 10^{-10}\qquad \text{(KNT based)}\,.
\end{align}
These results are smaller than, and show an important tension with, the most
precise pure lattice determinations of the same quantity by the
BMW~\cite{Borsanyi:2020mff}, the RBC/UKQCD~\cite{RBC:2024fic}, the
Mainz~\cite{Djukanovic:2024cmq}, and the
Fermilab/HPQCD/MILC~\cite{FermilabLatticeHPQCD:2024ppc} collaborations.

Access to the individual exclusive-mode spectra from the KNT data
compilation allowed the authors of
Refs.~\cite{Boito:2022rkw,Boito:2022dry,Benton:2023dci,Benton:2023fcv,Benton:2024kwp}
to obtain analogous results for the three RBC/UKQCD and other windows proposed
in the literature \cite{Aubin:2022hgm,Boito:2022njs}.
Results for the three RBC/UKQCD windows can be found in
Ref.~\cite{Benton:2024kwp}. The pre-CMD-3, purely KNT-based versions
of these results are
\begin{align}
\amuSDud  &=  46.96(48) \times 10^{-10}\,,\notag\\
\amuWud  &=  199.0(1.1) \times 10^{-10}\,,\notag\\
\amuLDud  &=  389.9(1.7) \times 10^{-10}\,.
\end{align}
An interesting exercise, performed to explore the potential impact of
the new CMD-3 $\pi\pi$ data on the data-driven determinations of the various
isospin-limit quantities, was the replacement of the combined $\pi^+\pi^-$
data of the KNT compilation with CMD-3 data alone, in the region between
0.33 and 1.2 GeV covered by the CMD-3 data.
Of course, as this exercise ignores all other $\pi^+\pi^-$ data in this
region, these ``CMD-3'' results are purely exploratory. The
results of this exploration are shown in \cref{fig:results-lqc-CMD3},
together with recent lattice results for the same quantities.

In Ref.~\cite{Benton:2023fcv} also a KNT-based data-driven result for
the $1.5$ to $1.9$~fm window ``$W2$'' of Ref.~\cite{Aubin:2022hgm}
was obtained. The value, $93.75(36)\times 10^{-10}$, is again
significantly lower than recent lattice values obtained in
Refs.~\cite{Aubin:2022hgm,FermilabLatticeHPQCD:2023jof,Boccaletti:2024guq}.
This discrepancy also vanishes if one replaces the KNT $\pi\pi$
combination with CMD-3 results in the CMD-3 data region.

\subsubsection{Strange + light-quark disconnected contributions}
\label{sec:strange+light}

The isospin-limit data-driven $s+\text{disc}$ contribution $\amuHVPLOsdisc$ to
$\amuHVPLO$ was first discussed in Ref.~\cite{Boito:2022rkw} and again
slightly updated in Refs.~\cite{Benton:2023fcv,Benton:2024kwp}. The
methodology employed is much the same as for the $ud$ contribution, with
the exception of the very small EM corrections which can only be bounded
based on the results of Ref.~\cite{Borsanyi:2020mff}. The DHMZ and KNT
data-compilation based results are \cite{Benton:2023fcv,Benton:2024kwp}
\begin{align}
\amuHVPLOsdisc&=39.8(2.4) \times 10^{-10}\qquad\text{(DHMZ based)}\,,\notag\\
\amuHVPLOsdisc&=40.7(1.7) \times 10^{-10}\qquad\text{(KNT based)}\,.
\end{align}
For the three RBC/UKQCD window quantities only results based on the KNT
compilation have been obtained~\cite{Benton:2023fcv,Benton:2024kwp};
they read \cite{Benton:2024kwp}
\begin{align}
\amuSDsdisc &=  9.21(36) \times 10^{-10}\,,\notag \\
\amuWsdisc  &=  26.98(84) \times 10^{-10}\,,\notag\\
\amuLDsdisc  &=  4.53(73) \times 10^{-10}\,.
\label{eq:LDsplqd}
\end{align}
Results for the SD and intermediate windows, as well as the $s+\text{disc}$
part of the total HVP, are shown in \cref{fig:results-s+lqd-CMD3},
together with recent lattice results for the same quantities. The only
lattice determination of the $s+\text{disc}$ contribution to the LD window
to date, by the Mainz collaboration~\cite{Djukanovic:2024cmq}, is
$\amuHVPLOsdisc=1.3 (2.4) \times 10^{-10}$ and is compatible with
\cref{eq:LDsplqd} within $1.3\sigma$.

\subsection{Higher-order iterations of HVP}
\label{sec:NLO}

At the required level of precision, also higher-order HVP iterations, including both NLO~\cite{Calmet:1976kd} and NNLO~\cite{Kurz:2014wya} effects, need to be considered. While traditionally, in the timelike method, the contributions are expressed in terms of the $R$-ratio via the corresponding kernel functions, also spacelike representations are possible, in which case the integrands are related to the HVP function $\Pi(Q^2)$ and the Adler function $D(Q^2)$. 
Specifically, the relation between the spacelike kernel $K_{\Pi}(Q^2)$ in
terms of the timelike kernel $K_{R}(s)$ has independently been obtained in
Refs.~\cite{Nesterenko:2021byp, Balzani:2021del} and the corresponding explicit expressions calculated at NLO~\cite{Nesterenko:2021byp,
Balzani:2021del} and NNLO~\cite{Balzani:2021del}. Additionally, the
complete set of six relations that mutually express all three
kernels $K_{\Pi}(Q^2)$, $K_{D}(Q^2)$, and $K_{R}(s)$ in terms of each other
has been obtained in Ref.~\cite{Nesterenko:2021byp}, which also provides the
results for $K_{D}(Q^2)$ at NLO. These results are important for calculations in lattice QCD, see \cref{sec:latticeHVP}, and the HVP measurement in MUonE, see \cref{sec:MUonE}, as in both cases spacelike representations are needed. In particular, the time--momentum representation of the NLO spacelike kernel was provided in Ref.~\cite{Balzani:2024gmu}.

Based on these developments, improved higher-order HVP calculations will become available in lattice QCD over the next years, while for existing calculations~\cite{Chakraborty:2018iyb} the precision is not yet competitive with the data-driven approach despite the tensions among the $e^+e^-$ measurements. To define a conservative estimate, we consider two evaluations obtained in the context of Ref.~\cite{DiLuzio:2024sps}
\begin{equation}
\amuHVPNLO = \left \{ \begin{array}{lll}
                                           -9.83(4) \times 10^{-10}    &  \text{KNT19}  &\text{\cite{Keshavarzi:2019abf}} \, ,  \\
                                           -10.08(6)\times 10^{-10}    &  \text{KNT19/CMD-3} &\text{\cite{DiLuzio:2024sps}}\, ,\\
                                            \end{array}  \right .
\end{equation}
where the second variant corresponds to the compilation from Ref.~\cite{Keshavarzi:2019abf} with the low-energy $\pi^+\pi^-$ contribution replaced by the CMD-3 measurement~\cite{CMD-3:2023alj,CMD-3:2023rfe} where applicable, and assign the mean and spread as our value\footnote{The same procedure for $\amuHVPLO$ would produce an uncertainty of $\pm 10.9\times 10^{-10}$, which roughly reflects the spread observed in \cref{fig:summary_plot_ee}, so that the uncertainty assignment in \cref{HVPNLO} should reasonably cover all presently observed tensions in the $e^+e^-$ data.}
\begin{equation}
\label{HVPNLO}
   \amuHVPNLO=\amuHVPNLOresultsection\times 10^{-10}\,. 
\end{equation}
For the NNLO contribution, we continue to use~\cite{Kurz:2014wya} 
\begin{equation}
\label{HVPNNLO}
 \amuHVPNNLO= \amuHVPNNLOresultsection\times 10^{-10}\,,
\end{equation}
even though the uncertainty is likely somewhat underestimated.
The impact of mixed leptonic and hadronic corrections at NNLO was estimated as $\lesssim 0.1\times 10^{-10}$ in Ref.~\cite{Hoferichter:2021wyj}, while NNLO hadronic corrections due to virtual photons emitted and reabsorbed by the HVP insertion were estimated to be of size ${\mathcal O}(10^{-12})$~\cite{Balzani:2021del}.

\subsection{An alternative approach: MUonE}
\label{sec:MUonE}

\nocite{CarloniCalame:2015obs,MUonE:2016hru}

\begin{figure}[t]
    \centering
    \includegraphics[width=0.98\textwidth]{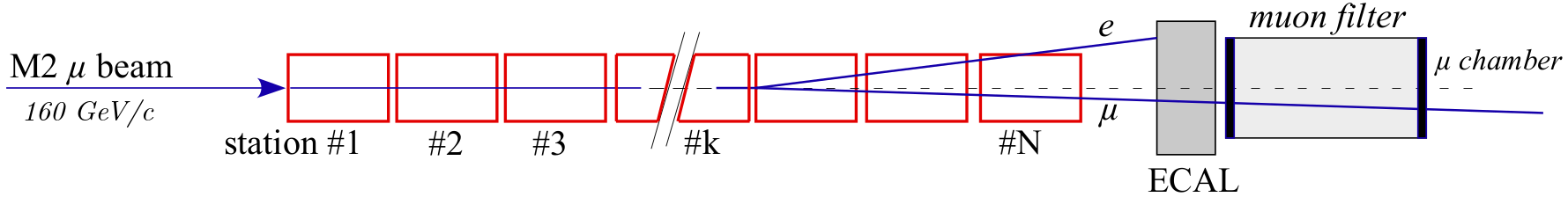}
    \caption{Schematic view of the MUonE experimental apparatus (not to scale). Figure taken from Ref.~\cite{MUonE:2019qlm}.}
    \label{fig:fig1}
\end{figure}

The MUonE experiment aims to provide an independent calculation of $\amuHVPLO$ using an innovative method based on the measurement of the hadronic contribution to the running of the EM coupling constant, $\Delta\alpha_\text{had}(t)$, for negative squared four-momentum transfers $t$~\cite{CarloniCalame:2015obs}. The hadronic running will be extracted from the shape of the $\mu$--$e$ elastic scattering differential cross section \cite{MUonE:2016hru}, which will be measured by colliding the $160\GeV$ muon beam available at CERN's M2 beam line with a thin low-$Z$ target. The detector will be segmented into 40 identical stations, each consisting of a target followed by a tracking system equipped with silicon strip sensors. The apparatus is completed by an EM calorimeter and a muon ID system placed downstream of the tracking stations, to improve event selection, and a spectrometer placed upstream to measure the beam momentum. A sketch of the experimental apparatus is shown in \cref{fig:fig1}.

The goal of MUonE is to determine $\amuHVPLO$ with a $\simeq 0.3\%$ statistical accuracy and comparable systematics. This poses several challenges not only on the experimental side~\cite{MUonE:2019qlm,Abbiendi:2022oks}, but also requires a huge effort to determine the higher-order radiative corrections to the $\mu$--$e$ scattering. The complete set of NLO electroweak, NNLO QED, and hadronic corrections have been calculated in Refs.~\cite{Alacevich:2018vez,Mastrolia:2017pfy,DiVita:2018nnh,Fael:2019nsf,Fael:2018dmz,CarloniCalame:2020yoz,Banerjee:2020tdt,Banerjee:2020rww,Bonciani:2021okt,Budassi:2021twh,Budassi:2022kqs,Broggio:2022htr}. The first steps towards an N3LO computation of the QED corrections were taken recently in Refs.~\cite{Fael:2022rgm,Fael:2022miw,Fael:2023zqr,Badger:2023xtl,Dave:2024ewl,Engel:2023ifn,Engel:2023rxp}, and atomic binding corrections to $\mu$--$e$ scattering were recently studied in Refs.~\cite{Plestid:2024xzh,Plestid:2024jqm}. The dominant NNLO QED and hadronic corrections are implemented in two different independent MC codes, \mesmer~\cite{CarloniCalame:2020yoz} and \mcmule~\cite{Banerjee:2020rww}. While the former includes also the main background processes~\cite{Budassi:2021twh,Budassi:2022kqs,Abbiendi:2024swt}, and is already integrated in the full detector simulations, the latter contains the complete NNLO photonic calculation~\cite{Broggio:2022htr}.
In addition to this, analytic expressions have been provided in Refs.~\cite{Balzani:2021del, Nesterenko:2021byp} to compute also $a_\mu^\text{HVP,NLO}$ and $a_\mu^\text{HVP,NNLO}$ in the spacelike region, see \cref{sec:NLO}, thus extending the capability of MUonE to determine the HVP contribution to $a_\mu$ to higher orders.
Finally, a different method has been proposed in Ref.~\cite{Ignatov:2023wma} to determine $\amuHVPLO$ from MUonE data in the spacelike region through the derivatives of $\Delta\alpha_\text{had}(t)$ at zero momentum transfer. Possible fit functions for $\Delta\alpha_\text{had}(t)$ are also discussed in Refs.~\cite{Greynat:2022geu, Boito:2024yat}.

\begin{figure}[t]
	\centering
	\includegraphics[width=.47\textwidth, keepaspectratio]{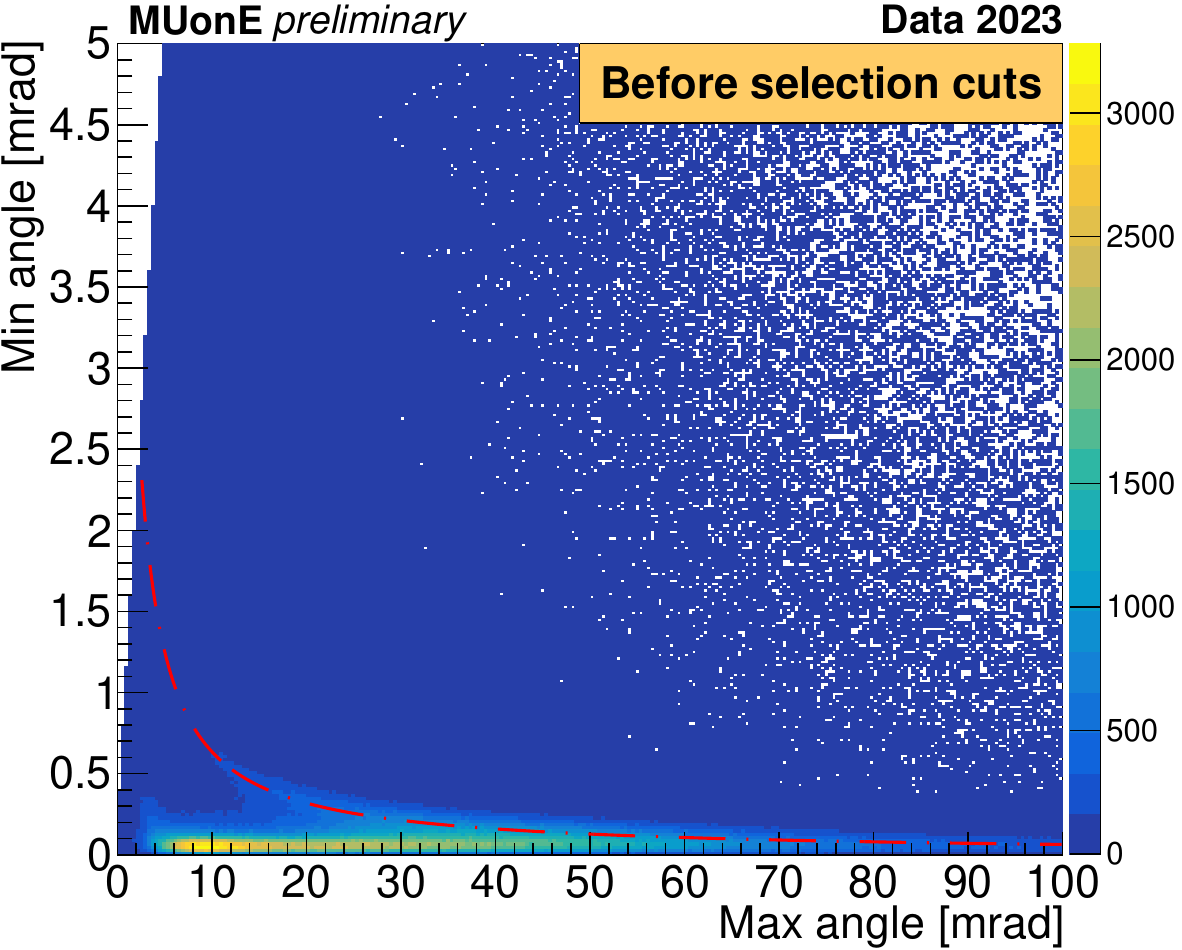}~\includegraphics[width=.47\textwidth, keepaspectratio]{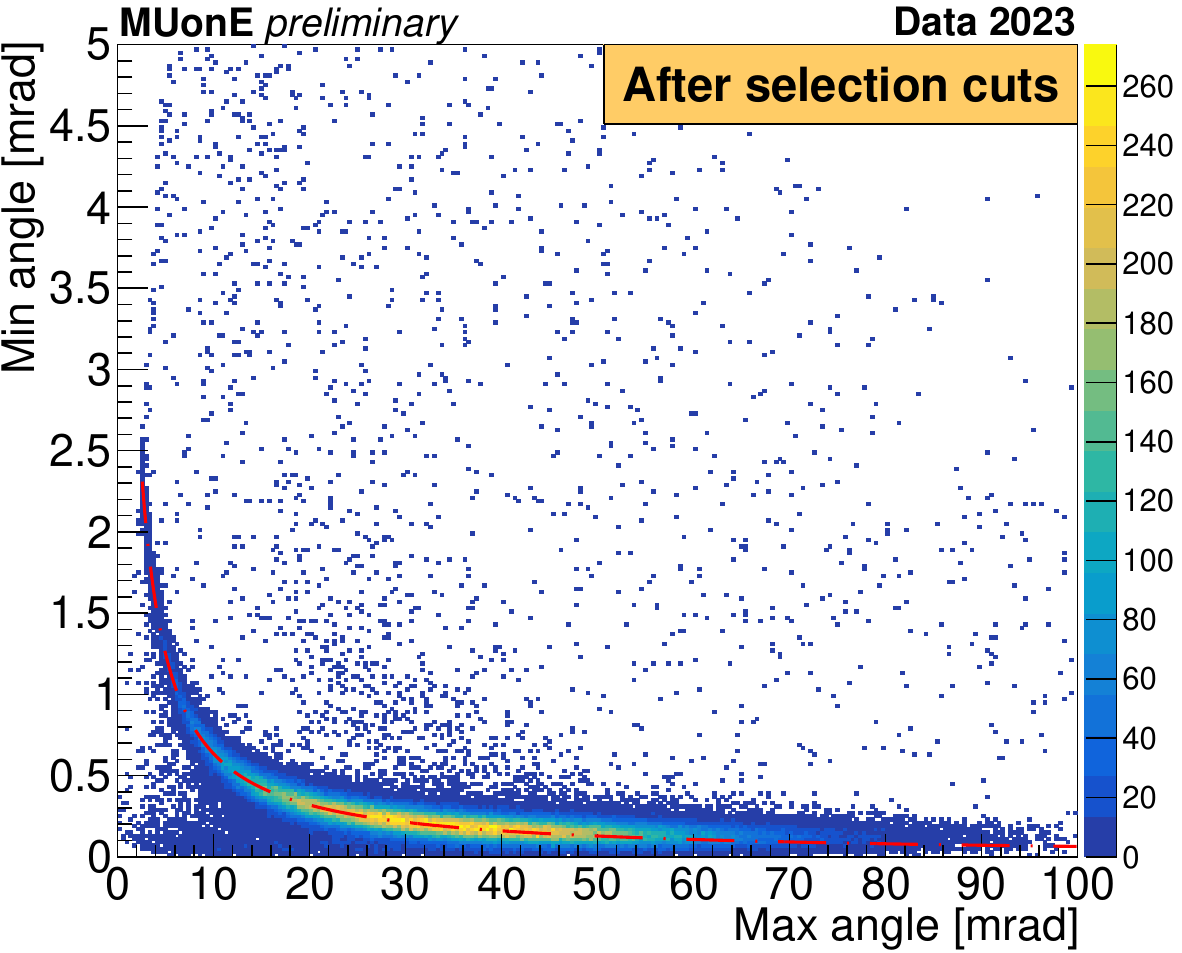}
	\caption{$(\theta_{max}, \theta_{min})$ distribution of elastic scattering candidates before (left) and after (right) the preliminary event selection described in the text. The red dashed line represents the expected elasticity curve for a $160\GeV$ muon beam. Figure adapted from Ref.~\cite{MUonE_Proposal}.}
	\label{fig:fig2}
\end{figure}

In the last few years, MUonE carried out a series of short beam tests at the M2 beam line with a first tracking station instrumented with prototype 2S modules, silicon strip sensors foreseen for the CMS Phase-2 Upgrade~\cite{CMS:2017lum}. This allowed one to commission the system mechanics and the data acquisition chain in view of a longer test performed in Summer 2023, whose goal was to demonstrate the ability to identify and reconstruct $\mu$--$e$ elastic events. The apparatus consisted of two tracking stations equipped with prototype 2S modules, followed by a prototype calorimeter. Graphite targets 2 or $3\,\text{cm}$ thick were installed between the two tracking stations, in order to evaluate multiple scattering effects and study background processes in different configurations.
Tracker data are currently being analyzed to optimize reconstruction algorithms, software alignment procedures and event selection. A sample of candidate elastic events is shown in \cref{fig:fig2}(left). The outgoing tracks are labeled according to the magnitude of their angles, denoted as $\theta_\text{max}$ and $\theta_\text{min}$, and a large fraction of background events is clearly visible at low $\theta_\text{min}$. \Cref{fig:fig2}(right) shows instead the effect of a preliminary event selection, including cuts on the total number of hits in the downstream station, on the acoplanarity of the event and on the vertex position, which is capable of rejecting most of the background. Residual background can be removed by cutting events with $\theta_\text{min} \leq 0.2\,\text{mrad}$. Further preliminary results on Test Run data analysis can be found in Refs.~\cite{MUonE_Proposal,Spedicato:2024lvp,Pilato:2024nvd}.

The MUonE Collaboration has recently submitted a proposal to the CERN SPS Committee~\cite{MUonE_Proposal} to run four weeks in 2025 with a small-scale version of the final apparatus, comprising three tracking stations, a calorimeter, a muon filter, and the beam spectrometer, obtaining positive recommendations. The Phase-1 MUonE Run will occur in the first half of 2025, allowing one to study signal and background under realistic conditions with a possible sensitivity to $\Delta\alpha_\text{had}(t)$. The 2025 data will also serve as a basis for a full-scale experiment proposal to be prepared during the CERN Long Shutdown 3 (2027--2029), aimed at taking data after 2030.

\subsection{HVP dispersive evaluation: comparison of experimental inputs and methods}
\label{sec:HVP_disp_summary}

\begin{figure}[t]
  \centering
  \includegraphics[width=0.8\linewidth]{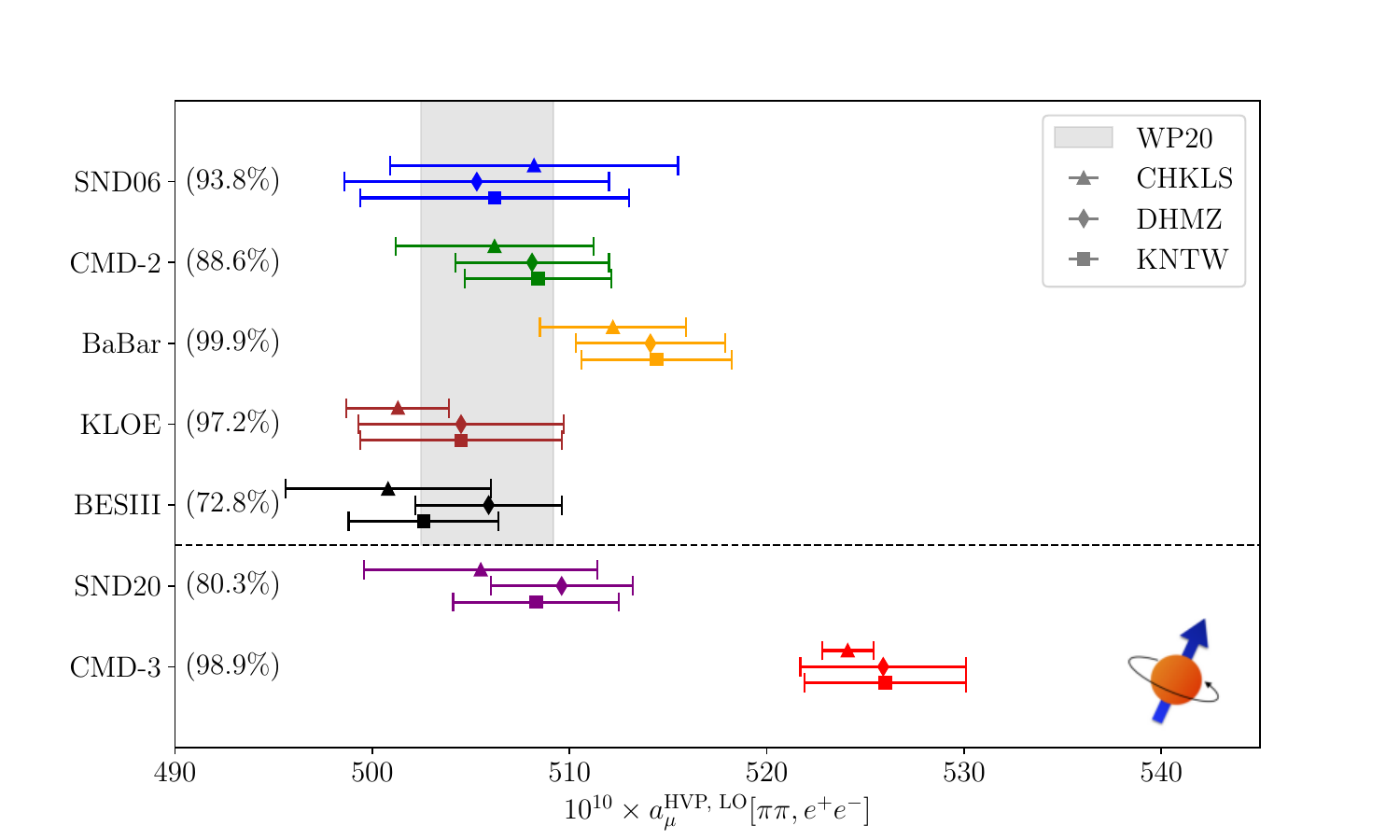}
  \caption{Dispersive theoretical predictions for $\amuHVPLO[\pi\pi]$, based on various measurements of  $e^+e^-\to\pi^+\pi^-$, fit/interpolated and complemented for the uncovered mass ranges~(percentages of the integral covered by each measurement are shown), for the three approaches ``CHKLS,'' ``DHMZ,'' and ``KNTW'' as detailed in the main text.
  The gray band indicates the result from WP20, including the error inflation due to the \babar--KLOE tension. The experiments above the dashed line entered the result for WP20, whilst those below are new measurements since then. The numerical values shown are reproduced in \cref{tab:summary_ee}.
}
\label{fig:summary_plot_ee}
\end{figure}

\begin{table}[t]
\small
	\centering	\renewcommand{\arraystretch}{1.1}
	\begin{tabular}{lrrrr}
	\toprule 
	Experiment & Percentage covered &       &      &     \\ \midrule 
        $e^+e^-$   &                    & CHKLS & DHMZ & KNTW\\ \midrule
	SND06 & $93.8\%$ & $508.2(7.3)$ & $505.3(6.7)$ & 	$506.2(6.8)$\\
	CMD-2 & $88.6\%$ & $506.2(5.0)$	& $508.1(3.9)$ &	$508.4(3.7)$\\
	BaBar & $99.9\%$ & $512.2(3.7)$ & $514.1(3.8)$ & 	$514.4(3.8)$\\
	KLOE & $97.2\%$ & $501.3(2.6)$ & $504.5(5.2)$ &	$504.5(5.1)$\\
BESIII & $72.8\%$ & $500.8(5.2)$ &	$505.9(3.7)$ &	$502.6(3.8)$\\
SND20 & $80.3\%$ & $505.5(5.9)$ & 	$509.6(3.6)$ & 	$508.3(4.2)$\\
CMD-3 & $98.9\%$ & $524.1(1.3)$ & $525.9(4.2)$ & 	$526.0(4.1)$\\\midrule
$\tau$ &&  \multicolumn{3}{c}{DHLMZ-23+LMR-24+WP25} \\ \midrule 
Belle+CLEO+ALEPH+OPAL & $100\%$ & \multicolumn{3}{c}{$517.2(5.8)$}\\
	 \bottomrule 
	\renewcommand{\arraystretch}{1.0}
	\end{tabular}
	\caption{$\amuHVPLO[\pi\pi,e^+e^-]$  up to $1.8\GeV$ (in units of $10^{-10}$) for the different $e^+e^-\to\pi^+\pi^-$ experiments, using the ``CHKLS,'' ``DHMZ,'' and ``KNTW'' approaches as detailed in the main text and illustrated in \cref{fig:summary_plot_ee}, where the second column indicates the percentage of the integral covered by each experiment. Depending on the method, a given experiment is either complemented by an average of others outside the covered mass range (DHMZ and KNTW) or extrapolated from the range in which data are taken (CHKLS), and also the treatment of the region above $1\GeV$ differs, as explained in the main text. We emphasize that these numbers are meant to illustrate the spread among the different experiments and analysis methods, and since the discrepancies are currently not understood, we do not attempt to derive a global $\amuHVPLO$ number based on $e^+e^-$ data. For comparison, the table also includes our estimate for $\amuHVPLO[\pi\pi,\tau]$ from \cref{eq:atau-ourestimate}.
	To obtain the corresponding ranges for $\amuSM$ shown in \cref{fig:summary_plot_data_hvp}, first the offset $187.3(2.2)\times 10^{-10}$ from WP20 is added to arrive at $\amuHVPLO$~(see main text). All other contributions are taken from WP25.}
 \label{tab:summary_ee}
\end{table}

\Cref{fig:summary_plot_ee} presents a summary of the dispersive theoretical predictions for $\amuHVPLO[\pi\pi]$, based on various measurements of $e^+e^-\to\pi^+\pi^-$, treated using either of the ``CHKLS'' (\cref{sec:disp_2pi}), ``DHMZ'' (\cref{Sec:DHLMZ}), and ``KNTW'' (\cref{Sec:KNTW}) approaches.
The various measurements cover  the range from threshold up to $1.8\GeV$ according to the percentages indicated in the figure.

Contributions from other channels and from the high mass region above 1.8~GeV are based on WP20.
It is, however, worth mentioning that significant progress has been made in these other channels since WP20 (see \cref{sec:e+e-}), e.g., next by uncertainty amplitude, the contribution from the $3\pi$ process was improved by more than a factor of two. 
However, tensions between the Belle II data and previous measurements are now visible. 
Other discrepancies are observed in a few other channels as well---see the corresponding discussion in \cref{Sect:th-summary} in the context of \cref{LOHVP_tau}---although their overall impact remains minor compared to the tension seen in the $2\pi$ channel. Continued progress in the overall $R_\text{had}(s)$ measurement program is essential for a future complete data-driven evaluation of $\amuHVPLO$.

The most important 
feature of \cref{fig:summary_plot_ee} concerns the strong tensions observed among the available $2\pi$ measurements, regardless of the
various methodologies employed for the data integration and extrapolation/completion to the full mass range.
The differences among the various methodologies are nevertheless also visible for the resulting central values and uncertainties, the latter differing up to a factor of two in some cases.
This comes from different treatment of correlations for fits/weight derivation, strategies to complete the data sets, and usage of additional constraints. The increasing precision of data and the larger tensions among experiments highlight the importance of a conservative treatment of experimental systematic uncertainties. The conventionally used practice to treat the provided experimental systematic errors, whose knowledge is inherently limited, is generally incomplete compared to what was assumed by the measured $e^+e^-$ data as discussed in Secs.\ 2.3.1--2.3.6 of WP20 and \cref{TI-CMD3,sec:disp_2pi} of WP25. In some cases, this leads to an underestimation of the evaluated integral errors due to assumptions made for the correlations of systematic uncertainties or the energy-dependent amplitude of uncertainties.

The ``CHKLS'' \cite{Colangelo:2018mtw,Colangelo:2022prz,Stoffer:2023gba,Leplumey:2025kvv} points are obtained using a dispersive representation to implement constraints from analyticity, unitarity, and crossing symmetry, so that the $\pi^+\pi^-$ cross sections can be analyzed in a global fit (with input for covariance matrices as provided by experiment), and thus the entire low-energy contribution can be evaluated for each experiment individually. In the figure, the variant of the dispersive approach without imposing the absence of zeros in the form factor is shown, see \cref{sec:disp_2pi} for details, evaluated up to $\sqrt{s}=1\GeV$, with the remainder taken from WP20.

``DHMZ'' perform a direct integration in the full range where a given
measurement is available, complemented with data combinations for the
rest, while ``KNTW'' limit their use of individual $\pi^+\pi^-$ data to the upper range of 1 GeV, using data combinations for the remaining values.
It is important to note that the latter combination-based integrals are by construction more precise than the integrals derived from one measurement alone (i.e., a measurement covering the low- and/or high-mass range can be misleadingly ``penalized'' when comparing the uncertainties).
A local $\chi^2$-based uncertainty rescaling is performed when necessary, although there are relatively little tensions in the relevant mass regions.

The ``KNTW'' numbers are solely based on the KNT19 combination~\cite{Keshavarzi:2018mgv, Keshavarzi:2019abf} and are, in particular, deliberately not including the CMD-3 data (see \cref{Sec:KNTW} for further details). The combination is fully model-independent, incorporates experimental correlation information in data fits for the full available energy range whilst simultaneously avoiding procedural systematic biases (such as the d'Agostini bias in correlated $\chi^2$ fits~\cite{DAgostini:1993arp,Blobel:2003wa}), and determines additional systematic uncertainties.

The ``DHMZ'' result employs a spline-based relatively-local averaging~(minimization of $\chi^2$ with correlations), with weight derivation accounting for different point-spacing/binning (see \cref{Sec:DHLMZ} for further details).
All existing data are used on the full mass ranges where combinations are employed. The combination procedure is also validated through a closure test.

The ``DHMZ'' and ``KNTW'' direct integration methods yield similar values in most cases~(differences not exceeding much more than 1 unit), except for the BESIII integral, where some difference~(of about 3 units) arises due to the larger range on which the combined data are employed and the differences of methodologies discussed above and in WP20. 
The ``CHLKS'' points show lower uncertainties in a few cases, since the additional theoretical constraints in a global fit reduce the uncertainties in the resulting integral. This reduction is most effective in cases in which the data are relatively scarce in some energy region---the extreme case concerning the ability to extrapolate in a robust manner into regions in which no data were taken---and in which the systematic uncertainties, most prominently from the truncation of the conformal expansion, are small. 
In the latter case, as for CMD-3, the limited knowledge of exact correlations of the systematic uncertainties in experiment becomes more critical, suggesting to somewhat enlarge the fit uncertainties, see \cref{TI-CMD3,sec:disp_2pi}.

\begin{figure}[t]
  \centering
  \includegraphics[width=0.8\linewidth]{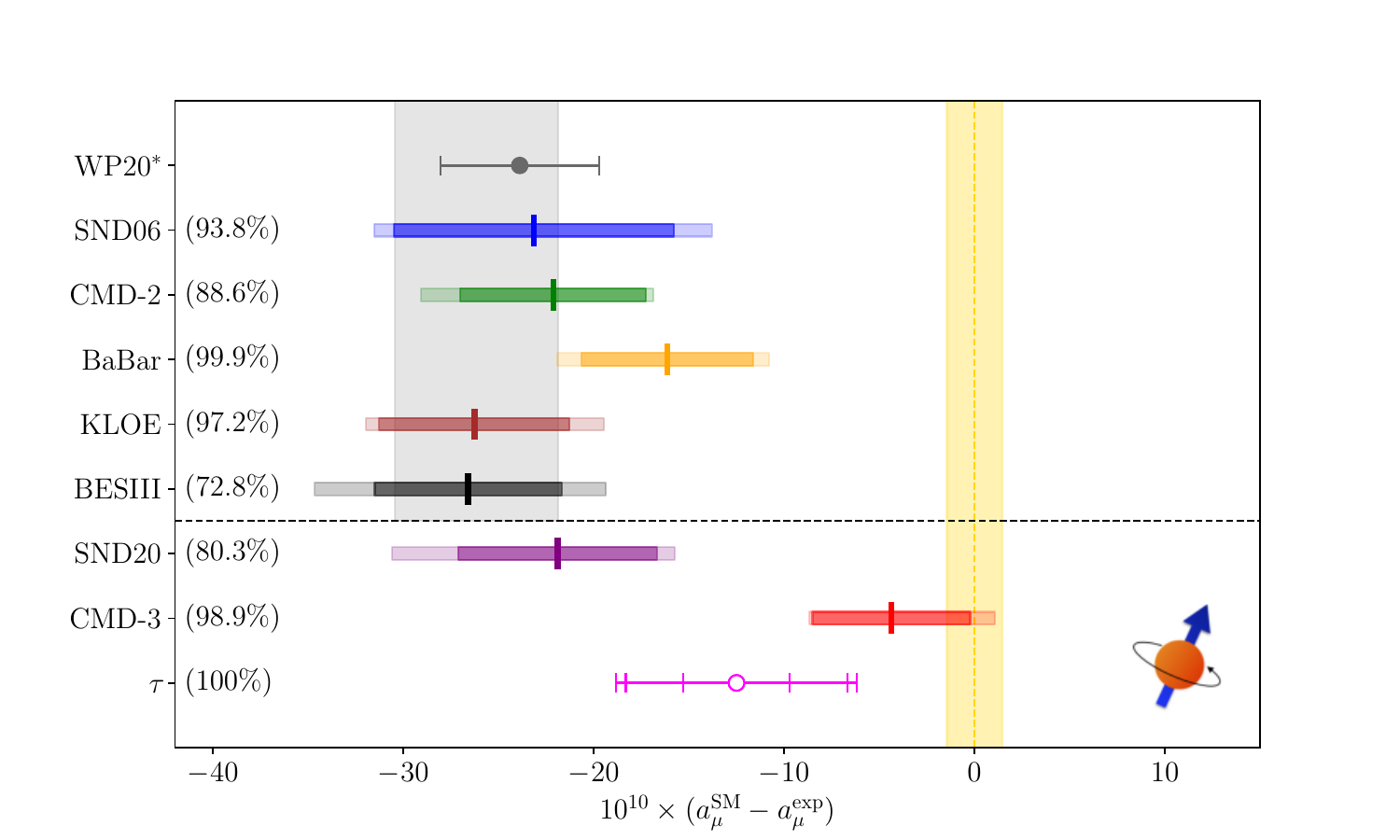}
  \caption{Summary of current data-driven evaluations of HVP, propagated to $\amuSM$ (the yellow band indicates $\amuexp$, the gray band the WP20 SM prediction based on the $e^+e^-$ data sets above the dashed line and the remainder from WP20, in particular, the WP20 HLbL value; the data point labeled WP20$^*$ indicates the shift upon using WP25 input for the other contributions besides LO HVP). The $\tau$ point corresponds to WP25 in \cref{fig:summary_plot_tau}, with the third, outmost error including the additional uncertainties beyond the $2\pi$ channel (the remainder of HVP is taken from WP20, the other contributions from WP25). The other points use input from the various $e^+e^-\to\pi^+\pi^-$ experiments according to \cref{fig:summary_plot_ee} (again with HVP remainder from WP20 and the other contributions from WP25), where for each experiment the central values are obtained as simple average of the three combination methods, the inner ranges as simple average of the uncertainties obtained in each method, and the outer ranges reflect the maximal range covered by all methods (the percentages indicate how much of the $2\pi$ contribution to the HVP integral is covered by each measurement). We emphasize that these ranges are merely meant to illustrate the current spread, they cannot be interpreted as uncertainties with a proper statistical meaning. The numerical values follow from \cref{tab:summary_ee,tab:summary}. The experimental world average has been updated including the final results from the Fermilab experiment.
}
\label{fig:summary_plot_data_hvp}
\end{figure}

\subsection{Summary and outlook}
\label{sec:HVP_disp_summary_outlook}

The unsatisfactory situation regarding the knowledge of the cross section for the process $e^+ e^- \rightarrow \pi^+ \pi^-$, which is known to contribute more than 70\% to the total HVP dispersion integral, presents a significant limitation to the data-driven evaluation of the HVP contribution and, consequently, to the SM prediction of the anomalous magnetic moment of the muon. This is illustrated in \cref{fig:summary_plot_data_hvp}. Since WP20, the introduction of the new CMD-3 measurement, which has a systematic uncertainty of 0.7\%, has changed the landscape of hadronic cross-section measurements. The CMD-3 result provides a significantly higher value for the HVP contribution compared to older evaluations.
Previously, for WP20, an averaging procedure with inflated uncertainties was adopted to accommodate the two most precise measurements from KLOE and \babar{}, which exhibited already some discrepancies. However, this approach is no longer appropriate given the current situation with the new CMD-3 data. Due to the significant spread of experimental results, with relative differences exceeding the claimed systematic uncertainties by substantial factors, it has been decided not to perform a new average for the two-pion channel and, consequently, the HVP contribution within WP25. Clarifying the current situation is of utmost importance, and the following developments from the recent past and the near future will be essential for achieving progress:

\begin{itemize}
    
    \item    
    The detailed understanding and control of radiative corrections in experimental analyses have become a focal point of research in recent years. The availability of high-precision event generators, along with a well-defined understanding of their uncertainties, is fundamentally important for the field of hadronic cross-section measurements, as well as for the QED normalization processes utilized by experiments. All ISR experiments (\babar\, BESIII, KLOE) use the NLO \phokhara{} event generator, which simulates hadronic and muonic events with one high-energy ISR (or FSR) photon representing the LO configuration. 
    The 
    dependence on \phokhara, 
    however, strongly depends on the event analysis.
    The \babar\ collaboration has conducted a detailed study on the fraction of (N)NLO photons and identified significant limitations of 
    \phokhara{} in simulating these higher-order corrections~\cite{BaBar:2023xiy}. They also found that the \babar\ measurement, due to its highly inclusive selection of (N)NNLO photons, remains unaffected.
    A subsequent publication by the DHMZ group~\cite{Davier:2023fpl} explored the potential impact of these shortcomings on existing ISR publications for the process $e^+ e^- \rightarrow \pi^+ \pi^-(\gamma)$. 
    Fast simulation by the DHMZ group suggested that the KLOE and BESIII measurements may have significantly underestimated the uncertainties attributed to radiative corrections.
    This latter claim holds especially for scenario 1 of the two scenarios considered in Ref.~\cite{Davier:2023fpl}. Further analysis by BESIII indicated that the large effect predicted for the BESIII measurement did not account for specific corrections implemented in the BESIII analysis~\cite{KEK24_Denig}. Moreover, both KLOE and BESIII demonstrated agreement at the 1\% level for the mass spectra for various event generators, which differ in their simulation of higher-order corrections~\cite{KEK24_Denig,gv_ti24}. 
    As a result of these studies, it was concluded that scenario 1 in Ref.~\cite{Davier:2023fpl} seems unlikely, and the shortcomings of \phokhara{} most likely do not explain the seen differences between different measurements.
    
    \item The {\it RadioMonteCarLow~2} initiative is committed to improving the theoretical predictions for hadron and lepton production at low-energy $e^+e^-$ colliders by bringing the available MC generators to NNLO+ precision. In the first phase, a review of the existing state-of-the-art  available generators has been concluded~\cite{Aliberti:2024fpq}. The comparison was performed by using approximated experimental selections as benchmarks. The study will continue by incorporating additional important selection variables. Moreover, Ref.~\cite{Aliberti:2024fpq} presents a detailed discussion of the different classes of higher-order radiative corrections for both direct-scan and ISR processes, identifying a critical class of virtual ISR corrections that, due to resonance enhancement, could contribute to the cross section at the relevant level (as first pointed out in Ref.~\cite{Abbiendi:2022liz}). A similar effect in the $C$-odd asymmetry in direct-scan experiments was observed by CMD-3, demonstrating that such structure-dependent radiative corrections can indeed far exceed estimates in a form-factor-times-sQED prescription~\cite{Ignatov:2022iou,Colangelo:2022lzg}.
    Actual improvements to the codes are expected to take place during the next Phase~II. To ultimately investigate the accuracy of event generators, high-quality codes with different approaches in simulating the radiative corrections are needed for cross-checks (both for energy scan and ISR experiments). The ongoing extension of the \babayaga, \phokhara, \mcmule, and additional codes are important steps in that context.

    \item 
    Fortunately, new precision measurements of the process $e^+e^- \rightarrow \pi^+\pi^-(\gamma)$ are being prepared with significant effort by several experimental collaborations. The CMD-3 and SND experiments will continue their campaigns to analyze the energy scans conducted at the VEPP-2000 collider, while the \babar{}, BESIII, and KLOE experiments will produce updated results using the ISR technique. For this purpose, considerably larger data sets will be utilized, and the event analyses will be designed to address possible limitations arising from event generators. In the case of \babar\, a new analysis strategy based on angular fits will be employed. The Belle-II collaboration has already presented a preliminary precision ISR analysis for the $3\pi$ channel~\cite{Belle-II:2024msd} and will soon contribute to the leading two-pion channel as well.
    
   \item Theory developments since WP20 include the derivation and improvement of unitarity and analyticity constraints for the $2\pi$, $3\pi$, $\pi^0\gamma$, and $\bar K K$ channels, with applications to chiral extrapolations, the determination of IB contributions, and structure-dependent radiative corrections. In addition, methods were developed to extract quark-flavor-specific results from the data. These efforts provide consistency checks on the data sets, establish correlations with other (low-energy) observables, and facilitate detailed comparisons to lattice-QCD calculations of HVP. Examples include the observed tensions with the dispersive constraints in the case of SND20 ($2\pi$)---while all other $2\pi$ measurements pass the consistency check---and Belle II ($3\pi$); spread in the phase of the $\rho$--$\omega$ mixing parameter and the charge radius among the $2\pi$ experiments; predictions for IB in the long-distance tail of the QED correction due to the pion mass difference (as well as other IB effects); the derivation of theoretically robust amplitudes for structure-dependent radiative corrections; and the determination of data-driven comparison values for light-quark-connected and strange + disconnected Euclidean windows.
    
    \item 
    The phenomenological analysis groups have made significant progress since the release of WP20. The DHMZ group has conducted detailed investigations to quantify the tensions among the various data sets, while the KNTW group has implemented the first fully blind data-driven HVP analysis and is in the process of scrutinizing, improving, and modernizing their data combination procedure. All these efforts over different groups are expected to lead to improved averaging procedures in the near future when the new data sets become available.
    
   \item   The hadronic $\tau$ decay $\tau \to \pi \pi \nu_\tau$ provides an alternative path to the dominant  two-pion contribution to  HVP. The  data sets from the LEP experiments, CLEO,  and Belle are consistent and provide a competitive uncertainty, with excellent prospects for further precision improvements with Belle-II measurements.  The current uncertainty in the $\tau$-based determination of HVP is dominated by challenging  IB corrections.  The two  analyses by the DHLMZ and LMR groups have been combined to obtain the WP25 estimate:  while the central value fully reflects the published results, the uncertainty has been reassessed to account for sources of IB that are not yet fully addressed in the literature. Work to tackle  structure-dependent radiative corrections and IB effects in the ratio of EM and  weak form factors is ongoing and new information is expected in the near future from  dispersive methods,  lattice-QCD,  and data-driven constraints.

\end{itemize}

As an outlook, it can be stated that the coming years will bring a series of new experimental results for the dominant hadronic channels used to evaluate the HVP contribution to the anomalous magnetic moment of the muon. Both the BESIII and \babar\ collaborations anticipate presenting new results for the dominant $e^+e^- \rightarrow \pi^+\pi^-(\gamma)$ channel already in 2025. The increased sensitivity to the treatment of higher-order radiative corrections in the experimental analyses, along with the development of new event generators to support these analyses, marks a significant improvement and is expected to help clarify the current situation. 
Furthermore, upcoming new measurements of hadronic $\tau$ decays with unprecedented statistics will be
performed at Belle II and will also be crucial to further improve the precision of the SM prediction of $a_\mu$. 
In the more distant future, the MUonE experiment aims to provide an independent and inclusive approach to determining the HVP contribution by measuring the spacelike running of HVP. Additionally, new analyses and highly complementary analysis techniques will be developed, such as BESIII's attempt to measure the inclusive hadronic cross section at low energies via ISR.

\FloatBarrier

\clearpage

\section{Lattice-QCD calculations of HVP}
\label{sec:latticeHVP}

\noindent
\begin{flushleft}
\emph{T.~Blum, M.~Bruno, M.~C\`e, D.~Clarke, M.~Della Morte, C.~DeTar, A.~X.~El-Khadra, R.~Frezzotti, G.~Gagliardi, A.~G\'erardin, D.~Giusti, S.~Gottlieb, V.~G\"ulpers, S.~Kuberski, S.~Lahert, C.~Lehner, L.~Lellouch, M.~K.~Marinkovi\'c, H.~Meyer, E.~Neil, J.~Parrino, A.~Portelli, J.~Sitison, J.~T.~Tsang, R.~Van de Water, G.~Wang, H.~Wittig}
\end{flushleft}

\subsection{Introduction}
\label{sec:introduction}

This section reviews lattice calculations of the HVP
contributions to the muon $g-2$. We start with an introduction summarizing standard
definitions and the outcome of WP20.
We adopt a Euclidean metric, which is the
appropriate setup for lattice formulations. The position-space HVP tensor is given by the QCD and QED expectation value
\begin{equation}
  C_{\mu\nu}(x)=\braket{j_{\mu}(x)j_{\nu}(0)}\,,
\end{equation}
where $j_{\mu}$ is the quark EM current
\begin{equation}
  j_{\mu}(x)=\sum_{f=1}^{N_f}Q_f\,\bar{q}_f(x)\gamma_{\mu}q_f(x)\,,
\end{equation}
where, for a flavor $f$, $q_f$ is the quark field and $Q_f$ is the electric
charge in units
of the elementary charge. The current above is conserved in continuous space--time, and its
implementation on a discrete space--time might differ in lattice-QCD calculations. The
momentum-space HVP tensor is then given by the Fourier transform
\begin{equation}
  \Pi_{\mu\nu}(Q)=\int d^4x\,C_{\mu\nu}(x)\,e^{-iQ\cdotp x}\,.
\end{equation}
The tensor above is known to be transverse as a consequence of gauge invariance and
Ward--Takahashi identities, and can be described through a single form factor $\Pi(Q^2)$:
\begin{equation}
  \Pi_{\mu\nu}(Q)=(Q_{\mu}Q_{\nu}-\delta_{\mu\nu}Q^2)\,\Pi(Q^2)\,.
\end{equation}
The function $\Pi(Q^2)$ is known to be UV divergent, and is conventionally regularized
by using the subtracted function $\hat{\Pi}(Q^2)=4\pi^2[\Pi(Q^2)-\Pi(0)]$, which fixes the residue of the photon propagator to one. Finally, with the definitions above, the
LO HVP contribution to the muon $g-2$ is given by
\begin{equation}
  \amuHVPLO=\left(\frac{\alpha}{\pi}\right)^2\int_{0}^{\infty}dQ^2\,f(Q^2)\hat{\Pi}(Q^2)\,,
  \label{eq:amuhvpdef}
\end{equation}
where the kernel function $f(s)$ is defined by
\begin{equation}
\label{def_f}
  f(s)=\frac{r\big[Z(r)\big]^3}{m_\mu^2}\frac{1-rZ(r)}{1+r\big[Z(r)\big]^2}\,,\qquad\text{with}\qquad
  Z(r)=-\frac{r-\sqrt{r^2+4r}}{2r}\quad\text{and}\quad
  r=\frac{s}{m_{\mu}^2}\,.
\end{equation}
\Cref{eq:amuhvpdef} gives the spacelike analog of the timelike master formula \cref{amu_HVP_master}, see also \cref{sec:windows_data_driven}.
In practice, it has become more standard in lattice
calculations to evaluate the integral above on the time variable, for zero spatial
momentum, as summarized below in~\cref{sec:methods}. Furthermore, we define the
LO HVP contribution, denoted by $\amuHVPLO$ in
\cref{eq:amuhvpdef}, in terms of the current--current correlator $C_{\mu\nu}$ restricted to
one-photon-irreducible contributions.

We conclude this introduction by reviewing the status of lattice-QCD calculations of the
LO HVP contribution to the anomalous magnetic moment of the muon at the time of WP20. Further details can be found in WP20 and references therein.
Additionally, the definitions of the various components of $\amuHVPLO$ are summarized in
the next section. In WP20, six groups quoted complete results for the total LO HVP and were
included in a world lattice average. Notice, however, that the disconnected contribution
was computed on the lattice by only three groups, and some groups used
phenomenological inputs in order to estimate the strong and QED isospin-breaking (IB)
contributions. There was a relatively large spread among lattice results, but all
calculations were in agreement, at the $1.5\sigma$ level. The final estimate was
\begin{equation}
  \amuHVPLO(\text{WP20}) =  711.6(18.4) \times 10^{-10} \, ,
  \label{eq:hvp_lat}
\end{equation}
with an uncertainty of 2.6\%. As a consequence, the lattice world average was consistent
with both the dispersive data-driven estimate and the ``no new physics'' scenario. The
uncertainty was a factor of 4.5 larger than that of the data-driven estimate, and the
lattice average was not included in the final SM value for $\amuHVPLO$. We also note that
the BMW-20~\cite{Borsanyi:2020mff} lattice calculation, which reported a precision of
0.8\%, was posted on arXiv after the WP20 deadline and published in April 2021. 
Consequently, it was not included in the WP20 lattice world average of \cref{eq:hvp_lat}. 

Since none of the results retained in WP20 for the lattice HVP average of \cref{eq:hvp_lat} included complete nonperturbative calculations of all components, it was
decided to perform averages on individual contributions. In a second step, the errors of
the individual contributions were added conservatively, assuming 100\% correlation on the
uncertainties, leading to the final result given by \cref{eq:hvp_lat}. The total
uncertainty was dominated by the isospin-symmetric connected light-quark contribution,
followed by strong IB (SIB) and QED corrections and the quark disconnected
contribution. For the dominant light-quark contribution itself, the total error was
dominated by the uncertainties associated with finite-volume (FV) corrections, statistics,
and the continuum extrapolation. The uncertainties on the charm and strange quark
contributions were already at or below the permil level relative to $\amuHVPLO$.

In order to achieve the permil-level precision needed to match the final Fermilab precision, the community identified a number of challenges early on. Within this section, we present in detail the progress made on them since WP20. Briefly, the main challenges and the progress made to address them so far are:
\begin{itemize}
  \item FV corrections are significant, even on $6~\mathrm{fm}$ lattices, and are among
    the largest contributions to the total error quoted in \cref{eq:hvp_lat}. They can be estimated using EFTs or EFT-inspired models, including NNLO ChPT \cite{Bijnens:2017esv,Aubin:2015rzx,Aubin:2019usy,Aubin:2020scy}, the Chiral Model~\cite{Chakraborty:2016mwy}, the Meyer--Lellouch--L\"uscher (MLL) formalism with a Gounaris--Sakurai parameterization of
the timelike pion form factor (MLLGS)~\cite{Meyer:2011um, Francis:2013fzp, DellaMorte:2017dyu, Gounaris:1968mw}, and the Hansen and Patella formalism \cite{Hansen:2019rbh,Hansen:2020whp}. However, there are now several direct FV studies also at the physical point \cite{Borsanyi:2020mff,RBC:2024fic,Djukanovic:2024cmq} that guide these extrapolations and provide well-quantified control over the associated uncertainties. 
  \item 
    The lattice scale uncertainty is amplified in $\amuHVPLO$  by close to a factor of two \cite{DellaMorte:2017dyu}, due to its dependence on the muon mass, and, hence must itself be determined with permil-level precision. This is part of the program of every lattice collaboration engaged in an effort to compute HVP at permil-level precision. 
  \item A permil determination of $\amuHVPLO$ requires the full inclusion of IB effects even for quark-disconnected contributions. In principle, these effects also enter the determination of the lattice spacing. 
    In WP20 only partial calculations of these effects were available, while now results for almost all IB effects are available from several groups  \cite{RBC:2018dos,Giusti:2019dmu,Borsanyi:2020mff,Djukanovic:2024cmq,MILC:2024ryz}, except for a small fraction associated with sea--sea and sea--valence QED corrections, for which only one calculation exists \cite{Borsanyi:2020mff}. The long-distance (LD) QED contributions, which amount to about 0.3\% of the total HVP, are, however, still a challenge. 
  \item Many of these systematics (e.g., cutoff effects) are discretization dependent, so
    it is imperative to have results from different lattice actions. The results presented in this review employ gauge field ensembles generated independently by several lattice groups with improved gauge actions and a variety of different fermion actions, including Wilson-Clover, twisted-mass Wilson, Domain Wall, HISQ, and stout-smeared staggered fermion actions. 
\end{itemize}
The rest of this section is organized as follows: 
\Cref{sec:Breakdown} details the standardized lattice breakdown of $\amuHVPLO$
into the isospin-symmetric flavor components and IB corrections.
\Cref{sec:methods} is devoted
to the discussion of some special features in the analysis of lattice
calculations of HVP that are common to all calculations (e.g., blinding).
\Cref{sec:windows} describes the calculation of time windows, which allow
sharper and more detailed comparisons between lattice QCD calculations.
\Cref{sec:flavor_decomposition} discusses the isospin-symmetric flavor
contributions to the full Euclidean-time integration region. In
\cref{sec:totalHVP}, the results from the different
collaborations are combined in order to produce a final estimate. The rest of the section
describes the IB correction to the data from $\tau$ decays and the hadronic contribution
to the running of $\aem$. Those are presented in \cref{sec:otherobs}. Finally, possible
cross-checks resulting from the use of alternative approaches such as the covariant
coordinate-space representation are introduced in \cref{sec:furtherchecks}.

\subsection{Breakdown of the total HVP contribution to
  \texorpdfstring{$\amuHVPLO$}{amuHVPLO} into
its various contributions}\label{sec:Breakdown}

IB corrections to $\amuHVPLO$ due to the mass difference between up and
down quarks, as well as EM effects, are expected to be an $\mathcal{O}(1\%)$
correction relative to the total $\amuHVPLO$. Therefore, it is desirable to write
\begin{equation}
  \amuHVPLO =  \amuHVPLOiso + \deltaHVPLO\,,
  \label{eq:HVP_IB}
\end{equation}
where $\amuHVPLOiso$ denotes the isospin-symmetric component and $\deltaHVPLO$ the
first-order IB corrections, which are sufficient in the present context.
However, the separation above is not prescription independent. Indeed, since all quarks interact
strongly and electromagnetically, any experimental input used to renormalize QCD and QED
will only define the sum of the two terms above. Defining the individual terms requires
additional conditions which define what physics is kept constant while taking $\aem$ to
zero. This issue was discussed in the most recent FLAG review (cf.\ Sec.~3 in
Ref.~\cite{FlavourLatticeAveragingGroupFLAG:2024oxs}), where a prescription
referred to as the \emph{Edinburgh Consensus} was agreed upon after consultation with the
community. This scheme is the recommended prescription for future calculations. However,
most calculations in the present review were initiated before the Edinburgh Consensus was
agreed upon, and we will use a slightly modified prescription. We define
\emph{isospin-symmetric QCD (isoQCD)} to be the $\alpha=0$ theory where quark masses are tuned to
reproduce the following complete set of inputs
\begin{equation}
  M_{\pi^+}=135.0~\mathrm{MeV}\,, \quad M_{K^0}=M_{K^+}=494.6~\mathrm{MeV}\,, \quad
  M_{D_s^+}=1967~\mathrm{MeV}\,,
  \quad\text{and}\quad w_0= 0.17236~\mathrm{fm}\,,\label{eq:scheme-wp25}
\end{equation}
where $w_0$ is the Wilson flow scale introduced in Ref.~\cite{BMW:2012hcm}. We
call this prescription the \emph{WP25 scheme}. The only
difference between the WP25 scheme and the Edinburgh Consensus is the use of $w_0$ to
set the QCD scale instead of the pion decay constant. Recent results estimated that this
change does not generate significant deviations within the current level of uncertainty
(cf.~comparison in \cref{sec:LDwin}), although the matching between the two prescriptions
will require more scrutiny in the future. Additionally, the variables used in the
prescription above can be changed to other hadronic inputs while keeping the scheme fixed,
although this change will generate an additional matching
uncertainty in actual numerical calculations. Another set of variables introduced originally by the BMW
collaboration~\cite{Borsanyi:2020mff} are the connected meson masses $M_{qq}$, where for a
quark flavor $q$, $M_{qq}$ is the mass of the pseudoscalar meson $\bar{q}q$ considering only
quark-connected contributions. These masses, although purely theoretical,
are well-defined quantities that can be matched to the physical inputs above. The scheme
employed in BMW-20~\cite{Borsanyi:2020mff} resulted in isoQCD defined by
\begin{equation}
  M_{uu}=M_{dd}=M_{\pi^0}^{\mathrm{PDG}}\,,\quad M_{ss}=689.89(49)~\mathrm{MeV}\,,
  \quad M_{D_s^+}=1967~\mathrm{MeV}\,,
  \quad\text{and}\quad w_0=0.17236(70)~\mathrm{fm}\,,\label{eq:scheme-bmw20}
\end{equation}
where $M_{\pi^0}^{\mathrm{PDG}}=134.9768(5)~\mathrm{MeV}$ is the current world average of
the $\pi^0$ mass experimental measurements~\citep{ParticleDataGroup:2024cfk}. It was
shown in BMW/DMZ-24~\cite{Boccaletti:2024guq} that \cref{eq:scheme-bmw20} matches
\cref{eq:scheme-wp25} up to uncertainties small enough to be discarded in the context of
the present review. Because of this, \cref{eq:scheme-wp25}, or
\cref{eq:scheme-bmw20} with discarded uncertainties, may be referred to in the
literature as \emph{BMW-20} or \emph{BMW scheme}. In summary, we consider that
it is well motivated to average results
produced with either prescription, and that propagation of matching uncertainties is
unnecessary at the current level of precision.

Furthermore, it is convenient to decompose $\amuHVPLOiso$ into quark-disconnected and
single-flavor quark connected contributions:
\begin{equation}
  \amuHVPLOiso=\amuHVPLOud+\amuHVPLOs+\amuHVPLOc+\amuHVPLOb+\amuHVPLOdisc\,.
  \label{eq:HVP_flavor}
\end{equation}
For a quark flavor $q$, $\amuHVPLO(q)$ in the equation above is obtained from
\cref{eq:amuhvpdef} by considering only the quark-connected contribution for flavor $q$ in
the current--current correlator $C_{\mu\nu}$, and $\amuHVPLOdisc$ is the sum for all
flavors of the remaining quark-disconnected contributions. We now discuss standard
methodologies to predict the quantities defined in this introduction from lattice
simulations.

\subsection{Methodology}
\label{sec:methods}

Here, we briefly discuss some of the methodology that is common to the lattice-HVP
calculations presented in more detail below.
\begin{itemize}
  \item \emph{Time--momentum representation (TMR):}
    The TMR has become the standard method for estimating
    the LO HVP contribution using lattice methods~\cite{Bernecker:2011gh}. In this
    approach, \cref{eq:amuhvpdef} is re-written
    \begin{equation}
      \amuHVPLO = \left(\frac{\alpha}{\pi}\right)^2 \int_0^\infty d x_0\,
      C(x_0) \tilde{f}(x_0) \, ,
      \label{eq:tmramu}
    \end{equation}
    with the zero-momentum time correlator
    \begin{equation}
      C(x_0) = -\frac{1}{3}\sum_{k=1}^3 \int d^3x \, C_{kk}(x_0,\mathbf{x})\; ,
      \label{eq:C_t}
    \end{equation}
    and the kernel function
    \begin{equation}
      \tilde{f}(x_0) = 8\pi^2\int_0^\infty \frac{d \omega}{\omega} f(\omega^2)
      \left[\omega^2x_0^2-4\sin^2\left(\frac{\omega x_0}{2}\right)\right]\,.
      \label{eq:tmrkernel}
    \end{equation}
    In the equations above, $C_{\mu\nu}$ and $f$ are defined in \cref{sec:introduction}. The function $\tilde{f}$ can be expressed in terms of a modified Bessel
    function of the second kind and Meijer's $G$ function~\cite{DellaMorte:2017dyu} as
    \begin{equation}
      \tilde{f}(x_0) = \frac{2\pi^2}{m_\mu^2}\left[-2+8\gamma_{\rm E}
        + \frac{4}{\hat{t}^2} + \hat{t}^2-\frac{8}{\hat{t}}K_1(2\hat{t})
        +8\log\hat{t}+G^{2,1}_{1,3}
        \left(\left.
          \begin{array}{c}\frac{3}{2}\\
            0,1,\frac{1}{2}
      \end{array}\right|\hat{t}^2\right)\right]\, ,
      \label{eq:tmrexplicitkernel}
    \end{equation}
    where $\hat{t}=m_\mu x_0$. The nonperturbative input computed on the lattice is the
    time correlator $C$. Other approaches, based on a direct evaluation of the HVP tensor,
    $\Pi_{\mu\nu}(Q)$ (see, for example, Ref.~\cite{Aubin:2012me}), or on the method of time moments introduced in Ref.~\cite{chakraborty:2014mwa} have been superseded by the TMR.

  \item \emph{Error treatment in HVP calculations:}  Lattice-QCD calculations yield \textit{ab
    initio} theoretical predictions that are systematically improvable.  Since lattice
    QCD is a MC method, statistical analysis is a crucial part of producing
    physical results.  In addition to statistical errors from the MC process,
    systematic errors can enter in a variety of forms, for example, through truncating the functional form that guides continuum limit extrapolations of lattice data obtained at finite sets of lattice spacings.  
    While the details may vary from group to group, all lattice analyses described
    below include estimates of systematic errors based on studies of variations of the functional forms of the underlying EFT expansions, e.g.,
    consideration of various functional forms for continuum extrapolation, that allow the
    assessment of systematic errors associated with the respective extrapolation (or interpolation).  More detailed
    discussions of specific types of systematic errors, for which the dominant
    uncertainties may be different for different windows, will be provided in
    context below.

  \item \emph{Blinding:}  Most of the HVP results presented below, including (for example)
    all three calculations discussed for the dominant LD light-quark connected
    contribution \cite{RBC:2024fic,Djukanovic:2024cmq,FermilabLatticeHPQCD:2024ppc}, are based on
    blinded analyses.  Blinding has been used widely in lattice QCD for many years, to
    prevent the introduction of bias in calculations of quantities whose values are known
    either from experiment or from previous calculations.  To conduct a blind analysis, a small
    but unknown blinding factor is introduced into the analysis, modifying the final HVP
    result.  Different lattice groups introduce the blinding in different places, but most
    commonly the blinding is applied as part of the analysis code, which has the advantage
    that different blinding factors can be easily applied for different windows. As the
    precision of lattice calculations improves, the importance of blinding will only
    increase, and we encourage all groups to adopt blind analyses going forward.

  \item \emph{Treatment of correlations using FLAG procedure:}  Except where explicitly noted
    otherwise, to combine results from different lattice calculations, we adopt the
    procedure used by the FLAG group for averaging
    \cite{FlavourLatticeAveragingGroupFLAG:2024oxs}.  Central values in this procedure
    amount to uncorrelated weighted averages of individual central values, with
    statistical and systematic errors added in quadrature. Error bars for averages are
    constructed using a procedure by which 100\% correlation can be assumed between
    specific contributions to individual error estimates, as described in Eqs.~(7)--(10)
    of the reference.  In cases where the overall $\chi^2$ per degree of freedom implied by the average over
    individual results is poor (i.e., $\chi^2/{\rm dof} > 1$), a rescaling to increase the
    error bars by $\sqrt{\chi^2/{\rm dof}}$ is performed to compensate.

    In some cases, results for certain quantities from lattice groups have been superseded
    by updated calculations.  Such superseded results are shown in tables and figures
    below, but are excluded from averages.  Depending on the quantity considered,
    different sources of uncertainty are taken to conservatively be 100\% correlated.
    This may include statistical uncertainties in cases where the two calculations share
    the same gauge configurations, or systematic uncertainties for shared FV
    correction schemes, valence quark discretization, or other sources of error as described below.

\end{itemize}

\subsection{Window contributions}
\label{sec:windows}
Dividing the Euclidean time integral into different ranges has proved to be a very
valuable tool in comparing the results of various collaborations and also for comparison
with calculations of HVP based on dispersion relations (where time ranges correspond
to energy ranges).  After defining the three windows most commonly used, we consider
details of the results for each window.

\subsubsection{Definitions}
\label{subsec:defs_win}
Breaking the time integral~\cref{eq:tmramu} into three components allows for a better
separation of the different sources of uncertainty~\cite{RBC:2018dos}. The window
contributions are defined by
\begin{align}
  \amuHVPLO &= \amuSD + \amuW + \amuLD \,,\label{amu:sum}\\
  \amuSD &=  \left(\frac{\alpha}{\pi}\right)^2 \int_0^\infty d x_0\, C(x_0)
  \tilde{f}(x_0) [1-\Theta(x_0,t_0,\Delta)]\,, \label{eq:amusd}\\
  \amuW &= \left(\frac{\alpha}{\pi}\right)^2 \int_0^\infty d x_0\, C(x_0)
  \tilde{f}(x_0)
  [\Theta(x_0,t_0,\Delta)-\Theta(x_0,t_1,\Delta)]\,, \label{eq:winW}\\
  \amuLD &= \left(\frac{\alpha}{\pi}\right)^2 \int_0^\infty d x_0\, C(x_0)
  \tilde{f}(x_0) \Theta(x_0,t_1,\Delta)\,, \label{eq:win}
\end{align}
where
\begin{equation}
\label{def_wind}
  \Theta(t, t', \Delta) = \frac12 + \frac12\tanh\left(\frac{t-t'}{\Delta}\right)\,,
\end{equation}
defines the window as a smooth function in Euclidean time. The width parameter $\Delta$,
in particular, smooths out the window on both sides of the interval.
The community
advocated the use of the widely studied choice $\Delta = 0.15$\,fm,  $t_0 = 0.4$\,fm, and $t_1 = 1.0$\,fm \cite{RBC:2018dos} as benchmark
parameters. Each window can then be treated independently and extrapolated to the
continuum and infinite-volume limits. With this choice, the \emph{intermediate window (W)}
$\amuW$ is less sensitive to discretization effects than the \emph{short-distance (SD) window}
$\amuSD$. It is also designed to be less sensitive to LD effects than the
\emph{long-distance window} $\amuLD$ and hence much less affected by the signal-to-noise
problems at large $x_0$.

Finally, it is understood that all the window quantities can be decomposed as
per~\cref{eq:HVP_IB,eq:HVP_flavor}, using the same definitions and conventions as in the
case of $\amuHVPLO$. We now turn to results for each of the windows.

\subsubsection{Short-distance window}
\label{sec:SDwin}
\begin{figure}[!t]
  \centering
  \includegraphics[width=\linewidth]{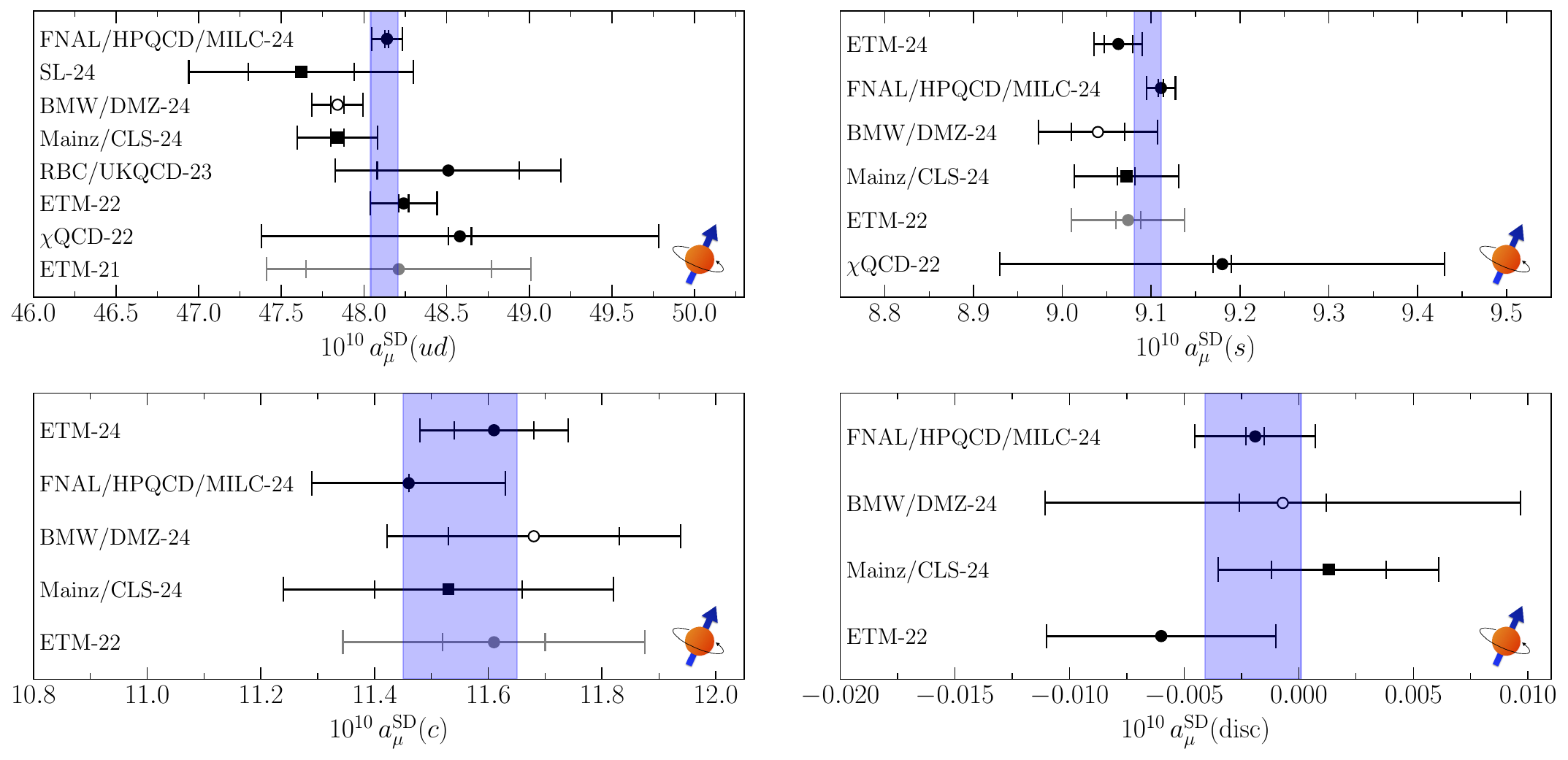}
  \caption{\small Comparisons of lattice results for flavor-specific contributions to $\amuSD$. Upper-Left: Light-quark-connected contribution $\amuSDud$. Upper-Right: Strange-quark-connected contribution $\amuSDs$. Lower-Left: Charm-quark-connected contribution $\amuSDc$. Lower-Right: Quark-disconnected contribution $\amuSDdisc$. We show in gray the results that have been superseded by new determinations. 
  Since the BMW/DMZ-24 results have not yet been published, they are excluded from the averages as denoted by the unfilled symbols. The plotting symbol indicates the number of sea quarks in the gauge ensembles employed in the underlying calculation. Circles correspond to ``$N_f = 2+1+1$'' ensembles with up, down, strange, and charm quarks in the sea, while squares denote ``$N_f = 2+1$'' in which charm quarks are omitted in the sea. The inner error bars, when displayed, represent the statistical uncertainties, while the outer error bars show the total uncertainties given by adding the statistical and systematic errors in quadrature. The lattice averages of~\cref{tab:amuSDflavsummary} are displayed here as light blue bands at the level of one standard deviation. See~\cref{tab:SD_fbf} for further details.}
  \label{fig:SD_fbf}
\end{figure}

We begin by reviewing the SD window observable, denoted as $\amuSD$ and
defined in~\cref{eq:amusd}. As in the case of the total LO HVP correction, also in the
SD window region the most important hadronic contribution is the one due to
the light $u$- and $d$-quark-connected contractions, $\amuSDud$, representing $\simeq
70\%$ of $\amuSDiso$, while the strange- and charm-quark-connected terms, $\amuSDs$ and
$\amuSDc$, constitute $\simeq 13\%$ and $\simeq 17\%$ of the total SD
contribution, respectively. In~\cref{tab:SD_fbf} and~\cref{fig:SD_fbf}, we collect results
for the flavor-specific contributions, namely, $\amuSDud$, $\amuSDs$, and $\amuSDc$, as
well as for the subleading correction $\amuSDdisc$ reported by the various lattice QCD
groups.
\begin{table}[!t]
\small
	\centering	\renewcommand{\arraystretch}{1.1}
    \begin{tabular}{lcllll}
      \toprule
      Collaboration&\hspace{-0mm} $N_f$&\hspace{-0mm} $\amuSDud$&\hspace{-0mm} $\amuSDs$&\hspace{-0mm} $\amuSDc$&\hspace{-0mm} $\amuSDdisc$\\
      \midrule
      ETM-24~\cite{ExtendedTwistedMass:2024nyi}&\hspace{-0mm}
      2+1+1&\hspace{-0mm} --&\hspace{-0mm} $9.063(16)(22)$&\hspace{-0mm}
      $11.61(07)(11)$&\hspace{-0mm} --\\
      FNAL/HPQCD/MILC-24~\cite{MILC:2024ryz}&\hspace{-0mm}
      2+1+1&\hspace{-0mm} $48.139(11)(91)$&\hspace{-0mm} $9.111(03)(16)$&\hspace{-0mm}
      $11.46(00)(17)$&\hspace{-0mm} $-0.0019(04)(26)$\\
      BMW/DMZ-24~\cite{Boccaletti:2024guq}&\hspace{-0mm} 2+1+1&\hspace{-0mm}
      $47.84(04)(15)$&\hspace{-0mm} $9.04(03)(06)$&\hspace{-0mm}
      $11.68(15)(21)$&\hspace{-0mm} $-0.0007(019)(102)$\\
      RBC/UKQCD-23~\cite{RBC:2023pvn}&\hspace{-0mm} 2+1(+1)&\hspace{-0mm}
      $48.51(43)(53)$&\hspace{-0mm} --&\hspace{-0mm} --&\hspace{-0mm} --\\
      ETM-22~\cite{ExtendedTwistedMass:2022jpw}&\hspace{-0mm} 2+1+1&\hspace{-0mm}
      $48.24(03)(20)$&\hspace{-0mm} $9.074(14)(62)^{\ast}$&\hspace{-0mm}
      $11.61(09)(25)^{\ast}$&\hspace{-0mm} $-0.006(5)$\\
      $\chi$QCD-22~\cite{Wang:2022lkq}&\hspace{-0mm} 2+1(+1)&\hspace{-0mm}
      $48.58(07)(1.2)$&\hspace{-0mm} $9.18(01)(25)$&\hspace{-0mm} --&\hspace{-0mm} --\\
      ETM-21~\cite{Giusti:2021dvd}&\hspace{-0mm} 2+1+1&\hspace{-0mm}
      $48.21(56)(57)^{\ast}$&\hspace{-0mm} --&\hspace{-0mm} --&\hspace{-0mm} --\\
      \midrule
      SL-24~\cite{Spiegel:2024dec}&\hspace{-0mm} 2+1/0&\hspace{-0mm}
      $47.62(32)(60)$&\hspace{-0mm} --&\hspace{-0mm} --&\hspace{-0mm} --\\
      Mainz/CLS-24~\cite{Kuberski:2024bcj}&\hspace{-0mm} 2+1&\hspace{-0mm}
      $47.84(04)(24)$&\hspace{-0mm} $9.072(10)(58)$&\hspace{-0mm}
      $11.53(13)(26)$&\hspace{-0mm} $0.0013(25)(41)$\\
      \bottomrule
    \end{tabular}
    \caption{Compilation of lattice results for flavor-specific contributions to $\amuSD$, in units of $10^{-10}$.
      The $N_f = 2+1$ Mainz/CLS calculation includes charm contributions from first
      principles in the valence sector, but not in the sea. Values marked with an asterisk
      are older results superseded by more recent determinations by the same
      collaboration. When results are displayed with two errors, the first is the
      statistical uncertainty and the second the systematic one. With only one quoted
    error, the statistical and systematic uncertainties are combined.}
    \label{tab:SD_fbf}
 \renewcommand{\arraystretch}{1.0}
\end{table}

The calculations in BMW/DMZ-24~\cite{Boccaletti:2024guq},
Fermilab/HPQCD/MILC-24~\cite{MILC:2024ryz}, and
ETM-22/24~\cite{ExtendedTwistedMass:2022jpw,ExtendedTwistedMass:2024nyi}
were performed on $N_f = 2+1+1$ ensembles. Those in Mainz/CLS-24~\cite{Kuberski:2024bcj}
and SL-24~\cite{Spiegel:2024dec}  are based on $N_f = 2+1$
ensembles. The former allow SD matching to the full SM at renormalization
scales equal to $m_b$, up to $1/m_b^2$ corrections, while that scale is $m_c$ and
corrections are of order $1/m_c^2$ for calculations based on $N_f = 2+1$ ensembles. The
Mainz/CLS-24~\cite{Kuberski:2024bcj} determination of $\amuSD$, however, includes the
leading correction due to the quenching of sea-charm quarks, estimating it to be at the
level of $0.02\times 10^{-10}$ based on perturbation theory~\cite{Hoang:1994it}. The
RBC/UKQCD calculation (RBC/UKQCD-23~\cite{RBC:2023pvn}) addresses that subleading
systematic effect from first principles, by generating gauge ensembles with dynamical
charm quarks matched to those with $N_f = 2+1$ sea quarks, and finds it to be small
compared to the quoted uncertainties. The gauge- and fermion-action dependence of the
SD window is studied by the $\chi$QCD group ($\chi$QCD-22~\cite{Wang:2022lkq})
using the overlap valence fermion action on ensembles with either the domain-wall fermion
(and the Iwasaki gauge action) $N_f = 2+1$-flavor sea from the RBC/UKQCD collaboration or
the highly-improved-staggered-quark (with the Symanzik gauge action) $N_f = 2+1+1$-flavor
sea from the MILC collaboration. Further details of the lattice formulations adopted for
the QCD action by the various groups can be found in the corresponding references
indicated in~\cref{tab:SD_fbf}.

The SD window observable can be obtained very precisely in lattice-QCD
calculations as far as statistical errors are concerned, not being affected by the
signal-to-noise problems present at large Euclidean time distances. FV effects,
which typically represent a source of uncertainty relevant for the calculation of
$\amuLD$, play hardly any role in $\amuSD$. Quantitatively, direct lattice evaluations,
estimates based on ChPT and data-driven determinations point to
sub-permil FV corrections for $L \simeq 6\, {\rm fm}$ and physical
$m_{ud}$~\cite{Boccaletti:2024guq}. As observed in
Refs.~\cite{RBC:2023pvn,Kuberski:2024bcj,MILC:2024ryz} the dependence of
$\amuSD$ on quark masses is mild as well. Also, the sensitivity to the lattice scale is
found to be subdominant~\cite{ExtendedTwistedMass:2022jpw,Kuberski:2024bcj}. On the
contrary, the SD window suffers from large discretization effects, associated
to the behavior of the lattice vector correlator at small-time distances, which are
expected to be logarithmically enhanced with respect to the naive $a^2$ scaling typical of
${\mathcal O}(a)$-improved calculations (compared to the usual case with on-shell
observables), as discussed in
Refs.~\cite{DellaMorte:2008xb,Ce:2021xgd,Sommer:2022wac,ExtendedTwistedMass:2022jpw}. Even
though this may present a significant challenge, the comparison among the results obtained
with different lattice regularizations provides the opportunity to test the robustness of
the continuum-limit extrapolation.

\begin{figure}[!t]
  \centering
  \raisebox{0.5cm}{\includegraphics[scale=0.13]{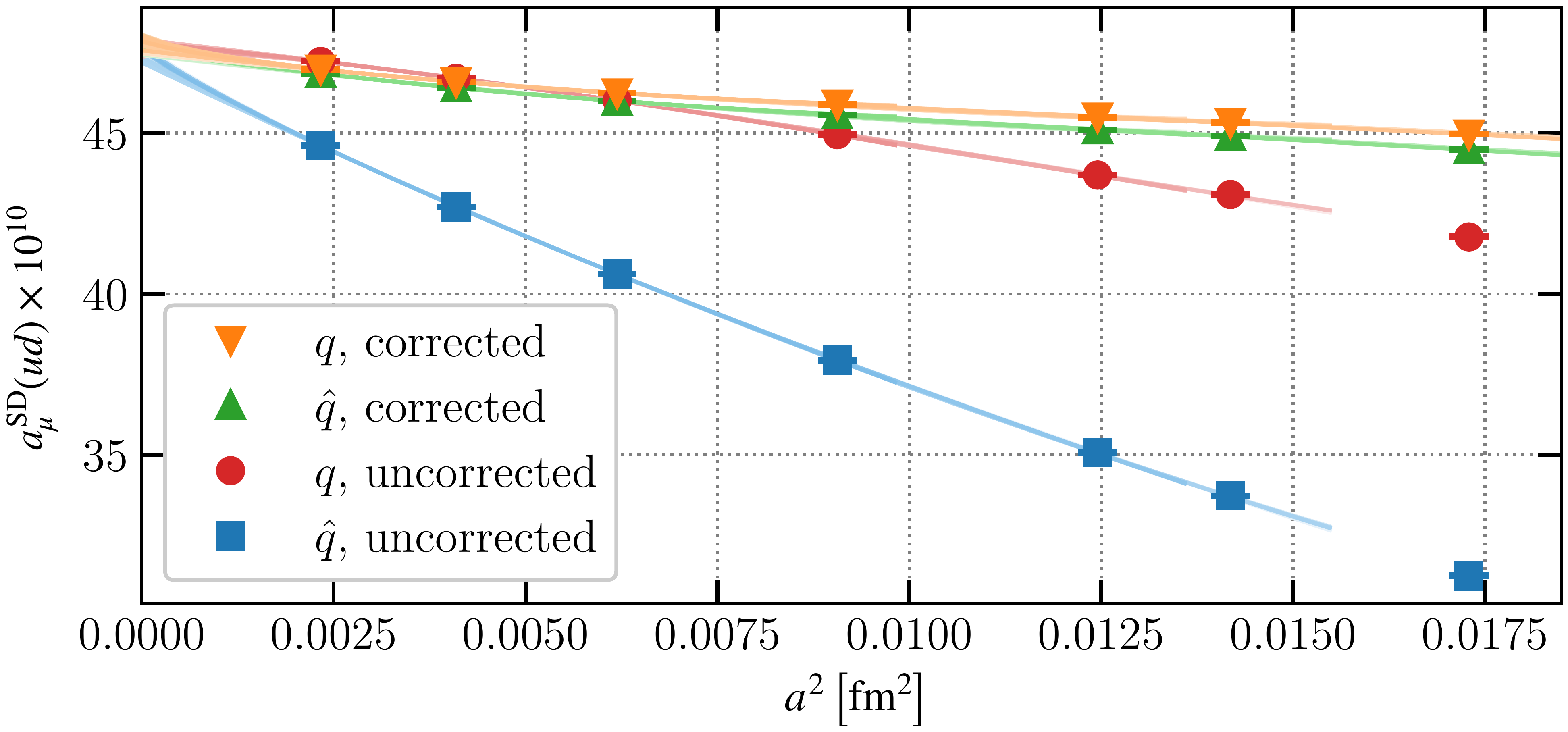}}
  \hspace{0.2cm} \includegraphics[scale=0.14]{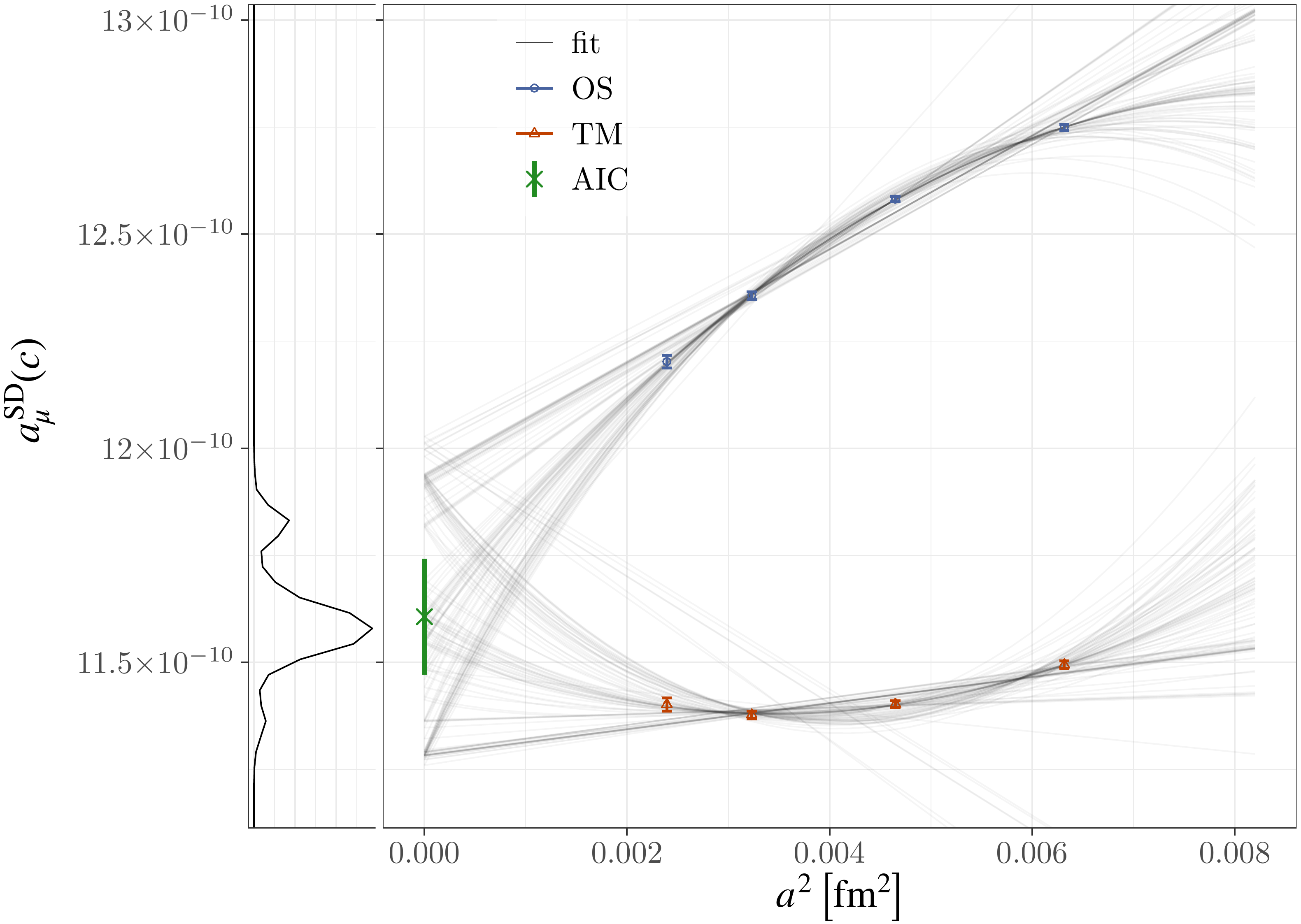}
  \caption{\small Quality of the extrapolations to the continuum limit for the light- and
    charm-quark connected contributions to $\amuSD$ performed in BMW/DMZ-24 (left panel) and
    ETM-24 (right panel), respectively. BMW/DMZ-24 uses data at seven lattice
    spacings and four
    variants of the observable of interest, while ETM-24 adopts two different
    valence-quark
    regularizations at four lattice spacings. Tree-level perturbative cutoff effects on
    lattice correlators are subtracted from the nonperturbative data in order
    to reduce the
    impact of dangerous ${\mathcal O}(a^2 \log a)$ artifacts. Figures adapted from
    Refs.~\cite{Boccaletti:2024guq,ExtendedTwistedMass:2024nyi}.}
  \label{fig:SD_continuum}
\end{figure}

The QCD action and the vector correlator employed in the calculations listed
in~\cref{tab:SD_fbf} are ${\mathcal O}(a)$-improved and all groups use data with at least
three lattice spacings (or two lattice spacings and two or more lattice regularizations of
the observable of interest) to control the extrapolation to the continuum limit.
Furthermore, different strategies are adopted to reduce the impact of (${\mathcal O}(a^2 \log
a)$) lattice artifacts from very short Euclidean distances: in
Refs.~\cite{ExtendedTwistedMass:2022jpw,RBC:2023pvn,Boccaletti:2024guq,Kuberski:2024bcj,
ExtendedTwistedMass:2024nyi}, tree-level perturbative cutoff effects on
lattice correlators are subtracted from the nonperturbative data; the use of a suitably
subtracted SD TMR kernel function together with
perturbative computations at very high momenta has been devised by the Mainz
group~\cite{Kuberski:2024bcj}; SL scrutinize the continuum limit of the
SD window using purely lattice QCD methods, without perturbative input, via a
combination of a quenched ($N_f = 0$) continuum extrapolation with a separate one of the
dynamical sea-quark corrections~\cite{Spiegel:2024dec}; while the Fermilab/HPQCD/MILC
collaboration adopts a highly-improved vector-current
discretization~\cite{MILC:2024ryz}. In the case of the staggered formulations
adopted in Refs.~\cite{Boccaletti:2024guq,MILC:2024ryz}, additional lattice
artifacts that generate the taste splittings among the pions are heavily suppressed in the
SD window region and found to be negligible at the level of precision reached.
The quality of the extrapolations to the continuum limit is shown
in~\cref{fig:SD_continuum} through the examples of the computations of $\amuSDud$ by
BMW/DMZ-24, who use the largest number (seven) of lattice spacings, and of $\amuSDc$ by
ETM-24, who quote the smallest relative uncertainty for that flavor-specific
contribution ($1.2\%$).

All results presented in~\cref{tab:SD_fbf} are extrapolated to the continuum and
infinite-volume limits and interpolated or extrapolated to the physical point. The quoted
errors in all lattice results include statistical and systematic uncertainties, where the
latter estimates effects from scale setting, input parameters, continuum extrapolation
(which represents the main source of uncertainty in the case of $\amuSD$), infinite-volume
extrapolation, and chiral interpolations/extrapolations. Typically, these systematic
errors are estimated by varying the chiral, continuum, or FV fit functions,
including adding higher-order terms in the corresponding EFT
expansions, or varying which lattice data are included, among other things. Overall, agreement is observed among the various (seven), currently available and more
recent determinations of the $ud$-quark connected contribution, shown in the upper-left
panel of~\cref{fig:SD_fbf}, with the most pronounced differences occurring between
the calculations in Mainz/CLS-24~\cite{Kuberski:2024bcj} and
BMW/DMZ-24~\cite{Boccaletti:2024guq}
on the one hand and those in ETM-22~\cite{ExtendedTwistedMass:2022jpw} and
Fermilab/HPQCD/MILC-24~\cite{MILC:2024ryz} on the other, at most at
the level of about $1.7\sigma$.
As far as the $s$-quark connected contribution is concerned, except for a
$1.5\sigma$ tension between the
ETM-24~\cite{ExtendedTwistedMass:2024nyi} and
Fermilab/HPQCD/MILC-24~\cite{MILC:2024ryz} determinations, the five
most up-to-date results,
listed in~\cref{tab:SD_fbf} and collected in the upper-right panel of~\cref{fig:SD_fbf},
exhibit good compatibility. The four results for $\amuSDc$ and for $\amuSDdisc$
(lower-left and -right panels of~\cref{fig:SD_fbf}, respectively) from
Refs.~\cite{ExtendedTwistedMass:2022jpw,Boccaletti:2024guq,Kuberski:2024bcj,
MILC:2024ryz,ExtendedTwistedMass:2024nyi} are nicely
consistent. For the $b$-quark connected contribution the ETM, Mainz, and
Fermilab/HPQCD/MILC groups provide the following estimates: $0.32\times
10^{-10}$~\cite{ExtendedTwistedMass:2022jpw} (based on perturbation theory);
$0.29(03)\times 10^{-10}$~\cite{Kuberski:2024bcj} (given by a combination of
  phenomenological/perturbative
  estimates~\cite{Kallen:1955fb,Chetyrkin:1995ii,Chetyrkin:1996ela,Erler:2021bnl}, and a
nonrelativistic-QCD-based lattice calculation~\cite{Colquhoun:2014ica}); and
$0.296(15)\times 10^{-10}$~\cite{MILC:2024ryz} (from the direct lattice
determination in Ref.~\cite{Hatton:2021dvg} and pQCD inputs), respectively.
Finally, the challenging nonperturbative calculation of the subleading IB
contributions $\delta \amuSD$ has been performed by only two collaborations so far. The
Mainz group performs an explicit calculation of the IB corrections to
the connected light- and strange-quark contributions due to unequal up- and down-quark
masses (SIB) and electric charges~\cite{deDivitiis:2011eh,deDivitiis:2013xla}, while the
QED correction to the charm-quark contribution is estimated in perturbation theory.
IB effects in the lattice scale and in the quark sea, as well as
quark-disconnected contributions in the calculation of the relevant correlation functions
are neglected. The total quoted correction is $0.095(95)\times
10^{-10}$~\cite{Kuberski:2024bcj}. The Fermilab/HPQCD/MILC collaboration computes from
first principles the LO SIB corrections for both the connected- and
disconnected-quark contractions. QED corrections are estimated in perturbation theory.
Higher-order contributions, including QED corrections to disconnected diagrams are
neglected. 
The resulting $-0.0049(35)\times 10^{-10}$ SIB connected, $0.015(12)\times 10^{-10}$ SIB disconnected and $0.028(28)\times 10^{-10}$~\cite{MILC:2024ryz} QED perturbative estimates yield a total SIB+QED contribution equal to $0.038(31)\times
10^{-10}$.

\begin{figure}[!t]
  \begin{center}
    \includegraphics[scale=0.5]{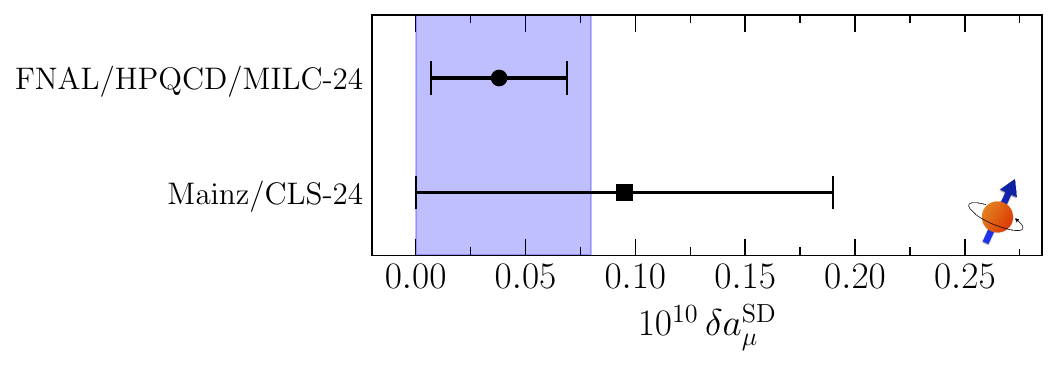}
    \caption{\small Comparison of the available lattice results for the SIB+QED IB corrections to the SD window observable, $\deltaSD$, displayed along with the corresponding average given in~\cref{tab:amuSDflavsummary}. Error ticks and plotting symbols and colors have the same meaning as in~\cref{fig:SD_fbf}.}
    \label{fig:SD_IB}
  \end{center}
\end{figure}

To average the lattice results for $\amuSDud$, $\amuSDs$, $\amuSDc$, and $\amuSDdisc$ given
in~\cref{tab:SD_fbf}, we use the strategy of the FLAG group described in
Ref.~\cite{FlavourLatticeAveragingGroupFLAG:2024oxs}. Any result that has been updated or
superseded by a more recent determination by the same lattice group is excluded from the
average. In~\cref{tab:SD_fbf} superseded results are marked with an asterisk. In addition, the results from BMW/DMZ-24 are not yet published, hence are also excluded from the following average. We treat the statistical and systematic errors in $\chi$QCD-22~\cite{Wang:2022lkq} and
RBC/UKQCD-23~\cite{RBC:2023pvn}, $\chi$QCD-22~\cite{Wang:2022lkq} and
Fermilab/HPQCD/MILC-24~\cite{MILC:2024ryz}, as well as in
RBC/UKQCD-23~\cite{RBC:2023pvn} and
SL-24~\cite{Spiegel:2024dec} as 100\% correlated since these calculations are based on
overlapping MILC and/or RBC/UKQCD gauge configurations, (partially) on the same
formulation of the QCD action, on common ensembles for the evaluation of systematic errors
(such as FV corrections) and/or on common scale uncertainty. All the other
determinations are treated as uncorrelated. Since sea-charm-quark effects on $\amuSD$ are
found to be below the permil level (see the above estimate in QCD perturbation
theory~\cite{Kuberski:2024bcj}), we can combine $N_f = 2+1$ and $N_f = 2+1+1$ results
in~\cref{tab:SD_fbf}. Note that $b$-quark effects are even more subleading and, therefore,
can be neglected here. For the valence-bottom-quark-connected contribution, $\amuSDb$, we
consider the uncertainties of the estimates mentioned above as 100\% correlated, since the
methods used by different teams to get them are similar. We provide an estimate for the
SIB+QED IB corrections, $\deltaSD$, by combining the two previously
described results given in Mainz/CLS-24~\cite{Kuberski:2024bcj} and
Fermilab/HPQCD/MILC-24~\cite{MILC:2024ryz} (see~\cref{fig:SD_IB}).
Since both determinations
come from first-generation calculations of some of those effects, we choose the
conservative approach of averaging Mainz/CLS-24 and Fermilab/HPQCD/MILC-24
under the assumption that the
uncertainties of the two calculations are 100\% correlated. IB effects in
the lattice scale and in the quark sea, as well as quark-disconnected QED contributions
neglected in those calculations are expected to be small and, thus, covered by the quoted
conservative error. The averages carry a $\chi^2/{\rm dof}<1$ in all cases.

Our averages for the individual flavor contributions to $\amuSD$, in the isospin limit and
for the leading SIB and QED corrections, are summarized in~\cref{tab:amuSDflavsummary}.
\begin{table}[!t]
\renewcommand{\arraystretch}{1.1}
  \begin{center}
    \small
    \begin{tabular}{cccccc}
      \toprule
      $\amuSDud$ & $\amuSDs$ & $\amuSDc$ & $\amuSDb$ & $\amuSDdisc$ & $\deltaSD$\\
      \midrule
      $48.123(83)$ & $9.096(15)$ & $11.55(10)$ & $0.30(02)$ & $-0.0020(21)$ & $0.04(04)$\\
      \bottomrule
    \end{tabular}
    \caption{\label{tab:amuSDflavsummary}Summary results for the individual flavor
      contributions to $\amuSDiso$ (in the isospin limit, $m_u=m_d$ and
      $\alpha=0$) and
    for the SIB and QED corrections, in units of $10^{-10}$.}
  \end{center}
  \renewcommand{\arraystretch}{1.0}
\end{table}
It should be stressed that the separation into isospin-symmetric flavor terms ($m_u =
m_d$ and $\alpha=0$) and IB corrections is scheme dependent, while the sum of
those contributions is unambiguous. We refer the reader to the various references given
in~\cref{tab:SD_fbf} for details on the schemes adopted by the lattice groups. Here it is
sufficient to note that ambiguities related to the different prescriptions used to define
isoQCD with respect to the full QCD+QED theory are negligible for the
SD window at the current permil precision level, as can be inferred from the
very modest impact of variations in the lattice-scale setting~\cite{ExtendedTwistedMass:2022jpw,FermilabLatticeHPQCD:2024ppc} and
quark-mass tuning~\cite{RBC:2023pvn,Kuberski:2024bcj,MILC:2024ryz} on $\amuSD$.

The contributions in~\cref{tab:amuSDflavsummary} can now be combined to obtain average
lattice numbers for the SD window contribution to LO HVP in the isospin
limit, $\amuSDiso$. Adding to this result the SIB and QED corrections yields the LO
HVP correction $\amuSD$, which can be compared with the phenomenological determinations of
$\amuSD$ (see also \cref{fig:results-lqc-CMD3,fig:results-s+lqd-CMD3} below). A final choice has to be made on how to combine uncertainties in adding up the
individual flavor and SIB+QED contributions. The uncertainty associated with the
determination of the overall scale is 100\% correlated among these contributions in a same
calculation. Also, in a calculation of the various contributions performed on the same set
of ensembles, statistical errors will be correlated. Thus, we choose the conservative
approach of adding the errors of the individual contributions linearly in our final
results. This procedure leads the following world average for the isospin-symmetric
SD window contribution to the total LO HVP correction:
\begin{equation}
  \label{eq:amuSD_iso_lat}
  \amuSDiso = 69.06(22)\times 10^{-10}\,.
\end{equation}
Adding to this result the average SIB+QED correction, we obtain the following lattice
world average for the total SD, LO HVP contribution to the muon $(g-2)$:
\begin{equation}
  \label{eq:amuSD_IB_lat}
  \amuSD = 69.10(26) \times 10^{-10}\,,
\end{equation}
where we include subleading digits at intermediate stages before rounding to the digits
shown. This result, which is roughly 10\% of the total LO HVP contribution, has a relative
uncertainty of about 0.4\%, and it is compatible with the estimate $68.4(5) \times
10^{-10}$ from $R$-ratio data given in Ref.~\cite{Colangelo:2022vok} in the pre-CMD-3
scenario. The absence of significant tensions in comparing lattice and $R$-data-driven
results for $\amuSD$ is consistent with what is found for the HVP $\Pi(Q^2)$ at large
Euclidean momentum $Q^2$, as well as with the constraints from EW precision tests of the
SM.

Another way of getting an average value for $\amuSD$, though with a less conservative
error, is to use directly the results quoted by a few groups, specifically $69.26(25)
\times 10^{-10}$ by the ETM collaboration
\cite{ExtendedTwistedMass:2022jpw,ExtendedTwistedMass:2024nyi},
$68.85(45) \times 10^{-10}$ by Mainz~\cite{Kuberski:2024bcj}, $69.05(21) \times
10^{-10}$ by Fermilab/HPQCD/MILC~\cite{MILC:2024ryz}. The average {\it \`a la}
FLAG of those independent determinations, treating the errors as uncorrelated across
calculations, amounts to $69.10(15) \times 10^{-10}$ which, remarkably, is in excellent agreement with our final value given in~\cref{eq:amuSD_IB_lat}.

\subsubsection{Intermediate window}
\label{sec:IDwin}

The standard intermediate window observable $\amuW$ contributing to the LO HVP term of
$a_\mu$ is defined in \cref{subsec:defs_win}, \cref{eq:winW}. Over the past few years
the intermediate window has played an important role, since it allows for a high-precision
comparison between SM predictions and experimental data for the $e^{+}e^{-}$
annihilation cross section into hadrons in the CM energy range around $1~{\rm
GeV}$. Indeed, from the perspective of lattice QCD, the computation of the intermediate
window is significantly less challenging than that of the full HVP contribution. The LD tail of the vector correlator is exponentially suppressed
by the integration kernel of $\amuW$, thereby mitigating both the signal-to-noise ratio
issues that affect the  correlator at large times and the large finite-size effects
associated with two-pion states. Moreover, unlike the SD window, $\amuW$ does
not suffer from large cutoff effects. Altogether these features enable very precise
lattice calculations of $\amuW$.

We first review results for $\amuW$ that are given in isoQCD. Since tiny
differences in the scheme defining isoQCD are present in the results obtained by the
various groups we adopt the WP25 scheme as a reference scheme prior to building averages
(see details below). Then we consider determinations of $\amuW$ in QCD+QED as quoted by
individual groups or coming from our average in the WP25-scheme
isoQCD and from
available estimates of IB effects (see \cref{tab:W_tot_comp}).

\begin{figure}[t!]
  \begin{center}
    \includegraphics[width=\linewidth]{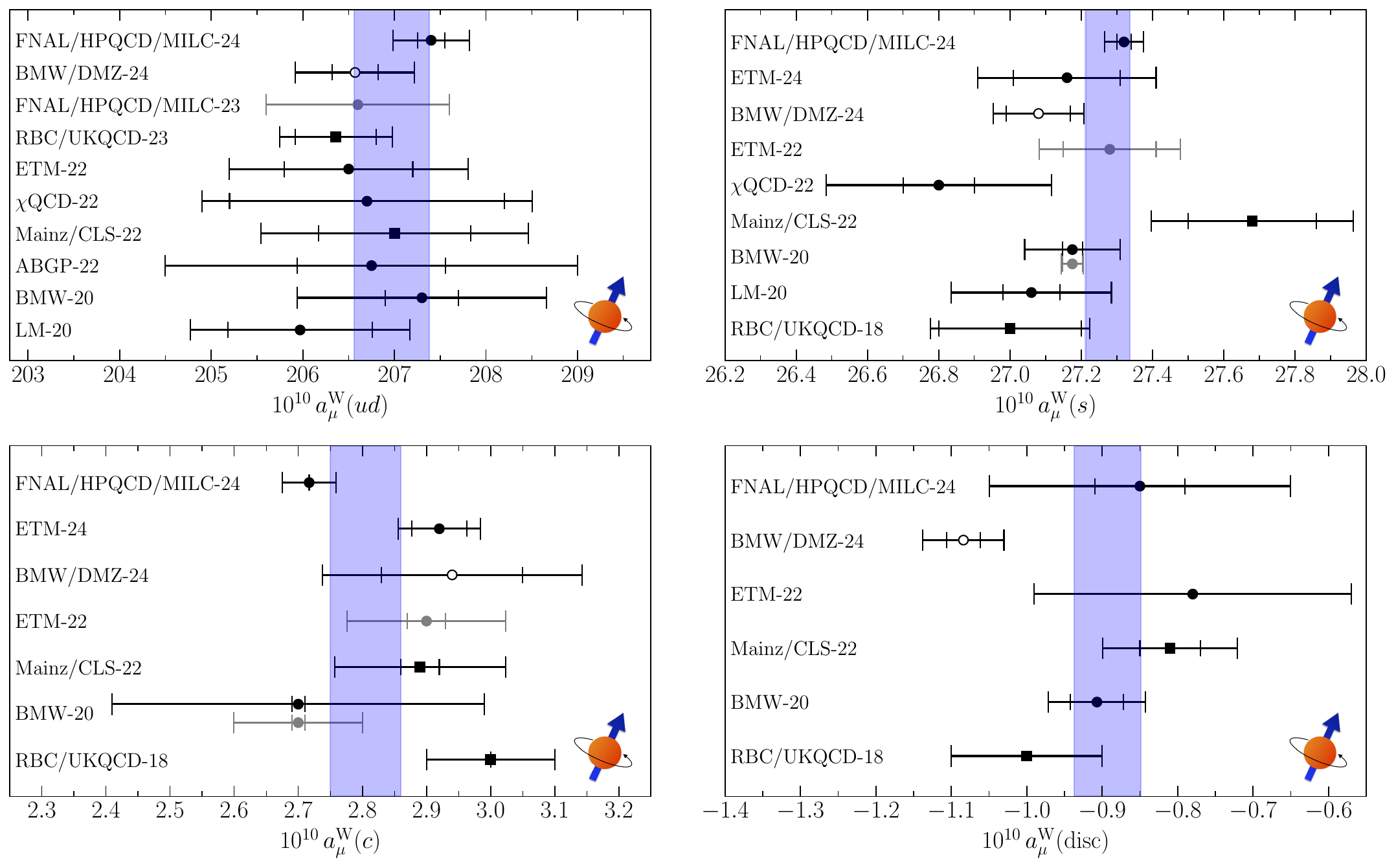}
    \caption{\small Compilation of lattice results for the partial flavor contributions to $\amuW$. Upper-Left: $ud$-quark connected, $\amuWud$. Upper-Right: $s$-quark connected, $\amuWs$. Lower-Left: $c$-quark connected, $\amuWc$. Lower-Right: quark disconnected, $\amuWdisc$.  Error ticks and plotting symbols and colors have the same meaning as in~\cref{fig:SD_fbf}.  For the BMW-20 results for $\amuWs$ and $\amuWc$, both the original published values and those including the additional systematic uncertainty explained in the caption of \cref{tab:W_comp} are shown in gray and black, respectively. The light blue band in each panel corresponds to the average $\pm 1 \sigma$ obtained using the procedure described in the text.}
    \label{fig:Wpartialcomp}
  \end{center}
\end{figure}

\begin{sloppypar}
We begin by separately discussing the connected light, strange, and charm quark
contributions to $\amuW$ in isospin-symmetric lattice QCD, denoted as $\amuWud$, $\amuWs$,
and $\amuWc$, respectively. For the connected $ud$-quark term, $\amuWud$, which is by
far the numerically dominant contribution to $\amuW$, we consider here all the lattice
results obtained using data with at least one ensemble at the physical pion mass point,
three lattice spacings (or two lattice spacings and two or more lattice regularizations of
the observable), and $M_{\pi}L \geq 3$. This criterion is relaxed for the subdominant
contributions, where results based on two lattice spacings are also included. Since the
deadline for inclusion in WP20, several lattice results have appeared:
LM-20~\cite{Lehner:2020crt}, BMW-20~\cite{Borsanyi:2020mff}, ABGP-22~\cite{Aubin:2022hgm},
Mainz/CLS-22~\cite{Ce:2022kxy}, $\chi$QCD-22~\cite{Wang:2022lkq},
ETM-22~\cite{ExtendedTwistedMass:2022jpw}, RBC/UKQCD-23~\cite{RBC:2023pvn},
Fermilab/HPQCD/MILC-23~\cite{FermilabLatticeHPQCD:2023jof},
BMW/DMZ-24~\cite{Boccaletti:2024guq}, and
Fermilab/HPQCD/MILC-24~\cite{MILC:2024ryz}. Concerning the lattice formulation
adopted, LM-20, ABGP-22, Fermilab/HPQCD/MILC-23, and Fermilab/HPQCD/MILC-24 employ the HISQ
staggered fermion discretization, BMW/DMZ-24 (BMW-20) use a staggered quark action with stout
smearing (4stout), RBC/UKQCD use domain-wall fermions, ETM-22 use Wilson-Clover
twisted-mass fermions, $\chi$QCD-22 use overlap valence fermions on both domain-wall and
HISQ sea, and Mainz/CLS-22 use ${\mathcal O}(a)$-improved Wilson fermions. These determinations of
the light-connected term, $\amuWud$ are collected in \cref{tab:W_comp}, and graphically
shown in the upper-left panel of \cref{fig:Wpartialcomp}. Within the quoted (statistical
plus systematic) errors a striking agreement is observed among all of them. The somewhat
older and since superseded result from RBC/UKQCD-18~\cite{RBC:2018dos} is also included in
the table for completeness.
\end{sloppypar}

For the $s$- and $c$-quark connected contributions,  $\amuWs$ and $\amuWc$, new results
have recently been presented by $\chi$QCD-22 ($s$-quark only), Mainz/CLS-22, ETM-22,
BMW/DMZ-24, and Fermilab/HPQCD/MILC-24, with an additional update from the ETM collaboration
(ETM-24~\cite{ExtendedTwistedMass:2024nyi}). All currently available data for these
contributions to the intermediate window are compiled in \cref{tab:W_comp} and displayed
in the upper-right and lower-left panels of \cref{fig:Wpartialcomp}. Overall, good agreement is observed
among the various determinations of the $s$-quark connected contribution, with the
Mainz/CLS-22 result lying slightly above the other ones. For the $c$-quark connected
contribution, a somewhat larger spread is evident, with the most pronounced differences
occurring between Fermilab/HPQCD/MILC-24 and RBC/UKQCD-18, and between
Fermilab/HPQCD/MILC-24 and ETM-24, each at the level of about $(2.6\text{--}2.7)\sigma$.

Turning to the quark-disconnected contribution to $\amuW$ in isoQCD, which
we denote as $\amuWdisc$, six lattice results are available:
RBC/UKQCD-18 \cite{RBC:2018dos}, BMW-20 \cite{Borsanyi:2020mff}, Mainz/CLS-22 \cite{Ce:2022kxy}, ETM-22 \cite{ExtendedTwistedMass:2022jpw}, BMW/DMZ-24 \cite{Boccaletti:2024guq}, and
Fermilab/HPQCD/MILC-24 \cite{MILC:2024ryz}. These are listed in \cref{tab:W_comp} and shown in the lower-right
panel of \cref{fig:Wpartialcomp}. The disconnected term gives only a tiny contribution to
the total $\amuW$, of the order of $0.05\%$, which is similar in size to the uncertainty
on the dominant light-connected term. A moderate compatibility is observed overall among
the different lattice results: the largest discrepancy, about $2.5\sigma$, arises between
the new BMW/DMZ-24 result and Mainz/CLS-22. We note that the BMW/DMZ-24 result is $20\%$ larger in
magnitude than BMW-20 (which had a relative uncertainty of about $7\%$).

In order to average the lattice results for
$
\amuWud,\,
\amuWs,\,
\amuWc,\,
\amuWdisc
$,
we follow the procedure established by the FLAG
group~\cite{FlavourLatticeAveragingGroupFLAG:2024oxs}.  We exclude from the average any
result that has been superseded by a more recent determination by the same collaboration, with the only exception of the BMW/DMZ-24 results, which have not yet been published.  Consequently in our averages we use the BMW-20 determinations,
supplementing them, when necessary, as specified in the caption of \cref{tab:W_comp},
with an additional systematic uncertainty.
In \cref{fig:Wpartialcomp}, superseded results appear in gray, while in \cref{tab:W_comp}
they are indicated by an asterisk. The final central values are determined via an
uncorrelated weighted average of the results, as described in Eqs.~(3)--(6) of
Ref.~\cite{FlavourLatticeAveragingGroupFLAG:2024oxs}. To estimate the final uncertainties,
we apply the method summarized in Eqs.~(7)--(10) of
Ref.~\cite{FlavourLatticeAveragingGroupFLAG:2024oxs}, which takes into account
correlations among the various results. As a conservative choice, in averaging we assume a
$100\%$ correlation in the statistical errors for groups that fully or partially share
gauge configurations, and a $100\%$ correlation in the systematic errors for groups using
the same discretization in the valence sector.

As anticipated, the average values for $\amuW$ partial and total results in
isoQCD are given in the reference WP25 scheme (see
\cref{sec:Breakdown}). Besides the BMW-20 and BMW/DMZ-24 results, the one of RBC/UKQCD-23 on $\amuWud$, and the ones by
Fermilab/HPQCD/MILC-24,
 the other intermediate-window results are given by the various authors in
slightly different schemes: we do not alter their central values and errors in the tables
and plots of this subsection, but in order to account for this systematic effect, prior to
and only for the sake of averaging, we add in quadrature to the results not in the WP25
scheme a $0.2\%$ systematic uncertainty for $\amuWud$, $\amuWc$, $\amuWdisc$, and a
$0.3\%$ systematic uncertainty for $\amuWs$. In almost all cases, this leads to a very
small increase of the total errors and thus to a correspondingly slight decrease of the
weight of these results in the average.

\newcommand{\emptycell}{--}
\begin{table}[t!]
\renewcommand{\arraystretch}{1.1}
  \begin{center}
    \small
    \begin{tabular}{lcllll}
      \toprule
      &\hspace{-0mm} $N_f$&\hspace{-0mm} $\amuWud$&\hspace{-0mm}
      $\amuWs$&\hspace{-0mm} $\amuWc$&\hspace{-0mm}
      $\amuWdisc$\\
      \midrule
      FNAL/HPQCD/MILC-24~\cite{MILC:2024ryz}&\hspace{-0mm}
      2+1+1&\hspace{-0mm}
      $207.40(15)(39)$&\hspace{-0mm} $27.32(2)(5)$&\hspace{-0mm}
      $2.717(0)(42)$&\hspace{-0mm}
      $-0.85(6)(19)$\\

      ETM-24~\cite{ExtendedTwistedMass:2024nyi}&\hspace{-0mm} 2+1+1&\hspace{-0mm}
      \emptycell&\hspace{-0mm} $27.16(15)(20)$&\hspace{-0mm} $2.920(43)(48)$&\hspace{-0mm}
      \emptycell\\

      BMW/DMZ-24~\cite{Boccaletti:2024guq}&\hspace{-0mm} 2+1+1&\hspace{-0mm}
      $206.57(25)(60)^{\#}$&\hspace{-0mm} $27.08(9)(9)^{\#}$&\hspace{-0mm}
      $2.94(11)(17)^{\#}$&\hspace{-0mm}
      $-1.084(22)(49)^{\#}$\\

      FNAL/HPQCD/MILC-23~\cite{FermilabLatticeHPQCD:2023jof}&\hspace{-0mm}
      2+1+1&\hspace{-0mm} $206.6(1.0)^{*}$&\hspace{-0mm} \emptycell&\hspace{-0mm}
      \emptycell&\hspace{-0mm} \emptycell\\

      RBC/UKQCD-23~\cite{RBC:2023pvn}&\hspace{-0mm} 2+1&\hspace{-0mm}
      $206.36(44)(43)$&\hspace{-0mm} \emptycell&\hspace{-0mm} \emptycell&\hspace{-0mm}
      \emptycell\\

      ETM-22~\cite{ExtendedTwistedMass:2022jpw}&\hspace{-0mm} 2+1+1&\hspace{-0mm}
      $206.5(7)(1.1)$&\hspace{-0mm} $27.28(13)(15)^{*}$&\hspace{-0mm}
      $2.90(3)(12)^{*}$&\hspace{-0mm} $-0.78(21)(0)$\\

      $\chi$QCD-22~\cite{Wang:2022lkq}&\hspace{-0mm} 2+1(+1)&\hspace{-0mm}
      $206.7(1.5)(1.0)$&\hspace{-0mm} $26.8(1)(3)$&\hspace{-0mm} \emptycell&\hspace{-0mm}
      \emptycell\\

      Mainz/CLS-22~\cite{Ce:2022kxy}&\hspace{-0mm} 2+1&\hspace{-0mm}
      $207.0(8)(1.2)$&\hspace{-0mm} $27.68(18)(22)$&\hspace{-0mm}
      $2.89(3)(13)$&\hspace{-0mm}
      $-0.81(4)(8)$\\

      ABGP-22~\cite{Aubin:2022hgm}&\hspace{-0mm} 2+1+1&\hspace{-0mm}
      $206.75(81)(2.10)$&\hspace{-0mm} \emptycell&\hspace{-0mm} \emptycell&\hspace{-0mm}
      \emptycell\\

      BMW-20~\cite{Borsanyi:2020mff}&\hspace{-0mm} 2+1+1&\hspace{-0mm}
      $207.3(4)(1.3)$&\hspace{-0mm} $27.175(28)(13)$&\hspace{-0mm}
      $2.70(1)(10)$&\hspace{-0mm} $-0.907(35)(54)$\\

      LM-20~\cite{Lehner:2020crt}&\hspace{-0mm} 2+1+1&\hspace{-0mm}
      $205.97(79)(90)$&\hspace{-0mm} $27.06(8)(21)$&\hspace{-0mm} \emptycell&\hspace{-0mm}
      \emptycell\\

      RBC/UKQCD-18~\cite{RBC:2018dos}&\hspace{-0mm} 2+1&\hspace{-0mm}
      $202.9(1.4)(0.3)^{*}$&\hspace{-0mm} $27.0(2)(1)$&\hspace{-0mm}
      $3.0(0)(1)$&\hspace{-0mm}
      $-1.0(1)(0)$\\

      \midrule
      Average &\hspace{-0mm} & \hspace{0mm}$206.97(41)$ & \hspace{0mm}$27.274(62)$ &
      \hspace{0mm}$2.805(55)$ & \hspace{0mm}$-0.893(44)$ \\
      \bottomrule
    \end{tabular}
    \caption{Single-flavor and disconnected contributions to $\amuW$ in
      isoQCD, in units of $10^{-10}$.
      In many cases, the results of the various groups correspond to slightly different
      definitions of isoQCD. To account for the impact of this
      ambiguity (even
      if small), we have added, prior to averaging, an additional systematic error to the
      published results, as described in the text. Values marked with an
      asterisk have been
      superseded by more recent results by the same collaboration and are hence
      excluded from
    the average that is given in the last row. Since the BMW/DMZ-24 results, which are indicated with a hash in the table, have not yet been published, we use the values quoted in the BMW-20 paper for the averages. For the specific contributions $\amuWs$ and $\amuWc$, however, the BMW/DMZ-24 analysis reports substantially larger systematic uncertainties. To remain conservative, in our averages we keep the BMW-20 central values while replacing their systematic errors with those quoted by BMW/DMZ-24 and adding the absolute difference between the BMW-20 and BMW/DMZ-24 determinations as an additional systematic uncertainty. This procedure yields $\amuWs=27.175(28)(90)(95)[134]$ and   $\amuWc =2.70(1)(17)(24)[29]$, where the error in square brackets is the total error.}
    \label{tab:W_comp}
  \end{center}
  \renewcommand{\arraystretch}{1.0}
\end{table}

The value of the systematic uncertainty added in quadrature to the results not in the
WP25 scheme can be justified as follows. Among the four renormalization inputs (for
scale, $ud$-, $s$-, $c$-quark mass) that specify an isoQCD scheme, all
lattice groups employ $M_{\pi}\simeq 135.0$~MeV to fix the $ud$-quark mass and $M_{D_s}
\simeq 1968.4$~MeV or a similar closely related input (e.g., $m_c/m_s \simeq 11.8$) to fix
the $c$-quark mass (which moreover has a tiny impact on the total $\amuW$). One is thus
only left with the ambiguities coming from use of different inputs as compared to the
WP25 scheme about the scale and the $s$-quark mass.

\begin{sloppypar}
The scale input in the WP25 scheme, which originally in the BMW-20 results is
given by $w_0\simeq 0.1724(7)$~fm (updated to
$0.1725(5)$~fm by BMW/DMZ-24), is commonly replaced by $M_\Omega \simeq 1672.5$~MeV (used,
e.g., by RBC/UKQCD-23 and Fermilab/HPQCD/MILC-24) or by the FLAG value of the pion decay
constant $f_\pi=130.5$~MeV (used, e.g., by ETM-22/24, Mainz/CLS-22, and
Fermilab/HPQCD/MILC-24). From RBC/UKQCD-23 we know that using the WP25 or the $M_\Omega$
scale input leads to $\lesssim 0.05\% $ central value shift in $\amuWud$. From the FLAG24
review~\cite{FlavourLatticeAveragingGroupFLAG:2024oxs} one learns that using $f_\pi$
as scale input leads to values of $w_0$ that are within 0.5\% of the WP25 value.
This can be seen in Table~77 of Ref.~\cite{FlavourLatticeAveragingGroupFLAG:2024oxs} by
comparing the results for $w_0$ from ETM-21~\cite{ExtendedTwistedMass:2021qui},
MILC-15~\cite{MILC:2015tqx}, HPQCD13A~\cite{Dowdall:2013rya}, and
CLS-21~\cite{Strassberger:2021tsu} (the last one inferred by comparing the value of
$\sqrt{t_0}$ to that of ETM-21). Using the estimate $d \log \amuWud / d \log a \simeq
0.4$ (see, e.g., ETM-22, Eq.~(A.20)) implies a systematic uncertainty on $\amuWud$ not
larger than  $0.2\%$. Using analogous arguments and information on the spectral density of
the current--current correlator in each flavor channel, a similar relative uncertainty due
to the $w_0$-scale ambiguity is estimated or expected for $\amuWs$ and $\amuWdisc$, and a
substantially smaller one for $\amuWc$.
\end{sloppypar}

The systematic error we add to non-WP25 scheme results for $\amuWs$ is 0.3\% because it
amounts to the sum in quadrature of 0.2\% arising from the scale ambiguity and another
0.2\% due to the difference in $s$-quark mass input for isoQCD as compared
to the WP25 one. Indeed, from an unpublished analysis of ETM-24 data at four lattice
spacings and physical pion mass we infer that the WP25 scheme input, which is the
fictitious strange-strange quark-connected meson mass $M_{ss}= 689.9$~MeV, leads to a
value of the kaon mass $M_K$ which is 0.7~MeV larger than $M_K({\rm FLAG})=494.6$~MeV
(used, e.g., by several lattice groups) and only 0.4~MeV smaller than $M_K({\rm
RBC/UKQCD})=495.7$~MeV. Combining the fact that the $M_K$ input spread compared to WP25
is about 0.2\% with the estimate of $d \log \amuWs / d \log M_K^2 = -0.41(6)$, which is
provided by Mainz/CLS-22 (see Eq.~(B.26) of Ref.~\cite{Ce:2022kxy}) and fully confirmed by
ETM-24 unpublished data, we estimate a systematic uncertainty of less than 0.2\% on
$\amuWs$ due to ambiguity in the $s$-quark input for isoQCD.

The averages of the single-flavor and disconnected contributions obtained in this way are
provided in the last row of \cref{tab:W_comp} and illustrated by the colored vertical
bands in \cref{fig:Wpartialcomp}. In the case of   $\amuWc$, the reduced
$\chi^{2}$ of the fit is approximately 3 and, due to the lower
level of compatibility among the lattice results in this case, we inflate the final
uncertainty by a factor of $\sqrt{\chi^{2}/\mathrm{dof}}$.\footnote{For $\amuWs$, the
  reduced $\chi^{2}$ of the fit is slightly larger than one (about 1.3), too, and we
inflated the final error by a factor $\sqrt{\chi^{2}/\mathrm{dof}}$.}

\begin{figure}[t!]
  \begin{center}
    \includegraphics[width=\linewidth]{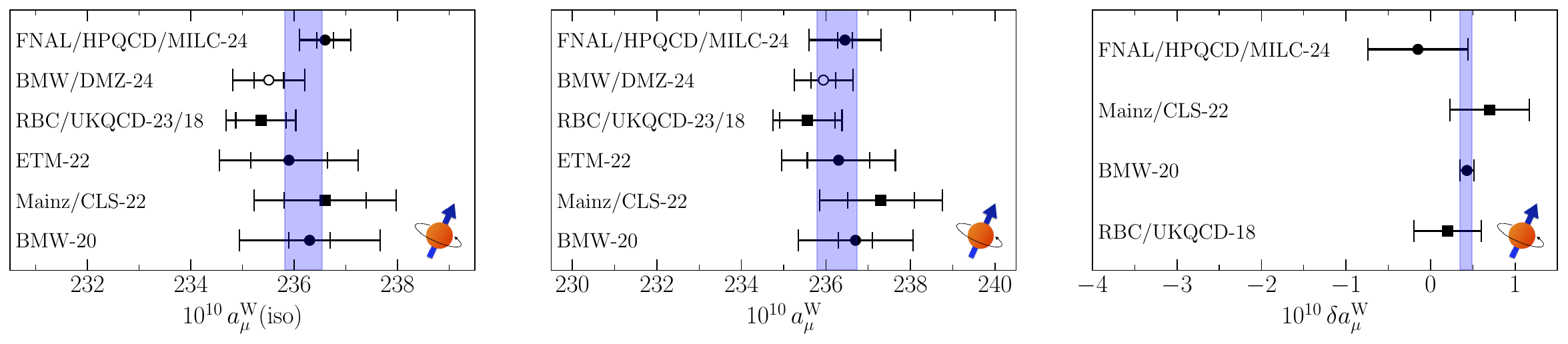}
    \caption{\small Compilation of lattice results for the total $\amuW$ in isospin-symmetric QCD (left), QCD+QED (center), and the available independently estimated IB corrections (right). Error ticks and plotting symbols and colors have the same meaning as in~\cref{fig:SD_fbf}. The light blue band in each panel denotes the average $\pm 1 \sigma$ using the procedure described in the text.}
    \label{fig:Wtotcomp}
  \end{center}
\end{figure}

We now turn to the results for $\amuWiso$ in isoQCD. There are five groups
who provided (or provided sufficient evidence to infer) results on $\amuWiso$:
Fermilab/HPQCD/MILC, BMW, RBC/UKQCD, ETM, and Mainz. The results are summarized in
\cref{tab:W_tot_comp} and shown in the left panel of \cref{fig:Wtotcomp}. For BMW/DMZ-24, we
sum the single-flavor and disconnected contributions, combining their errors in
quadrature. Instead, ETM-22, Mainz/CLS-22, and Fermilab/HPQCD/MILC-24 directly quote the value of
$\amuWiso$. For RBC/UKQCD-23/18, we begin with the value of $\amuWud$ quoted in the WP25
scheme~\cite{RBC:2023pvn} and then add the $s$- and $c$-quark connected contributions, as well
as the disconnected contribution given in Ref.~\cite{RBC:2018dos}. Because these three
contributions are not originally provided in the WP25 isoQCD scheme, we assign
systematic errors of $0.3\%$, $0.2\%$, and $0.2\%$, respectively, before summation. The
resulting RBC/UKQCD value and error are then treated as if given in the WP25 scheme. As
for BMW-20, we subtract an IB correction $\deltaW =
0.43(8) \times 10^{-10}$ from the total $\amuW$ computed in QCD+QED.

\begin{table}[t!]
\renewcommand{\arraystretch}{1.1}
  \begin{center}
    \small
    \begin{tabular}{lclll}
      \toprule
      &\hspace{-0mm} $N_f$&\hspace{-0mm} $\amuWiso$&\hspace{-0mm}
      $\amuW$&\hspace{-0mm} $\deltaW$\\
      \midrule
      FNAL/HPQCD/MILC-24~\cite{MILC:2024ryz}&\hspace{-0mm}
      2+1+1&\hspace{-0mm}
      $236.60(16)(47)$&\hspace{-0mm} $236.45(17)(83)$ 
      &\hspace{-0mm}$-0.15(59)$\\

      BMW/DMZ-24~\cite{Boccaletti:2024guq}&\hspace{-0mm} 2+1+1&
      \hspace{-0mm}
      $235.51(29)(63)^{\#}$ &\hspace{-0mm} $235.94(29)(63)^{\#}$
      &\hspace{-0mm}~\cite{Borsanyi:2020mff}\\

      RBC/UKQCD-23/18~\cite{RBC:2023pvn, RBC:2018dos}&\hspace{-0mm} 2+1&\hspace{-0mm}
      $235.36(49)(46)$ &\hspace{-0mm} $235.56(65)(50)$ &\hspace{-0mm}$0.2(4)$\\

      ETM-22~\cite{ExtendedTwistedMass:2022jpw}&\hspace{-0mm} 2+1+1&\hspace{-0mm}
      $235.90(74)(1.12)$ &\hspace{-0mm} $236.30(74)(1.12)$
      &\hspace{-0mm}~\cite{Borsanyi:2020mff}\\

      Mainz/CLS-22~\cite{Ce:2022kxy}&\hspace{-0mm} 2+1&\hspace{-0mm}
      $236.60(79)(1.13)$&\hspace{-0mm} $237.30(79)(1.22)$ &  $0.70(47)$ 
      \hspace{-0mm}\\

      BMW-20~\cite{Borsanyi:2020mff}&\hspace{-0mm} 2+1+1&\hspace{-0mm} $236.3(4)(1.3)$
      &\hspace{-0mm} $236.7(4)(1.3)$ &\hspace{-0mm}$0.43(08)$\\

      \bottomrule
      Average &\hspace{-0mm} & \hspace{0mm}$236.18(36)$ & \hspace{0mm}$236.26(47)$ &
      \hspace{0mm}$0.42(07)$ \\
      \bottomrule
    \end{tabular}
    \caption{Results for $\amuW$, in units of $10^{-10}$. We quote in different columns: the number of active quark flavors; 
    the isoQCD values discussed in this subsection (see \cref{eq:ave_iso} and related text), $\amuWiso$, with the average in last line given in the WP25 scheme (see text about how we treat the ETM-22 and Mainz/CLS-22 results that are quoted in a slightly different scheme); 
    the QCD+QED values based on the papers cited in each line, $\amuW$, with their average given in last line; 
    the IB contribution, $\deltaW$, with respect to
    isoQCD in the WP25 scheme (see text for details),
     and its average in the last line. As in the case of the single-flavor and disconnected contributions, since the BMW/DMZ-24 results have not yet been published (they are indicated with a hash in the table), we use the values quoted in the BMW-20 paper for the averages.  }    
    \label{tab:W_tot_comp}
  \end{center}
  \renewcommand{\arraystretch}{1.0}
\end{table}

In analogy to the procedure above, the average of results for $\amuWiso$ is also given
in the WP25 isoQCD scheme. In order to account for the slightly different definitions of
the isoQCD entering the results by ETM and Mainz in
\cref{tab:W_tot_comp}, we add, prior to averaging, a systematic uncertainty of $0.2\%$
(i.e., the same systematic error we attributed previously on $\amuWud$) to all results not
quoted in the WP25 isoQCD scheme. The average, performed according to the FLAG procedure
just as for the partial single-flavor and disconnected contributions, yields\footnote{As in the case of the single-flavor and disconnected contributions, since the BMW/DMZ-24 results have not yet been published, we base our averages on the BMW-20 determinations (see also the caption of \cref{tab:W_tot_comp}).}
\begin{align}
  \label{eq:ave_iso}
  \amuWiso = 236.18(36) \times 10^{-10}\,,
\end{align}
with an uncertainty of about 0.15\%.

An alternative procedure to obtain $\amuWiso$ is to sum the (average of) single-flavor and
disconnected contributions listed in the last row of \cref{tab:W_comp}. Noting that the
quark-connected correlators for different flavors and the quark-disconnected correlators
all have very different statistical fluctuations, reflecting the completely different
physics of the corresponding intermediate-state channels, we expect their statistical
correlations to be tiny already within the results of a single group and further diluted
by the average leading to last line of \cref{tab:W_comp}, and hence safely negligible. For
this reason we proceed by adding the uncertainty of these contributions in quadrature to
obtain
\begin{align}
  \amuW(\mathrm{iso}, \mathrm{single\text{-}flav\text{-}sum})
  = 236.16(42) \times 10^{-10}\,. \label{eq:amu-iso-flasum}
\end{align}
This result is in excellent agreement with that of \cref{eq:ave_iso}. If one were to treat
the individual flavor contributions as fully correlated one would obtain a result with a
36\% larger uncertainty, namely
\begin{align}
  \amuW(\mathrm{iso}, \mathrm{single\text{-}flav\text{-}sum\text{-}correlated})
  = 236.16(57) \times 10^{-10}\,. \label{eq;amu-iso-flasum-lin}
\end{align}
For the reasons outlined above, we consider this latter value to be overly conservative
and advocate the value given in \cref{eq:amu-iso-flasum} as our central result for the
isospin-symmetric window contribution $\amuWiso$ in the WP25 scheme.

To provide a QCD+QED result with a conservative error for $\amuW$ we add the average of IB
effects $\deltaW$ to the result \cref{eq:amu-iso-flasum} obtaining the value
\begin{align}
  \label{eq:ave_amu_QCD_QED}
  \amuW = 236.58(43) \times 10^{-10}\,. 
\end{align}
The estimates of the IB effect $\deltaW$ with respect to isoQCD in the WP25 scheme are given for each group in the last column of \cref{tab:W_tot_comp}. For the cases where the value of $\deltaW$ is not taken from the BMW-20 paper, we proceed as follows: for Fermilab/HPQCD/MILC-24 and RBC/UKQCD-23/18 we obtain it by combining the published information on the correlated difference in $\amuW(\mathrm{iso})$ between the used scheme and the WP25 scheme with the requirement that the sum $\amuWiso +\deltaW$ be scheme independent; for Mainz/CLS-22 we keep the published central value of $\deltaW$ in the scheme of choice, assuming that the scheme dependence is accounted for in the estimated systematic error. 

Another way of obtaining a final result for $\amuW$ is to use directly the results quoted in QCD+QED by a few groups, see the
corresponding column of \cref{tab:W_tot_comp}. An average \emph{\`a la} FLAG leads to $\amuW =
236.26(47) \times 10^{-10}$, which is reassuringly in very good agreement with the value
in \cref{eq:ave_amu_QCD_QED}.

Finally, we stress that the result of \cref{eq:ave_amu_QCD_QED} is potentially of high
relevance for the experimental collaborations measuring $e^+ e^- \to \text{hadrons}$ and
for precision tests of the SM. Indeed, the quantity $\amuW$ can be accessed experimentally
and lattice QCD+QED provides for it a pure SM prediction with an accuracy of about 0.18\%.

\subsubsection{Long-distance window}
\label{sec:LDwin}

A precise estimation of the LD contribution is crucial for achieving high
precision in $\amuHVPLO$, as it dominates both the central value and the uncertainty in
lattice-QCD calculations. The latter arises primarily from the exponential signal-to-noise
problem, which results in larger statistical uncertainties compared to $\amuSD$ and
$\amuW$. Additionally, significant FV effects pose challenges for
conventionally sized lattices with $M_\pi L \leq 5$ at the physical value of the pion
mass. Moreover, when the staggered fermion formulation is used to simulate quark fields,
substantial taste-breaking effects introduce deviations from the continuum theory, further
complicating the continuum extrapolation.

Four groups have performed dedicated computations of contributions to the LD window $\amuLD$. RBC/UKQCD-24~\cite{RBC:2024fic},
Mainz/CLS-24~\cite{Djukanovic:2024cmq}, and Fermilab/HPQCD/MILC-24~\cite{FermilabLatticeHPQCD:2024ppc}
have all evaluated the dominant light-quark connected contribution to $\amuLD$. In addition,
Mainz/CLS-24 provide results for the remaining flavor contributions, yielding a complete
determination of the LD window in isoQCD. The ETM collaboration
has computed the subleading strange- and charm-connected contribution to $\amuLD$ in
ETM-24~\cite{ExtendedTwistedMass:2024nyi}.

The RBC/UKQCD collaboration employs simulations with $2+1$ flavors of domain-wall fermions
at three lattice spacings, down to $0.073\,\text{fm}$. Three ensembles at physical quark
masses are complemented by six simulations with heavier-than-physical pion masses. A
spectral reconstruction of the vector correlation function from two-pion states is used to
reduce statistical noise in the LD regime. The Mainz calculation is based
on $2+1$ flavors of $\mathcal{O}(a)$-improved Wilson fermions at six lattice spacings, with
the finest reaching $0.039\,\text{fm}$. The analysis includes 34 gauge ensembles with pion
masses ranging from $420\,\text{MeV}$ down to $132\,\text{MeV}$, including three ensembles
with physical quark masses. Low-mode averaging is supplemented by a spectral reconstruction on
selected ensembles to compute the vector correlation function. Fermilab/HPQCD/MILC use
four ensembles with $2+1+1$ flavors of highly improved staggered quarks, all at physical
quark masses. The finest ensemble corresponds to a lattice spacing of $0.06\,\text{fm}$.
Low-mode averaging is used for the vector correlation function, and EFT-based
taste-breaking corrections are applied alongside finite-size corrections to mitigate
cutoff effects specific to the staggered formulation. The ETM collaboration computes
$s$- and $c$-quark connected correlation functions on six physical-mass ensembles
with $2+1+1$ flavors of twisted-mass fermions at four lattice spacings, with the finest
being $0.049\,\text{fm}$.

Currently, no lattice calculation exists for IB effects in the pure
LD regime, restricting the averaging of results to the isospin-symmetric case.
Given the enhanced scale dependence of $\amuLD$ compared to $\amuSD$ and $\amuW$, and the
fact that IB corrections primarily affect large Euclidean distances, a
significant dependence on the scheme used for isoQCD cannot be excluded at
present. To mitigate this and enable meaningful comparisons and averaging of lattice-QCD
results, groups computing the light-connected contribution to $\amuLD$ have performed
their analyses in multiple schemes.

RBC/UKQCD-24 quote results in the RBC/UKQCD-18 scheme~\cite{RBC:2018dos} as well as in
the WP25 scheme, both of them using the $\Omega$ baryon mass
$M_\Omega$ to set the scale. Mainz/CLS-24 provide results in a scheme defined in
Ref.~\cite{Ce:2022kxy}, which employs a combination of the pion and kaon decay constants,
$f_\pi$ and $f_K$, to set the scale, as well as in the WP25 scheme. Similarly,
Fermilab/HPQCD/MILC-24 quote results in two schemes that differ only in the scale-setting
quantity, either $M_\Omega$ or $f_\pi$, and also in the WP25 scheme. ETM-24 employ the
scheme of the Edinburgh consensus based on $f_\pi$ to set the scale.

\begin{figure}[!t]
  \centering
  \includegraphics[scale=0.5]{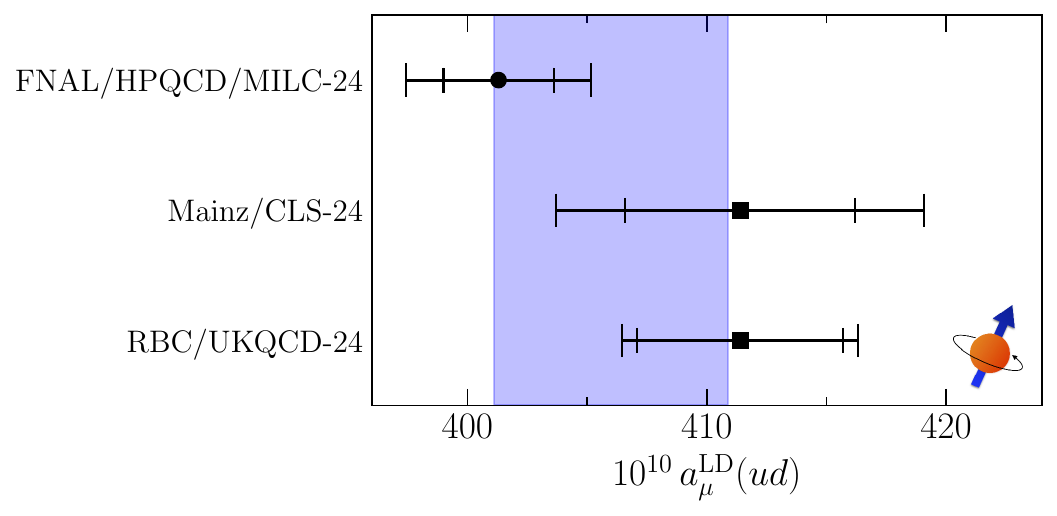}
  \caption{\small Compilation of lattice results for $\amuLDud$ as listed in \cref{tab:LD_comp}. The light blue band corresponds to the average given in \cref{e:hvp_lat_amuLD_I1_av}. Error ticks and plotting symbols and colors have the same meaning as in~\cref{fig:SD_fbf}.}
  \label{fig:comp_LD}
\end{figure}

While the values provided by the collaborations in the WP25 scheme are expected to match within the quoted uncertainties, the degree to which the collaboration-specific $M_\Omega$- and $f_\pi$-based schemes differ has not yet been systematically studied.  
We recognize that the spread of the $f_\pi$-based
schemes between Fermilab/HPQCD/MILC-24 and Mainz/CLS-24 is significant and needs to be scrutinized in future work.  To account for this effect in WP25, we consider additional averages in the WP25 scheme with Fermilab/HPQCD/MILC-24 and Mainz/CLS-24 excluded, respectively.  We then take half of the difference of the central value of those two averages as an additional uncertainty that is added in quadrature, following the approach adopted in WP20 for the \babar{}--KLOE tension of the data-driven HVP evaluation.
As a result, an average of the three results for the light-quark connected contribution can be
robustly performed in the WP25 scheme. The results across all schemes are
summarized in \cref{tab:LD_comp} and shown in \cref{fig:comp_LD} for the common WP25 scheme.

\begin{table}[!t]
\renewcommand{\arraystretch}{1.1}
  \begin{center}
    \small
    \begin{tabular}{lclll}
      \toprule
      &\hspace{-0mm} $N_f$&\hspace{-0mm} WP25&\hspace{-0mm} Own scheme ($M_\Omega$) &\hspace{-0mm}
      Own scheme ($f_\pi$) \\
      \midrule
      FNAL/HPQCD/MILC-24~\cite{FermilabLatticeHPQCD:2024ppc} &
      \hspace{-0mm}
      2+1+1 &
      \hspace{-0mm}
      $401.3(2.3)(3.1)$ &
      \hspace{-0mm}
      $400.2(2.3)(3.7)$ &
      \hspace{-0mm}
      $396.6(2.2)(3.3)$ \\

      Mainz/CLS-24~\cite{Djukanovic:2024cmq} &
      \hspace{-0mm}
      2+1 &
      \hspace{-0mm}
      $411.4(4.8)(6.0)$ &
      \hspace{0mm}
      -- &
      \hspace{-0mm}
      $420.8(4.1)(3.5)$ \\

      RBC/UKQCD-24~\cite{RBC:2024fic} &
      \hspace{-0mm}
      2+1 &
      \hspace{-0mm}
      $411.4(4.3)(2.4)$ &
      \hspace{-0mm}
      $413.6(6.0)(2.9)$ &
      \hspace{0mm}
      -- \\

      \bottomrule
    \end{tabular}
    \caption{%
      Results for the light-connected contribution $\amuLDud$ in units of
      $10^{-10}$, obtained
      by three groups using different schemes for isoQCD. The third column
      lists results computed in the WP25 scheme. The remaining two columns
      show results in each collaboration's own scheme labeled by
      the quantity used for scale setting (see main text for details). We note that (except for WP25) schemes with the same scale setting used by different collaborations do not match exactly. Uncertainties are given in parentheses, where the first
      denotes statistical errors and the second systematic errors. }
    \label{tab:LD_comp}
  \end{center}
  \renewcommand{\arraystretch}{1.0}
\end{table}

While all three results are statistically independent, they adopt similar approaches to
correct for finite-size effects. RBC/UKQCD-24 apply the method developed by Hansen and
Patella~\cite{Hansen:2019rbh,Hansen:2020whp}. Mainz/CLS-24 employ the same method for
short Euclidean distances, complemented at larger distances by the
Meyer--Lellouch--L\"uscher (MLL) formalism, which uses a Gounaris--Sakurai parameterization of
the timelike pion form factor~\cite{Meyer:2011um, Francis:2013fzp, Gounaris:1968mw}.\footnote{A recent study finds that the MLL formalism is valid at all Euclidean distances and reproduces the result of Hansen--Patella by including sufficiently high-energy two-pion states \cite{Itatani:2024fpr}.} Both studies
explicitly validate these corrections through measurements in multiple volumes with
identical Lagrangian parameters. Fermilab/HPQCD/MILC-24, on the other hand, perform a
model average over three different approaches to account for FV effects,
including the MLL method. RBC/UKQCD-24, Mainz/CLS-24, and Fermilab/HPQCD/MILC-24 estimate
the uncertainty for the light-connected contribution $\amuLDud$ arising from FV
corrections to be 0.8, 1.7, and 1.4 units of $10^{-10}$, respectively. 
To ensure a conservative assessment, we adopt these values as a correlated contribution to the
systematic uncertainty across all three results.

When averaging according to the FLAG procedure, we apply a slight $\sqrt{\chisq}$ rescaling of
$1.6$ to account for the fact that the result from Fermilab/HPQCD/MILC-24 is
significantly smaller than the other two. The $p$-value of the average is $0.2$. The outcome of the partly correlated average, combined with the additional systematic uncertainty of $3.1 \times 10^{-10}$, is
\begin{align} \label{e:hvp_lat_amuLD_I1_av}
  \amuLDud = 406.0( 4.9) \times 10^{-10}\,,
\end{align}
where the contribution of the uncertainty of $w_0$ that is part of the scheme definition
is not included. This uncertainty originates from the determination of the physical value
of the gradient flow scale $w_0 = 0.17236(70)$\,fm, based on $M_\Omega$ in
Ref.~\cite{Borsanyi:2020mff}. The average is displayed as a blue band alongside the three
inputs in \cref{fig:comp_LD}.

To compute the full LD window observable, the averaged result for the isovector
contribution can be combined with the Mainz/CLS-24 determination of the isoscalar
contribution within the WP25 scheme. The strange-connected contribution is included in
the isoscalar part. Its determination in Mainz/CLS-24 is slightly larger than, but
statistically compatible with, the result from ETM-24, where the two employed schemes are
similar but not identical. The charm-connected contribution, computed in Mainz/CLS-24 and
ETM-24, is numerically negligible at the current level of precision and has not been
obtained in the WP25 scheme in either work.
We note that this procedure uses the LD disconnected and strange contributions from only Mainz/CLS-24.

By adding the averaged isovector contribution, $\amuLD(\mathrm{I1}) = \frac{9}{10}
\amuLDud$, to the isoscalar contribution, $\amuLD(\mathrm{I0}) =
42.5(1.8)_\mathrm{stat}(1.5)_\mathrm{syst}$~\cite{Djukanovic:2024cmq}, one obtains
\begin{align}
  \amuLD(\mathrm{iso}) = 407.9( 5.0) \times 10^{-10}\,,
  \label{eq:amuLDiso}
\end{align}
for the LD window observable in isoQCD within the WP25 scheme,
again without scale uncertainty.

\subsection{Flavor decomposition}\label{sec:flavor_decomposition}

In this section, we summarize the results for flavor specific contributions to
$\amuHVPLO$. Following the breakdown in \cref{eq:HVP_flavor}, we consider the
light-, strange-, and charm-quark connected contributions, followed by a
discussion of the disconnected contribution. These results are obtained by
considering the individual flavor contributions over the entire integration
range. In principle, this corresponds to
performing the integral of \cref{eq:tmramu}. However, in practice, the results are now
more commonly obtained from summing the contributions to the three windows as in
\cref{amu:sum}. This summed approach leverages the fact that each window
contribution has a tailored analysis, resulting in a more robust determination,
once inter-window correlations are taken into account~\cite{RBC:2024fic,Djukanovic:2024cmq,FermilabLatticeHPQCD:2024ppc}. Final averages are given
in the WP25 isospin-symmetric scheme of \cref{eq:scheme-wp25}. Where possible, results in figures are also presented in this scheme.

\begin{table}[!t]
\renewcommand{\arraystretch}{1.1}
  \begin{center}
    \small
    \begin{tabular}{lcllll}
      \toprule
      &\hspace{-0mm} $N_f$&\hspace{-0mm} $\amuHVPLOud$&\hspace{-0mm}
      $\amuHVPLOs$&\hspace{-0mm} $\amuHVPLOc$&\hspace{-0mm}
      $\amuHVPLOdisc$\\
      \midrule
      FNAL/HPQCD/MILC-24~\cite{FermilabLatticeHPQCD:2024ppc}&\hspace{-0mm} 2+1+1 &\hspace{-0mm}
      $655.2(2.3)(3.9)$ &\hspace{-0mm}\emptycell &\hspace{-0mm}\emptycell
      &\hspace{-0mm}\emptycell\\
      ETM-24~\cite{ExtendedTwistedMass:2024nyi}&\hspace{-0mm}
      2+1+1&\hspace{-0mm}\emptycell &\hspace{-0mm} 53.57(41)(48)&\hspace{-0mm}
      14.56(10)(9)&\hspace{-0mm}\emptycell\\
      Mainz/CLS-24~\cite{Djukanovic:2024cmq}&\hspace{-0mm} 2+1&\hspace{-0mm}
      675.7(4.1)(3.7) &\hspace{-0mm} 54.5(3)(3)&\hspace{-0mm} 14.4(2)(2)&\hspace{-0mm}
      $-16.1(1.2)(1.2)$\\
      RBC/UKQCD-24~\cite{RBC:2024fic}&\hspace{-0mm} 2+1&\hspace{-0mm} 668.7(6.1)(2.9) 
      &\hspace{-0mm}\emptycell&\hspace{-0mm}\emptycell&\hspace{-0mm}\emptycell\\
      ABGP-22~\cite{Aubin:2022hgm}&\hspace{-0mm} 2+1+1&\hspace{-0mm} 646 (11.4)(7.8) &
      \emptycell&\hspace{-0mm}\emptycell&\hspace{-0mm}\emptycell\\
      HPQCD-20~\cite{Hatton:2020qhk}&\hspace{-0mm} 2+1+1&\hspace{-0mm}\emptycell
      &\hspace{-0mm}\emptycell &\hspace{-0mm} $14.606(47)$ &\hspace{-0mm}\emptycell\\
      LM-20~\cite{Lehner:2020crt}&\hspace{-0mm} 2+1+1&\hspace{-0mm} 657(26)(12)
      &\hspace{-0mm} 52.83(22)(65)&\hspace{-0mm}\emptycell&\hspace{-0mm}\emptycell\\
      BMW-20~\cite{Borsanyi:2020mff}&\hspace{-0mm} 2+1+1&\hspace{-0mm} 652.4(2.0)(5.3)
      &\hspace{-0mm} 53.393(89)(68)&\hspace{-0mm} 14.6(0)(1)&\hspace{-0mm}
      $-15.4(1.2)(1.4)$\\
      ETM-18/19~\cite{Giusti:2018mdh,Giusti:2019hkz}&\hspace{-0mm} 2+1+1&\hspace{-0mm}
      629.1(13.7)&\hspace{-0mm} 53.1(1.6)(2.0)$^{*}$&\hspace{-0mm}
      14.75(42)(37)$^{*}$&\hspace{-0mm}\emptycell\\
      Mainz/CLS-19~\cite{Gerardin:2019rua}&\hspace{-0mm} 2+1&\hspace{-0mm}
      674(12)(5)$^{*}$ &\hspace{-0mm} 54.5(2.4)(0.6)$^{*}$&\hspace{-0mm}
      14.66(45)(6)$^{*}$&\hspace{-0mm} $-23.2(2.2)(4.5)^{*}$\\
      ABGP-19~\cite{Aubin:2019usy}&\hspace{-0mm} 2+1+1&\hspace{-0mm} 659(22)$^{*}$
      &\hspace{-0mm}\emptycell&\hspace{-0mm}\emptycell&\hspace{-0mm}\emptycell\\
      FNAL/HPQCD/MILC-19~\cite{Davies:2019acq}&\hspace{-0mm} 2+1+1&\hspace{-0mm}
      637.8(8.8)$^{*}$
      &\hspace{-0mm}\emptycell&\hspace{-0mm}\emptycell&\hspace{-0mm}\emptycell\\
      PACS-19~\cite{Shintani:2019wai}&\hspace{-0mm} 2+1&\hspace{-0mm} 673(9)(11)
      &\hspace{-0mm} 52.1(2)(5)&\hspace{-0mm} 11.7(0.2)(1.6)&\hspace{-0mm}\emptycell\\
      RBC/UKQCD-18~\cite{RBC:2018dos}&\hspace{-0mm} 2+1&\hspace{-0mm}
      649.7(14.2)(4.9)$^{*}$ &\hspace{-0mm} 53.2(4)(3)&\hspace{-0mm}
      14.3(0)(7)&\hspace{-0mm} $-11.2(3.3)(2.3)$\\
      BMW-17~\cite{Budapest-Marseille-Wuppertal:2017okr}&\hspace{-0mm} 2+1+1&\hspace{-0mm}
      647.6(7.5)(17.7)$^{*}$ &\hspace{-0mm} 53.73(4)(49)$^{*}$&\hspace{-0mm}
      14.74(4)(16)$^{*}$&\hspace{-0mm} $-12.8(1.1)(1.6)^{*}$\\
      HPQCD-14~\cite{chakraborty:2014mwa}&\hspace{-0mm} 2+1(+1)&\hspace{-0mm}
      \emptycell&\hspace{-0mm} 53.41(0)(59)&\hspace{-0mm}
      14.42(0)(39)$^{*}$&\hspace{-0mm}\emptycell\\
      \bottomrule
    \end{tabular}
    \caption{Single-flavor and disconnected contributions to $\amuHVPLO$, see also
      \cref{fig:comp_full}, in units of $10^{-10}$. All results are given in the isospin-symmetric scheme of the individual groups, information on these schemes can be found in the associated references. Results which have been superseded by new calculations by the
      same collaboration are denoted with an asterisk. }
    \label{tab:amu_fbf}
  \end{center}
  \renewcommand{\arraystretch}{1.0}
\end{table}

For the dominant light-quark connected component $\amuHVPLOud$, which accounts for
$\simeq88\%$ of the total HVP contribution, there have been six
new results since the release of WP20: BMW-20~\cite{Borsanyi:2020mff},
LM-20~\cite{Lehner:2020crt}, ABGP-22~\cite{Aubin:2022hgm},
RBC/UKQCD-24~\cite{RBC:2024fic}, Mainz/CLS-24~\cite{Djukanovic:2024cmq}, and
Fermilab/HPQCD/MILC-24~\cite{FermilabLatticeHPQCD:2024ppc}. The first three are obtained
from an analysis of the full integral, whereas the latter three come from
summing the windows. These six new results are collected in the third column of
\cref{tab:amu_fbf}, along with previous determinations that were included in
WP20. All these new results (aside from LM-20) are updates on previous
determinations, which we denote with an asterisk in the table. As discussed in
\cref{sec:LDwin}, a central component of these newer calculations (aside from
LM-20) is the use of exact low-modes of the Dirac operator, which is the
primary reason for the dramatic improvement in statistical precision over older
determinations. The four most precise determinations are all also available in the WP25 isospin-symmetric scheme, which we collect in \cref{tab:amu_fbf_wp25}.

For the strange and charm connected components, which are
$\simeq7\%$ and $\simeq2\%$ of the total HVP contribution, respectively,
there are four new determinations of $\amuHVPLOs$, from
BMW-20~\cite{Borsanyi:2020mff}, LM-20~\cite{Lehner:2020crt},
Mainz/CLS-24~\cite{Djukanovic:2024cmq}, and
ETM-24~\cite{ExtendedTwistedMass:2024nyi}, and four new
determinations of $\amuHVPLOc$, from BMW-20~\cite{Borsanyi:2020mff},
HPQCD-20~\cite{Lehner:2020crt}, Mainz/CLS-24~\cite{Djukanovic:2024cmq}, and
ETM-24~\cite{ExtendedTwistedMass:2024nyi}, since WP20. These
are given in the fourth and fifth columns of \cref{tab:amu_fbf}. Again, aside
from LM-20, these supersede previous determinations from the same groups.
Finally, for the disconnected contribution $\amuHVPLOdisc$, given in the last
column of \cref{tab:amu_fbf}, there have been two new results,
BMW-20~\cite{Borsanyi:2020mff} and Mainz/CLS-24~\cite{Djukanovic:2024cmq}, both
updates on previous determinations.

\begin{table}[!t]
\renewcommand{\arraystretch}{1.1}
  \begin{center}
    \small
    \begin{tabular}{lcllll}
      \toprule
      &\hspace{-0mm} $N_f$&\hspace{-0mm} $\amuHVPLOud$&\hspace{-0mm}
      $\amuHVPLOs $&\hspace{-0mm} $\amuHVPLOc$&\hspace{-0mm}
      $\amuHVPLOdisc$\\
      \midrule
      FNAL/HPQCD/MILC-24~\cite{FermilabLatticeHPQCD:2024ppc}&\hspace{-0mm} 2+1+1 &\hspace{-0mm}
      $656.9(2.3)(3.2)$ 
      &\hspace{-0mm}\emptycell&\hspace{-0mm}\emptycell&\hspace{-0mm}\emptycell\\
      Mainz/CLS-24~\cite{Djukanovic:2024cmq}&\hspace{-0mm} 2+1&\hspace{-0mm}
      666.2(4.9)(6.1) &\hspace{-0mm} 53.56(21)(33)&\hspace{-0mm}\emptycell&\hspace{-0mm}
      $-16.3(1.7)(1.7)$\\
      RBC/UKQCD-24~\cite{RBC:2024fic}&\hspace{-0mm} 2+1&\hspace{-0mm} 666.2(4.3)(2.5)
      &\hspace{-0mm}\emptycell&\hspace{-0mm}\emptycell&\hspace{-0mm}\emptycell\\
      BMW-20~\cite{Borsanyi:2020mff}&\hspace{-0mm} 2+1+1&\hspace{-0mm} 652.4(2.0)(5.3)
      &\hspace{-0mm} 53.393(89)(68)&\hspace{-0mm} 14.6(0)(1)&\hspace{-0mm}
      $-15.4(1.2)(1.4)$\\
      \bottomrule
    \end{tabular}
    \caption{Single-flavor and disconnected results of \cref{tab:amu_fbf} in the WP25 isospin-symmetric scheme, see \cref{eq:scheme-wp25}, in units of $10^{-10}$. The BMW-20 results are duplicated for clarity.}
    \label{tab:amu_fbf_wp25}
  \end{center}
  \renewcommand{\arraystretch}{1.0}
\end{table}

\begin{figure}[t!]
  \begin{center}

    \vspace{-0.2cm}
    \includegraphics[width=\linewidth]{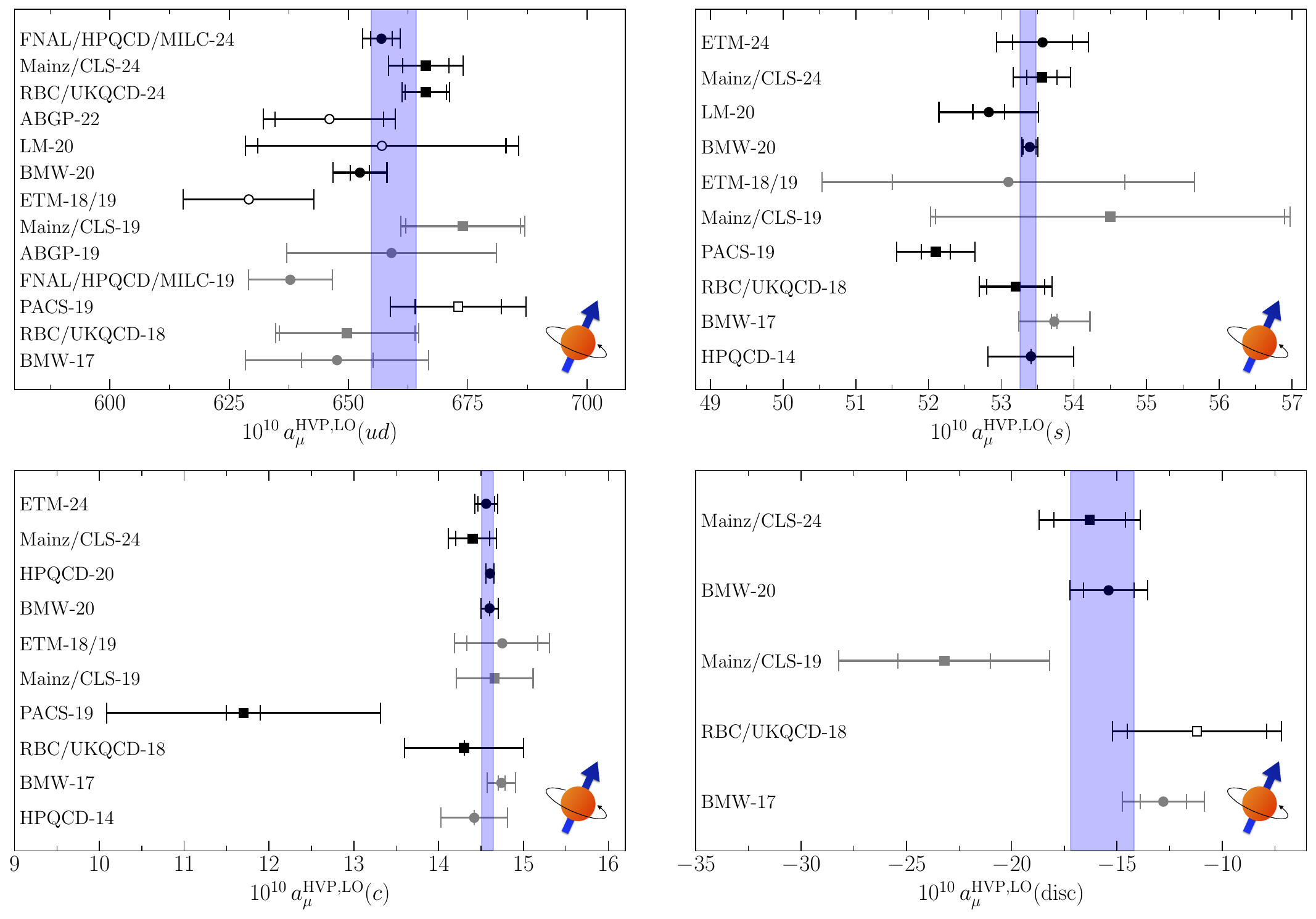}
    \caption{\small Compilation of lattice results for the flavor contributions to $\amuHVPLO$. Upper-Left: light-quark connected $\amuHVPLOud$. Upper-Right: strange-quark connected $\amuHVPLOs$. Lower-Left: charm-quark connected $\amuHVPLOdisc$. Lower-Right: quark disconnected $\amuHVPLOdisc$. Where possible, we display results in the WP25 isospin-symmetric scheme (\cref{eq:scheme-wp25}), corresponding to the results in \cref{tab:amu_fbf_wp25}.
    The light blue bands correspond to ``Avg.~A'' in the first row of \cref{tab:full_avgs}. Results not included in the average are denoted by unfilled symbols.
    Error ticks and plotting symbols and colors have the same meaning as in~\cref{fig:SD_fbf}.}
    \label{fig:comp_full}
  \end{center}
\end{figure}

The results for each contribution are compared in \cref{fig:comp_full}, where
we show superseded results with a gray marker.\footnote{The results of Refs.~\cite{Chakraborty:2016mwy,DellaMorte:2017dyu} are not included in the comparison, because they were already superseded in WP20 by newer results.} We stress here, as in
\cref{sec:LDwin}, that these determinations, in particular $\amuHVPLOud$ and
$\amuHVPLOs$, are sensitive to how one defines the isospin-symmetric pure QCD
scheme. Hence, when comparing results directly one should be aware of the
particular scheme definitions, which are described in the caption and
references of \cref{tab:amu_fbf,tab:amu_fbf_wp25}.

We perform the world averages for each of the flavor contributions using two
independent approaches. The first approach makes direct use of the
results in \cref{tab:amu_fbf,tab:amu_fbf_wp25}. For this, we follow the FLAG averaging procedure discussed at the end of \cref{sec:methods}. In particular, results that have been superseded are not included, as well as results based on less than three lattice spacings for dominant contributions and less than two for subdominant ones. Before performing the averages, the results that are not in the preferred isospin-symmetric scheme of \cref{eq:scheme-wp25} must be adjusted. To do this we follow the strategy laid out in \cref{sec:IDwin}.

\begin{table}[!t]
\renewcommand{\arraystretch}{1.1}
  \begin{center}
    \small
    \begin{tabular}{lccccc}
      \toprule
      &$\amuHVPLOud$ & $\amuHVPLOs$ & $\amuHVPLOc$ & $\amuHVPLOdisc$ & $\amuHVPLOiso$ \\
      \midrule
      Avg.~A &$659.5(4.7)$ & $53.37(11)$ & $14.576(68)$& $-15.7(1.5)$ & $712.0(4.9)$  \\
      Avg.~B &$661.1(5.0)$ & $53.18(28)$ & $14.37(11)$ &
      $-16.4(2.3)$ & $712.5(5.4)$\\
      \bottomrule
    \end{tabular}
    \caption{Results for the individual flavor contributions to $\amuHVPLO$ (in the
      isospin-symmetric scheme of \cref{eq:scheme-wp25} in units of $10^{-10}$). Also given is the total isospin-symmetric contribution to $\amuHVPLO$, which includes the $\amuHVPLOb$
      contribution from \cref{tab:amuSDflavsummary}. The first row (Avg.~A) corresponds to directly averaging over the results of \cref{tab:amu_fbf,tab:amu_fbf_wp25} using the FLAG
    procedure. The second row (Avg.~B) corresponds to combining the window averages of \cref{sec:SDwin,sec:IDwin,sec:LDwin}.}\label{tab:full_avgs}
  \end{center}
  \renewcommand{\arraystretch}{1.0}
\end{table}

For $\amuHVPLOud$, the results from Fermilab/HPQCD/MILC-24~\cite{FermilabLatticeHPQCD:2024ppc},
RBC/UKQCD-24 \cite{RBC:2024fic}, and BMW-20 \cite{Borsanyi:2020mff} are
available in the preferred scheme, which relies on $M_\Omega$ to determine
$w_0$ in fm. The result from Mainz/CLS-24 \cite{Djukanovic:2024cmq} is not;
however, the corresponding determination of $\amuLDud$ is available in the
WP25 scheme, which can be combined with the relatively scale-insensitive
short- and intermediate-distance contributions. All these WP25 determinations are given in \cref{tab:amu_fbf_wp25}. The remaining results, ABGP-22
\cite{Aubin:2022hgm}, LM-20 \cite{Lehner:2020crt}, ETM-18/19
\cite{Giusti:2018mdh,Giusti:2019hkz}, and PACS-19 \cite{Shintani:2019wai}, all
have significantly larger uncertainties and use $f_\pi$ directly to set the
scale or to determine $w_0$ in~fm. Hence, for this average, we include only the four sub-percent results in the WP25 scheme. We account for correlations due to the shared approaches to correct for FV effects in the same fashion as in \cref{sec:LDwin}. In addition, we correlate the systematic uncertainty due to taste-breaking effects from BMW-20 and Fermilab/MILC/HPQCD-24, corresponding to 3.7 and 0.9 in units of $10^{-10}$, respectively. The additional uncertainty of $3.1\times 10^{-10}$ from the spread of the $f_\pi$-based LD window results in Sec.~\ref{sec:LDwin} is also applied to $\amuHVPLOud$ here. 

For $\amuHVPLOs$, there are only two\footnote{The Mainz/CLS-24 $\amuHVPLOs$ result in the WP25 scheme is constructed in the same fashion as described for the $\amuHVPLOud$ result, with an additional 0.3\% relative uncertainty applied to the $\amuWs$ component. This additional 0.3\% uncertainty is estimated in \cref{sec:IDwin}.} results available in the WP25 scheme (see \cref{tab:amu_fbf_wp25}), hence,  following \cref{sec:IDwin}, for results not in this scheme we include an additional 0.9\% and 0.2\% systematic uncertainty (in quadrature) due to scale setting input and the spread in input values for $M_K$. These relative uncertainties are estimated using the derivatives $d \log \amuHVPLOs / d \log w_0 \simeq 1.8$ and $d \log \amuHVPLOs / d \log M^2_K \simeq
0.6$ (Eq.~(B.29) of Ref.~\cite{Ce:2022kxy}). As the charm contribution is dominated by $\amuSDc$ and $\amuWc$ we assign just the $0.2\%$ systematic uncertainty obtained in \cref{sec:IDwin}.\footnote{For $\amuHVPLOc$ results obtained using $M_{J/\psi}$ as opposed to $M_{D_S}$ as in \cref{eq:scheme-wp25}, namely HPQCD-20, we include a 0.5\% systematic uncertainty, estimated similarly as described for $\amuHVPLOs$.} Finally, for the disconnected contribution, there are only two qualifying results obtained from a lattice calculation with more than one lattice spacing, BMW-20~\cite{Borsanyi:2020mff} and Mainz/CLS-24~\cite{Djukanovic:2024cmq}. The BMW-20 result is already in the
preferred scheme, whereas the Mainz/CLS-24 result can be obtained by the same procedure as described above for $\amuHVPLOud$. This first set of flavor averages (Avg.~A), as well as the corresponding result for
$\amuHVPLOiso=\amuHVPLOud+\amuHVPLOs+\amuHVPLOc+\amuHVPLOb+\amuHVPLOdisc$, is collected in the first row of \cref{tab:full_avgs}. We note here that the $\amuHVPLOs$ average is largely dominated by the BMW-20 result, with a weight of 0.84. This is due to it being one of only two results already in the WP25 scheme, and already having smaller errors than all other determinations.

The second approach for obtaining world averages is to make use of the windowed averages obtained in \cref{sec:SDwin,sec:IDwin,sec:LDwin}. For the short- and intermediate-distance windows, the separate flavor averages are all performed in the WP25 scheme. For the LD window, the
light-quark contribution $\amuLDud$ of \cref{e:hvp_lat_amuLD_I1_av} is also
obtained in the WP25 scheme. Due to the sensitivity of $\amuLDs$ to the choice of scale-setting and strange-quark mass input, we take the sole determination in the WP25 scheme from Mainz/CLS-24~\cite{Djukanovic:2024cmq} as our window-``average'' for this quantity.\footnote{The other determination of this quantity from ETM-24~\cite{ExtendedTwistedMass:2024nyi} uses $f_\pi$ to set the scale as opposed to $w_0$ via $M_\Omega$ as in the WP25 scheme \cref{eq:scheme-wp25}. The shift in $\amuLDs$ from the two scale-setting choices can be estimated using the two results for $\amuLDs$ in Ref.~\cite{Djukanovic:2024cmq}(Table 3 and Eq.~(A.9)), giving $\Delta \amuLDs \simeq 0.92\times 10^{-10}$. Including this shift as an additional source of uncertainty on the result of Ref.~\cite{ExtendedTwistedMass:2024nyi} dramatically inflates the overall error and renders a FLAG average of the two results pointless.} 
The LD charm contribution, $\amuLDc$, is almost negligible since it contributes less than $0.1\%$ to the charm total.
In addition, the sensitivity of $\amuLDc$ to the choice of scale setting is suppressed by the large value of the ratio $m_c/m_\mu$.
To obtain $\amuLDc$, we perform a FLAG average of the two
determinations of this quantity from Mainz/CLS-24~\cite{Djukanovic:2024cmq} and
ETM-24~\cite{ExtendedTwistedMass:2024nyi}. For the LD disconnected contribution $\amuLDdisc$, we take the only available determination of this quantity from Mainz/CLS-24~\cite{Djukanovic:2024cmq}.

With all these results in hand, the sum in \cref{amu:sum} can be performed for each flavor. Given the conservative choices made to account for correlations between groups in \cref{sec:SDwin,sec:IDwin}, we perform an uncorrelated sum of the windows to obtain the window-summed averaged values for $\amuHVPLOs$ and
$\amuHVPLOc$. For $\amuHVPLOud$ and $\amuHVPLOdisc$, a systematic uncertainty that is
common to all determinations of these quantities in the W and
LD windows concerns the EFT-based approaches to correcting for FV effects. To account for this, we assume 100\% correlation of FV uncertainty between the intermediate and LD windows. To obtain a correlation coefficient that corresponds to this assumption, we examine results from the three groups that have both $\amuWud$ and $\amuLDud$ determinations~\cite{Ce:2022kxy,RBC:2023pvn,RBC:2024fic,Djukanovic:2024cmq,MILC:2024ryz,FermilabLatticeHPQCD:2024ppc}. In particular, we compute the following correlation coefficient $\rho =\sigma_{\mathrm W}^\text{FV} \sigma_{\mathrm{LD}}^\text{FV} / (\sigma_{\mathrm W} \sigma_{\mathrm{LD}})$, where $\sigma^\text{FV}$ is the uncertainty due to FV, for each case. A weighted average of $\rho=0.1$ is obtained using the weights from the FLAG average in \cref{e:hvp_lat_amuLD_I1_av}. This value is used for $\amuHVPLOdisc$ as well. The final window-summed averaged results (Avg.~B) are given in the
second row of \cref{tab:full_avgs}. We note here that the Avg.~B determination of $\amuHVPLOiso$ neglects the nontrivial anti-correlations between the Mainz/CLS-24 results for $\amuLDud$ and $\amuLDdisc$, which arise when obtaining these flavor contributions from linear combinations of $\amuLD(\mathrm{I0})$ and $\amuLD(\mathrm{I1})$ as in Ref.~\cite{Djukanovic:2024cmq}. Therefore, the uncertainty on this determination of $\amuHVPLOiso$ is less well estimated than that of \cref{eq:amuIso_avg2} discussed in \cref{sec:totalHVP}.

We observe that all values from this second average are broadly consistent with the averages from the first approach. The central value of $\amuHVPLOud$ from Avg.~A is slightly lower than Avg.~B and its uncertainty is smaller. This is due to the LD contribution from BMW-20 playing a part in the first average but not in the second. For the same reason, the uncertainty in $\amuHVPLOdisc$ is significantly smaller in the first average. 

Comparing the two averages further, we note that every flavor-specific average that enters Avg.~A is based on three or more inputs, while Avg.~B takes as inputs for $\amuLDs$ and $\amuLDdisc$ a result from a single lattice group. However, Avg.~A excludes some of the most recent results for the $\amuSD$ and $\amuW$ contributions, particularly for the $(ud)$ and $(c)$ flavor contributions. We can consider a variation to Averages A and B, where (almost) all recent results are included and where all contributions are based on two or more inputs, by combining $\amuHVPLOs$ and $\amuHVPLOdisc$ from Avg.~A with $\amuHVPLOud$ and $\amuHVPLOc$ from Avg.~B. This yields the value $\amuHVPLOiso = 713.5(5.2)$, which is virtually identical to Avg.~B and to the average in \cref{eq:amuIso_avg2} in \cref{sec:lattice_HVP_world_average}. Other variations yield similar results.

\subsection{Total HVP}
\label{sec:totalHVP}

\newcommand{\avgone}{4}
\newcommand{\avgtwo}{1}
\newcommand{\avgthree}{3}
\newcommand{\avgfoura}{2A}
\newcommand{\avgfourb}{2B}
In this section, we present our averages for the total HVP
contribution to $a_\mu$ computed in lattice QCD. When combined with QED,
EW, and the remaining hadronic contributions (also computed using lattice
methods), these averages yield purely theoretical SM predictions
for $a_\mu$. We emphasize that most of the values from individual groups that comprise our averages have
used blinding procedures, and we use the FLAG averaging procedure throughout.

Since the pioneering works of Blum~\cite{Blum:2002ii} and Aubin and
Blum~\cite{Aubin:2006xv}, and the seminal paper of ETM in 2011~\cite{Feng:2011zk} in
two-flavor QCD, many lattice calculations of $\amuHVPLO$ have been
performed~\cite{Burger:2013jya,Chakraborty:2016mwy,Budapest-Marseille-Wuppertal:2017okr,DellaMorte:2017dyu,
RBC:2018dos,Giusti:2018mdh,Giusti:2019hkz,Gerardin:2019rua,FermilabLattice:2019ugu,
Shintani:2019wai,Borsanyi:2020mff,Boccaletti:2024guq,Djukanovic:2024cmq,RBC:2023pvn,
RBC:2024fic,Aubin:2022hgm}. In addition to these, there are now many works giving the individual windows and other contributions as discussed in the previous two sections~\cite{Lehner:2020crt,Wang:2022lkq,Ce:2022kxy,ExtendedTwistedMass:2022jpw,Kuberski:2024bcj,Spiegel:2024dec,ExtendedTwistedMass:2024nyi,MILC:2024ryz,FermilabLatticeHPQCD:2024ppc}. Hence, to obtain a final lattice prediction for $\amuHVPLO$, there are a number of ways to combine
the lattice results from the various groups. These differ in the contributions to $\amuHVPLO$ that are averaged before being combined. 
Since groups do not necessarily compute all contributions, it matters in which order they are averaged and added up.
Here we consider four different combinations:
\begin{enumerate}
  \item \emph{Total $\amuHVPLO$ from a combination of the averages of all
      available lattice
      results for the SD, W, and LD windows in the isospin-symmetric limit
      plus the average
    of the total IB corrections:} 
   The advantage of this approach is that there are many results for some flavor components of the isospin-symmetric
    windows (see \cref{tab:W_comp}). This is particularly true for the $ud$ contribution to
    the intermediate-distance window for which there are eleven results and,
    more generally, for all flavor
    contributions to that window. While this large number of computations allows a
    particularly robust estimate of the corresponding systematic errors, this
    intermediate-distance window
    contributes only about 35\% to the total. There is also a fair number of computations
    of the SD window (see \cref{tab:SD_fbf}). For the LD window, which
    contributes 55\%, the
    situation is less favorable. There are only three results~\cite{Djukanovic:2024cmq,RBC:2024fic,FermilabLatticeHPQCD:2024ppc}, of which only the one
    from the Mainz group~\cite{Djukanovic:2024cmq} includes all flavor and
    IB
    contributions, as shown in \cref{tab:LD_comp}. More generally, the situation for the
    IB corrections to all windows is quite spotty. In addition,
    this averaging
    approach requires a common definition of the isoQCD limit
    for each of the
    windows and flavor contributions. Indeed, for the most important $ud$ LD window, Mainz~\cite{Djukanovic:2024cmq}, RBC/UKQCD-\cite{RBC:2024fic}, and
    Fermilab/HPQCD/MILC~\cite{FermilabLatticeHPQCD:2024ppc} have gone the extra mile to convert their
    results to the common WP25 scheme. We note that the shift required for results that are not available in the WP25 scheme includes an additional uncertainty, as discussed in \cref{sec:windows,sec:flavor_decomposition}.   

\item \emph{ Total $\amuHVPLO$ as a combination of lattice results for
      isospin-symmetric $\amuHVPLOiso$, obtained via flavor decomposition, with an average of the total IB
    correction:}
    Here $\amuHVPLOiso$ is obtained by summing the results of determinations of
    individual flavor contributions to that quantity.
    \Cref{sec:flavor_decomposition} considers two ways of obtaining those
    flavor contributions.
    In the first (approach 2A), the flavor components of $\amuHVPLOiso$ are obtained from averages of the corresponding results by the different collaborations. 
    That procedure may appear to be a rearrangement of approach 3 described below.
    However, for approach 3 there are only three computations of the total isospin-symmetric
    contribution to $\amuHVPLO$, while there are many more of the various flavor
    components of $\amuHVPLOiso$ (see \cref{tab:amu_fbf}).
    The second way (approach 2B) of obtaining the flavor components of $\amuHVPLOiso$ is by summing
    the averages of the individual flavor contributions to the SD, W, and LD windows.
    As explained in \cref{sec:flavor_decomposition}, this procedure, which uses the same inputs as approach 1, described above, and hence is very closely related to it, yields a slightly less-well quantified uncertainty. 
    
    From the two determinations of $\amuHVPLOiso$ obtained in \cref{sec:flavor_decomposition}, two values of $\amuHVPLO$
    are obtained by adding the average SIB and QED contributions to that quantity obtained in \cref{sec:lattice_HVP_world_average}.

  \item \emph{Total $\amuHVPLO$ as a combination of an average of lattice results for
      isospin-symmetric $\amuHVPLOiso$ and for the corresponding IB
    corrections:}
    This is similar to the first combination described in the preceding paragraph except that the isospin-symmetric average is
    taken over the totals from each group without averaging flavor-specific contributions as an intermediate step.
    Thus, one of the drawbacks compared to approach \avgfoura{} is that there are fewer computations of the total isospin-symmetric contribution to $\amuHVPLO$ than there are of some
    individual flavor-specific contributions. However, there are no more
    computations of the $ud$
    contribution to the isospin-symmetric LD window~\cite{Djukanovic:2024cmq,RBC:2024fic,FermilabLatticeHPQCD:2024ppc}, which overwhelms the
    uncertainty on $\amuHVPLO$, than of the total isospin-symmetric contribution~\cite{Djukanovic:2024cmq,RBC:2018dos,RBC:2024fic,Borsanyi:2020mff}. Indeed,
    here the full isospin-symmetric result from BMW-20~\cite{Borsanyi:2020mff}
    re-enters the
    average. Thus, the uncertainties in this averaging approach will not necessarily be
    larger than in approach 1. It is worth noting that the groups who have computed the
    total isospin-symmetric value are the same as those who have computed the complete $\amuHVPLO$,
    so that the overlap in input between this approach and approach 4 (below) is significant.

\item\emph{Averaging lattice results for total $\amuHVPLO$:} The advantage of this
    approach is that all ambiguity surrounding different conventions used in
    the separation
    into strictly QCD results and QED and strong IB corrections
    is absent. Each
    collaboration can choose its preferred decomposition scheme or way of fixing
    the scale and
    tuning quark masses. The result obtained corresponds to the same quantity:
    $\amuHVPLO$ is
    a well-defined physical observable. This makes averaging results straightforward. The
    downside is that very few collaborations have computed all necessary
    contributions. The
    BMW~\cite{Borsanyi:2020mff} and Mainz~\cite{Djukanovic:2024cmq}
    collaborations have done
    so. It is possible to obtain such a result from the RBC/UKQCD collaboration publications~\cite{RBC:2018dos,RBC:2023pvn,RBC:2024fic}.  We note the
    latter two employed
    blinded analyses, at least for part of their calculation.

\end{enumerate}

In addition to the advantages and disadvantages of the averaging methods discussed above,
other features can play an important role in the final uncertainties. In particular,
because we perform the canonical PDG rescaling of uncertainties by
$\sqrt{\chi^2/\text{dof}}$ in averages of results with slight tensions, for which
$\chi^2/\text{dof}>1$, the order in which different contributions are averaged and
combined may significantly impact the final uncertainty. Such effects will be found below.

\subsubsection{Result for \texorpdfstring{$\amuHVPLO$}{amuHVPLO} from the sum of averaged
windows}
\label{sec:lattice_HVP_world_average}

Here we simply rely on~\cref{sec:windows}, which provides the following averages
for the SD, W, and LD windows in the isospin-symmetric limit: $\amuSDiso = 69.06(22)\times
10^{-10}$, $\amuWiso = 236.16(42) \times 10^{-10}$, and $\amuLDiso =
407.9(5.0)$. These results are to be understood as
given in the WP25 scheme for the isospin limit. Adding the contributions together
without any correlations, we obtain the following result for $\amuHVPLO$ in the WP25
isospin limit:
\begin{equation}
  \label{eq:amuIso_avg2}
  \amuHVPLOiso\Big|^\text{lat}_\text{Avg.~\avgtwo} = 713.1(5.0)\times 10^{-10}
  \,.
\end{equation}

To that result, we have to add QED and SIB effects. We average the results of
BMW-20 \cite{Borsanyi:2020mff}, RBC/UKQCD-18 \cite{RBC:2018dos}, ETM-19~\cite{Giusti:2019xct}, and
Mainz/CLS-24 \cite{Djukanovic:2024cmq}. We must first account for the following modifications
to those effects and also consider that the IB corrections computed by Mainz/CLS-24, ETM-19, and RBC/UKQCD-18 are not given in the WP25 scheme.

To begin with, it should be noted that many of the correlation functions that enter the computation of those effects are very noisy at distances above $t\simeq 2\,\mathrm{fm}$, with details depending on the specific effect considered or on the approach used. 
Thus, all collaborations make assumptions about the behavior of those correlators at long distances. 
At such
separations, the QED corrections are dominated by the $\pi^\pm$--$\pi^0$ mass difference.
There are also small contributions from $\pi\pi\gamma$ and $\pi^0\gamma$ states as well as
from $\rho$--$\omega$ mixing. 
For Euclidean times $t\ge 2.8\fm$, those effects integrate to
$-2.0(1)\times 10^{-10}$ according to the phenomenological evaluation of
Ref.~\cite{Hoferichter:2023sli}.\footnote{The breakdown of the different sources of IB is provided in \cref{tab:IB_channels}, in particular, the dependence of the dominant LD effect due to the pion mass difference ($\Delta_\pi$) on the cut in Euclidean time $t_2$ is shown in \cref{fig:pion_mass_difference}. The full estimate for $t_2=2.8\fm$ roughly decomposes as $-2.0\simeq -2.5_{\Delta_\pi}+0.1_{\pi^0\gamma}+0.3_{\text{FSR } (2\pi)}+0.1_{\rho\text{--}\omega}$.} 
That time range approximately corresponds to the region
above which the QED and SIB correlators are set to zero in Ref.~\cite{Borsanyi:2020mff}. 
More generally, it is important to account for the uncertainties associated with the assumptions made by the various collaborations.
To do so we will add the absolute value of this phenomenological estimate to the uncertainty on the average of the determinations, without shifting the central value. 
Since that correction is only an estimate, we choose to add it linearly.
It should be noted that this effect is not relevant for the hybrid BMW/DMZ-24 
result~\cite{Boccaletti:2024guq} discussed in \cref{sec:hybrid}, because in that calculation
$e^+e^-$ and $\tau$ spectral functions are used to compute the contribution for $\amuHVPLO$ for $t\ge 2.8\fm$.

\begin{table}[!t]
  \centering
  \renewcommand{\arraystretch}{1.1}
  \small
  \begin{tabular}{cccc}
    \toprule
    Collaboration & $\deltaHVPLO$ &
    $\deltaHVPLO(\text{WP25})\vert_{\text{corr}}$ & References\\
    \midrule
    Mainz/CLS-24 & $-4.1(4.4)$ & $5.2(4.4)$ & \cite{Djukanovic:2024cmq} \\
    BMW-20 & $-0.2(1.3)$ & $-0.2(1.4)$ & \cite{Borsanyi:2020mff} \\
    ETM-19 &  $7.1(2.6)$ &   $-3.7(4.3)$ & \cite{Giusti:2019xct}\\
    RBC/UKQCD-18 & $9.5(10.4)$ & $6.9(10.4)$ & \cite{RBC:2018dos}\\
    \bottomrule
  \end{tabular}
  \caption{IB contributions to $\amuHVPLO$ in units of
    $10^{-10}$. The values
    in the second column are as published (in the collaboration's scheme), while
    the third is
    given in the WP25 scheme. They also include corrections for possibly omitted
  contributions but not the estimate, $2\times 10^{-10}$, of the common uncertainty associated with the assumptions made about the LD behavior of the IB-correction correlators. As described in the main text, the latter is added linearly to the average IB contribution.}
  \label{tab:IBcorrAvg}
  \renewcommand{\arraystretch}{1.0}
\end{table}

In Ref.~\cite{Boccaletti:2024guq} a small mistake in the sea--sea QED contribution to $\amuHVPLOud$ of BMW-20 was found, as described in Sec.~7.3 of that reference. 
The corresponding correction and its uncertainty are tiny compared to the total $5.5\times 10^{-10}$
uncertainty quoted for $\amuHVPLO$ in Ref.~\cite{Borsanyi:2020mff}. 
Once propagated in a correlated manner to the total QED contribution, its net effect is to reduce the central value of the BMW-20 result for that contribution by $0.64\times 10^{-10}$.
It would also reduce its squared uncertainty by $0.45\times 10^{-21}$. 
However, because BMW/DMZ-24 is not yet published, we add to the uncertainty the absolute value of the central-value shift in quadrature.
Taking all of those changes into account, the sum of the total QED and SIB corrections for BMW-20 that we use here
is $-0.2(1.5)\times 10^{-10}$.

Turning to RBC/UKQCD-18, the total QED and SIB correction given in
Ref.~\cite{RBC:2018dos} is $9.5(10.4)\times 10^{-10}$ in their isospin-decomposition
scheme. Many QED and SIB corrections are computed, except for the quark-disconnected SIB one and for a number of quark-disconnected and all sea QED
ones. BMW~\cite{Borsanyi:2020mff} calculated these to be $-4.67(54)(69)\times 10^{-10}$ for the former and $-0.40(80)\times
10^{-10}$ for the latter (including the small correction and the error increase mentioned
above).
The quark-disconnected SIB result is also consistent with an estimate from
partially quenched ChPT of $-6.9(3.5)\times 10^{-10}$
\cite{Lehner:2020crt}.
We note that the suppression of the disconnected SIB is not as pronounced as in the isospin-symmetric case. This can be understood from partially quenched ChPT~\cite{Lehner:2020crt}, where the cancellation of two-pion connected and disconnected SIB contributions is manifest at LO. The partially quenched ChPT argument, together with the significant cancellation of quark-connected and quark-disconnected SIB contributions calculated on the lattice~\cite{Borsanyi:2020mff,Parrino:2025afq}, stresses the importance of quark-disconnected SIB for future computations of $\amuHVPLO$.
In Ref.~\cite{RBC:2018dos}, an error of $1.1\times 10^{-10}$ is ascribed for
the neglect of the SIB correction and of
$0.3\times 10^{-10}$ for the QED ones. While the combination of these errors may appear quite generous, it does not cover the corresponding large shift in central value. Thus, we add to RBC/UKQCD's result the
shift $-5.1(1.2)\times 10^{-10}$ as well as the difference
between the central values of the $ud$ contribution in the WP25 and
RBC/UKQCD-18 schemes, i.e., $2.5\times 10^{-10}$~\cite{RBC:2024fic}. Finally, we arrive at $\deltaHVPLO(\text{RBC/UKQCD-18})=6.9(10.4)\times 10^{-10}$.

We include in the average the ETM-19 calculation of IB
corrections~\cite{Giusti:2019xct}, where quark connected SIB and QED
contributions are computed for light, strange, and charm quarks in a quenched QED
setup. ETM-19 includes an extra uncertainty of $1.2\times 10^{-10}$ for the
missing disconnected SIB and QED contributions, as well as the sea QED contributions. 
According to the BMW-20-based determination discussed above, these neglected corrections amount to $-5.6(1.2)\times 10^{-10}$ and are not covered by ETM-19's extra uncertainty. Thus, after removing that uncertainty, 
we add to ETM-19's result, $7.1(2.6)\times
10^{-10}$, the determination of the neglected contribution and combine uncertainties in quadrature.
While Refs.~\cite{DiCarlo:2019thl,Boccaletti:2024guq} provide sufficient
evidence that the differences between the Gasser--Rusetsky--Scimemi (GRS)~\cite{Gasser:2003hk} and WP25 schemes are negligible
as far as the hadronic inputs to extract quark masses are concerned, ETM-19's
GRS scheme is implemented employing the pion decay constant to set the scale.
To account for the scheme mismatch, we use the Fermilab/HPQCD/MILC-24~\cite{FermilabLatticeHPQCD:2024ppc} computation of the correlated difference of the
isospin-symmetric total HVP~\cite{MILC:2024ryz,FermilabLatticeHPQCD:2024ppc}
between a scheme based on $f_\pi$, close to the ETM one, and the one used for
our averages. 
This difference amounts to $-5.2(3.3)\times 10^{-10}$, which we also add to the
ETM-19 estimate, combining uncertainties in quadrature. All the above
corrections yield $\smash{\deltaHVPLO(\text{ETM-19})}=-3.7(4.4)\times 10^{-10}$.

For Mainz/CLS-24, we get the total QED plus SIB correction in their scheme directly
from Ref.~\cite{Djukanovic:2024cmq}.
It is $-4.1(4.4)\times 10^{-10}$.
Most QED and SIB corrections were computed, except for
a number of quark-disconnected and sea QED ones, which were either taken into
account using scalar QED~\cite{Djukanovic:2024cmq,Parrino:2025afq} or included
via an additional systematic error estimate. It is worth noting that Mainz does
not account for the renormalization of the
lattice spacing due to QED effects. This effect is believed to be negligible compared to
their quoted total error.
We must now convert this result to the WP25 scheme. 
We do the latter by
adding the difference between the central values of the LD window in the WP25
and Mainz/CLS-24 schemes, i.e., $9.3\times 10^{-10}$~\cite{Djukanovic:2024cmq}. 
Thus, in the WP25 scheme we obtain
$\smash{\deltaHVPLO(\text{Mainz/CLS-24})}=5.2(4.4)\times 10^{-10}$.

\begin{figure}[!t]
  \centering
  \includegraphics[scale=0.6]{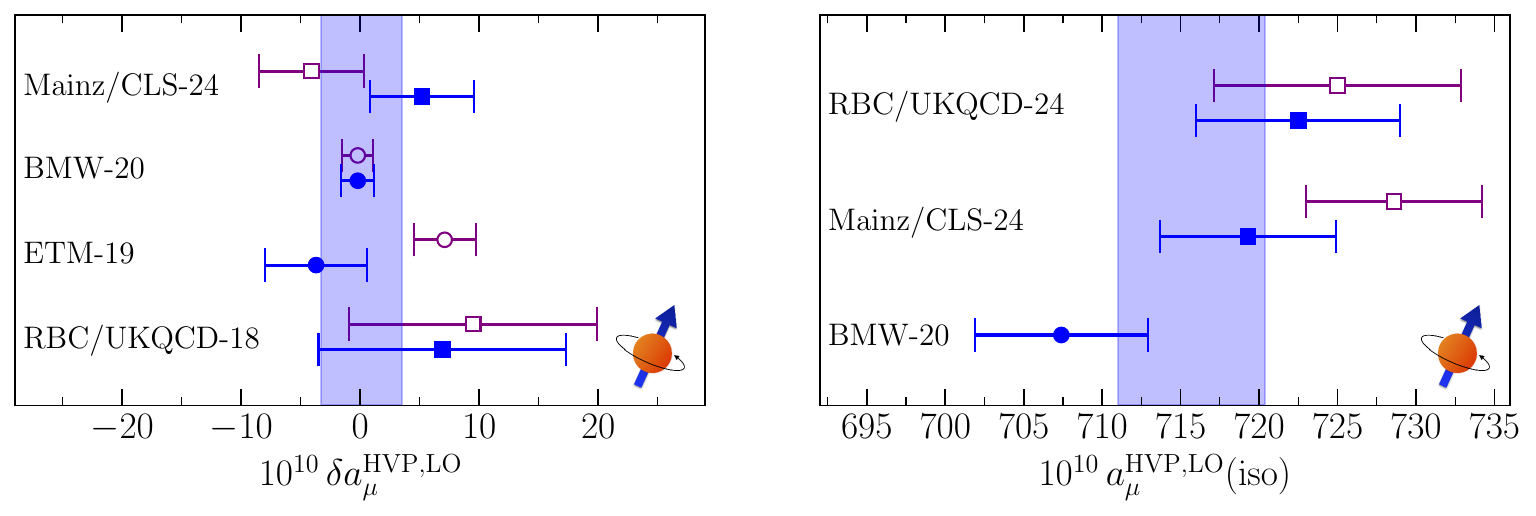}
  \caption{\small Total IB (left) and isospin-symmetric (right) contributions to $\amuHVPLO$. On the left both corrected (blue, filled symbols) and uncorrected (purple, unfilled symbols) results, as described in the text, are shown. The blue band corresponds to the average of the corrected values in the WP25 scheme. On the right results are shown in original (purple, unfilled symbols) and WP25 schemes (blue, filled symbols). The light blue band is the average in the WP25 scheme. Plotting symbols have the same meaning as in~\cref{fig:SD_fbf}.}
  \label{fig:IB and IS Avg}
\end{figure}

Before averaging we have to account for correlations between BMW-20, ETM-19, and RBC/UKQCD-18 which arise through the corrections applied to the latter two for missing QED and SIB contributions. The uncertainty
associated with the SIB contribution is $0.9\times 10^{-10}$ for BMW-20 and $1.1\times
10^{-10}$ for RBC/UKQCD-18. 
For the sea QED corrections, the value is $0.8\times 10^{-10}$
for the two results. 
For ETM-19 the uncertainty is $1.2\times
10^{-10}$ for both.
We can now average the total QED plus SIB correction from Mainz/CLS-24, BMW-20,
ETM-19, and RBC~18.
These are summarized in \cref{tab:IBcorrAvg} and shown in the left panel of \cref{fig:IB
and IS Avg}. 
We find $0.1(1.4)\times 10^{-10}$ with a $\chi^2/\text{dof}=2.6/3$.
As discussed above, to account for the uncertainties associated with assumptions made concerning the LD behavior of lattice IB-correction correlators, we add a $2.0\times 10^{-10}$ uncertainty linearly to that average.
Thus, we obtain
\begin{equation}
  \label{eq:amuhvpIB_avg2}
  \deltaHVPLO\Big|^\text{lat}_\text{Avg.} =
  0.1(3.4)\times 10^{-10}
  \,.
\end{equation}
Adding these IB corrections to the
isospin-symmetric results
of~\cref{eq:amuIso_avg2}, we obtain:
\begin{equation}
  \label{eq:amuhvp_avg2}
  \amuHVPLO\Big|^\text{lat}_\text{Avg.~\avgtwo} =
  713.2(6.1)\times 10^{-10}\,.
\end{equation}

In addition to the IB estimates averaged here,
we refer to an independent estimate of the SIB for the SD and W contributions
recently provided by Fermilab/HPQCD/MILC-24~\cite{MILC:2024ryz} and
their ongoing calculation of the QED effects~\cite{Ray:2022ycg}. Alternative
formulations beyond
QED$_L$~\cite{Hayakawa:2008an} and
QED$_\infty$~\cite{Green:2015mva,RBC:2018dos,Feng:2018qpx} implemented in the estimates
discussed in detail here, include a novel IR-improved QED action
QED$_{r}$~\cite{Davoudi:2018qpl,DiCarlo:2025uyj}, simulating at nonzero
photon mass $m_\gamma$ and taking the limit
$m_\gamma\rightarrow 0$~\cite{Endres:2015gda, Bussone:2017xkb,Clark:2022wjy}, and a
local, gauge invariant formulation with $C^{\star}$ boundary
conditions~\cite{Lucini:2015hfa}, which has recently been used to compute HVP in an
unphysical setup~\cite{Altherr:2025kqw}.

\subsubsection{Result for \texorpdfstring{$\amuHVPLO$}{amuHVPLO} from the sum of the
  individual flavor components of \texorpdfstring{$\amuHVPLOiso$}{amuHVPLO(iso)}
and the average isospin-breaking corrections}
\label{sec:FbyF_combinations}

Here we rely on the results of \cref{sec:flavor_decomposition} for the
determination of $\amuHVPLOiso$, in the WP25 scheme, from averages of its
individual flavor components.
These are obtained in two ways.
In the first, the flavor components of $\amuHVPLOiso$ are obtained from averages of the corresponding results by the different collaborations. 
The results that enter these averages are collected in \cref{tab:amu_fbf} and plotted in \cref{fig:comp_full}.
Their averages are summarized in the first row of \cref{tab:full_avgs}.
In that same row is given the result $\amuHVPLOiso=712.0(4.9)\times
10^{-10}$, obtained by combining those averages.

\begin{table}[!t]
\renewcommand{\arraystretch}{1.1}
  \begin{center}
    \small
    \begin{tabular}{lccccccc}
      \toprule
Contribution &$(ud)$ & $(s)$ & $(c)$ & $(\mathrm{disc})$ & $(\mathrm{iso})$ & Total \\
      \midrule
      $\amuSD$ {\scriptsize [BMW-20$\to$BMW/DMZ-24]} &$48.063(73)$ & $9.094(14)$ & $11.57(10)$  & $-0.0020(21)$ & $ 69.02(20)$ & $69.06(24)$\\
      $\amuW$ {\scriptsize [BMW-20$\to$BMW/DMZ-24]} &$206.89(37)$ & $27.261(68)$ & $2.810(56)$  & $-0.995(59)$  & $235.97(38)$  & $ 236.39(39)$\\
      $\amuHVPLO\big|^\text{lat}_\text{Avg.~\avgtwo}$ {\scriptsize [BMW-20$\to$BMW/DMZ-24]} &\hspace{-0mm} \emptycell & \hspace{-0mm} \emptycell & \hspace{-0mm} \emptycell & \hspace{-0mm} \emptycell & $712.9(5.0)$   & $713.0(6.1)$\\
      $\amuHVPLO\big|^\text{lat}_\text{Avg.~\avgfourb}$ {\scriptsize [BMW-20$\to$BMW/DMZ-24]} &$661.0(5.0)$ & $53.16(28)$ & $14.39(11)$ & $-16.5(2.3)$ & $712.3(5.4)$   & $712.4(6.4)$\\
      \bottomrule
    \end{tabular}
\caption{\label{tab:amuSummaryWithBMWDMZ24} Values corresponding to the world averages in \cref{tab:amuSDflavsummary,eq:amuSD_iso_lat,eq:amuSD_IB_lat,tab:W_comp,eq;amu-iso-flasum-lin,eq:ave_amu_QCD_QED,tab:full_avgs,eq:amuIso_avg2,eq:amuIso_avg2,eq:amuhvp_avg2,eq:amuhvp_avg4a} with the inclusion of the as-of-yet-unpublished BMW/DMZ-24~\cite{Boccaletti:2024guq} results.}
  \end{center}
  \renewcommand{\arraystretch}{1.0}
\end{table}

The second way of obtaining the flavor components of $\amuHVPLOiso$ is by summing averages of the individual flavor contributions to the SD, W, and LD windows computed in \cref{sec:SDwin,sec:IDwin,sec:LDwin}.
This procedure may appear to be merely a rearrangement of that of approach \avgtwo{} 
above, where $\amuHVPLOiso$ is obtained by first summing the flavor contributions
window-by-window and then adding the three windows together.
They are clearly closely related.
However, there are some differences.
These arise from taking as only result for $\amuLDs$ the one from
Ref.~\cite{Djukanovic:2024cmq}, from including a very small charm contribution
to $\amuLD$ obtained by averaging the results of
Mainz/CLS-24~\cite{Djukanovic:2024cmq} and
ETM-24~\cite{ExtendedTwistedMass:2024nyi}, and from considering the
disconnected contribution to $\amuLDiso$ instead of the iso-singlet one.
Nevertheless, because those differences are small, it is not surprising that
the resulting value for $\amuHVPLOiso$ given in the third row of
\cref{tab:full_avgs}, $712.5(5.4)\times 10^{-10}$, is close to the
result in \cref{eq:amuIso_avg2}.
We note that the central values of the two averages are very close to each other. 

Now we combine each of these determinations of $\amuHVPLOiso$ from averages of
its individual flavor components with the average of the IB
corrections computed in \cref{sec:lattice_HVP_world_average} and given in \cref{eq:amuhvpIB_avg2}.
Using the first determination we obtain
\begin{equation}
  \label{eq:amuhvp_avg4a}
  \amuHVPLO\Big|^\text{lat}_\text{Avg.~\avgfoura} = 712.1(6.0)\times 10^{-10}
  \,,
\end{equation}
and for the second,
\begin{equation}
  \label{eq:amuhvp_avg4b}
   \amuHVPLO\Big|^\text{lat}_\text{Avg.~\avgfourb} = 712.6(6.4)\times 10^{-10}
  \,.
\end{equation}
These two results are virtually identical to $\amuHVPLO\big|^\text{lat}_\text{Avg.~\avgtwo}$, which is also true of the alternate average considered in \cref{sec:flavor_decomposition}. 

As noted in \cref{sec:windows,sec:flavor_decomposition}, because Ref.~\cite{Boccaletti:2024guq} is, at the time of the writing of WP25, not yet published, the BMW/DMZ-24 results obtained therein are not included in any of the averages reported there, and hence also not included in the averages of this section and \cref{sec:lattice_HVP_world_average}. For completeness, we list in \cref{tab:amuSummaryWithBMWDMZ24} values for the averages from \cref{sec:windows,sec:flavor_decomposition,sec:lattice_HVP_world_average} as well as this section that include the corresponding BMW/DMZ-24 results. 
In particular, \cref{tab:amuSummaryWithBMWDMZ24} lists BMW/DMZ-24-inclusive averages for the $\amuSD$ and $\amuW$ observables of \cref{tab:amuSDflavsummary,eq:amuSD_iso_lat,eq:amuSD_IB_lat,tab:W_comp,eq;amu-iso-flasum-lin,eq:ave_amu_QCD_QED} and the $\amuHVPLO$ results of \cref{tab:full_avgs,eq:amuIso_avg2,eq:amuhvp_avg2,eq:amuhvp_avg4a}. The corresponding BMW-20~\cite{Borsanyi:2020mff} results are then excluded as inputs as they are then superseded by BMW/DMZ-24.

\subsubsection{Result for \texorpdfstring{$\amuHVPLO$}{amuHVPLO} from the sum
  of the averages for
the isospin-symmetric total and isospin-breaking contributions}
\label{sec:avg3}

\cref{tab:totamu 2} lists the results for the total isospin-symmetric contribution to
$\amuHVPLO$ that we consider here. These are also shown in the right panel of \cref{fig:IB
and IS Avg}. 
These are from the BMW, Mainz, and RBC/UKQCD collaborations whose results enter the average of
approach \avgone{} and whose calculations are discussed in more detail in \cref{sec:totalHVPav}. 
We perform a weighted average of those values in the WP25
scheme, while accounting for the correlations among the FV
corrections in the Mainz/CLS-24 and RBC/UKQCD-24 computations. We obtain
\begin{equation}
  \label{eq:amuhvp_iso_avg3}
  \amuHVPLOiso\Big|^\text{lat}_\text{Avg.~\avgthree} = 715.7(4.7)\times 10^{-10}\,,
\end{equation}
with $\chi^2/\text{dof}=3.8/2$, so the quoted error has been enlarged by a factor of
$1.4$.
Taking the IB correction from \cref{eq:amuhvpIB_avg2} computed for Avg.~1, we arrive at
\begin{equation}
  \label{eq:amuhvp_avg3}
  \amuHVPLO\Big|^\text{lat}_\text{Avg.~\avgthree} = 715.8(5.8)\times 10^{-10}\,,
\end{equation}
which is slightly larger, but consistent, with the averages in the previous two subsections.

\begin{table}[!t]
\renewcommand{\arraystretch}{1.1}
  \centering
  \small
  \begin{tabular}{cccc}
    \toprule
    Collaboration & $\amuHVPLOiso$  & $\amuHVPLO(\text{iso, WP25})$ & References\\
    \midrule
    BMW-20 & -- & 707.4 (5.5) & \cite{Borsanyi:2020mff} \\
    Mainz/CLS-24 & 728.6 (5.6) & 719.3(5.6) &
    \cite{Djukanovic:2024cmq,Kuberski:2024bcj,Ce:2022kxy} \\
    RBC/UKQCD-24 & 725.0 (7.9) & 722.5 (6.5) &
    \cite{RBC:2018dos,RBC:2023pvn,RBC:2024fic}\\
    \bottomrule
  \end{tabular}
  \caption{Total isospin-symmetric contribution to $\amuHVPLO$  in units of $10^{-10}$ in
    the collaboration's original isospin-decomposition scheme (second column) and in the
  WP25 scheme (third column). }
  \label{tab:totamu 2}
  \renewcommand{\arraystretch}{1.0}
\end{table}

\begin{table}[!t]
\renewcommand{\arraystretch}{1.1}
  \centering
  \small
  \begin{tabular}{cccc}
    \toprule
    Collaboration & $\amuHVPLO$ & $\amuHVPLO\vert_{\text{corr}}$ & References\\
    \midrule
    BMW-20 & 707.5 (5.5) & 707.2 (5.5) & \cite{Borsanyi:2020mff} \\
    Mainz/CLS-24 & 724.5 (7.1) & $-$ &
    \cite{Djukanovic:2024cmq,Kuberski:2024bcj,Ce:2022kxy} \\
    RBC/UKQCD-24 & 734.5 (13.0) & 729.4 (13.0) &
    \cite{RBC:2018dos,RBC:2023pvn,RBC:2024fic}\\
    \bottomrule
  \end{tabular}
  \caption{ Total HVP contribution to $a_\mu$ in units of $10^{-10}$.
    The values in
    the second column are as published, while the third includes corrections for omitted
    contributions as described in the text, except for
    Ref.~\cite{Djukanovic:2024cmq} that is left unchanged. 
    Not included is the estimate, $2\times 10^{-10}$, of the common uncertainty associated with the assumptions made about the LD behavior of the IB-correction correlators. As described in \cref{sec:lattice_HVP_world_average}, the latter is added linearly to the average of corrected results.
    }
  \label{tab:totamu}
  \renewcommand{\arraystretch}{1.0}
\end{table}

\subsubsection{Averaging lattice results for total \texorpdfstring{$\amuHVPLO$}{amuHVPLO}}
\label{sec:totalHVPav}

\begin{figure}[!t]
  \centering
  \includegraphics[scale=0.6]{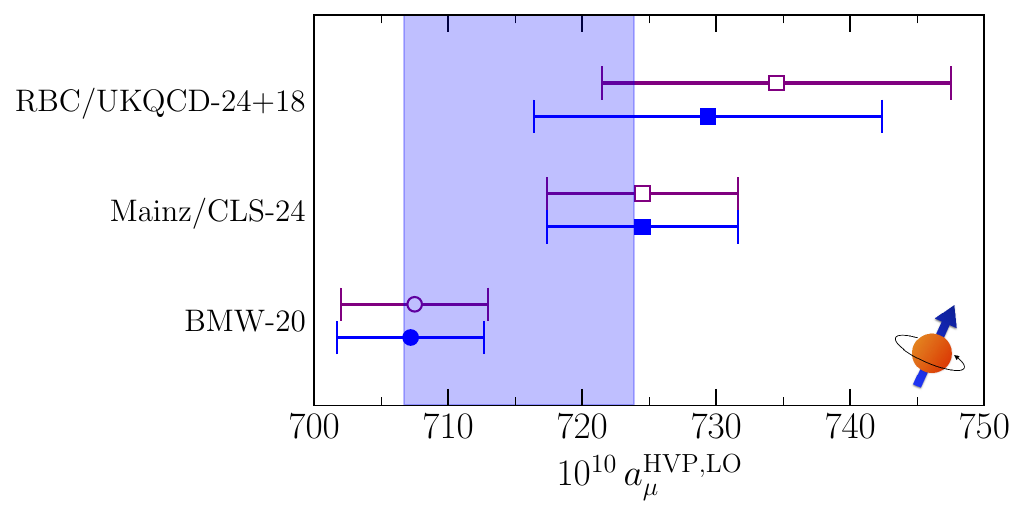}
  \caption{\small LO HVP contribution to $a_\mu$. Results shown are published (purple, unfilled symbols) and with corrections (blue, filled symbols) as explained in the text. The vertical band displays the weighted average of the corrected results as given in \cref{eq:amuhvp_avg1} (note that the Mainz/CLS-24 result has not been corrected). Plotting symbols have the same meaning as in~\cref{fig:SD_fbf}.}
  \label{fig:tot-amu}
\end{figure}

Here we focus on the computations that include all flavor contributions and at least the
dominant QED and SIB corrections (for details about the less recent ones see
\cref{sec:flavor_decomposition} and Ref.~\cite{Aoyama:2020ynm}). We also require
that the precision of the isoQCD contribution be around one percent. 
This makes them suitable for a comparison with data-driven determinations of
$\amuHVPLO$ and, as input to the SM prediction, with the measurement of
$a_\mu$~\cite{Muong-2:2006rrc,Muong-2:2021ojo,Muong-2:2023cdq}. We now discuss the three
calculations that satisfy those criteria and that we use to obtain our final lattice
average of $\amuHVPLO$. These are presented in \cref{tab:totamu} and shown in
\cref{fig:tot-amu}.

The BMW collaboration was the first to reach the precision goal in 2020, quoting 
a 0.8\% total
uncertainty~\cite{Borsanyi:2020mff}. The 31 gauge ensembles used are obtained with
$N_f=2+1+1$ flavors of stout-smeared staggered fermions, quark masses closely bracketing
their physical values, and six lattice spacings. The calculation includes all LO QED and SIB corrections. 
FV corrections are determined via dedicated simulations on $L=11\fm$ lattices. 
In addition to the modifications to the QED and SIB corrections computed in BMW-20 and discussed 
in~\cref{sec:lattice_HVP_world_average}, the subtraction of the very small one-photon-reducible contribution, $0.321(11)\times 10^{-10}$, performed in BMW-20 was not required and thus should be removed.

Recently, Mainz reported results for the LD window~\cite{Djukanovic:2024cmq}, which
they combined with earlier results for the SD~\cite{Kuberski:2024bcj} and
intermediate~\cite{Ce:2022kxy} windows to give a final prediction for $\amuHVPLO$. They quote 
a prediction for $\amuHVPLO$ with a total relative error of $1.0\%$. Their results are
obtained from 34 ensembles with $N_f=2+1$ flavors of $\mathcal{O}(a)$-improved fermions, 
four of which have near-physical light quark masses. Their simulations provide them with
six lattice spacings. QED and SIB corrections are explained in~\cref{sec:lattice_HVP_world_average}.

The third collaboration, whose work alone can provide a determination of $\amuHVPLO$, is
RBC/UKQCD. They were the first collaboration to present a determination of the connected
$ud$ contribution to the LD window in the isospin-symmetric limit~\cite{RBC:2024fic}. By
combining that result with an earlier determination of the connected $ud$ contribution to
the SD and intermediate windows~\cite{RBC:2023pvn}, they obtain a prediction for that contribution
to the full $\amuHVPLO$~\cite{RBC:2024fic}, denoted here $\amuHVPLOud$. In turn, that
prediction can be combined with RBC/UKQCD's 2018 computation of the quark-disconnected,
strange, charm, QED, and SIB contributions~\cite{RBC:2018dos} to obtain a result for
$\amuHVPLO$. In that process, the QED and SIB corrections of RBC/UKQCD-18 must be supplemented by missing contributions, as discussed in \cref{sec:lattice_HVP_world_average}.
Their prediction for $\amuHVPLOud$ is based on ten $N_f=2+1$ domain-wall
fermion ensembles at three lattice spacings, of which four are generated with masses of
light quarks around their physical values~\cite{RBC:2023pvn,RBC:2024fic}. Their results
for the strange-quark connected contribution are obtained from two ensembles with
different lattice spacings~\cite{RBC:2018dos} and, for the charm, eight ensembles with
three spacings. Both determinations involve two ensembles with physical pion masses. For
the light- and strange-quark disconnected contributions~\cite{Blum:2015you}, as well as for
QED and SIB corrections, one ensemble with the larger lattice spacing and physical masses
is used.

These three results are given in \cref{tab:totamu} and shown in \cref{fig:tot-amu}, with
and without the applied correction factors discussed in~\cref{sec:lattice_HVP_world_average}.
We are now almost ready to combine the corrected results of BMW-20, Mainz/CLS-24, and
RBC/UKQCD-24. First, we must discuss the correlations among them. Mainz/CLS-24 and
RBC/UKQCD-24 employ very similar models for FV corrections. By far the largest
contribution to those corrections concerns the $I=1$ contribution to the LD window. The way
in which the models are used differs in the two calculations, but both their physical-mass
simulations are performed on $L\simeq 6\fm$ lattices. Thus, we fully correlate the
uncertainties on the collaborations' estimates of FV corrections for the LD window, i.e.,
$1.5\times 10^{-10}$ for Mainz and $0.8\times 10^{-10}$ for RBC/UKQCD. Correlations arising from IB corrections between BMW-20 and RBC/UKQCD-24 are also included, see~\cref{sec:lattice_HVP_world_average}.

Putting all of those ingredients together, we obtain the average $715.3(6.6)\times 10^{-10}$ for $\amuHVPLO$. This average carries a
$\chi^2/\text{dof}=5.0/2$, leading to a rescaling of its error by $1.6$. While this
rescaling factor is not excessively large, it applies to the total uncertainty on $\amuHVPLO$, 
instead of being applied to only the sub-components that are in tension, 
as in the three other combination procedures described above.
Since the uncertainty on the total $\amuHVPLO$ is obviously larger than those on its sub-components, a similar 
rescaling factor has a larger impact, in absolute terms, on the final uncertainty of this particular
combination of results than on the others.
For example, the current average makes use of the same set of results as approach \avgthree, except for the addition of ETM-19~\cite{Giusti:2019dmu} in the determination of $\deltaHVPLO\big|^\text{lat}_\text{Avg.}$ in \cref{eq:amuhvpIB_avg2}. 
The lesser precision here comes from the fact that the tensions between different lattice inputs, 
which are strongest between respective isospin-symmetric results, also induce a rescaling of
the large uncertainties associated with IB corrections, while they only induce a rescaling
of the uncertainty on the isospin-symmetric averages in approach \avgthree.
This is an example that shows that the order in which averages are combined
can have a significant impact on the final uncertainty.

Now, the uncertainty on the above average does not take into account the $2.0\times 10^{-10}$ uncertainty associated with the assumptions made on the LD behavior of the IB-correction correlators discussed in \cref{sec:lattice_HVP_world_average}. 
Following that discussion, we add this uncertainty linearly to the one of the average, leading to:
\begin{equation}
  \label{eq:amuhvp_avg1}
  \amuHVPLO\big|^\text{lat}_\text{Avg.~\avgone} = 715.3(8.6)\times 10^{-10}
  \,.
\end{equation}
This result also carries a significantly larger uncertainty than those of the first two approaches, 
described in \cref{sec:lattice_HVP_world_average,sec:FbyF_combinations}, which is due to the fact they include many more lattice inputs.  

\begin{table}[!t]
\renewcommand{\arraystretch}{1.1}
\small
  \centering
  \begin{tabular}{ccc}
    \toprule
    Avg.~ID & $\amuHVPLO\big|^\text{lat}_\text{Avg.~ID}$ & \\
    \midrule
    \avgtwo & $713.2(6.1)$ & \cref{eq:amuhvp_avg2}\\
    \avgfoura &  $712.1(6.0)$ & \cref{eq:amuhvp_avg4a}\\
    \avgfourb & $712.6(6.4)$ & \cref{eq:amuhvp_avg4b}\\
    \avgthree & $715.8(5.8)$ & \cref{eq:amuhvp_avg3}\\
    \avgone & $715.3(8.6)$ & \cref{eq:amuhvp_avg1}\\
    \bottomrule
  \end{tabular}
  \caption{Summary of lattice weighted averages for $\amuHVPLO$ in units of
    $10^{-10}$ ; ``\avgtwo'' from the sum of
    averages of the isospin-symmetric SD, intermediate, and LD windows from
    \cref{sec:SDwin,sec:IDwin,sec:LDwin} plus our average of IB
    corrections;
    ``\avgfoura'' using averages of
    flavor components of $\amuHVPLOiso$ obtained in \cref{sec:flavor_decomposition} plus
    that same IB correction average; ``\avgfourb'' using a flavor decomposition of
    the windows implemented in \cref{sec:flavor_decomposition} plus the same IB correction;
    ``\avgthree'' from our average of the total $\amuHVPLOiso$ values plus the
    IB correction average; and 
    ``\avgone''
    is our average of total $\amuHVPLO$ values.}
  \label{tab:LatTotHVPavgs}
  \renewcommand{\arraystretch}{1.0}
\end{table}

\subsubsection{Summary of lattice world averages for
\texorpdfstring{$\amuHVPLO$}{amuHVPLO}}

\begin{figure}[!t]
  \centering
  \includegraphics[scale=0.65]{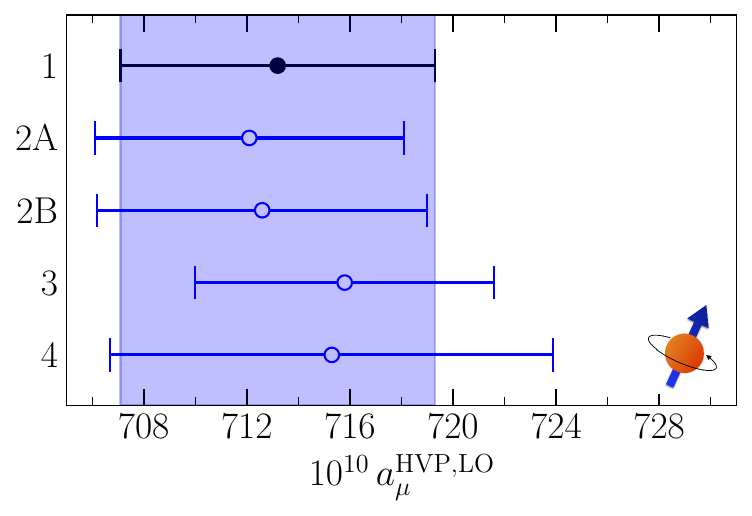}
  \caption{\small Weighted averages of the total HVP contribution to $a_\mu$ for the various averaging procedures described in the text, see~\cref{tab:LatTotHVPavgs} for more details. The final quoted value for $\amuHVPLO$ is denoted by the black, filled circle; results for the other averaging procedures are shown as blue, unfilled symbols.}
  \label{fig:amu lat Avg summary}
\end{figure}

\cref{tab:LatTotHVPavgs} collects the main results of this ``Total HVP''
section, and they are also shown in \cref{fig:amu lat Avg summary}.
While a few of the lattice inputs enter all five averages, the overlap among the inputs used in the different approaches varies somewhat. Averages~\avgtwo{} and ~\avgfourb{} use the exact same lattice inputs, while the overlap among the inputs used in Avgs.~\avgtwo{}, \avgfourb{}, and \avgfoura{} is considerable. Similarly, Avgs.~\avgthree{} and \avgone{} are based on almost the same lattice inputs. 
We note that the small tensions observed among results from different
collaborations for some components of $\amuHVPLO$ contribute differently
depending on the order in which they are combined.
Nevertheless, these different averages and combinations are remarkably
similar, well within quoted uncertainties, demonstrating the general consistency of the lattice results.

We note that Avgs.~\avgtwo{} and \avgfourb{} include all of the most recent lattice inputs and thus information about $\amuHVPLO$, where the windowed sum of Avg.~\avgtwo{} yields a slightly more reliable uncertainty estimate. Average \avgfoura{}, obtained from a flavor decomposition, includes some different inputs that yield better consolidation for some sub-components, see \cref{sec:flavor_decomposition}. We also note that \cref{sec:flavor_decomposition} discusses an alternate average that mixes the inputs in Avgs.~\avgfoura{} and \avgfourb{} to yield a virtually identical result to Avg.~\avgtwo{}. 
Incidentally, the Avg.~\avgtwo{} central value and error are close to being medians for the five results listed in \cref{tab:LatTotHVPavgs}. 
In summary, we take $\amuHVPLO\big|^\text{lat}_\text{Avg.~\avgtwo}$ as our final result for the WP25 consolidated lattice HVP average. Hence, quoting \cref{eq:amuhvp_avg2}, we obtain 
\begin{equation}
  \amuHVPLO\Big|^\text{lat}_\text{WP25} = \amuHVPLOresultsection\times 10^{-10}\,,
  \label{eq:amuLOHVWP25}
\end{equation}
which has a precision of 0.9\%.
This result is a testimony to the significant progress accomplished by lattice
computations of $\amuHVPLO$ since WP20~\cite{Aoyama:2020ynm}.

\subsection{Other observables}
\label{sec:otherobs}

\subsubsection{Isospin-breaking corrections for \texorpdfstring{$\tau$}{tau} data}
\label{sec:tau_IB_lattice}

Dispersive determinations of $\amuHVPLO$ based on $\tau$ data,
specifically on vector-mediated decays of the $\tau$ lepton into final hadronic states,
have been proposed long ago~\cite{Alemany:1997tn}. In the limit of exact isospin symmetry
the underlying charged spectral density agrees with the corresponding neutral one,
obtained from $e^+e^-$ data, but given the high-precision target of $\amuHVPLO$, accurate experimental determinations of the differential rates together
with a precise control of IB effects are necessary. These aspects are
discussed, from a phenomenological point of view, in \cref{sec:tau} with a
specific focus on the dominant $2\pi$ channel. In this problem, lattice QCD+QED
simulations can play an important role in providing solid first-principle results as
proposed in Ref.~\cite{Bruno:2018ono}, where the difference of the (fully) inclusive
spectral densities in the charged and neutral channels has been first examined, see also \cref{sect:tau-lattice}.

Contrary to the calculation of IB effects for HVP, the presence of electric charge
in initial and final states severely complicates the problem. Setting aside the
technicalities on the definition of QED on a torus, an important challenge arising here is
in the IR divergences. By studying the contributions to the decay rate, rather than
the individual Feynman diagrams, it is possible to isolate three classes of IR safe
contributions, as reported in Ref.~\cite{Bruno:2023}, amounting to factorized
initial- and final-state effects, and mixed nonfactorizable contributions (where a photon
is exchanged between the lepton and the hadronic system). For the first, thanks to the
perturbative nature of QED, pure analytic control is achieved
\cite{Cirigliano:2001er,Cirigliano:2002pv}, see also \cref{sec:tau}, and progress in defining a factorized
correction has been recently reported~\cite{Bruno:2024}. 

The focus of Ref.~\cite{Bruno:2018ono} was instead on the final-state effects, with the
intent to study them in a fully inclusive manner by re-using QED and SIB correlation
functions, already generated for the study of IB effects of the (Euclidean
lattice) HVP contribution. 
By splitting the $ud$ EM current $j_\mu$ into its two isospin components, 
$j_k^I$ (with $I = 0,1$ and $I_3 = 0$), we introduce the Euclidean-time correlators $G_{II'}(x_0)$ 
involving two isospin-projected currents~\cite{Bruno:2018ono}
\begin{equation}
  G_{II'}^\gamma(x_0) \equiv -\frac{1}{3} \sum_{k=1}^3 \int d^3x \,
  \braket{j_k^{I} (x_0, \mathbf{x} ) j_k^{I'} (0)}
  \,,
\end{equation}
such that the two-point function used for the leading lattice HVP calculation obeys the
relation
\begin{equation}
  C(x_0) \equiv G_{00}^\gamma(x_0) + 2 G_{01}^\gamma(x_0) + G_{11}^\gamma(x_0) \,.
\end{equation}
In the isospin limit, $G_{01}^\gamma(x_0)$ vanishes and $G_{11}^\gamma(x_0)$ reduces to the
so-called Wick-connected diagram. A Laplace transform relates the latter to the isovector
spectral density, which is dominated at low-energies by the two-pion channel. The same
spectral density appears in hadronic decays of the $\tau$ lepton and the corresponding
Euclidean correlator is given by
\begin{equation}
  G^W_{11}(x_0) \equiv \, - \frac13 \sum_{k=1}^3 \int d^3x \, \bra{0}
  j^{1,+}_k(x_0,\mathbf{x}) j^{1,-}_k(0) \ket{0}\,,
\end{equation}
with $j^{1,+}_k$ the (Euclidean) charged $ud$ vector current. By turning on QED and SIB
effects, both $G_{01}^\gamma$ and the difference $G_{11}^W-G_{11}^\gamma$ acquire a
nonzero value. The correlator $G_{11}^W$, being charged, is evidently gauge dependent,
and requires an appropriate scale- and scheme-dependent renormalization. It is therefore
defined only as an intermediate quantity, which may nevertheless be used to extract some
model-dependent parameters, such as the $\rho$ mass shift induced by EM interactions,
where details of the operators do not matter. While eventually a model-independent calculation of the matrix elements is required, at present the phenomenological estimate is informed by IB in the $\rho$ parameters, see \cref{Sect:th-summary}. 

At present, a possible strategy to calculate the IB corrections required
for a $\tau$-based inclusive determination of the HVP contribution to $(g-2)_\mu$ has been
reported in several talks~\cite{Bruno:2023,Bruno:2024}. Some additional developments are, however, still
required. One is the need to define the matching between a
renormalization scheme suitable for a lattice calculation of $G_{11}^W$ and the
$W$-regularization or $\overline{\mathrm{MS}}$ schemes used to determine the
SD matching with the SM. The mixed terms where
the photon is exchanged between the $\tau$ and the hadronic system constitute an
additional problem: the strategy advocated so far in Ref.~\cite{Bruno:2023} consists of
using the available knowledge on the LD effects of these terms from ChPT~\cite{Cirigliano:2001er, Cirigliano:2002pv}, which is, however, limited to the
two-pion intermediate channel. To make further progress one should consider matching 
these terms to the other phenomenological descriptions of the LD effects 
(including their scheme dependence) discussed in \cref{sec:tau}. 
The restriction to the two-pion channel generates an additional systematic error if
one is interested in a fully inclusive approach, and eventually one should consider the
complete calculation of these mixed nonfactorizable effects from lattice QCD+QED
simulations, possibly solving the corresponding inverse problem.

In conclusion, while further work is still required to properly address all systematic
errors of a $\tau$-based prediction of $\amuHVPLO$ using lattice QCD+QED inputs, it is
nevertheless worth remarking that several of these systematic effects are expected to be
mitigated in the intermediate window contribution $\amuW$, where high-multiplicity
channels are suppressed, a quantity that should likely be the target of a first lattice
QCD+QED prediction from $\tau$ data.

\subsubsection{Running of the fine-structure constant}
\label{sec:running_alpha}

The dependence of the QED coupling $\alpha$ on the interaction energy can be parameterized
in the on-shell scheme as $\alpha(q^2)=\alpha/[1-\Delta\alpha(q^2)]$, where the $q^2=0$
value $\alpha=1/137.035999\dots$ is the precisely-measured fine-structure constant. The
uncertainty in the running $\Delta\alpha$ is dominated by the hadronic contribution
$\Delta\alpha_{\text{had}}$.
For spacelike momenta, $q^2=-Q^2<0$, it is proportional to the subtracted HVP
function defined in \cref{sec:introduction}, i.e.,
\begin{equation}
  \Delta\alpha_{\text{had}}(-Q^2) = \frac{\alpha}{\pi}\,\hat{\Pi}(Q^2) ,
  \qquad\text{with}\qquad
  \hat{\Pi}(Q^2) =4\pi^2[\Pi(Q^2) - \Pi(0)]\,.
\end{equation}
Thus, $\Delta\alpha_{\text{had}}(-Q^2)$ is intimately related to the HVP contribution $\amuHVPLO$ and
can be computed from first principles on the lattice. First, we
discuss results at a spacelike $Q^2$ of a few $\GeV^2$.
These constitute yet another set of quantities that are directly accessible by lattice
computations and data-driven methods alike (see also \cref{sec:corr_obs,sec:comparison}) and can be used as an alternative to $\amuHVP$
and the window observables to compare independent determinations. In particular, if $R_\text{had}(s)$
is split in three intervals of $\sqrt{s}$ below $600\MeV$, above $900\MeV$ and between,
the relative weight of the three contributions is similar in
$\Delta\alpha_{\text{had}}(-1\GeV^2)$ and $\amuW$~\cite{Ce:2022kxy}. Following pioneering
works by ETM \cite{Burger:2015lqa} and Mainz \cite{Francis:2015grz}, the BMW collaboration provided values for
$\Delta\alpha_{\text{had}}(-Q^2)$ at $Q^2=1.0$, $2.0$, \dots, $5.0\GeV^2$ in the
supplementary material of Ref.~\cite{Budapest-Marseille-Wuppertal:2017okr}, albeit not including the
FV corrections. The first complete lattice calculation of the five-flavor
$\Delta\alpha_{\text{had}}^{(5)}(-Q^2)$ including contributions up to the bottom quark and
all known systematic effects is given in Ref.~\cite{Borsanyi:2020mff} at $Q^2=1\GeV$,
$\Delta\alpha_{\text{had}}^{(5)}(-1\GeV^2)=0.003 784(22)$, and in terms of the difference
with $Q^2=10\GeV$,
$\Delta\alpha_{\text{had}}^{(5)}(-10\GeV^2)-\Delta\alpha_{\text{had}}^{(5)}(-1\GeV^2)
=0.004867(32)$. In Ref.~\cite{Ce:2022eix} Mainz published their dedicated study of the
hadronic contribution to
the running of $\alpha$ up to a scale of $7\GeV^2$. The result at $Q^2=1\GeV^2$ is $0.003
864(32)$, larger than the corresponding BMW result. These results can be directly
compared with the data-driven estimates of
Refs.~\cite{Davier:2019can,Keshavarzi:2018mgv,Jegerlehner:2019lxt}, which are in agreement
among themselves. For instance, using the data from Ref.~\cite{Keshavarzi:2018mgv} one
obtains $0.003 682(14)$ at $Q^2=1\GeV^2$. One concludes that lattice results are in
significant tension with data-driven estimates obtained prior to 
\mbox{CMD-3's} publication of their data in the $\pi^+\pi^-$ channel, even if only the lower BMW result is considered. 
This persists at larger values of $Q^2$: Ref.~\cite{Ce:2022eix} obtains
$\Delta\alpha_{\text{had}}^{(5)}(-5\GeV^2)=0.007 16(9)$, that is $3.5\%$ larger and in a
$2.6\sigma$ tension with the data-driven result of $0.006
915(33)$~\cite{Keshavarzi:2018mgv}. In general, the discrepancy observed in
Refs.~\cite{Budapest-Marseille-Wuppertal:2017okr,Borsanyi:2020mff,Ce:2022eix} between the lattice results and
the data-driven estimates is consistent with the scenario suggested by the lattice $\amuW$
results.

In terms of phenomenological impact, the most relevant quantity is the running
contribution to the $Z$ pole, $\Delta\alpha_{\text{had}}(M_Z^2)$, which is an input to
global EW fits. Current lattice simulations cannot access spacelike $Q^2$
much larger than a few $\GeV^2$, beyond which cutoff effects are not under control. A
solution is writing the running to the $Z$ pole using the so-called Euclidean split
technique (also known as Adler function
approach)~\cite{Eidelman:1998vc,Jegerlehner:2008rs} as the sum of three terms
\begin{equation}
  \label{eq:euclidead_split_technique}
  \Delta\alpha_{\text{had}}^{(5)}(M_Z^2) =
  \Delta\alpha_{\text{had}}^{(5)}(-Q_0^2) +
  [\Delta\alpha_{\text{had}}^{(5)}(-M_Z^2) - \Delta\alpha_{\text{had}}^{(5)}(-Q_0^2)] +\
  [\Delta\alpha_{\text{had}}^{(5)}(M_Z^2) - \Delta\alpha_{\text{had}}^{(5)}(-M_Z^2)]\,,
\end{equation}
where the first term on the RHS at an intermediate spacelike scale $Q_0^2$ of the
order of a few $\GeV^2$ has been computed on the lattice. The second term on the RHS
of \cref{eq:euclidead_split_technique} can either be estimated with data-driven
methods, or, if one is after a result independent of $R$-ratio data, it can be obtained
by integrating the Adler function $D(Q^2)$ that, for sufficiently large $Q^2$, is
calculable in pQCD plus minor nonperturbative
corrections~\cite{Eidelman:1998vc,Chetyrkin:1996cf,Erler:2023hyi,Davier:2023hhn}. The third term on the RHS of
\cref{eq:euclidead_split_technique}, which provides the link between the spacelike
and timelike regions at $M_Z$, has been computed in pQCD and has a negligible
error compared to the other two terms~\cite{Jegerlehner:2019lxt}.

In Ref.~\cite{deBlas:2021wap} the Euclidean split technique is applied combining the
lattice results of Ref.~\cite{Budapest-Marseille-Wuppertal:2017okr} with the pQCD running at
$Q_0^2=4\GeV^2$, obtaining $\Delta\alpha_{\text{had}}^{(5)}(M_Z^2)=0.027 66(10)$. This fit
prior results in a small pull of $1.3\sigma$ (or $1.1\sigma$ in a conservative scenario),
leading the authors of Ref.~\cite{deBlas:2021wap} to conclude that there is no
inconsistency between this lattice evaluation and the EW fit. In Ref.~\cite{Ce:2022eix},
the result of the combination of the lattice calculation with the pQCD
expansion of the Adler function is $\Delta\alpha_{\text{had}}^{(5)}(M_Z^2)=0.027 73(15)$,
substantially constant when varying $Q_0^2$ in the $3$ to $7\GeV^2$ range. While this
result is larger than a pure data-driven determination, e.g.,
$\Delta\alpha_{\text{had}}^{(5)}(M_Z^2)=0.027 61(11)$ from Ref.~\cite{Keshavarzi:2019abf},
the clear tension in the running up to $Q_0^2$ is hidden by the relatively large error of
the higher-energy part of the running corresponding to the second term in
\cref{eq:euclidead_split_technique}. Indeed, the result of Ref.~\cite{Ce:2022eix} is
little more than $1\sigma$ larger than the indirect determinations from most EW global
fits~\cite{Haller:2018nnx,Keshavarzi:2020bfy,Malaescu:2020zuc,deBlas:2021wap},
except for the global fit in Ref.~\cite{Crivellin:2020zul} (based on Ref.~\cite{DeBlas:2019ehy}) that obtains a slightly
smaller value with a smaller error hinting at a larger tension.
When interpreting these findings, one must keep in mind that differences among global fits can originate from the input values for $m_t$ and $M_H$, as well as the treatment of $M_W$.

Similar connections exist between $\hat{\Pi}(q^2)$ and the running of the weak mixing angle, whose nonperturbative contributions arise from a different flavor combination of the same correlator.
Reference~\cite{Ce:2022eix} provides the first lattice calculation thereof, including the SU(3) breaking correlator $\bar\Pi^{08}$, which removes the SU(3) assumptions required in previous phenomenological determinations~\cite{Marciano:1983ss,Jegerlehner:1985gq,Jegerlehner:1990uiq,Marciano:1993jd}.
In particular, the evolution of the weak mixing angle to $q^2=M_Z^2$ can be better controlled~\cite{Erler:2024lds}, which reduces the uncertainty in the $\gamma$--$Z$ correlator $\bar\Pi^{\gamma Z}(-M_Z^2)$, required for the prediction of $\amuEW$, see \cref{sec:EW} and Ref.~\cite{Hoferichter:2025yih}.

\subsection{Further cross-checks}
\label{sec:furtherchecks}

Given the high precision that is being aimed at, it is important to provide stringent
cross-checks of the lattice calculations. Reaching consistency among the results of
different lattice collaborations is a crucial validation step, in particular as a test of
universality.  In this subsection, we discuss further cross-checks aimed at eliminating
potentially underestimated common sources of systematic error.

In recent years, all lattice collaborations have used the TMR, which means that the states created by the EM current are at rest. An
alternative method, which is particularly natural on physically large lattices, is the
covariant coordinate-space (CCS) representation~\cite{Meyer:2017hjv},
\begin{equation}
  \amuHVPLO =\int d^4z \; H_{\lambda\sigma}(z)\,\braket{ j_{\sigma}(z)
  j_\lambda(0)}\,,
\end{equation}
with $H_{\mu\nu}(x)$ a known tensorial kernel. The latter can be chosen to be, e.g.,
transverse, traceless, or proportional to $x_\mu x_\nu$, by exploiting the conservation of
the vector current. A recipe to translate a quantity defined in the TMR representation to
the CCS representation is provided in Ref.~\cite{Chao:2022ycy}. Consistency between
results obtained from the TMR and the CCS representation are ultimately a test of the
restoration of Lorentz symmetry in the continuum and infinite-volume limit. In particular,
finite-size effects are different for the two representations, and therefore the
comparison TMR vs.\ CCS provides a test of our understanding of these effects.

The CCS representation with a traceless kernel has been applied
recently~\cite{Chao:2022ycy} to compute the intermediate window quantity, thereby
providing a cross-check for the Mainz calculation~\cite{Ce:2022kxy} at the point
$M_\pi=350\,$MeV, $M_K=450\,$MeV with a precision of 1.1\%. Good consistency was found in
the continuum, even though the discretization errors are rather different for the lattice
correlator based on one local and one conserved vector current.

The CCS representation can also be used to compute the bare EM correction to
the HVP contribution~\cite{Biloshytskyi:2022ets}
\begin{equation}
  a_\mu^{1\gamma^*}(\Lambda) =  -\frac{e^2}{2}\delta_{\mu\nu}\int
  d^4x\,d^4y\,d^4z\, H_{\lambda\sigma}(z)\,
  G_0(\Lambda;y-x)
  \, \braket{j_\sigma(z)j_\nu(y)j_\mu(x) j_\lambda(0)}\,,
  \label{eq:CCSmaster}
\end{equation}
where $G_0$ is the massless scalar propagator UV-regulated at scale $\Lambda$, either
using the same lattice as the QCD degrees of freedom, in which case $\Lambda=1/a$, or at a
scale $\Lambda$ well below the inverse lattice spacing, for instance \`a la Pauli--Villars.
In the latter case, the expression has a similar structure as the coordinate-space
expression for $\amuHLbL$. The continuum limit of $ a_\mu^{\rm 1\gamma^*}(\Lambda)$ can be
taken at a fixed value of $\Lambda$. It is entirely determined by the forward
light-by-light amplitude, and a Cottingham-like formula~\cite{Biloshytskyi:2022ets} allows
for an evaluation of $ a_\mu^{\rm 1\gamma^*}(\Lambda)$ in the continuum. Quark-contraction
diagrams in which the photon propagator connects two distinct quark loops are UV-finite
and may be evaluated without regularization~\cite{Parrino:2025afq}.

If $\mathbf{M}$ denotes an $(N_{\rm f}+1)$-component vector of stable hadron masses
defining the parameters of QCD in the presence of dynamical photons, the master equation
for the correction to HVP to be added to its value in isoQCD reads
\begin{equation}\label{eq:d_amuhvp_master}
  \deltaHVPLO= \lim_{\Lambda\to\infty}\left[a_\mu^{1\gamma^*}(\Lambda)
  - \nabla_{\mathbf{M}} \amuHVPLO \cdot  \mathbf{M}^{\rm self}(\Lambda)\right]
  + \nabla_{\mathbf{M}} \amuHVPLO \cdot \delta \mathbf{M}\,.
\end{equation}
The term containing the bare EM self-energies $\mathbf{M}^{\rm
self}(\Lambda)$ of the reference hadrons corresponds to subtracting from the bare
correction $ a_\mu^{1\gamma^*}(\Lambda)$ the effect that merely comes from an
EM shift of the reference hadron masses. The last term corresponds to a shift
in $\amuHVPLO$ due to the fact that at the isoQCD expansion point, not all reference
hadron masses are at their physical values, $\delta \mathbf{M} = \mathbf{M}^{\rm phys}-
\mathbf{M}^{\rm isoQCD}$. \Cref{eq:d_amuhvp_master} is also a rigorous starting point for
computing $\deltaHVPLO $ in the continuum, since $\mathbf{M}^{\rm self}(\Lambda)$ can be
evaluated using Cottingham formulae for each reference hadron (see, e.g.,
Ref.~\cite{Gasser:2020hzn}). Thus, the correction $\deltaHVPLO $ can ultimately be
cross-checked by continuum evaluations, for which one might expect a precision
of $2\times 10^{-10}$ to be achievable, by analogy with $\amuHLbL$.

\subsection{Conclusions and outlook}

At the time of WP20 lattice-QCD calculations were not precise enough to impact the SM prediction, and, as a result, the value for HVP in the SM prediction of WP20 was entirely based on data-driven analyses of hadronic $e^+ e^-$ cross-section data.
Since then a great deal of progress has been achieved, thanks to dedicated efforts by the world-wide lattice field theory community, allowing for a precise and robust first-principles calculation of the HVP contribution from a variety of lattice fermion actions including Wilson-Clover and twisted-mass Wilson, domain wall, highly improved staggered quark (HISQ), and other implementations of staggered quark actions.

A key ingredient for the observed consolidation of lattice results was the introduction of window observables and their adoption by the lattice community as a standard for cross-checking and benchmarking lattice calculations for sub-contributions to HVP.
The SD, intermediate, and LD windows amplify or suppress technical challenges related to discretization effects, FV corrections, and statistical noise, allowing for tailored approaches for each of them.
The second important development since WP20 is the widespread adoption of blinding procedures by the lattice community to avoid confirmation bias. This has proved instrumental for establishing the reliability of the observed consolidation among independent lattice-QCD results and, in turn, for producing a robust and precise average for LO HVP.

The intermediate-distance window was the first quantity for which a consolidated set of lattice results became available around 2022, hinting at a significant tension with the corresponding data-driven estimates based on $e^+e^-$ data prior to CMD-3's 2023 result.
The currently available set of results for the light-quark, isospin-symmetric short- and intermediate-distance window observables shows a remarkable degree of consistency and precision among different lattice collaborations. Fewer results are available for the LD window or for unwindowed contributions, but the results for the latter are in good agreement with sums of windows.

As the precision of lattice results improves, the focus is gradually shifting towards the calculation of IB corrections.
Only a small fraction associated with sea--sea and sea--valence QED corrections, which amounts to 
approximately 0.06\% of the total HVP contribution, has been calculated by a single lattice collaboration so far.
For all other contributions, at least three independent lattice results exist.  The LD QED contributions, which amount to about 0.3\% of the total HVP correction, are particularly challenging.  For these contributions additional cross-checks against phenomenology have been performed.

The review of lattice-QCD results in WP25 is based on seventeen different papers from eight independent lattice-QCD collaborations~\cite{ExtendedTwistedMass:2024nyi, MILC:2024ryz, Spiegel:2024dec, Boccaletti:2024guq, Kuberski:2024bcj, ExtendedTwistedMass:2022jpw, Wang:2022lkq, RBC:2023pvn, Borsanyi:2020mff, Ce:2022kxy, Aubin:2022hgm, Lehner:2020crt, RBC:2018dos, Djukanovic:2024cmq, FermilabLatticeHPQCD:2024ppc, RBC:2024fic, Giusti:2019xct}, including three
lattice calculations of the entire LO HVP contribution~\cite{Borsanyi:2020mff,RBC:2024fic,Djukanovic:2024cmq}. 
We use five approaches for obtaining averages for $\amuHVPLO$, which are in
excellent agreement.
As our final SM prediction of $\amuHVPLO$ we take Avg.~1 from \cref{eq:amuhvp_avg2}, which includes the maximum number of independent lattice results from Refs.~\cite{ExtendedTwistedMass:2024nyi, MILC:2024ryz, Spiegel:2024dec, Boccaletti:2024guq, Kuberski:2024bcj, ExtendedTwistedMass:2022jpw, Wang:2022lkq, RBC:2023pvn, Borsanyi:2020mff, Ce:2022kxy, Aubin:2022hgm, Lehner:2020crt, RBC:2018dos, Djukanovic:2024cmq, FermilabLatticeHPQCD:2024ppc, RBC:2024fic, Giusti:2019xct}.
This consolidated average of lattice-QCD results provides a reliable determination of the LO HVP contribution to the SM prediction of $a_\mu$. The total error of $\pm6.1\times10^{-10}$ is  larger compared to that of the data-driven estimate quoted in WP20.
Compared with WP20, the relative precision for $\amuHVPLO$ obtained from lattice QCD has improved by a factor of approximately three. 

This section also describes
how lattice QED+QCD can provide model-independent
results for the isospin rotation needed to extract the spectral function from hadronic
$\tau$ decays.
In addition to considering the LO HVP contribution, we have also reviewed lattice calculations of the closely related hadronic running of the EM coupling $\alpha$, which, unsurprisingly, shows similar tensions between lattice results and dispersive analyses based on $e^+ e^-$ cross-section data prior to the results published by CMD-3.
Finally, a valuable consistency check of the entire lattice approach to HVP can be performed by employing the CCS representation as an alternative to the standard TMR.

Continuing efforts by the world-wide lattice community are expected to yield further significant improvements in precision and, hopefully, even better consolidation thanks to a diversity of methods. We expect, in particular, more precise evaluations of IB effects and the noisy contributions at long distances.

\FloatBarrier

\clearpage

\section{Interplay of lattice-QCD and data-driven evaluations of HVP}
\label{sec:comparison}

\noindent
\begin{flushleft}
\emph{D.~Boito, A.~X.~El-Khadra,  M.~Golterman, M.~Hoferichter, C.~Lehner, L.~Lellouch, T.~Leplumey, B.~Malaescu}
\end{flushleft}

\subsection{Data-driven evaluations of Euclidean windows}
\label{sec:windows_data_driven}

For the full HVP contribution, timelike~\cite{Bouchiat:1961lbg,Brodsky:1967sr,Lautrup:1969fr,Gourdin:1969dm} and spacelike~\cite{Lautrup:1971jf,Blum:2002ii,Bernecker:2011gh} representations 
\begin{equation}
\label{amu_HVP}
 \amuHVPLO=\bigg(\frac{\alpha m_\mu}{3\pi}\bigg)^2\int_{s_\text{thr}}^\infty ds \frac{\hat K(s)}{s^2}R_\text{had}(s)=\left(\frac{\alpha}{\pi}\right)^2 \int_0^\infty d x_0\,
      C(x_0) \tilde{f}(x_0)  \, ,
\end{equation}
are connected via change of the corresponding kernel functions $\hat K(s)$ and $\tilde f(x_0)$, 
where $\hat K(s)$ and $R_\text{had}(s)$ are defined in \cref{sec:intro_data_HVP}, 
$\tilde f(x_0)$ and $C(x_0)$ in \cref{sec:methods}.
 Ultimately, the two formulations are related by a change of kernel functions, $\hat K(s)$ vs.\ $\tilde f(x_0)$. 
In particular, this implies that additional weight functions introduced in the integrals in \cref{amu_HVP} in one approach can be translated to the other accordingly,\footnote{In a perturbative approach, also the effect of quark-flavor thresholds needs to be taken into account~\cite{Nesterenko:2023bdc}.} e.g., the spacelike weight functions $\Theta(x_0)$ introduced in Ref.~\cite{RBC:2018dos}, see \cref{amu:sum,eq:amusd,eq:amusd,eq:winW,eq:win,def_wind}, translate to timelike ones $\tilde \Theta(s)$ via
\begin{align}
 \tilde \Theta(s)=\frac{3 s^{5/2}}{m_\mu^2 \hat K(s)}\int_0^\infty dx_0\, \Theta(x_0)e^{-x_0\sqrt{s}} \int_0^\infty \frac{d\omega}{\omega} f\big(\omega^2\big)\bigg[\omega^2x_0^2-4\sin^2\bigg(\frac{\omega x_0}{2}\bigg)\bigg] \,,
\end{align}
with $f(s)$ defined in \cref{def_f}.
The possibility of evaluating the Euclidean window quantities defined by $\Theta(x_0)$ also with data-driven techniques renders them an invaluable diagnostic tool to compare lattice-QCD and data-driven evaluations of HVP in more detail, and, ultimately, to combine the two approaches via hybrid evaluations, using either method in the parameter space where it is most precise~\cite{RBC:2018dos}. 

\begin{figure}[t!]
\begin{center}
\includegraphics[scale=0.5]{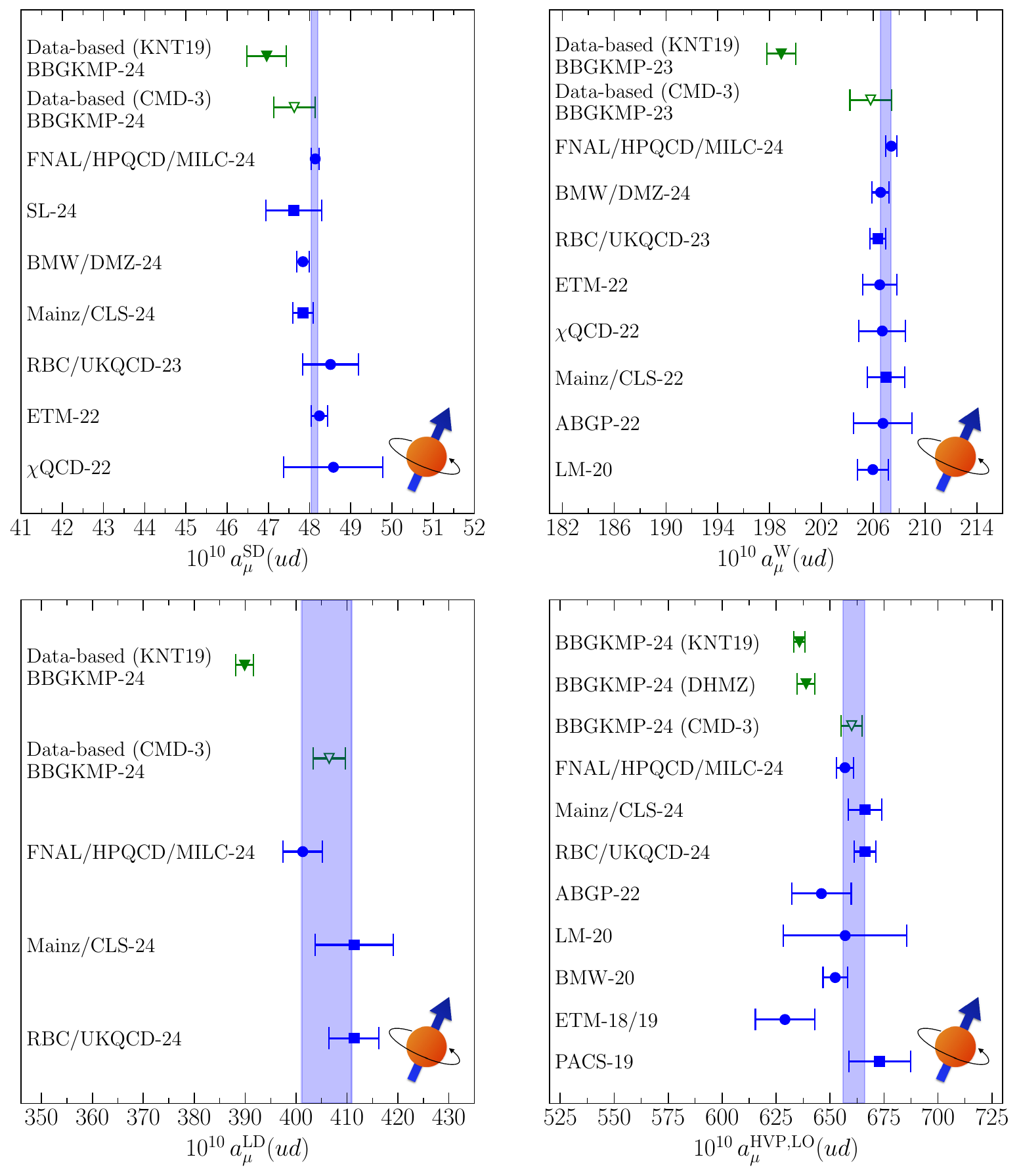}
\caption{Data-driven results for
the light-quark connected ($ud$) components of the RBC/UKQCD windows compared with isospin-symmetric
lattice-QCD determinations of the same quantities.
The label ``BBGKMP-24(23)''
refers to Refs.~\cite{Boito:2022dry,Benton:2023dci,Benton:2023fcv,Benton:2024kwp}.
The bands show averages for these quantities taken from  \cref{sec:latticeHVP}.
Upper-Left:
Data-driven $\amuSDud$ compared with results from
Refs.~\cite{ExtendedTwistedMass:2022jpw,RBC:2023pvn,Kuberski:2024bcj,Boccaletti:2024guq,Spiegel:2024dec,MILC:2024ryz,Wang:2022lkq}.
Upper-Right: Data-driven $\amuWud$ compared with results from
Refs.~\cite{Boccaletti:2024guq,Lehner:2020crt,Wang:2022lkq,Aubin:2022hgm,Ce:2022kxy,ExtendedTwistedMass:2022jpw,RBC:2023pvn,MILC:2024ryz}.
Lower-Left: Data-driven $\amuLDud$ compared with the recent
results of Refs.~\cite{RBC:2024fic,Djukanovic:2024cmq,FermilabLatticeHPQCD:2024ppc}.
Lower-Right: Data-driven $\amuHVPLOud$ compared with results from
Refs.~\cite{Aubin:2022hgm,Borsanyi:2020mff,Lehner:2020crt,Giusti:2018mdh,Giusti:2019hkz,Shintani:2019wai,RBC:2024fic,FermilabLatticeHPQCD:2024ppc,Djukanovic:2024cmq}.
Entries marked with ``CMD-3'' in each figure show the exploration of the impact of the
replacement of KNT $2\pi$ data for energies between 0.33 and 1.2 GeV with
CMD-3 data (see main text).}
\label{fig:results-lqc-CMD3}
\end{center}
\end{figure}

In practice, the comparison is currently performed at the level of the spacelike windows, as the inversion to the timelike domain amounts to an ill-posed inverse Laplace transform. Optimizing the resolution in $\sqrt{s}$ that can be achieved in such an inversion therefore requires precise knowledge of the covariances within a given lattice-QCD calculation, and several studies already explored the consequences for detailed comparisons between phenomenological and lattice-QCD evaluations~\cite{Lehner:2020crt,DeGrand:2022lmc,Boito:2022njs,Colangelo:2022vok,Davier:2023cyp}, including the development of strategies to extract the maximum amount of information on the timelike domain from a given set of lattice calculations for $\amuHVPLO$, Euclidean windows, and $\Delta \alpha_\text{had}(-Q^2)$. In addition to comparing the full HVP contribution in a given window~\cite{Borsanyi:2020mff,Colangelo:2022vok,Davier:2023cyp}, also comparison quantities for specific parts of a complete lattice calculation have been extracted with data-driven techniques, such as IB contributions, see \cref{sec:isospin_breaking}, the light-quark-connected contribution, see \cref{sec:lqc}, and the strange+disconnected contribution, see \cref{sec:strange+light}.  

\begin{figure}[t!]
\begin{center}
\includegraphics[scale=0.5]{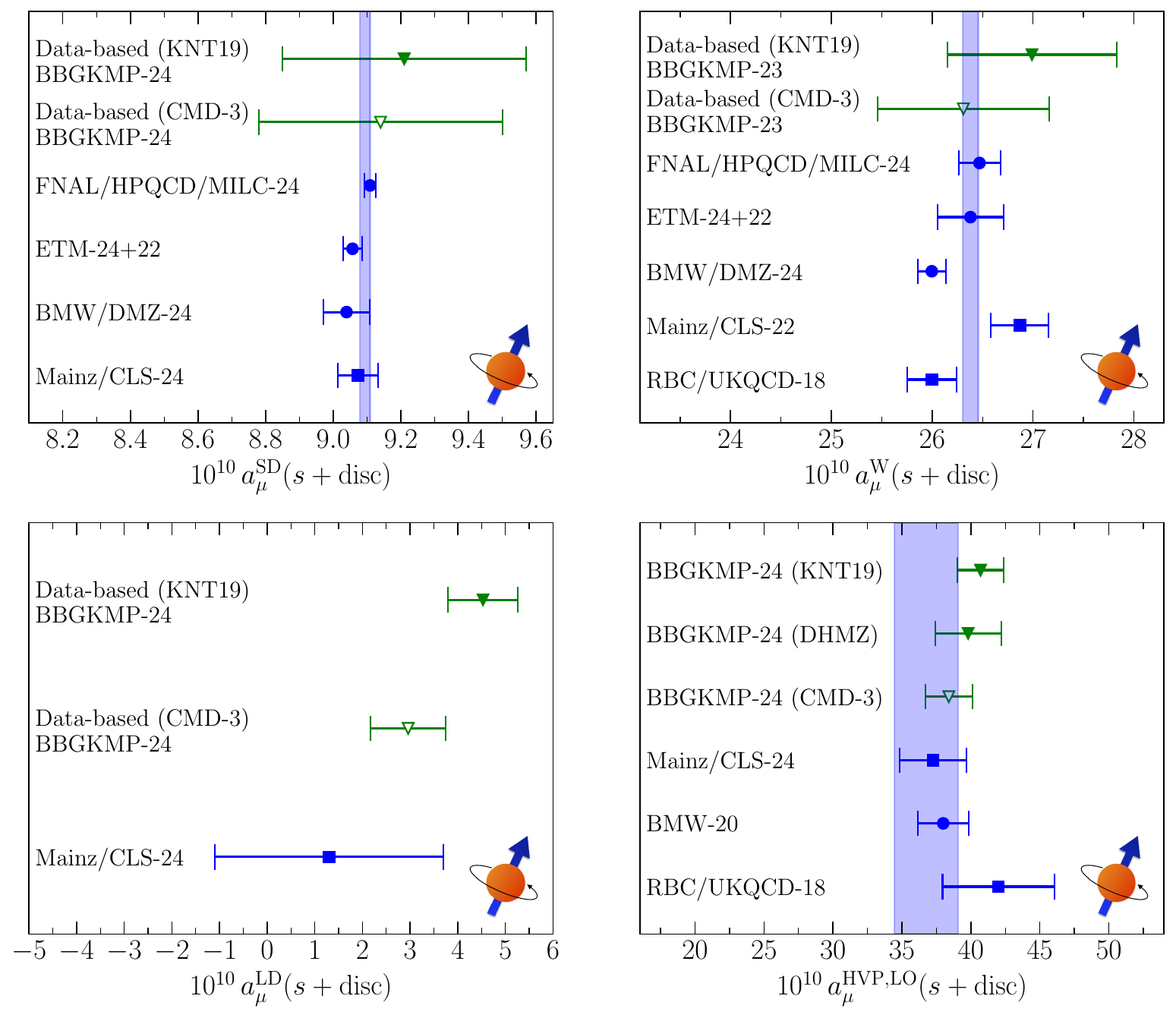}
\caption{Data-driven results for the $s$+disc components of the RBC/UKQCD
 windows compared with isospin-symmetric lattice
determinations of the same quantities obtained by adding strange-connected and disconnected contributions given in \cref{sec:latticeHVP}. The label  ``BBGKMP 24(23)'' refers to
Refs.~\cite{Boito:2022rkw,Benton:2023dci,Benton:2023fcv,Benton:2024kwp}.
The bands show averages for these quantities obtained from  \cref{sec:latticeHVP}
by adding the averages of $s$  and disc contributions, see main text.
Upper-Left: Data-driven results for $\amuSDsdisc$ compared with
results from
Refs.~\cite{MILC:2024ryz,Kuberski:2024bcj,Boccaletti:2024guq,ExtendedTwistedMass:2024nyi,ExtendedTwistedMass:2022jpw}.
Upper-Right: Data-driven
results for $\amuWsdisc$ compared with results from
Refs.~\cite{MILC:2024ryz,Boccaletti:2024guq,RBC:2018dos,Ce:2022kxy,ExtendedTwistedMass:2022jpw,ExtendedTwistedMass:2024nyi}. 
Lower-Left: Data-driven
results for $\amuLDsdisc$ compared with the result from Ref.~\cite{Djukanovic:2024cmq}  (see also Ref.~\cite{Benton:2024kwp}).
Lower-Right: Data-driven results for $\amuHVPLOsdisc$ compared
with results from Refs.~\cite{RBC:2018dos,Djukanovic:2024cmq,Borsanyi:2020mff}.
Entries marked with ``CMD-3'' in each figure  show the exploration of the impact of
the replacement of KNT $2\pi$ data for energies between 0.33
and 1.2 GeV with CMD-3 data (see main text).}
\label{fig:results-s+lqd-CMD3}
\end{center}
\end{figure}

\subsection{Hybrid calculations}
\label{sec:hybrid}

The Euclidean windows defined in Ref.~\cite{RBC:2018dos}
are well-defined individually and are independent of the regulator.  They can therefore also be computed with different methods and then combined to obtain the total HVP value.  This was already recognized
in Ref.~\cite{RBC:2018dos}, where the SD window and the LD window were obtained from a data-driven approach and the intermediate-distance window was obtained using the lattice method.  This requires that the lattice calculation be complete, i.e., include all quark flavors and also QED and SIB corrections.  Since Ref.~\cite{RBC:2018dos} appeared, tensions in the data-driven sector have been identified that make the original hybrid result of RBC/UKQCD-18 no longer appropriate to use since it included the LD contribution from $t_1=1.0$ fm from the data-driven method.  Variations of this idea can, however, be useful with well-consolidated results.

BMW/DMZ-24 builds on this idea, strongly limiting the use of the data-driven component to a LD window beyond $t_1=2.8$~fm, see \cref{eq:win}~\cite{Boccaletti:2024guq}.
Thus, more than $95\%$ of their result is obtained with a lattice calculation and only 
the remaining $\lesssim5$\% utilizes a subset of the measured $e^+e^-\to \text{hadrons}$ (and $\tau\to\nu_\tau+\pi^-\pi^0$) spectral function discussed in \cref{sec:dataHVP}.
More importantly, this LD contribution is predominantly determined by the low-mass region of the spectrum where all measurements agree within uncertainties, below the $\rho$-peak region that displays the tensions discussed in \cref{sec:dataHVP}.
Even such a small replacement brings a significant reduction in overall uncertainty.
As discussed in \cref{sec:latticeHVP}, the statistical noise of light-quark contributions increases rapidly with distance when present-day lattice methods are utilized.
The same is true for effects associated with the finite volume of the lattice.
However, those challenges are circumvented when the measured spectra are used to determine LD contributions instead.
The authors find that for the choice of $t_1$ considered in Ref.~\cite{Boccaletti:2024guq}, the resulting uncertainty on this LD part becomes a negligible contribution to the one on the final result.
It is worth noting that investigations of the dependence of $\amuHVPLO$ on $t_1$ have been performed in Refs.~\cite{RBC:2018dos,Davies:2024pvv}.
The result of Ref.~\cite{Boccaletti:2024guq} for $\amuHVPLO$ is not yet published and the authors plan to complete the documentation of the lattice-QCD calculations that enter the hybrid evaluation in the published version.  
Moreover, it is important that the different components, which enter that result, be computed independently by other groups in view of providing a consolidated combination with similar or even improved precision in the future.
We note that once the discrepancies among experimental $e^+e^-$ data are resolved and if data-driven and lattice results are found to be in good agreement, the hybrid method has the potential to yield valuable additional insights, including evaluations of HVP with improved precision, to fully leverage the Fermilab experiment's measurement of the muon $g-2$.

\subsection{Comparison}
\label{comp}

\begin{figure}[t!]
       \begin{center}
\includegraphics[width=0.85\textwidth]{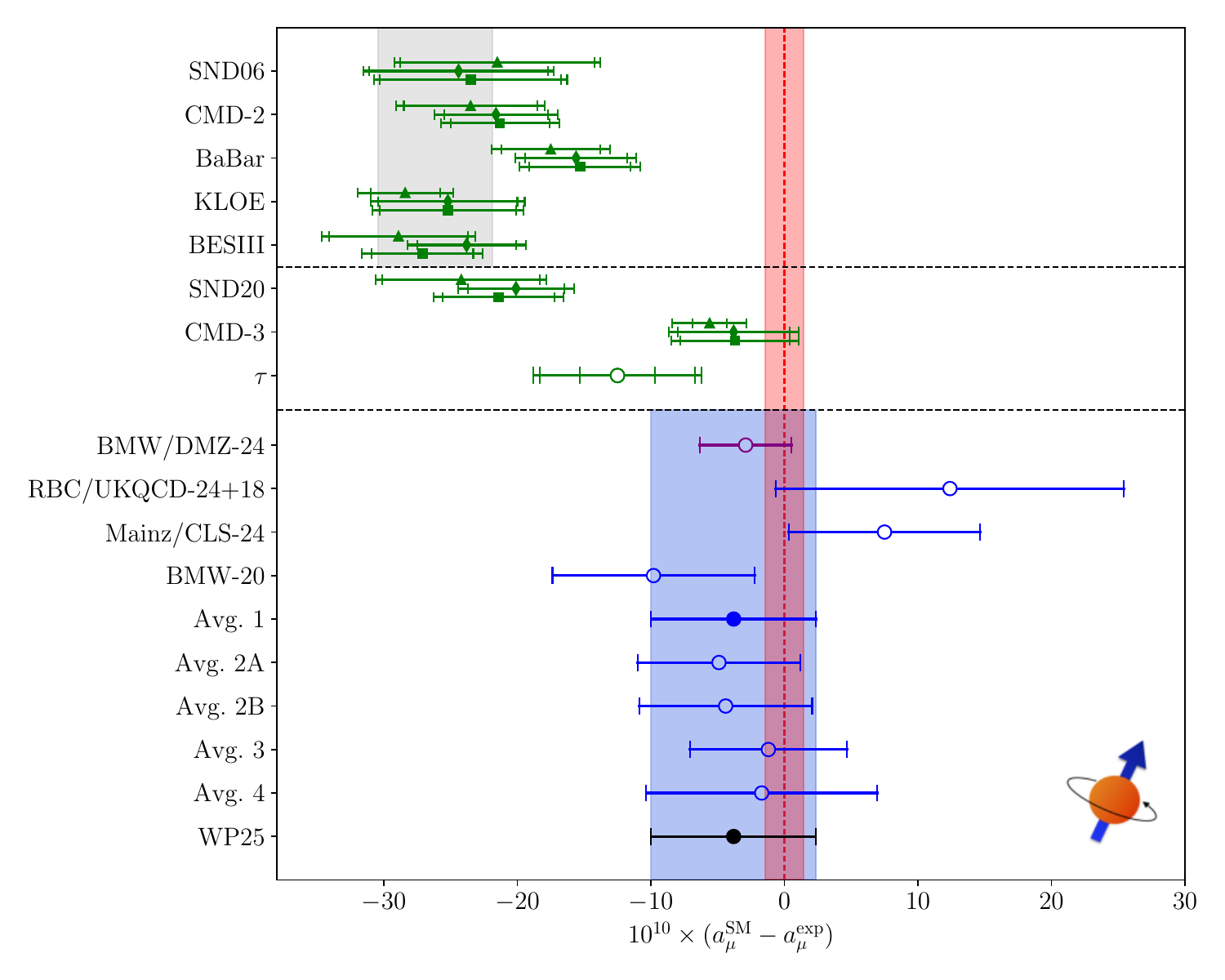}
\caption{Final summary of various determinations of $\amuHVPLO$ discussed in \cref{sec:dataHVP,sec:latticeHVP}, propagated to $\amuSM$. The first two panels refer to data-driven determinations, where the three points for each $e^+e^-$ experiment reflect the ``CHKLS,'' ``DHMZ,'' and ``KNTW'' methods, see \cref{fig:summary_plot_ee,fig:summary_plot_data_hvp} for more details. The gray band indicates the WP20 result, based on the $e^+e^-$ experiments above the first dashed line. The $\tau$ point corresponds to \cref{LOHVP_tau}. 
The last panel summarizes lattice-QCD determinations, including the hybrid evaluation~\cite{Boccaletti:2024guq}, the three  individual lattice-QCD calculations shown in \cref{fig:tot-amu}, and the five lattice HVP averages from \cref{fig:amu lat Avg summary}. The blue band refers to the final WP25 result, which coincides with ``Avg.~1.''
In all cases, except for the gray WP20 band, the remaining contributions to $\amuSM$  beyond $\amuHVPLO$ are taken from WP25, as given in \cref{tab:summary}.
The red band denotes the experimental world average, which has been updated including the final results from the Fermilab experiment.}
\label{fig:finalHVPsummary}
\end{center}
\end{figure}

The results of the analysis in \cref{sec:lattice_comparison} are shown in 
\cref{fig:results-lqc-CMD3} for isospin-symmetric light-quark connected quantities, and in \cref{fig:results-s+lqd-CMD3} for isospin-symmetric $s$+disc  quantities, comparing the data-driven results with those obtained from the lattice. Since the analyses in question~\cite{Boito:2022rkw,Boito:2022dry,Benton:2023dci,Benton:2023fcv,Benton:2024kwp} use phenomenological input for IB corrections in the $2\pi$ and $3\pi$ channels~\cite{Hoferichter:2023sli,Colangelo:2022prz,Hoferichter:2023bjm}, which to good approximation is scheme independent once the pion mass is identified with the mass of the neutral pion, and 
BMW-20~\cite{Borsanyi:2020mff} input for small additional QED IB corrections, it is justified to compare with lattice-QCD results in the BMW/WP25 scheme, see also \cref{sec:isospin_breaking}.  
The lattice averages shown in \cref{fig:results-lqc-CMD3}
are taken from \cref{sec:latticeHVP}, while those in \cref{fig:results-s+lqd-CMD3}
are obtained by adding strange-connected and disconnected averages from \cref{sec:latticeHVP} (adding errors in quadrature, as suggested therein), using the fact that charm-disconnected contributions are very small, and can thus
be neglected.   The three lattice points shown in the lower-right panel of \cref{fig:results-s+lqd-CMD3}
average to a value larger than the band shown because the band includes more strange-connected
results from \cref{tab:amuSDflavsummary,{tab:W_comp}}.
For the single lattice point in the lower-left panel, see Ref.~\cite{Benton:2024kwp}.
For the $ud$ and $s$+disc  HVP lattice averages, the line ``Avg.~B'' in \cref{tab:full_avgs} has been used.

\Cref{fig:results-lqc-CMD3,fig:results-s+lqd-CMD3} show
that for the LD and intermediate light-quark connected RBC/UKQCD window quantities,
there are significant discrepancies between the KNT-compilation-based
data-driven and the lattice-based estimates, which lead to a significant
discrepancy in the total  $\amuHVPLOud$ when
comparing with the most precise lattice determinations. In contrast, for 
$\amuSDud$ and for all the $s$+disc window quantities, there
are no discrepancies (though the data-driven errors for the latter
are relatively large). Moreover, the exploratory exercise of
replacing the $\pi^+\pi^-$ KNT-compilation data in the interval between
$0.33$ and $1.2$ GeV with the CMD-3 $\pi^+\pi^-$ data suggests
that these discrepancies could be due to discrepancies in the experimental
data for the $\pi^+\pi^-$ component of $R_\text{had}(s)$ in the region around the
$\rho$ peak. With this replacement, the discrepancies in the light-quark connected results
are eliminated without disturbing the good agreement for the $s$+disc and
the light-quark connected SD parts. The $\pi^+\pi^-$ channel is responsible for 72\%
of the data-driven $a_\mu^{W}(ud)$ result and for 88\% of the
data-driven $\amuLDud$, but only 32\% of the $\amuSDud$ value and
only very small fractions of the $s$+disc results. 
These conclusions for the quark-flavor-specific contributions agree with the findings from Ref.~\cite{Davier:2023cyp}, obtained for the full contributions, including all flavors and all IB corrections. Confirming the original findings of Ref.~\cite{Borsanyi:2020mff}, the authors observe a significant discrepancy between the pre-CMD-3 data-driven and lattice results for the full intermediate-window contribution and a large, but less statistically significant one, for the total HVP contribution to $a_\mu$. In addition they show that there is relatively good agreement for $\Delta \alpha_\text{had}(-10\GeV^2)-\Delta \alpha_\text{had}(-1\GeV^2)$. 
Using a new approach that allows one to investigate how the experimentally measured spectral functions would have to be modified to reconcile the data-driven results with the lattice ones, they show that an enhancement of that function in any interval of CM energy that includes the $\rho$ peak could explain the observed disagreement pattern. 
In particular, this pattern can be explained by a rescaling by $5\%$ of the contribution to each of those observables from the $\rho$-peak region, defined via the interval $\sqrt{s}\in [0.63,\,0.92]$. 
Of course, such a rescaling is significantly larger than the uncertainties quoted in pre-CMD-3 combinations of the $e^+e^-$ spectra.
This study included statistical-and-systematic-uncertainty correlations among the observables computed on the lattice and found the conclusions to be stable with respect to them. It was also performed with an initial blinding on the data-driven results.

A priori, it is not evident that significant modifications to the $2\pi$ spectral function can be introduced without violating QCD constraints from analyticity and unitarity, but it was shown in Ref.~\cite{Colangelo:2020lcg} that this is possible. One option is a modification of the $\pi\pi$ phase shift, which would, however, induce rather large changes in the cross section concentrated around the $\rho$ resonance. The second possibility concerns changes in the inelastic contributions, which lead to largely uniform relative shifts of the cross section in the whole low-energy region; the CMD-3 measurement amounts to a realization of this latter scenario. Indeed, the CMD-3 data pass the tests of unitarity and analyticity~\cite{Stoffer:2023gba,Leplumey:2025kvv}, see \cref{sec:corr_obs}. However, inverting the Laplace transform to translate from Euclidean windows back to CM energies is inherently ill-posed, where one possible avenue concerns optimizing linear combinations of a number of Euclidean windows~\cite{Lehner:2020crt,DeGrand:2022lmc,Colangelo:2022vok}, e.g., Ref.~\cite{Colangelo:2022vok} explicitly constructs such linear combinations for localization in CM energy, leading to strong oscillations in Euclidean time, and also proposes to isolate a given hadronic channel via suitable linear combinations of Euclidean windows. It is emphasized  that such optimizations require a precise knowledge of the full covariance matrix in lattice-QCD calculations. Finally, in Ref.~\cite{ExtendedTwistedMassCollaborationETMC:2022sta} it is proposed to calculate a version of the $R$-ratio convolved with Gaussian smearing kernels in lattice QCD using the method from Ref.~\cite{Hansen:2019idp} to extract smeared spectral densities from Euclidean correlators, including  a proof-of-concept calculation with resolutions between $0.44$ and $0.63\GeV$. 

With the new and upcoming lattice results available for the SD, intermediate, and LD windows and for other HVP related quantities those studies will have to be significantly updated.
This is all the more important given the tensions now observed among the different $e^+e^-\to\text{hadrons}$ measurements, as well as with those of hadronic $\tau$ decays. 
Such studies should offer interesting perspectives in the comparison of lattice and data-driven results for $\amuHVPLO$.
It is important that those studies account for all correlations in lattice and $R$-ratio observables, as well as for uncertainties on these~\cite{Davier:2023cyp}.
Double-blinding, for observables not yet computed, is also important.

Finally, \cref{fig:finalHVPsummary} presents a detailed overview of the current situation for HVP, comparing the data-driven, lattice, and hybrid results discussed in \cref{sec:dataHVP,sec:latticeHVP}, as well as \cref{sec:hybrid}.
We note that the good agreement between the lattice and CMD-3-based results does not imply that the CMD-3 result is to be preferred. The discrepancy between the CMD-3 and other experimental measurements remains an independent puzzle. This issue, along with the well-known \babar--KLOE discrepancy, will require further experimental investigations and detailed studies of the MC generators and other tools used by the different experiments in their data analyses.

In contrast, the complete lattice HVP results from Refs.~\cite{Borsanyi:2020mff,RBC:2018dos,RBC:2023pvn,RBC:2024fic,Djukanovic:2024cmq} show good agreement, including with the hybrid result from Ref.~\cite{Boccaletti:2024guq} and the consolidated lattice averages. These averages are based on different combinations of a large number of results from well-defined sub-contributions, obtained independently by various lattice-QCD collaborations, as described in \cref{sec:latticeHVP}.

\FloatBarrier

\clearpage

\section{Data-driven and analytic approaches to HLbL}
\label{sec:dataHLbL}

\noindent
\begin{flushleft}
\emph{J.~Bijnens, L.~Cappiello, G.~Eichmann, C.~S.~Fischer, S.~Gonz\`alez-Sol\'is, N.~Hermansson-Truedsson, M.~Hoferichter, B.-L.~Hoid, S.~Holz, B.~Kubis, A.~Kup\'s\'c, P.~Masjuan, M.~Procura, A.~Rebhan, C.~F.~Redmer, A.~Rodr\'\i guez-S\'anchez, P.~Roig, P.~S\'anchez-Puertas, P.~Stoffer,  M.~Zillinger}
\end{flushleft}

\subsection{Introduction}
\label{Section:HLbLIntro}

\begin{figure}[t]
    \begin{center}
        \includegraphics{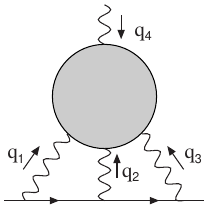}
    \end{center}
    \caption{The HLbL contribution depicted graphically. The bottom line is the muon. The blob indicates all possible hadronic interactions. The $q_4$ leg indicates the external magnetic field. Figure taken from Ref.~\cite{Bijnens:2020xnl}.}
    \label{fig:hlblgeneral}
\end{figure}

The HLbL contribution to the muon anomalous magnetic moment is depicted in \cref{fig:hlblgeneral}.\footnote{We have kept the $q_4$ sign different in different sections to keep with the notation in the original references.} For a very long time it was considered impossible to estimate it with any reliability. This changed during the 90s when a number of theoretically reasonably well argued models were used, final numbers can be found in Refs.~\cite{Bijnens:2001cq,Hayakawa:2001bb}. However, it was realized much more recently that a proper well-defined separation of different hadronic contributions was possible~\cite{Colangelo:2015ama}. This provided together with a proper definition of short-distance contributions (SDCs) from QCD~\cite{Melnikov:2003xd,Bijnens:2019ghy} the basis for a full evaluation of the HLbL contribution in the previous white paper of the muon $g-2$ theory initiative~\cite{Aoyama:2020ynm}. A proper description of the large amount of work done earlier can also be found there. The Fermilab experiment expects to release a new result for their total data set in the near future. We therefore need to update the HLbL result from Ref.~\cite{Aoyama:2020ynm}. The main improvements are that the dispersive framework has been improved to deal with spin-1 hadrons in a singularity-free fashion and first steps towards dealing with higher spin resonances have been done. The short-distance results have also been significantly improved. The model and phenomenological work has also been updated. This is described in the following sections.

The overall framework is discussed in \cref{sec:framework}. Improvements in the experimental inputs that are needed in the various approaches are discussed in \cref{sec:experiment}.
The short-distance part is improved in various ways, by including higher orders in the operator product expansion (OPE) as well as gluonic corrections, both for the general case and for the case with only two internal photons far off-shell. This is described in \cref{sec:SDC}. 
The improved dispersive framework is used for determining the $\eta,\eta'$ and axial-vector contributions as well as an update on scalars and a first estimate for tensors. This is
described in \cref{sec:Pseudoscalars}--\cref{sec:ResonanceContributions}. The method to perform a dedicated matching to the short-distance QCD results is explained in~\cref{sec:MatchingToSDCs}. Holographic QCD (hQCD) is a model for QCD that incorporates a lot of restrictions from full QCD. The HLbL contribution can be calculated in various versions of this class of models and the results are discussed in detail in \cref{Section:HolographicModels}. In particular we have used this to estimate errors on tensor and missing resonances from the results in \cref{sec:hQCDscalartensor}. In WP20 the rational approximants were used as a cross-check for the dispersive result for the $\pi^0$
 pole and as input for the $\eta,\eta'$ pole. These together with the newer analyses using the resonance-chiral-theory (R$\chi$T) approach for these contributions are described in \cref{sec:CApoles}. An improved phenomenological treatment of axial vectors and how they are used together with the short-distance results can be found in \cref{sec:Regge}. Another approach to QCD are the functional methods using Dyson--Schwinger (DSE) and Bethe--Salpeter (BSE) equations. The approximations needed to be able to solve the equations make this approach a QCD-based model, but it can capture different aspects than the phenomenological or holographic approaches. The most recent results relevant for HLbL are given in \cref{sec:functional}. A few quantities can be easily compared between the various approaches. This comparison is performed in \cref{sec:formfactors}.
The combination of all the results into a new number for the HLbL is presented in \cref{sec:finalnumber}.
Finally, prospects for further improvements on the theory side are given in \cref{sec:TriangleDR} and a wish list and expected future experimental results in \cref{sec:ExpProsp}.

\subsection{Framework}
\label{sec:framework}

The HLbL tensor is defined as the hadronic Green's function of four EM currents in pure QCD
\begin{align}
	\label{eq:HLbLTensorDefinition}
	\Pi^{\mu\nu\lambda\sigma}(q_1,q_2,q_3) = -i \int d^4x \, d^4y \, d^4z \, e^{-i(q_1 \cdot x + q_2 \cdot y + q_3 \cdot z)} \langle 0 | T \{ j_\mathrm{em}^\mu(x) j_\mathrm{em}^\nu(y) j_\mathrm{em}^\lambda(z) j_\mathrm{em}^\sigma(0) \} | 0 \rangle \, ,
\end{align}
which includes the contribution from the three lightest quarks:
\begin{align}
	j_\mathrm{em}^\mu := \bar q Q \gamma^\mu q \, , \qquad q = ( u , d, s )^T \, , \qquad Q = \mathrm{diag}\left(\frac{2}{3}, -\frac{1}{3}, -\frac{1}{3}\right) \, .
\end{align}
Contracting $\Pi^{\mu\nu\lambda\sigma}$ with polarization vectors yields the hadronic contribution to the helicity amplitudes for off-shell photon--photon scattering, $\gamma^*(q_1,\mu) \gamma^*(q_2,\nu) \to \gamma^*(-q_3,\lambda) \gamma^*(q_4,\sigma)$. The starting point of a dispersive approach to HLbL is the decomposition of the tensor $\Pi^{\mu\nu\lambda\sigma}$ into a generating set of Lorentz structures $T_i^{\mu\nu\lambda\sigma}$ that are individually gauge invariant, with scalar coefficient functions $\Pi_i$ free of kinematic singularities. Such a decomposition was constructed in~\cite{Colangelo:2015ama} following the recipe by Bardeen, Tung, and Tarrach (BTT)~\cite{Bardeen:1969aw,Tarrach:1975tu}, which leads to a redundant set of 54 structures with manifest crossing symmetry,
\begin{equation} \label{eq:BTT}
\Pi^{\mu\nu\lambda\sigma} = \sum_{i=1}^{54} T_i^{\mu\nu\lambda\sigma}\; \Pi_i\,.
\end{equation}
Although the scalar functions $\Pi_i$ are free of kinematic singularities, which is a prerequisite for a dispersive representation, the redundancy in the decomposition needs to be addressed: a fully off-shell basis contains only 41 elements, given by the number of independent helicity amplitudes.

The HLbL contribution to $a_\mu$ can be derived from $\Pi^{\mu\nu\lambda\sigma}$ using Dirac-space projection operator techniques~\cite{Aldins:1970id} in the limit $q_4 \to 0$. This leads to the following master formula, which is well-suited for a numerical evaluation:\footnote{Different variables for the integration can be used, some more examples are given in Ref.~\cite{Bijnens:2024jgh}.}
\begin{align}
	\label{eq:MasterFormula}
	a_\mu^{\rm HLbL} = \frac{\alpha^3}{432 \pi^2} \int_0^\infty d\Sigma\, \Sigma^3 \int_0^1 d r\, r \sqrt{1-r^2} \int_0^{2\pi} d \phi \sum_{i=1}^{12} T_i(\Sigma, r, \phi)\, \bar{\Pi}_i(Q_1^2, Q_2^2, Q_3^2)\,,
\end{align}
with known kernel functions $T_i(\Sigma, r, \phi)$~\cite{Colangelo:2017fiz} and Euclidean photon virtualities $Q_i^2 \equiv -q_i^2$ given by
\begin{align}
Q_1^2  &=  \frac{\Sigma}{3}\left(1 - \frac{r}{2} \cos\phi - \frac{r}{2} \sqrt{3}\sin \phi\right)\,, \nonumber \\
Q_2^2 & =  \frac{\Sigma}{3}\left(1 - \frac{r}{2} \cos\phi + \frac{r}{2} \sqrt{3}\sin \phi\right)\,, \nonumber \\
Q_3^2 & =  \frac{\Sigma}{3}\left(1 + r \cos\phi\right)\,.
\end{align}
The functions $\bar \Pi_i$ are linear combinations of the $\Pi_i$ introduced above and fully parameterize the hadronic content in $a_\mu^\text{HLbL}$. The ambiguities in the redundant set of functions $\Pi_i$ cancel in the linear combinations $\bar\Pi_i$ in the limit $q_4\to0$, hence each of the 12 terms in the master formula \cref{eq:MasterFormula} is a well-defined quantity.

Unitarity rigorously determines the absorptive part of the HLbL tensor, i.e., the residues of dynamical single-particle poles and the discontinuities of multi-particle branch cuts of the HLbL tensor in the different kinematic channels and relates them to various sub-processes with on-shell intermediate states.
In order to reconstruct the dispersive (real) part of the HLbL tensor, one makes use of analyticity to write dispersion relations for the BTT scalar functions in a given set of kinematic variables. This requires the scalar functions to be free of ambiguities, i.e., the redundancy in \cref{eq:BTT} needs to be resolved without introducing kinematic singularities in the dispersed variables. This has been achieved in the original approach of Refs.~\cite{Colangelo:2015ama,Colangelo:2017fiz} in four-point kinematics in the limit of an external on-shell photon, $q_4^2 = 0$, i.e., dispersion relations have been derived in terms of the Mandelstam variables for $\gamma^*\gamma^*\to\gamma^*\gamma$ scattering with fixed photon virtualities, with the limit $q_4 \to 0$ relevant for $a_\mu^\text{HLbL}$ taken only afterwards. Although the derived scalar functions are free of singularities in the dispersed Mandelstam variable, they contain kinematic singularities in the fixed virtualities $q_i^2$. In the static limit $q_4 \to 0$, their residues vanish due to a set of sum rules, hence $a_\mu^\text{HLbL}$ remains unaffected. These sum rules ensure that $a_\mu^\text{HLbL}$ is independent of the choice of the tensor basis. However, the fulfillment of the sum rules typically involves a cancellation between different hadronic intermediate states, which leads to complications in practice. In Ref.~\cite{Hoferichter:2024fsj}, an optimized basis has been constructed that limits the appearance of spurious kinematic singularities to intermediate states of spin $\ge2$.

In Ref.~\cite{Ludtke:2023hvz}, an alternative dispersive framework has been introduced, formulated directly for the functions $\bar\Pi_i$ in the static limit $q_4\to0$. In this approach, all redundancies and kinematic singularities are manifestly absent, but the dispersion relations in the photon virtualities require new sub-processes as input, see \cref{sec:TriangleDR}.

In the dispersive framework in four-point kinematics, the contributions from one- and two-meson intermediate states have been explicitly accounted for as discussed in the next sections: pseudoscalar-pole and -box contributions, scalars, axial-vector contributions, as well as partial contributions from tensor mesons. These intermediate-state contributions are expressed in terms of physical transition form factors (TFFs) and helicity amplitudes, which can either be reconstructed dispersively, obtained from lattice QCD, or taken as input from hadronic models. Since the dispersive approach relates all contributions to on-shell intermediate states, these observables serve as the experimental input for the dispersive data-driven determination of $a_\mu^\text{HLbL}$.

The dispersive treatment of exclusive intermediate states is feasible only at sufficiently low energies. Therefore, this description has to be complemented by matching to SDCs, which stem from the OPE and pQCD, see \cref{sec:SDC}.

\subsection{Experimental results}
\label{sec:experiment}

Experimental results can serve as input to the calculation of the contributions to HLbL in a direct and a more indirect way.
Direct input to the pseudoscalar-meson pole contributions are TFFs $F_{P\gamma^*\gamma^*}(q_1^2,q_2^2)$, which can be measured in different regions of momentum transfer $q_i^2$. The momentum transfer dependence of the relevant cross sections and decay rates due to the TFFs makes the experimental access to information at arbitrary values of $q_i^2$ challenging. Similar limitations hold for experimental information on multi-meson production in two-photon collisions, which can provide corresponding input for the $\pi\pi$ contribution or contributions of heavier scalar, axial-vector, and tensor resonances.
In a more indirect way, experimental information on other hadronic and radiative processes is needed to determine missing direct information on TFFs from dispersion relations, as successfully demonstrated for the $\pi^0$ pole contribution in WP20.

With respect to the available experimental information and the possible approaches, priorities for new experimental input have been formulated in WP20.
In the following an overview on new experimental results beyond those listed there is provided. An outlook on planned measurements and data to become available in the near future is given in \cref{sec:ExpProsp}.

\subsubsection{Meson transition form factors}
Experimental methods to obtain information on meson TFFs comprise Primakoff production of mesons $P$ to address the normalization at $F_{P\gamma^*\gamma^*}(0,0)$, meson production in two-photon scattering at $e^+e^-$ machines to address the spacelike momentum dependence of $F_{P\gamma^*\gamma^*}(q_1^2,q_2^2)$, and radiative meson production at $e^+e^-$ colliders as well as meson Dalitz decays to test the timelike momentum dependence of the respective TFFs.
Furthermore, rare decays of pseudoscalar mesons into lepton pairs can be related to TFFs.

New measurements of Dalitz decays of $\eta$ and $\eta^\prime$ involving a single off-shell photon are provided by the BESIII collaboration~\cite{BESIII:2024pxo} using the radiative decays of a total of 10 billion inclusively recorded $J/\psi$ decays as source of the mesons. The invariant mass distribution $M_{ee}^2 = q^2$ of the lepton pairs is investigated to determine the TFFs, which are parameterized with respect to the slope at $q^2=0$.
The slope parameter $\Lambda^2$ is obtained by fitting the data with the model parameterization\footnote{The pseudoscalar TFF $F_{P\gamma^*\gamma^*}$ is formally defined in \cref{eq:def-PTFF}.}
\begin{equation}
   |F_{P\gamma^*\gamma^*}(q_1^2,0)|^2=
   \frac{\Lambda^2(\Lambda^2+\gamma^2)}{(\Lambda^2-q^2)^2+\Lambda^2\gamma^2}\,, \label{eq:TFF1}
\end{equation}
where $\Lambda$ and $\gamma$ can be interpreted as the mass and width of a contributing effective vector meson, in the context of a vector-meson-dominance (VMD) picture. In case of the $\eta^\prime$, the full parameterization is needed to properly describe the data with $\Lambda^2_{\eta^\prime} = 0.6432(56)(64)\GeV^2$ and $\gamma^2_{\eta^\prime} = 0.0128(11)(2)\GeV^2$, which is a four-times improved accuracy compared to the previously reported first observation~\cite{BESIII:2015zpz}. The slope parameter of the $\eta$ TFF is obtained by setting $\gamma_\eta=0$ as $\Lambda^2_\eta = 0.561(20)(5)\GeV^2$. Compared to the previously reported results of the A2~\cite{Adlarson:2016hpp} and NA60~\cite{NA60:2016nad} collaborations, the BESIII result tends towards smaller values, but it is compatible within errors. Currently, a new measurement of the $\eta$ TFF is prepared at BESIII using $\eta^\prime$ decays as source of $\eta$ mesons, which are more copiously produced. The different environment allows for a cross-check with similar statistical, but different systematic uncertainties. It should be mentioned that the timelike TFF has a cusp at the opening of the $\pi^+\pi^-$ channel corresponding to $q^2=4M_\pi^2$. The effect of the cusp is not included in \cref{eq:TFF1}, which is a source of bias in the determination of the slopes.
The same data set of inclusive $J/\psi$ decays also allowed one to observe the double Dalitz decay of the $\eta^\prime$ at BESIII with a significance of $5.7\sigma$ and to determine the branching fraction for the first time~\cite{BESIII:2022cul}.

The CMS collaboration reported the branching ratio of the double Dalitz decay into muon pairs normalized to the decay into a single muon pair $\Gamma(\eta\to2\mu^+2\mu^-)/\Gamma(\eta\to\mu^+\mu^-)=0.86(14)(12)\times10^{-3}$~\cite{CMS:2023thf}. The result can be related to the double off-shell $\eta$ TFF~\cite{Petri:2010ea}.

Additional information on doubly-virtual TFFs through the investigation of dilepton decays of mesons is provided by the BESIII collaboration in the measurement of the direct production of the axial-vector state $\chi_{c1}$ in $e^+e^-$ annihilation~\cite{BESIII:2022mtl}. The meson production is established by exploiting the interference pattern of the decay particles and the radiative dimuon continuum predicted in Ref.~\cite{Czyz:2016xvc} with a $5.1\sigma$ significance. The interference method allowed for the first measurement of nonvector meson production at $e^+e^-$ colliders with an electronic width of $\chi_{c1}$ of $\Gamma_{ee}=(0.12^{+0.13}_{-0.08})\,\text{eV}$, and is currently tested to provide equivalent information on further mesons.

\subsubsection{Further results from meson decays}

The decay of $\eta^\prime$ into four pions was prioritized as one of the required inputs for the development of a dispersive framework of $\eta$ and $\eta^\prime$ TFFs in WP20. The BESIII collaboration reported not only an upper limit of $\eta^\prime\to4\pi^0$, but also differential information of the decay $\eta^\prime\to2\pi^+2\pi^-$~\cite{BESIII:2023ceu}. An amplitude analysis is performed based on the work of Ref.~\cite{Guo:2011ir} and finds the coupling constants to be in agreement with their assumptions.

Similarly relevant to the construction of a dispersive framework for the $\eta^\prime$ TFF is the decay into a pion pair and a lepton pair. First studies reported by the BESIII collaboration~\cite{BESIII:2020otu,BESIII:2020elh} were updated to the fully available data set of $J/\psi$ decays~\cite{BESIII:2024awu}. Using different VMD models, the slope parameter of the TFF is determined from the differential decay distributions as $\Lambda^2_{\eta^\prime} = 0.77(11)\GeV^2$. Furthermore, a test for unconventional $CP$ violation is performed based on the angular distributions of the pion and lepton decay planes.

\subsubsection{Hadronic cross sections}
Apart from information on meson decays also  hadronic cross sections in $e^+e^-$ collisions are relevant for the development of dispersive frameworks of the pseudoscalar meson TFFs. Predominantly in the context of the HVP contribution new measurements of the total cross section of $e^+e^-\to\pi^+\pi^-\pi^0$ using initial state radiation are reported by the \babar{} and Belle-II collaborations~\cite{BABAR:2021cde,Belle-II:2024msd}. Additional information on intermediate states are reported from a partial-wave analysis performed by the BESIII collaboration on high-statistics data points of an energy scan measurement at CM energies between $2.0\GeV$ and $3.08\GeV$~\cite{BESIII:2024okl}. The three-pion system is found to be dominated by $\rho\pi$, $\omega\pi$, $\rho(1450)\pi$, $\rho(1700)\pi$, and $\rho_3(1690)\pi$.

An equivalent study of the $e^+e^-\to\pi^+\pi^-\eta$ cross section is performed on the same data~\cite{BESIII:2023sbq}. The intermediate states are described by dominant contributions of $\rho\eta$ and $a_2(1320)\pi$. Further significant contributions of higher resonant states are identified.
Additional results on the total cross sections of the related processes $e^+e^-\to\omega \pi^0/\eta^{(\prime)}$ and $e^+e^-\to\pi^+\pi^-\eta^\prime$ are also reported based on the same data~\cite{BESIII:2020xmw,BESIII:2024qjv,BESIII:2020kpr}.

\subsection{Short-distance constraints}
\label{sec:SDC}

SDCs can be derived to address the HLbL contributions to $a_\mu$ in kinematical regions where two or three photon virtualities $Q_{1}^2$, $Q_{2}^2$, and $Q_{3}^2$ become large. These constraints are obtained by applying OPEs to the EM quark currents that define the HLbL tensor~\cite{Melnikov:2003xd,Bijnens:2019ghy,Bijnens:2020xnl,Bijnens:2021jqo,Bijnens:2022itw,Bijnens:2024jgh}. In this section we review these constraints, in particular what has been done since WP20.

In the $a_\mu$ kinematics, the external momentum $q_4$ is always soft, but the other three vary in the integration over virtualities. SDCs can be obtained in two regions
\begin{enumerate}
    \item Pure short-distance region: $Q_{i}^2\gg \Lambda^2 _{\text{QCD}}$,
    \item Mixed, corner, or Melnikov--Vainshtein (MV), region: $Q_{i}^2,Q_{j}^2\gg Q_{k}^2,\Lambda _{\text{QCD}}^2$.
\end{enumerate}
In the underlying correlation function with four EM currents, the above limits respectively correspond to the cases where three and two currents are close in position space, which is how the OPE is applied.
Here it should be noted that there is no inherent hierarchy on $Q_{k}^2$ and $\Lambda _{\text{QCD}}^2$ in the mixed region, but if one imposes $Q_{k}^2\gg \Lambda _{\text{QCD}}^2$, there is partial overlap between the two regions. Below we discuss them separately. Additionally, other kinds of SDCs also exist, such as the Brodsky--Lepage constraint on form factors~\cite{Lepage:1979zb,Brodsky:1981rp}. These are not further discussed in this section.

\subsubsection{Pure short-distance region}
\begin{figure}[t]
\begin{center}
\includegraphics[width=0.18\textwidth]{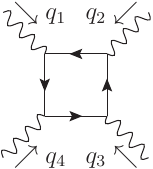}
\caption{The perturbative quark loop. Figure taken from Ref.~\cite{Bijnens:2022itw}.}
\label{fig:pertQLoop}
\end{center}
\end{figure}

Whenever the rest of Euclidean momenta entering into the problem are large, the external soft photon at $q_4\rightarrow 0$ can be treated as a background field for the OPE~\cite{IOFFE1984109,Balitsky:1983xk}. This was used for the electroweak contribution to $a_\mu$ in Ref.~\cite{Czarnecki:2002nt}. In Refs.~\cite{Bijnens:2019ghy,Bijnens:2020xnl} it was shown that the HLbL tensor in $g-2$ kinematics can be directly related to a three-point function with the external field in the background captured in an external state,
\begin{equation}
\label{eq:backdyson1}
\Pi ^{\mu_{1} \mu_{2} \mu_{3} }(q_{1},q_{2}) = -\frac{1}{e}\int\frac{d^4 q_{3}}{(2\pi)^4}  \left(\prod_{i=1}^{3}\int d^{4}x_{i}\, e^{-i q_{i} x_{i}}\right)  \langle 0 | T\left(
\prod_{j=1}^{3}J^{\mu_{j}}(x_{j})
\right)
| \gamma(q_4) \rangle \, .
\end{equation}
One may perform a well-defined and systematic OPE at this stage~\cite{Bijnens:2019ghy}. The leading operator is $F_{\mu\nu}$. Its contribution corresponds to the massless quark loop depicted in \cref{fig:pertQLoop}, where $q_4$ corresponds to the soft photon, $q_4\to 0$.

The leading power correction, suppressed by the quark mass, was addressed in Ref.~\cite{Bijnens:2019ghy}. In Ref.~\cite{Bijnens:2020xnl} the OPE was further developed to study the subleading contributions beyond the perturbative quark loop, through orders $g_{s}\,\frac{\Lambda_{\text{QCD}}^4}{Q^4}$, $\frac{m_q^2}{Q^2}$, $g_s^2 \,\frac{\Lambda_{\text{QCD}}^4}{Q^4}$,  $m_q\frac{\Lambda_{\text{QCD}}}{Q^2}$, $m_q\frac{\Lambda_{\text{QCD}}^3}{Q^4}$, and $m_q^3\frac{\Lambda_{\text{QCD}}}{Q^4}$. The appearing operators are
\begin{align}
Q_{1,\,\mu\nu}& =  e\,   e_{q}  F_{\mu\nu} \,  ,
\qquad
Q_{2,\,\mu\nu} =   \bar{q}\sigma_{\mu\nu}q   \,
\qquad
Q_{3,\,\mu\nu}=  i \,  \bar{q} G_{\mu\nu}q  \, ,
\qquad
Q_{4,\,\mu\nu}=  i \, \bar{q} \bar{G}_{\mu\nu}\gamma_{5} q \, ,\notag
\\
Q_{5,\,\mu\nu}& =   \bar{q} q\; e\,   e_{q}F_{\mu\nu} \, ,
\qquad
Q_{6,\,\mu\nu}  = \frac{\alpha_{s}}{\pi}\, G_ {a}^{\alpha\beta}G^{a}_{\alpha\beta}\; e\,   e_{q}F_{\mu\nu}  \, ,\notag
\\
Q_{7,\,\mu\nu}&=  \bar{q}(G_{\mu\lambda}D_{\nu}+D_{\nu}G_{\mu\lambda})\gamma^{\lambda}q-(\mu\leftrightarrow\nu) \, ,
\qquad
Q_{\{8\},\,\mu\nu}= \alpha_{s}\, (\bar{q}\, \Gamma \,q \; \bar{q}\Gamma q)_{\mu\nu} \, .
\label{eq:HLbLoperators}
\end{align}
Here $\sigma _{\mu \nu} = i/2\, [\gamma _\mu , \gamma _\nu]$, $G_{\mu \nu}$ is the gluon field strength tensor, $\bar{G}_{\mu \nu}$ its dual and the $Q_{\{8\},\,\mu\nu}$ is a class of four-quark operators with Dirac matrix structure $\Gamma$. While the massless quark loop provides the leading contribution through the $F_{\mu\nu}$ operator, the leading quark-mass correction is not driven by the (massive) quark loop. Instead, it is linear in $m_q$ and mediated by the $\bar{q}\sigma^{\mu\nu}q$ operator. The operators in~\cref{eq:HLbLoperators} are defined in the $\overline{\text{MS}}$ scheme. The numerical estimates for their nonperturbative condensates labeled $X_i$ are based on various methods and approximations as discussed in Ref.~\cite{Bijnens:2020xnl} and given in~\cref{tab:numSD1}. 

With an OPE in place for the HLbL tensor in the corresponding kinematic regime, the scalar functions in its Lorentz decomposition, $\hat{\Pi}_i$, can be determined using projection techniques~\cite{Bijnens:2020xnl}.
One may then use them to compute the short-distance regions of the $a_\mu^{\text{HLbL}}$ integral, $Q_{1,2,3}>Q_{\text{min}}$. The numerical impact on $a_\mu^{\text{HLbL}}$ of the associated terms in the OPE for two values of $Q_{\textrm{min}}$ is shown in~\cref{tab:numSD1}. The massless perturbative quark loop corresponds to contribution $X_{1,0}$, and is clearly dominating over all other corrections. It should be emphasized that also those contributions that are not suppressed by the light-quark masses are negligible at the current level of precision. 

\begin{table}[t]
\begin{center}
\small
\renewcommand{\arraystretch}{1.1}
\begin{tabular}{llrr}\toprule
   Contribution
   & Inputs $[\text{GeV}]$
   & $Q_{\text{min}}=1 \GeV$
   & $Q_{\text{min}}=2 \GeV$
   \\ \midrule
   $X_{1,0}$
   &
   & $1.73\times 10^{-10}$
   & $4.35\times 10^{-11}$
   \\
   $X_{1,m^2}$
   &
   & $-5.7 \times 10^{-14}$
   & $-3.6 \times 10^{-15}$
   \\
   $X_{2,m}$
   & $X_2=-4\times 10^{-2}$
   & $-1.2\times 10^{-12}$
   & $-7.3\times 10^{-14}$
   \\
   $X_{2,m^3}$
   & $X_2=-4\times 10^{-2}$
   & $6.4\times 10^{-15}$
   & $1.0\times 10^{-16}$
   \\
   $X_{3}$
   & $X_3=3.51\times 10^{-3}$
   & $-3.0\times 10^{-14}$
   & $-4.7\times 10^{-16}$
   \\
   $X_{4}$
   &  $X_4=3.51\times 10^{-3}$
   & $3.3\times 10^{-14}$
   & $5.3\times 10^{-16}$
   \\
   $X_{5}$
   & $X_5=- 1.56 \times 10^{-2} $
   & $-1.8\times 10^{-13}$
   & $-2.8\times 10^{-15}$
   \\
   $X_{6}$
   & $X_6=2\times 10 ^{-2}$
   & $1.3 \times 10^{-13}$
   & $2.0\times 10^{-15}$
   \\
   $X_{7}$
   &  $X_7=3.33\times 10 ^{-3} $
   & $9.2\times 10^{-13}$
   & $1.5\times 10^{-14}$
   \\
   $X_{8,1}$
   & $\overline{X}_{8,1}=-1.44\times 10^{-4}$  & $3.0\times 10^{-13}$             &    $4.7\times 10^{-15}$         \\
$X_{8,2}$    & $\overline{X}_{8,2}=-1.44\times 10^{-4} $ &   $-1.3\times 10^{-13}$              &  $-2.0\times 10^{-15}$ \\
\bottomrule
\end{tabular}\end{center}
 \caption{\label{tab:numSD1}Numerical contribution to $a_{\mu}^{\text{HLbL}}$ for the indicated inputs. The $X_i$ correspond to the operators in \cref{eq:HLbLoperators} and the order of quark masses is indicated as well. The input values for the condensates are described in detail in Ref.~\cite{Bijnens:2020xnl}. Table from 
 Ref.~\cite{Bijnens:2020xnl}.}
 \renewcommand{\arraystretch}{1}
\end{table}

\begin{figure}[t]
\begin{center}
\includegraphics[width=0.45\textwidth]{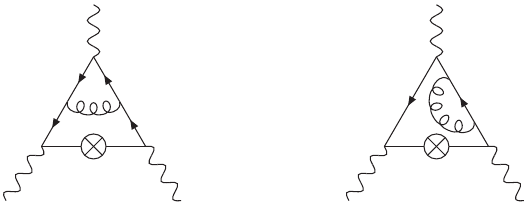}
\caption{Gluonic corrections to the perturbative quark loop, with the soft external field denoted by a crossed vertex. Figures taken from Ref.~\cite{Bijnens:2021jqo}.}
\label{fig:pertGluons}
\end{center}
\end{figure}

\begin{figure}[t]
\begin{center}
\includegraphics[width=0.49\textwidth]{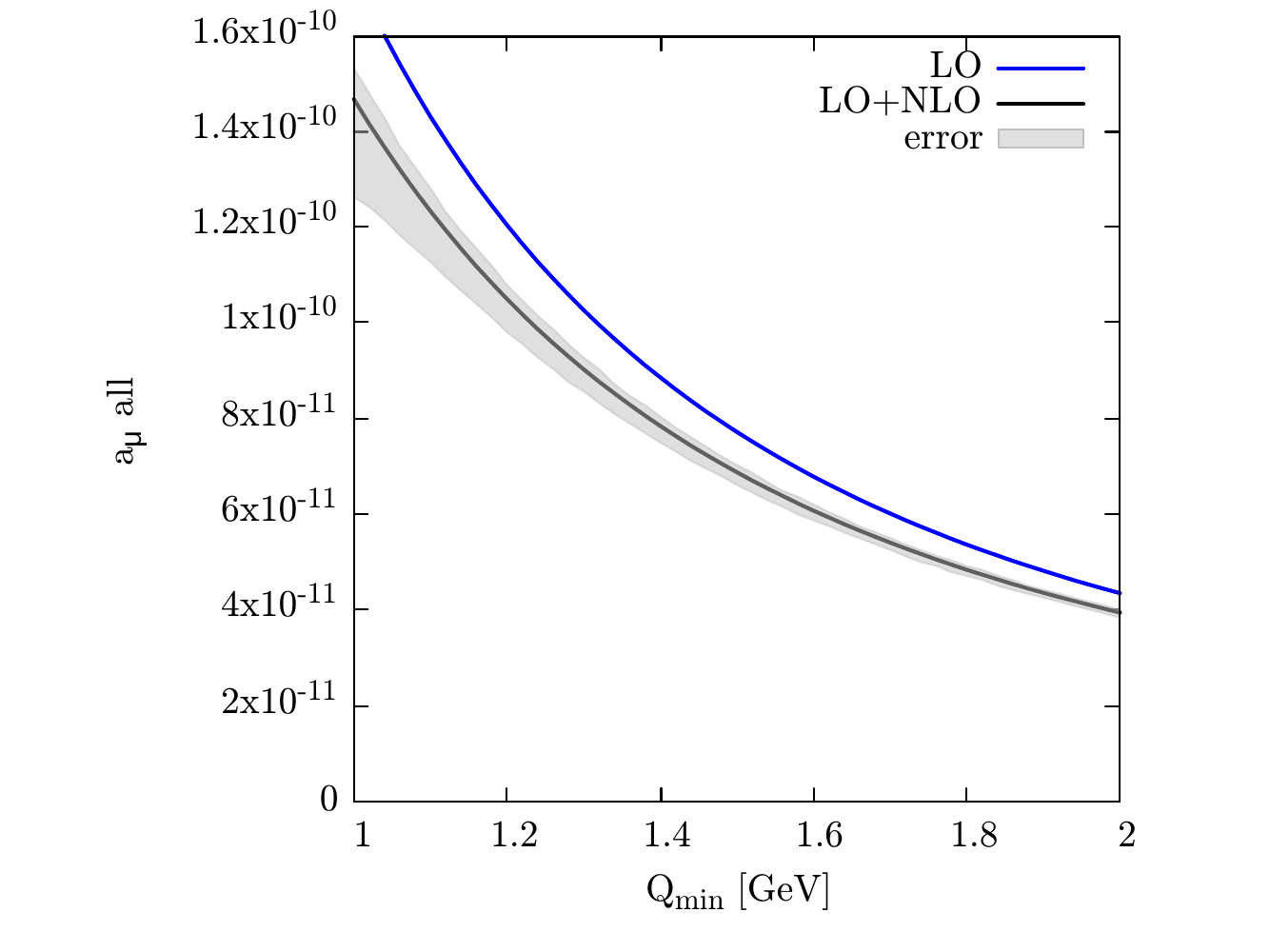}\includegraphics[width=0.49\textwidth]{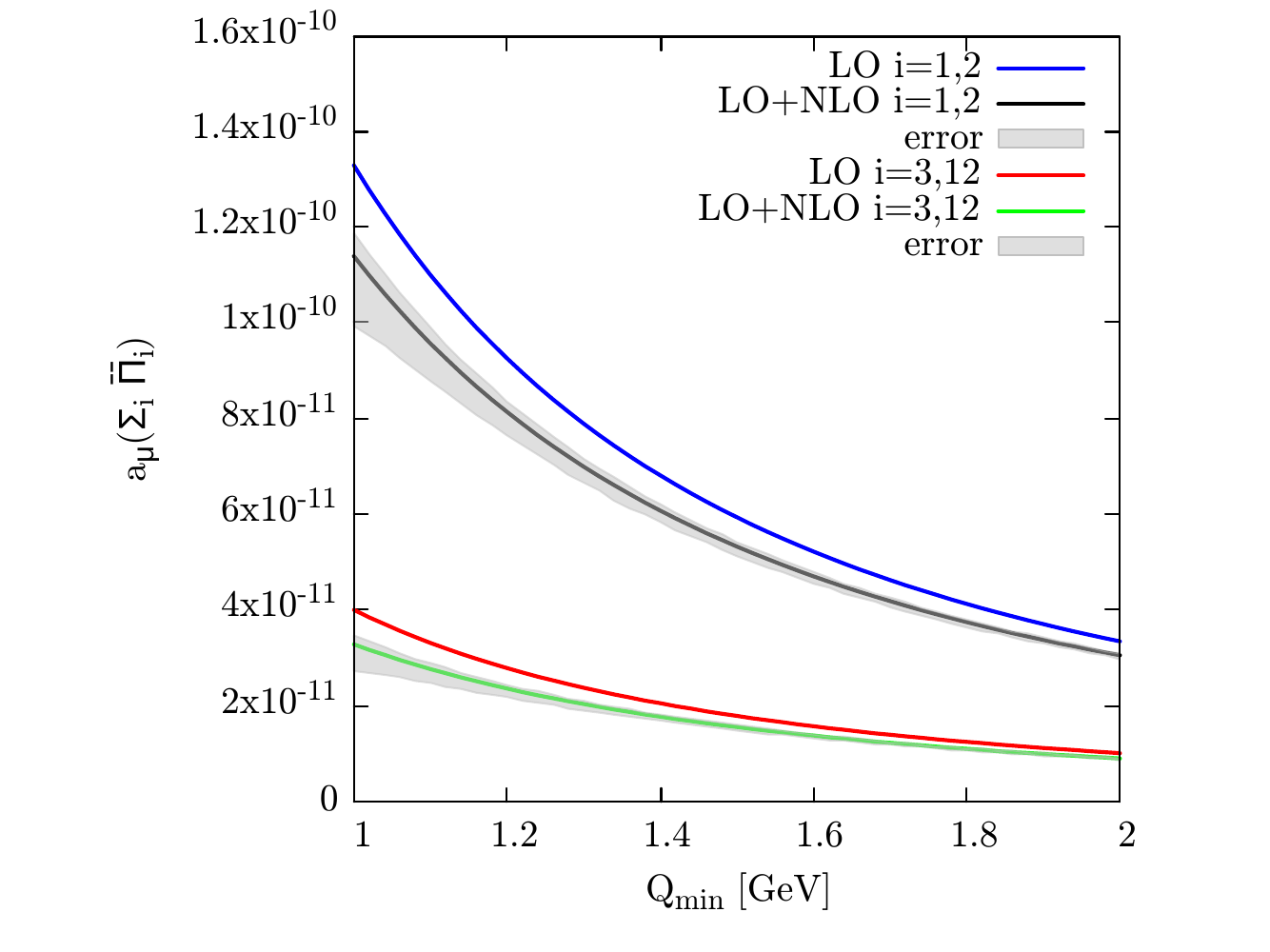}
\caption{Short-distance contribution to HLbL as a function of the cutoff on the three virtualities. Left: the full contribution. Right: separated in the longitudinal $\overline\Pi_{i=1,2}$ and transverse $\overline\Pi_{i=3,12}$ amplitudes. The blue line is the massless quark loop (LO). The full line is the quark loop including gluonic corrections (LO+NLO) and the shaded region the uncertainty due to varying the scale of evaluating $\alpha_s$.
We used the value $\alpha_s(1.5\GeV)=0.3501$ and varied the scale of $\alpha_s$ from $Q_\text{min}/\sqrt{2}$ to $\sqrt{2} Q_\text{min}$. The corresponding error becomes large below about $1.2\GeV$ and may be slightly overoptimistic for low $Q_\text{min}$ values depending on the size of higher-order corrections.
}
\label{fig:HLbLSDgluon}
\end{center}
\end{figure}

Another possibility for large corrections are the gluonic corrections to the massless quark loop, i.e., the leading term in the OPE. These have been worked out fully analytically in Ref.~\cite{Bijnens:2021jqo}. This perturbative two-loop calculation consists in adding all possible gluonic corrections to the massless perturbative quark loop as in \cref{fig:pertGluons}, and reducing them to a set of six known master integrals~\cite{Usyukina:1994iw,Chavez:2012kn,Bijnens:2021jqo}.
The main conclusion is that the gluonic corrections are of the order of $-10\%$ depending on the value of $\alpha_s$ used. In \cref{fig:HLbLSDgluon} we show the massless quark loop contribution and the gluonic corrections as a function of the lower cutoff $Q_{\text{min}}$ on the three virtualities, $Q_{i}>Q_{\text{min}}$. The uncertainty indicated is from varying $\alpha_s$~\cite{Bijnens:2021jqo}.

The gluonic corrections have been partially included in model implementations. Especially in the holographic method as discussed in \cref{Section:HolographicModels} it allowed for a significantly improved matching.
In WP20 it was only known that the perturbative quark loop was the leading contribution and the directly subleading term was negligible, as had been shown in Ref.~\cite{Bijnens:2019ghy}. From Refs.~\cite{Bijnens:2020xnl,Bijnens:2021jqo} it has now been established that the perturbative quark loop is a good representation at the current level of precision up to $10\%$ corrections from perturbative gluonic corrections.

\begin{figure}[t]
\begin{center}
\includegraphics[width=0.8\textwidth]{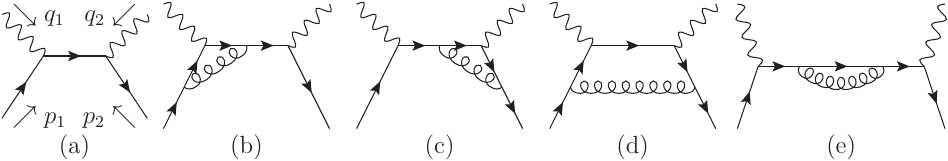}
\caption{Diagrammatic depiction of the two-current OPE in the mixed MV kinematical region. Here the particular case $Q_1^2,Q_2^2\gg Q_3^2,\Lambda ^2_{\text{QCD}}$ was chosen. Figures taken from Ref.~\cite{Bijnens:2024jgh}.}
\label{fig:HLbLMVOPE}
\end{center}
\end{figure}

\subsubsection{Mixed region (Melnikov--Vainshtein)}
In the mixed kinematical region there are two currents close in the underlying HLbL function. This momentum configuration was first studied in Ref.~\cite{Melnikov:2003xd}, and after WP20 this work was extended in Refs.~\cite{Bijnens:2022itw,Bijnens:2024jgh}. Following the notation of the latter references, for the specific case $Q_1^2,Q_2^2\gg Q_3^2,\Lambda ^2_{\text{QCD}}$ the HLbL tensor can be written as
\begin{equation}\label{eq:twopoint}
\Pi^{\mu_{1}\mu_{2}\mu_3\mu_4}=
\sum_{j,k}\frac{i e_{q_j}e_{q_k}}{e^{2}}\int \frac{d^{4}q_4}{(2\pi)^4}\int d^{4}x_1\int d^{4}x_2\, e^{-i(q_1 x_1+q_2 x_2)}
\langle 0 |T\{ J_j^{\mu_1}(x_1)J_k^{\mu_2}(x_2)\}|\gamma^{\mu_3}(q_3)\gamma^{\mu_4}(q_4) \rangle \, .
\end{equation}
The OPE between the two above currents at tree level can be diagrammatically depicted as in \cref{fig:HLbLMVOPE}(a), and including gluonic corrections as in \cref{fig:HLbLMVOPE}(b)--(e). Additional topologies corresponding to operators made out of photons and gluons start at $D=4$.
The large Euclidean scale in the expansion is $\hat{Q} = \sqrt{-\hat{q}^2}$, where $\hat{q} = (q_1-q_2)/2$. The leading term (dimension $D=3$) studied in Ref.~\cite{Melnikov:2003xd} enters as $1/\hat{Q}^2$, and is directly related to the axial current with two nonperturbative form factors, $\omega _L$ and $\omega _T$. It was also pointed out that the nonrenormalization theorems for the axial anomaly have significant consequences for $g-2$. The implementation was discussed extensively in WP20 and also more recently, see Refs.~\cite{Colangelo:2019lpu,Colangelo:2019uex,Melnikov:2019xkq,Leutgeb:2019gbz,Ludtke:2020moa,Knecht:2020xyr,Masjuan:2020jsf,Cappiello:2021vzi,Colangelo:2021nkr,Danilkin:2021icn,Leutgeb:2021bpo,Leutgeb:2021mpu,Zanke:2021wiq,Leutgeb:2022lqw}. An update on the implementation in the different approaches can be found in the following, see \cref{Section:Dispersive}, \cref{Section:HolographicModels}, and \cref{Section:OtherApproaches}.

The extension of the leading-order result to dimension $D=4$ in Refs.~\cite{Bijnens:2022itw,Bijnens:2024jgh} leads to a better understanding of the power counting and of the structure of perturbative and nonperturbative corrections to the HLbL tensor in the studied kinematic regime. Additionally, new constraints on the scalar functions determining the HLbL tensor were derived. Through dimension $D=4$ the independent operators are~\cite{Bijnens:2022itw,Bijnens:2024jgh}
\begin{align}
			D=3: & \quad \mathcal{O}_{q3}^{\alpha \beta \gamma} =  \overline{q} \Big[
			\gamma ^{\alpha}\gamma ^{\beta} \gamma ^{\gamma}
			-
			\gamma ^{\gamma}\gamma ^{\beta} \gamma ^{\alpha}
			\Big] q
     \, ,\notag
	\\
			D=4: &  \quad \mathcal{O}_{q4}^{\alpha \beta} = \overline{q}
			\gamma ^{\beta}
			\Big[
			\overrightarrow D^\alpha
   -
   \overleftarrow D^\alpha
			\Big] q
     \, ,
			\notag\\
			&  \quad \mathcal{O}_{FF,1}^{\alpha  \beta} = F^{\alpha \gamma}F_{\gamma }^{\, \,  \beta}
     \, ,
            \qquad \mathcal{O}_{FF,2}^{\alpha  \beta} = F^{\gamma \delta }F_{\gamma \delta} \, g^{\alpha \beta}
     \, ,
            \qquad \mathcal{O}_{F} = F\times F
     \, ,
			\notag\\
			&
            \quad \mathcal{O}_{GG,1}^{\alpha  \beta} = G^{\alpha \gamma}G^{\, \, \beta}_{\gamma}
    \, ,
            \qquad \mathcal{O}_{GG,2}^{\alpha  \beta} = G^{\gamma \delta }G_{\gamma \delta} \, g^{\alpha \beta}
     \, ,
            \qquad \mathcal{O}_{G} = G\times G
   \, .
\end{align}
The operators $\mathcal{O}_{F}$ and $\mathcal{O}_G$ represent classes of operators with Lorentz structure different from $\mathcal{O}_{FF,i}^{\alpha \beta}$ and $\mathcal{O}_{GG,i}^{\alpha \beta}$~\cite{Bijnens:2024jgh}. In Ref.~\cite{Bijnens:2024jgh} also gluonic corrections to the OPE were included as well as the corresponding renormalization through dimension $D=4$. This in particular proved the conjecture of Ref.~\cite{Ludtke:2020moa} that at $D=3$ the gluonic corrections enter as $-\alpha_s/\pi$ times the original result of Ref.~\cite{Melnikov:2003xd}. It was also explicitly checked that in the overlap region between OPEs, where $Q_3^2\gg \Lambda _{\text{QCD}}^2$, there is analytic agreement between derived scalar functions $\hat{\Pi}_i$ through the relevant orders. The current limits of predictivity of the OPE in the three mixed regions for the $\hat{\Pi}_i$ are shown in \cref{tab:truncorder} (rewritten in terms of the corner variable $\overline{Q}_k = Q_i+Q_j $).

\begin{table}[t]\centering
\small
\renewcommand{\arraystretch}{1.1}
\begin{tabular}{cccc}
\toprule
& $\overline{Q}_1$ & $\overline{Q}_2$ & $\overline{Q}_3$
 \\
 \midrule
$\hat{\Pi}_1$  & 5 & 5 & 4 \\
$\hat{\Pi}_4$  & 3 & 3 & 5 \\
$\hat{\Pi}_7$  & 4 & 5 & 6 \\
$\hat{\Pi}_{17}$ & 5 & 5 & 5 \\
$\hat{\Pi}_{39}$ & 5 & 5 & 5 \\
$\hat{\Pi}_{54}$ & 5 & 5 & 5
\\ \bottomrule
\end{tabular}
\caption{\label{tab:truncorder}The leading power $n$ of $1/ \overline{Q}_k^{n}$, where the OPE through $D=4$ lacks prediction. Table from Ref.~\cite{Bijnens:2024jgh}.}
\renewcommand{\arraystretch}{1}
\end{table}

Although there in principle is a plethora of different nonperturbative form factors at dimension $D=4$, it was  found in Ref.~\cite{Bijnens:2024jgh} that, when incorporating the different components of the HLbL tensor into the $g-2$ integrand, a very strong cancellation occurs for all computed contributions. This suggests that, up to chiral corrections, the leading term of the integrand in the corresponding kinematic regime may be determined by the axial current form factors $\omega _L$ and $\omega _T$.

\begin{figure}[t]
\begin{center}
\includegraphics[width=0.6\textwidth]{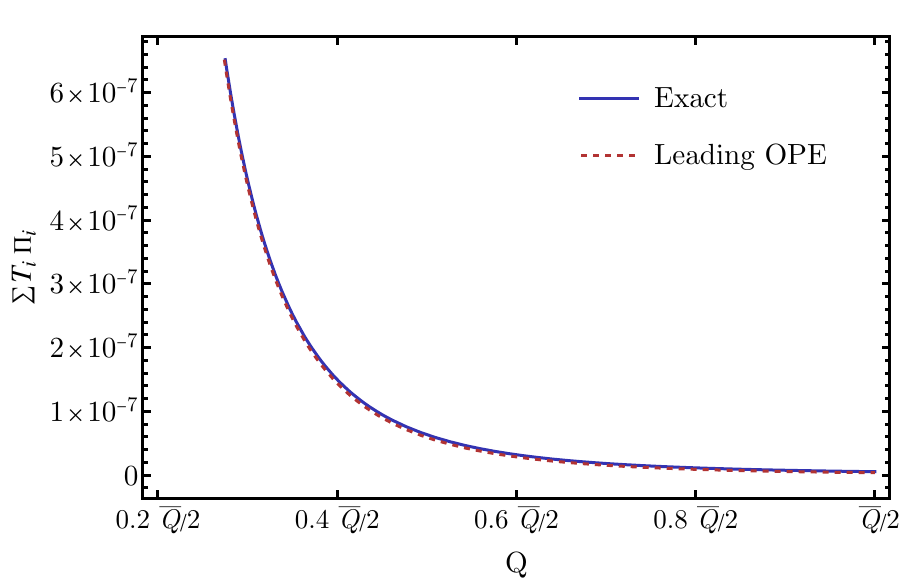}
\caption{Numerical comparison between the quark loop and OPE in the MV region. Figure adapted from Ref.~\cite{Bijnens:2024jgh}.}
\label{fig:MVOPENum}
\end{center}
\end{figure}

To assess how well the leading OPE contribution works, Ref.~\cite{Bijnens:2024jgh} compared the derived expansion through NLO for the sum $\sum T_{i}\bar{\Pi}_i$ in the $g-2$ integral with the LO one given by the quark loop. Since the NLO result depends on the nonperturbative form factors $\omega _L$ and $\omega _T$, the perturbative limit was taken where $2\, \omega _T = \omega _L = -2/Q_{i}^2$ in the region where $Q_i^2$ is the small scale. At fixed large scale $\bar{Q}_i=10\GeV$ and corner variable $y_{jk} = (Q_j-Q_k)/Q_i = 0$ one can study $\sum T_{i}\bar{\Pi}_i$ by varying the small scale $Q_i$ in units of $\bar{Q}_i/2$. This is shown in \cref{fig:MVOPENum} for $y_{jk}=0$. Overall, there is good agreement up to regions relatively close to the symmetric point, where $Q_i$ tends towards $\bar{Q}_i/2$.

\subsection{Dispersive approach}
\label{Section:Dispersive}

Within the dispersive framework~\cite{Colangelo:2014dfa,Colangelo:2014pva,Colangelo:2015ama,Colangelo:2017fiz,Hoferichter:2013ama,Pauk:2014rfa}, the different contributions to HLbL can be ordered according to the mass threshold in the unitarity relation. The lightest intermediate state is a single neutral pion, followed by two charged pions, etc. This ordering turns out to be roughly reflected in the numerical dominance of the individual contributions to $a_\mu^\text{HLbL}$. Single-pseudoscalar contributions are discussed in \cref{sec:Pseudoscalars} and two-particle intermediate states in \cref{sec:TwoMesonContributions}. A rigorous dispersive description of three-particle intermediate states would be challenging---this contribution is assumed to be negligible unless resonantly enhanced. It is therefore described in terms of single-resonance contributions in the narrow-width limit, see \cref{sec:ResonanceContributions}. The infinite sum over hadronic intermediate states needs to saturate the SDCs discussed in \cref{sec:SDC}. In practice, only a finite number of intermediate states can be related to data input, hence the remainder typically involves some modeling, which is however controlled by the matching to the SDCs, see \cref{sec:MatchingToSDCs}.

\subsubsection{Pseudoscalars}
\label{sec:Pseudoscalars}

In WP20~\cite{Aoyama:2020ynm}, the largest individual contribution to HLbL, the $\pi^0$ pole, was determined from three independent approaches: dispersively~\cite{Hoferichter:2018dmo,Hoferichter:2018kwz}, using Canterbury approximants (CA)~\cite{Masjuan:2017tvw}, and on the lattice~\cite{Gerardin:2019vio}.  For the subleading $\eta$- and $\eta'$-pole contributions, however, the consensus relied only on the determination from Canterbury approximants~\cite{Masjuan:2017tvw}, as dispersion-theoretical or lattice-QCD calculations had not been completed. Since then, progress has been made on both latter approaches, see \cref{sec:HLbLexcl_state_LAT} for that on the lattice. Meanwhile, a dedicated effort to evaluate the $\eta^{(\prime)}$-pole contributions with dispersion relations led to their first determination~\cite{Holz:2024lom,Holz:2024diw}, which requires the dispersive reconstruction of the pertinent doubly-virtual TFFs~\cite{Stollenwerk:2011zz,Hanhart:2013vba,Kubis:2015sga,Holz:2015tcg,Holz:2022hwz,Holz:2022smu} as input, extending a similar program for the $\pi^0$ TFF~\cite{Schneider:2012ez,Hoferichter:2012pm,Hoferichter:2014vra,Hoferichter:2018dmo,Hoferichter:2018kwz,Hoferichter:2021lct}. The pseudoscalar TFFs describe the decay $P(q_1+q_2)\to\gamma^*(q_1,\mu)\gamma^*(q_2,\nu)$, formally defined by the matrix elements
\begin{equation}
i \int  d^4x\, e^{iq_1 \cdot x} \langle 0 | T\{ j_{\mu}(x)j_{\nu}(0) \}  | P(q_1+q_2) \rangle =\epsilon_{\mu\nu\rho\sigma}q_1^{\rho}q_2^{\sigma} F_{P\gamma^*\gamma^*}(q_1^2,q_2^2)\,, \label{eq:def-PTFF}
\end{equation}
where the normalization $F_{P\gamma^*\gamma^*}(0,0)\equiv F_{P\gamma\gamma}$ is determined by the $P\to\gamma\gamma$ decay rate.

Such a reconstruction, based on the low-lying singularities, data input for $\eta^{(\prime)}\to\gamma\gamma$, $\eta^{(\prime)}\to\pi^+\pi^-\gamma$, $\eta^{(\prime)}\to2(\pi^+\pi^-)$, and $e^+ e^- \to e^+ e^- \eta^{(\prime)}$, as well as
an implementation of asymptotic constraints in the dispersive representations of $\eta$ and $\eta'$ transition form factors comparable to Refs.~\cite{Hoferichter:2018dmo,Hoferichter:2018kwz} was recently concluded~\cite{Holz:2024lom,Holz:2024diw}. This form factor representation reads
\begin{equation}
   F_{\eta^{(\prime)} \gamma^*\gamma^*} = F_{\eta^{(\prime)}}^{(I=1)} + F_{\eta^{(\prime)}}^{(I=0)} + F_{\eta^{(\prime)}}^{\text{eff}} + F_{\eta^{(\prime)}}^{\text{asym}}\,. \label{Eq:tff_compl}
\end{equation}
In contradistinction to the $\pi^0$ case, the final-state photons here carry either both isovector ($I=1$) or both isoscalar ($I=0$) quantum numbers, leading to the two independent terms in \cref{Eq:tff_compl}. The first of these, the isovector dispersive piece, dominates the TFFs at low energies and can be cast in the form of a double-spectral representation
\begin{equation}
    F_{\eta^{(\prime)}}^{(I=1)}(-Q_1^2,-Q_2^2) = \frac{1}{\pi^2} \int_{4M_\pi^2}^{\Lambda^2}d x\, d y\,  \frac{ \rho_{\eta^{(\prime)}}(x,y)}{(x+Q_1^2)(y+Q_2^2)} + (Q_1 \leftrightarrow Q_2)\,, \label{Eq:TFF_I1}
\end{equation}
with the double-spectral density
\begin{equation}
     \rho_{\eta^{(\prime)}}(x,y) = \frac{x \sigma_\pi^3(x)}{192 \pi} \operatorname{Im} \Big\{\big[F_\pi^V(x)\big]^*  \mathcal{F}_{\eta^{(\prime)} \pi\pi\gamma^*}(x,y)\Big\}\,,
\end{equation}
where $\sigma_\pi(s)=\sqrt{1-4M_\pi^2/s}$, and $F_\pi^V$ is the pion vector form factor (VFF).
$\mathcal{F}_{\eta^{(\prime)} \pi\pi\gamma^*}(t,k^2)$ is the $P$-wave amplitude for $\eta^{(\prime)} \gamma^*(k^2) \to \pi^+\pi^-$. It is constructed based on dispersive representations of $\eta^{(\prime)} \to 2(\pi^+\pi^-)$ amplitudes (cf.\ Ref.~\cite{Guo:2011ir}) with the approximation of taking only pairwise rescattering of $\pi^+\pi^-$ into account. In addition, factorization-breaking effects are included by considering the exchange of $a_2(1320)$ tensor mesons as a left-hand-cut contribution, which leads to a description by means of an inhomogeneous Omn\`es representation. The subtraction constants of these representations, which are subject to constraints from ChPT, are eventually fixed by fits to $\eta^{(\prime)} \to \pi^+ \pi^- \gamma$ decay data~\cite{KLOE:2012rfx,BESIII:2017kyd}. For the description of $F_\pi^V$ an Omn\`es representation~\cite{Omnes:1958hv} is utilized. Due to technical reasons~\cite{Gasser:2018qtg}, in the solution of the inhomogeneous Omn\`es problem,
a representation of the $\pi\pi$ $P$-wave phase shift based on unitarized ChPT~\cite{Dobado:1989qm,Truong:1991gv,Dobado:1992ha,Niehus:2020gmf} is employed. The resulting representation of $F_\pi^V$ is fixed by fitting to data of $\tau^- \to \pi^- \pi^0 \nu_\tau$~\cite{Belle:2008xpe}.

The isoscalar channel, on the other hand, is dominated by the narrow $\omega$ and $\phi$ resonances in the low-energy regime. Their (small) contribution is parameterized in a VMD ansatz,
 \begin{align}
     &F_{\eta^{(\prime)}}^{(I=0)}(-Q_1^2,-Q_2^2) = \sum\limits_{V\in \lbrace \omega,\phi\rbrace} \frac{w_{\eta^{(\prime)} V \gamma} F_{\eta^{(\prime)} \gamma \gamma}  M_V^4}{(M_V^2+Q_1^2)(M_V^2+Q_2^2)}\,, \label{Eq:TFF_I0}
 \end{align}
 with $M_V$ as mass parameters and weight factors $w_{\eta^{(\prime)} V \gamma}$~\cite{Hanhart:2013vba,Gan:2020aco} determined from the respective decays widths of $\omega\to \eta \gamma$, $\eta' \to \omega \gamma$, $\phi\to\eta^{(\prime)} \gamma$, and $\omega,\, \phi \to e^+ e^-$~\cite{ParticleDataGroup:2024cfk}.

 The third term in \cref{Eq:tff_compl}, the effective-pole piece, parameterizes higher intermediate states. It serves two purposes: it ensures that a sum rule for the form factor normalization is exactly fulfilled, and additionally, helps to describe high-energy, singly-virtual data on $e^+ e^- \to e^+ e^- \eta^{(\prime)}$. Two distinct parameterizations have been implemented,
  \begin{align}
     F_{\eta^{(\prime)}}^{\text{eff}\,(A)}(-Q_1^2,-Q_2^2) &= \frac{g_\text{eff} F_{\eta^{(\prime)} \gamma \gamma}   M_{\text{eff}}^4}{(M_{\text{eff}}^2+Q_1^2)(M_{\text{eff}}^2+Q_2^2)}\,,\notag \\
     F_{\eta^{(\prime)}}^{\text{eff}\,(B)}(-Q_1^2,-Q_2^2) &= \sum\limits_{V\in \lbrace \rho',\rho''\rbrace}\frac{g_V F_{\eta^{(\prime)} \gamma \gamma} M_V^4}{(M_V^2+Q_1^2)(M_V^2+Q_2^2)}\,,
     \label{Eq:TFF_eff}
 \end{align}
 where in the variant $(A)$ the effective coupling $g_\text{eff}$ is used to restore the normalization. For both $\eta$ and $\eta'$, $|g_\text{eff}|$ is found in a range up to about $10\, \%$. In this variant, the effective mass parameter is fit to singly-virtual TFF data~\cite{CELLO:1990klc,CLEO:1997fho,L3:1997ocz,BaBar:2011nrp} and is found in the range $(1.3$--$2.2)\GeV$. On the other hand, variant $(B)$ describes the two poles of $\rho(1450) \equiv \rho'$ and $\rho(1700)\equiv \rho''$ resonances, with their mass parameters taken from Ref.~\cite{ParticleDataGroup:2024cfk}. One resonance coupling $g_V$ is used to fulfill the normalization sum rule, while the other is fit to the singly-virtual TFF data.

 The last term in \cref{Eq:tff_compl} is used to implement constraints from pQCD~\cite{Lepage:1979zb,Lepage:1980fj,Brodsky:1981rp}.
 As $\eta$ and $\eta'$ are much more massive than the $\pi^0$,
 the masses are explicitly accounted for in this representation. Starting from an asymptotic piece in the massless limit~\cite{Hoferichter:2018dmo,Hoferichter:2018kwz} based on a dispersive reformulation~\cite{Khodjamirian:1997tk} of the leading term of the light-cone sum rule expansion evaluated with asymptotic distribution amplitudes~\cite{Lepage:1979zb,Lepage:1980fj,Brodsky:1981rp,Novikov:1983jt,Nesterenko:1982dn,Gorsky:1987mu}, a form including mass effects can be written as~\cite{Zanke:2021wiq,Holz:2024lom,Hoferichter:2024bae,Holz:2024diw}
 \begin{align}
     &F_{\eta^{(\prime)}}^{\text{asym}}(q_1^2,q_2^2) = \frac{- \bar{F}_{\text{asym}}^{\eta^{(\prime)}}}{M_{\eta^{(\prime)}}^4} \int_{2 s_m}^{\infty}dv\, \bigg[ \frac{q_2^2}{v-q_1^2} \bigg(\frac{1}{v-q_1^2-q_2^2} - \frac{1}{q_1^2-q_2^2} \bigg) f^{\text{asym}}_{\eta^{(\prime)}}(v,q_1^2) + (q_1^2 \leftrightarrow q_2^2) \bigg]\,, \notag \\
     &f^{\text{asym}}_{\eta^{(\prime)}}(v,q^2) = \frac{(v-2q^2)^2 - M_{\eta^{(\prime)}}^2 v}{\sqrt{(v-2q^2)^2 - 2 M_{\eta^{(\prime)}}^2 v +M_{\eta^{(\prime)}}^4}} +2 q^2 - v\,,\label{Eq:TFF_asym}
 \end{align}
 which does not contribute to the singly-virtual direction, but maintains the correct doubly-virtual behavior, especially the limit
 \begin{equation}
     \lim_{Q^2 \to \infty} Q^2 F_{\eta^{(\prime)} \gamma^*\gamma^*}(-Q^2,-Q^2) = \frac{1}{3} \bar{F}_{\text{asym}}^{\eta^{(\prime)}}\,. \label{eq:DVlimit}
 \end{equation}
 The asymptotic coefficient $\bar{F}_{\text{asym}}^{\eta^{(\prime)}}$ is extracted from data utilizing the singly-virtual limit
  \begin{equation}
     \lim_{Q^2 \to \infty} Q^2 F_{\eta^{(\prime)} \gamma^*\gamma^*}(-Q^2,0) = \bar{F}_{\text{asym}}^{\eta^{(\prime)}}\,, \label{eq:SVlimit}
 \end{equation}
of the isovector, isoscalar, and effective-pole pieces in the TFF representation.

\begin{figure}[t]
     \centering
     \begin{minipage}[c]{0.495\textwidth}
        \centering
        \includegraphics[width=\linewidth]{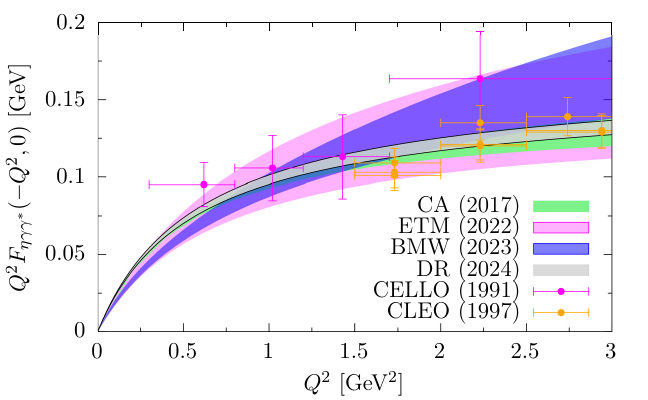}
     \end{minipage}
    \begin{minipage}[c]{0.495\textwidth}
        \centering
        \includegraphics[width=\linewidth]{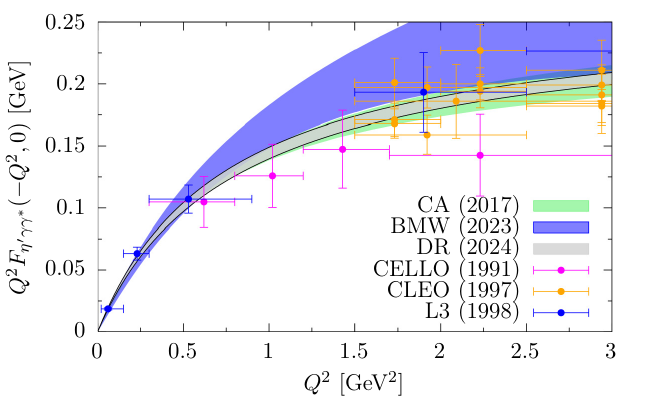}
     \end{minipage}\\
    \begin{minipage}[c]{0.495\textwidth}
        \centering
        \includegraphics[width=\linewidth]{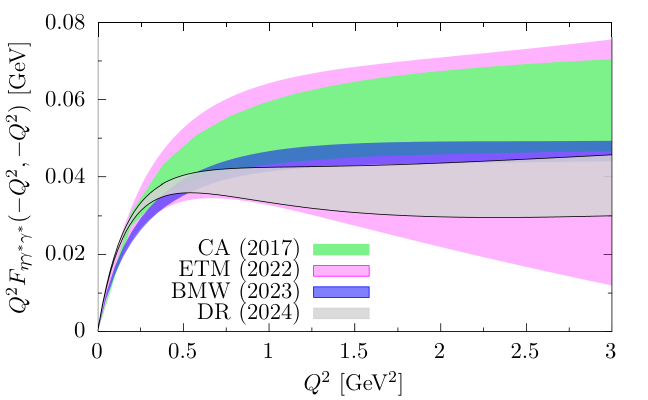}
     \end{minipage}
    \begin{minipage}[c]{0.495\textwidth}
        \centering
        \includegraphics[width=\linewidth]{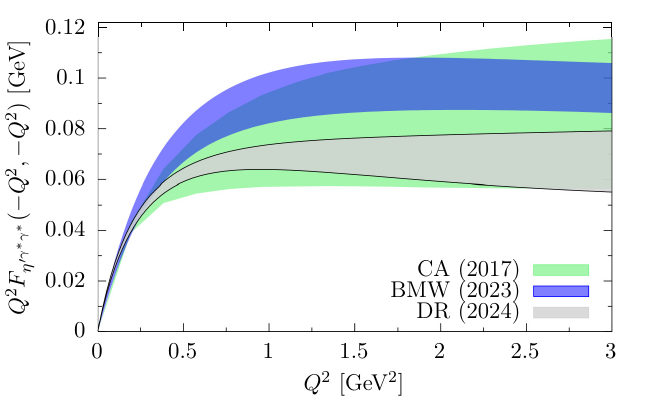}
     \end{minipage}
     \caption{Comparison of the singly-virtual (top) and doubly-virtual (bottom) dispersive $\eta$ (left) and $\eta'$ (right) TFFs (DR)~\cite{Holz:2024lom,Holz:2024diw} with the results of CA~\cite{Masjuan:2017tvw}, the lattice-QCD approach by ETM~\cite{ExtendedTwistedMass:2022ofm} and BMW~\cite{Gerardin:2023naa}, and singly-virtual experimental data from CELLO~\cite{CELLO:1990klc}, CLEO~\cite{CLEO:1997fho}, and L3~\cite{L3:1997ocz}.}
     \label{Fig:dispersive_tffs}
\end{figure}

The pole contributions to HLbL can be evaluated according to the well-known expression
\begin{align}\label{Eq:amu-P-pole}
  a_{\mu}^{P\text{-pole}} = \left(\frac{\alpha}{\pi}\right)^3\int dQ_1dQ_2 d\tau\Big[&w_1(Q_1,Q_2,\tau) F_{P\gamma^*\gamma^*}(-Q_1^2,-Q_3^2)F_{P\gamma^*\gamma^*}(-Q_2^2,0)  \notag\\
     &+w_2(Q_1,Q_2,\tau) F_{P\gamma^*\gamma^*}(-Q_1^2,-Q_2^2)F_{P\gamma^*\gamma^*}(-Q_3^2,0)\Big]\,,
\end{align}
where $Q_3^2\equiv Q_1^2+Q_2^2+2\tau Q_1Q_2$, and
the explicit form of the weight functions $w_{1/2}(Q_1,Q_2,\tau)$ is given, e.g., in Refs.~\cite{Jegerlehner:2009ry,Nyffeler:2016gnb,Masjuan:2017tvw,Hoferichter:2018kwz}.
Obviously, \cref{Eq:amu-P-pole} requires the spacelike singly- and doubly-virtual TFFs as input.  In \cref{Fig:dispersive_tffs}, we therefore compare the dispersively reconstructed transition form factors $F_{\eta^{(\prime)} \gamma^*\gamma^*}(-Q^2,0)$ and $F_{\eta^{(\prime)} \gamma^*\gamma^*}(-Q^2,-Q^2)$ to the previous approach based on CA~\cite{Masjuan:2017tvw} adopted in WP20~\cite{Aoyama:2020ynm}, as well as to lattice-QCD calculations by ETM~\cite{ExtendedTwistedMass:2022ofm} and BMW~\cite{Gerardin:2023naa}.  In general, we find very good agreement between the different approaches, as well as with the singly-virtual data~\cite{CELLO:1990klc,CLEO:1997fho,L3:1997ocz} at low energies that the dispersive representation has not included in the fit.  The narrowness of the uncertainty band, in particular in the singly-virtual direction, demonstrates the usefulness of the many constraints combined in the dispersive reconstruction of the TFFs.  In the doubly-virtual direction, there is a tendency to approach the asymptotic behavior dictated by \cref{eq:DVlimit} slightly more slowly than found in Ref.~\cite{Masjuan:2017tvw}.

Numerically, the $\eta$- and $\eta'$-pole contributions are determined as
\begin{align}
   a_\mu^{\eta\text{-pole}}&=14.72(56)_\text{norm}(32)_\text{disp}(23)_\text{BL}(54)_\text{asym}\times 10^{-11} = 14.72(87)\times 10^{-11}\,,\notag\\
   a_\mu^{\eta'\text{-pole}}&=13.50(48)_\text{norm}(15)_\text{disp}(20)_\text{BL} (48)_\text{asym}\times 10^{-11}=13.50(72)\times 10^{-11}\,,
   \label{Eq:amu}
 \end{align}
 where the different uncertainties arise from the following sources: (i) ``norm'': the normalization uncertainty reflecting the uncertainty on the average of experimental determinations on the $\Gamma(\eta^{(\prime)} \to \gamma \gamma)$ decay widths; (ii) ``disp'': the dispersive uncertainty, where different variants on the underlying representations of the isovector piece as well as different cutoff values for the dispersive integrals have been taken into account; (iii) ``BL'': the fit to high-energy singly-virtual TFF data approaching the Brodsky--Lepage limit, as well as the difference of the two variants in \cref{Eq:TFF_eff}; (iv) ``asym'': the variation of the onset of the asymptotic piece through variation of $s_{\text{m}}$ in \cref{Eq:TFF_asym} as well as variation of the asymptotic coefficient between the data-driven determination, by means of the $(I=1)$, $(I=0)$, and effective pieces, and lattice-QCD determinations~\cite{Ottnad:2017bjt,Bali:2021qem}. Furthermore, the total uncertainty refers to the quadratic sum.

We mention two more data-driven consistency checks on the dispersive reconstruction of the $\eta$ and $\eta'$ transition form factors.
The doubly-virtual $\eta$ transition form factor is closely linked to the $\gamma^{(*)} \to \eta\pi^+\pi^-$ amplitude.  In Ref.~\cite{Holz:2015tcg}, $\pi^+\pi^-$ spectra in $e^+e^-\to\eta\pi^+\pi^-$ data~\cite{BaBar:2007qju,BaBar:2018erh} have been analyzed for consistency with those measured in the corresponding real-photon decay $\eta\to\pi^+\pi^-\gamma$~\cite{KLOE:2012rfx}.  It was found that the dependence on the $\pi^+\pi^-$ invariant mass factorizes to a larger degree from that on the photon virtuality than what a data-driven model of singularities in the crossed channel induces~\cite{Kubis:2015sga}, already suggesting rather mild breaking of factorization in the doubly-virtual $\eta$ TFF at low-to-medium energies.
For the $\eta'$ singly-virtual TFF, a dispersive analysis~\cite{Holz:2022hwz} similarly relies on input for $\eta'\to\pi^+\pi^-\gamma$~\cite{BESIII:2017kyd} and the pion VFF (see Refs.~\cite{Dai:2017tew,Benayoun:2023dkl} for related works).  As both display a prominent isospin-breaking signal due to $\rho$--$\omega$ mixing~\cite{Hanhart:2016pcd}, in Ref.~\cite{Holz:2022hwz}, these were combined in a consistent coupled-channel framework.  While such effects are of relevance for high-precision data on $\eta'\to\ell^+\ell^-\gamma$ in the future, the impact of these isospin-breaking corrections in the spacelike region is rather negligible.

\subsubsection{Two-meson contributions: boxes and \texorpdfstring{$S$}{}-wave rescattering}
\label{sec:TwoMesonContributions}

The next-lightest intermediate state after a single pion is given by the two-pion state. It can be split further according to intermediate states in the crossed channels, resulting in the dispersively defined pion-box contribution~\cite{Colangelo:2017fiz,Colangelo:2017qdm}
\begin{align}
	a_\mu^\text{$\pi$-box} = -15.9(2) \times 10^{-11}
\end{align}
and two-pion rescattering contributions. The first only depends on the pion VFF, whereas the rescattering contribution can be expanded into partial waves and related to $\gamma^*\gamma^*\to\pi\pi$ helicity partial waves~\cite{Colangelo:2017fiz}. Both contributions have been discussed in detail in WP20.

Similarly to two-pion intermediate states, the dispersive analysis of two-meson intermediate states can be extended to $K\bar K$ and $\pi\eta$ intermediate states.
Due to the heavier mass, the contribution of the kaon box is small.\footnote{Similarly, the proton box contribution ($F_1$ form factor) is very small, $0.18(1)\times10^{-11}$~\cite{Estrada:2024rfi}.} It was previously evaluated based on model input for the kaon VFF~\cite{Aoyama:2020ynm,Eichmann:2019bqf}. In the meantime, a dispersive analysis of the kaon VFFs has become available~\cite{Stamen:2022uqh}, leading to
\begin{align}
	a_\mu^{K^\pm\text{-box}} = -0.48(1) \times 10^{-11} \, , \quad a_\mu^{K^0\text{-box}} = -0.5(4) \times 10^{-15}\,,
\end{align}
and confirming the evaluations based on model input.

The rescattering contributions are expanded into partial waves, with the $S$-waves providing a model-independent description of scalar resonances. While the low-energy contribution of the scalar $f_0(500)$ resonance was already treated dispersively in terms of $\pi\pi$ rescattering~\cite{Colangelo:2017fiz,Colangelo:2017qdm}, WP20 combined additional scalar and tensor contributions together into $a_\mu^\text{scalars+tensors} = -1(3) \times 10^{-11}$~\cite{Aoyama:2020ynm}. In Ref.~\cite{Danilkin:2021icn}, the $S$-wave rescattering contribution was extended by using as input the helicity partial waves for $\gamma^*\gamma^*\to\pi\pi / K\bar K_{I=0}$~\cite{Garcia-Martin:2010kyn,Hoferichter:2011wk,Moussallam:2013una,Danilkin:2018qfn,Hoferichter:2019nlq,Danilkin:2019opj}, derived from a modified coupled-channel Muskhelishvili--Omn\`es analysis~\cite{Danilkin:2019opj}. The input for the coupled-channel hadronic rescattering $\pi\pi/K\bar{K}_{I=0}$ was taken from the dispersive analysis of Ref.~\cite{Danilkin:2020pak}.
This covered the region of the $f_0(980)$ resonance within the dispersive rescattering formalism, updating the isospin $I=0$ contribution. Together with the $I=2$ estimates of Refs.~\cite{Colangelo:2017fiz,Colangelo:2017qdm}, an $S$-wave rescattering contribution of
\begin{align}
	a_\mu^\mathrm{HLbL}[\pi\pi/K\bar K_{I=0}, \text{$S$-waves}] = -8.7(1.0) \times 10^{-11}
\end{align}
was obtained. By investigating the line shape of the resonance, an $f_0(980)$ contribution of $a_\mu^\text{HLbL}[f_0(980)]|_\text{rescattering} = -0.2(1) \times 10^{-11}$ was isolated. In addition, Ref.~\cite{Danilkin:2021icn} performed a comparison of the dispersive resonance contribution in terms of two-particle $S$-wave rescattering with a narrow-width approximation, using the same tensor decomposition of the dispersive framework. The dispersive definition of narrow-width contributions eliminates model-dependent nonpole contributions, but the use of a consistent framework, in particular the choice of the tensor basis, is crucial for a meaningful combination of different hadronic contributions: the independence of the dispersive result for HLbL on the choice of the tensor basis only follows from the requirement that a set of sum rules be fulfilled by the scalar functions~\cite{Colangelo:2017fiz}. While the pseudoscalar poles and the pion box fulfill the sum rules individually, this is no longer the case for rescattering contributions or narrow resonances. This implies that these individual contributions depend on the choice of basis, with only the sum over all hadronic states being unambiguous.

In Ref.~\cite{Danilkin:2021icn}, it was observed that similarly to the single-channel solution of~\cite{Colangelo:2017fiz,Colangelo:2017qdm} the coupled-channel solution for $\gamma^*\gamma^*\to\pi\pi/K\bar K_{I=0}$ only leads to small violations of the sum rule relevant for $S$-waves. Therefore, the obtained rescattering contribution is essentially basis independent.
Differences in the narrow-width estimate of the $f_0(980)$ contribution between Ref.~\cite{Danilkin:2021icn} and the earlier estimate of Ref.~\cite{Knecht:2018sci} were attributed mainly to the propagator definition of the scalar resonance, which only corresponds to a dispersive pole in a different tensor basis and therefore indicates sum-rule violations of the narrow resonance. To a lesser degree, the differences are related to the input for the transition form factors~\cite{Schuler:1997yw,Hoferichter:2020lap}. The comparison to Ref.~\cite{Pauk:2014rta} was complicated by the fact that this estimate was based on a single helicity amplitude, which suffers from kinematic singularities that had to be removed by hand via angular averages.

Based on the narrow-width approximation, Ref.~\cite{Danilkin:2021icn} also evaluated the contribution of the isospin $I=1$ scalar $a_0(980)$. In Ref.~\cite{Deineka:2024mzt}, the $S$-wave contribution of $\pi\eta/K\bar K_{I=1}$ intermediate states has been evaluated dispersively in terms of helicity partial waves. By incorporating precise experimental data from two-photon processes \cite{Belle:2009xpa, Belle:2013eck} and utilizing the modified Muskhelishvili--Omn\`es formalism, the dispersive analysis
provides
\begin{align}
	a_\mu^\text{HLbL}[\pi\eta/K\bar K_{I=1}, \text{$S$-waves}] = -0.44(5) \times 10^{-11} \, .
\end{align}
This result represents an order of magnitude improvement in precision compared to the previous narrow-width estimate and is of the same order as the charged kaon box contribution in HLbL scattering.
In total, the $I=0,1,2$ rescatterings (including the effects of the $f_0(980)$ and $a_0(980)$ resonances) amount to a scalar contribution of
\begin{equation}
	\label{eq:RescatteringScalarContribution}
	a_\mu^\text{HLbL}[\text{scalars}] = -9.1(1.0) \times 10^{-11}\,.
\end{equation}

In Ref.~\cite{Danilkin:2021icn}, the contribution of the even heavier scalars $f_0(1370)$ and $a_0(1450)$ was estimated to be around $-1\times 10^{-11}$, but it was pointed out that the two-photon couplings and TFFs of these states are highly uncertain. In Ref.~\cite{Hoferichter:2024vbu,Hoferichter:2024bae}, these heavy-scalar contributions below a matching scale $Q_0=1.5\GeV$ were estimated as
\begin{align}
	a_\mu^\text{HLbL}[\text{heavy scalars}, Q_i < 1.5\text{ GeV}] = -0.7(3) \times 10^{-11} \, ,
\end{align}
using a simple quark model for the TFFs~\cite{Schuler:1997yw}
\begin{align}
	\frac{\F_1^S(q_1^2,q_2^2)}{\F_1^S(0,0)}=\frac{\Lambda_S^2(3\Lambda_S^2-q_1^2-q_2^2)}{3(\Lambda_S^2-q_1^2-q_2^2)^2} \, , \qquad
	\frac{\F_2^S(q_1^2,q_2^2)}{\F_1^S(0,0)}=-\frac{2\Lambda_S^4}{3(\Lambda_S^2-q_1^2-q_2^2)^2} \, ,
\end{align}
with the normalization $\F_1^S(0,0)$ determined from the two-photon widths $\Gamma_{\gamma\gamma}$ as discussed in Ref.~\cite{Danilkin:2021icn}. The scale is set to $\Lambda_{S}=M_{\rho}$, based on the observation that for the scalar mesons $f_0(980)$ and $a_0(980)$, adopting the expected VMD scale results in better agreement with the explicit calculations using helicity partial waves~\cite{Danilkin:2021icn,Deineka:2024mzt}.

\subsubsection{Axial-vector and tensor contributions}
\label{sec:ResonanceContributions}

As already discussed in WP20, the issue of sum-rule violations is more severe for the contribution of resonances beyond spin zero. At the time of WP20, the inclusion of axial-vector states or tensor resonances was not possible due to the presence of kinematic singularities. In the case of axial-vector states, this problem has been solved in Ref.~\cite{Hoferichter:2024fsj} with the construction of an optimized basis that allows for the inclusion of axial-vector states while leaving previously evaluated contributions like scalar resonances in terms of $S$-wave rescattering unchanged. The contribution of pseudoscalar poles and the scalar-QED pion box are left unchanged, as they do not depend on the choice of basis: while pseudoscalar poles do not contribute to the set of sum rules, the box contributions fulfill them individually. Rescattering contributions or narrow resonances do not fulfill the sum rules exactly, which implies that these individual contributions depend on the choice of basis, with only the sum over all hadronic states being unambiguous.

The new basis optimized for axial-vector states has been tested by analyzing the convergence behavior of the partial-wave expanded pion box, observing an even faster convergence when summing up higher partial waves than in the original basis~\cite{Hoferichter:2024fsj}.
The new basis does not fully solve the issue of kinematic singularities, but significantly simplifies the remaining singularity structure even for tensor resonances. The tensor-meson contribution can in general be described in a narrow-width approximation in terms of five TFFs~\cite{Hoferichter:2020lap}. In the new basis, no kinematic singularities show up in the tensor-resonance contribution if only the TFF $\F_1^T$ is retained, as well as in the special cases when only $\F_{1,3}^T$ or $\F_{2,3}^T$ are present. In Refs.~\cite{Hoferichter:2024vbu,Hoferichter:2024bae}, the tensor contributions are estimated using the simple quark model of Ref.~\cite{Schuler:1997yw}, which
indeed only gives a contribution to $\F_1^T$ of the form
\begin{align}
	\label{eq:TFF_tensor}
	\frac{\F_1^T(q_1^2,q_2^2)}{\F_1^T(0,0)}=\bigg(\frac{\Lambda_T^2}{\Lambda_T^2-q_1^2-q_2^2}\bigg)^2\,,\qquad \F_{2\text{--}5}^T(q_1^2,q_2^2)=0\,.
\end{align}
For a more realistic description at low energies,
the contribution of the tensor mesons $f_2(1270)$, $a_2(1320)$, and $f_2'(1525)$ below the matching scale $Q_0 = 1.5\GeV$ are calculated for
$\Lambda_T = M_\rho$
\begin{equation}
	a_\mu^\text{HLbL}[\text{tensors}, Q_i < 1.5\text{ GeV}] = -2.5(3) \times 10^{-11} \, ,
\end{equation}
where the uncertainty is propagated from the two-photon widths only. This reflects the expectation that the vector-meson scale should be important at low energies both because the $f_2(1270)$ arises primarily from the unitarization of vector-meson left-hand cuts and because the momentum dependence of the respective TFFs is again largely determined by vector-meson physics~\cite{Garcia-Martin:2010kyn,Hoferichter:2011wk,Moussallam:2013una,Hoferichter:2013ama,Danilkin:2018qfn,Hoferichter:2019nlq,Danilkin:2019opj}.  Further arguments to consider only $\F_1^T$ as a first estimate originate from the high-energy limit, since $\F_1^T$ gives the dominant contribution in the light-cone expansion~\cite{Hoferichter:2020lap}, and from the low-energy behavior, as the leading term in a chiral-Lagrangian approach only generates $\F_1^T$~\cite{Bellucci:1994eb}. Finally, up to a small correction from $\F_2^T$, it is $\F_1^T$ that determines the on-shell two-photon decay width.\footnote{The hQCD scenario described in \cref{sec:hQCDscalartensor}, featuring $\F_{1,3}^T$, can also be evaluated in the optimized basis from Ref.~\cite{Hoferichter:2024fsj}, but the appearance of $\F_3^T$ is rather unexpected from the arguments given here, e.g., in contrast to hQCD the Lagrangian realization in Ref.~\cite{Mathieu:2020zpm} violates gauge invariance. However, given the limited information on tensor TFFs presently available, for the final compilation in \cref{sec:finalnumber} both the $\F_1^T$ and $\F_{1,3}^T$ scenarios will be taken into account.}

The tensor-resonance estimate of Ref.~\cite{Danilkin:2016hnh} included in WP20, $a_\mu^\text{tensors} = 0.9(1) \times 10^{-11}$, was based on a model-dependent propagator definition that resulted in the need to average over kinematic singularities~\cite{Pauk:2014rta}, see Ref.~\cite{Danilkin:2021icn} for further discussion. It can now be replaced by an estimate within the same basis as other contributions. However, we emphasize that
even in the new basis, tensor-meson contributions beyond the effect of $\F_1^T$ still suffer from kinematic singularities. This issue has been solved within a new dispersive framework in triangle kinematics~\cite{Ludtke:2023hvz}, see \cref{sec:TriangleDR}. The new framework requires further sub-processes as input, but opens a path towards a model-independent evaluation of tensor contributions in terms of two-meson rescattering~\cite{Hoferichter:2019nlq,Danilkin:2019opj}. As in the case of heavy scalars, the final contribution of tensor states should be studied in combination with the SDCs, see below.

In contrast to the rather small tensor contributions, model estimates point to a substantial contribution of axial-vector mesons as discussed in \cref{Section:HolographicModels} and \cref{Section:OtherApproaches}. These contributions are tightly connected to the saturation of SDCs, see \cref{sec:SDC}, \cref{sec:MatchingToSDCs}, and \cref{sec:finalnumber}. With the solution of the problem of kinematic singularities for axial contributions in Ref.~\cite{Hoferichter:2024fsj}, the problem is reduced to the determination of the input for the axial-vector TFFs, which at present are poorly constrained from data. This is related to the fact that the Landau--Yang theorem forbids the decay of an axial-vector meson into two on-shell photons, hence constraining the TFFs requires data from processes involving at least one off-shell photon. For the three physical TFFs that enter the dispersive representation, the asymptotic behavior has been worked out in Ref.~\cite{Hoferichter:2020lap} using a light-cone expansion. In Ref.~\cite{Zanke:2021wiq, Hoferichter:2023tgp}, a VMD representation was used to combine all available experimental constraints on the TFFs of the $f_1(1285)$.
The minimal particle content that is necessary to fulfill asymptotic constraints as well as reproducing data from $e^{+}e^{-}\to f_{1}\pi^{+}\pi^{-}$ requires the inclusion of at least three multiplets of vector mesons. In Ref.~\cite{Hoferichter:2023tgp}, the $\rho=\rho(770)$, $\rho'=\rho(1450)$, and $\rho''=\rho(1700)$ were used for the dominant isovector piece and $\omega=\omega(782)$, $\omega'=\omega(1420)$, $\omega''=\omega(1650)$ as well as $\phi=\phi(1020)$, $\phi'=\phi(1680)$, $\phi''=\phi(2170)$ for the isoscalar contributions.

\begin{figure}[t]
	\centering
	\includegraphics[width=0.496\textwidth]{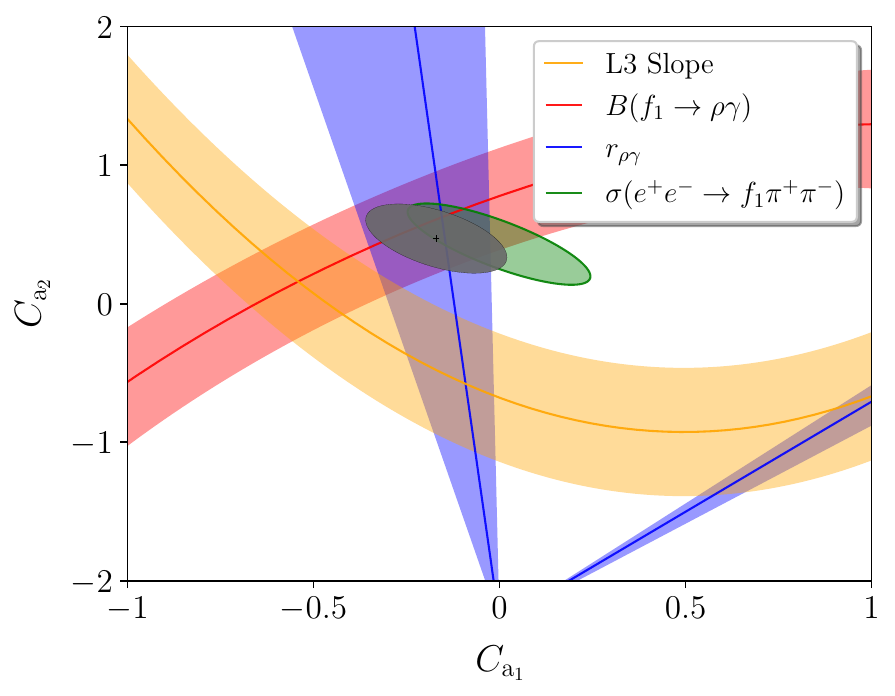}
	\includegraphics[width=0.496\textwidth]{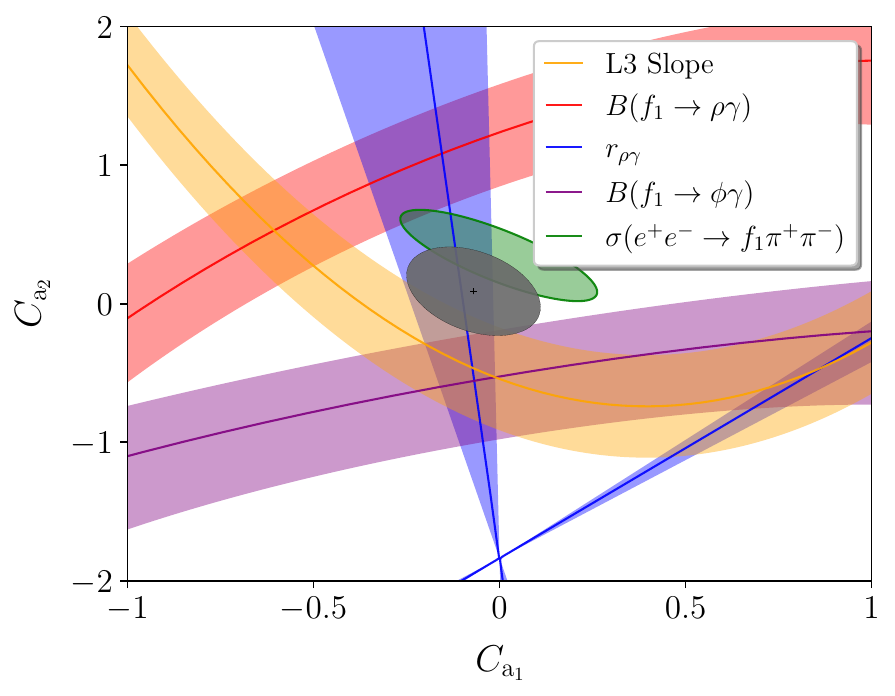}
	\caption{Constraints on the normalizations of the two antisymmetric $f_1$ TFFs, $C_{\text{a}_1}$ and $C_{\text{a}_2}$, excluding (left) and including (right) a possible constraint from $f_1\to\phi\gamma$. The bands come from
the L3 normalization and slope, branching and helicity fraction for $f_1\to \rho\gamma$, and $\sigma(e^+ e^- \to f_1 \pi^+ \pi^-)$, while the gray ellipse indicates the global fit. Figures taken from Ref.~\cite{Hoferichter:2023tgp}.}
	\label{fig:solutions_global_fit_pipi}
\end{figure}

Experimental input concerns the tree-level processes $e^+e^-\to e^+e^-f_1$~\cite{L3:2001cyf}, which mainly determines the equivalent two-photon decay width and constrains the symmetric TFF rather well; the decay $f_1\to\rho\gamma$; and, for the subleading isoscalar part, $f_1\to\phi\gamma$.  The decay $f_1\to4\pi$ does not provide information on the TFFs, as it is likely dominated by the decay chain $f_1\to a_1\pi\to4\pi$. In addition, the loop induced decay $f_1\to e^+e^-$, measured for the first time in inverse kinematics by SND~\cite{SND:2019rmq}, could provide valuable information on all three TFFs. Unfortunately, the data from this decay are not yet at the required level of precision to have a meaningful impact on the global fit. Instead, the scattering process $e^{+}e^{-}\to f_{1}\pi^{+}\pi^{-}$~\cite{BaBar:2007qju,BaBar:2022ahi} was considered in Ref.~\cite{Hoferichter:2023tgp}, as it provides further sensitivity also to the antisymmetric TFFs, see \cref{fig:solutions_global_fit_pipi}. However, even combining all currently available data, at best the normalizations of the $f_1$ can be determined, while $f_{1}'$ and $a_{1}$ are largely unconstrained. In this situation, estimates for their TFFs can be obtained assuming U(3) symmetry, since the corresponding mixing angle $\theta_A$ between $f_{1}$ and $f_{1}'$ was measured by L3 via the ratio of the respective equivalent two-photon decay widths $\tilde\Gamma_{\gamma\gamma}$ yielding $\theta_{A}=62(5)\degree$~\cite{L3:2001cyf,L3:2007obw}.

\begin{figure}[t]
    \centering
    \includegraphics[width=0.6\linewidth]{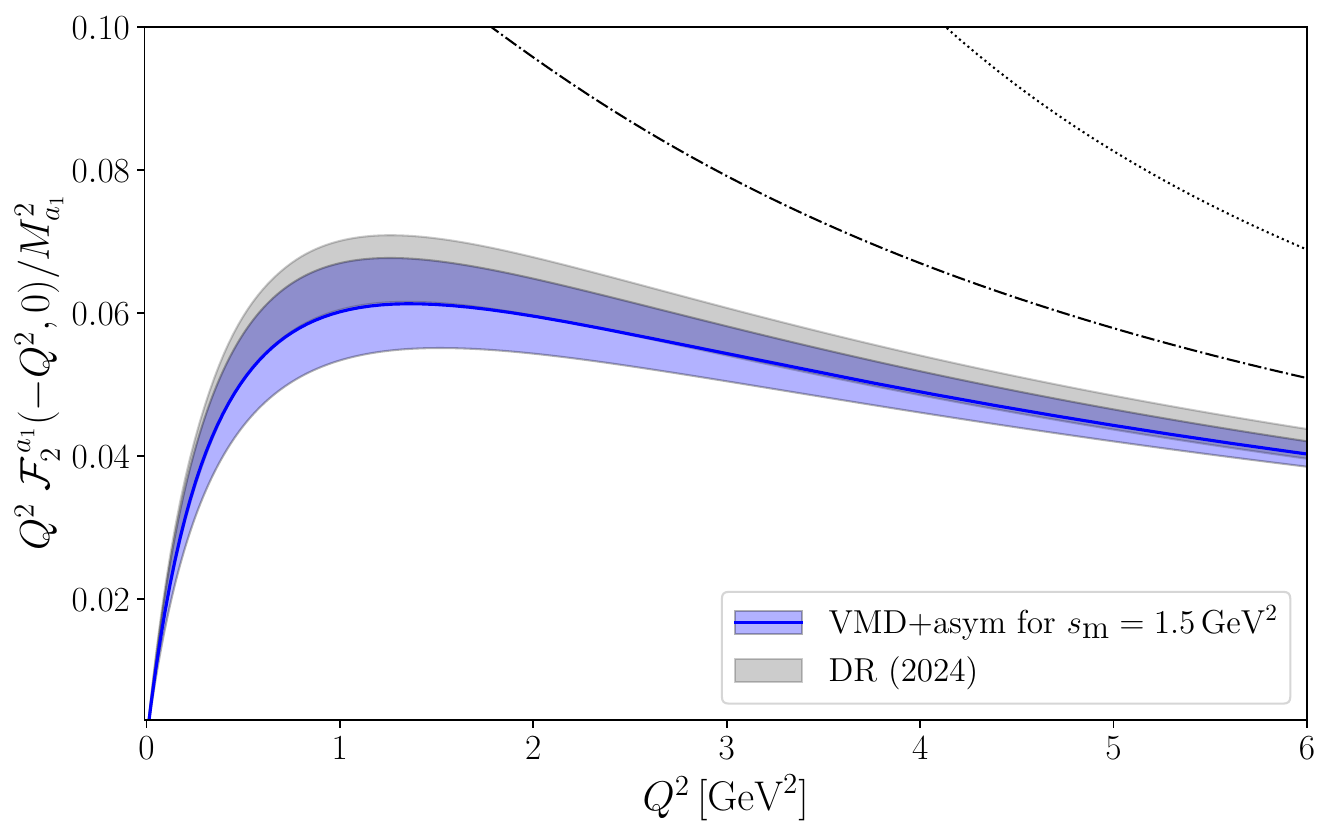}
    \caption{Comparison of the $a_1$ singly-virtual TFF $\mathcal{F}_2^{a_1}(-Q^2,0)$ from the VMD representation and U(3) symmetry (blue band)~\cite{Hoferichter:2023tgp,Hoferichter:2024bae}, in comparison with the dispersive result~\cite{Ludtke:2024ase} (gray band).}
    \label{fig:a1VMD}
\end{figure}

While full determinations of all three doubly-virtual TFFs are thus difficult to obtain, as reflected by the challenges to extract even the three couplings arising in a VMD-motivated representation for the $f_1$ from data, in the special case of the singly-virtual $a_1$ TFF $\F_2^{a_1}$ a cross-check is possible against a recent dispersive description derived in the context of the $VVA$ correlator~\cite{Ludtke:2024ase}, see \cref{fig:a1VMD}. This comparison shows that the assumption of U(3) symmetry
leads to a result for the $a_1$ in surprisingly good agreement.  Moreover, both results for $\F_2^{a_1}$ agree well with hQCD models, see \cref{sec:AxialTFFhQCD}.

Finally, the VMD representation taking into account three multiplets of vector mesons does not reproduce the correct asymptotic scaling in the doubly-virtual direction, and the coefficient of the singly-virtual limit is also not reproduced exactly, making it necessary to supplement the VMD representation with an asymptotic piece. The asymptotic contribution takes a form that is an appropriate extension of the one established for pseudoscalars, see \cref{sec:Pseudoscalars}. In Ref.~\cite{Hoferichter:2024vbu,Hoferichter:2024bae}, a suitable representation was found that incorporates relevant mass effects, but at the same time leaves unaltered the low-energy properties such as normalization and slope, which are already determined by the fit of the VMD representation to data. The additional asymptotic form introduces a parameter $s_m$ that describes the transition to the asymptotic regime of the corresponding form factor. Good agreement with the dispersive evaluation of the form factor $\mathcal{F}_{2}^{a_{1}}(q_{1}^{2},q_{2}^{2})$ of Ref.~\cite{Ludtke:2024ase} was found for $s_{m}=1.5\GeV^{2}$, the typical matching scale expected from light-cone sum-rule calculations~\cite{Khodjamirian:1997tk,Agaev:2014wna}.

As the inclusion of axial-vector states is still affected by a basis dependence, their contribution needs to be studied in combination with the matching to SDCs, see \cref{sec:MatchingToSDCs}. In this context, the tails of the pseudoscalar-pole contributions need to be subtracted, while the other contributions summarized in
\cref{tab:disp_summary} have negligible overlap with the mixed and short-distance regions.

\begin{table}[t]
\small
	\centering
	\renewcommand{\arraystretch}{1.1}
	\begin{tabular}{lrr}
	\toprule
	 Contribution & $a_\mu [10^{-11}]$ & References\\\midrule
	 $\pi^0$, $\eta$, $\eta'$ poles & $91.2^{+2.9}_{-2.4}$ & \cite{Hoferichter:2018dmo,Hoferichter:2018kwz,Holz:2024lom,Holz:2024diw}\\
	 $\pi^\pm$ box & $-15.9(2)$ & \cite{Colangelo:2017qdm,Colangelo:2017fiz}\\
	 $K^\pm$ box & $-0.5(0)$ & \cite{Stamen:2022uqh}\\
	 $S$-wave rescattering & $-9.1(1.0)$ & \cite{Colangelo:2017qdm,Colangelo:2017fiz,Danilkin:2021icn,Deineka:2024mzt}\\\midrule
	 Sum & $65.7^{+3.1}_{-2.6}$ &\\
\bottomrule
	\renewcommand{\arraystretch}{1.0}
	\end{tabular}
	\caption{Summary of previously evaluated contributions in dispersion theory. The $S$-wave rescattering subsumes effects that correspond to the light scalar resonances $f_0(500)$, $f_0(980)$, and $a_0(980)$. Table from Ref.~\cite{Hoferichter:2024vbu}.}
	\label{tab:disp_summary}
\end{table}

\subsubsection{Matching to short-distance constraints}
\label{sec:MatchingToSDCs}

Matching the sum of hadronic intermediate states to the SDCs of the HLbL tensor is most conveniently organized in terms of a matching scale $Q_{0}$ that separates the low-energy regime from the high-energy part.  Moreover, a parameter $r\in[0,1]$ has to be introduced in order to keep track of the applicability of the OPE constraints in the mixed region. In the following, we describe the matching strategy from Refs.~\cite{Hoferichter:2024vbu,Hoferichter:2024bae}.

\begin{figure}[t]
     \centering
     \includegraphics[width=0.7\linewidth]{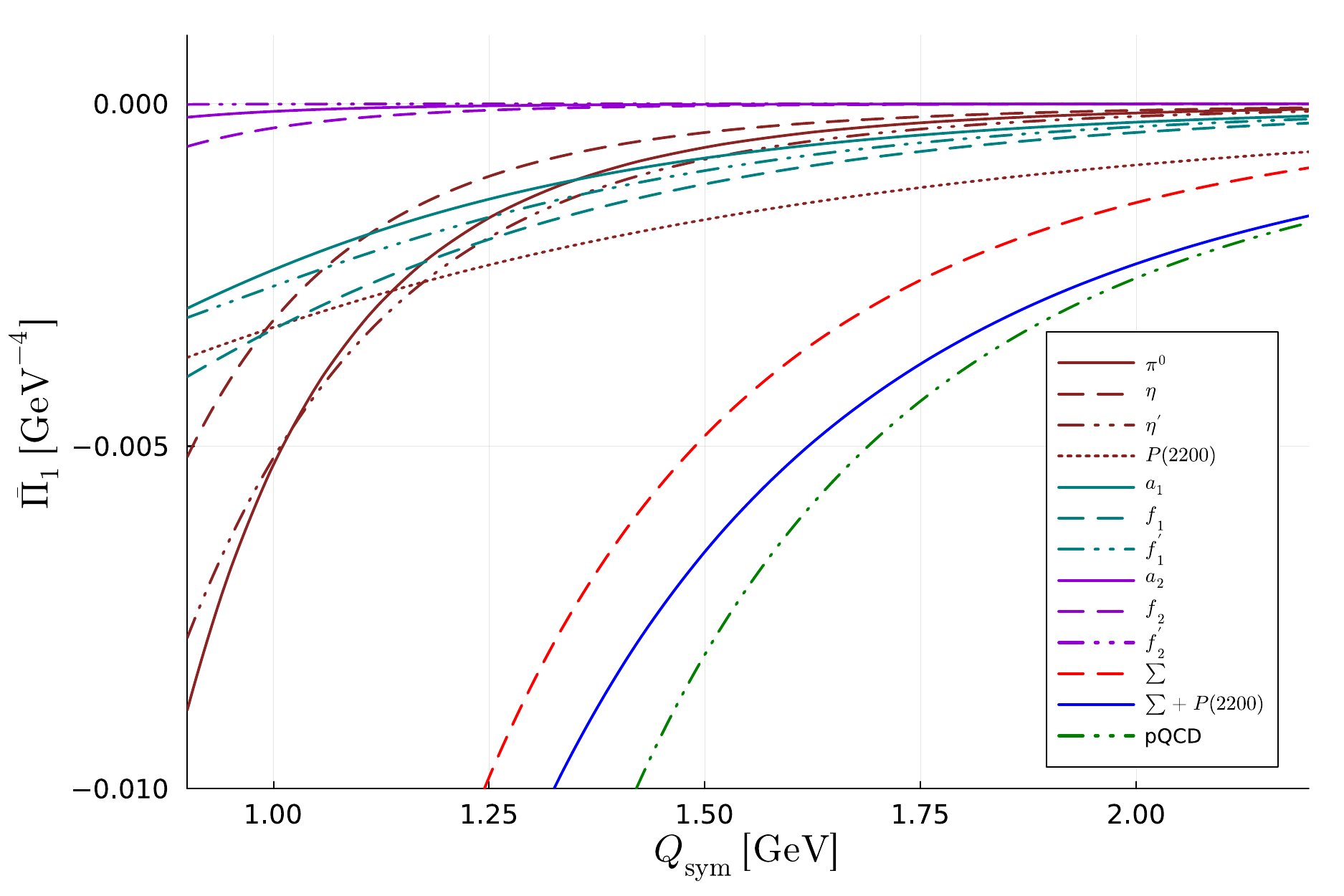}
     \caption{Matching between the sum of hadronic states, $\sum=\pi^0+\eta+\eta'+a_1+f_1+f_1'+a_2+f_2+f_2'$, in $\bar\Pi_1$ as a function of $Q_\text{sym}\equiv Q_1=Q_2=Q_3$. $P(2200)$ denotes the contribution of the effective pseudoscalar pole. The maroon solid, dashed, and dot-dashed lines refer to the lowest-lying pseudoscalar triplet, the turquoise ones to the lowest axial-vector triplet, and the violet lines to the lowest tensor triplet. The red dashed line represents the sum of all these contributions, the solid blue line the sum including, in addition, the effective pseudoscalar pole, to be compared to the dot-dashed green curve representing pQCD. Figure taken from Ref.~\cite{Hoferichter:2024vbu}.}
     \label{fig:matching}
 \end{figure}

With a finite number of hadronic states, e.g., including the pseudoscalar poles, the heavy scalars $f_{0}(1370)$ and $a_{0}(1450)$, the first axial-vector multiplet $f_{1}(1285)$, $f_{1}'(1420)$, and $a_{1}(1260)$, and the first tensor multiplet $f_2(1270)$, $f_2'(1525)$, and $a_2(1320)$, the matching cannot be expected to be smooth in all possible directions of the HLbL tensor. In \cref{fig:matching}, the scalar function $\bar{\Pi}_{1}$ for the symmetric limit $Q_{\text{sym}}=Q_{1}=Q_{2}=Q_{3}$ is shown. It displays the remaining discrepancy between the sum of hadronic states and pQCD in the regime above $1\GeV$. Therefore, in order to estimate the remaining uncertainty from the matching procedure, the relevant scales $Q_{0}$ and $r$ are varied in suitable intervals.

\begin{figure}[t]
     \centering
     \includegraphics[width=0.7\linewidth]{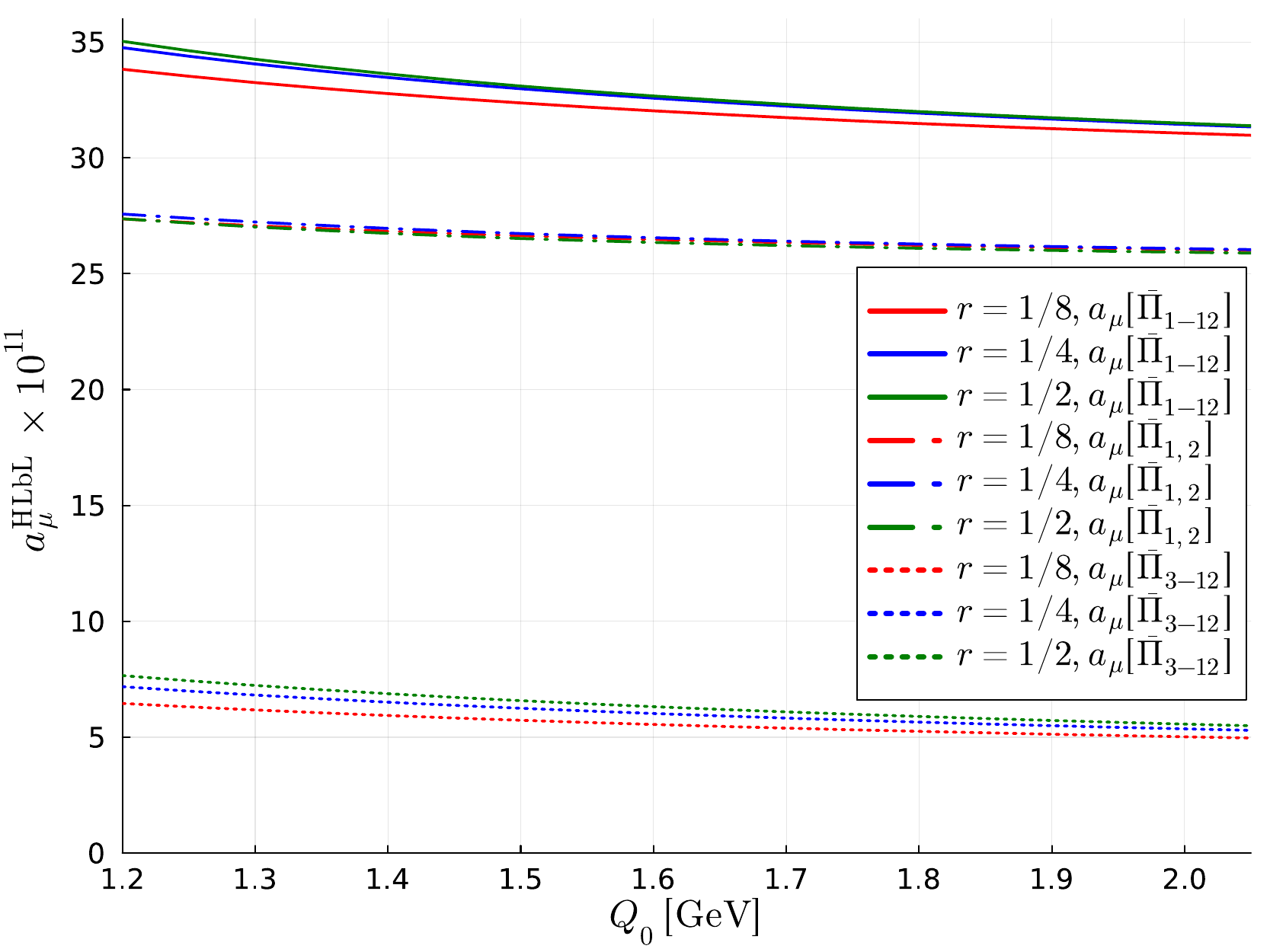}
     \caption{Stability of the $a_\mu$ integral under variation of $Q_0$, $r$. Figure taken from Ref.~\cite{Hoferichter:2024vbu}.}
     \label{fig:Q0r}
 \end{figure}

In the region in which all $Q_{i}<Q_{0}$ are small, the sum of hadronic intermediate states is used, and in the opposite case, when all $Q_{i}$ are larger than $Q_0$, the
pQCD result including $\alpha_s$ corrections evaluated at scale $\mu=Q_0$. The central value is chosen for $Q_0=1.5\GeV$, with scale varied between $1.2\GeV$, the lower boundary at which $\alpha_s$ corrections can still be controlled, and $2\GeV$, the upper limit at which a hadronic description should still be meaningful.
The mixed region can be further divided into the part with $Q_{1}>Q_{0}$ and $Q_{2},Q_{3} < Q_{0}$ and crossed versions thereof. In this regime, no OPE constraint applies and hence the sum of hadronic intermediate states is used. Instead, for
\begin{equation}
\label{mixed_OPE}
Q_3^2\leq r \frac{Q_1^2+Q_2^2}{2}\,,\qquad Q_1^2\geq Q_0^2\,,\qquad Q_2^2\geq Q_0^2\,,\qquad Q_{3}^{2}\leq Q_{0}^{2}\,,
\end{equation}
and crossed versions thereof, the HLbL tensor is related to the $VVA$ correlator, leading to the identification
\begin{align}
\label{hatPi_wL_wT}
 \hat{\Pi}_1&=-\frac{1}{\pi^2\hat q^2}\sum_{a=0,3,8}C_a^2w_L^{(a)}(q_3^2)\,,\notag\\
 \hat{\Pi}_5&=\hat{\Pi}_6=-\hat q^2 \hat{\Pi}_{10}=-\hat q^2 \hat{\Pi}_{14}=\hat q^2 \hat{\Pi}_{17}=\hat q^2 \hat{\Pi}_{39}=2\hat q^2 \hat{\Pi}_{50}=2\hat q^2 \hat{\Pi}_{51}
 =-\frac{2}{3\pi^2\hat q^2}\sum_{a=0,3,8}C_a^2w_T^{(a)}(q_3^2)\,,
\end{align}
in terms of longitudinal and transverse form factors $w_{L,T}(q^{2})$.\footnote{These form factors are identical to $\omega_{L,T}(q^2)$ introduced in \cref{sec:SDC}, up to an overall factor of $N_c=3$, $w_{L,T}(q^{2})=N_c \omega_{L,T}(q^2)$.} In particular, it was shown in Ref.~\cite{Bijnens:2024jgh} that certain corrections at higher orders in the OPE accidentally cancel at the level of the $a_\mu$ integration, making the constraint \cref{hatPi_wL_wT} more robust.  For the numerical analysis, the dedicated dispersive analysis of the hadronic $VVA$ correlator from Ref.~\cite{Ludtke:2024ase} is used for the triplet component $a=3$, reproduced in Ref.~\cite{Hoferichter:2024fsj} in a simplified  set-up that can be readily generalized to $a=0,8$.
For the central value $r=1/4$ is used, varied by a factor of $2$ in both directions in order to assess the uncertainty introduced by this specific matching parameter. The stability of $a_{\mu}$ by varying the scales $Q_{0}$ and $r$ can be seen in \cref{fig:Q0r}.

The mismatch between the sum of hadronic intermediate states and pQCD in the region between $1.2\GeV$ and $2.0\GeV$ could have an impact on the low-energy region and the parts of the mixed region for which no OPE constraint applies. When including more states or even an infinite tower of states in order to saturate the asymptotic behavior and OPE constraints of the HLbL tensor, a potential nonnegligible contribution might arise at low energies as well, and this uncertainty would not be captured by the variation of the matching scales $Q_{0}$ and $r$. Therefore,  effective poles are introduced that saturate the prescribed asymptotic behavior of the HLbL tensor. As a single state or a finite number of states in four-point kinematics cannot exactly reproduce the required behavior, the dispersive framework in triangle kinematics proves beneficial~\cite{Ludtke:2023hvz}, introducing a corresponding pseudoscalar pole for the longitudinal components and an axial-vector state for the transverse ones, e.g.,
\begin{align}
\hat \Pi_1^\text{eff}&=\frac{F_{P\gamma^*\gamma^*}(q_1^2,q_2^2)F_{P\gamma^*\gamma^*}(M_P^2,0)}{q_3^2-M_P^2}\,,\notag\\
\hat \Pi_4^\text{eff}
&=\frac{(q_1^2+q_3^2-M_A^2)\F_2^A(M_A^2,0)\big[2\F_1^A(q_1^2,q_3^2)+\F_3^A(q_1^2,q_3^2)\big]}{2M_A^4(q_2^2-M_A^2)}+\big(q_1^2\leftrightarrow q_2^2\big)\,.
\end{align}
Note that the axial-vector state in triangle kinematics does not contribute to the longitudinal part in contrast to the dispersive framework in four-point kinematics.
The overall couplings are determined from the short-distance matching
in the symmetric limit $Q_{\text{sym}}=Q_1=Q_2=Q_3$, while the asymmetric matching is included in the error estimate. The mass scales are taken as $M_{P}^{\text{eff}}=2.2\,\text{GeV}$ and $M_{A}^{\text{eff}}=1.7\,\text{GeV}$, motivated by the phenomenology of excited pseudoscalar and axial-vector states~\cite{Hoferichter:2024bae}, and the TFF scale is varied in the same range as $Q_0$, with central value at $1.5\,\text{GeV}$.
The various contributions in each region are summarized in \cref{tab:subleading}, displaying results for the central values and those errors directly derived from experiment (and $\alpha_s$).

\begin{table}[t]
\small
	\centering
	\renewcommand{\arraystretch}{1.1}
	\begin{tabular}{llrrr}
	\toprule
	 Region & & $a_\mu[\bar\Pi_{1,2}] \ [10^{-11}]$ & $a_\mu[\bar\Pi_{3\text{--}12}] \ [10^{-11}]$ & Sum $[10^{-11}]$\\\midrule
\multirow{4}{*}{$Q_i<Q_0$} & $A=f_1,f_1',a_1$ &  $7.2(1.4)_\text{exp}$ & $5.0(1.0)_\text{exp}$ & $12.2(2.3)_\text{exp}$\\
& $S=f_0(1370),a_0(1450)$ & -- & $-0.7(3)_\text{exp}$ & $-0.7(3)_\text{exp}$\\
& $T=f_2,a_2,f_2'$ & $2.6(3)_\text{exp}$ & $-5.1(7)_\text{exp}$ & $-2.5(3)_\text{exp}$\\
& Effective poles & $2.5$ & $-0.4$ & $2.0$\\\midrule
\multirow{3}{*}{Mixed} & $A,S,T$ & $2.5(7)_\text{exp}$ & $1.3(3)_\text{exp}$ & $3.8(1.0)_\text{exp}$\\
& OPE & $6.3$ & $4.7$ & $10.9$\\
&Effective poles & $1.1$ & $0.1$& $1.2$\\\midrule
$Q_i>Q_0$ & pQCD & $4.8^{+0.1}_{-0.2}$ & $1.6^{+0.0}_{-0.1}$ & $6.3^{+0.2}_{-0.3}$\\\midrule
Sum & & $26.9(2.1)_\text{exp}(3.7)_\text{sys}(3.2)_\text{eff}$ & $6.3(1.5)_\text{exp}(0.2)_\text{sys}(2.2)_\text{eff}$ & $33.2(3.3)_\text{exp}(4.6)_\text{sys}(3.9)_\text{eff}$\\
\bottomrule
	\renewcommand{\arraystretch}{1.0}
	\end{tabular}
	\caption{Summary of the various subleading contributions considered in Refs.~\cite{Hoferichter:2024vbu,Hoferichter:2024bae}, at the matching scale $Q_0=1.5\GeV$ and with the OPE applied for $Q_3^2<r(Q_1^2+Q_2^2)/2$, $r={1/4}$. In the regions in which OPE and pQCD are used, the tails of the pseudoscalar poles are subtracted to avoid double counting. The effective poles are determined from the matching in the symmetric asymptotic limit, with TFF scale $1.5\GeV$. The errors are labeled as in \cref{result_subleading}, see main text for details,
    while the matching uncertainties from the variation of $Q_0$, $r$ are not yet included. The errors for the pQCD contribution are propagated from $\alpha_s$. Table from Ref.~\cite{Hoferichter:2024bae}.}
	\label{tab:subleading}
\end{table}

In total, the following results for the subleading HLbL contributions are obtained
\begin{align}
\label{result_subleading}
 a_\mu[\bar\Pi_{1,2}]&=26.9(2.1)_\text{exp}(1.0)_\text{match}(3.7)_\text{sys}(3.2)_\text{eff}[5.4]_\text{total}\times 10^{-11}\,,\notag\\
 a_\mu[\bar\Pi_{3\text{--}12}]&=6.3(1.5)_\text{exp}(1.4)_\text{match}(0.2)_\text{sys}(2.2)_\text{eff}[3.0]_\text{total}\times 10^{-11}\,,\notag\\
 a_\mu[\bar\Pi_{1\text{--}12}]&=33.2(3.3)_\text{exp}(2.2)_\text{match}(4.6)_\text{sys}(3.9)_\text{eff}[7.2]_\text{total}\times 10^{-11}\,,
\end{align}
where the ``exp'' error refers to the two-photon couplings of the heavy scalars and tensor mesons as well as the axial-vector TFFs; the ``match'' error indicates the maximal variation under $Q_0\in[1.2,2.0]\text{GeV}$ and $r\in[1/8,1/2]$; the ``sys'' error subsumes various systematic effects (a $30\%$ uncertainty on the hadronic contributions to reflect U(3) assumptions for the axial-vector TFFs~\cite{Zanke:2021wiq,Hoferichter:2023tgp} and the simplified form of the tensor TFFs; and a $100\%$ uncertainty for the total tensor contribution to $a_\mu[\bar\Pi_{1\text{--}12}]$ to reflect the strong cancellations observed among the different scalar functions); the ``eff''  error estimates the uncertainties in the effective-pole contributions (variation of TFF scale and symmetric vs.\ asymmetric matching), where the uncertainties for the pseudoscalar and axial-vector poles are considered separately and added in quadrature in the end, again due to a cancellation observed between them.  Combined with the dispersively evaluated contributions from \cref{tab:disp_summary} one finds a total HLbL contribution
$a_\mu^\text{HLbL}=98.9(7.9)\times 10^{-11}$ (excluding the charm loop).

\subsection{Holographic approach}
\label{Section:HolographicModels}

In WP20, results from simple (chiral) holographic models, which had been shown to naturally implement the MV SDC through their infinite towers of axial-vector mesons~\cite{Leutgeb:2019gbz,Cappiello:2019hwh}, were observed to be roughly consistent with the combined estimate of the contributions of axials and SDCs, however, with numerical results around the upper value of the estimated error bar $a_\mu^{\text{axials+SDC}}=21(16)\times 10^{-11}$. In the meantime, more realistic models including finite quark masses as well as the anomalous mass of the $\eta'$ have been worked out that make a numerical comparison with the results from the updated data-driven approach as well as with lattice results even more interesting. Before doing so, it is useful to recall the theoretical basis of the holographic approach.

Holographic QCD (hQCD) refers to hadronic models of QCD in the large-$N_c$ limit
constructed in analogy to the conjectural but well-established AdS/CFT correspondence
\cite{Maldacena:1997re,Aharony:1999ti}, where a conformally invariant supersymmetric Yang--Mills theory in the limit of infinite 't Hooft coupling can be
mapped to a five-dimensional classical gravity theory on anti-de Sitter (AdS) space.
The extra dimension turns out to correspond to the energy scale of the holographically
dual quantum field theory living on the conformal boundary of AdS space.

A ``top-down'' construction of a holographic dual to low-energy large-$N_c$ QCD with
chiral quarks has been achieved by Sakai and Sugimoto~\cite{Sakai:2004cn,Sakai:2005yt},
building upon earlier work by Witten~\cite{Witten:1998zw}, where supersymmetry and conformal invariance is
broken by an additional Kaluza--Klein compactification. While remarkably
successful for a number of low-energy observables, this model does not make
contact to QCD at large energies and momenta. However, there are by now a number
of phenomenologically interesting ``bottom-up'' models of hadron physics that
combine salient features of the top-down construction with a simpler geometry
that is asymptotically AdS$_5$ and therefore can be matched to the asymptotic
conformal symmetry of QCD in the UV. Confinement and chiral symmetry breaking
is implemented either by a hard-wall (HW) cutoff~\cite{Erlich:2005qh,DaRold:2005mxj,Hirn:2005nr}
with appropriate boundary conditions or by a soft wall (SW) provided by
a nontrivial dilaton~\cite{Ghoroku:2005vt,Karch:2006pv,Kwee:2007dd,Gursoy:2007cb,Gursoy:2007er,Colangelo:2008us,Gherghetta:2009ac,Branz:2010ub,Colangelo:2011xk}, which has similarities with light-front hQCD~\cite{Brodsky:2014yha}.

Already the simplest hQCD models have proved to provide good qualitative and
oftentimes quantitative predictions of observables in hadron physics with
typical errors of (sometimes less than) 10 to 30\% with a minimal set of free parameters~\cite{Erlich:2005qh,DaRold:2005mxj,Hirn:2005nr}.
While this is clearly much too crude to be of help with the HVP
contribution to $a_\mu$, bottom-up hQCD models provide interesting predictions for the
HLbL amplitude, where they naturally satisfy the longitudinal SDC
after matching leading-order OPE results for the vector correlation function~\cite{Leutgeb:2019gbz,Cappiello:2019hwh,Leutgeb:2021mpu}.
In this context, it is useful to recall that hQCD relies on the so-called
holographic dictionary underlying the original conjectured AdS/CFT duality, where for
each gauge-invariant quantum operator $\mathcal{O}_\Delta(x)$ of the 4D gauge theory with scaling dimension $\Delta$ one has a corresponding 5D field $\phi(x,z)$, whose value
on the UV boundary at $z=0$ is identified (modulo a certain power of $z$) with the 4D source of the operator, $J(x)\propto\phi(x,0)$, and whose 5D mass is determined by $\Delta$ and form degree $p$ through
$m_\phi^2=(\Delta-p)(\Delta+p-4)$.

The generating functional of the 4D theory is computed from the
5D action evaluated on-shell,
\begin{equation}
    \exp(iW[J])\equiv \langle 0 | T \exp \left(i\int d^4x J(x)\mathcal{O}_\Delta(x)\right) | 0 \rangle_{4D} = \exp \left( i S_{5D}^\text{on-shell}[J] \right)\,,
\end{equation}
so that $n$-point functions of the 4D theory
$\langle 0| T\left\{ \mathcal{O}_\Delta(x_1)\mathcal{O}_\Delta(x_2)\cdots\mathcal{O}_\Delta(x_n)\right\} |0\rangle$
can be obtained from variational derivatives $\delta^n S_{5D}^\text{on-shell}/\delta J(x_1)\delta J(x_2)\cdots \delta J(x_n)$.
The calculation of correlation functions of quantum operators thus amounts to computing
tree-level (Witten) diagrams on the gravity side, where the external sources reside on
the conformal boundary of AdS space, which connect with bulk-to-boundary propagators
to interactions in the bulk of the 5D theory, involving also bulk-to-bulk propagators
when $n\ge 4$ as in the case of the HLbL amplitude. In this case $\mathcal{O}_\Delta$
involves the vector current $\bar q \gamma^\mu t^a q$ with $\Delta=3$, corresponding
to a massless gauge field in the bulk. The global chiral symmetry U($N_f$)$_L\times$ U($N_f$)$_R$
thus turns into a (flavor) gauge theory in the bulk with 5D action
\begin{equation}\label{S5D}
S_\text{YM}
=-\frac{1}{4g_5^2}
\;\text{tr}\int d^4x \int_0^{z_0} dz\,e^{-\Phi(z)}\sqrt{-g}\, g^{PR}g^{QS}
\left(\mathcal{F}^{(L)}_{PQ}\mathcal{F}^{(L)}_{RS}
+\mathcal{F}^{(R)}_{PQ}\mathcal{F}^{(R)}_{RS}\right)\,,
\end{equation}
where $P,Q,R,S=0,\dots,3,z$ and $\mathcal{F}_{MN}=\partial_M \mathcal{B}_N-\partial_N \mathcal{B}_M-i[\mathcal{B}_M,\mathcal{B}_N]$. Here $z_0$ can either be a
finite upper value, when confinement is implemented by a hard wall,
or $z_0=\infty$ when confinement is due to a nontrivial dilaton field $\Phi(z)$.

\begin{figure}[t]
\centering
\includegraphics[width =.7\textwidth,clip]{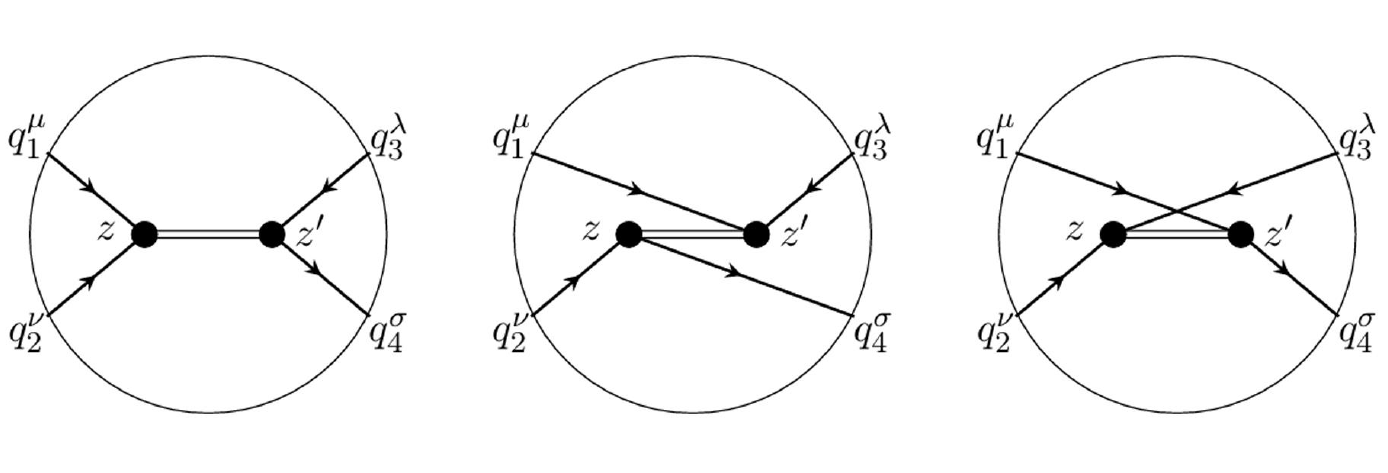}
\caption{5D Witten diagrams contributing to the HLbL tensor in hQCD models, generalizing the one-particle exchange diagrams in 4D.  The AdS$_5$ space is represented as the disk, with $z$ as the holographic radial coordinate and ordinary 4D spacetime as its boundary.  Solid lines represent the vector bulk-to-boundary propagators, depending on the 4D external momenta. Double lines denote the Green function of a 5D field, containing the whole tower of massive resonances in the given channel (pseudoscalars, axial vectors etc.) in the form $G(q,z,z')=\sum_n \varphi_n(z)\varphi_n(z')/(q^2-m_n^2)$.  Dots represent interaction vertices
derived from the 5D action.
}
\label{fig:WittenDiagram}
\end{figure}

It is a striking feature of hQCD models that the correlation functions are saturated
by infinite sums of narrow 4D resonances of increasing mass, as is expected in the large-$N_c$ limit of QCD. In particular, the normalizable modes of the 5D Yang--Mills fields
$\mathcal{B}^{L,R}_{M}=\mathcal{B}^{V}_{M}\mp \mathcal{B}^{A}_{M}$ correspond to an infinite tower
of massive vector and axial-vector mesons.
VMD is naturally built in through the bulk-to-boundary propagators
for EM currents, which can be expressed in terms of mode functions of the massive vector mesons.
Typically, hQCD models provide closed-form analytic or semi-analytic expressions summing up the contributions of those infinite towers of resonances.
Moreover, hQCD models with (asymptotically) AdS metric reproduce the approximately conformal behavior of QCD at short distances, permitting in many cases a matching of SDCs.

Chiral symmetry is broken either by introducing an extra bifundamental
scalar field $X$ dual to quark bilinears $\bar q_L q_R$ as in the original HW model~\cite{Erlich:2005qh,DaRold:2005mxj} (denoted by HW1 in the following) and in the SW model~\cite{Karch:2006pv}, or
through different IR boundary conditions for vector and axial vector fields
at $z_0$ as in the inherently chiral Hirn--Sanz (HW2) model~\cite{Hirn:2005nr} (as is also the case in
the top-down Sakai--Sugimoto model~\cite{Sakai:2004cn,Sakai:2005yt}). Versions of the HW1 model with the latter boundary conditions have been advocated in Ref.~\cite{Domenech:2010aq} and further studied in Ref.~\cite{Leutgeb:2021mpu} (denoted by HW3 therein). With the exception of the HW2 model, these bottom-up models have just enough parameters to match $F_\pi$, the $\rho$ meson mass, and the leading OPE term of the vector--vector correlation function; the simpler HW2 model can match only two of those ingredients at once.

In all these models, flavor anomalies follow uniquely
from 5D Chern--Simons terms $S_{\rm CS}^L-S_{\rm CS}^R$, where (in
differential-form notation)
\begin{equation}\label{SCS}
            S_\text{CS}=\frac{N_c}{24\pi^2}\int\text{tr}\left(\mathcal{B}\mathcal{F}^2-\frac{i}2 \mathcal{B}^3\mathcal{F}
            -\frac1{10}\mathcal{B}^5\right)\,.
\end{equation}
In particular, this determines the anomalous interactions of photons with pseudoscalars
and axial-vectors which appear together in the mode expansion of $\mathcal{B}_M^A
=(\mathcal{B}_M^R-\mathcal{B}_M^L)/2$.

Although suppressed at large $N_c$, a Witten--Veneziano mechanism for the $\eta_0$ mass through the U(1)$_A$ anomaly is included in the Sakai--Sugimoto model~\cite{Sakai:2004cn} and yields satisfactory numerical results at $N_c=3$ \cite{Brunner:2015yha}.
Extensions of the bottom-up models that implement the U(1)$_A$ anomaly have been proposed in Refs.~\cite{Katz:2007tf,Casero:2007ae},
and have been employed in the context of HLbL scattering in Refs.~\cite{Leutgeb:2022lqw,Colangelo:2023een}.

\subsubsection{Pseudoscalar TFFs in hQCD}
\label{section:Pseudoscalar}

In a mode decomposition of the bulk-to-bulk propagator in the 5D Witten diagrams for the HLbL tensor shown in \cref{fig:WittenDiagram}, integration over the holographic coordinate of an internal vertex yields
the TFF of the exchanged meson. The internal vertex is
provided by the Chern--Simons action \cref{SCS}, yielding
\begin{equation}\label{pi0TFF}
            F_{\pi^0\gamma^*\gamma^*}(Q_1^2,Q_2^2)=
            \frac{N_c}{12\pi^2 F_\pi}\int_{0}^{z_{0}}dz\,\mathcal{J}(Q_1,z)\mathcal{J}(Q_2,z)
            \Psi(z)\,,
\end{equation}
where $\mathcal J(Q,z)$ is the bulk-to-boundary propagator of a photon with virtuality $Q^2=-q^2$ and $\Psi(s)$ a holographic pion profile function.
The authors of Refs.~\cite{Grigoryan:2007wn,Grigoryan:2008up,Grigoryan:2008cc} were the first to notice that in hQCD models with asymptotic AdS$_5$ geometry the expression  in \cref{pi0TFF} reproduces the asymptotic momentum dependence obtained by BL in  QCD~\cite{Brodsky:1981rp,Lepage:1979zb,Lepage:1980fj},
\begin{equation}
                F_{\pi^0\gamma^*\gamma^*}(Q_1^2,Q_2^2)
                \to C \frac{2 F_\pi}{Q^2}\left[ \frac1{w^2}-\frac{1-w^2}{2w^3}\log\frac{1+w}{1-w} \right]\,,
                \label{pionTFFas}
\end{equation}
with $C=g_5^2N_c/(12\pi^2)$, $Q^2=(Q_1^2+Q_2^2)/2\to\infty$, $w=(Q_1^2-Q_2^2)/(Q_1^2+Q_2^2)$.
The correct functional dependence on $w$
is due to the infinite sum of vector resonances
contained in $\mathcal J$, and cannot be obtained in
VMD models with a finite number of resonances.
When the vector--vector correlator is
matched to the leading-order OPE result, one has
$C=1$, in exact agreement with the BL limit.
This can be achieved in the bottom-up hQCD models with asymptotic AdS geometry
except for the simple HW2 model, when the latter matches
both $F_\pi$ and $M_\rho$, which reduces
$C$ to 0.616 (fitting the asymptotic limit would instead lead to an overweight $\rho$ meson with mass of $987\MeV$).

\paragraph{Pion-pole contribution}

First estimates of the corresponding holographic results for the pion pole contribution $a_\mu^{\pi^0}$ were obtained in Ref.~\cite{Hong:2009zw,Cappiello:2010uy} and more fully in  Ref.~\cite{Leutgeb:2019zpq}.
The experimental data on the pion TFF turned out to be bracketed by the HW1 and HW2 results when these models are fit to $F_\pi$ and $M_\rho$. Since the HW2 model then undershoots the asymptotic limit of \cref{pionTFFas}
by $38\%$, its prediction of $a_\mu^{\pi^0}=56.9\times 10^{-11}$ is correspondingly on the low side. The HW1 model, which exactly satisfies the asymptotic limit \cref{pionTFFas}
and which can also accommodate finite quark masses~\cite{Leutgeb:2021mpu}, predicts $a_\mu^{\pi^0}=(66.0\text{--}66.6)\times 10^{-11}$ depending on the precise boundary conditions~\cite{Leutgeb:2021mpu}.
This is consistent with, but slightly larger than the dispersive result $63.0^{+2.7}_{-2.1}\times 10^{-11}$~\cite{Hoferichter:2018dmo,Hoferichter:2018kwz,Aoyama:2020ynm}.

\begin{figure}[t]
\centerline{\includegraphics[width=0.495\textwidth]{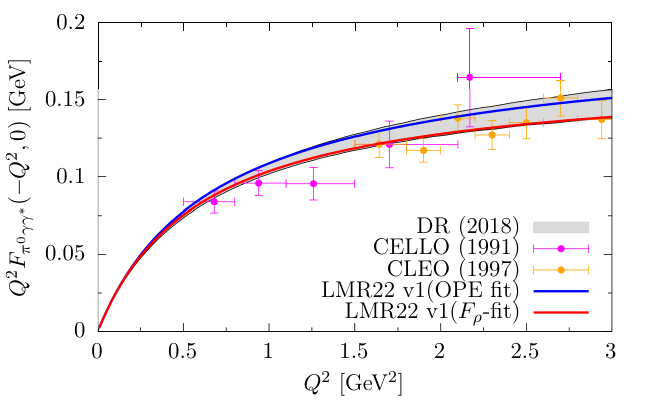}
\includegraphics[width=0.495\textwidth]{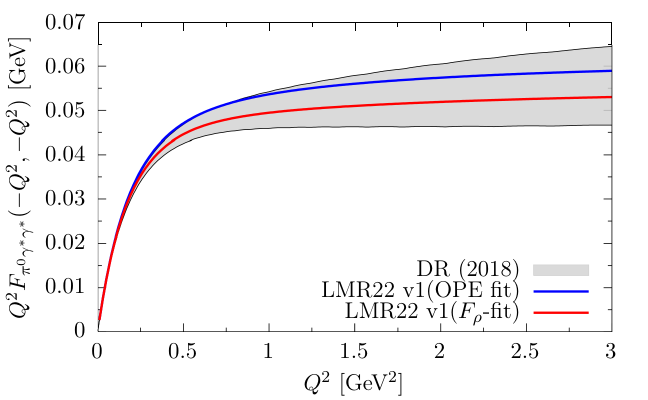}
}
\caption{Comparison of the hQCD results for the pion TFF of Ref.~\cite{Leutgeb:2022lqw} (LMR22) with
OPE fit of $g_5^2$ (blue lines) and a $10\%$ reduced coupling $g_5^2$ (red lines)
from fitting $F_\rho$ instead, compared to the dispersive results of
Ref.~\cite{Hoferichter:2018kwz} and experimental data.
}
\label{fig:pionTFFHW1}
\end{figure}

At high but still phenomenologically relevant energy scales, gluonic corrections modify the large-$Q^2$ behavior of the TFF  by about  $10\%$~\cite{Melic:2002ij,Bijnens:2021jqo}. Holographic QCD models with a simple AdS$_5$ background have no running coupling constant and thus
approach the asymptotic limit too quickly. Similar corrections, but with opposite sign, appear in the vector correlator determining also HVP. Indeed, the HW1 model with a complete UV fit underestimates the HVP contribution. Reducing $g_5^2$ by $10\%$ (incidentally corresponding to an exact fit of the decay constant of the $\rho$ meson) brings
HVP in line with the dispersive result (to within $5\%$)~\cite{Leutgeb:2022cvg}. It also makes the HW1 result with $a_\mu^{\pi^0}=63.4\times 10^{-11}$ almost completely
coincident with the dispersive result~\cite{Leutgeb:2021mpu,Leutgeb:2022lqw}.
In \cref{fig:pionTFFHW1} the HW1 result (with finite quark masses) is compared with experimental data and the dispersive result of Ref.~\cite{Hoferichter:2018kwz}
for the two choices of $g_5$.

\begin{figure}[t]
\centerline{
\includegraphics[width=0.495\textwidth]{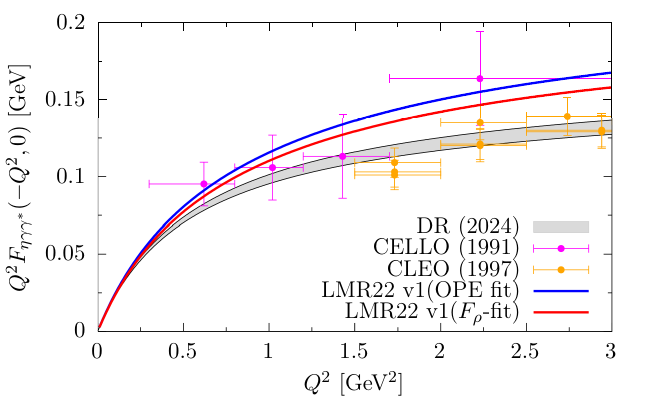}
\includegraphics[width=0.495\textwidth]{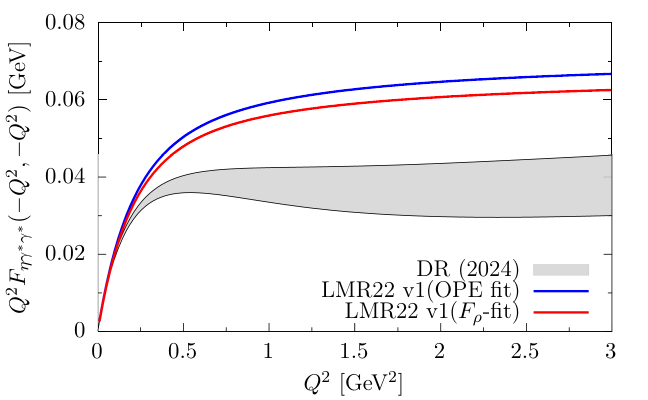}
}
\centerline{
\includegraphics[width=0.495\textwidth]{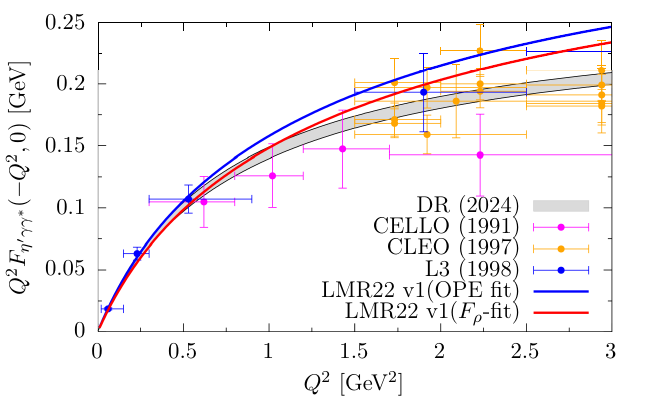}
\includegraphics[width=0.495\textwidth]{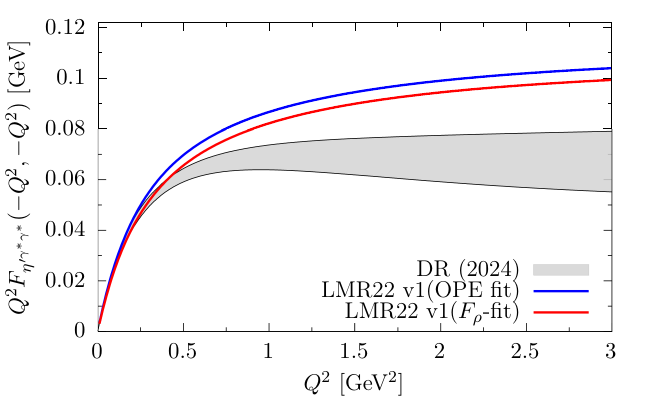}
}
\caption{Comparison of the hQCD results for the $\eta$ and $\eta'$ TFFs of Ref.~\cite{Leutgeb:2022lqw} (LMR22) with
OPE fit of $g_5^2$ (blue lines) and a $10\%$ reduced coupling $g_5^2$ (red lines)
from fitting $F_\rho$ instead, compared to the dispersive results of
Refs.~\cite{Holz:2024lom,Holz:2024diw} and experimental data.}
\label{fig:etasTFFHW}
\end{figure}

More recently, the pseudoscalar TFFs and their contribution to $a_\mu$ have been evaluated also in the SW model~\cite{Colangelo:2023een}. In this model,
the pseudoscalar TFFs turn out to approach the BL limit from above, and the good agreement with data obtained in the HW1 model is lost. The result for the pion contribution is $a_\mu^{\pi^0}=75.2\times 10^{-11}$, significantly larger than the results from data-driven and lattice approaches. A drawback of the SW model as set up in Ref.~\cite{Colangelo:2023een} is that the background of the bifundamental scalar, which implements chiral symmetry breaking, is taken over from the HW models but is not a solution of the field equations in the modified geometry. On the other hand, self-consistent solutions would make the chiral condensate proportional to quark masses~\cite{Karch:2006pv}, so that some modifications are needed which presumably would change the behavior at larger momenta. Improved hQCD models with nontrivial dilaton background exist in the literature~\cite{Gursoy:2007er,Gursoy:2007cb,Jarvinen:2011qe}, but they involve a rather large parameter space; so far they have not been worked out with regard to the HLbL amplitude.

\paragraph{Excited pions}

With the exception of the inherently chiral HW2 model, the hQCD models also have excited pseudoscalars.
Those decouple from the anomaly in the chiral limit, where $F_{\pi_n}\to0$ for $n>1$, but even then have nonvanishing two-photon decay amplitudes. Away from the chiral limit, one has the sum rule~\cite{Leutgeb:2021mpu}
\begin{equation}
    \sum_{n=1}^\infty F_{\pi_n} F_{\pi_n\gamma\gamma}=\frac{N_c}{12\pi^2}
\end{equation}
and asymptotic behavior according to the BL result given in \cref{pionTFFas}.

In the massive HW models, the first excited pion has a mass of $1.9\GeV$ (HW1) or $1.7\GeV$ (HW3), contributing $a_\mu^{\pi^{0*}}=0.7\times 10^{-11}$ or $0.8\times10^{-11}$, respectively~\cite{Leutgeb:2021mpu}. In Ref.~\cite{Domenech:2010aq} it was noted that with a different scaling dimension for the bifundamental operator one can adjust the HW3 model such that the mass of the $\pi(1300)$ resonance is matched. This increases the contribution to $a_\mu^{\pi^{0*}}=1.5\times10^{-11}$; however,
the predicted decay amplitude $|F_{\pi^{0*}\gamma\gamma}|=0.206\GeV^{-1}$ is in conflict with the estimated experimental upper bound \cite{Colangelo:2019uex}
$|F_{\pi(1300)\gamma\gamma}|<0.0544(71) \GeV^{-1}$.
Without this modification, the contribution from the collection of excited pions with $n\ge2$ is $\sum_{n\ge2}a_\mu^{\pi_n}\simeq 0.8\times10^{-11}$ in the HW models. In the SW model of Ref.~\cite{Colangelo:2023een}, where the first excited pion has a mass of about $2.1\GeV$, already this mode would contribute $1.68\times10^{-11}$.
In contrast to the HW model, the contributions from higher excited pions do not fall off quickly, despite their more rapidly growing masses~\cite{LMR-SW}.

\paragraph{$\eta$ and $\eta'$}

In order to describe the $\eta$ and $\eta'$ pseudoscalars consistently, the larger strange quark mass as well as the anomaly induced mass of the $\eta_0$ singlet has to be taken into account. In the chiral models studied initially in Refs.~\cite{Cappiello:2010uy,Leutgeb:2019zpq,Leutgeb:2019gbz,Cappiello:2019hwh},
the decay constants and masses were adjusted manually, whereas in the early HW1 study
of Ref.~\cite{Hong:2009zw} with finite quark masses the Chern--Simons action was implemented incorrectly~\cite{Leutgeb:2021mpu}. From the chiral HW2 model, Ref.~\cite{Cappiello:2019hwh} obtained the ranges $a_\mu^{\eta}=(14\ldots21)\times 10^{-11}$
and $a_\mu^{\eta'}=(10\ldots16)\times 10^{-11}$, while from the chiral HW1 model Ref.~\cite{Leutgeb:2019zpq} estimated $a_\mu^{\eta}=18.2\times 10^{-11}$
and $a_\mu^{\eta'}=15.6\times 10^{-11}$.

In the hQCD models with bifundamental scalar field, quark masses can be introduced by the nonnormalizable modes of the latter, while the anomalous mass of the $\eta_0$ singlet can be obtained by a Witten--Veneziano mechanism. This permits to predict masses, decay constants, and TFFs of $\eta$ and $\eta'$ when the quark masses are fixed to reproduce $\pi$ and $K$ masses.
The best result so far was obtained in Ref.~\cite{Leutgeb:2022lqw} by
using the Katz--Schwartz model~\cite{Katz:2007tf} for implementing the U(1)$_A$ anomaly in a HW AdS background, but allowing for a gluon condensate (referred to as model v1 in LMR22 \cite{Leutgeb:2022lqw}).
This reproduced $\eta$ and $\eta'$ masses and
$F_{\eta^{(\prime)}\gamma\gamma}(0,0)$ to within a few percent.\footnote{The SW model of Ref.~\cite{Colangelo:2023een} with U(1)$_A$ anomaly also achieves this, but the TFFs of $\eta$ and $\eta'$ approach the BL limit from above like the SW result for $\pi^0$, leading to rather poor agreement with data, in particular in the case of $\eta$.}

The HW results for the singly and doubly virtual TFFs are displayed in \cref{fig:etasTFFHW}. While they are fully compatible with the
CA results used in WP20, they are significantly
above the new dispersive results~\cite{Holz:2024lom,Holz:2024diw}.
Correspondingly the results for $a_\mu^{\eta}$ and $a_\mu^{\eta'}$ listed in
\cref{tab:HWamuHLbL}, which would agree with WP20 within errors, are larger than those of Refs.~\cite{Holz:2024lom,Holz:2024diw}, by $3\sigma$ and $2\sigma$, respectively.

\nocite{Leutgeb:2024rfs}

\subsubsection{Axial-vector TFFs in hQCD}
\label{sec:AxialTFFhQCD}

All hQCD models involve an infinite tower of axial vector mesons.
Their coupling to two photons is determined
by the same action \cref{SCS} that gives rise to the anomalous TFFs of
the pseudoscalars. It has the form \cite{Leutgeb:2019gbz,Cappiello:2019hwh}
\begin{equation}
\label{calMa}
            \mathcal{M}_{\mathcal{A}_n\gamma^*\gamma^*}=i\frac{N_c}{4\pi^2}\mathrm{tr}(\mathcal{Q}^2 t^a)\,\epsilon_{(1)}^\mu \epsilon_{(2)}^\nu
            \epsilon_\mathcal{A}^{*\rho} \epsilon_{\mu\nu\rho\sigma}
            \left[q_{(2)}^\sigma Q_1^2 A_n^a(Q_1^2,Q_2^2)-q_{(1)}^\sigma Q_2^2 A_n^a(Q_2^2,Q_1^2)\right]\,,
\end{equation}
involving an asymmetric structure function
        \begin{equation}
            A_n(Q_1^2,Q_2^2) = \frac{2g_5}{Q_1^2} \int_0^{z_0} dz \left[ \frac{d}{dz} \mathcal{J}(Q_1,z) \right]
            \mathcal{J}(Q_2,z) \,\psi^A_n(z)\,,
        \end{equation}
where $\psi^A_n(z)$ is the holographic profile function of the $n$th axial vector meson. The Landau--Yang theorem~\cite{Landau:1948kw,Yang:1950rg}, which forbids the decay of an axial vector meson into two real photons, is satisfied because $\partial_z\mathcal{J}(0,z)\equiv0$.

In the notation of Refs.~\cite{Hoferichter:2020lap,Zanke:2021wiq,Hoferichter:2023tgp},
$A_n(Q_1^2,Q_2^2)\propto \mathcal{F}_2(q_1^2,q_2^2)\equiv -\mathcal{F}_3(q_2^2,q_1^2)$. The most general decay amplitude of axial-vector mesons would permit one further structure function~\cite{Pascalutsa:2012pr,Roig:2019reh,Zanke:2021wiq},
denoted as $\mathcal{F}_1\equiv \mathcal{F}_{\mathrm{a}_1}$ in Refs.~\cite{Hoferichter:2020lap,Zanke:2021wiq,Hoferichter:2023tgp}.
The holographic prediction that this vanishes is consistent with the results of the study of all available experimental data on $f_1(1285)$ in Ref.~\cite{Hoferichter:2023tgp}.

\begin{figure}[t]
\centerline{\includegraphics[width=0.6\textwidth]{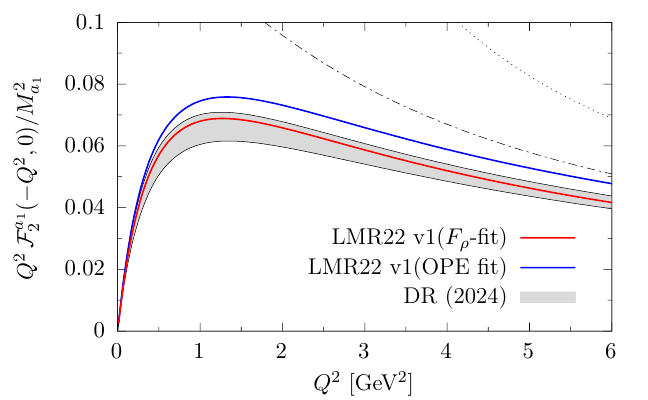}
}
\caption{Comparison of the hQCD results for the $a_1$ TFF of Ref.~\cite{Leutgeb:2022lqw} (LMR22) with
OPE fit of $g_5^2$ (blue lines) and a 10\% reduced coupling $g_5^2$ (red lines)
from fitting $F_\rho$ instead, compared to the dispersive result of
Ref.~\cite{Ludtke:2024ase}. The BL limit~\cite{Hoferichter:2020lap} is shown as a dotted (dash-dotted) gray line without (with) mass corrections.
}
\label{fig:a1TFFHW1}
\end{figure}

In all hQCD models that are asymptotically AdS, the asymptotic form of $A_n(Q_1^2,Q_2^2)$ is given by \cite{Leutgeb:2019gbz}
\begin{equation}
                A_n(Q_1^2,Q_2^2) \to C
                \frac{12\pi^2 F^A_{n}}{N_c Q^4}
                \frac1{w^4}\left[
                w(3-2w)+\frac12 (w+3)(1-w)\log\frac{1-w}{1+w}
                \right]\,,
\end{equation}
for $Q^2=(Q_1^2+Q_2^2)/2\to\infty$, $w=(Q_1^2-Q_2^2)/(Q_1^2+Q_2^2)$,
which agrees with the result obtained by light-cone expansions in Ref.~\cite{Hoferichter:2020lap}, where in fact $\mathcal{F}_1$
does not contribute terms of order $1/Q^{4}$.
Again, $C=g_5^2N_c/(12\pi^2)=1$ when $g_5$ is fixed such
that the asymptotic behavior of the
vector correlator matches the leading-order OPE result in QCD.

In \cref{fig:a1TFFHW1} the result obtained for the singly-virtual $a_1$ TFF in the HW model
of Ref.~\cite{Leutgeb:2022lqw} (LMR22) is compared to the recent dispersive result of
Ref.~\cite{Ludtke:2024ase}. With $g_5$ fixed by fitting $F_\rho$ instead of the
OPE result, the BL limit is reached asymptotically only at the level of
$89\%$, but for virtualities at moderate values of $Q^2$ a perfect agreement with the dispersive
result is obtained.

\subsubsection{Short-distance constraints in hQCD}

In the BTT basis of the HLbL four-point function~\cite{Colangelo:2015ama}, the longitudinal SDC of MV~\cite{Melnikov:2003xd} in the region
$Q_1^2 \simeq Q_2^2\gg Q_3^2 \gg M_\rho^2$ and $Q_4=0$, which is
governed by the chiral anomaly and protected by its nonrenormalization
theorem,
reads \cite{Melnikov:2003xd,Bijnens:2019ghy,Colangelo:2019lpu,Colangelo:2019uex}
\begin{equation}\label{MVSDC}
\lim_{Q_3\to\infty} \lim_{Q\to\infty} Q^2 Q_3^2 \bar\Pi_1(Q,Q,Q_3)=-\frac{2}{3\pi^2}
\end{equation}
for $N_c=N_f=3$, while the symmetric limit is 2/3 of this,
\begin{equation}
\label{symSDC}
\lim_{Q\to\infty} \lim_{Q\to\infty} Q^4 \bar\Pi_1(Q,Q,Q)=-\frac{4}{9\pi^2}\,.
\end{equation}
The short-distance behavior of the form factors of both pseudoscalars
and axial-vector mesons implies that each individual meson gives a pole contribution
with $\bar\Pi_1(Q,Q,Q_3)\simeq Q^{-2} Q_3^{-4}$.
However, in Ref.~\cite{Leutgeb:2019gbz,Cappiello:2019hwh} it was shown that in hQCD a summation over the
infinite tower of axial-vector mesons changes this. The infinite sum yields
\begin{equation}
    \bar{\Pi}_1^{AV}=-\frac{2C}{\pi^2 Q_3^2} \int_0^{z_0} dz \int_0^{z_0} dz' \mathcal{J}'(Q,z) \mathcal{J}(Q,z) \mathcal{J}'(Q_3,z') G_A(0;z,z')\,, \label{HWPi1}
\end{equation}
where $G_A$ is the Green function for the axial-vector mode equation
at $q^2=0$.
For large $Q,  Q_3\gg M_\rho$, \cref{HWPi1} is dominated by $z, z'\ll z_0$, where all hQCD models with (at least asymptotic) AdS geometry have
$\mathcal{J}(Q,z)\to Qz K_1(Qz)$, and
\begin{equation}\label{GAsmallz}
G_A(0,z,z') = \frac12 \left(\mathrm{min}(z,z')\right)^2\left(1+\mathcal{O}(Q^{-n})+
\mathcal{O}(Q_3^{-n})\right)\,,\quad n>0\,,
\end{equation}
when $z=\xi/Q$ and $z'=\xi'/Q_3$.
For $Q\gg Q_3\gg M_\rho$ this leads to
\begin{equation}
\label{Pi1AVlimit}
\lim_{Q_3\to\infty} \lim_{Q\to\infty} Q^2 Q_3^2 \bar\Pi_1^\mathrm{AV}(Q,Q,Q_3)
=\frac{C}{2\pi^2} \int_0^\infty d\xi
\xi^2\frac{d}{d\xi}[\xi K_1(\xi)]^2=
-\frac{2}{3\pi^2}C\,,
\end{equation}
reproducing \cref{MVSDC} for $C=1$, i.e., whenever the
leading-order OPE behavior of the vector correlator is matched.
As shown in Ref.~\cite{Leutgeb:2021mpu}, an infinite tower of excited
pseudoscalars, which is present in the hQCD models apart from the HW2 model,
does not contribute to \cref{Pi1AVlimit}.
An analogous analysis of the symmetric limit $Q_1^2 \simeq Q_2^2\simeq Q_3^2 \gg M_\rho^2$ yields
\begin{equation}
\label{Pi1AVsymlimit}
\lim_{Q\to\infty} Q^4 \bar\Pi_1^\mathrm{AV}(Q,Q,Q)
=-\frac{C}{\pi^2} \int_0^\infty d\xi
\,\xi [\xi K_1(\xi)]^3=
-0.812\frac{4}{9\pi^2}C\,,
\end{equation}
which reproduces the correct power behavior,
but for $C=1$ is only $81\%$ of the result \cref{symSDC}.

\subsubsection{Complete axial-sector HLbL contributions to \texorpdfstring{$a_\mu$}{} in hQCD}

The virtue of bottom-up hQCD models is to provide simple, self-contained
hadronic models for the HLbL tensor covering all virtualities, including the
consequences of the axial anomaly for the longitudinal SDC in full, while also capturing $81\%$ of the transverse SDC.
Because the pseudoscalar and axial-vector contributions are coupled, complete results
for the HLbL contribution in a given hQCD model require that both be fully evaluated.
So far this has been done in the chiral HW models studied in Ref.~\cite{Leutgeb:2019gbz,Cappiello:2019hwh}, flavor-symmetric HW models with
a quark mass that matches $M_\pi$ in Ref.~\cite{Leutgeb:2021mpu}, and more recently in an $N_f=2+1$ model including the anomalous $\eta'$ mass in Ref.~\cite{Leutgeb:2022lqw}.
The resulting values for $a_\mu^\text{HLbL}$ are summarized in \cref{tab:HWamuHLbL}.
There CCDGI(set1-set2) refers to the results for the HW2 model from Ref.~\cite{Cappiello:2019hwh}, which
made an estimate for the $N_f=2+1$ case by using the chiral TFFs with physical pseudoscalar masses and by introducing a different decay constant
for $\eta'$, with two sets of parameters which either
undershoot the BL limit by a factor of $0.616$, or overshoot the $\rho$ mass,
but give pseudoscalar TFFs that bracket the experimental results.
LMR22 refers to the HW model Ref.~\cite{Leutgeb:2022lqw} with
quark masses and an $\eta_0$ mass contribution from the U(1)$_A$ anomaly. The latter entails a pseudoscalar glue ball ($G/\eta''$) that mixes with $\eta$ and $\eta'$ and also couples to two photons.
Most recently, variations of such hQCD models have been considered in
Ref.~\cite{Leutgeb:2024rfs} with scalar-extended Chern--Simons actions, motivated
by string-theoretic constructions~\cite{Casero:2007ae}. One of those
named CS''($F_\rho$-fit)
is included in \cref{tab:HWamuHLbL}, which has the virtue of
reproducing quite well the experimentally observed~\cite{L3:2001cyf,L3:2007obw}
equivalent photon rates and mixing of
$f_1^{(\prime)}$ axial-vector mesons, albeit at the expense of a less good
fit of $\eta'$ data. Moreover, the $\pi^0$ TFF appears to be overestimated
at nonzero virtualities. However, the total sum of all contributions
turns out to be similar to the LMR22 result due to larger contributions
from the tower of excited pseudoscalars and axials, which compensate for the
smaller result from ground-state axial-vector mesons.
Reference~\cite{Leutgeb:2024rfs} still considers the LMR22 v1($F_\rho$-fit)
as the best-guess hQCD result, with the variation obtained in
the CS$''$ (and the partially scalar-extended CS$'$) models providing
estimates for inherent errors.

A comparison of these results with other approaches that determine
the pole contributions from single meson exchanges is made
in \cref{sec:formfactors,sec:finalnumber} below, 
and reasonable agreement is obtained within estimated systematic errors.\footnote{An exception in recent work concerns Refs.~\cite{Dorokhov:2011zf,Radzhabov:2023odj,Radzhabov:2025pyx}, where significantly smaller results,
$a_\mu^{\pi^0+\eta+\eta'}=58.5(8.7)\times10^{-11}$ and
$a_\mu^{a_1+f_1+f_1'}=3.6(1.8)\times10^{-11}$, are obtained in
a nonlocal quark model. However, these results and the other contributions in those papers do not correspond to pure pole contributions, and it is thus not clear how to compare to the results discussed here.} 

\begin{table}[t]  
\centering
\small
\renewcommand{\arraystretch}{1.1}
\begin{tabular}{lcccc}
\toprule
$a_\mu\times 10^{11}$ & CCDGI(set1-set2) & LMR22 v1(OPE fit) & LMR22 v1($F_\rho$-fit) & CS$''$($F_\rho$-fit)\\
 \midrule
 $\pi^0$ & $57\text{--}75$ & $66.1$ &  $63.4$ & $68.8$\\
 $\eta$ & $14\text{--}21$ & $19.3$ &  $17.6$ & $17.2$\\
 $\eta'$ & $10\text{--}16$ & $16.9$ &  $14.9$ & $12.2$\\
 $G/\eta''$ & -- & $0.2$ &  $0.2$ & $2.6$\\
 $\sum_{PS^*}$ & -- & $1.6$ & $1.4$ & $3.2$ \\
 \midrule
 $\pi^0+\eta+\eta'$ & $81\text{--}112$ & $102.3$ & $95.9$ & $98.2$ \\
 PS poles & $81\text{--}112$ & $104$ &  $97.5$ & $104$\\\midrule
 $a_1$& -- & $7.8$ &  $7.1$ & $4.9$ \\
 $f_1+f_1'$ & --  & $20.0$ &  $17.9$ & $12.0$ \\
 $\sum_{a_1^*}$ & -- & $2.5$ &  $2.6$ & $2.3$\\
 $\sum_{f_1^{(\prime)*}} $ & -- & $4.0$ &  $3.5$ & $7.9$\\
 \midrule
 AV+LSDC & $28(2)$ & $34.3$ &  $31.1$ & $27.0$ \\
 AV+P$^*$+LSDC & $28(2)$ & $36.0$ & $32.7$ & $32.8$ \\
 \midrule
 Total& $110\text{--}140$ & $138$ & $129$ & $131$ \\
 \bottomrule
\end{tabular}
\renewcommand{\arraystretch}{1.0}
\caption{Breakdown of the axial-sector contributions to $a_\mu^\text{HLbL}$
    in $N_f=2+1$ hQCD models, where the HLbL contributions of pseudoscalars and axial-vector mesons have been evaluated completely. LMR22 v1($F_\rho$-fit)
    represents the best-guess model of Ref.~\cite{Leutgeb:2022lqw}. The table gives
     a partial summary of the results from Refs.~\cite{Cappiello:2019hwh,Leutgeb:2022lqw,Leutgeb:2024rfs} for the different contributions to $a_\mu^\text{HLbL}$.
    }
    \label{tab:HWamuHLbL}
\end{table}

\subsubsection{Scalar and tensor contributions in hQCD}
\label{sec:hQCDscalartensor}
 In WP20 the broad resonance $f_0(500)$ was covered by the $S$-wave $\pi\pi$ scattering contribution and largely dominates those of heavier scalars, which  have been estimated in the narrow-width approximation.  The contributions of the lightest tensor mesons $f_2(1270)$, $f_2(1565)$, $a_2(1320)$ and $a_2(1700)$, were evaluated in the narrow-resonance approximation in Refs.~\cite{Pauk:2014rta,Danilkin:2016hnh}. The total contribution from scalar and tensor resonances with masses greater that $1\GeV$ was estimated to be $a_\mu^\text{scalar+tensors}=-1(3)\times10^{-11}$.
There were no dispersive estimate for the single tensor resonance exchanges due to some spurious dependence on the result on the tensor basis. Recently, this problem has been overcome in Ref.~\cite{Hoferichter:2024fsj} allowing for the dispersive evaluation reported in Ref.~\cite{Hoferichter:2024vbu}

The hQCD models which involve a bifundamental scalar field for
implementing chiral symmetry breaking also feature nonets of
scalar mesons and their excitations. However, in the minimal
setup of Refs.~\cite{Erlich:2005qh,Karch:2006pv,DaRold:2005mxj},
they do not give rise to a coupling to photons.
Such a coupling is only naturally present for the flavor-singlet
dilaton, which corresponds to a scalar glue ball, and metric
fluctuations, corresponding to a tensor glue ball.
In Ref.~\cite{Hechenberger:2023ljn}, this was considered
within the Witten--Sakai--Sugimoto model with two-photon decay
rates in the keV range, but the resulting $a_\mu$ contributions
have been found to be negligible.

In Ref.~\cite{Cappiello:2021vzi}, the minimal HW1 model of
Ref.~\cite{Erlich:2005qh,Karch:2006pv,DaRold:2005mxj} was extended by adding, in the 5D Lagrangian, two new interaction terms  quadratic in both the 5D scalar field and  tensor gauge fields in \cref{S5D}. Chiral symmetry breaking generated  scalar--photon--photon interaction vertices and hence a nonvanishing scalar TFF.  Only the chiral limit was considered in that model. The contribution of the full tower of scalar resonances in each channel was obtained. The final result for the total contribution of $\sigma(500)$, $a(980)$, and $f_0(980)$ was $a_\mu^S=-9(2) \times 10^{-11}$, with roughly $90\%$ of it coming from $\sigma(500)$, being compatible with the estimates in WP20. A shortcoming of the model is that the asymptotic behavior of the scalar TFF does not match  the one expected from a BL analysis~\cite{Hoferichter:2020lap} in QCD for large photon virtualities. In Ref.~\cite{Leutgeb:2024rfs} it has been shown that this mismatch persists also beyond the chiral limit, although consistency with the OPE limit is achieved.

\begin{figure}
\centering
\includegraphics[width=.66\textwidth]{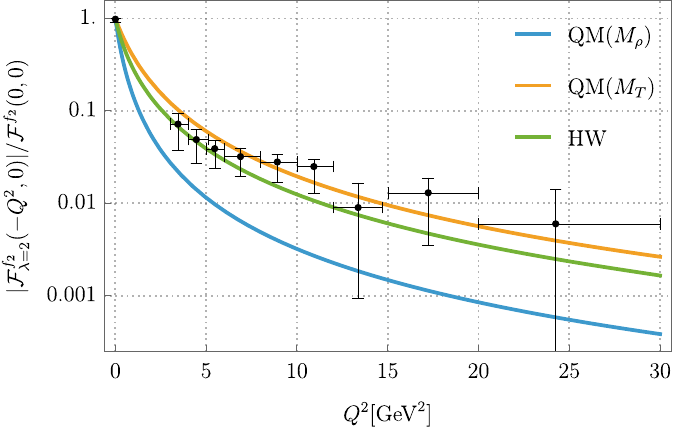}
\caption{Comparison of hQCD result (HW) and quark model
ansatz with $\Lambda=M_\rho$~\cite{Hoferichter:2024bae} and $M_T$~\cite{Hoferichter:2020lap,Schuler:1997yw} for the singly-virtual tensor TFF for helicity $\lambda=2$ with Belle data~\cite{Belle:2015oin} for the $f_2(1270)$, normalized to
$\F^{f_2}(0,0)\equiv
\sqrt{5 \Gamma_{\gamma\gamma}/(\pi\alpha^2 M_T)}$.
Figure adapted from Ref.~\cite{Cappiello:2025fyf}.}
\label{F1T-Belle-comparison}
\end{figure}

In Ref.~\cite{Colangelo:2023een}
a holographic description of tensor mesons was considered
which postulated two-photon vertices as obtained from
metric fluctuations, analogously to tensor glue balls, but
with phenomenologically determined coupling strengths.
In this paper a result of $a_\mu^{f_2(1270)}\simeq +0.6\times 10^{-11}$ was reported for both HW and SW models, when the experimental decay width is matched,
however, in a simplified and
thus incomplete evaluation.
Recently, in Ref.~\cite{Cappiello:2025fyf},
a full evaluation of the tensor meson pole contribution has been performed
for the HW models.\footnote{This contribution is the same
in all the HW models considered above, depending only on $z_0$, which
is matched to $M_\rho$.}
This involves two structure functions,
$\F^T_{1}$ and $\F^T_{3}$ in the notation of Ref.~\cite{Hoferichter:2024fsj},
which have the correct $Q^2$ scaling expected from the analysis
of Ref.~\cite{Hoferichter:2020lap}, but not the same asymmetry functions.
The normalized TFF of singly-virtual tensor decays involves only
$\F^T_{1}$. This turns out to agree well with available Belle data, see \cref{F1T-Belle-comparison}, in particular at the lowest available
$Q^2$ values. Using this TFF together with $\F^T_{3}$, which only contributes
in the doubly-virtual case, the pole contribution as defined by
the dispersive approach in the optimized basis of Ref.~\cite{Hoferichter:2024fsj}
yields a positive $a_\mu$ result of 
$3.4(4)\times 10^{-11}$ when
refit to the lowest $f_2$, $a_2$, $f_2'$ multiplet ($3.2(4)\times 10^{-11}$
from the integration region $Q_i<1.5\GeV$).
This is in stark contrast to the quark model, where similarly large but
negative results are obtained~\cite{Hoferichter:2024bae}.
In fact, setting $\F^T_{3}$ to zero in the
holographic result would reverse the sign and increase the absolute value
to $-4.6(6)\times 10^{-11}$ for $a_\mu^{f_2+a_2+f_2'}$ (of which
$-4.5(6)\times 10^{-11}$ is due to $Q_i<1.5\GeV$), see \cref{tab:HWtensorresults}.

In Ref.~\cite{Mager:2025pvz} it was moreover shown that summing over the
infinite tower of tensor mesons gives a contribution to the symmetric
longitudinal SDC, where axial vector mesons give only $81\%$ of the OPE result,
see \cref{Pi1AVsymlimit}, without modifying the result for the MV
constraint \cref{Pi1AVlimit}. Fixing the tensor normalization by the
condition that tensor and axial-vector contributions together match
both longitudinal SDCs leads to a somewhat larger result than the original
prescription of Ref.~\cite{Katz:2005ir} and it also restores the correct $N_c$ dependence
of tensor meson couplings.

In contrast to the case of pseudoscalar and axial-vector contributions,
evaluating the complete contribution of tensor mesons to the
HLbL amplitude in the HW model does give
a different result than the pole contribution as defined by
the dispersive approach.
Instead of $3.17\times 10^{-11}$ ($2.93\times 10^{-11}$ from the IR region), the lowest tensor mode
as given by the model contributes~\cite{Cappiello:2025fyf,Mager:2025pvz}
\begin{equation}\label{eq:hQCDT1}
a_\mu^{T,n=1}=8.3\times 10^{-11}\,,\qquad
a_\mu^{T,n=1}|_\text{IR}=7.4\times 10^{-11}\,,
\qquad\text{($F_\rho$-fit)}
\end{equation}
which increases to a value of
\begin{equation}\label{eq:hQCDTall}
a_\mu^{T}=11.1\times 10^{-11}\,,\qquad
a_\mu^{T}|_\text{IR}=8.5\times 10^{-11}\,,
\qquad\text{($F_\rho$-fit)}
\end{equation}
when the
whole tower of tensor meson modes is included.
From the perspective of the holographic model, which
is self-contained and does not need matching with pQCD
results except for the parameters, it is
more natural to take the complete contribution into account, which
is independent of the choice of basis.
Remarkably, the sum over individual pole contributions from the tensor tower
converges much more slowly than the sum over the complete contributions, while
both sums tend to equally large totals~\cite{Cappiello:2025fyf}.

In \cref{tab:Qregions} below, the contribution
in hQCD that is attributed to the ground-state tensor mesons, however,
refers only to the $n=1$ pole contribution as defined in the dispersive approach;
the remaining contribution from the tower of tensor modes is included under
``other'' contributions, together with the contributions from the axial sector
as given by the LMR22 v1($F_\rho$-fit) result of \cref{tab:HWamuHLbL}.
However, a larger value is obtained than in all previous estimates,
with a positive sign due to the presence of $\F^T_3$.
Unfortunately, at present there are no experimental data available to test
the hQCD prediction of a nonzero $\F^T_{3}$ as this requires
double virtuality.\footnote{A nonzero $\F^T_{3}$ also appears in
so-called minimal models of tensor mesons~\cite{Mathieu:2020zpm}, which contributes even at vanishing virtualities due to poles at $Q_i^2=0$, thereby violating gauge invariance,
in contrast to the hQCD case; see \cref{{sec:ResonanceContributions}} for a related discussion of the tensor TFFs.}

\begin{table}
    \centering
    \small
    \renewcommand{\arraystretch}{1.1}
    \begin{tabular}{ccccc}
    \toprule
    $M_T$ [GeV] & $\Gamma_{\gamma \gamma}$ [keV] &IR &Mixed  & $a_\mu\  [10^{-11}]$\\
    \midrule
       $1.235$  & $2.3+0.8+0.2$  & $9.61-6.68=2.93$ & $0.55-0.32=0.23$   &   $3.17$ \\
    \midrule
    $1.2754(8)$ & $2.65(45)$ & $6.6-4.3=2.3$ & $0.4-0.2=0.2$ &  $2.4(4)$ \\
    $1.3182(6)$ & $1.01(9)$ & $2.2-1.3=0.9$ & $<0.1$ &  $0.9(1)$ \\
    $1.5173(24)$ & $0.08(2)$   & $0.1-0.04=0.06$ & $<0.01$ & $0.06(2)$ \\
    \midrule
    $f_2+a_2+f_2'$ & & $8.9-5.7=3.2$ & $0.5-0.3=0.2$ & $3.4(4)$ \\
    $f_2+a_2+f_2'$ w/o $\F^T_3$ & &
    $4.6-9.1=-4.5$ & $0.2-0.3=-0.1  $ &  $-4.6(6)$ \\
    \bottomrule
    \end{tabular}
    \renewcommand{\arraystretch}{1.0}
    \caption{$a_\mu$ results for the tensor meson pole contributions, in units of $10^{-11}$, obtained by inserting the hQCD results for $\F^T_{1,3}$ in the formulae obtained in Ref.~\cite{Hoferichter:2024fsj}, where
    the first line corresponds to the lowest HW tensor mode normalized as in Ref.~\cite{Cappiello:2025fyf,Mager:2025pvz}  ($F_\rho$-fit)
     and the remaining lines to a refit to experimental data.
    The IR region is defined by $Q_i\le Q_0=1.5\GeV$ for all $i=1,2,3$, the mixed region by having only one or two below $Q_0$. The first term of each sum is from $\overline{\Pi}_{1,2}$, while the second one is from the other $10$ structure functions. Table adapted from Ref.~\cite{Cappiello:2025fyf}.}
    \label{tab:HWtensorresults}
\end{table}

\subsubsection{Holographic QCD summary}

The strongest point of hQCD models so far has been
clearly the ability to match
anomalies in the hadron sector in a way that is consistent with
asymptotic conformal invariance and corresponding SDCs.
Comparatively minimal models succeed in reproducing the
masses and photon couplings of $\pi^0$, $\eta$, and $\eta'$ rather well,
which makes their predictions for the experimentally less well-known
axial-vector sector that is responsible for the saturation of the
MV SDC also quantitatively interesting.

In principle, such models would also permit the study of $\pi^\pm$ and
$K^\pm$ box contributions, by evaluating one-loop Witten diagrams, but
this has not been carried out so far. However, such contributions
are anyway determined with negligible errors by the dispersive approach.
On the other hand,
the similarly important $S$-wave rescattering contribution,
corresponding to light scalar resonances, depends on phenomenological
data for which hQCD models do not provide particular clues.
However, using available data on scalars as input, the hQCD model calculation
of Ref.~\cite{Cappiello:2021vzi} obtained a result
that completely agrees with the dispersive approach, with
somewhat enlarged error.

It thus seems more pertinent to compare only the available hQCD results
for the axial sector and the tensor sector
with corresponding results from other approaches,
or to combine it with the
contribution from pseudoscalar boxes,
which in the dispersive approach~\cite{Colangelo:2017fiz,Colangelo:2017qdm,Stamen:2022uqh} amounts to
$-16.4(2)\times 10^{-11}$
together with the $S$-wave rescattering contribution from either
the dispersive approach or the hQCD model calculation~\cite{Cappiello:2021vzi}, the latter amounting to $-9(2)\times 10^{-11}$.
The sum total given in \cref{tab:HWamuHLbL} for the best-fitting
LMR22 v1($F_\rho$-fit) model,
combined with the relatively large positive contribution 
from the tower of tensor modes~\cite{Cappiello:2025fyf,Mager:2025pvz},
$11.1^{+1.3}_{-3.0}\times 10^{-11}$,
would then correspond to
\begin{equation}
    a_\mu^\text{HLbL}\Big|_\text{hQCD\; completed}= 114_{-4}^{+10}\times 10^{-11}\,,
\end{equation}
where the estimate for an (hQCD-systematic) error accommodates the range of the
($F_\rho$-fit) models considered in Ref.~\cite{Leutgeb:2024rfs}
(including also the OPE-fit models would enlarge the upwards error to +17 points).

\subsection{Rational approximants, resonance chiral theory, and Regge evaluations}
\label{Section:OtherApproaches}

\subsubsection{\texorpdfstring{$\pi^0,\eta,\eta'$}{} poles}
\label{sec:CApoles}

Until recently, the estimate for the $\eta$, $\eta'$ poles was obtained from CA alone, which provides us with a mathematical framework to approximate the TFFs in a systematically improvable way, while other approaches available at the time were deemed insufficient to meet the quality criteria outlined in WP20~\cite{Aoyama:2020ynm}.  As such, new alternative evaluations are necessary to better establish their current estimate. In particular, while for the singly-virtual case a wealth of experimental data confirms the idea that TFFs are reasonably described by VMD, the doubly-virtual case stands on a different footing.
More specifically, while the singly-virtual behavior is dominated by the physics of the light vector resonances (and any approach faithfully describing them should provide an excellent parameterization), for the doubly-virtual case the lightest vector mesons cannot fulfill pQCD asymptotics. Instead, an infinite number of them, or some alternative mechanism supplying the missing physics, is required.
While guidance can be found from pQCD at asymptotically large energies, 
the fact that fundamental physics is missing at intermediate energies calls for a critical inspection of this region and potential systematic effects therein.
Advances along these lines have been undertaken in different theoretical approaches, see \cref{sec:Pseudoscalars,section:Pseudoscalar,sec:functional}.

Concerning data, the very useful singly-virtual (spacelike) measurements from the \babar~\cite{BaBar:2009rrj, BaBar:2011nrp}, Belle~\cite{Belle:2012wwz}, CELLO~\cite{CELLO:1990klc},  CLEO~\cite{CLEO:1997fho}, and LEP~\cite{L3:1997ocz} collaborations are complemented, for the doubly-virtual case, only by five data points published by \babar{} in the $\eta^\prime$ channel~\cite{BaBar:2018zpn}. Therefore, it is natural to wonder if $\eta'$ data could shed light on the $\pi^0,\eta$ cases based on U(3)$_F$ symmetry (and its breaking) arguments. In this context, Ref.~\cite{Estrada:2024cfy} explores this possibility incorporating, also, lattice QCD results for double virtuality \cite{Gerardin:2023naa, ExtendedTwistedMass:2022ofm, ExtendedTwistedMass:2023hin}, that improves notably the control on this difficult dependence of the lightest pseudoscalar TFFs and has a nonnegligible impact on $a_\mu^\text{PS-poles}$~\cite{Estrada:2024cfy}.
To such endeavor, Ref.~\cite{Estrada:2024cfy} uses the framework of R$\chi$T, incorporating resonances along similar lines as developed in Refs.~\cite{Ecker:1988te, Ecker:1989yg} in a large-number of colors expansion~\cite{tHooft:1973alw} and accounting for flavor-breaking\footnote{It improves over and/or extends earlier descriptions in this framework~\cite{Kampf:2011ty,Roig:2014uja,Guevara:2018rhj,Kadavy:2020hox}.} and includes all operators with resonance fields that---after being integrated out---saturate the $\mathcal{O}(p^6)$ chiral low-energy constants~\cite{Kampf:2011ty}. In this approach, the effect of the infinite tower of vector and pseudoscalar resonances is accounted for with a second (effective) meson multiplet and contact terms---of ChPT origin---that, together with the first multiplet, comply with the singly- and doubly-virtual leading asymptotic behaviors predicted by pQCD~\cite{Nesterenko:1983ef,Novikov:1983jt,Brodsky:1973kr,Lepage:1980fj}\footnote{The subleading correction to the doubly-virtual result, computed within the OPE, is also fit, with its coefficient determined from QCD sum rules~\cite{Novikov:1983jt}.} and order by order with the chiral-symmetry-breaking expansion. The relations found among the Lagrangian couplings agree with the consistent SDCs in the odd-intrinsic parity sector~\cite{Roig:2013baa}. In addition to the singly- and doubly-virtual experimental and lattice data quoted above,\footnote{Reference~\cite{Estrada:2024cfy} explains that only three lattice data points per channel can be used, to represent faithfully the lattice information.} Ref.~\cite{Estrada:2024cfy} also fits the $P\to\gamma\gamma$ decay widths~\cite{ParticleDataGroup:2022pth} and uses the $\eta$--$\eta^\prime$ mixing parameters~\cite{Feldmann:1998vh} as stabilization points.
It is instructive~\cite{Estrada:2024cfy} to relate this R$\chi $T approach to the CA: the resulting $\pi^0$ TFF corresponds to a $C_2^2$ and the $\eta^{(\prime)}$ to $C_4^4$ CA\footnote{Note that coefficients must be chosen to reproduce the appropriate high-energy behavior. Further, it must be noted that, by contrast to the most general CA (such as in Ref.~\cite{Masjuan:2017tvw}) the denominator reduces to a product of poles, which would relate to what is known in the literature as Pad\'e-type approximants, with different convergence properties.} (for which clearly there is not enough data to determine all their unknown coefficients), although in the chiral limit they correspond as well to a $C_2^2$. Even though Ref.~\cite{Estrada:2024cfy} considers data fits using independent $C_2^2$ for the three $P$ channels in which the OPE is constrained, it is concluded that these would require more doubly-virtual data (either from experimental measurements or from lattice QCD) to prevent overfitting. 
Interestingly, R$\chi$T allows one to predict from \babar{} $\eta^\prime$ doubly-virtual measurements the $\pi^0$ and $\eta$ corresponding dependence due to the built-in U(3)$_F$ relations, though lattice results (as future measurements on any of the $P$ channels would also do) improve this knowledge. This is not possible in CAs that, in addition, provide limited ability to extract information from lattice QCD data due to the absence of U(3)$_F$ relations~\cite{Estrada:2024cfy}. 
Concluding, the R$\chi$T model predicts
\begin{align}
a_{\mu}^{ \{\pi^0,\eta,\eta' \} } &=  \Big\{
61.9(0.6)(^{+2.4}_{-1.5})\,,
15.2(0.5)(^{+1.1}_{-0.8})\,,
14.2(0.7)(^{+1.4}_{-0.9})
\Big\}\times 10^{-11}\,, \notag\\
a_\mu^\text{PS-poles}&=91.3(1.0)(^{+3.0}_{-1.9})\times 10^{-11} \, ,
\end{align}
where the first uncertainty is statistical and the second one is the systematic theory error. In order of importance, the sources of the main systematic uncertainties are: the effect of modeling the $n=2,\dots,\infty$ multiplets by a second effective one, subleading corrections in $1/N_c$, and the combination of experimental and lattice data. In turn, they contribute by $\lbrace+1.8,\pm1.5,+0.4\rbrace,$ $\lbrace+1.0,\pm0.5,-0.6\rbrace,$ $\lbrace+1.4,\pm0.3,-0.8\rbrace$ to the error for the $\lbrace\pi^0,\eta,\eta^\prime\rbrace$ channels (see Ref.~\cite{Estrada:2024cfy} for a detailed discussion, including smaller uncertainties as well). This result is fully compatible for the $\pi^0$, compared to the estimate based on DR or CA, and the $\eta$, $\eta'$ ones agree well with previous estimates based on CA ($a_\mu^{\eta} = 16.3(1.0)(0.5)(0.9)\times10^{-11}$, $a_\mu^{\eta'} = 14.5(0.7)(0.4)(1.7) \times10^{-11}$).

\begin{figure}[t]
    \centering
    \includegraphics[width=\linewidth]{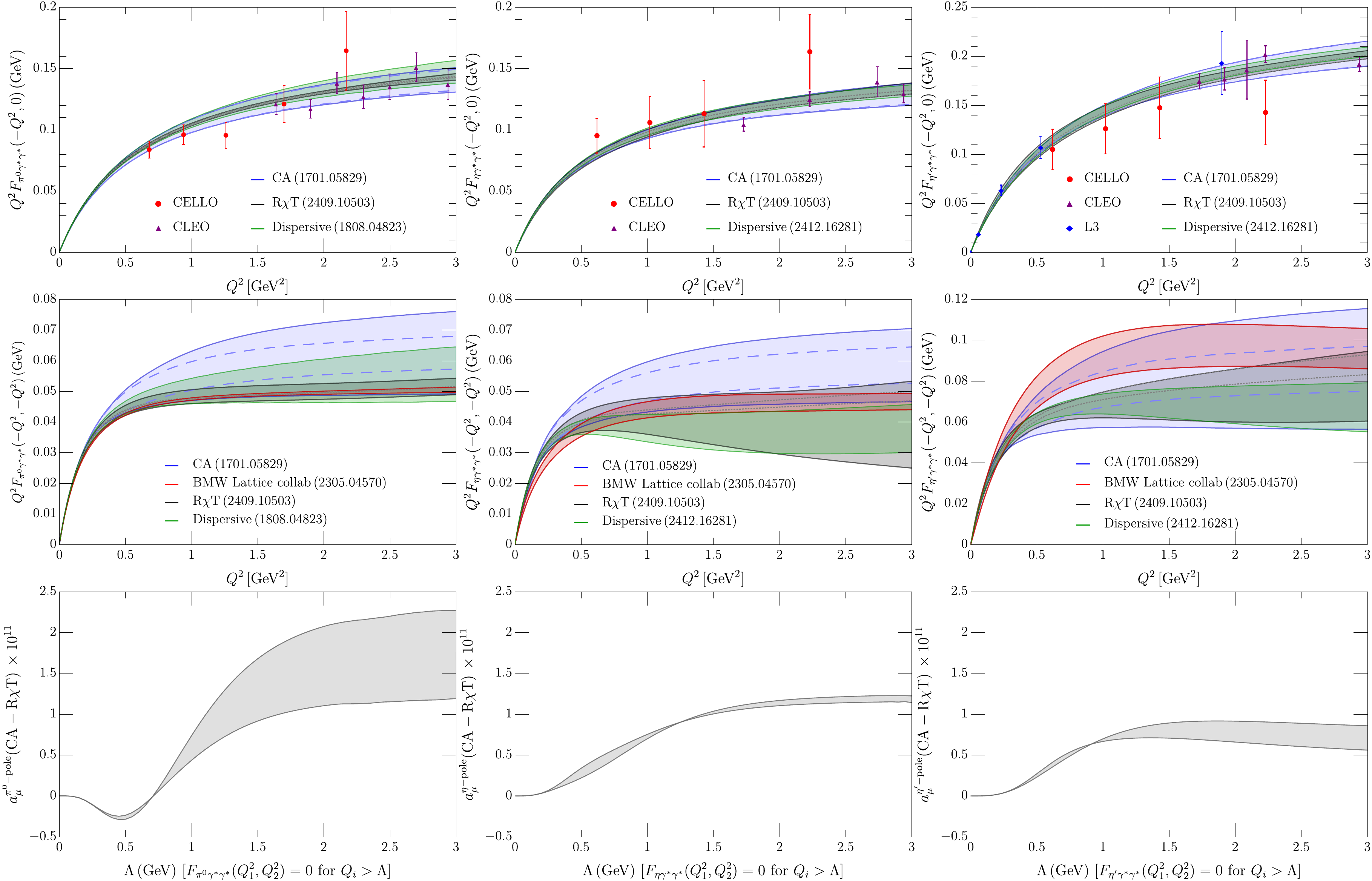}
    \caption{First row: the singly-virtual TFFs. The inner dashed-blue (dotted-black) band is the CA (R$\chi$T) result with statistical uncertainties, whereas the full ones include a combination in quadrature of statistics and systematic uncertainties. The green band is the dispersive results from Refs.~\cite{Holz:2024lom,Holz:2024diw} (cf.\ \cref{sec:Pseudoscalars}). The red band stands for the $z$-expansion fit to lattice data~\cite{Gerardin:2023naa} (no systematics). Second row: the results analogous to the first row for doubly-virtual TFFs. Third row: the difference of the two approaches to the partial contribution to $a_\mu^{\text{HLbL}, P}$ up to a given scale.}
    \label{fig:CaRcLcomp}
\end{figure}

The results are compared to CAs for the singly and doubly-virtual TFFs in \cref{fig:CaRcLcomp} together with a $z$-expansion fit to lattice data from Ref.~\cite{Gerardin:2023naa}. Good agreement is found---even within statistical uncertainties---for the singly-virtual TFFs between R$\chi$T and CA. In turn, differences appear in doubly-virtual kinematics for the $\pi^0$ and $\eta$ case, yet results are fully consistent if systematic uncertainties are accounted for (for CA this is estimated as the difference among the $C^1_2$ and $C^0_1$ approximants). In this respect, it would be interesting to better constrain the doubly-virtual behavior in the region relevant for the HLbL contribution. In particular, in the last row from \cref{fig:CaRcLcomp} we show the difference of the contributions among CA and R$\chi$T to $a_\mu^{\text{HLbL}}$ if the TFFs are cut off at an upper scale $\Lambda$ (the band stands for the central one obtained with CA, but lacks any kind of uncertainties). As can be observed, the main differences arise in the energy region up to $Q_i^2\simeq 2\GeV^2$, well below \babar's data for the doubly-virtual $\eta'$ case. This region could be accessed at BESIII~\cite{Anderson:2024hdr} and could help improve the uncertainties above provided the experimental errors are below the plotted bands.

As a final remark, the size of doubly-virtual systematics emphasizes the relevance of less-controlled intermediate physics effects in both approaches. This can be better appreciated in CA, where no data was employed to constrain the doubly-virtual TFFs. There, one observes that singly-virtual data is vastly dominated by a single hadronic scale in the spacelike region, that relates to the success of VMD. In turn, the slower convergence for the doubly-virtual one is related to the existence of further scales that require one to go beyond the simplest $C^0_1$ approximant or naive VMD approaches, requiring further input to determine the additional parameters in CA and R$\chi$T approaches.

\subsubsection{Other mesons and short-distance constraints}
\label{sec:Regge}
\begin{sloppypar}
Previous results on axial-vector-meson contributions scattered over a large range~\cite{Aoyama:2020ynm} due to differences in their implementation. While some of them may be ascribed to an incorrect high-energy doubly-virtual behavior~\cite{Pauk:2014rta}, additional ambiguities arise from incorrect claims on the Landau--Yang theorem, kinematic zeros~\cite{Jegerlehner:2017gek}, or the treatment of the axial-vector meson propagator~\cite{Jegerlehner:2017gek,Roig:2019reh}, that clearly required an improved understanding in Lagrangian/resonance-exchange models. Along these lines, recent evidence has accumulated that axial-vector mesons play a leading role in fulfilling SDCs~\cite{Cappiello:2019hwh,Leutgeb:2019gbz,Masjuan:2020jsf}.
\end{sloppypar}

In particular, Ref.~\cite{Masjuan:2020jsf} makes use of the important connection of SDCs in the MV limit~\cite{Melnikov:2003xd} of the HLbL tensor to the $\langle VVA \rangle$ Green's function and the anomaly. The central point is to advocate for a tensor basis free of kinematic singularities that allows for a transparent identification of the role of intermediate states.  This contrasts to the standard $\langle VVA \rangle$ basis employed in earlier discussions of the MV limit, which basis naturally splits longitudinal, $w_L$, from transverse contributions at the cost of introducing kinematic singularities. More specifically, borrowing the notation from Ref.~\cite{Aoyama:2020ynm}, Eq.~(4.76), one can show that ($q_{12}=q_1+q_2$)~\cite{Knecht:2020xyr,Masjuan:2020jsf}
\begin{equation}\label{eq:VVALl}
    \frac{w_L(q_1^2, q_2^2, q_{12}^2)}{8\pi^2} = \frac{\tilde{w}_0(q_1^2, q_2^2, q_{12}^2)}{8\pi^2} -\frac{(q_1^2 +q_2^2)}{q_{12}^2} \frac{w_T^+(q_1^2, q_2^2, q_{12}^2)}{8\pi^2}
   + \frac{(q_1^2 -q_2^2)}{q_{12}^2} \frac{w_T^-(q_1^2, q_2^2,q_{12}^2)}{8\pi^2}\,.
\end{equation}
The form factors on the RHS ($\tilde{w}_0,w_T^{\pm}$) contain only physical singularities, but $w_L$ clearly contains, in general, kinematic ones.
More specifically, pseudoscalar poles exclusively appear in $w_L$, $\tilde{w}_0$, whereas $w_T^{\pm}$ contain axial-vector meson poles. Clearly, the only physical pole at $q_{12}^2\to0$ (in the chiral limit) identifies with the $\pi^0$ one, and its residue relates to its TFF, that only fulfills the anomaly ($w_L =2N_c/q_{12}^2$) for real photons. For virtual photons, the anomaly can only be fulfilled if transverse contributions in \cref{eq:VVALl} are included, showing that transverse physics cannot be excluded when accounting for the anomaly. It is also along these lines that Ref.~\cite{Knecht:2020xyr} discusses the $(g-2)_\mu$ soft-photon limit. Identifying the form factors there~\cite{Knecht:2020xyr}, $-8\pi^2\{ w_0 +w_2, w_1, w_2, w_3 \} = \{ \tilde{w}_0, w_T^-, w_T^+ , \tilde{w}_T^- \}$, one obtains $-w_L(q_1^2,0,q_1^2)/8\pi^2 = w_0(q_1^2,0,q_1^2) +w_1(q_1^2,0,q_1^2)$. Only the full tensor is free from unphysical poles.
This is the principal line of thought in Ref.~\cite{Masjuan:2020jsf}, which investigates axial-vector meson contributions to the HLbL tensor and $\langle VVA \rangle$. There it is shown how (an infinite number of) axial-vector mesons, that introduce pole-free contributions to $w_L$ in \cref{eq:VVALl}, together with the $\pi^0$ pole, can fulfill the anomaly. In particular, this fixes all the ambiguities with axial-vector-meson propagators previously discussed that, together with high-energy constraints, fix the way in which Lagrangian models/resonance-exchange axial contributions should be accounted for. Note also that similar findings have been obtained recently albeit in a dispersive approach for the $\langle VVA \rangle$ function~\cite{Ludtke:2024ase}. In addition, the discussion explains why R$\chi$T calculations of axial exchanges---that are necessarily transverse---cannot reproduce SDCs with axial-vector mesons alone, and contact terms would be required. This is nevertheless standard in R$\chi$T, but discourages its use in the context of axial contributions to HLbL scattering.

\begin{figure}[t]
    \centering
    \includegraphics[width=0.5\linewidth]{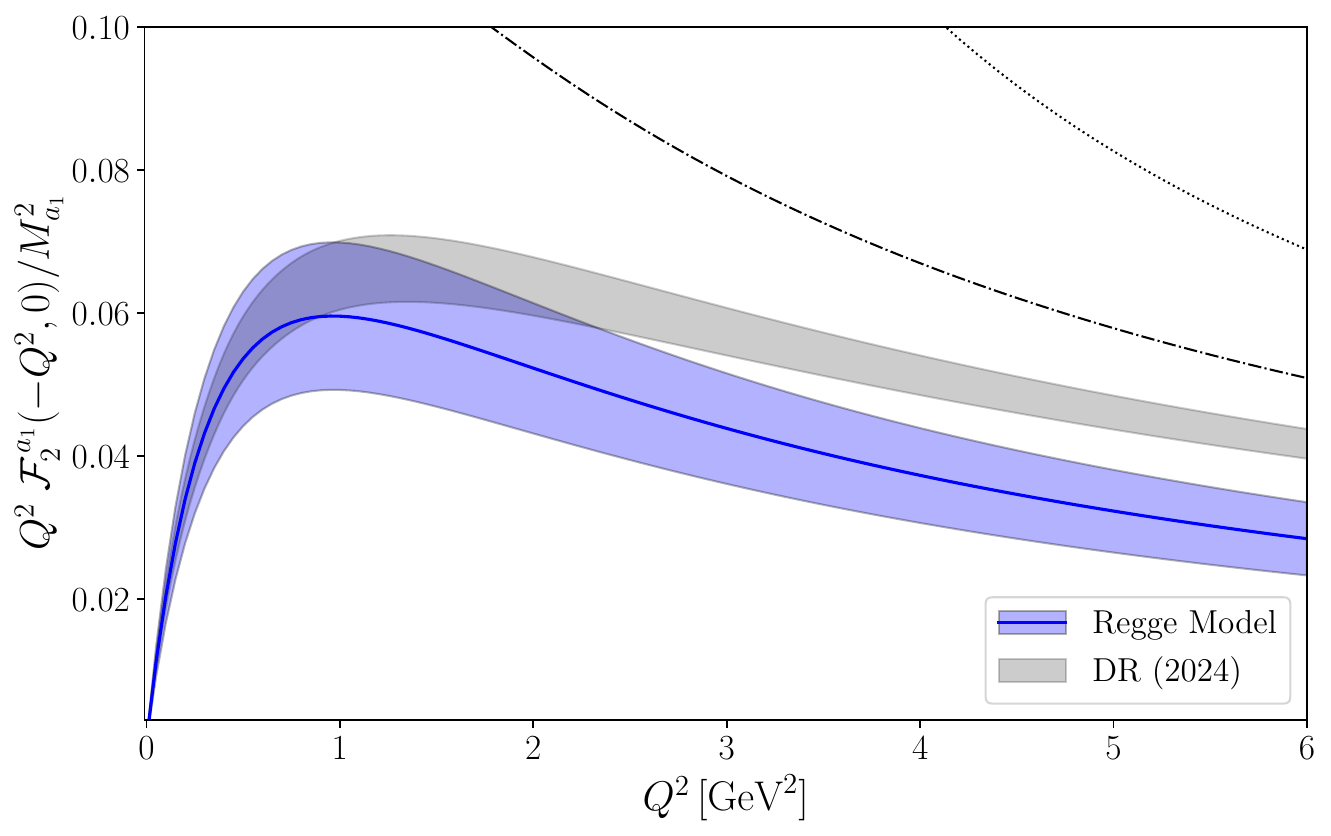}
    \caption{The ground state $a_1$ meson singly-virtual TFF $M_{a_1}^{-2}\mathcal{F}_2^{a_1}(-Q^2,0)$. The results from the Regge model~\cite{Masjuan:2020jsf} identify with \cref{eq:AxReggeModel} and are shown in blue, with uncertainties from $F_{a_1}$ and the half-width rule applied to $M_{a_1}$. The dispersive result~\cite{Ludtke:2024ase} is shown as a gray band and the BL limit~\cite{Hoferichter:2020lap} is shown as a dotted (dash-dotted) gray line without (with) mass corrections.}
    \label{fig:a1Regge}
\end{figure}

Accounting for all these subtleties and focusing on the symmetric form factor exclusively, Ref.~\cite{Masjuan:2020jsf} finds, for the ground state multiplet,
\begin{equation}
\label{amu_axial_RchT}
a_{\mu}^{a_1} +a_{\mu}^{f_1} +a_{\mu}^{f_1'} = [ 5.4(^{+3.7}_{-3.3}) +8.3(^{+3.4}_{-2.9}) +2.3(^{+1.1}_{-0.9}) ] \times 10^{-11}  =16.0(^{+5.1}_{-4.5}) \times 10^{-11}\,,
\end{equation}
where the symmetric form factor is analogous in functional form to that in \cref{eq:AxReggeModel}. Note, however, that for a complete evaluation also the two asymmetric TFFs need to be taken into account, as their numerical contribution might not be negligible and only ${\mathcal F}_1$ is suppressed asymptotically~\cite{Hoferichter:2020lap}, see \cref{sec:ResonanceContributions,sec:AxialTFFhQCD}. Still, in this framework, this suffices to fulfill the SDCs for the HLbL tensor in the MV regime when an infinite number of axial-vector mesons are considered (see further comments in this respect below). 
The $a_1$ input used for \cref{amu_axial_RchT} was obtained from the R$\chi$T estimate in Ref.~\cite{Roig:2019reh} under U(3)$_F$ relations and the assumption that $f_1$, $f_1'$ are, analogously to the vector case, mostly light and strange mesons, respectively.
This seems reasonable from $\text{BR}[f_1\to\gamma \rho]/\text{BR}[f_1\to\gamma \phi]$ decays, and the equivalent $2\gamma$ decay widths for $f_1$, $f_1'$ are in line with experiment due to their R$\chi$T scaling $\propto M_A/\Lambda_H^3$ differing from the standard assumption $\propto M_A^{-2}$~\cite{L3:2007obw} (in holographic models $\propto \Lambda_H^{-2}$), outlining once more the need for experimental input for the axials.\footnote{The mixing angle of $f_1$, $f_1'$ can be estimated as well from the equivalent two-photon decay widths~\cite{L3:2001cyf,L3:2007obw} under U(3) assumptions. Apart from data on axial-vector TFFs directly, these flavor assumptions can also be tested in radiative decays to vector mesons. For a more complete discussion, see Refs.~\cite{Zanke:2021wiq,Hoferichter:2023tgp} and \cref{tab:exp_prio}.} The estimates $B_{2S}^{a_1}(0,0)=0.245(63)\GeV^{-2}$ (in agreement with $0.252(30)\GeV^{-2}$~\cite{Ludtke:2024ase} or $0.230\GeV^{-2}$~\cite{Leutgeb:2024rfs}) and $\Lambda_{a_1} = 1.0(1)\GeV$ lead in addition to a doubly-virtual high-energy behavior in agreement with the OPE, $\lim_{Q^2\to\infty} Q^4 B_{2S}(-Q^2,-Q^2) = \sum_a M_A F_A^a \text{Tr}\big(\mathcal{Q}^2\lambda^a\big)$~\cite{Masjuan:2020jsf}.
Further, in order to fulfill the MV SDCs, Ref.~\cite{Masjuan:2020jsf} introduces a Regge-like model for an infinite tower of axial-vector meson resonances and their form factors where
\begin{equation}\label{eq:AxReggeModel}
    B_{2S}^{A_n^a} = \frac{4F_A \text{Tr}(\mathcal{Q}^2 \lambda^a) M_{A_n^a}}{[q_1^2 +q_2^2 -(M_a^2 +n\Lambda_V^2)]^2}\,, \quad M_{A_n^a}^2 = M_{A_0^a}^2 +n\Lambda_A^2\,,\qquad  a=\{ 3,q,s \}\,,
\end{equation}
with $A_n^a$ the $n$th axial-vector resonance with isospin index $a$ and decay constant $F_A= 140(10)\MeV$. MV SDCs, the anomaly, and the asymptotic behavior of the doubly-virtual $\pi^0$ TFF completely fix the parameter for the form factors to $\Lambda_V = 4\pi F_A/\sqrt{N_c}$ and $M_3^2 = (8\pi^2/3)[F_A^2 +2F_{\pi}^2]$, predicting $B_{2S}^{a_1}(0,0)=0.246\GeV^{-2}$ and a slope for the $\pi^0$ TFF in agreement with recent estimates~\cite{Masjuan:2017tvw,Hoferichter:2018kwz}, that warrants then the appropriate low-energy behavior for the transverse part of the $\langle VVA\rangle$ Green's function (find similar discussions in Ref.~\cite{Ludtke:2024ase}). This reflects the connection between the pseudoscalar and axial-vector sectors in the model. To bypass a detailed analysis of $\eta$--$\eta'$ mixing, the parameters $M_{q,s}$ in \cref{eq:AxReggeModel} were fixed to reproduce the experimental $2\gamma$ equivalent decay widths for the ground state, and $\Lambda_A^2$ was taken to reproduce the Regge trajectories. As a result, the estimate for the tower of axial-vector mesons was found to be
\begin{equation}
a_{\mu}^{\text{HLbL}; \{ a_1,f_1,f_1'\} } = \{ 5.89, 10.52, 1.97 \}\times 10^{-11}\,,
\qquad \sum _{n=1}^{99} a_{\mu}^{\text{HLbL}; \{ a_1,f_1,f_1'\} }  = \{ 3.67, 8.46, 0.90 \}\times 10^{-11}\,.
\end{equation}
While not discussed in Ref.~\cite{Masjuan:2020jsf}, by choosing $\Lambda_A = \Lambda_V$, it is possible to satisfy the MV SDC for the transverse part as well ($w_L(Q^2) = 2w_T(Q^2)$ asymptotically~\cite{Melnikov:2003xd}).
In the following, we provide an update for the tower of excited axial-vector meson states if $\Lambda_A =\Lambda_V$ is taken, including as well uncertainties from the half-width rule~\cite{Masjuan:2012gc}, $F_A$ ~\cite{Nugent:2013hxa}, and experimental uncertainties on $B_{2S}^{f_1,f_1'}(0,0)$\footnote{We note that changes are minor, and differences are mostly due to the updated numerical integration recipe Divonne from CUBA library~\cite{Hahn:2004fe}.}
\begin{equation}
a_{\mu}^{\{ a_1,f_1,f_1'\} }  = \{ 6.0(0.2), 10.5(1.5), 2.0(0.5) \}\times 10^{-11}\,, \qquad
\sum _{n=1}^{99} a_{\mu}^{\{ a_1,f_1,f_1'\} }  =  \{ 3.6(0.4), 8.6(0.7), 0.91(0.09) \}\times 10^{-11} \,,
\end{equation}
but still lacking a genuine systematic uncertainty from the model itself. In order to compare against other results, it is interesting to provide their contributions divided in the different regions introduced in Refs.~\cite{Hoferichter:2024bae,Hoferichter:2024vbu} for $Q_0=1.5\GeV$ (cf.\ \cref{sec:ResonanceContributions}). In particular, $ a_{\mu}^{ a_1,f_1,f_1'}\vert_{Q_i <Q_0} = 10.9(1.0)\times 10^{-11}$, $ a_{\mu}^{ \sum A^*}\vert_{Q_i <Q_0} = 3.2(6)\times 10^{-11}$, $a_{\mu}^{ \sum A}\vert_{\text{Mixed}} = 12.8(5)\times 10^{-11}$, $a_{\mu}^{ \sum A}\vert_{Q_i>Q_0} = 4.8(1)\times 10^{-11}$,  $a_{\mu}^{ \sum A} = 31.7(1.6)\times 10^{-11}$, where $\sum A^*$ refers to the sum of excited axial-vector multiplets and $\sum A$ to the full tower (including the ground state), cf.\ \cref{tab:Qregions}. In addition, we show the prediction of the model above for the ground state $a_1$ meson $M_{a_1}^{-2}\mathcal{F}_2^{a_1}(-Q^2,0)$ TFF compared to the dispersive result in \cref{fig:a1Regge}. Remarkably, the plot shows a nice agreement at low energies. Note, in particular, that the model above predicts $M_{a_1}^{-2}\mathcal{F}_2^{a_1}(0,0)=0.25(5)\GeV^{-2}$, to be compared with the dispersive result $0.25(2)\GeV^{-2}$. At high energies some differences appear, that are expected from the simplistic model and the omission of the antisymmetric form factors. More specifically, the latter seem necessary, in connection with the anomaly, to correctly match the singly-virtual $\pi^0$ TFF asymptotic behavior. Note that, still, the pQCD relation $w_L(Q^2)=2w_T(Q^2)$ holds.

Finally, also the tensor contributions $a_\mu^{ (a_2+f_2+f_2')\text{-pole}}$ have recently been computed in R$\chi$T~\cite{Estrada:2025bty}. Using a large-$N_c$ expansion, working in the chiral limit, and including two vector resonance multiplets to fulfill the SDCs, this contribution reads $\big(-4.5^{+0.5}_{-0.3}\big)\times 10^{-11}$ (cutting off the integrals above $Q_0=1.5\GeV$). This coincides with the result given in the last line of table \ref{tab:HWtensorresults}, corresponding to the fact that in this approach the only nonvanishing TFF is $\mathcal{F}_1^T$, in line with the discussion in \cref{sec:ResonanceContributions}, and its shape turns out similar to hQCD, see \cref{sec:hQCDscalartensor}. The numerical effect is therefore covered by the uncertainty estimates discussed in \cref{sec:finalnumber}.

\subsection{Functional methods}
\label{sec:functional}

\nocite{Huber:2020keu,Eichmann:2021zuv,Aguilar:2024dlv,Eichmann:2016yit,Ding:2022ows,Raya:2024ejx}

\begin{figure}[!t]
	\centering%
	\includegraphics[width=0.28\linewidth]{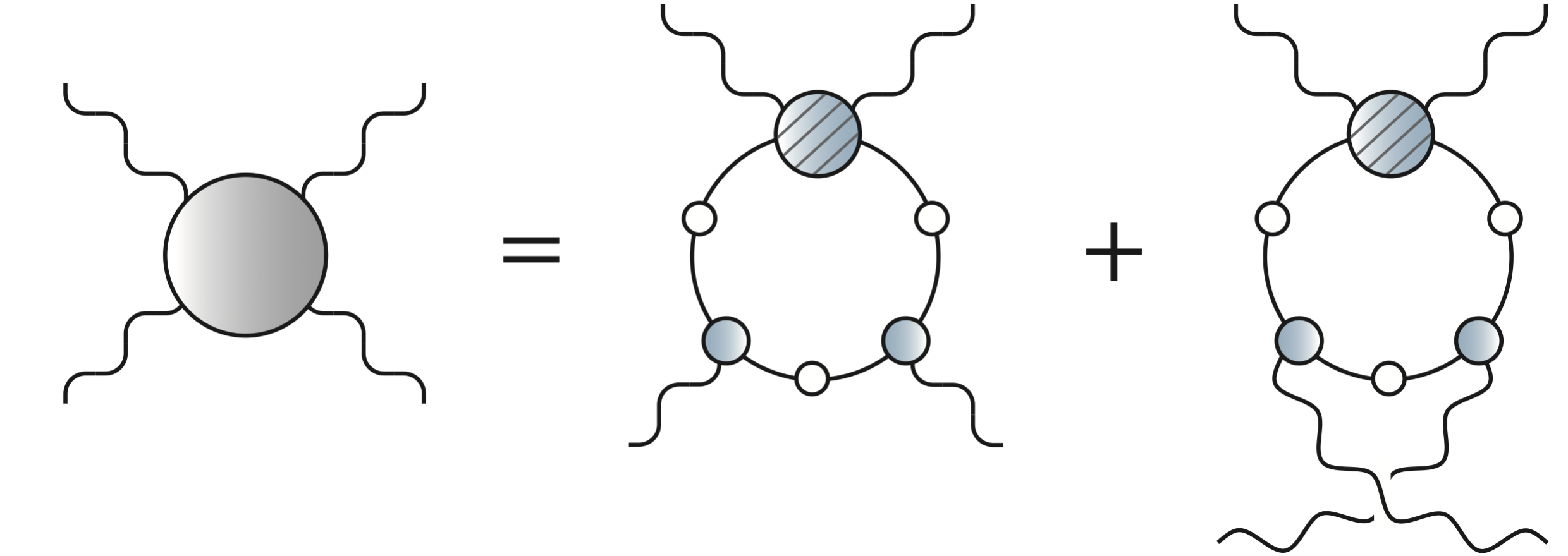} \qquad\quad	\includegraphics[width=0.65\linewidth]{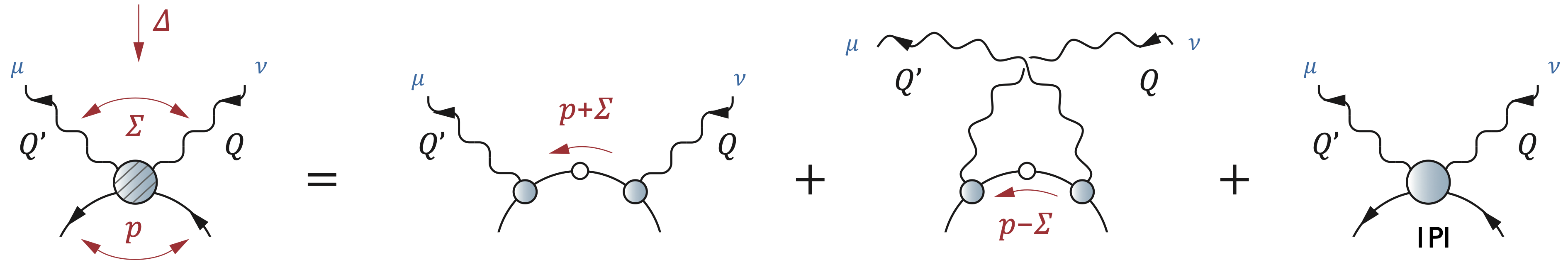}%
	\caption{\label{fig:DSE1}%
		HLbL in terms of a quark Compton vertex including contributions from the dressed quark loop and meson exchange and box
		diagrams, as well as other contributions subsumed in the 1PI part of the Compton vertex. Figures taken from Ref.~\cite{Eichmann:2012mp}.
	}
\end{figure}

Another approach to the muon $g-2$ problem employs functional methods, in particular a  formalism based on coupled Dyson--Schwinger equations (DSEs) and Bethe--Salpeter equations (BSEs)  in QCD.
The DSEs are the quantum equations of motion which relate QCD's $n$-point correlation functions with each other, including the dressed quark and gluon propagators, the
quark--gluon and three-gluon vertices, etc. These are exact equations in QCD, but in practice they require truncations to obtain numerical solutions.
The status quo in the DSE sector is by now quite advanced and arriving at a point where ab-initio solutions,
whose only approximations amount to neglecting higher $n$-point functions, have become feasible~\cite{Huber:2020keu,Eichmann:2021zuv,Aguilar:2024dlv}.
On the other hand, its implementation in hadronic BSEs
is technically challenging and has so far generally been limited to the rainbow-ladder (RL) truncation, where the quark--(anti-)quark interaction is modeled by an effective gluon exchange~\cite{Eichmann:2016yit}.
This is the leading order in a systematic expansion of the kernel,
which satisfies chiral symmetry constraints by construction, including its spontaneous breaking which is responsible for hadron mass generation.
Over the past decades this approach has been employed in studying a wide range of hadron properties from light- and heavy-meson spectroscopy to baryon spectroscopy,
four-quark states, EM elastic and TFFs, parton distributions and various scattering amplitudes; see Refs.~\cite{Eichmann:2016yit,Ding:2022ows,Raya:2024ejx} for reviews.

In principle, the HLbL contribution within the DSE/BSE formalism can be evaluated through two different frameworks:
(i) directly and (ii) indirectly via the dispersive framework (see below).
First, in the direct approach, the HLbL amplitude can be systematically expressed in terms of the underlying
correlation functions like the quark propagator, the quark--photon vertex and quark--two-photon vertex, which
are then computed through their DSEs and BSEs~\cite{Eichmann:2016yit}. This yields a decomposition of the
amplitude in terms of a dressed quark loop plus diagrams with intermediate $q\bar{q}\to q\bar{q}$ four-point
functions, which automatically contain all possible meson poles. The latter can be rearranged such that the
central ingredient is the quark--two-photon vertex, see Fig.~\ref{fig:DSE1}, which has been calculated in the
context of nucleon Compton
scattering~\cite{Eichmann:2012mp}.
Therefore, this establishes an independent approach to compute HLbL scattering that is complementary to lattice QCD
and dispersion relations.

The full HLbL scattering amplitude is electromagnetically gauge invariant and thus transverse. This is preserved by
the RL truncation but does not hold for the individual diagrams contributing to the amplitude like
the dressed quark loop. The main obstacle so far is the lack of a complete basis for the HLbL amplitude that is
``minimal,'' i.e., free of kinematic constraints. The amplitude depends on 136 linearly independent tensors, of which
41 are purely transverse and 95 nontransverse~\cite{Eichmann:2014ooa,Eichmann:2015nra}. A minimal basis for the
transverse part alone is known~\cite{Colangelo:2014dfa,Eichmann:2015nra} but its analog for the nontransverse
part is still missing. However, the latter is needed for a projection onto all 136 tensors. This allows one to remove
gauge artifacts because any further approximation (in addition to the RL) would break gauge invariance and
induce kinematic singularities, which impedes an extraction of $a_\mu^\text{HLbL}$. An analogous but simpler example
was recently discussed in the context of axial-vector-meson TFFs contributing to HLbL scattering:
In that case, three tensors are transverse and three are nontransverse; the full RL calculation is gauge
invariant, but any further approximation would introduce gauge artifacts that can only be removed by projecting onto
the complete basis~\cite{Eichmann:2024glq}. The construction of a complete basis for HLbL scattering is ongoing work.

\begin{table}[t]
	\centering
    \small
	\renewcommand{\arraystretch}{1.1}
	\begin{tabular}{ccc} \toprule
		&  DSE/BSE          & WP20   \\ \midrule
		$\pi^0$ exchange                  & $62.6(1)(1.3)$     & $63.0^{+2.7}_{-2.1}$                 \\
		$\eta$ exchange                 & $15.8(2)(3)(1.0)$  & $16.3(1.4)$                         \\
		$\eta'$ exchange                & $13.3(4)(3)(6)$   & $14.5(1.9)$                         \\
		$\pi^0$, $\eta$, $\eta'$ exchange    & $91.6(1.9)$        & $93.8^{+4.0}_{-3.6}$                 \\\midrule
		$\pi$ box                       &$-15.7(2)(3)$    &$-15.9(2)$                        \\
		$K$   box                       &$-0.48(2)(4)$    &$-0.46(2)$                        \\
		$\pi$, $K$ boxes/loops    		&$-16.2(5)$       &$-16.4(5)$                        \\\midrule
		$S$-wave $\pi\pi$ rescattering	& --   &$-8.0(1.0)$                        \\
		higher scalar exchange         	&$-1.6(5)$        &$-2.0(2.0)$                        \\
		AV exchange (single)      		& $17.4(6.0)$        &   $6.0(6.0)$                      \\
		AV exchange (tower) + SDC   	& $24.8(6.1)$        &  $21.0(16.0)$                     \\\bottomrule
	\end{tabular}
	\renewcommand{\arraystretch}{1.0}
	\caption{DSE/BSE results for various contributions to $a_\mu^\text{HLbL}$ in units of $10^{-11}$ compared to the values quoted in WP20~\cite{Aoyama:2020ynm}.
		The results for pseudoscalar exchange and pseudoscalar-box contributions are from Refs.~\cite{Eichmann:2019tjk,Eichmann:2019bqf}
		and those from  axial-vector and scalar exchange from Ref.~\cite{Eichmann:2024glq}.}
	\label{tab:DSEBSE}
\end{table}

The second approach to HLbL scattering with functional methods has been advocated in
Refs.~\cite{Eichmann:2019tjk,Eichmann:2019bqf,Eichmann:2024glq}. Here the meson EM and transition form
factors entering in the HLbL amplitude are calculated separately in the DSE/BSE framework and then inserted into the
dispersive formalism described in \cref{Section:Dispersive}. In this case the systematics of meson-exchange and meson-box diagrams
is identical to the dispersive approach, while the effects from the quark loop must be implemented through SDCs, see \cref{sec:SDC}. Once again, all ingredients in the form factor calculations (quark propagators, quark--photon
vertices and meson amplitudes) are determined in the RL truncation without further approximations. In particular,
the quark--photon vertex dynamically develops vector-meson poles so that the form factors in the timelike region automatically
capture the physics of VMD.

\begin{figure}[t]
\centerline{\includegraphics[width=0.46\textwidth]{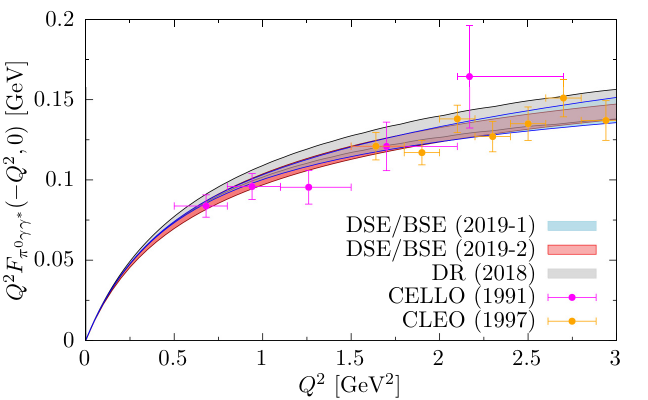}
\includegraphics[width=0.46\textwidth]{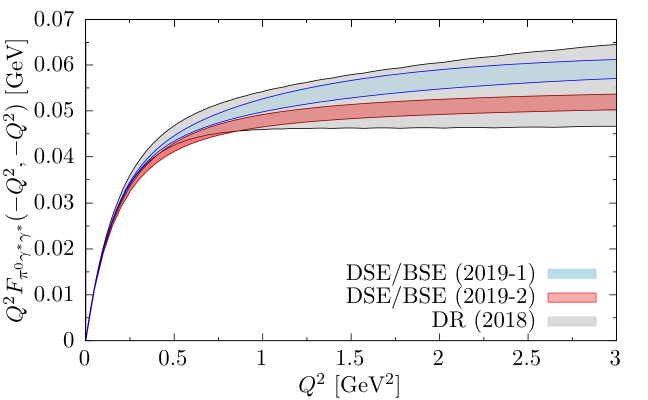}}
\centerline{\includegraphics[width=0.46\textwidth]{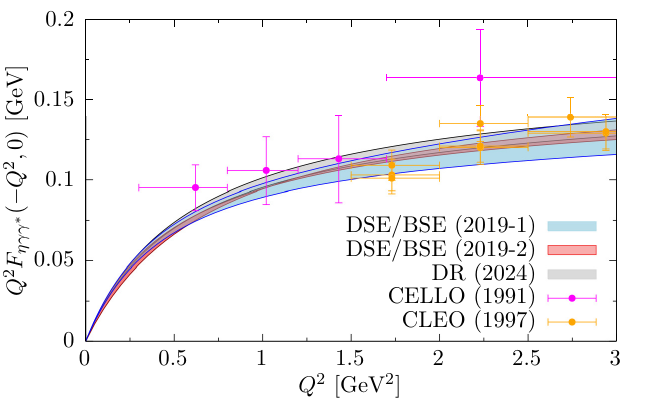}
\includegraphics[width=0.46\textwidth]{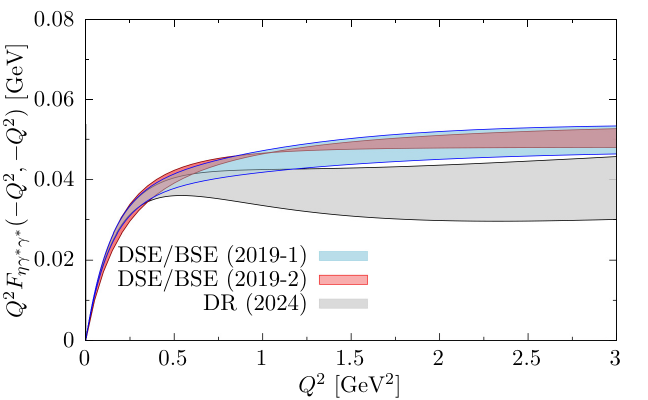}}
\centerline{\includegraphics[width=0.46\textwidth]{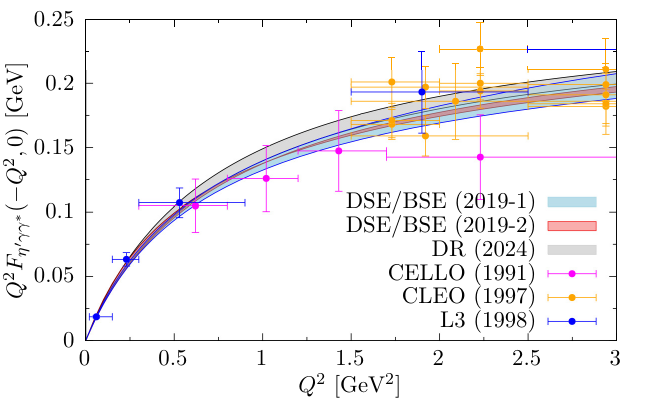}
\includegraphics[width=0.46\textwidth]{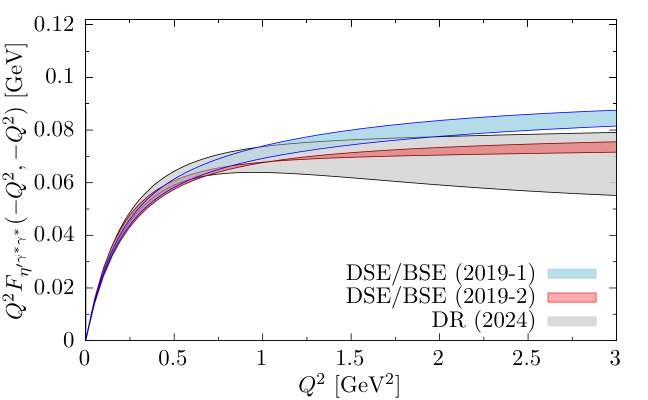}}
\caption{Results for the singly- (left) and doubly-virtual (right) TFFs of $\pi^0$ (first row), $\eta$ (second row),
and $\eta'$ (third row) from the DSE/BSE approach of Ref.~\cite{Eichmann:2019tjk} (2019-1) and
Ref.~\cite{Raya:2019dnh} (2019-2) compared to dispersive results~\cite{Hoferichter:2018dmo,Hoferichter:2018kwz,Holz:2024lom,Holz:2024diw} and experimental data~\cite{CELLO:1990klc,CLEO:1997fho,L3:1997ocz}.
}
\label{fig:TFF_DSE}
\end{figure}

The results obtained in the DSE/BSE framework are collected in \cref{tab:DSEBSE} and compared to the values quoted
in WP20.\footnote{The DSE/BSE framework also delivers results for the HVP contribution in the ballpark
	of dispersive and lattice results, although the error of roughly 3\% is much too large to make any claims for this quantity~\cite{Goecke:2011pe}.}
The sum of intermediate $\pi^0$, $\eta$, and $\eta'$ single-meson pole contributions amounts
to $a_\mu^\text{PS-poles} = 91.6(1.9) \times 10^{-11}$, where the error contains the variation of the parameter in the
RL effective interaction, the numerical error, and the uncertainties in the $\eta$ and $\eta'$ mixing
parameters~\cite{Eichmann:2019tjk}. The corresponding ${\pi^0, \eta, \eta'}\to \gamma^{(\ast)} \gamma^{(\ast)}$
TFFs, displayed in \cref{fig:TFF_DSE}, are calculated in the spacelike domain including the doubly- and singly-virtual kinematic
limits~\cite{Eichmann:2017wil,Weil:2017knt,Eichmann:2019tjk}. Similar results in a slightly different framework have
been found in Ref.~\cite{Raya:2019dnh}. The sum of the pion and kaon box diagrams, which depend
on the pion and kaon EM form factors, yields $a_\mu^\text{PS-box} = -16.2(5) \times 10^{-11}$~\cite{Eichmann:2019tjk,Eichmann:2019bqf}.
All these results are in good agreement with results from other approaches and the WP20 estimate.

\begin{figure}[t]
\centerline{\includegraphics[width=0.495\textwidth]{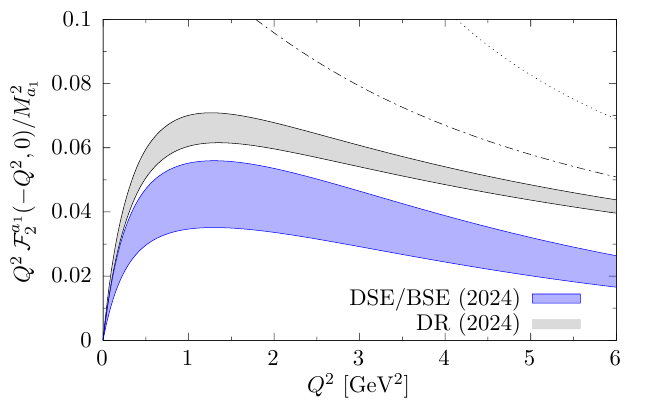}}
\caption{Results for the axial-vector TFF $\mathcal{F}_2$
from the DSE/BSE approach of Ref.~\cite{Eichmann:2024glq} (2024)
compared to the dispersive result of Ref.~\cite{Ludtke:2024ase}. The momentum dependency of the form factors agrees nicely; differences in an overall factor can be mostly attributed to the assumption of ideal mixing in the DSE/BSE result. The
BL limits \cite{Hoferichter:2020lap} are shown as dash-dotted/dotted
lines similar to previous plots.}
\label{fig:axTFF_DSE}
\end{figure}

Since WP20, three new issues have been addressed in the DSE/BSE approach. First, the authors of
Ref.~\cite{Miramontes:2021exi} have determined corrections to the pseudoscalar pion and kaon
box contributions beyond RL (BRL), see also Ref.~\cite{Raya:2022ued} for a summary.
 To this end they determined the (RL and BRL) EM form factors of the pion and the kaon in the space- and timelike region~\cite{Miramontes:2021xgn,Miramontes:2022uyi,Miramontes:2023ivz}. A major result is the
remarkable robustness of the RL truncation in the spacelike region, leading to the observation that BRL pion and kaon
box contributions to HLbL are safely contained within the error bars of previous RL results.
Second, after calculating the
EM form factor of the first radial excitation of the pion, $\pi_1$, its box contribution to the HLbL component of the $a_{\mu}$ has been computed in Ref.~\cite{Miramontes:2024fgo}, producing, respectively, $a_{\mu}^{\pi_1-\text{box}}(\text{RL}) = -2.03(12) \times 10 ^{-13}$ and
 $a_{\mu}^{\pi_1-\text{box}}(\text{BRL}) = -2.02(10) \times 10 ^{-13}$ in the RL and BRL truncation schemes.
Third, the contributions from intermediate axial-vector and
scalar mesons have been calculated~\cite{Eichmann:2024glq}. Their corresponding TFFs have
been determined in the very same setup used previously to determine pseudoscalar meson exchange
($\pi^0$, $\eta$ and $\eta'$) as well as meson ($\pi$ and $K$) box contributions. The corresponding results are also
shown in \cref{tab:DSEBSE}. In the scalar meson sector, contributions from $f_0(980)$, $a_0(980)$, $f_0(1370)$,
and $a_0(1450)$ have been combined into a single number, $a_\mu^{\text{HLbL}}[\text{scalar}] = -1.6(5)  \times 10^{-11}$.
In the axial-vector sector, the result $a_\mu^{\text{HLbL}}[\text{AV exchange (single)}] = 17.4(6.0)  \times 10^{-11}$
combines contributions from the lowest lying multiplet of $a_1$, $f_1$, and $f_1'$ states, whereas the result
$a_\mu^{\text{HLbL}}[\text{AV-tower+SDC}] = 24.8(6.1)  \times 10^{-11}$ combines contributions from a whole tower
of $a_1$, $f_1$, and $f_1'$ multiplets including short-distance contributions in the form of a quark loop using the matching
procedure outlined above and in Refs.~\cite{Colangelo:2019uex,Colangelo:2021nkr}. The latter is necessary because, even though a whole tower of axial-vector contributions has been taken into account, one cannot show
analytically whether an infinite resummation of such a tower would satisfy the SDCs as it does in hQCD~\cite{Leutgeb:2019gbz,Cappiello:2019hwh,Leutgeb:2021mpu,Leutgeb:2022lqw}. In contrast to hQCD, the TFFs in the DSE/BSE
framework are given numerically, and much more refined numerical methods would be needed to extract the necessary information. A comparison of the $\F_2$ TFF to the dispersive result is shown in \cref{fig:axTFF_DSE}.

In summary, systematic cross-checks and comparisons between the dispersive approach,
lattice QCD, hQCD, and the approach to QCD via functional methods certainly has the potential to further decrease the
spread of results and consequently the error bar of the axial-vector and other contributions to HLbL in the future.

\subsection{Form factor comparisons}
\label{sec:formfactors}

\begin{table}[t]
\small
    \renewcommand{\arraystretch}{1.1}
\begin{center}
\begin{tabular}{cccccc}
\toprule
 & Dispersive~\cite{Hoferichter:2018dmo,Hoferichter:2018kwz,Holz:2024lom,Holz:2024diw} & CA~\cite{Masjuan:2017tvw}  & R$\chi$T~\cite{Estrada:2024cfy} & hQCD \cite{Leutgeb:2022lqw} & DSE/BSE~\cite{Eichmann:2019tjk} \\
 \midrule
 $b_{\pi^0}~[\text{GeV}^{-2}]$ & $1.73(5)$ 
 & $1.76(10)$ & $1.74(1)$ 
 & $1.68(8)$ & $1.71(1)$\\
 $b_\eta ~[\text{GeV}^{-2}]$ & $1.83(4)$
 & $1.91(3)$ & $2.00(3)$ & $1.53(3)$ & $1.87(1)$\\
 $b_{\eta^\prime}~[\text{GeV}^{-2}]$ & $1.49(3)$ 
 & $1.43(3)$ & $1.37(2)$ & $1.31(2)$ & $1.54(3)$\\
 \midrule
 $\bar{F}^{\pi^0}_{\text{asym}}$ & $2F_{\pi}$ & $2F_{\pi}$ & $2F_{\pi}$ & $2F_{\pi}[\times0.9]$ & $2.6(4)F_{\pi}$\\
 $\bar{F}^{\eta}_{\text{asym}}~[\text{GeV}]$ & $0.186(13)$ & $0.180(12)$ & $0.174(3)$ & $0.194(14)$ & $0.21(2)$\\
 $\bar{F}^{\eta'}_{\text{asym}}~[\text{GeV}]$ & $0.264(13)$ & $0.255(4)$ & $0.260(4)$ & $0.32(4)$ & $0.36(4)$\\
 \bottomrule
\end{tabular}
\end{center}
\renewcommand{\arraystretch}{1.0}
\caption{The slope and the asymptotic value of the pseudoscalar singly-virtual TFFs from the different approaches.}
\label{tab:pseudoscalarcomparison}
\end{table}

For the pseudoscalar TFFs, several cross-checks between the various approaches have already been discussed and illustrated throughout the various subsections.  Here, we add a broader comparison of all phenomenological theoretical calculations, specifically for the TFF slopes and asymptotic behavior.
We define the slope parameters for the pseudoscalar TFFs according to
\begin{equation}
    b_P \equiv \frac{1}{F_{P\gamma\gamma}} \frac{\partial}{\partial q^2}  F_{P \gamma^* \gamma^*}(q^2,0) \big|_{q^2=0}\,.
\end{equation}
In the simplest VMD model, the slopes are expected to be given approximately by $b_P \simeq M_\rho^{-2} \simeq 1.66\GeV^{-2}$ for all three pseudoscalars.
These slope parameters, as well as the singly-virtual asymptotic values
\begin{equation}
\lim\limits_{Q^2\to\infty} Q^2 F_{P \gamma^* \gamma^*}(-Q^2,0) = \bar{F}^{P}_{\text{asym}}\,,
\end{equation}
are collected, for various theoretical approaches, in \cref{tab:pseudoscalarcomparison}.
We find that the slopes  $b_{\pi^0}$ of the $\pi^0$ TFF agree within uncertainties for all calculations.  The hQCD results for $b_\eta$ and $b_{\eta'}$ are smaller than all others, with the most significant discrepancy for $b_\eta$, while the R$\chi$T number for $b_\eta$ deviates somewhat towards a larger value.  Dispersive, CA, and DSE/BSE slopes are well consistent with each other also for $\eta$ and $\eta'$.  The results for the singly-virtual asymptotic behavior are largely consistent with each other within uncertainties for all TFFs, with discrepancies at most at the $2\sigma$ level.

\begin{table}[t]
    \renewcommand{\arraystretch}{1.1}
\begin{center}
\small
\begin{tabular}{ccccccccc}
\toprule
& Dispersive & VMD, U(3) & Regge & R$\chi$T & hQCD & DSE/BSE \\
& \cite{Ludtke:2024ase} & \cite{Hoferichter:2020lap,Hoferichter:2023tgp} & \cite{Masjuan:2020jsf} & \cite{Roig:2019reh,Masjuan:2020jsf} & \cite{Leutgeb:2022lqw} & \cite{Eichmann:2019tjk,Eichmann:2019bqf} \\
 \midrule
$\mathcal{F}_2(0,0)/M_{a_1}^2$ & $0.25(2)$ & $0.23(3)$ & $0.25(5)$ & $0.25(6)$ & $0.265(20)$ & $0.18(3)$ \\
$M_{a_1}$ & $1.23(4)$ & $1.23(4)$  & $1.23(21)$ & $1.23(21)$ & $1.28(8)$ & $1.23(4)$\\
 \bottomrule
\end{tabular}
\end{center}
\renewcommand{\arraystretch}{1.0}
\caption{The normalization of the $a_1$ TFF $\mathcal{F}_2(0,0)/M_{a_1}^2$ and $M_{a_1}$ used from the different approaches in units of $\GeV^{-2}$ and $\GeV$, respectively.}
\label{tab:a1TFFNormalizations}
\end{table}

In \cref{tab:a1TFFNormalizations}, we compare the different models for the TFF normalization $\mathcal{F}_2(0,0)/M_{a_1}^2$ for the $a_1(1260)$ axial-vector meson, again finding excellent agreement among all of them.  Note that this only constrains one out of the in total three axial-vector TFFs, cf.\ the discussion in \cref{sec:ResonanceContributions}.

\subsection{A new final number for analytic HLbL}\label{sec:finalnumber}

\begin{table}
\small
    \renewcommand{\arraystretch}{1.1}
\begin{center}
\begin{tabular}{cccccc}
\toprule
 & Dispersive~\cite{Hoferichter:2018dmo,Hoferichter:2018kwz,Holz:2024lom,Holz:2024diw} & CA~\cite{Masjuan:2017tvw}   & R$\chi$T~\cite{Estrada:2024cfy} & hQCD~\cite{Leutgeb:2022lqw} & DSE/BSE~\cite{Eichmann:2019tjk} \\
 \midrule
 $\pi^0$ & $63.0^{+2.7}_{-2.1}$ & $63.6(2.7)$ & $61.3^{+2.5}_{-1.6}$ & $63.4(2.7)$ & $62.6(1.3)$ \\
 $\eta$ & $14.7(9)$ 
 & $16.3(1.4)$ & $15.2^{+1.2}_{-0.9}$ & $17.6(1.7)$ & $15.8(1.1)$ \\
 $\eta^\prime$ & $13.5(7)$
 & $14.5(1.9)$ & $14.2^{+1.6}_{-1.1}$ & $14.9(2.0)$ & $13.3(8)$ \\
 \midrule
 Sum & $91.2^{+2.9}_{-2.4}$ & $94.4(3.6)$ & $91.3^{+3.2}_{-2.1}$ & $95.9(3.8)$ & $91.6(1.9)$\\
 \bottomrule

\end{tabular}
\end{center}
\renewcommand{\arraystretch}{1.0}
\caption{The pseudoscalar pole contributions to $a_\mu$ in units of $10^{-11}$ from the different approaches.
}
\label{tab:pseudoscalar}
\end{table}

We now have a fully dispersive evaluation of the $\pi^0$, $\eta$, and $\eta^\prime$ contributions, which agrees well with the other approaches based on phenomenological analyses, CA and R$\chi$T modeling as well as the more QCD modeling approaches hQCD and the functional approaches. The comparison of these is shown in \cref{tab:pseudoscalar}.
This contribution is improved over WP20~\cite{Aoyama:2020ynm} of $93.8^{+4.0}_{-3.6}\times 10^{-11}$ to
\begin{equation}
a^\text{PS-poles}_\mu = 91.2^{+2.9}_{-2.4}\times 10^{-11}\,.
\end{equation}
The next contributions are those from the pion and kaon box and rescattering in the $S$-wave channel. The dispersive box contributions for the pion have not changed compared to WP20.
The kaon box is now evaluated with dispersive methods with a very small change compared to WP20. Both contributions are now evaluated as well by functional methods in good agreement with the dispersive results. The sum is unchanged giving
\begin{equation}
    a^{\pi,K\text{-box}} = -16.4 (2)\times 10^{-11}\,.
\end{equation}
The $S$-wave rescattering has been updated and now includes $I=0,1,2$ $\pi\pi$, $\bar KK$, $\pi\eta$ rescattering, leading to
\begin{equation}
 a^\text{S-waves}_\mu = -9.1(1.0)\times 10^{-11}\,.
\end{equation}
The above contributions are well established and agree well among the different approaches. They sum to
\begin{equation}
\label{eq:disp-low}
a_\mu^\text{disp-low} = 65.7^{+3.1}_{-2.6}\times 10^{-11}\,.
\end{equation}
This is slightly lower than the WP20 result of $69.4(4.1)\times 10^{-11}$ and a somewhat smaller error. This is mainly due to the improved estimates of the $\eta$, $\eta^\prime$ contributions from dispersive methods compared to the CA results used previously.

The main source of the error in WP20 came from the short- and intermediate-distance domain with both missing resonances and issues of double counting. The SDCs have been improved very much as described in \cref{sec:SDC}. The remaining contributions are split in different integration regions which we treat differently. This allows one to deal with the double counting issue. The integration regimes are split by a scale $Q_0=1.5\GeV$ in a purely short-distance regime with all three $Q_i > Q_0$, a purely long-distance regime with all three $Q_i<Q_0$, and a mixed regime.

The purely short-distance regime can be calculated by using the QCD results discussed in \cref{sec:SDC} with contributions already included in \cref{eq:disp-low} subtracted. Similarly, in the phenomenological approaches all other contributions can be included. This is shown in \cref{tab:Qregions} in the row $Q_i>Q_0$. The agreement from the QCD result with the other approaches is reasonable with the possible exception of the functional approach. For this region we will use the QCD result with
\begin{equation}
a^{\text{SD-QCD}}_\mu = 6.2^{+0.2}_{-0.3}\times 10^{-11}\,.
\end{equation}
The error is mainly from the uncertainty in $\alpha_s$.

The next part is the mixed region where at least one of the $Q_i$ is larger than $Q_0$ and one smaller. In this regime there are known short-distance results~\cite{Melnikov:2003xd,Knecht:2020xyr,Bijnens:2022itw,Bijnens:2024jgh}. These constraints have been included to a large extent in all the modeling as well for direct evaluation of some parts of this region. For the modeling approaches the results have been calculated for the sum. The dispersive results follow from \cref{tab:subleading}. The uncertainty on the hadronic part has a 30\% error added quadratically to the input uncertainties added. The results from the four approaches are in reasonable agreement but the dispersive is somewhat larger due to a large contribution from the OPE domain. The error on the OPE domain has been determined from varying $r$, defined in \cref{mixed_OPE}, and shows that there is some remaining mismatch even after adding the effective poles. For this part we will use
\begin{equation}
 a^\text{Mixed}_\mu = 15.9(1.7)\times 10^{-11}\,.
\end{equation}
The remaining error we expect to be covered by the $Q_0$ variation discussed below.

\begin{table}[t]
\small
    \renewcommand{\arraystretch}{1.1}
\begin{center}
\begin{tabular}{llccccc}
\toprule
Region & & Dispersive & hQCD & Regge & DSE/BSE \\
\midrule
$Q_i>Q_0$ & & $6.2^{+0.2}_{-0.3}$ & $6.3(7)$ & $4.8(1)$ & $2.3(1.5)$\\
\midrule
Mixed & $A,S,T$ & $3.8(1.5)$ & & \\
 & OPE   & $10.9(0.8)$ & & \\
 & Effective pole & $1.2$ & & \\
 & Sum & $15.9(1.7)$ & $13.5(2.4)$ 
            & $12.8(5)$ & $10.1(3.0)$\\
\midrule
$Q_i<Q_0$ & $A=f_1,f_1',a_1$ & $12.2(4.3)$ & $13.1(1.5)$ & $10.9(1.0)$ & $8.6(2.6)$\\
    & $S=f_0(1370),a_0(1450)$ & $-0.7(4)$ & & &$-0.8(3)$\\
    & $T=f_2,a_2$ & $-2.5(8)$ & 
    $2.9(4)$ \\ 
    & Other & $2.0$ & $8.0(9)$
       & $3.2(6)$ & $2.8(6)$\\
    & Sum & $11.0(4.4)$ & $24.0(2.8)$  & $14.1(1.2)$ & $10.6(2.7)$ \\
\midrule
Sum & & $33.2(4.7)$ &$43.8(5.9)$ & $31.7(1.6)$ & $23.0(7.4)$\\
\bottomrule
\end{tabular}
\end{center}
\renewcommand{\arraystretch}{1.0}
\caption{The contributions in units of $10^{-11}$ from the different regions with the parts included in $a_\mu^\text{disp-low}$ removed. Dispersive results from \cref{tab:subleading}, with a $30\%$ error added for the hadronic contributions; hQCD, Regge, and DSE/BSE results are calculated from the models described in \cref{sec:Regge}, \cref{Section:HolographicModels}, and \cref{sec:functional} with this value of $Q_0$ and numbers can thus be slightly different from those quoted there. The row ``other'' includes all contributions not specified explicitly, in particular from excited axial-vector mesons, and for hQCD also from excited tensors and tensor nonpole contributions (the latter two give $5.6$ out of the $8.0$).
In the row labeled ``Sum'' the errors for hQCD have been added linearly.}
\label{tab:Qregions}
\end{table}

The last remaining contribution is the low-energy one with all three $Q_i<Q_0$. The contribution from the lowest axial-vector nonet has been much clarified compared to WP20 and all approaches are now in acceptable agreement. The heavy scalars are estimated both in the functional and dispersive approach and are in good agreement with each other as well as with estimates for the scalars used in WP20, but now with a smaller error. For the tensor contributions the error estimate used in WP20 was shown to not be reliable. Here the dispersive approach uses a phenomenological estimate including only one of the five TFFs, as argued from the quark model. The hQCD estimate includes one more and finds a different sign due to this; with only the same single TFF as in the dispersive calculation the result for hQCD for the tensors is somewhat more negative than the dispersive estimate.
 We will use as the tensor contribution $-1(4)$ covering all available results. With this and the dispersive results agreeing well with the others we will use for this contribution
\begin{equation}
    a^\text{Low}_\mu = 12.5(5.9)\times 10^{-11}\,.
\end{equation}

When putting all together there are still some remaining uncertainties. These we estimate by varying $Q_0$ as described in \cref{sec:MatchingToSDCs} and Ref.~\cite{Hoferichter:2024bae} to be an error of $1.9$ and the estimate of missing contributions by varying how matching to the SDCs is done. This is estimated by varying how the matching with the effective poles is done and adds an error of $3.9$. However, while the estimate of ``Other'' from the dispersive estimates agrees well with the Regge and DSE/BSE results, the hQCD estimate of the remaining contributions is higher, we therefore increase the error for this uncertainty to $5.0$.

Finally we use the same estimates for the NLO HLbL charm contribution  as in WP20:
\begin{equation}
    a^\text{charm}_\mu = 3(1)\times 10^{-11}\,.
\end{equation}
Adding all the errors quadratically and the contributions gives a total (LO) HLbL contribution of
\begin{equation}
\label{eq:amuHLbLdata}
  \amuHLbL = \amuHLbLdataresult\times 10^{-11}.
\end{equation}
This should be compared with the WP20 result of $92(19)\times10^{-11}$.
Our new number is perfectly compatible with the previous number within errors and has a significantly improved precision due to improvements of many hadronic contributions  as well as improvements in short-distance calculations and matching.

Finally, there is the estimate of the NLO HLbL contribution from Ref.~\cite{Colangelo:2014qya}, which we update along the lines of  Ref.~\cite{Aoyama:2020ynm}:
\begin{equation}
\label{eq:amuHLbLNLOdata}
    \amuHLbLNLO = \amuHLbLNLOdataresult\times 10^{-11}\,.
\end{equation}

\subsection{Prospects for future improvements}

\subsubsection{Theory}
\label{sec:TriangleDR}

The original approach of Refs.~\cite{Colangelo:2015ama,Colangelo:2017fiz} is based on dispersion relations for the HLbL tensor in general four-point kinematics, which can be derived from the Mandelstam double-spectral representation. The photon virtualities are treated as fixed external variables, while dispersion relations are written in terms of the Mandelstam variables for off-shell photon--photon scattering. A disadvantage of this approach is that, even in the optimized BTT decomposition of Ref.~\cite{Hoferichter:2024fsj}, the cancellation of kinematic singularities from the contribution of intermediate states of spin equal to 2 or larger is not manifest. This is due to the fact that in this approach---based on the decomposition of the HLbL tensor in terms of a redundant set of BTT structure---kinematic singularities in the photon virtualities are eliminated by sum rules that involve a tower of intermediate states in the unitarity relation. This issue is solved within a novel approach in triangle kinematics~\cite{Ludtke:2023hvz}, in which the external photon is taken soft prior to setting up the dispersion relations for the scalar functions entering the master formula in \cref{eq:MasterFormula}. In this framework the original cuts in the Mandelstam variables and in the photon virtualities are no longer separated, which leads to more complicated unitarity relations and the fact that the dependence of pole contributions on the photon virtualities is not automatically resummed as in the original approach. In the new framework, a reshuffling of intermediate-state contributions takes place and further sub-processes enter the two-pion unitarity relations, e.g., $\gamma^*\gamma^*\gamma\to\pi\pi$. These can be dispersively reconstructed without introducing kinematic singularities nor ambiguities~\cite{Ludtke:2023hvz}. A comparison and a suitable combination of the two dispersive approaches to HLbL~\cite{Ludtke:2023hvz} will, in the future, provide optimal control over residual uncertainties in the matching between hadronic and asymptotic continuum contributions. A detailed study of the relationship between the two types of dispersion relations and the reshuffling of different hadronic intermediate-state contributions has been recently performed in Ref.~\cite{Ludtke:2024ase} for the simpler case of the VVA correlator.

Ultimately, one would like to have in addition a data-driven approach given in terms of a single observable, as is done for the HVP contribution. Such an alternative approach is, in principle, provided by the Schwinger sum rule~\cite{Schwinger:1975ti}. It expresses the anomalous magnetic moment via an integral of a polarized photoabsorption cross section as follows:
\begin{equation}
a_\mu = \frac{m_\mu^2}{\pi^2\alpha}\int_{\nu_0}^\infty \dd\nu \left[\frac{\sigma_{LT}(\nu,Q^2)}{Q}\right]_{Q^2\rightarrow 0}\,,
\label{eq:SchwingerSR}
\end{equation}
where $Q^2$ is the photon virtuality, $\nu$ is the photon lab frame energy, and $\nu_0$ is the first inelastic particle-production threshold. The polarized photoabsorption cross section $\sigma_{LT}$ is the observable quantity that corresponds to the sum of the muon spin structure functions, $g_1(x,Q^2)+g_2(x,Q^2)$.
In this approach the HVP and HLbL contributions appear on the same footing; see Ref.~\cite{Hagelstein:2017obr} for more details. Work in this direction is ongoing~\cite{Hagelstein:2019tvp,VBthesis,SSRtoAppear}.

\subsubsection{Experiment}
\label{sec:ExpProsp}
\begin{sloppypar}
Experimental input for a data-driven evaluation of the HLbL contribution requires measurements of TFFs $F_{P\gamma^*\gamma^*}(Q_1^2,Q_2^2)$ of pseudoscalar mesons $P$ at arbitrary virtualities $Q_i^2$ to improve on the pseudoscalar-pole contributions.
The normalization of the TFF at $Q_i^2=0$ is given by the radiative width $\Gamma(P\to\gamma\gamma)$ of the respective pseudoscalar meson. A new measurement of the radiative width of $\eta$ using Primakoff-type meson production is announced by the PrimEx-\textit{eta} collaboration at JLab~\cite{primexeta:2009}. A first, preliminary evaluation without considering systematic effects yields $\Gamma(\eta\to\gamma\gamma) = 0.499(36)\,\text{keV}$~\cite{Smith:2024wvr}, which is in agreement with the current value $\Gamma(\eta\to\gamma\gamma) = 0.515(18)\,\text{keV}$~\cite{ParticleDataGroup:2024cfk} listed by the PDG. As the PDG only considered results from $e^+e^-$ experiments in their average and fits, the agreement suggests that the previously observed discrepancy with results from Primakoff-type experiments is resolved.
\end{sloppypar}

With the proposed upgrade of JLab to $22\GeV$ beam energy significant improvements on the accuracy of the radiative width measurements of $\pi^0$, $\eta$, and $\eta^\prime$ will be possible~\cite{Accardi:2023chb}. While the Primakoff-type measurements for $\eta$ and, for the first time, also for $\eta^\prime$, will be performed on nuclear targets with projected accuracies of $2\%$ and $3.5\%$, respectively, the measurement for $\pi^0$ can be performed off atomic electrons, which will allow one to avoid uncertainties of nuclear effects and achieve sub-percent accuracy on the radiative width. The Primakoff program is to be extended to measure the momentum dependence of the meson TFFs in a range of $0.001\GeV^2$ up to $0.3\GeV^2$, which is complementary to the ongoing investigations at $e^+e^-$ colliders.

The preliminary result of the BESIII collaboration on the singly virtual spacelike TFF of the $\pi^0$~\cite{Redmer:2019zzr} has already demonstrated the potential of the $e^+e^-$ experiment to contribute high-precision data in the most relevant region of momentum transfer around $1\GeV^2$. In spring 2024 the collaboration finalized the data taking of a $20\,\text{fb}^{-1}$ data set at $\sqrt{s}=3.773\GeV$, which improves on the statistics of the preliminary result by a factor seven and will be the new standard sample for two-photon fusion reaction studies at BESIII. First results on singly-virtual TFFs of $\pi^0$, $\eta$, and $\eta^\prime$ based on the complete data set are expected in 2025, providing information for virtualities from $Q^2\geq0.1\GeV^2$ to $5\GeV^2$. A comparison with the results of \babar{} and Belle at lower values $Q^2$ should be possible. The results will consider the full radiative corrections at NLO, as provided by the {\sc Ekhara 3.0} event generator~\cite{Czyz:2018jpp} without applying the previously accepted approximations~\cite{Druzhinin:2014sba,Ong:1988kg}. The size of the data set will also allow one to provide information on the doubly-virtual spacelike TFF of the lightest pseudoscalar mesons.

The data will also be used to provide information on the partial waves  in $\gamma^*\gamma^{(*)}\to \pi^+\pi^- / \pi^0\pi^0$. Results with a single off-shell photon are reported to be provided for invariant masses from the two-pion threshold up to $2.0\GeV$ at $0.2\GeV^2\leq Q^2 \leq 3.0\GeV^2$ and the full coverage of the pion helicity angle. Furthermore, a revised analysis scheme allows for doubly-virtual information in a limited region of the second virtuality~\cite{Lellmann:2024rbg}.

The analysis of single-virtual production of higher meson multiplicities will allow to provide results on the TFFs of the axial-vector meson $f_1(1285)$ at virtualities in the region at $0.2\GeV^2\leq Q^2 \leq 3.0\GeV^2$. Here, a partial-wave analysis is performed to separate the $a_0(980)\pi$ and $f_0\eta$ contributions in the $\pi^+\pi^-\eta$ final state. Further investigations of higher-multiplicity pion final states as well as final states involving kaons are being prepared to provide information on further axial and tensor states.

Complementary measurements at $Q_i^2>3\GeV^2$ are to be expected from the Belle-II collaboration as a continuation of the successful activities at Belle~\cite{Belle-II:2018jsg}.

In addition to the ongoing investigation \cref{tab:exp_prio} summarizes recommendations for intensified experimental activities to improve on HLbL.

\begin{table}[t]
\small
    \centering
    \renewcommand{\arraystretch}{1.1}
    \begin{tabular}{ll}
    \toprule
         &  Experimental input\\
    \midrule
    \multirow{3}{*}{Axial-vector TFFs}& $e^+e^-\to e^+e^- A$, $A=f_1, f_1', a_1$\\
    &  Radiative decays $A\to V\gamma$, $V=\rho,\omega,\phi$\\
         & Dilepton decays $A\to e^+e^-$\\\midrule
Scalar and    tensor TFFs&  $\gamma^*\gamma^{(*)}\to\pi\pi, \pi\eta,\bar K K,\pi\pi\pi$
     \\  \midrule
    \multirow{3}{*}{Pseudoscalar TFFs} & $\gamma\gamma\to\eta,\eta'$ \\
    & $e^+e^-\to e^+e^-(\gamma^*\gamma^{(*)}\to\pi^0,\eta,\eta')$ \\
        & $e^+e^-\to e^+e^-(\gamma\gamma\to P)$, $P=\pi(1300),\eta(1295),\eta(1405)$ \\
    \bottomrule
    \end{tabular}
    \renewcommand{\arraystretch}{1.0}
    \caption{Examples of useful experimental inputs related to the exclusive hadronic channels.}
    \label{tab:exp_prio}
\end{table}

\subsubsection{Monte-Carlo event generators}

The measurement of multi-particle final states produced in two-photon fusion reactions with virtual photons beyond the individual decay modes of $\eta$ and $\eta^\prime$ mesons requires new event generators to evaluate detection efficiencies.

The {\sc Ekhara 3.0} generator~\cite{Czyz:2018jpp} has been extended to provide a data-driven description of the resonance contributions in $\gamma^{(*)}\gamma^{(*)}\to\pi^+\pi^-/\pi^0\pi^0$ up to invariant pion pair masses of $2\GeV$ and virtualities $Q_{1,2}^2\leq 4\GeV^2$. The dominant resonances $f_0(500)$, $f_0(980)$, and $f_2(1270)$ are included based on Refs.~\cite{Danilkin:2018qfn,Danilkin:2019mhd}.

The development of {\sc Ekhara} will unfortunately not continue. To provide the MC simulations necessary in experimental investigations for efficiency studies and partial-wave analyses, a new event generator project has been started~\cite{Lellmann:2025}. The focus is on multi-particle final states, which are relevant for measurements of axial-vector and tensor mesons, but the design allows for flexibility in the generated final states. The generator determines the cross section observable at $e^+e^-$ machines as the product of the luminosity function~\cite{Budnev:1975poe,Pascalutsa:2012pr} and two-photon cross-sections. The latter depend on the polarization of the virtual photons and can be taken from theory calculations, e.g., for two-pion~\cite{Danilkin:2018qfn, Danilkin:2019opj} or $f_1(1285)\to\pi^+\pi^-\eta$~\cite{Ren:2024uui} production in two-photon collisions, or can be modeled from existing experimental results. The ongoing measurements of $\gamma\gamma^*\to\pi^0\pi^0$ and $\gamma\gamma^*\to f_1(1285)$ at BESIII are the first applications of the new event generator.

\FloatBarrier

\clearpage

\section{Lattice approaches to HLbL}
\label{sec:latticeHLbL}

\noindent
\begin{flushleft}
\emph{X.~Feng, A.~G\'erardin,  L.~Jin, T.~Lin, H.~Meyer}
\end{flushleft}

\subsection{Introduction}

The idea to compute the HLbL contribution to the muon $(g-2)$ in lattice QCD was pioneered in Ref.~\cite{Hayakawa:2005eq}.
The relevant diagram is illustrated on the RHS of \cref{fig:QEDker}.
A first peer-reviewed publication on the subject based on these methods appeared in 2014~\cite{Blum:2014oka}.
Substantial improvements to the methodology~\cite{Blum:2015gfa} led to a lattice calculation of $\amuHLbL$ at physical pion mass~\cite{Blum:2016lnc}. 
An important feature of this methodology is the calculation of lepton and photon propagators on the same (periodic) lattice as the QCD degrees of freedom.
By the time of WP20, one complete lattice calculation, by the RBC/UKQCD collaboration~\cite{Blum:2019ugy}, had been published; it entered the WP20 average.
Starting in 2015, in a series of conference proceedings~\cite{Green:2015mva,Asmussen:2016lse,Asmussen:2017bup,Asmussen:2018lcw,Asmussen:2019act} and papers~\cite{Asmussen:2022oql}, the Mainz/CLS group developed a method treating the QED parts of the diagram in continuum and infinite volume coordinate-space perturbation theory. This type of method, sometimes referred to as QED$_{\infty}$, was also investigated by the RBC/UKQCD collaboration starting with Ref.~\cite{Blum:2017cer}, providing some important refinements. 
Since WP20, three collaborations (Mainz/CLS, RBC/UKQCD, and BMW) have produced lattice results for $\amuHLbL$ based on QED$_{\infty}$.
Their methodology and results are reviewed in \cref{sec:direct-lattice-hlbl}, which concludes with a lattice average for $\amuHLbL$.

A separate line of approach to HLbL scattering in the muon $(g-2)$ is to provide the relevant hadronic input to compute individual contributions as defined in a given dispersive framework.
The most important case is that of the pseudoscalar meson exchanges.
Starting with Ref.~\cite{Gerardin:2016cqj}, lattice calculations have aimed at providing the $F_{P\gamma^*\gamma^*}(q_1^2,q_2^2)$ TFFs ($P=\pi^0,\eta,\eta'$), or else directly evaluating the relevant integral in coordinate space~\cite{Lin:2024khg}. By now, several evaluations have appeared for the $\pi^0$ pole contribution and first corresponding calculations for the $\eta$ are available as well. All these are reviewed in \cref{sec:HLbLexcl_state_LAT}.

\subsection{Direct lattice calculations of the hadronic light-by-light contribution}
\label{sec:direct-lattice-hlbl}

\begin{figure}
\centerline{\includegraphics[width=0.32\textwidth]{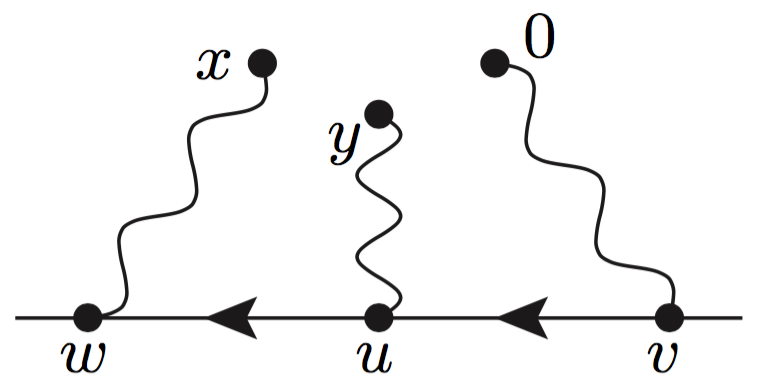}~~~~~~~~~~~~\includegraphics[width=0.36\textwidth]{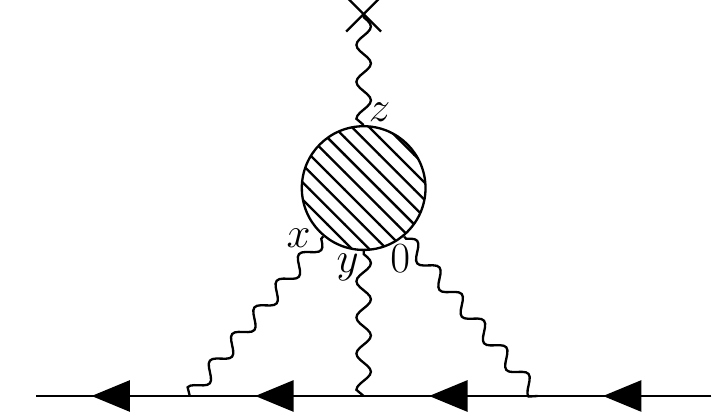}}
\caption{\label{fig:QEDker} Left: QED part of the HLbL Feynman diagram, with the muon in- and outgoing with a definite momentum $p$. 
The position-space  QED kernel $ {\cal L}_{[\rho,\sigma];\mu\nu\lambda}(p,x,y)$ results from integrating over the internal vertices $w,u,v$. Note that the represented amplitude is translationally invariant, hence the origin 0 can be chosen to coincide with one of the photon end-points without loss of generality. Figure taken from Ref.~\cite{Asmussen:2018oip}. Right: the full amplitude yielding $\amuHLbL$, with the blob corresponding to the HLbL scattering amplitude. Figure taken from Ref.~\cite{Asmussen:2022oql}.}
\end{figure}

The starting point for computing $\amuHLbL$ in lattice QCD using Euclidean position-space techniques can be written as~\cite{Green:2015mva}
\begin{align}
\amuHLbL &= \frac{m_\mu e^6}{3} \int d^4x\;d^4y\;
{\cal L}_{[\rho,\sigma];\mu\nu\lambda}(p,x,y)\;
i\widehat\Pi_{\rho;\mu\nu\lambda\sigma}(x,y)\,,
\label{eq:master_formula} \\
i\widehat \Pi_{\rho;\mu\nu\lambda\sigma}( x, y)  &=
-\int d^4z\; z_\rho\, \Big\langle\,j_\mu(x)\,j_\nu(y)\,j_\sigma(z)\,
j_\lambda(0)\Big\rangle_{\rm QCD}\,.
\label{eq:iPihat_def}
\end{align}
The nonperturbative physics of the strong interaction enters via the QCD four-point function \cref{eq:iPihat_def}
of the EM current carried by the quarks, $j_\mu=\frac{2}{3}\bar u \gamma_\mu u - \frac{1}{3} \bar d \gamma_\mu d - \dots$
The QED kernel
\begin{align}
\label{eq:kerHLbL_def}
  {\cal L}_{[\rho,\sigma];\mu\nu\lambda}(p,x,y) &= \frac{1}{16m_\mu^2}
  \int d^4u\,d^4v\,d^4w\, G(w-x)\, G(u-y)\, G(v)\, e^{-ip\cdot(w-v)}
 \notag\\ 
 & \times {\rm Tr}\left\{[\gamma_\rho,\gamma_\sigma]\;
 (-i\slashed{p}+m_\mu)
 \gamma_\mu S(w-u) \gamma_\nu S(u-v) \gamma_\lambda
(-i\slashed{p}+m_\mu)
 \right\}
\end{align}
represents the photon and muon elements
of the HLbL diagram, which are depicted in the left diagram of 
\cref{fig:QEDker}.
In \cref{eq:kerHLbL_def}, $G(x)=1/(4\pi^2x^2)$ is the massless scalar propagator, 
\begin{equation}
    S(x)=\int \frac{d^4p}{(2\pi)^4}\, \frac{-i\slashed{p}+m_\mu}{p^2+m_\mu^2}\, e^{ip\cdot x}
\end{equation} is the muon propagator, and the notation is fully Euclidean, in particular for the space--time scalar products, and the (Hermitian) Dirac matrices obey
$\{\gamma_\mu,\gamma_\nu\}=2\delta_{\mu\nu}$.
The kernel is dimensionless as well as infrared finite thanks to the trace of \cref{eq:kerHLbL_def}, which projects onto the anomalous magnetic moment, and obeys the properties
\begin{align}\label{eq:LisOdd}
{\cal L}_{[\rho,\sigma];\mu\nu\lambda}(-p,-x,-y) =&
- {\cal L}_{[\rho,\sigma];\mu\nu\lambda}(p,x,y) \,,
\notag\\
{\cal L}_{[\rho,\sigma];\mu\nu\lambda}(-p^*,x,y) =&
~~{\cal L}_{[\rho,\sigma];\mu\nu\lambda}(p,x,y)^*\,,
\notag\\
{\cal L}_{[\rho,\sigma];\lambda\nu\mu}(p,x,x-y) =&
 - {\cal L}_{[\rho,\sigma];\mu\nu\lambda}(-p^*,x,y)^*\,.
\end{align}
While in earlier approaches the QED kernel used to be ``immersed'' in the lattice formulation of $\amuHLbL$, recent calculations have worked with the kernel in the continuum and infinite-volume theory. A reliable calculation of \cref{eq:kerHLbL_def}, or its momentum-averaged version $\bar{\cal L}_{[\rho,\sigma];\mu\nu\lambda}$ defined below in \cref{eq:kerHLbL_paver} is a demanding task in itself~\cite{Asmussen:2016lse,Blum:2017cer,Asmussen:2022oql}. A code with auxiliary look-up files providing $\bar{\cal L}_{[\rho,\sigma];\mu\nu\lambda}$ is publicly available \cite{Asmussen:2022oql}.\footnote{\tt https://github.com/RJHudspith/KQED}

As for the muon momentum, one option, adopted by the RBC/UKQCD collaboration~\cite{Blum:2017cer,Blum:2023vlm}, is to choose the rest frame of the muon, \emph{de facto} averaging over the Euclidean momenta $p=(\pm im_\mu,\boldsymbol{0})$; see \cref{sec:HLbL_RBC} for more details. The option followed by all other collaborations~\cite{Chao:2021tvp,Fodor:2024jyn,Kalntis:2024dyd} so far is to employ a full average of the QED kernel over the direction of the muon momentum, thus making the following substitution~\cite{Green:2015mva} in \cref{eq:master_formula},
\begin{equation}
{\cal L}_{[\rho,\sigma];\mu\nu\lambda}(p,x,y)
\rightarrow
 \bar {\cal L}_{[\rho,\sigma];\mu\nu\lambda}(x,y)
\equiv \frac{1}{2\pi^2}\int d\Omega_{\hat\epsilon} \;
{\cal L}_{[\rho,\sigma];\mu\nu\lambda}(p=im_\mu\hat\epsilon,x,y)\,.
\label{eq:kerHLbL_paver}
\end{equation}
Here $\hat\epsilon$ is understood to be a real, four-component unit vector, and $d\Omega_{\hat\epsilon} $ the corresponding solid-angle differential. An advantage of using $\bar{\cal L}$ is that, after contraction of the Lorentz indices and integration (say) over $x$ in \cref{eq:master_formula},
the remaining integral over $y$ reduces to a one-dimensional integral over $|y|$ that can be performed reliably with $\Order(20)$ points.

Irrespective of the choice made for handling the muon momentum, a significant amount of flexibility remains to optimize the QED kernel for numerical purposes. First, there is a freedom to add total-divergence terms---sufficiently well-behaved  not to generate boundary terms upon partial integration---to the kernel, due to the conservation of the vector currents~\cite{Blum:2017cer,Asmussen:2019act}.
A proposal that has been adopted by several collaborations~\cite{Chao:2021tvp,Fodor:2024jyn,Kalntis:2024dyd} is~\cite{Asmussen:2019act}
\begin{equation}\label{eq:lamsub}
\bar{\cal L}^{(\Lambda)}_{[\rho,\sigma];\mu\nu\lambda}(x,y) = \bar{\cal L}_{[\rho,\sigma];\mu\nu\lambda}(x,y)
-\partial_\mu^{(x)} (x_\alpha e^{-\Lambda m_\mu^2 x^2/2}) \bar{\cal L}_{[\rho,\sigma];\alpha\nu\lambda}(0,y) 
- \partial_\nu^{(y)} (y_\alpha e^{-\Lambda m_\mu^2 y^2/2})\bar{\cal L}_{[\rho,\sigma];\mu\alpha\lambda}(x,0)\,,
\end{equation}
where the value 0.4 was chosen for the free parameter $\Lambda$.
The rationale behind this choice is to make the integrand neither too long-range, nor too short-distance dominated, as this leads to smaller overall uncertainties in the lattice calculation.
Second, there is the option of symmetrizing the kernel with respect to the Bose symmetries of $\widehat\Pi$, for instance under the exchange of the pairs $(x,\mu)\leftrightarrow(y,\nu)$.
Thirdly, for each individual Wick contraction contributing to $\widehat\Pi$, translation invariance can be
exploited~\cite{Blum:2015gfa,Blum:2016lnc,Chao:2020kwq} so as to minimize the overall computational cost.

The notation used in \cref{eq:master_formula} is the one introduced by the Mainz/CLS group. The connection to the notation used by the RBC/UKQCD collaboration~\cite{Blum:2023vlm}, provided here for convenience, is
 \begin{align}
    \frac{i}{4}\; {\rm Tr}\{[\gamma_\rho,\gamma_\sigma]
    \mathfrak{G}^{\rm RBC}_{\mu\nu\lambda}(x,y,z)\}
    &=   {\cal L}_{[\rho,\sigma];\mu\nu\lambda}(p=(im_\mu,\boldsymbol{0}),x-z,y-z)\,,
\notag \\
  -6 \int d^4x_{\rm op}\; (x_{\rm op} - x_{\rm ref})_j \; {\cal H}^{\rm RBC}_{k,\rho,\sigma,\lambda}(x_{\rm op},x,y,z)
  &= i\widehat\Pi_{j;\rho\sigma\lambda k}(x-z, y-z)\,.
  \label{eq:iPihat_RBC}
  \end{align}
A further redundancy in the master formula \cref{eq:master_formula} for $\amuHLbL$ exploited in Ref.~\cite{Blum:2023vlm} is that in the definition of $i\widehat\Pi$, \cref{eq:iPihat_def}, the factor $z_\rho$ could be replaced by\footnote{A more general class of subtractions to the integrand $j_{\sigma}(z)\,z_{\rho}$ is $j_\alpha(z)\,\partial^{(z)}_\alpha f_{\rho\sigma}(x,y,z)$, where the special case is recovered for $f_{\rho\sigma}(x,y,z) \equiv z_\sigma\,z_\rho^{\rm ref}(x,y)$.} $(z_\rho-z^{\rm ref}_\rho)$---as made explicit in \cref{eq:iPihat_RBC}, where in RBC/UKQCD notation the external vertex is written $x_{\rm op}$ instead of $z$.

\begin{figure}
\centerline{\includegraphics[width=0.28\textwidth]{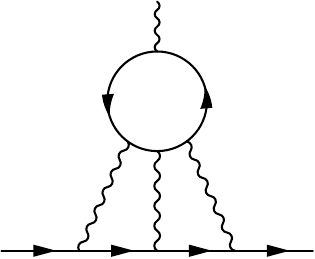}~~~~~~~~~~~~~\includegraphics[width=0.28\textwidth]{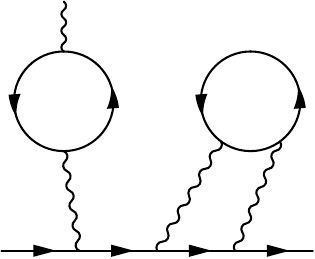}}
\caption{Illustration of the connected and (2+2) disconnected Wick-contraction diagrams for the quarks. The latter are not SU(3)$_{F}$ suppressed and turn out to be the dominant class of disconnected diagrams. Figures taken from Ref.~\cite{Blum:2023vlm}.
\label{fig:ConnLeadDiscDiag}}
\end{figure}

The flavor structure of $i\widehat\Pi$ plays an important role.
In terms of the isospin decomposition $j_\mu = j_\mu^3 + j_\mu^8/\sqrt{3}$ of the EM current in the $(u,d,s)$ sector, $i\widehat\Pi$ can be decomposed into three terms, involving either zero, two, or four isovector currents $j_\mu^3$, which lead respectively to the contributions $a_\mu^{{\rm HLbL}\{j^8\}}$, $a_\mu^{{\rm HLbL}\{j^3,j^8\}}$, and $a_\mu^{{\rm HLbL}\{j^3\}}$.
On the other hand, for a given quark current there are five classes of  Wick contractions, the connected (4) as well as the (2+2), the (3+1), the (2+1+1), and the (1+1+1+1) disconnected diagrams. For instance, we denote the (2+2) contribution to $\amuHLbL$ from the $u,d$ quarks
 $a_\mu^{\rm HLbL,(2\ell+2\ell)}$.
In particular, the contribution involving four isovector currents $j_\mu^3$ corresponds to a linear combination of the two leading quark-level contractions, $a_\mu^{\rm HLbL,(4)\ell}$ and $a_\mu^{\rm HLbL,(2+2)\ell}$.
We write
\begin{equation}
\amuHLbL = a_\mu^{\rm HLbL,\ell}
+ a_\mu^{\rm HLbL,s}
+ a_\mu^{\rm HLbL,c}
+ a_\mu^{\rm HLbL,rest}\,,
\end{equation}
where
\begin{align}
a_\mu^{\rm HLbL,\ell} &\equiv 
a_\mu^{\rm HLbL,(4\ell)}+a_\mu^{\rm HLbL,(2\ell+2\ell)}\,,
\notag\\
a_\mu^{\rm HLbL,s} &\equiv  a_\mu^{\rm HLbL,(4s)}
+ a_\mu^{\rm HLbL,(2s+2\ell)}
+ a_\mu^{\rm HLbL,(2s+2s)} \,,
\notag\\
a_\mu^{\rm HLbL,c} &\equiv  a_\mu^{\rm HLbL,(4c)}+a_\mu^{\rm HLbL,(2c+2\ell)}+a_\mu^{\rm HLbL,(2c+2c)}\,,
\notag\\
a_\mu^{\rm HLbL,rest} &\equiv  a_\mu^{\rm HLbL,(3\ell+1\ell s)}
+ a_\mu^{\rm HLbL,(2\ell+1\ell s+1\ell s)}
+ a_\mu^{\rm HLbL,(1\ell s+1\ell s+1\ell s+1\ell s)}
+ \dots\,,
\end{align}
where we will neglect the contributions represented by the ellipsis.
The notation $1\ell s$ refers to the difference of a light- and a strange-quark loop with a single, vector-current insertion.
We begin by reviewing the calculations by three different collaborations of the sum $a_\mu^{\rm HLbL,\ell}$ of the connected and leading disconnected light-quark contributions illustrated in \cref{fig:ConnLeadDiscDiag}.
Subsequently, we treat the strange, as well as the charm-quark contribution,
and finally we present the status of the subleading contributions in the $(u,d,s)$ quark sector.
The term $a_\mu^{\rm HLbL,\ell}$ by far dominates the total $\amuHLbL$, and the other contributions are found to be smaller than the current uncertainty of $a_\mu^{\rm HLbL,\ell}$.

\subsubsection{\texorpdfstring{$a_\mu^{\rm HLbL,\ell}$}{}: the Mainz/CLS calculation}

\begin{figure}[t]
  \includegraphics[width=0.5\textwidth]{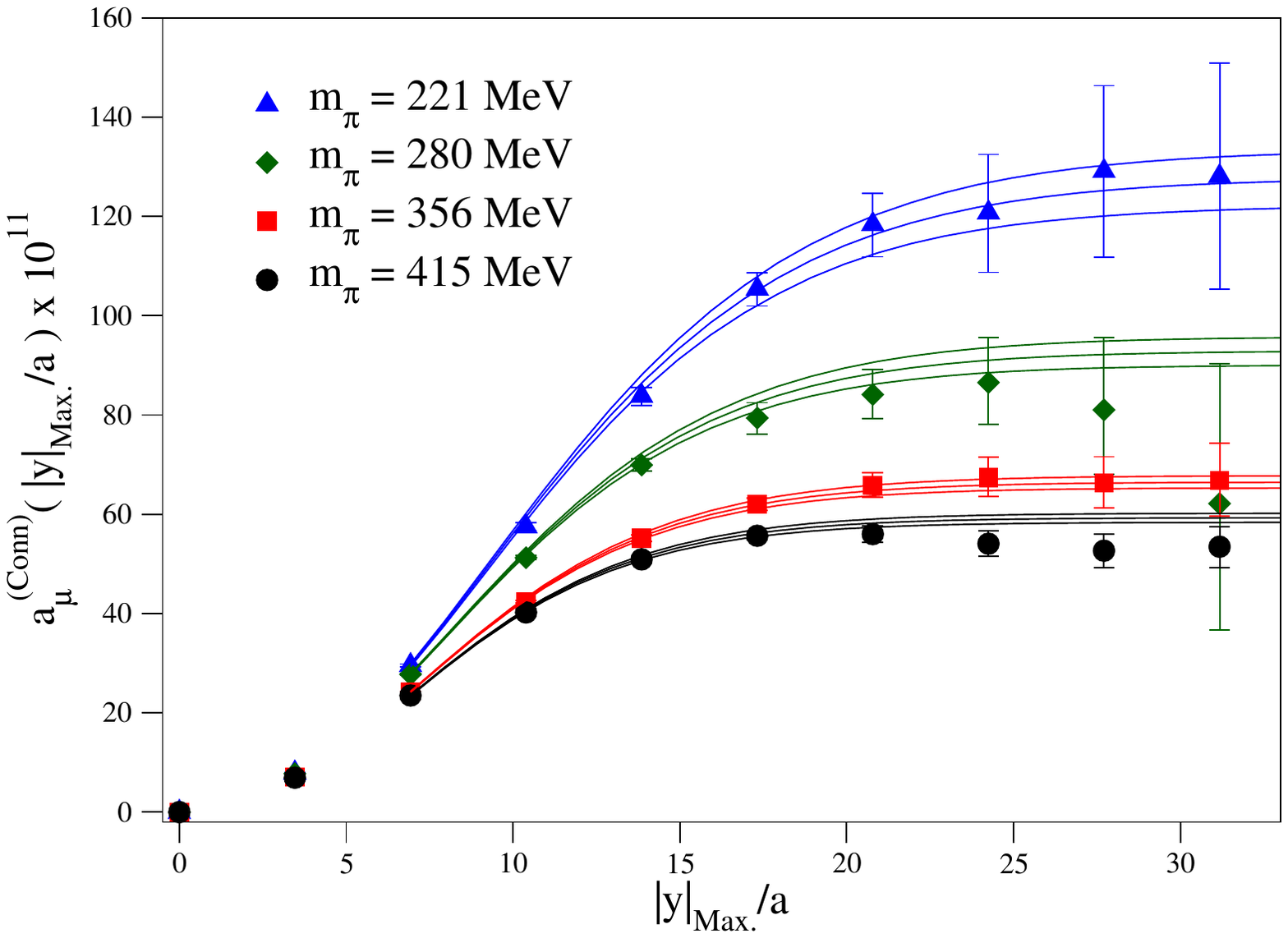}
  \hspace{-8pt}
  \includegraphics[width=0.5\textwidth]{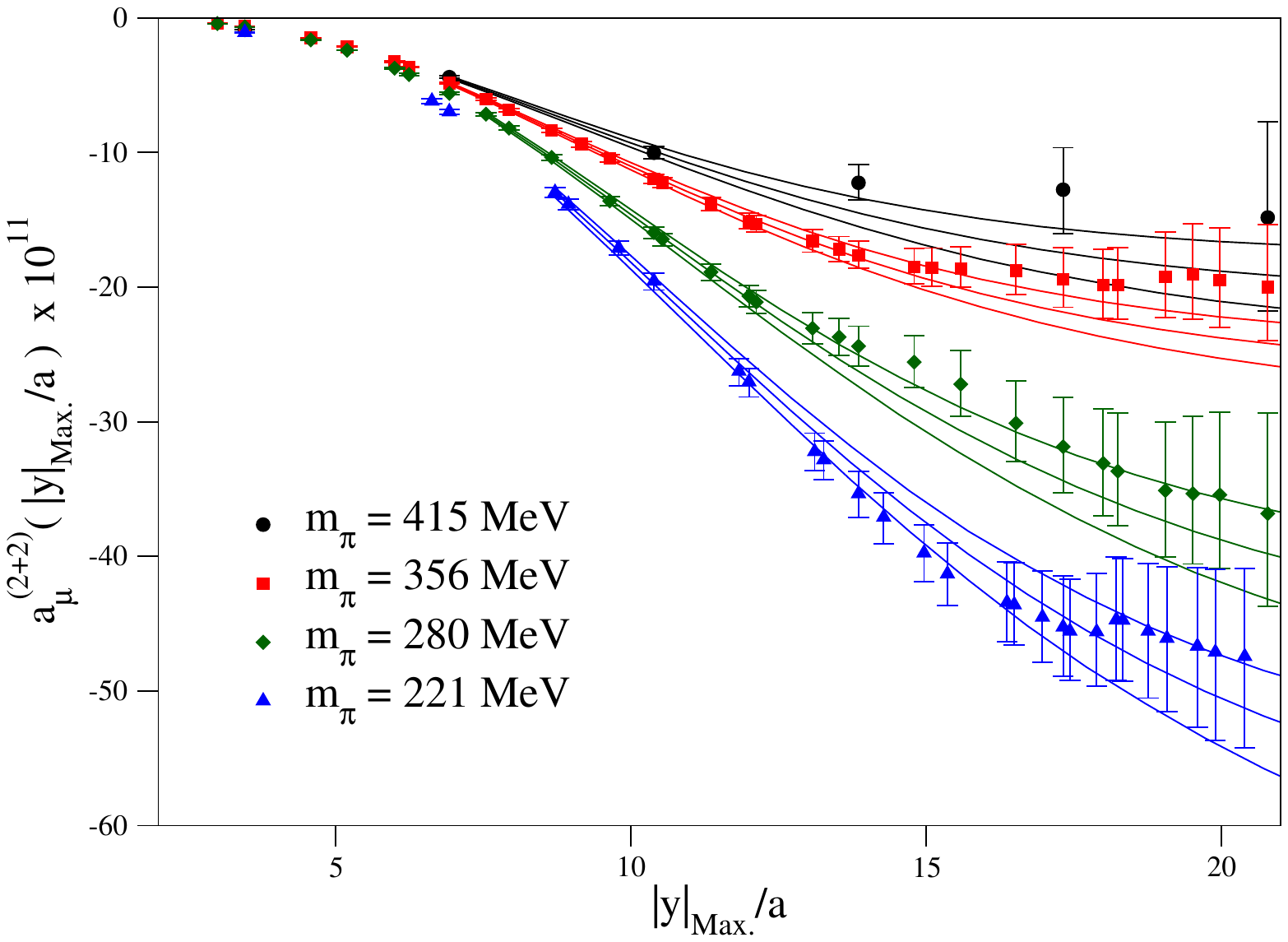}
  \caption{Partially-integrated light-quark connected (left) and (2,2) disconnected contribution to $\amuHLbL$ versus $|y|_{\rm Max.}/a$ for four ensembles which have a broad range of pion masses but the same lattice spacing ($a=0.086\,$fm) and similar volumes ($4\lesssim M_\pi L\lesssim 6$). The points are the numerically integrated lattice data and the curves result from switching the integrand to a fit above a certain distance.  Figures taken from Ref.~\cite{Chao:2021tvp}. }\label{fig:plotsIntgd}
\end{figure}

A calculation of $\amuHLbL$ by the Mainz group~\cite{Chao:2021tvp}  was published in 2021.
In this calculation, the kernel employed is $\bar{\cal L}^{(\Lambda=0.4)}_{[\rho,\sigma];\mu\nu\lambda}(x,y)$
as given in \cref{eq:lamsub}.
The lattice formulation consists of a nonperturbatively $\Order(a)$-improved Wilson fermion action with an $\Order(a^2)$ tree-level improved L\"uscher--Weisz gauge action,
and the calculations are performed on ensembles generated as part of the Coordinated Lattice Simulations (CLS) initiative~\cite{Bruno:2014jqa}.
The Mainz group follows the strategy of performing
the $x$ and $z$ integrals as explicit four-dimensional sums over the
lattice, whereas the final integral over $y$ is performed in spherical
coordinates, by sampling $\Order(20)$ values of $|y|$, see \cref{fig:plotsIntgd}.

\paragraph{A pilot study at $M_\pi=M_K\simeq 415\MeV$}

The Mainz group initially performed a pilot study~\cite{Chao:2020kwq} at
the SU(3)$_{F}$-symmetric point $M_\pi=M_K\simeq 415\MeV$.  The enlarged flavor symmetry leads
to a simplification, namely only the fully-connected and the $(2+2)$
topologies contribute.
For the connected contribution, two methods were explored:  the first based on computing all three connected Wick-contractions of
the HLbL amplitude called ``Method~1.'' This requires the calculation of a
large number of (sequential) propagators and is therefore rather
expensive. The second by contrast, ``Method~2,'' exploits the symmetries of the
tensor $i\widehat\Pi$ to reduce the connected contribution to
$\amuHLbL$ to a single Wick contraction. This can then be computed
with far fewer propagator calculations. ``Method-2'' was found to broaden the integrand somewhat, and the subtraction scheme of \cref{eq:lamsub} was necessary to have it peak in a range suitable for the lattice. In the end, the two methods led to the
same result for $\amuHLbL$ within uncertainties.
For the $(2+2)$ disconnected contribution, only one method was used:
the contraction in which a quark loop connects the origin to point $y$
was handled by interchanging the integration variables $x$ and $y$ in \cref{eq:master_formula}.

One challenge in the calculation is the control over long-distance
effects.  This includes finite-size effects on the final $|y|$
integrand, as well as the extension of this integrand to arbitrarily
long distances, since the statistical noise is difficult to tame in
this regime. Both issues were handled using a coordinate-space
calculation of the $\pi^0$ exchange in finite volume, based on a
VMD TFF with parameters
determined from a direct lattice calculation of the TFF on the same
ensemble~\cite{Gerardin:2019vio}.  A 25\% uncertainty was assigned to
the total long-distance correction applied to the data.

The other major challenge is to control cutoff effects.  For the (positive) connected contribution at the SU(3)$_{F}$-symmetric point, these were found to lead to a 15\% (upward) correction between the lattice data at the coarsest lattice spacing and the continuum result.
Discretization errors on the (negative) disconnected contribution
turned out to have the same sign as for the connected. Thus, while a
strong cancellation takes place when adding connected and disconnected
contributions in the continuum, no such cancellation occurs for the corresponding
discretization errors. At the SU(3)$_{F}$-symmetric
point~\cite{Chao:2020kwq}, this means that the continuum result for
$\amuHLbL$ is a factor of about 1.6 larger than $\amuHLbL$ at the coarsest lattice spacing used, $a=0.086$\,fm.  The
uncertainty of the continuum extrapolation is made larger by the
expected presence of $\Order(a)$, as opposed to $\Order(a^2)$ discretization
errors, due to contact terms in the vector four-point function and the use of unimproved vector currents.
The relative size of the $\Order(a)$ and the $\Order(a^2)$ discretization errors
is not known in the range of lattice spacings employed (0.049 to 0.086\,fm)
and is difficult to determine from the lattice data.
Nevertheless, applying various fit ans\"atze reflecting this uncertainty, the result
obtained in \cite{Chao:2020kwq} was:
\begin{equation}
\amuHLbL = 65.4 (4.9)( 6.6) \times 10^{-11} \qquad \quad {\rm at}~ M_\pi=M_K\simeq 415\MeV\,,
\end{equation}
with the first error resulting from the uncertainties on the individual
gauge ensembles, and the second from the systematic error of the continuum extrapolation.

\paragraph{Lattice result extrapolated to physical $(u,d,s)$ quark masses}

The calculation of Ref.~\cite{Chao:2020kwq} was extended away from the $\text{SU}(3)_{F}$-symmetric point
towards physical quark masses in Ref.~\cite{Chao:2021tvp}. The lattice ensembles used Ref.~\cite{Bruno:2014jqa} keep the sum of
the $(u,d,s)$ quark masses constant, and the smallest pion mass reached
was around $200\MeV$.  They found a strong increase in the value of the light-quark
fully-connected contribution to $\amuHLbL$ with decreasing pion mass. Similarly, the $(2+2)$ disconnected contribution was found to increase in
magnitude in this limit. The absolute amount of cancellation between the two
contributions thus strongly increases as the pion mass is lowered, and the authors concluded that adding the two contributions prior to the continuum extrapolation was beneficial.
Furthermore, three more Wick-contraction topologies appear away from
the SU(3)$_{F}$-symmetric point. They have all been computed
within Ref.~\cite{Chao:2021tvp} and are discussed in \cref{subsec:HLBL_rest_contrib}.

In Ref.~\cite{Chao:2021tvp}, the tail of the integrand and the finite-size
correction, though still based on the $\pi^0$ exchange, were handled
differently: first, an ansatz describing the final integration in the
variable $|y|$ faithfully for the $\pi^0$ exchange was fit to the
lattice data in order to extend the integration up to $|y|=\infty$.
Secondly, a finite-size correction term of the form $\exp(-M_\pi L/2)$
with a fit coefficient was included in the final extrapolation to
the physical point. This alternative, more data-driven procedure was
shown to yield results consistent with those of
Ref.~\cite{Chao:2020kwq} at the SU(3)$_{F}$ symmetric point.

\begin{figure}[t!]
\centerline{  \includegraphics[width=0.6\textwidth]{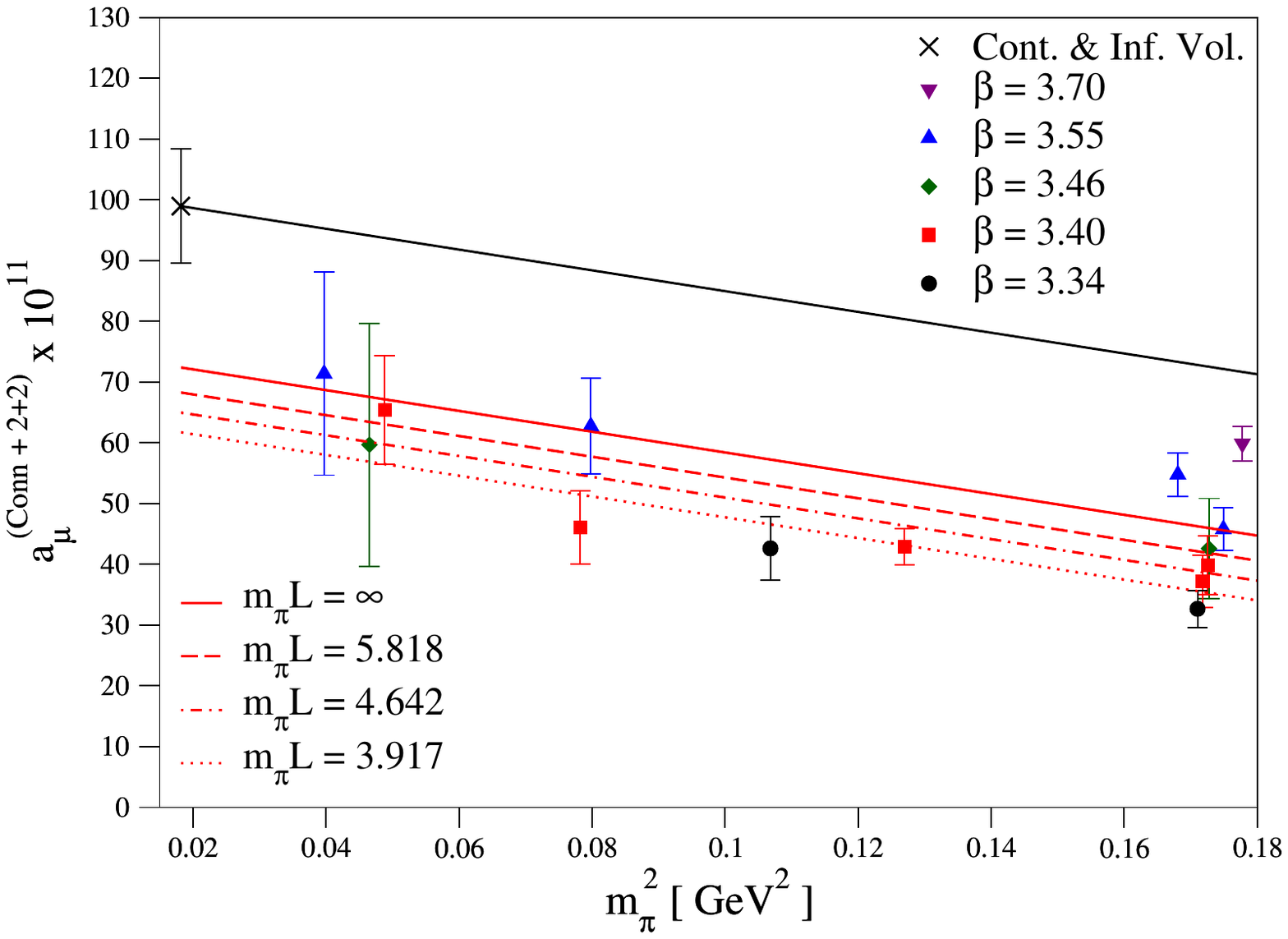}}
\caption{Example of a chiral, continuum, and infinite-volume extrapolation of the sum of the light-quark, fully-connected, and $(2+2)$ contributions to $\amuHLbL$. For this fit ansatz, the discretization errors are assumed to be linear in $a^2$ and $M_\pi$-independent, and the chiral dependence to be linear in $M_\pi^2$.
  The red straight lines labeled by an $M_\pi L$ value correspond to the lattice spacing $a=0.086\,$fm. The infinite-volume, continuum behavior
inferred from the fit is shown as the black, uppermost straight line. Figure taken from Ref.~\cite{Chao:2021tvp}.}\label{fig:sumplot}
\end{figure}

The final value of $\amuHLbL$ at the physical point is the
result of an extrapolation to infinite volume, zero lattice spacing,
and physical pion mass, see \cref{fig:sumplot}. Various fit ans\"atze as well as cuts in the
pion mass and lattice spacing were used to assess the systematic uncertainty.
The final result reads
\begin{equation}\label{eq:amuHLbLellMainz}
a_\mu^{\rm HLbL,\ell} = 107.4 ( 15.8) \times 10^{-11}\,,
\end{equation}
where the indicated total error of 15.8 consists of the statistical error (11.3), of the systematic
error from varying the fit ansatz and data cuts (9.2), which is dominated by the difference
in the results obtained from extrapolations linear in $a$ and in $a^2$; and finally, of an error
associated with the chiral extrapolation quantified as (6.0). The different components were then combined in quadrature.
In Ref.~\cite{Chao:2022zqs}, a modified strategy for the chiral extrapolation was investigated, namely by first
subtracting the $\pi^0$ exchange contribution from $\amuHLbL$ computed on each gauge ensemble,
and adding back the $\pi^0$ contribution at the every end. This procedure leads to a very flat
chiral extrapolation, with a result entirely consistent with \cref{eq:amuHLbLellMainz}.

\subsubsection{\texorpdfstring{$a_\mu^{\rm HLbL,\ell}$}{}: the RBC/UKQCD calculation \label{sec:HLbL_RBC}}

By the time of WP20, the RBC/UKQCD collaboration had published a lattice calculation~\cite{Blum:2019ugy} with both QCD and QED in a finite-size box using the so-called QED$_L$~\cite{Hayakawa:2008an} scheme.
The finite volume error of HLbL with QED$_L$ scales with $\mathcal O(1/L^2)$. Lattices with different sizes ranging from $L=4.67~\mathrm{fm}$ up to $L=9.33~\mathrm{fm}$ were used to extrapolate to the infinite-volume limit.
Two different lattices $a=0.114~\mathrm{fm}$ and $a=0.084~\mathrm{fm}$ are used for extrapolating to the continuum limit.
The results obtained are listed below:
\begin{align}
  a_\mu^{\rm HLbL,\ell} + a_\mu^{\rm HLbL,(2s+2\ell)} + a_\mu^{\rm HLbL,(2s+2s)} &=
  78.7(30.6)_\text{stat}(17.7)_\text{syst}~[35.4]\times 10^{-11}
  \,, \notag\\
  a_\mu^{\rm HLbL,(4\ell)} &=
  241.6(23.0)_\text{stat}(51.1)_\text{syst}~[56.0] \times 10^{-11}
  \,,\notag\\
  a_\mu^{\rm HLbL,(2\ell+2\ell)} + a_\mu^{\rm HLbL,(2s+2\ell)} + a_\mu^{\rm HLbL,(2s+2s)}  &=
  -164.5(21.3)_\text{stat}(39.9)_\text{syst}~[45.2]\times 10^{-11}
  \,.
\end{align}
This publication does not quote $a_\mu^{\rm HLbL,\ell}$ or $a_\mu^{\rm HLbL,(2\ell+2\ell)}$ alone.
For the purpose of comparison, we subtract the result $a_\mu^{\rm HLbL,(2s+2\ell)} + a_\mu^{\rm HLbL,(2s+2s)} = -3.6 (2.2)_\text{stat}(0.3)_\text{syst}\times 10^{-11}$ calculated in RBC/UKQCD's new publication~\cite{Blum:2023vlm}, and obtain the following results:
\begin{align}
  a_\mu^{\rm HLbL,\ell} &=
  82.3(30.7)_\text{stat}(17.7)_\text{syst}~[35.4]\times 10^{-11} 
  \,, \notag\\
  a_\mu^{\rm HLbL,(2\ell+2\ell)}  &=
  -160.9(21.4)_\text{stat}(39.9)_\text{syst}~[45.3]\times 10^{-11}
  \,.
\end{align}

The remaining part of this subsection is dedicated to the new RBC/UKQCD calculation~\cite{Blum:2023vlm} with QED$_\infty$. In this calculation, the muon momentum $p$ is set to its rest frame value, $p^2=-m^2_\mu$ with $\mathbf{p}=\boldsymbol{0}$. Note that the time direction is identified with the axis of the hypercubic lattice that has the longest extent. A first observation is that due to the property  $\widehat\Pi_{\rho;\mu\nu\lambda\sigma}(-x,-y)= - \widehat\Pi_{\rho;\mu\nu\lambda\sigma}(x,y)$, the kernel ${\cal L}$ can be replaced by its odd part,
\begin{equation}\label{eq:LI}
{\cal L}^{{\rm I}}_{[\rho,\sigma];\mu\nu\lambda}(p,x,y)
= {\textstyle\frac{1}{2}}\left( {\cal L}_{[\rho,\sigma];\mu\nu\lambda}(p,x,y)
- {\cal L}_{[\rho,\sigma];\mu\nu\lambda}(p,-x,-y)\right)\,.
\end{equation}
This modification, which by property \cref{eq:LisOdd} is equivalent to averaging the kernel over the two values $(\pm im_\mu,\boldsymbol{0})$ of the muon momentum, has the technical advantage that  ${\cal L}^{{\rm I}}$ computed using \cref{eq:kerHLbL_def} is infrared finite prior to taking the Dirac trace.
The kernel ${\cal L}^{{\rm I}}$ is then fed into the linear combination
\begin{equation}\label{eq:LII}
    {\cal L}^{{\rm II}}_{[\rho,\sigma];\mu\nu\lambda}(p,x,y)
    =  {\cal L}^{{\rm I}}_{[\rho,\sigma];\mu\nu\lambda}(p,x,y)
    - {\cal L}^{{\rm I}}_{[\rho,\sigma];\mu\nu\lambda}(p,y,y)
     - {\cal L}^{{\rm I}}_{[\rho,\sigma];\mu\nu\lambda}(p,x-y,0)\,,
\end{equation}
which has the property of vanishing for $y=0$ and for $x=y$.
The second term in \cref{eq:LII} does not contribute to $\amuHLbL$ (in infinite volume) due to the property $\partial_\mu^{(x)}\widehat\Pi_{\rho;\mu\nu\lambda\sigma}(x,y)=0$, and neither does the third term due to the conservation of the EM current at the origin, which implies $(\partial_\lambda^{(x)}+\partial_\lambda^{(y)})\widehat\Pi_{\rho;\mu\nu\lambda\sigma}(x,y)=0$.
The main motivation for this subtraction is to reduce the discretization error.
Compared with the scheme adopted by the Mainz group proposed in Ref.~\cite{Asmussen:2019act},
this choice of the subtraction scheme introduced in Ref.~\cite{Blum:2017cer} has more contributions from the long-distance region,
and needs a dedicated study for this region as described below.

Subsequently, the kernel ${\cal L}^{{\rm II}}$ is replaced by the version ${\cal L}^{{\rm II, sym}}$, which is symmetrized under the permutation group of the three photon end-points $(x,\mu)$, $(y,\nu)$ and $(0,\lambda)$.
Exploiting the invariance of ${\cal L}^{{\rm II, sym}}$ under permutation of the internal photon vertices, the calculation of the connected contribution is reduced (from three) to a single Wick contraction with quark flow in both directions. The selected Wick contraction is such that it would lead to an overall planar quark--photon--muon diagram if combined with (the unsymmetrized) ${\cal L}^{{\rm II}}$.
Similarly, using ${\cal L}^{{\rm II, sym}}$ the three disconnected diagrams of topology (2+2) are traded for the single diagram in which the origin is connected by quark lines to the external vertex.

The subtraction vector $z^{\rm ref}$ (introduced below \cref{eq:iPihat_RBC}) can depend on $x,y$.
In the treatment of the (2+2)-type diagrams,
$z^{\rm ref}$ is set to be located at the internal vertex on the quark loop that couples to the external photon.
With the reduction of the Wick contractions due to the symmetrization in ${\cal L}^{{\rm II, sym}}$, this choice of $z^{\rm ref}$ simply means $z^{\rm ref}=0$.
This choice of $z^{\rm ref}$ suppresses the (2+2)-type diagram contribution when the external photon vertex is close to 
the internal vertex belonging to the same quark loop, whereby one expects milder cutoff effects.
The choice of $z^{\rm ref}$ is legitimate due to the classical current conservation condition being satisfied at every vertex in each individual (2+2)-type diagram.

For the connected diagrams, the same choice of $z^{\rm ref}$ as the (2+2)-type diagrams does not apply, since all vertices belong to the same quark loop. Therefore, a slightly different choice is made which (a) maintains the symmetry among the three internal vertices and (b)
mimics the choice for the (2+2)-type diagrams in the long-distance region:  $z^{\rm ref}$ is set to coincide with the internal vertex that is opposite to the shortest side of the triangle formed by the internal vertices $(0, x, y)$.
Indeed, in the long-distance region, the $\pi^0$-exchange contribution dominates, and the two vertices coupled to one end of the $\pi^0$ propagator are relatively close to each other, but far away from the other two vertices, which are coupled to the other end of the $\pi^0$ propagator. Furthermore, in the (2+2)-type diagram case, the two vertices on the same quark loop are coupled to the same end of the $\pi^0$ propagator. Therefore the integrand of the (2+2)-type diagram in which the external vertex and the vertex at $z^{\rm ref}$ are in the same quark loop is dominated precisely by the constellation where $z^{\rm ref}$ is far away from the other two internal vertices.

\begin{figure}[t]
\centerline{\includegraphics[width=0.44\textwidth]{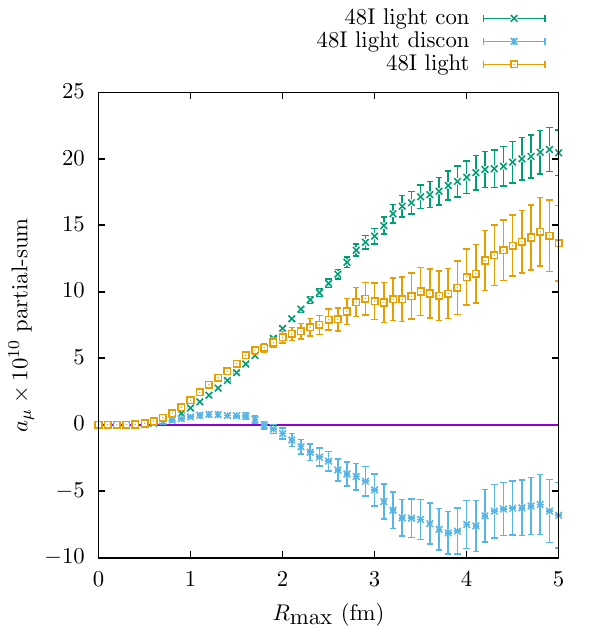}\includegraphics[width=0.44\textwidth]{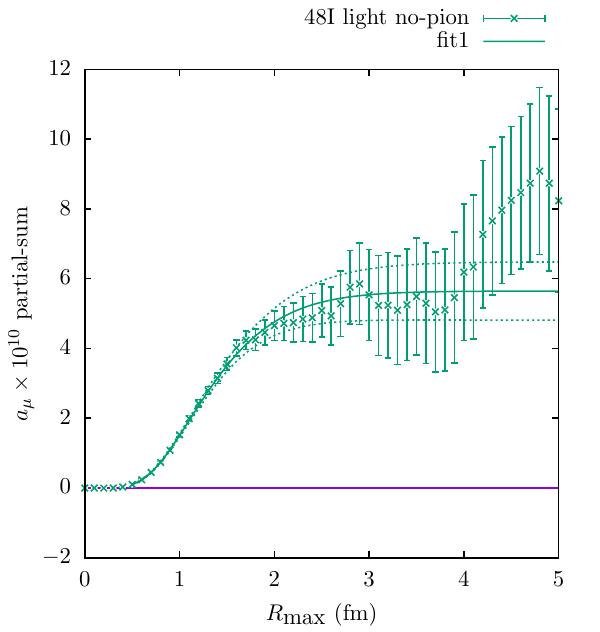}}
\caption{Left: partially integrated light-quark contributions computed on a M\"obius domain-wall fermion ensemble of size $48^3\times96$ with $M_\pi=139\MeV$, $a=0.114\,$fm from the connected diagrams, the (2+2) disconnected diagrams, and the total.
Right: linear combination of light-quark connected and (2+2) disconnected diagrams canceling the $\pi^0$ exchange contribution from long distances.
The variable $R_{\rm max}$ is the longest side of the triangle formed by the internal vertices.
Figures taken from Ref.~\cite{Blum:2023vlm}.}\label{fig:RBC_sumplot}
\end{figure}

The position-space integral as a function of its upper boundary is illustrated in \cref{fig:RBC_sumplot} for the connected and (2+2) disconnected diagrams, where $R_{\rm max}$ represents the longest side of the triangle formed by the internal vertices.
Reading off the values from the left plot in \cref{fig:RBC_sumplot} for contributions within $R_{\rm max} < 4~\mathrm{fm}$:
\begin{align}
  a_\mu^{\rm HLbL,\ell}(R_{\rm max} < 4~\mathrm{fm})  &= 111.1(21.1)_\text{stat}\times 10^{-11}\,,\notag\\
  a_\mu^{\rm HLbL,(4\ell)}(R_{\rm max} < 4~\mathrm{fm})  &= 186.1(12.2)_\text{stat}\times 10^{-11}\,,\notag\\
  a_\mu^{\rm HLbL,(2\ell+2\ell)}(R_{\rm max} < 4~\mathrm{fm})  &= -74.9(18.2)_\text{stat}\times 10^{-11}\,.
\end{align}
These results are obtained with the 48I MDWF ensemble from RBC/UKQCD collaborations.
The lattice spacing is 0.114~fm. Due to the subtraction of the QED kernel described
in \cref{eq:LII}, the discretization error is expected to be smaller than the
48I results obtained from the previous QED$_L$ calculation.
Due to the lack of light-quark data from a second lattice spacing, the discretization error of the above
results were estimated based on the continuum extrapolation for the strange-quark-connected contribution,
in which case the continuum limit is 8\% larger than the 48I result.
Therefore, RBC/UKQCD estimate 8\% discretization systematic error for the total light quark contribution $(8.3)_\text{syst}\times 10^{-11}$.

The observed cancellation between connected and disconnected contributions motivates the partitioning of the dominant HLbL contribution into a term proportional to the contribution from the purely isovector part $j_\mu^3=(\bar u\gamma_\mu u - \bar d\gamma_\mu d)/2$ of the photon coupling, $a_\mu^{{\rm HLbL}\{j^3\}}$, and a remaining fraction of the light-connected contribution,
\begin{equation}\label{eq:alt_lincomb}
    a_\mu^{\rm HLbL,\ell} = \underbrace{\frac{100}{81} a_\mu^{{\rm HLbL}\{j^3\}}}_{= a_\mu^{\rm no\,pion}} + \frac{9}{34} a_\mu^{\rm HLbL,ud\;conn}\,.
\end{equation}
The integrand for the first term, denoted $a_\mu^{\rm no\,pion}$ in Ref.~\cite{Blum:2023vlm}, being free of the pion-pole contribution~\cite{bijnens:2016hgx}, is expected to be of shorter range.
On the other hand, $a_\mu^{{\rm HLbL}\{j^3\}}$ is expected to contain the entire charged-pion-loop contribution to $\amuHLbL$, which is negative.
Clearly, the positive result shown on the right panel of 
\cref{fig:RBC_sumplot} indicates that other effects, such as the $\eta$ and $\eta'$ exchanges and shorter-distance contributions, dominate over the charged pion loop.
To take advantage of the shorter-range property of $a_\mu^{\rm no\,pion}$,
this contribution is included directly from a lattice calculation up to $R_\text{max} < 2.5~\mathrm{fm}$, which gives
$50.9(7.6)_\text{stat}\times 10^{-11}$.
For the region where $R_\text{max} > 2.5~\mathrm{fm}$, a fit is performed
\begin{align}
    \label{eq:fit-form}
f(R_\text{max}) = A / \text{fm}^4 \frac{R_\text{max}^6}{R_\text{max}^3 + (C~\text{fm})^3} e^{-B R_\text{max} / (\text{fm}\times\text{GeV})}\,,
\end{align}
as illustrated in the right plot of \cref{fig:RBC_sumplot}. This tail contribution is $3.1 (2.2)_\text{stat}(3.1)_\text{syst} \times 10^{-11}$. This is the only part of the entire calculation that is based on a fit. Many different fitting forms are studied and 100\% systematic error is assigned to this contribution.

\begin{figure}[t]
\begin{center}
\includegraphics[width=0.4\textwidth]{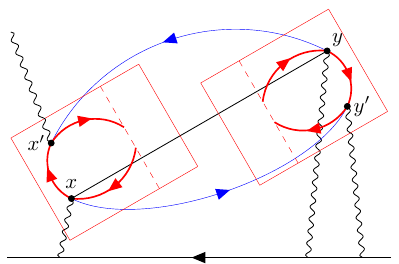}
\caption{The long-distance HLbL contribution to the muon $g-2$ associated with $\pi^0$ exchange. The amplitudes inside the boxes are calculated in lattice QCD, while the pion propagator linking them (solid lines) is given by the analytic, infinite-volume, continuum expression. Figure taken from Ref.~\cite{Blum:2023vlm}.}
\label{fig:long-distance-pi0}
\end{center}
\end{figure}

As mentioned above, the long-distance $\pi^0$ exchange contribution is calculated separately as illustrated in \cref{fig:long-distance-pi0} based on the following long-distance approximation:
\begin{equation}
\label{eq:pi0-exch-approx}
\langle T  j_{\mu'}(x') j_{\mu}(x)  j_{\nu'}(y') j_{\nu}(y) \rangle
\simeq
D_{\pi^0}(x - y)
\mathcal{F}_{\mu',\mu}\Big(x'-x, i M_\pi \frac{x-y}{|x-y|}\Big)
\mathcal{F}_{\nu',\nu}\Big(y'-y, i M_\pi \frac{y-x}{|y-x|}\Big)\,,
\end{equation}
where the $\pi^0$ TFF $\mathcal{F}_{\mu',\mu}(x, i M_\pi \hat n )$ is directly calculated on the lattice with
\begin{equation}
    \label{eq:pi0-tff}
\mathcal{F}_{\mu,\nu} (x, i M_\pi \hat t)  = \langle 0 | T j_{\mu}(x) j_{\nu}(0) | \pi^0(\mathbf{p}=\boldsymbol{0}) \rangle
\end{equation}
and a proper Euclidean space--time rotation.
Here, the long-distance contribution includes regions with $R_\text{max} > 4~\mathrm{fm}$.
This contribution is calculated to be $20.0(1.1)_\text{stat}(2.8)_\text{syst}\times 10^{-11}$.

The finite-volume correction for the contribution in the region $R_\text{max} < 4~\mathrm{fm}$ is calculated based on the
coordinate space formulation of the $\pi^0$-pole contribution in finite volume, with the $\pi^0$ transition form factor given by the LMD model. The correction is $-4.7(1.1)_\text{syst}\times 10^{-11}$. This model can also be used to calculate the above long-distance contribution
and is consistent with the above lattice results based on the long-distance approximation.

The last correction is from a small mismatch of the physical pion mass $135~\mathrm{MeV}$ and the pion mass used in the above lattice calculation, $139~\mathrm{MeV}$. The correction is based on a separate lattice calculation at $M_\pi=340~\mathrm{MeV}$.
The correction is calculated to be $3.5(0.7)_\text{stat}(1.7)_\text{syst}\times 10^{-11}$.

The final value of $a_\mu^{\rm HLbL,\ell}$ after applying the corrections from finite volume, small mismatch of pion mass is
\begin{equation}
  a_\mu^{\rm HLbL,\ell}  = 122.0(10.1)_\text{stat}(9.5)_\text{syst}~[13.8] \times 10^{-11}\,.
\end{equation}
The leading source of systematic error is from the estimation of the discretization effects,
which is estimated to be $(8.3)_\text{syst}\times 10^{-11}$. A future calculation at a finer lattice spacing can help reduce this systematic error.
Results for the individual contributions are also obtained
\begin{align}
  a_\mu^{\rm HLbL,(4\ell)} &= 257.0(13.3)_\text{stat}(19.9)_\text{syst}~[23.9]\times 10^{-11}\,,\notag\\
  a_\mu^{\rm HLbL,(2\ell+2\ell)} &= -135.0(13.6)_\text{stat}(12.1)_\text{syst}~[18.2]\times 10^{-11}\,,\notag\\
  a_\mu^{\rm no\,pion} &=\frac{25}{34}a_\mu^{\rm HLbL,(4\ell)} + a_\mu^{\rm HLbL,(2\ell+2\ell)}
  =  54.0(9.4)_\text{stat}(5.3)_\text{syst}~[10.8]\times 10^{-11}\,.
\end{align}

\subsubsection{\texorpdfstring{$a_\mu^{\rm HLbL,\ell}$}{}: the BMW calculation}

In Ref.~\cite{Fodor:2024jyn}, the staggered-fermion formulation of QCD is used with $N_f=2+1+1$ dynamical quarks, whose action involves four steps of stout smearing. The pion and kaon masses are very close to their physical values.
The light-quark contribution is computed on seven ensembles at three values of the lattice spacing, 0.132, 0.112 and 0.095\,fm. The linear extent of three ensembles is just above $L=6\,{\rm fm}$, one has $L=4.2\,{\rm fm}$, and the other three, which do not enter the final continuum extrapolation, have $L\simeq3.1\,{\rm fm}$.
The conserved vector current, which does not receive any multiplicative renormalization, is used throughout.

The starting point for the integral representation chosen of $a_\mu^{\rm HLbL,\ell}$ is the kernel $\bar{\cal L}^{(\Lambda)}$ defined in \cref{eq:lamsub}, for which the same numerical implementation \cite{Asmussen:2022oql} is used as in the Mainz/CLS calculation.
In contrast to the latter, however, the kernel $\bar{\cal L}^{(\Lambda)}$ is then entirely symmetrized with respect to the three internal vertices. The $x$ and $z$ integrals are performed as sums over the lattice, while the remaining $y$ integral is handled as a one-dimensional integral over $|y|$ thanks to the use of spherical coordinates.
The position vector $y$ of the vertex is chosen to be an integer multiple of (1,1,1,1) or (3,1,1,1).
In fact, this vertex is spread over the eight nearest neighbors of its nominal position in order to suppress oscillating lattice artifacts.

For the connected contribution $a_\mu^{\rm HLbL,(4\ell)}$, the calculation is reduced to the quark-contraction diagram in which the vertex at the origin is connected to the internal vertex at $x$ and to the external vertex at $z$. Due to the use of translation invariance to perform this reduction to one diagram, effectively a subtraction term $z_\rho^{\rm ref}=-x_\rho/3$ appears (see the comment below \cref{eq:iPihat_RBC}).
The chosen representation is validated by reproducing the known contribution of a lepton loop  to $a_\mu^{\rm LbL}$ on the lattice, for a lepton twice as heavy as the muon.

\begin{figure}[t]
\centerline{\includegraphics[width=0.495\textwidth]{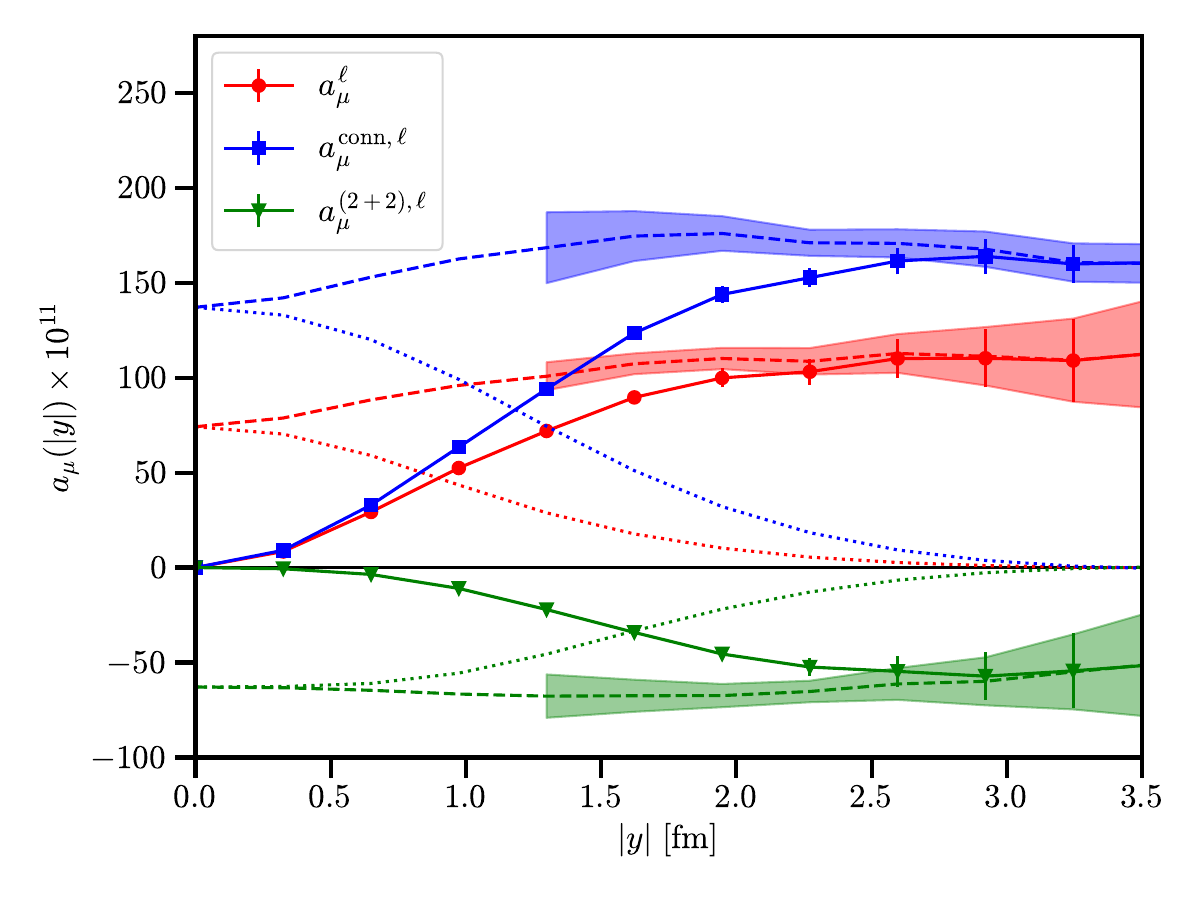}\includegraphics[width=0.495\textwidth]{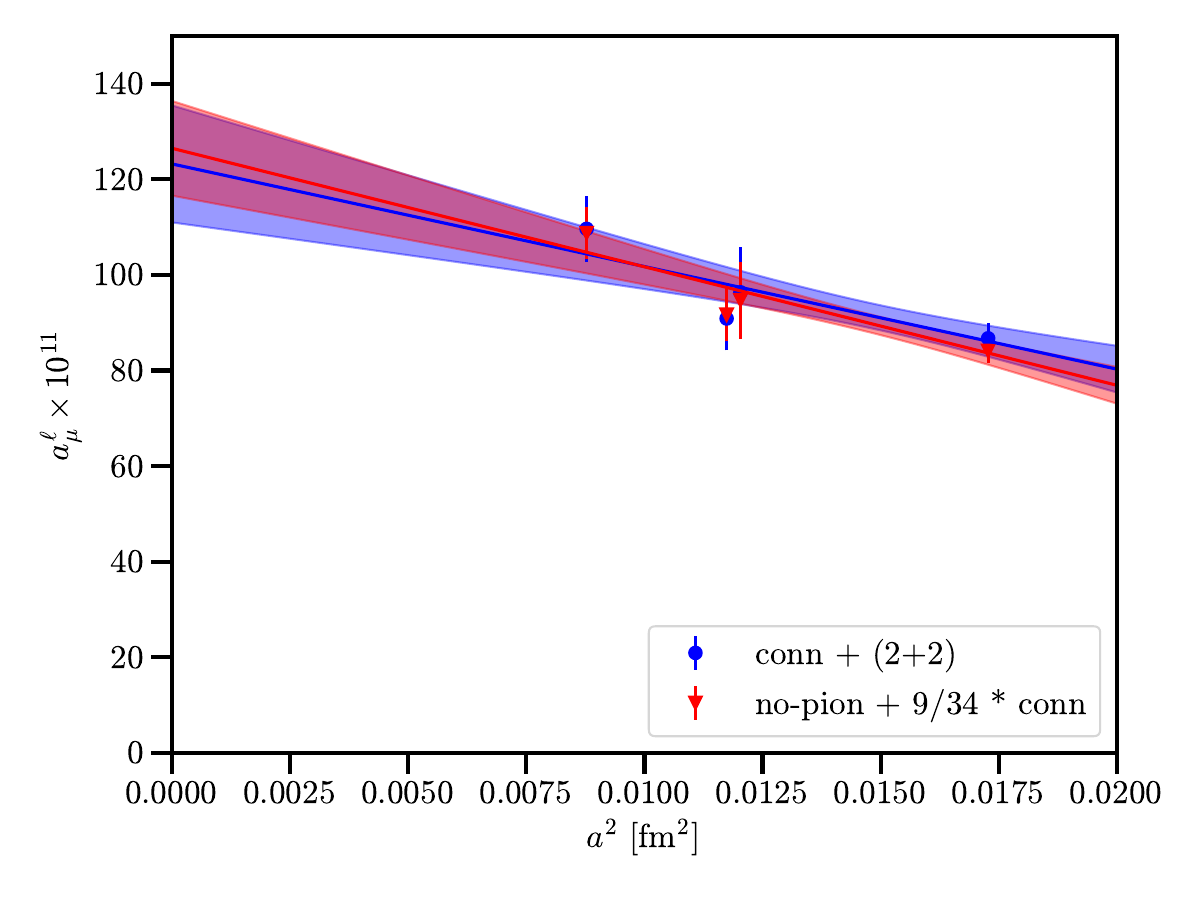}}
\caption{
 Left: partially integrated light-quark contributions computed on a staggered-quark, physical-mass ensemble of size $64^3\times96$ with  $a=0.095\,$fm from the connected diagrams, the (2+2) disconnected diagrams, and the total.
 Right: continuum extrapolation of $a_\mu^{\rm HLbL,\ell}$, either based on the direct calculation of the sum of the connected and disconnected diagram, or based on combining the $a_\mu^{\rm no\,pion}$ with the connected contribution.
Figures taken from Ref.~\cite{Fodor:2024jyn}.}\label{fig:BMW_sumplot}
\end{figure}

For the disconnected contribution $a_\mu^{\rm HLbL,(2\ell,2\ell)}$, there are again three Wick contractions, which can all be expressed in terms of the diagram in which 0 is connected (by two quark lines) with $x$, or in terms of the diagram where 0 is connected with $z$.
Each of these two possibilities leads to a valid expression for $a_\mu^{\rm HLbL,(2\ell,2\ell)}$, and the final estimator is the average of these two expressions.
This choice has the effect of giving the pseudoscalar-pole contribution to the connected diagram and the two disconnected diagrams precisely the same weight function, so that the $\pi^0$ contribution can be made to cancel at the \emph{integrand} level in the linear combination $a_\mu^{\rm no\,pion}$.
The final $|y|$ integral as a function of its upper bound is displayed in the left panel of \cref{fig:BMW_sumplot} for $a_\mu^{\rm HLbL,(4\ell)}$, $a_\mu^{\rm HLbL,(2\ell,2\ell)}$ and for their sum, $a_\mu^{\rm HLbL,\ell}$.

The pion-pole contribution is computed in position space and in finite volume in order to extend the $|y|$ integrand at long distances and to correct for finite-size effects.
The switch from lattice data to the pion-pole prediction is done at $y_{\rm cut}\simeq2.3$\,fm.
It is also used to assess how soon the $x$ integral can be truncated once $|y|>1.5\,$fm, a technique that is used to reduce statistical fluctuations. The parameterizations used for the pion TFF are the VMD and LMD ones. The latter is found to provide a satisfactory description of finite-size effects, while the prediction using the VMD form factor yields corrections that are 20 to 25\% smaller.
Overall, the corrections applied to the raw lattice data summed up to $y_{\rm cut}$ are on the order of 4 to $6.5\times10^{-11}$ for $a_\mu^{\rm HLbL,\ell}$.

The continuum extrapolation of $a_\mu^{\rm HLbL,\ell}$ is shown in the right panel of \cref{fig:BMW_sumplot}.
The alternative route, which consists in working with the linear combination \cref{eq:alt_lincomb} of the shorter-range $a_\mu^{\rm no\,pion}$ and the connected part, gives consistent results.
The final results are collected in the rightmost column of \cref{tab:amuHLbL_light}. No continuum-extrapolated result is quoted for $a_\mu^{\rm no\,pion}$, however, the results at finite lattice spacing are in the range 49 to $59\times10^{-11}$, increasing toward the continuum. This contribution would end up being higher than the central value obtained by the RBC/UKQCD collaboration (penultimate column of \cref{tab:amuHLbL_light}), but still compatible given the uncertainties.

\begin{table}[t]
\centering
\small
\renewcommand{\arraystretch}{1.1}
    \begin{tabular}{cc}
    \toprule
    $ [10^{-11}]$  & RBC/UKQCD QED$_L$~\cite{Blum:2019ugy} \\
         \midrule
    $a_\mu^{\rm HLbL,(4\ell)} \phantom{^{\big|}}$ & $241.6(23.0)_\text{stat}(51.1)_\text{syst}~[56.0]$  \\
    $a_\mu^{\rm HLbL,(2\ell+2\ell)}\phantom{\big|} $  & $-160.9(21.4)_\text{stat}(39.9)_\text{syst}~[45.3]$  \\
    \midrule
 $a_\mu^{\rm HLbL,\ell}\phantom{^{\big|}}$  &  $82.3(30.7)_\text{stat}(17.7)_\text{syst}~[35.4]$  \\
    \bottomrule
    \end{tabular}
    \\~\\~\\
    \begin{tabular}{cccc}
    \toprule
    $ [10^{-11}]$    & Mainz/CLS~\cite{Chao:2021tvp}  &  RBC/UKQCD~\cite{Blum:2023vlm} &  BMW~\cite{Fodor:2024jyn} \\
         \midrule
    $a_\mu^{\rm HLbL,(4\ell)} \phantom{^{\big|}}$ & see text     & $257.0(13.3)_\text{stat}(19.9)_\text{syst}~[23.9]$ &  $220.1(13.0)_\text{stat}(3.8)_\text{syst}$ \\
    $a_\mu^{\rm HLbL,(2\ell+2\ell)}\phantom{\big|} $  & see text  & $-135.0(13.6)_\text{stat}(12.1)_\text{syst}~[18.2]$  &  $-101.1(12.4)_\text{stat}(3.2)_\text{syst}$ \\
    $a_\mu^{\rm no\,pion}\phantom{\big|}$ & $-$                  & $54.0(9.4)_\text{stat}(5.3)_\text{syst}~[10.8]$ &  see text \\
    \midrule
 $a_\mu^{\rm HLbL,\ell}\phantom{^{\big|}}$  & $107.4(11.3)(9.2)(6.0)[15.8]$  & $122.0(10.1)_\text{stat}(9.5)_\text{syst}~[13.8]$ & $122.6(11.5)_\text{stat}(1.8)_\text{syst}$  \\
    \bottomrule
    \end{tabular}
    \renewcommand{\arraystretch}{1}
    \caption{Results for the light-quark contributions in various linear combinations, as well as the total $a_\mu^{\rm HLbL,\ell}$.}
    \label{tab:amuHLbL_light}
\end{table}

\subsubsection{\texorpdfstring{$a_\mu^{\rm HLbL,\ell}$}{}: our average}

Before proceedings to average $a_\mu^{\rm HLbL,\ell}$, we briefly compare the results for the connected and the disconnected contributions; see \cref{tab:amuHLbL_light}.
 The Mainz/CLS publication does not quote a final result for the two topologies separately, but provides the results of four different fit ans\"atze with regards to the chiral extrapolation. This results in a fairly large spread. Nevertheless, all three collaborations obtain central values in the range $200$ to $260$ for the connected, while the disconnected part is in the range $-145$ to $-100$. Clearly, the absolute size of systematic effects is larger when the connected and the disconnected contributions are handled separately, due to the enhanced weight with which the $\pi^0$ exchange contributes in these. For that reason, we refrain for now from presenting an average for $a_\mu^{\rm HLbL,(4\ell)}$ and $a_\mu^{\rm HLbL,(2\ell+2\ell)}$ separately. The decomposition into connected and $a_\mu^{\rm no\,pion}$ is more favorable in that respect.

We now proceed to perform an average of the light-quark contribution $a_\mu^{\rm HLbL,\ell}$.
The four calculations were done with three qualitatively different quark actions on independent ensembles; therefore the statistical errors are fully uncorrelated (the two RBC/UKQCD calculations are performed with the same quark action but with two almost independent data sets).
We adopt the approach of taking into account correlations among the systematic errors conservatively.
In the Mainz/CLS case, the systematic uncertainty of $6.0\times10^{-11}$ associated with the chiral extrapolation will be treated as being uncorrelated to the other two calculations, since the latter are performed directly at (near-)physical quark masses.

Now treating the systematic errors $\{17.7,9.2,9.5,1.8\}\times10^{-11}$ of the four calculations as being 100\% correlated (that is the systematic errors from different results added linearly in the average), we construct the covariance matrix of the results and perform a fit to a constant.
\footnote{In the weighted average we always add linearly the absolute value of the weighted systematic errors. The fit is performed to find the weights $w_i$ (the weights should satisfy $\sum_i w_i = 1$) that minimize the total error of the weighted average.
The formula for the final error is $\sqrt{\sum_i w_i^2 e_{\text{stat},i}^2 + (\sum_i|w_i e_{\text{syst},i}|)^2 + (\sum_i|w_i e_{\text{chiral-extrap},i}|)^2}$, where $e_{\text{stat},i}$ and $e_{\text{syst},i}$ are the statistical and systematic errors for each lattice calculations, $e_{\text{chiral-extrap},i}$ are the errors for the chiral extrapolations, which are equal to zero for calculations performed at physical quark masses.
Note that, with the above minimization procedure, the final weights should always be nonnegative.
} The $\chi^2/{\rm dof}$ is 0.69 and the result is
\begin{equation}\label{eq:amuHLbL_light_aver}
 a_\mu^{\rm HLbL,\ell}= 119.6(7.1)_\text{stat}(5.5)_\text{syst}[9.0]_\text{tot} \times 10^{-11}
 \qquad \textrm{(our~average)}\,.
\end{equation}
If we treated the systematic errors as being uncorrelated, the result would change to $117.1(6.5)_\text{stat}(4.0)_\text{syst}[7.6]_\text{tot}\times 10^{-11}$. We have checked that interpolating between these two results with a scale factor $\xi\in[0,1]$ multiplying the off-diagonal components of the covariance matrix, the error of the average increases monotonously towards the fully correlated $\xi=1$ case.

\subsubsection{The strange contribution \texorpdfstring{$a_\mu^{\rm HLbL,s}$}{}}
\label{subsec:HLBL_strange_contrib}

\begin{figure}
    \centering
    \includegraphics[width=0.495\linewidth]{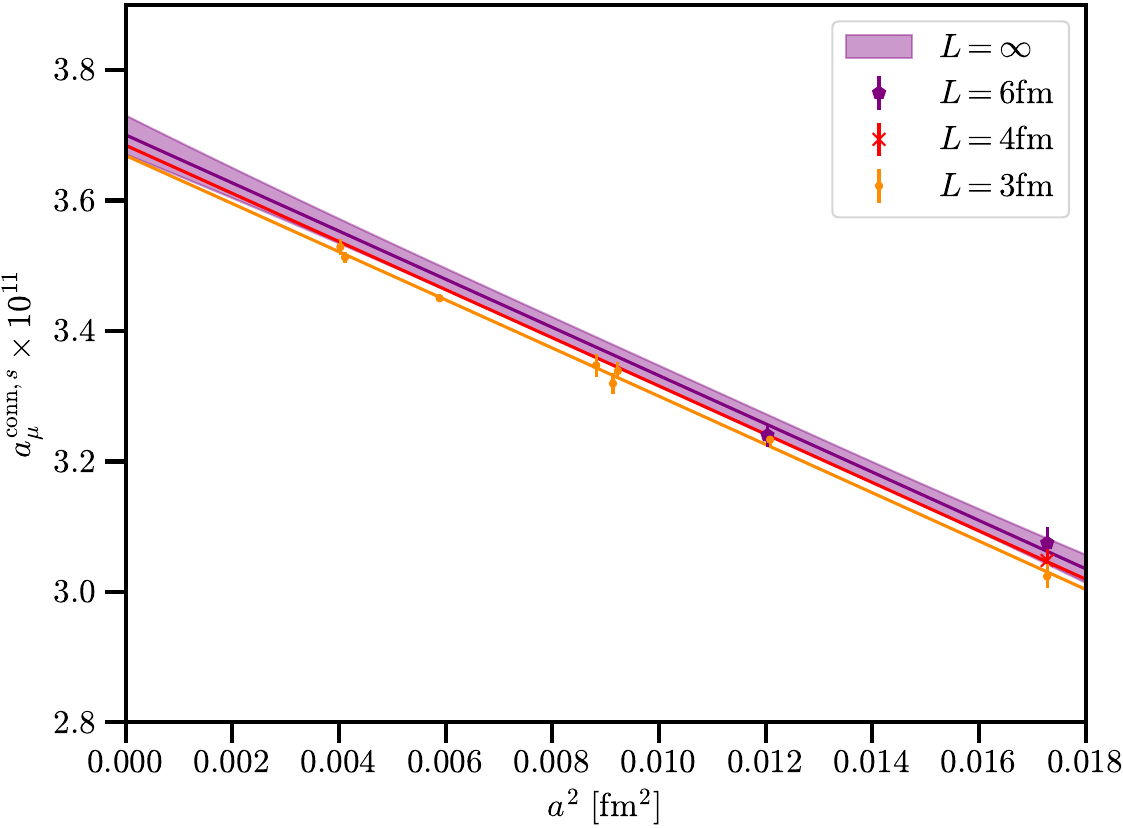}
    \includegraphics[width=0.495\linewidth]{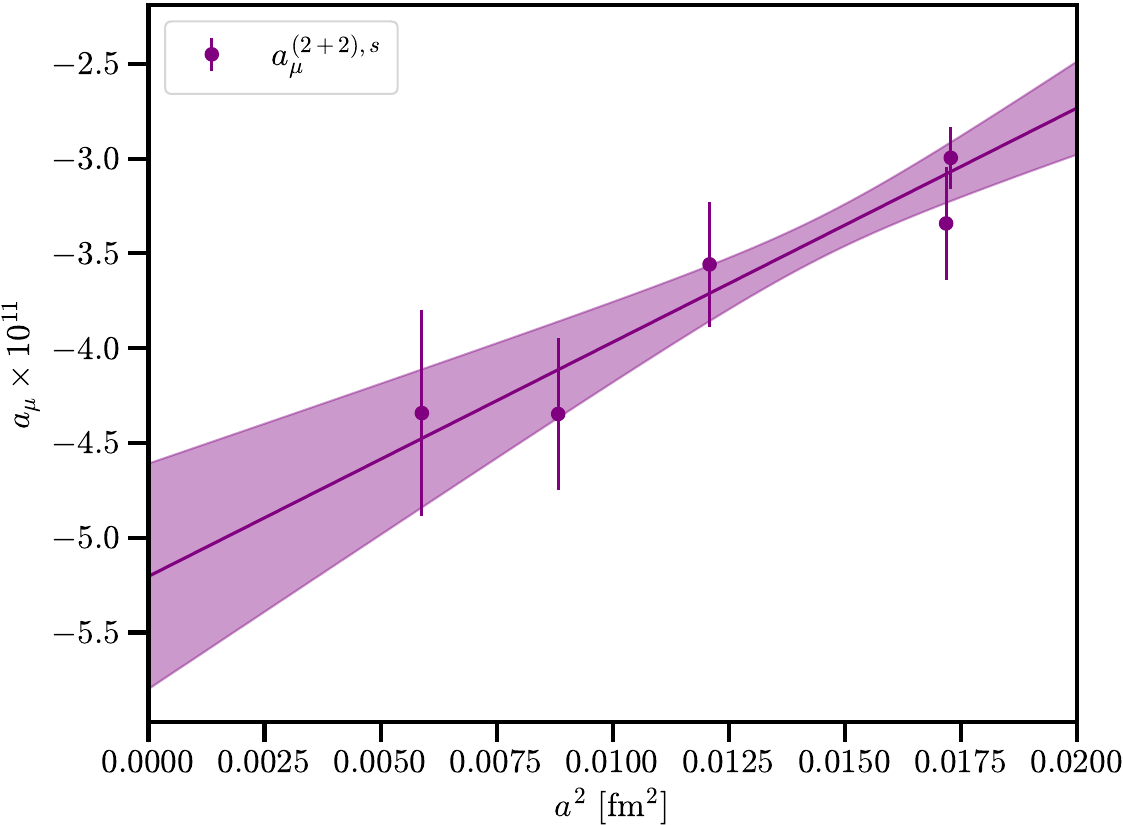}
    \caption{Extrapolation of the strange connected (left, $a_\mu^{\rm HLbL,(4s)}$) and disconnected contribution (right, $a_\mu^{\rm HLbL,(2l+2s)}+a_\mu^{\rm HLbL,(2s+2s)}$) in the calculation by the BMW collaboration~\cite{Fodor:2024jyn} based on ensembles with dynamical up/down, strange, and charm  quarks tuned to their physical masses. Figures taken from Ref.~\cite{Fodor:2024jyn}.}
    \label{fig:extrap_strange}
\end{figure}

 Three collaborations~\cite{Chao:2021tvp,Blum:2023vlm,Fodor:2024jyn} have produced results for $a_\mu^{\rm HLbL,s}$. We begin by discussing the connected part alone.
The strange connected contribution $a_\mu^{\rm HLbL,(4s)}$ is an excellent benchmark quantity to compare lattice calculations, since the signal is of good quality, not too long range, and finite-size effects are expected to be suppressed. We quote the results
\begin{equation}
   a_\mu^{\rm HLbL,(4s)} = \left\{
\begin{array}{ll}
    3.530(70)_{\rm stat} \times 10^{-11} & {\rm RBC/UKQCD}\phantom{_\big|}
\\
    3.694(17)_{\rm stat}(18)_{a}(8)_{\rm syst} \times 10^{-11} & {\rm BMW}
    
\end{array}\right.
    \,.
\end{equation}
Both calculations are performed directly at (practically) physical quark masses.
The BMW calculation uses five lattice spacings in the range 0.064 to 0.132\,fm, see \cref{fig:extrap_strange}, and the statistical precision was sufficiently high that finite-size effects could be resolved---and corrected for. The cutoff effect on the coarsest lattice spacing is about 16\% of the continuum value.
The RBC/UKQCD calculation uses two lattice spacings, 0.114 and 0.084\,fm to obtain the continuum limit, relative to which the data on the coarser ensemble is 8\% lower.
Here we also mention the preliminary results of the ETM collaboration \cite{Kalntis:2024dyd}, that obtains continuum-extrapolated values somewhat lower than RBC/UKQCD or BMW, with an error still to be quantified.

The Mainz/CLS publication~\cite{Chao:2021tvp} does not quote a result for $a_\mu^{\rm HLbL,(4s)}$ alone. For the purpose of a comparison with the two results above, we produce an estimate from (a) the continuum-extrapolated result at the $M_\pi=M_K\simeq 415\MeV$ point, Eq.~(22)~\cite{Chao:2020kwq} with the charge factor adjust from $18/81$ to $1/81$ in order to isolate the strange contribution of $5.49(14)\times 10^{-11}$, and (b) by fitting the $a_\mu^{\rm HLbL,(4s)}$ values obtained on
three ensembles at a fixed lattice spacing $a=0.086\,$fm and $M_K=415,\;461$ and $487\MeV$
 in order to correct to the physical $M_K=494.6\MeV$.
 Indeed, the data for $a_\mu^{\rm HLbL,(4\ell)}$ suggests that the cutoff effects on the valence-quark mass dependence is weak.
 We do so using a quadratic fit in $M_K^2$ and obtain $a_\mu^{\rm HLbL,(4s)}= 3.64(14)_{{\rm SU(3)}_{F}}(9)_{M_K\,{\rm fit}}\times 10^{-11}$, where we have assigned a 5\% uncertainty to the $M_K$ dependence.

 We conclude that the agreement between the three calculations of $a_\mu^{\rm HLbL,(4s)}$ is satisfactory. At the same time, there is evidence that the quoted errors are somewhat underestimated. A straightforward fit to a constant leads to $\chi^2/{\rm dof}=1.8 $;  the error inflated by the square-root of this quantity would be $0.034 \times 10^{-11}$.  For simplicity, we then quote
 \begin{equation}
  a_\mu^{\rm HLbL,(4s)}= 3.68(4)\times 10^{-11}\qquad {\rm (our~average)}\,.
 \end{equation}

After this important consistency check,
we quote the following results for the full contribution $a_\mu^{\rm HLbL,s}$,
\begin{equation}
\label{eq:amuHLBL_strange_list}
   a_\mu^{\rm HLbL,s} = \left\{
\begin{array}{ll}
-0.6 (2.0)_{\rm stat}\times 10^{-11} & {\rm Mainz/CLS}
\\
   -0.0 (2.2)_{\rm stat}(0.3)_{\rm syst}\times 10^{-11}  & {\rm RBC/UKQCD}\phantom{\Big|}
\\
   -1.7 (8)_{\rm stat}(3)_{\rm syst}\times 10^{-11} & {\rm BMW}
\end{array}
    \right.
    \,.
\end{equation}
Due to, in part, the much larger statistical errors of the disconnected diagrams, these results are entirely consistent with each other. Performing a straightforward average, we again slightly inflate the error
and obtain
\begin{equation}\label{eq:amuHLBL_strange_aver}
  a_\mu^{\rm HLbL,s}= -1.4(8)\times 10^{-11}\qquad {\rm (our~average)}\,.
 \end{equation}

\subsubsection{The charm contribution \texorpdfstring{$a_\mu^{\rm HLbL,c}$}{}}
\label{subsec:HLBL_charm_contrib}

\begin{figure}
  \centerline{\includegraphics[width=0.52\textwidth]{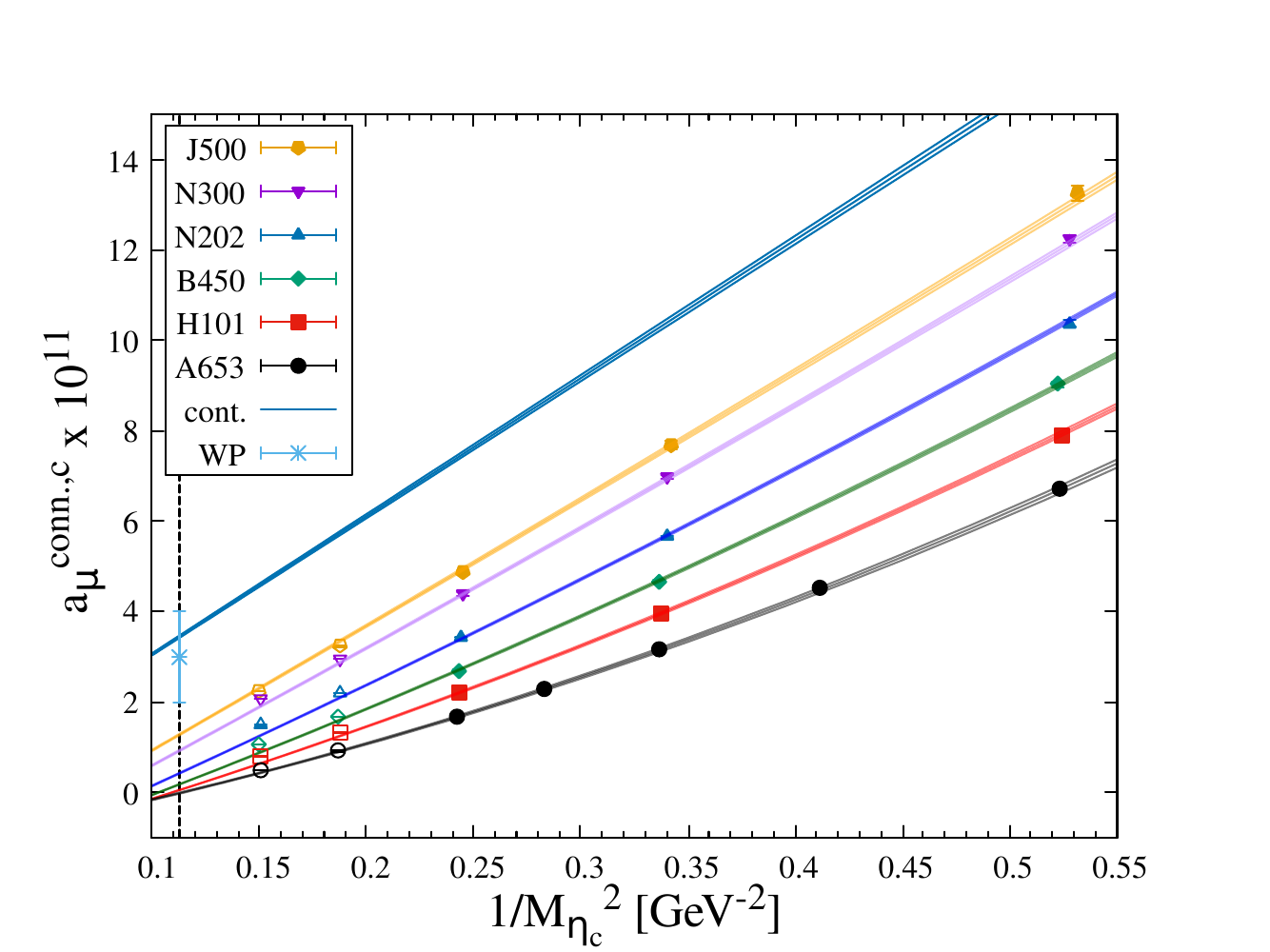}\includegraphics[width=0.47\textwidth]{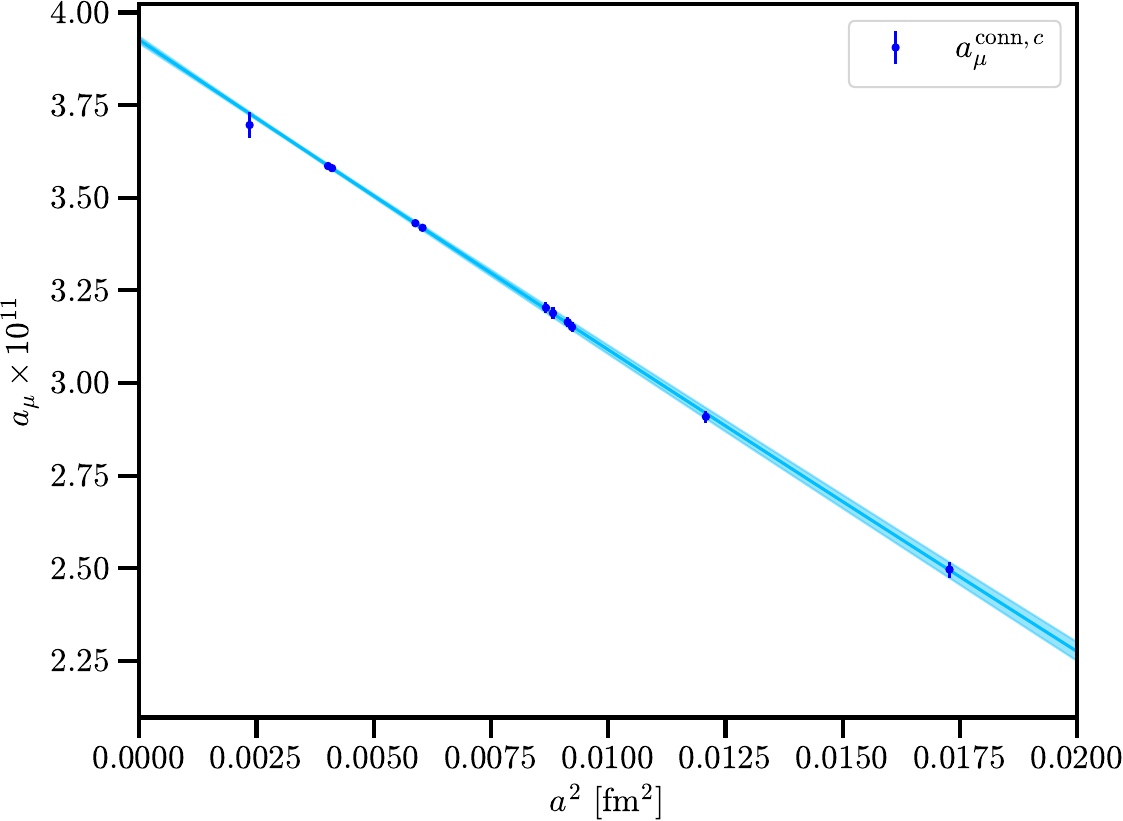}}
  \caption{\label{fig:charmHLbL} Extrapolations of the connected charm contribution. Left: One representative fit to the lattice data of the Mainz/CLS collaboration~\cite{Chao:2022xzg}
    at different charm-quark masses and lattice spacings. The $\eta_c$ meson mass is used as a proxy for the charm mass.
    The continuum limit of the lattice data is represented by the blue, uppermost curve.
    For comparison, the WP20 estimate of $a_\mu^{\rm HLbL,c}=3(1)\times 10^{-11}$ is displayed at the physical $\eta_c$ mass. Figure taken from Ref.~\cite{Chao:2022xzg}. Right: Representative continuum extrapolation at physical charm quark mass by the BMW collaboration~\cite{Fodor:2024jyn}. Figure taken from Ref.~\cite{Fodor:2024jyn}.}
\end{figure}

The Mainz/CLS and the BMW collaboration have performed dedicated lattice calculations of the
charm-quark contribution $a_\mu^{\rm HLbL,c}$~\cite{Chao:2022xzg,Fodor:2024jyn}, see \cref{fig:charmHLbL},
 and preliminary results by the ETM collaboration are available \cite{Kalntis:2024dyd}.
Due to the large mass scale involved, the main challenge in this case
is to control the cutoff effects.

In the Mainz/CLS calculation, based on quarks with an $\Order(a)$-improved Wilson action, the adopted strategy consisted in using very fine lattice spacings
(down to 0.039\,fm) and lighter-than-physical charm-quark masses. These calculations were followed by a simultaneous extrapolation in both variables to the continuum and physical charm quark mass, respectively, to obtain the final result.  To make this strategy
computationally affordable, the calculation was performed on ensembles
with dynamical $(u,d,s)$ quarks at the SU(3)$_{F}$-symmetric point
with $M_\pi=M_K\simeq415\MeV$. Indeed, neither the charm-quark loop,
nor the $\eta_c$ exchange is expected to depend strongly on the
light-quark masses.

An important question is then what fit ans\"atze to employ in order to
describe the charm mass and lattice spacing dependence of $a_\mu^{\rm HLbL,c}$.
Being a heavy degree of freedom, general arguments~\cite{Beresetskii:1956,Cowland:1958}
strongly suggests a leading $1/m_c^2$ dependence. As for the
lattice-spacing dependence, both $\Order(a)$ and $\Order(a^2)$ effects are
present in the Mainz setup. The (dominant) connected charm
contribution and its uncertainty were estimated so as to cover the
continuum results emerging from all fits with good $\chi^2/{\rm dof}$.

The correlation between a light-quark and a charm-quark two-vector-insertion loop
(2+2 Wick-contraction topology), expected to be the next most important contribution,
was computed as well and found to amount to a $-10\%$ correction to the connected piece.
The corresponding (2+2) diagram involving two charm loops was found to be negligible.
Altogether, the final result of Ref.~\cite{Chao:2022xzg} was
\begin{equation}\label{eq:charm_Mz}
a_\mu^{\rm HLbL,c} = 2.8(5) \times 10^{-11} \qquad \textrm{(Mainz/CLS)}\,,
\end{equation}
where the uncertainty is entirely systematics dominated.
It also contains an estimate of the effect of not performing the calculation at physical
$(u,d,s)$ quark masses.

In the BMW calculation~\cite{Fodor:2024jyn}, which is performed directly at physical quark masses in the staggered-quark formulation, the continuum extrapolation also represented a major source of uncertainty.
Additively implemented tree-level improvement of the observable $a_\mu^{\rm HLbL,c}$ was applied to data obtained from six lattice spacings in the range 0.048 to 0.132\,fm.
Two directions were used for the vertex $y$ integrated over last, $(1,1,1,1)$ and $(0,1,1,1)$, leading to separate continuum extrapolations. The difference in results was taken as an additional  systematic error.
Additional checks were performed, by using lighter-than-physical charm quark masses as well as by reproducing the result for a lepton loop in the continuum. As a result of the spread obtained in the latter exercise, a sizable uncertainty ($0.25\times 10^{-11}$) was associated with the numerical implementation of the kernel.

The leading disconnected diagrams involving a charm-quark loop, namely those of topology (2+2), were computed at four values of the lattice spacing, extrapolated to the continuum and found to be negative. The final integrand is more extended than for the connected diagram, reaching its maximum size around $|y|=0.6\,$fm and is found to yield a contribution on the order of $-0.2\times 10^{-11}$ after continuum extrapolation. All in all,
the total charm contribution was found to be
\begin{equation}\label{eq:charm_BMW}
a_\mu^{\rm HLbL,c} = 3.73(26) \times 10^{-11} \qquad \textrm{(BMW)}\,,
\end{equation}
where the error is entirely dominated by the systematic uncertainty associated with the short-distance part of the kernel, see the previous paragraph, and the continuum extrapolation of the connected part.

The ETM collaboration has performed preliminary calculations of the connected charm contribution \cite{Kalntis:2024dyd} at physical quark masses with twisted-mass Wilson sea quarks, using three lattice spacings in the range 0.057 to 0.080\,fm. While an upward trend is observed in the $a_\mu^{\rm HLbL,c}$ results as the continuum is approached, pointing to a similar order of magnitude as found by Mainz/CLS and BMW, it is too early to derive a quantitative result.

We thus proceed to average the Mainz/CLS and BMW results.
The BMW result \cref{eq:charm_BMW} exhibits a mild $1.7\sigma$ tension\footnote{Recall that the statistically precise strange connected contribution agrees well between the two collaborations.} with the Mainz result \cref{eq:charm_Mz}. Therefore, after fitting the results of the two collaborations to a constant, we need to inflate the error. The scale factor obtained from $\chi^2/\text{dof}$ would lead to an error of $0.4\times10^{-11}$. We make a slightly more conservative choice for the latter and thereby obtain for the charm light-by-light contribution
\begin{equation}\label{eq:charm_av}
a_\mu^{\rm HLbL,c} = 3.5 (5) \times 10^{-11} \qquad \textrm{(our average)}\,.
\end{equation}

\subsubsection{Subleading disconnected diagrams, \texorpdfstring{$a_\mu^{\rm HLbL,rest}$}{}}
\label{subsec:HLBL_rest_contrib}

For those topologies (X+1) containing one or more quark loops consisting of a single vector insertion, $L^f_\alpha(x) = -{\rm Tr}\{\gamma_\alpha S_f(x,x)\}$, where $S_f(y,x)$ denotes the quark propagator of flavor $f$, it is helpful to include all three flavors, $\sum_{f=u,d,s} {\cal Q}_f L^f_\alpha(x) $, exploiting the fact that $\sum_{f=u,d,s} {\cal Q}_f =0$.
The latter property implies the vanishing of these diagrams at the SU(3)$_{F}$-symmetric point, $m_l=m_s$ and therefore a suppression at short distances.

\begin{figure}
\centerline{\includegraphics[width=0.21\textwidth]{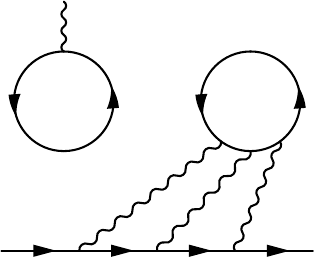}~~~~~~~~~~~~~~\includegraphics[width=0.24\textwidth]{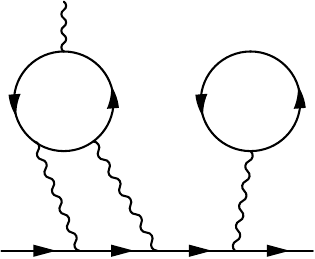}}
\caption{\label{fig:HLbLsublDiags} The two quark-contraction diagrams of the (3+1) topology class, up to permutations of the vertices on the muon propagator. Figures taken from Ref.~\cite{Blum:2023vlm}.}
\end{figure}

The topology (3+1), see \cref{fig:HLbLsublDiags}, consists of two quark loops and is therefore of the same order in $1/N_c$ counting as the (2+2) topology. Also, the pion loop contributes to this topology with a significant weight. Therefore, a direct lattice calculation is called for. The Mainz/CLS calculation could not resolve a signal and yielded $a_\mu^{\rm HLbL,3+1}=0.0(6)\times 10^{-11}$.
The more recent BMW calculation \cite{Fodor:2024jyn} did obtain a positive signal,
\begin{equation}\label{eq:amu_3p1BMW}
     a_\mu^{\rm HLbL,3+1} = 0.82(18)_{\rm stat}(17)_{\rm syst}\times 10^{-11} \qquad
     \textrm{(BMW)}\,.
\end{equation}
Since the Mainz/CLS result only amounts to an upper bound on the magnitude of this contribution,  we will adopt the result \cref{eq:amu_3p1BMW} for our average.

\begin{figure}
\centerline{\includegraphics[width=0.24\textwidth]{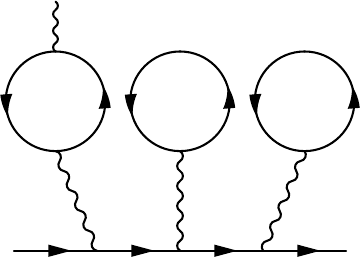}~~~~~~~~~~~~
\includegraphics[width=0.32\textwidth]{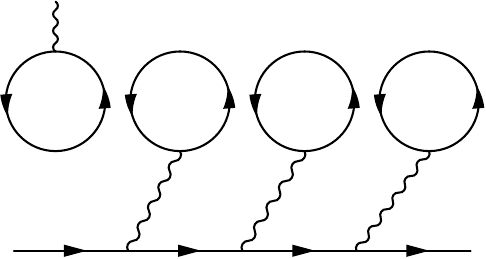}}
\caption{\label{fig:HLbLsubsublDiags} Representatives of the (2+1+1, left) and (1+1+1+1, right) class of diagrams. Figures taken from Ref.~\cite{Blum:2023vlm}.}
\end{figure}

Two collaborations~\cite{Chao:2022xzg,Fodor:2024jyn} have found the (2+1+1) and the (1+1+1+1) contributions (see \cref{fig:HLbLsubsublDiags}) to be entirely negligible, hence we will not go into any details. No statistically significant signal is obtained for these contributions. However, the lattice data allows one to bound them. The Mainz/CLS calculation arrives at
\begin{align}
     a_\mu^{\rm HLbL,(2+1+1)} &= 0.0(3)\times 10^{-11}\,,
\notag\\
a_\mu^{\rm HLbL,(1+1+1+1)} &= 0.0(1)\times 10^{-11}\,,
\end{align}
while the BMW collaboration quotes even tighter bounds, $0.04\times 10^{-11}$ and $0.0005\times 10^{-11}$, respectively. Therefore, we will neglect these contributions altogether.

\subsubsection{Final average for \texorpdfstring{$\amuHLbL$}{}}
\label{sec:amuHLbLlatticeaverage}

Adding the four contributions of \cref{eq:amuHLbL_light_aver,eq:amuHLBL_strange_aver,eq:charm_av,eq:amu_3p1BMW}, treating their uncertainties as being uncorrelated,
we arrive at our final average of lattice QCD results for $\amuHLbL$,
\begin{equation}\label{eq:amuHLbL_aver}
    \amuHLbL = 122.5(7.1)_\text{stat}(5.6)_\text{syst}\times10^{-11}=\amuHLbLlatticeresult\times 10^{-11}
    \qquad
    {\rm (lattice~QCD~average)}\,.
\end{equation}
Clearly, the uncertainty is entirely dominated by that of the light-quark contribution $a_\mu^{\rm HLbL,\ell}$. The entire remaining contribution is
\begin{equation}\label{eq:amuHLbL_s_c_rest_aver}
    a_\mu^{\rm HLbL,s} + a_\mu^{\rm HLbL,c} + a_\mu^{\rm HLbL,rest} = 2.92(98)\times10^{-11}\qquad
    {\rm (lattice~QCD~average)}\,.
\end{equation}
A summary of the status of the HLbL contribution is shown in \cref{fig:summary_plot_hlbl} in \cref{sec:conclusionsWP}.

\subsection{Exclusive state contributions \label{sec:HLbLexcl_state_LAT}}

In the dispersive framework, the dominant contribution to the HLbL diagram comes from the light pseudoscalar-poles according to \cref{Eq:amu-P-pole}.
The hadronic inputs are the TFFs $\FFP$ at spacelike virtualities with $P=\pi^0,\eta,\etap$.
The integrand in \cref{Eq:amu-P-pole} is peaked at low virtualities and the dominant contribution comes from the region $Q_{1,2}^2 < 3~\GeV^2$~\cite{Nyffeler:2016gnb} that can be reached by lattice calculations.

Regarding the determination of TFFs, two distinct approaches are currently employed. The first approach adopts the time--momentum representation, as proposed by Ref.~\cite{Ji:2001wha},
where the momenta for the pseudoscalar meson and photons are constrained to discrete values. This necessitates the use of parameterizations like the $z$-expansion for momentum interpolation.
This method was used in early calculations to evaluate the TFFs and the lifetime of the pion~\cite{Feng:2012ck} and has recently been adopted by three groups to determine the
pseudoscalar-pole contributions to HLbL, extending from the pion pole to the $\eta$- and $\eta^{\prime}$-poles~\cite{Gerardin:2016cqj,Gerardin:2019vio,Gerardin:2023naa,ExtendedTwistedMass:2022ofm,ExtendedTwistedMass:2023hin,Koponen:2023zle}.
The second approach, proposed very recently by the RBC/UKQCD collaboration~\cite{Lin:2024khg}, calculates TFFs with arbitrary photon momenta by introducing an appropriate coordinate-space weight function (coordinate-space representation)~\cite{Christ:2022rho,Meng:2021ecs}. To date, this method has only been applied to the pion-pole contribution, where the significant signal-to-noise problem in computing TFFs at large virtualities is mitigated by introducing a structure function. Below, we present the results obtained from both approaches.

\subsubsection{Time--momentum representation}

By the time of WP20, a single complete lattice calculation of the pion TFF, by the Mainz group~\cite{Gerardin:2019vio,Gerardin:2016cqj}, had been published. The TFF had been computed in the kinematic range relevant to the $(g-2)_{\mu}$ using $N_f=2+1$ Wilson-Clover quarks. Four lattice spacings in the range [0.050--0.086]~fm and pion masses down to $200\MeV$ were used to extrapolate the result to the physical point using a systematically improvable parameterization, inspired by the analysis of other hadronic form factors. This parameterization satisfies the short-distance constraints~\cite{Lepage:1979zb,Lepage:1980fj,Brodsky:1981rp,Nesterenko:1982dn,Novikov:1983jt} and can be further constrained by experimental inputs, such as the two-photon decay width.
In the first publication~\cite{Gerardin:2016cqj} the calculation was limited to the pion rest-frame. In Ref.~\cite{Gerardin:2019vio} it was realized that including data in a moving frame, where the pion carries one unit of momentum, allows a better coverage of the ($Q_1^2,Q_2^2$)-plane, especially when one of the virtualities tends to zero (single-virtual regime).
Using the parameterization of the form factor, the Mainz group was able to provide the first lattice calculation of the pion-pole contribution to the HLbL. The result is
\begin{equation}
a_{\mu}^{\pi^0\text{-pole}} = 59.7(3.4)(0.9)(0.5) \times 10^{-11} = 59.7(3.6) \times 10^{-11}\,,
\label{eq:ps-mainz}
\end{equation}
where the first error is statistical, the second one is the systematic error associated with the parameterization of the TFF, and the third one from the disconnected contribution. The disconnected contribution is $-1.0(5)\times 10^{-11}$.
Using, in addition, the experimental measurement of the two-photon decay width~\cite{PrimEx:2010fvg,PrimEx-II:2020jwd} as a constraint in the fit of the TFF, the authors quote $a_{\mu}^{\pi^0\text{-pole}} = 62.3(2.3) \times 10^{-11}$.

Since the publication of WP20, two additional groups have presented lattice calculations of the pion TFF in the time--momentum representation.
The ETM collaboration~\cite{ExtendedTwistedMass:2023hin} uses $N_f=2+1+1$ flavors of Wilson Clover twisted mass quarks.
The simulations are performed at the physical pion mass and at maximal twist~\cite{Frezzotti:2003ni,Frezzotti:2004wz}. This ensures automatic $\Order(a)$-improvement of the three-point function.
The latter is evaluated in the pion rest frame for three values of the lattice spacing in the range [0.057--0.080]~fm using lattices of spatial extents $L \simeq 5.5~$fm. The calculation includes both connected and disconnected contributions.
The form factor is extrapolated to the continuum limit using the parameterization introduced by the Mainz group~\cite{Gerardin:2019vio}.
The authors quote
\begin{equation}
a_{\mu}^{\pi^0\text{-pole}} = 56.7(3.1)(1.0) \times 10^{-11} = 56.7(3.2) \times 10^{-11}\,,
\label{eq:ps-etm}
\end{equation}
where the first error is statistical and the second is the systematic uncertainty.

The BMW collaboration~\cite{Gerardin:2023naa} uses $N_f = 2+1+1$ flavors of staggered quark fermions with four steps of stout smearing. The calculation is also performed at the physical pion mass and includes five values of the lattice spacing in the range [0.065--0.132]~fm. The spatial extents of the lattices are in the range [6.1--6.7]~fm.
In addition to the pion rest frame, the pion frame with one unit of momentum is also considered.
As for the other groups, both connected and disconnected contributions are included and the parameterization introduced in Ref.~\cite{Gerardin:2019vio} is used to extrapolate the TFF in the continuum limit. The result reads
\begin{equation}
a_{\mu}^{\pi^0\text{-pole}} = 57.8(1.8)(0.9) \times 10^{-11} = 57.8(2.0) \times 10^{-11}\,.
\label{eq:ps-bmw}
\end{equation}
where the disconnected contribution is $-1.33(19)\times 10^{-11}$.

Besides the pion, the $\eta$ and $\etap$-poles are the largest pseudoscalar-pole contributions.
Compared to the pion, the TFFs are much more challenging to compute on the lattice. The calculation requires additional quark-disconnected diagrams that involve a single pseudoscalar loop. This noisy disconnected contribution is large and of opposite sign as compared to the fully connected part. Thus, large statistics is needed to control this delicate cancellation. Because of the heavier pseudoscalar masses, the integrand in \cref{Eq:amu-P-pole} is also peaked at larger virtualities and the pole contribution is less sensitive to the normalization of the TFF~\cite{Nyffeler:2016gnb}.
The $\eta$ meson is the lowest-lying eigenstate with quantum numbers $I^G(J^{PC}) = 0^+(0^{-+})$ and the associated form factor can be extracted using the same strategy as the one used for the pion. In this case, the $\etap$ contribution is seen as an excited-state contribution that vanishes for sufficiently large time separations between the pseudoscalar insertion and the two vector currents. This is the strategy followed by the ETM collaboration in Ref.~\cite{ExtendedTwistedMass:2022ofm,Burri:2022gdg}. In principle, any pseudoscalar interpolating operator can be used as long as it overlaps with the physical $\eta$ meson.
For the $\etap$ meson, a proper treatment of the mixing between the flavor-octet and flavor-singlet states is required to isolate the eigenstate of the Hamiltonian.

\begin{figure}[t]
	\includegraphics*[width=0.47\linewidth]{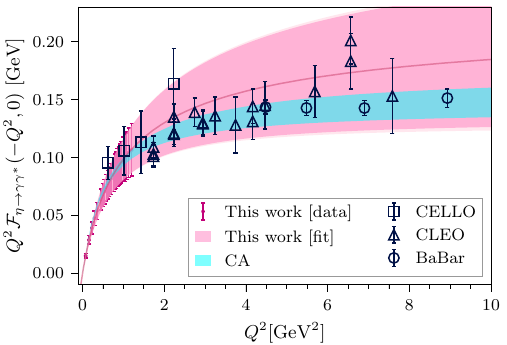}
	\includegraphics*[width=0.47\linewidth]{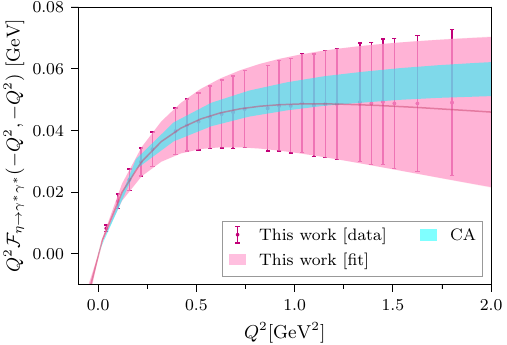}
	\caption{\label{fig:ETM-TFF}The $\eta$ TFF from the ETM collaboration in the single-virtual (left) and double-virtual (right) kinematics~\cite{ExtendedTwistedMass:2022ofm}. The result, obtained at a single lattice spacing, is compared with experimental results and the Canterbury estimate (cyan bands). Figures taken from Ref.~\cite{ExtendedTwistedMass:2022ofm}.}
\end{figure}

In Ref.~\cite{ExtendedTwistedMass:2022ofm}, the ETM collaboration has presented the first lattice calculation of the $\eta$ TFF, at the physical pion mass but at a single lattice spacing. The result for the double-virtual and single-virtual kinematics is shown in \cref{fig:ETM-TFF} and compared with experimental data and the CA estimate~\cite{Masjuan:2017tvw}. Using this TFF and the master formula \cref{Eq:amu-P-pole}, they quote
\begin{equation}
a_{\mu}^{\eta\text{-pole}} = 13.2(5.2)_{\rm stat}(1.3)_{\rm syst}[5.3]_{\rm tot} \times 10^{-11}\,,
\end{equation}
where the error is dominated by statistics. The main systematic error comes from varying the order of their $z$-expansions used to parameterize the virtuality dependence of the form factor. The diagrams that contain single vector loops are suppressed in the SU(3) flavor limit and have been neglected in this work.

\begin{figure}[t!]
	\includegraphics[width=0.49\linewidth]{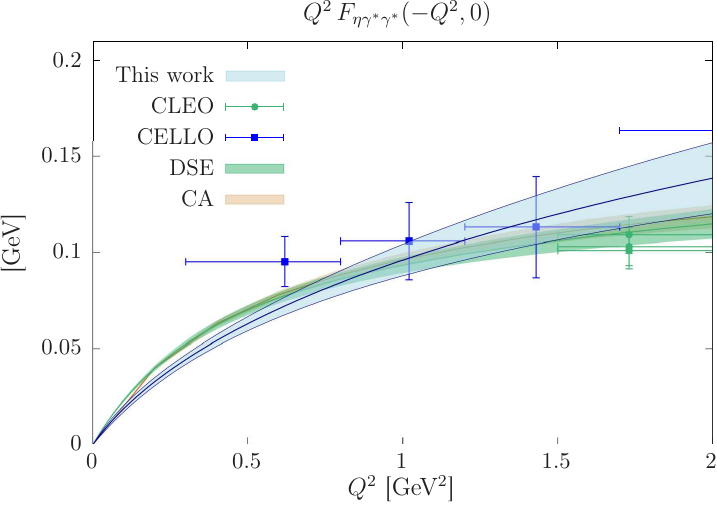}
	\includegraphics[width=0.49\linewidth]{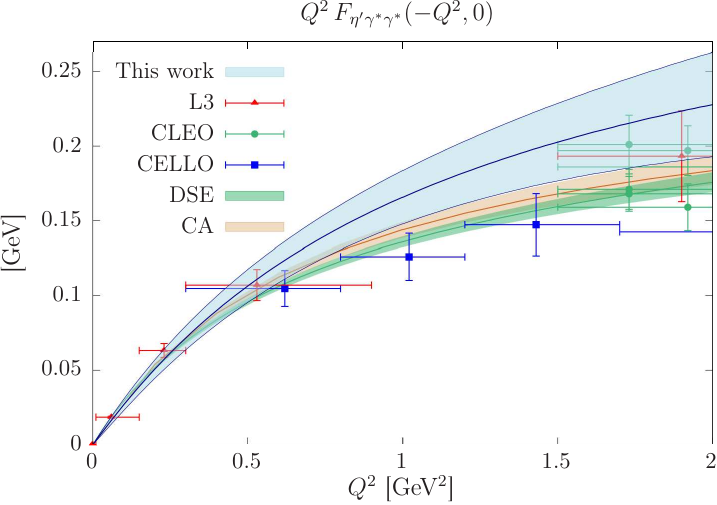}
	\caption{\label{fig:BMW-TFF}The $\eta$ (left) and $\eta^{\prime}$ (right) single-virtual TFFs from the BMW collaboration at the physical point and in the continuum limit. The Canterbury approximant (CA) result is extracted from Ref.~\cite{Masjuan:2017tvw} and the Dyson--Schwinger equation (DSE) result comes from~Ref.~\cite{Eichmann:2019tjk}. Measurements from CELLO~\cite{CELLO:1990klc}, CLEO~\cite{CLEO:1997fho}, and L3~\cite{L3:1997ocz} are shown for comparison. Figures taken from Ref.~\cite{Gerardin:2023naa}.}
\end{figure}

The BMW collaboration has presented a first calculation including both the $\eta$ and $\eta^{\prime}$ contributions. The results for the single-virtual regime are depicted in \cref{fig:BMW-TFF}.
All connected and disconnected contributions have been estimated. It has been shown that the disconnected diagrams that are suppressed in the SU(3) flavor limit are small and negligible at the current statistical precision. As for the pion, two pseudoscalar frames have been included for cross-checks.
This calculation includes a continuum extrapolation based on five lattice spacings.
The corresponding pseudoscalar-pole contributions read
\begin{align}
a_{\mu}^{\eta\text{-pole}} &= 11.6  (1.6)_{\rm stat} (0.5)_{\rm syst} (1.1)_{\rm FSE} \times 10^{-11}\,, \notag\\
a_{\mu}^{\eta^\prime\text{-pole}} &= 15.7 (3.9)_{\rm stat} (1.1)_{\rm syst} (1.3)_{\rm FSE} \times 10^{-11}\,.
\end{align}
The uncertainty is dominated by statistics, followed by the finite-volume effects.
The $\eta$ and $\etap$ contributions are comparable in size and their sum is about half of the pion-pole contribution.

The $\eta$-pole contribution is compatible within 1.4 standard deviations with the data-driven dispersive result presented in Ref.~\cite{Holz:2024diw}.
Good agreement is found for the $\eta^{\prime}$-pole contribution, although with larger uncertainties on the lattice side.

\subsubsection{Coordinate-space representation}

In the coordinate-space representation, the methodology is designed within the framework of infinite volume, where spatial momentum can take arbitrary values.
The TFFs can be extracted via the Fourier transform
\begin{equation}
\label{eq:F_munu}
\mathcal{F}_{\mu\nu}(Q,Q')= \int d^4x\, e^{-i\left(Q-\frac{Q_\pi}{2}\right)\cdot x}\mathcal{H}_{\mu\nu}(x)=i\varepsilon_{\mu\nu\alpha\beta}Q_{\alpha}Q_{\pi,\beta}F_{\pi^0\gamma^*\gamma^*}(-Q^2,-{Q'}^2)\,,
\end{equation}
with $Q'=Q_\pi-Q$ and $Q=(iE,\mathbf{q})$ parameterized in terms of the real-valued kinematic variables $E$ and $\mathbf{q}$. The  hadronic matrix element $\mathcal{H}_{\mu\nu}(x)\equiv \left\langle 0\left|T\left\{j_{\mu}\left(\frac{x}{2}\right)j_{\nu}\left(-\frac{x}{2}\right)\right\}\right|\pi(Q_\pi)\right\rangle$ is defined in Euclidean spacetime with $j_\mu(x)$ representing the EM current and $Q_\pi=(iE_\pi,\mathbf{q}_\pi)$ denoting the four-momentum of the on-shell pion. This matrix element can be decomposed as $\mathcal{H}_{\mu\nu}(x) = -\varepsilon_{\mu\nu\alpha\beta}x_{\alpha}Q_{\pi,\beta} H(x^2,Q_\pi\cdot x)$. Consequently, the TTFs can be determined through an integral~\cite{Lin:2024khg}
\begin{equation}
F_{\pi^0\gamma^*\gamma^*}(-Q^2,-Q'^2) = \int d^4x \,\omega(K,Q_\pi,x) H(x^2,Q_\pi\cdot x)\,,
\end{equation}
where the weight function $\omega(K,Q_\pi,x)$ is defined as
\begin{equation}
\omega(K,Q_\pi,x) \equiv i\,e^{-iK\cdot x} \frac{(K\cdot x) Q_\pi^2 - (K\cdot Q_\pi)(Q_\pi\cdot x)}{K^2Q_\pi^2-(K\cdot Q_\pi)^2}\,,
\end{equation}
with $K=Q-Q_\pi/2$. Unfortunately, the factor $e^{-iK\cdot x}=e^{(E-E_\pi/2)t}e^{-i(\mathbf{q}-\mathbf{q}_\pi/2)\cdot\mathbf{x}}$ grows rapidly when $E$ becomes large, leading to
a severe signal-to-noise problem, which is also common in the time--momentum representation. The challenge is addressed by introducing a pion structure function $\phi_{\pi}(x^2, u)$ and expressing $H(x^2, Q_\pi \cdot x)$ as
\begin{equation}
\label{eq:structure_func}
H(x^2, Q_\pi\cdot x) = \int_{0}^{1}du\, e^{i\left(u-\frac{1}{2}\right)Q_\pi\cdot x}\phi_{\pi}(x^2,u)H(x^2,0)\,.
\end{equation}
In this way, the TFFs can be obtained through
\begin{equation}
F_{\pi^0\gamma^*\gamma^*}(-Q^2,-Q'^2) = \int_0^1 du\int d^4x \,\omega(\bar{K},Q_\pi,x)\phi_{\pi}(x^2,u)H(x^2,0)\,,
\end{equation}
with $\bar{K}=Q-uQ_\pi$. Since $\phi_{\pi}(x^2,u)H(x^2, 0)$ is symmetric under SO(4) spacetime rotations, an SO(4) average can be performed for the weight function $\omega(\bar{K}, Q_\pi, x)$, yielding $\langle \omega(\bar{K},P,x)\rangle_\text{SO(4)}=2\frac{J_2(|\bar{K}||x|)}{\bar{K}^2}$, with $J_2(x)$ being the Bessel function.
As $J_2(|\bar{K}||x|)$ does not
exhibit exponential growth at large $|\bar{K}||x|$, there is no signal-to-noise problem in computing the TFFs. The integral can be further improved by replacing $H(x^2,0)$ with
$H(x^2,Q_\pi\cdot x)$ based on the relation \cref{eq:structure_func}.

\begin{figure}[t!]
\centering
\includegraphics[width=0.485\textwidth]{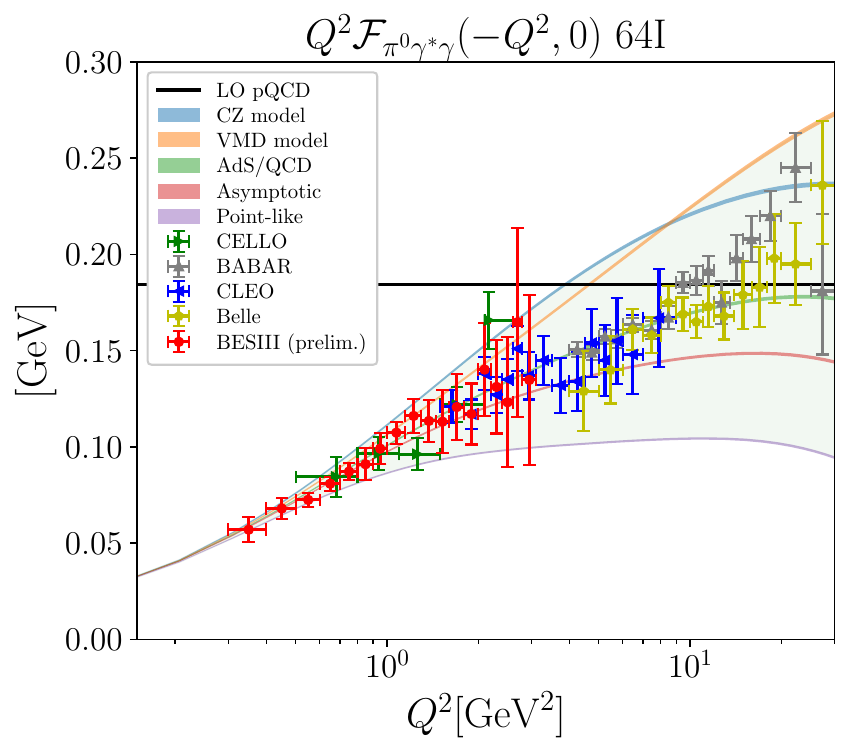}\quad
\includegraphics[width=0.485\textwidth]{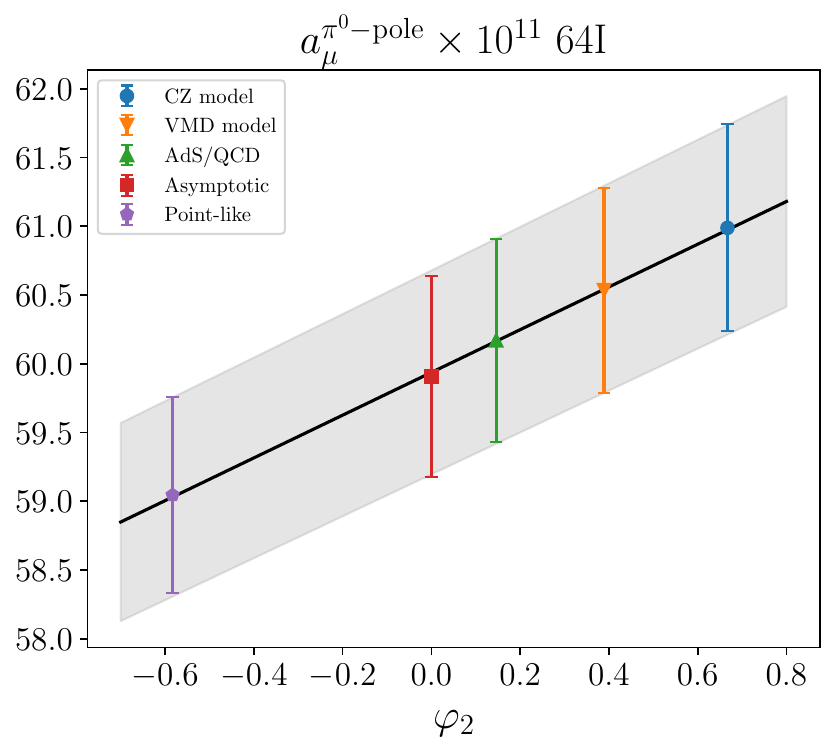}
\caption{Left: Comparison between experimental measurements of TFFs with a single virtual photon and lattice QCD calculations using different parameterizations of $\phi_\pi(x^2,u)$ as inputs. Right: $a_{\mu}^{\pi^0\text{-pole}}$ as a function of $\varphi_{2}(|x|)$. The results align closely with a linear form, indicating that higher-order effects are negligible. Figures taken from Ref.~\cite{Lin:2024khg}.}
\label{fig:Fq0_ex}
\end{figure}

The methodology involves a Gegenbauer expansion to express the pion structure function $\phi_{\pi}(x^2,u)$. The size of different Gegenbauer terms is evaluated by
incorporating lattice data, and the LO Gegenbauer term is demonstrated to be model independent.
Five parameterizations for $\phi_{\pi}(x^2,u)$, which span a wide range of large-momentum behaviors for the TFFs and cover all experimental data
as shown in \cref{fig:Fq0_ex}, are used to estimate the systematic effects induced by $\phi_{\pi}(x^2,u)$. These parameterizations yield the coefficients
for different Gegenbauer terms. The results of $a_{\mu}^{\pi^0\text{-pole}}$ align closely with a linear form of the NLO coefficient $\varphi_2(|x|)$
as shown in \cref{fig:Fq0_ex}, indicating that the higher-order effects are negligible. More details are presented in Ref.~\cite{Lin:2024khg}.

\begin{figure}[t!]
\centerline{\includegraphics[width=0.79\linewidth]{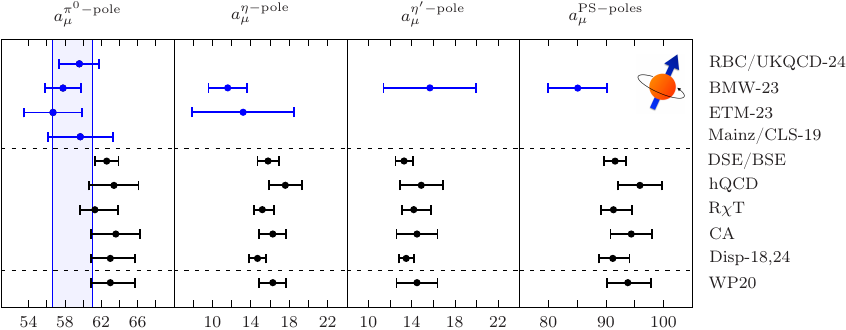}}
\caption{Summary of the lattice calculations (in blue) and of the approaches presented in \cref{sec:dataHLbL}, and comparison with the WP20 average. The blue band represents the lattice average for the pion-pole contribution.}
\label{fig:ps-pole-summary}
\end{figure}

Using eight gauge ensembles generated with 2+1 flavor domain wall fermions, incorporating multiple pion masses, lattice spacings, and volumes,
the RBC/UKQCD calculation provides the result\footnote{While this paper was under review, the final version of
Ref.~\cite{Lin:2024khg} appeared, $a_{\mu}^{\pi^0\text{-pole}} =
61.2(1.1)_{\mathrm{stat}}(1.0)_{\phi}(0.7)_{a}(0.4)_{\mathrm{FV}}\times
10^{-11}=61.2(1.7)\times10^{-11}$, with changes
to~\cref{eq:RBC-pion-pole} due to increased statistics. The average
in~\cref{eq:amuHLbLpi0lat} is based on \cref{eq:RBC-pion-pole}.}
\begin{equation}
\label{eq:RBC-pion-pole}
a_{\mu}^{\pi^0\text{-pole}} = 59.6(1.6)_{\mathrm{stat}}(1.0)_{\phi}(1.0)_{a}(0.4)_{\mathrm{FV}}\times 10^{-11}=59.6(2.2)\times10^{-11}\,.
\end{equation}
In this result, the first error is statistical, the second reflects the dependence on the structure function, the third represents lattice artifacts,
and the fourth accounts for residual finite-volume effects.
It is noteworthy that the finite-volume effects are estimated by assuming that the hadronic function $\mathcal{H}_{\mu\nu}(x)$
at long distances is dominated by the $\rho$ meson and $\pi\rho$ states. These finite-volume effects can reach $0.8\times10^{-11}$
for a lattice size of $L\simeq4.6$ fm and are quickly suppressed to $0.2\times10^{-11}$ at $L\simeq 5.5$ fm.
In \cref{eq:RBC-pion-pole}, the uncertainty $(0.4)_{\mathrm{FV}}\times10^{-11}$ indicates potential residual effects after the finite-volume correction is applied.
When using small lattice volumes, the finite-volume effects are not negligible, making corrections essential, particularly as lattice precision improves in future studies.
In the total contribution to $a_{\mu}^{\pi^0\text{-pole}}$, the connected diagrams account for $57.8(2.2)\times 10^{-11}$, while the disconnected pieces contribute $1.81(15)\times10^{-11}$, based on the results from the finest ensemble.

\subsubsection{Conclusion}

\begin{figure}[t!]
\centerline{\includegraphics[width=0.495\linewidth]{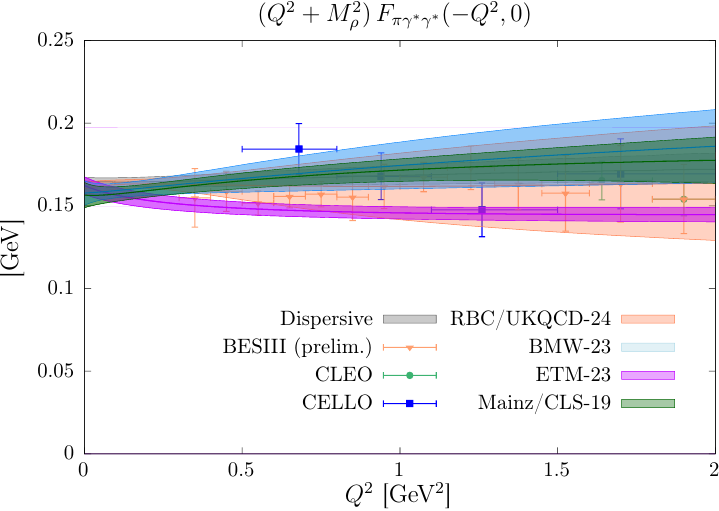}
\includegraphics[width=0.495\linewidth]{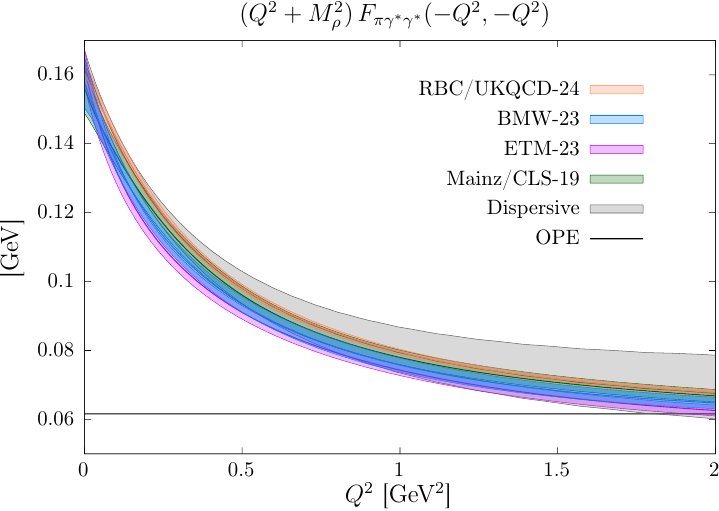}}
\centerline{\includegraphics[width=0.495\linewidth]{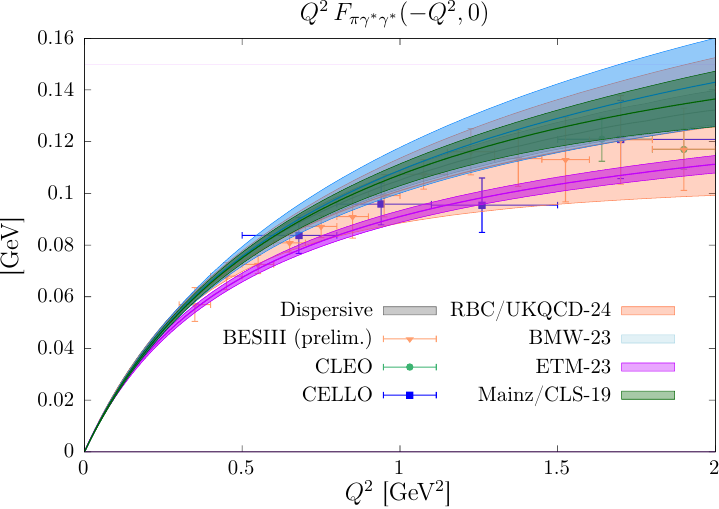}
\includegraphics[width=0.495\linewidth]{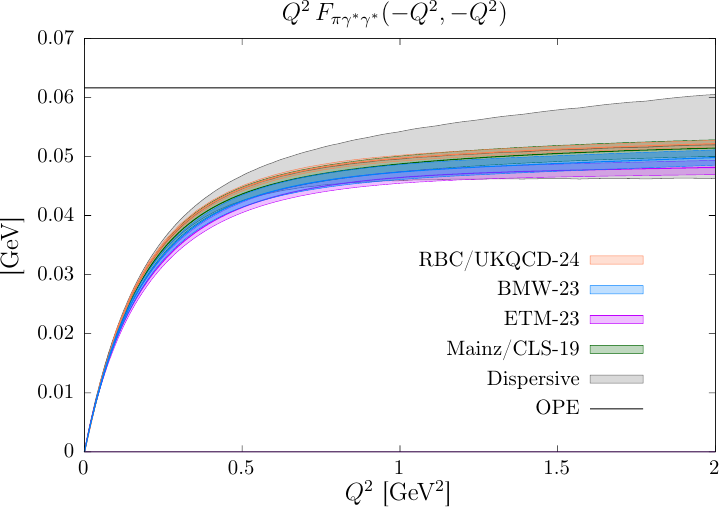}}
\caption{
Comparison of lattice calculations \cite{Gerardin:2019vio,ExtendedTwistedMass:2023hin,Gerardin:2023naa,Lin:2024khg}, the dispersive determination of Ref.~\cite{Hoferichter:2018dmo}, and experimental data  (Refs.~\cite{CELLO:1990klc,CLEO:1997fho} and preliminary BESIII data~\cite{Redmer:2018uew,Aoyama:2020ynm}) for the single-virtual and the double-virtual $\pi^0$ TFF.}
\label{fig:ps-TFF-compa}
\end{figure}

A summary of the lattice calculations of pseudoscalar-pole contributions is given in \cref{fig:ps-pole-summary}.
Four groups, using different lattice regularizations, have presented results on the pion-pole contribution, with precisions between 3.5 and 6\%. They agree within one standard deviation. However, the sign of the small disconnected contribution $a_{\mu}^{\pi^0\text{-pole},\mathrm{disc}}$ is found to be different among groups and needs further investigations.
At present, Mainz, BMW, and ETM find the disconnected contribution to be negative, and
the more recent RBC/UKQCD calculation finds it to be positive.
Also, Mainz, BMW, and RBC/UKQCD provided the results for
the disconnected contribution $a_{\mu}^{\pi^0\text{-pole},\mathrm{disc}}$, in which case, we can subtract it from the total and obtain
the connected-only contribution $a_{\mu}^{\pi^0\text{-pole},\mathrm{conn}}$.
The ETM group has only obtained the total contribution, which should equal to $a_{\mu}^{\pi^0\text{-pole},\mathrm{conn}} - |a_{\mu}^{\pi^0\text{-pole},\mathrm{disc}}|$ in their calculation. We can perform a global fit and obtain both the averaged connected-only contribution and the magnitude of the disconnected contribution.
We find good agreements in the connected contribution with $\chi^2/\text{dof}= 0.24$.
For the disconnected diagram, in addition to the sign disagreement, there is also about $2\sigma$ tension in the magnitude between the BMW and RBC/UKQCD results.
Therefore, we use the connected-only contribution as the central value and
the disconnected contribution is treated as an additional uncertainty.
The final lattice QCD average is\footnote{Using the final result of Ref.~\cite{Lin:2024khg}, this
average would change to $a_{\mu}^{\pi^0\text{-pole}} =
59.4(0.9)_\text{stat}(1.1)_\text{syst}(1.5)_{\rm disc}[2.1]_\text{tot}
\times 10^{-11}$.}  
\begin{equation}
a_{\mu}^{\pi^0\text{-pole}} = a_{\mu}^{\pi^0\text{-pole},\mathrm{conn}} \pm \big|a_{\mu}^{\pi^0\text{-pole},\mathrm{disc}}\big| = 58.8(1.1)_\text{stat}(1.1)_\text{syst}(1.6)_{\rm disc}[2.2]_\text{tot} \times 10^{-11}
\qquad {\rm (lattice~QCD~average)}
\,.
\label{eq:amuHLbLpi0lat}
\end{equation}
The lattice average is shown in \cref{fig:ps-pole-summary}, where it is compared to the data-driven approaches presented in \cref{sec:dataHLbL}. In particular, the value \cref{eq:amuHLbLpi0lat} is compatible with the data-driven result of Ref.~\cite{Hoferichter:2018dmo} within $1.3\sigma$. A more detailed comparison between the latter and lattice calculations, at the level of the TFF, is made in \cref{fig:ps-TFF-compa}. Both at single- and at double-virtual kinematics, one observes broad agreement between the different determinations. We note that the number of parameters used in parameterizing the TFF has a significant impact on the uncertainty bands.

Since WP20, first lattice calculations for the $\eta$ and $\eta^{\prime}$ have appeared. They suffer from the noisy disconnected contribution that largely cancels the connected contribution.
The ETM collaboration has published results for the $\eta$-pole contribution at the physical point using a single lattice spacing and the BMW collaboration has presented results for both the $\eta$- and $\eta^{\prime}$-pole contributions at the physical point and in the continuum limit.
Finally, it is worth noting that the light pseudoscalar TFFs can be valuable inputs in reducing the statistical and systematic errors in the direct HLbL calculation.

\begin{table}[t]
\small
\centering
\renewcommand{\arraystretch}{1.1}
\begin{tabular}{ccc}
\toprule
Contribution & Average $[10^{-11}]$ & Sources \\\midrule
$a_\mu^{\rm HLbL,\ell}$ & $119.6(7.1)_\text{stat}(5.5)_\text{syst}[9.0]_\text{tot}$ &
\cref{eq:amuHLbL_light_aver}, \cref{tab:amuHLbL_light} \\
$a_\mu^{\rm HLbL,s}$ & $-1.4(8)$ & \cref{eq:amuHLBL_strange_aver,eq:amuHLBL_strange_list} \\
$a_\mu^{\rm HLbL,c}$ & $3.5(5)$ & \cref{eq:charm_av,eq:charm_Mz,eq:charm_BMW} \\
$a_\mu^{\rm HLbL,3+1}$ & $0.82(25)$ & \cref{eq:amu_3p1BMW} \\
$\amuHLbL$ & $122.5(7.1)_\text{stat}(5.6)_\text{syst}[9.0]_\text{tot}$ & \cref{eq:amuHLbL_aver} \\
\bottomrule
\end{tabular}
\renewcommand{\arraystretch}{1}
\caption{Current lattice results for the direct calculation of $\amuHLbL$, where statistical and systematic errors have been added in quadrature.}
\label{tab:hlbl_lattice}
\end{table}

\subsection{Summary and future prospects}

\subsubsection{Summary of current knowledge from the lattice}

Great strides have been made in the lattice computation of the HLbL contribution to $(g-2)_\mu$ compared to WP20.
In WP20, the only determination was that of the RBC/UKQCD collaboration using finite-volume $\text{QED}_L$, but now an infinite-volume QED result has also been determined and the two are consistent.
Chiral-continuum determinations of all contributing diagrams using lattice QCD and infinite-volume QED have now been measured in the direct calculation~\cite{Chao:2021tvp}. It is shown that only the fully-connected and $(2+2)$-disconnected light quark contributions are significant to the overall value, and there are three independent lattice calculations of these contributions: those of the RBC/UKQCD collaboration~\cite{Blum:2019ugy,Blum:2023vlm}, the Mainz group~\cite{Chao:2021tvp}, and the BMW collaboration~\cite{Fodor:2024jyn}.

\Cref{tab:hlbl_lattice} summarizes various contributions to $\amuHLbL$ based on the averages of different lattice studies currently available.

\subsubsection{Final estimate and outlook}

Our estimate for the lattice determination of the HLbL contribution reads as follows:
\begin{equation}
a_\mu^{\text{HLbL}} = 122.5(7.1)_\text{stat}(5.6)_\text{syst}[9.0]_\text{tot} \times 10^{-11}
\qquad {\rm (lattice~QCD~average)}\,.
\end{equation}
The individual contributions of the determinations are listed in \cref{tab:hlbl_lattice}.
It has been shown that the current techniques and methods are sufficient for the expected precision on this quantity that are required. It is now evident that only two contributions are really relevant, the light-quark ($u$, $d$) fully-connected, and $(2+2)$ disconnected contributions, thus facilitating cheaper future studies of $\amuHLbL$. Future improvements in the lattice determination are possible with higher statistics and more gauge ensembles with different lattice spacings and volumes.

In addition, lattice QCD calculations have obtained the $\pi^0$-pole contribution in HLbL. The lattice average is
\begin{equation}
a_{\mu}^{\pi^0\text{-pole}} = 58.8(1.1)_\text{stat}(1.1)_\text{syst}(1.6)_{\rm disc}[2.2]_\text{tot} \times 10^{-11}
\qquad {\rm (lattice~QCD~average)}\,.
\end{equation}

\FloatBarrier

\clearpage

\section{The QED contributions to \texorpdfstring{$\boldsymbol{a_\mu}$}{}}
\label{sec:QED}

\noindent
\emph{T.~Aoyama, M.~Hayakawa, M.~Nio, S.~Volkov}
\vspace{\baselineskip}

No significant changes have arisen in the QED contribution to $a_\mu$ since WP20~\cite{Aoyama:2020ynm}.\footnote{Slight changes are due to the updated mass-independent tenth-order coefficient $A_1^{(10)}$ as detailed in \cref{subsec:QED10th} as well as the updated sixth-order coefficient $A_2^{(6)}(m_\mu/m_\tau)=0.000\,360\,601(83)$. The latter change does not affect any digits reported in this section.}
However, several remarkable new results were obtained in related studies.  They  are summarized in this section, and
their influences on the theoretical value of $a_\mu$ are discussed.

\subsection{The fine-structure constant \texorpdfstring{$\alpha$}{} from atom-interferometer experiments}

The EM coupling constant known as the fine-structure constant $\alpha$  is a fundamental and essential input for predicting the theoretical value of the lepton $g-2$.
The latest and most accurate values of $\alpha$ are provided in two methods.
One is the atom-mass measurement using an atom interferometer, and the other is the measurement of the electron's anomalous magnetic moment $a_e$
together with its theoretical prediction.

The atom-interferometer experiment determines the quotient of the Planck constant $h$ and the mass of an atom X $m_\text{X}$.
Since May 2019, the Planck constant $h$ is a defined constant, so that it becomes a direct measurement of the mass of a neutral atom.
 The cesium ($^{133}$Cs) atom mass \cite{Parker:2018vye} and the rubidium ($^{87}$Rb) atom mass  \cite{Morel:2020dww} are available for
 the determination of $\alpha$:
\begin{align}
h/m_{\text{Cs}} &=3.002\,369\,4721(12) \times 10^{-9} \,\text{m}^2\text{s}^{-1} \hspace{-4cm}&& [0.40 \ppb] \, ,
\notag
\\
h/m_{\text{Rb}}  &=4.591\,359\,258\,90(65) \times 10^{-9} \, \text{m}^2\text{s}^{-1} \hspace{-4cm}&& [0.14\ppb] \, .
\label{eq:h/m}
\end{align}
Both experiments are ongoing, aiming for an order-of-magnitude improvement.
Another atom-interferometer experiment with strontium (Sr) or ytterbium (Yb) has started also,
aiming at a relative precision $0.01\ppb$~\cite{Schelfhout:2024bvz}.
The  mass $m_\text{X}$ is converted to $\alpha$ with the help of more precisely determined physical constants:
\begin{equation}
\alpha (\text{X}) = \left [ \frac{2 R_\infty}{c} \frac{ A_r(\text{X})}{A_r(e)}\frac{h}{m_\text{X}} \right ]^{1/2}\, ,
\end{equation}
where
$R_\infty $ is the Rydberg constant, $c$ is the speed of light in vacuum,
and $A_r(\text{X})$ and $A_r(e)$ are the relative atomic mass of the atom X and that of the electron, respectively.

The Rydberg constant $R_\infty$  has been determined by combining many measurements on various transition frequencies of the hydrogen, deuterium, and muonic atoms from 1979 to 2022
and the theoretical QED predictions of these spectra.  The latest CODATA2022 recommended value of $R_\infty$  is \cite{Mohr:2024kco}
\begin{equation}
R_\infty = 10\, 973\, 731.568\, 157(12)\,\text{m}^{-1}\quad [1.1\ppt]\,.
\label{eq:R_infty}
\end{equation}
The relative atomic masses of Cs and Rb atoms were also updated in 2020~\cite{Huang:2021nwk, Wang:2021xhn}.
The relative atomic masses of the ions of these nuclei were measured and then converted to the neutral atom masses
 by adding the appropriate number of electron masses and subtracting the ionization energies.
 The latest values are
\begin{align}
A_r(^{133}\text{Cs}) & = 132.905\,451\,9585(86)  \hspace{-4cm}&& [0.065 \ppb] \, ,
\notag
 \\
A_r(^{87}\text{Rb}) &   =   86.909\,180\,5291(65)  \hspace{-4cm}&& [0.075 \ppb] \, .
\label{eq:A_r}
\end{align}
The relative atomic mass of the electron  $A_r(e)$
has been determined by measuring cyclotron and spin-precession frequencies of hydrogen-like ions of carbon $(^{12}\text{C}^{5+})$ and silicon $(^{28}\text{Si}^{13+})$ using a Penning trap, in conjunction with the QED predictions for the $g$ factor of a Coulomb-bound electron.
The CODATA2022  recommended value of $A_r(e)$ is \cite{Mohr:2024kco}
\begin{align}
A_r(e)=5.485\,799\,090\,441 (97) \times 10^{-4} \quad [0.018 \ppb]\, .
\label{eq:A_r(e)}
\end{align}
The values of the fine-structure constant $\alpha$ from the atom-interferometer experiments are then updated by using
\cref{eq:h/m,eq:R_infty,eq:A_r,eq:A_r(e)},
and the defined value $c=299\,792\,458\,\text{m}/\text{s}$
as
\begin{align}
\alpha^{-1}(\text{Cs}) &= 137.035\,999\,045(27) \hspace{-4cm}&& [0.20\ppb]\, ,     \label{eq:alpha_Cs} \\
\alpha^{-1}(\text{Rb}) &=  137.035\,999\, 2052(97)\hspace{-4cm}&&  [0.071\ppb]\, .      \label{eq:alpha_Rb}
\end{align}
The last digit of $\alpha^{-1}(\text{Cs})$ in Ref.~\cite{Parker:2018vye} is changed from 6 to 5, but the uncertainty is unchanged.
The updated value of
$\alpha^{-1}(\text{Rb})$ is  smaller than that in Ref.~\cite{Morel:2020dww} by $0.8\times10^{-9}$ and the uncertainty is
reduced from $11\times 10^{-9}$ to $9.7 \times 10^{-9}$.
The difference between \cref{eq:alpha_Cs,eq:alpha_Rb} is
\begin{align}
\alpha^{-1}(\text{Cs}) - \alpha^{-1}(\text{Rb}) = - 0.160 (29) \times 10^{-6} \, ,
\end{align}
corresponding to a significance of $5.5\sigma$.

\subsection{Measurement of the electron anomalous magnetic moment \texorpdfstring{$a_e$}{}}

The electron anomalous magnetic moment $a_e = (g_e-2)/2$ has been measured using a single electron trapped in EM fields known as a Penning trap.
The longstanding $a_e$ measurement at Harvard University (HV) in 2008 \cite{Hanneke:2008tm} was superseded in 2022 by a new measurement from the same team, now based at Northwestern University (NW) \cite{Fan:2022eto}.
The new measurement
\begin{align}
a_e^\text{exp}(\text{NW}22) =1\,159\,652\,180.59(13) \times 10^{-12} \quad [0.11 \ppb]
\label{eq:ae_NW}
\end{align}
demonstrates a 2.2-fold improvement in precision, with a shift $-0.14\times 10^{-12}$ from $a_e^\text{exp}(\text{HV}08)$, which remains within the assigned uncertainties.
As correlations between the two similar measurements are not obvious,
the authors of Ref.~\cite{Fan:2022eto} do not recommend averaging the two determinations.
Stabilizing the magnetic field of a Penning trap typically takes weeks, but the NW team developed a faster method.
This breakthrough enabled measurements at multiple magnetic field strengths for the first time.
The consistency of results across different magnetic field strengths significantly enhanced the reliability of the measurement.

\begin{figure}[t]
\centering
\begin{minipage}[b]{0.475\columnwidth}
\centering
    \includegraphics[width=0.98\columnwidth]{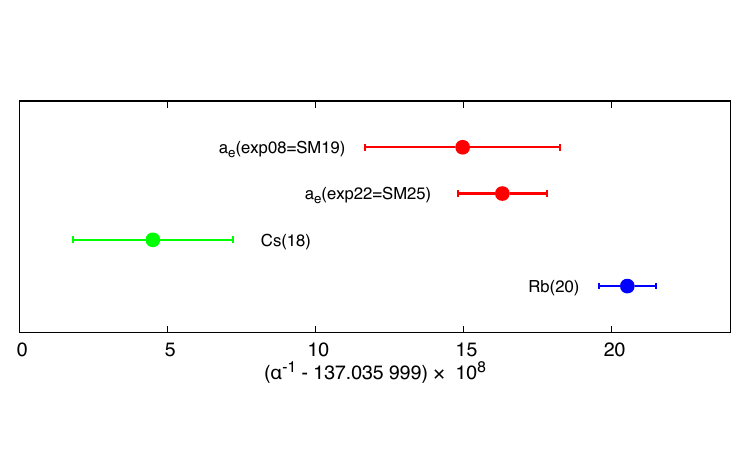}
 \end{minipage}
\hspace{3mm}
\begin{minipage}[b]{0.475\columnwidth}
\centering
    \includegraphics[width=0.98\columnwidth]{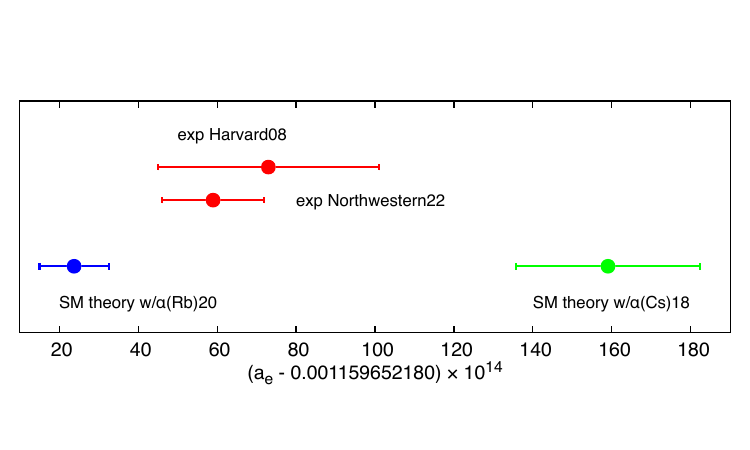}
\end{minipage}
\caption{The best three values of the fine-structure constant $\alpha$  (left) and a comparison of the measurement and SM prediction of $a_e$ (right).
The values of $\alpha$ are determined from the measurement and theory of $a_e$ and the atom-mass measurements of Cs and Rb with atom interferometers.
The two SM predictions for $a_e$  are based on two different values of $\alpha$.
These two figures present the three experimental results and one theoretical prediction from different perspectives.}
\label{fig:three_alphas}
\end{figure}

A value of the fine-structure constant $\alpha$ from $a_e$ can be obtained by treating $\alpha$ as an unknown
variable in the theoretical formula for $a_e$ and equating it to the measured value in \cref{eq:ae_NW}.
The derived value  is
\begin{align}
\alpha^{-1}(a_e)  =  137.035\,999\,163 (15)\quad [0.11\ppb]\,,
\label{eq:alpha_ae}
\end{align}
which is smaller than the value reported in Ref.~\cite{Fan:2022eto} by $ 3 \times 10^{-9}$. This shift, within the assigned uncertainties, is attributed to two factors. The first is a 13 \% reduction of the tenth-order QED term of $a_e$, as discussed in \cref{subsec:QED10th}.
The second is an updated evaluation of the HVP contribution, detailed in \cref{subsec:a_etheory}, which
affects the result in the opposite direction.
As a result, $\alpha^{-1}(a_e)$ remains nearly unchanged.
The total uncertainty is entirely driven by the measurement in \cref{eq:ae_NW},
while the uncertainties from the QED tenth-order term and the hadronic contributions are
$0.44 \times 10^{-9}$ and $0.36\times 10^{-8}$, respectively.

The differences between the values of $\alpha$ determined by the atom-interferometer experiments and that obtained from $a_e$
are
\begin{align}
\alpha^{-1}(\text{Cs}) - \alpha^{-1} (a_e) &=   -0.118 (31) \times 10^{-6} \,,\notag\\
\alpha^{-1}(\text{Rb}) - \alpha^{-1} (a_e) &=   +0.042 (18) \times 10^{-6}  \, ,
\end{align}
corresponding to $-3.8\sigma$ and $+2.3\sigma$, respectively.
The relationships among the three values of $\alpha$ in  \cref{eq:alpha_Cs,eq:alpha_Rb,eq:alpha_ae} are visualized in \cref{fig:three_alphas}.

\subsection{The QED tenth-order mass-independent and universal term}
\label{subsec:QED10th}

The QED contribution involves only four particles: the photon, electron, muon, and $\tau$ lepton.
Thus, the QED contribution to $a_\mu^\text{QED}$  can be divided according to the lepton-mass dependence as
\begin{equation}
a_\mu^\text{QED} = A_1 + A_2(m_\mu/m_e) + A_2(m_\mu/m_\tau) + A_3(m_\mu/m_e, m_\mu/m_\tau)\,,
\end{equation}
and each can be calculated by perturbation theory using Feynman diagrams as
\begin{equation}
   A_i = \sum_{n=1,2,\cdots}  \left(\frac{\alpha}{\pi}\right )^{n} A_i^{(2n)}\, .
\end{equation}
To match the current precision of  $a_\mu$, the QED contribution up to the tenth-order
of perturbation theory are required. The second-, fourth-, sixth-, and eighth-order terms are well established
and were reported in the QED section of
WP20~\cite{Aoyama:2020ynm}.

All Feynman diagrams contributing to $a_\mu$ of the tenth-order terms $A_1^{(10)}$, the two $A_2^{(10)}$, and $A_3^{(10)}$, as well as the  term $A_2^{(10)}(m_e/m_\mu)$ contributing to $a_e$ were numerically calculated by a single team AHKN \cite{Aoyama:2012wj,Aoyama:2012wk}.
The $\tau$-lepton contribution to $a_e$ of the tenth order is negligible.
The mass-independent and universal term $A_1^{(10)}$, which contributes equally to both $a_e$ and $a_\mu$,
receives contributions from 12,672 vertex Feynman diagrams.  Among these,  6,318 diagrams
that include at least one fermion loop also contribute to $A_2^{(10)}$.
Due to the complexity and enormity of the calculation, only relatively simple diagrams were independently cross-checked until recently~\cite{Laporta:1994md,Baikov:2013ula}.

In Ref.~\cite{Volkov:2019phy} and successively in Ref.~\cite{Volkov:2024yzc}, the contribution to $A_1^{(10)}$ from Set~V, which consists of 6,354 vertex diagrams without a fermion loop, was
obtained using numerical integration. The two results, both calculated by Volkov using different intermediate renormalization schemes, are consistent with each other.
The best estimate is then obtained by statistically combining both results~\cite{Volkov:2024yzc}:
\begin{equation}
A_1^{(10)}[\text{Set~V}:\text{Volkov}] = 6.828(60)\,.
\label{eq:A_1^(10)Volkov}
\end{equation}
This exhibited a $5\sigma$ discrepancy from the latest report $7.668(159)$ given in Ref.~\cite{Aoyama:2019ryr}
by AHKN, which was used to determine the QED contribution to  $a_\mu$ in WP20.

To resolve the discrepancy, AHKN performed a diagram-by-diagram numerical comparison with Volkov's results~\cite{Aoyama:2024aly}.
The 6,354 vertex diagrams were consolidated into 389 groups using the Ward--Takahashi identity.
Because of different choices of intermediate renormalization schemes, the integrated magnetic moment amplitudes of the same group
in Volkov's and AHKN's calculations show different values.
However,  the lower-order magnetic moment amplitudes and finite renormalization constants can account for this difference.
No apparent inconsistencies were found between Volkov's and AHKN's results for any individual group. However, the sum of the results from 98 groups---each containing a single second-order self-energy subdiagram and no other self-energy subdiagrams---was found to be responsible for the $5\sigma$ discrepancy.
The 98 integrals representing these 98 groups were then reevaluated by Monte-Carlo integration with substantially increased statistics.  Although the old and new results of individual integrals are consistent with each other, the sum of the 98 integrals becomes smaller than before. As a consequence, AHKN's tenth-order contribution has been updated to
\begin{equation}
A_1^{(10)}[ \text{Set~V}:\text{AHKN}] = 6.800(128)\,,
\label{eq:A_1^(10)AHKN}
\end{equation}
which is in good agreement  with \cref{eq:A_1^(10)Volkov}.

The contributions from the remaining 6,318 diagrams of the tenth order, which involve at least one fermion loop, were also calculated
and reported in Ref.~\cite{Volkov:2024yzc}:
\begin{equation}
A_1^{(10)}[\text{with fermion loops}: \text{Volkov}] = -0.9377(35)\, .
\label{eq:with_fermion_loop_Volkov}
\end{equation}
The corresponding AHKN value was first reported in Ref.~\cite{Aoyama:2012wj} and updated in Ref.~\cite{Aoyama:2017uqe}:
\begin{equation}
A_1^{(10)}[\text{with fermion loops}: \text{AHKN}] = -0.9304(36)\,,
\label{eq:with_fermion_loop_AHKN}
\end{equation}
which is larger than \cref{eq:with_fermion_loop_Volkov}  by $0.0073(50)$, indicating good agreement between the two.
The mass-independent and universal $A_1^{(10)}$ terms of Volkov and AHKN then become
\begin{align}
A_1^{(10)} [ \text{all}: \text{Volkov}] &= 5.891 (61)\, ,\notag\\
A_1^{(10)} [\text{all}: \text{AHKN}] & = 5.870 (128)\,.
\end{align}
Since they are in good agreement and independent, the weighted average of the two determinations gives the best estimate
\begin{align}
A_1^{(10)} = 5.887 (55)\, ,
\label{eq:A_1^(10)_2025}
\end{align}
which is used to derive the fine-structure constant $\alpha(a_e)$ in \cref{eq:alpha_ae}.

\subsection{Theory of the electron anomalous magnetic moment \texorpdfstring{$a_e$}{} }
\label{subsec:a_etheory}

To compare the measurement of $a_e$ to the theoretical prediction,  the SM contributions
to $a_e$ besides the QED contributions must be considered.
The HVP  contribution to the muon $g-2$, $a_\mu^\text{HVP, LO}$, has been under discussion in light of recent lattice-QCD calculations and new measurements of the $\pi^+\pi^-$ cross section from CMD-3. The details are described in \cref{sec:dataHVP,sec:latticeHVP}.
In connection with these new evaluations of HVP, the HVP contribution to $a_e$ was recalculated in Ref.~\cite{DiLuzio:2024sps}, replacing the $\pi^+\pi^-$ channel of KNT19~\cite{Keshavarzi:2019abf} by the corresponding CMD-3 data.
Additionally, lattice-QCD predictions of  $a_e^\text{HVP, LO}$ are available~\cite{Budapest-Marseille-Wuppertal:2017okr,Giusti:2019hkz,Giusti:2020efo}, though their precision is not yet comparable to data-driven calculations, whose values are, e.g.,
\begin{equation}
a_e^\text{HVP, LO} = \left \{ \begin{array}{lll}
                                           1.8608(66) \times 10^{-12}    &  \text{KNT19}  &\text{\cite{Keshavarzi:2019abf}} \, ,  \\
                                           1.920(9)\times 10^{-12}    &  \text{KNT19/CMD-3} &\text{\cite{DiLuzio:2024sps}}\, .\\
                                            \end{array}  \right .
\end{equation}
The difference between KNT19 and KNT19/CMD-3 corresponds to a $6.2\sigma$ discrepancy.

The magnitude of this difference  $0.059 \times 10^{-12}$  is still much smaller than the difference of $1.35\times 10^{-12}$ arising from the discrepancy between the two theoretical predictions of $a_e$ caused by the two values of $\alpha$, see \cref{eq:alpha_Cs,eq:alpha_Rb,eq:ae_SM}.
To take a conservative approach, we adopt the simple mean of KNT19 and KNT19/CMD-3 and assign  half of their difference as the uncertainty, in analogy to \cref{sec:NLO}:
\begin{align}
a_e^\text{HVP, LO} = 1.89(3) \times 10^{-12}\,,
\label{eq:aeHVPLO}
\end{align}
which should cover all realistic HVP evaluations for $a_e$ at the moment and 
is also consistent with the lattice-QCD predictions.
Other hadronic and EW contributions are summarized as follows:
\begin{align}
\begin{array}{llr}
a_e^\text{HVP, NLO}   &=                -0.2263(35) \times 10^{-12}     & \text{\cite{Keshavarzi:2019abf,DiLuzio:2024sps} } \,,\\
a_e^\text{HVP, NNLO} &=       0.027\,99(17) \times 10^{-12}  &\text{\cite{Jegerlehner:2017zsb} }\,,\\
a_e^\text{HLbL}            &=        0.0351(23) \times 10^{-12}           &\text{\cite{Hoferichter:2025fea} } \,, \\
a_e^\text{EW} &=        0.030\,53  (23) \times 10^{-12}   & \text{\cite{Jegerlehner:2017zsb} }  \,.
\end{array}
\label{eq:aeHVPNLOandweak}
\end{align}
Equations~\eqref{eq:aeHVPLO} and \eqref{eq:aeHVPNLOandweak} are used to derive the fine-structure constant $\alpha(a_e)$ in \cref{eq:alpha_ae}
as well as the SM prediction for $a_e$.
With QED perturbation theory up to the tenth order and  the hadronic and EW contributions, the SM predictions
$a_e^\text{SM}$ are determined using the fine-structure constant $\alpha$ of \cref{eq:alpha_Cs} or \cref{eq:alpha_Rb} as
\begin{align}
a_e^\text{SM}\big[\alpha(\text{Cs})\big]&=1\,159\,652\,181.59 (23) (0)(3)\times 10^{-12}\hspace{-3cm}&& [0.20 \ppb]\,,\notag
\\
a_e^\text{SM}\big[\alpha(\text{Rb})\big]&= 1\,159\,652\,180.238 (82)(4)(30)\times 10^{-12} \hspace{-3cm}&&  [0.075\ppb]\,,
\label{eq:ae_SM}
\end{align}
where the uncertainties, listed from left to right, correspond to the fine-structure constant $\alpha$, the tenth-order QED term $A_1^{(10)}$, and the hadronic contributions.
While the uncertainties in $\alpha$ still dominate, a more reliable HVP contribution remains essential for probing BSM physics through $a_e$.
The differences from the measurement \cref{eq:ae_NW} are
\begin{align}
a_e^\text{exp}(\text{NW}22) -a_e^\text{SM}\big[\alpha(\text{Cs}) \big]& =  -1.00 (26) \times10^{-12}   \, ,  \notag\\
a_e^\text{exp}(\text{NW}22)-a_e^\text{SM}\big[\alpha(\text{Rb}) \big]  & =  +0.35 (16) \times 10^{-12}   \, ,
\end{align}
corresponding to $-3.8\sigma$ and $+2.2\sigma$, respectively.
The comparison between the measurement and the SM predictions is shown in \cref{fig:three_alphas}.

\subsection{The QED contribution to the muon anomalous magnetic moment \texorpdfstring{$a_\mu$}{}}
\label{sec:amuQEDfinal}

The lepton-mass ratios necessary for $a_\mu^\text{SM}$ are unchanged since WP20 and they are~\cite{Mohr:2024kco}
\begin{align}
m_e/m_\mu &=4.836\,331\,70 (11) \times 10^{-3}\,,  \notag\\
m_e/m_\tau &=2.875\,85(19) \times 10^{-4}\, .
\end{align}
As discussed earlier, the three values of $\alpha$ differ from each other by $2.4\sigma$ to $5.6\sigma$, making it inappropriate to average them.
Accordingly, we present the three values of the QED contribution to $a_\mu^\text{SM}$ using three values of $\alpha$ in \cref{eq:alpha_Cs,eq:alpha_ae,eq:alpha_Rb}:
\begin{align}
\label{eq:amuQED}
a_\mu^\text{QED}\big[\alpha(\text{Cs})\big] &=    116\,584\,718.926 (23) (7)(17)(6)(100) [104] \times 10^{-11} \,,   \notag\\
a_\mu^\text{QED}\big[\alpha(a_e) \big]        &=    116\, 584\,718.825(13) (7) (17)(6)(100) [103]\times 10^{-11}\,,    \notag\\
a_\mu^\text{QED}\big[\alpha(\text{Rb}) \big]&=     116\, 584\,718.789(8) (7) (17)(6)(100) [102]\times 10^{-11}   \,,
\end{align}
where the uncertainties from left to right arise from  $\alpha$, the $\tau$-lepton mass, the QED eighth-order term, the QED tenth-order term,
the estimated QED twelfth-order term, and the total combined uncertainties~\cite{Aoyama:2020ynm}.
The difference between the largest and the smallest values of $a_\mu^\text{QED}$ is $0.137 \times 10^{-11}$,
which is of the same order of magnitude as the estimated QED twelfth-order term.
In view of \cref{eq:amuQED}, we use
\begin{equation}
\label{amuQEDfinal}
a_\mu^\text{QED} =    \amuQEDresult\times 10^{-11} \,,
\end{equation}
in the calculation of the complete SM prediction of $a_\mu$.

\FloatBarrier

\clearpage

\section{The electroweak contributions to \texorpdfstring{$\boldsymbol{a_\mu}$}{}}
\label{sec:EW}

\noindent
\emph{D.~St\"ockinger, H.~St\"ockinger-Kim}
\vspace{\baselineskip}

\begin{figure}[t]
\centerline{  \includegraphics[scale=.45]{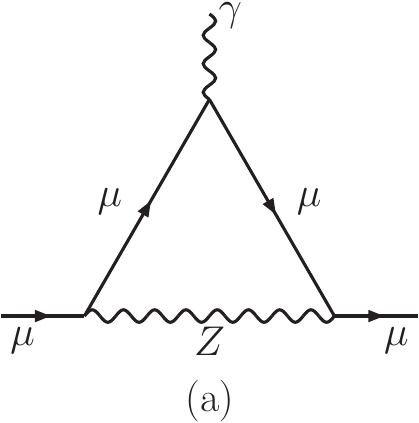}
  \includegraphics[scale=.45]{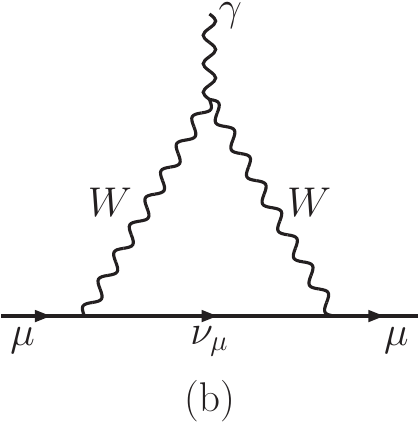}
  \includegraphics[scale=.45]{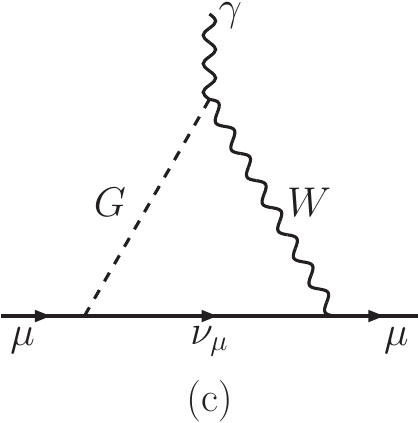}
  \includegraphics[scale=.45]{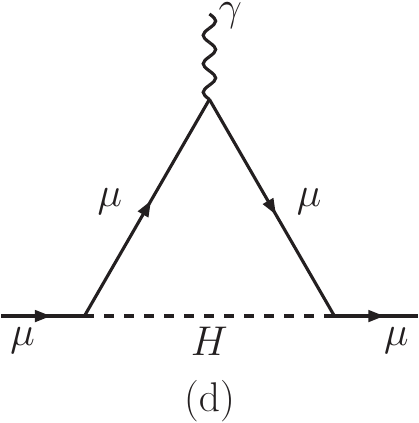}}
  \caption{One-loop Feynman diagrams contributing to $\amuEW$. Figures taken from Ref.~\cite{Aoyama:2020ynm}.}
  \label{fig:SMEW1}
\end{figure}

\begin{figure}[t]
\centerline{  \includegraphics[scale=.45]{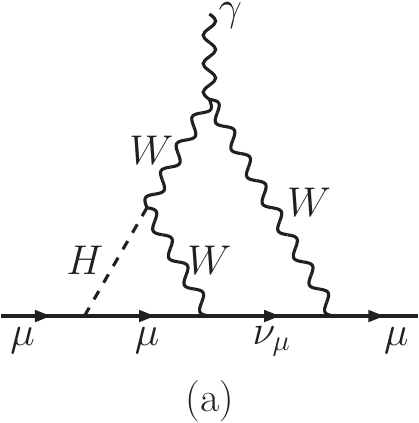}
\quad  \includegraphics[scale=.45]{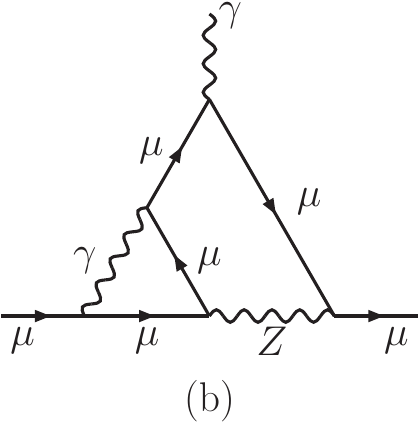}
\quad  \includegraphics[scale=.45]{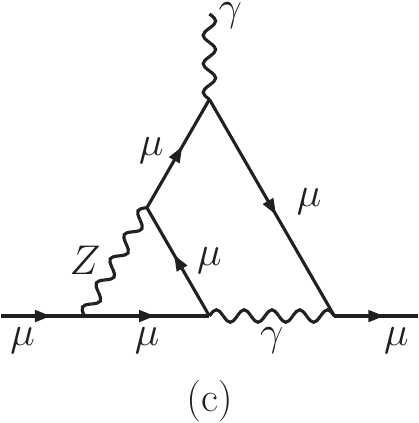}}
  \caption{Sample bosonic two-loop Feynman diagrams contributing to $\amuEW$. Figures taken from Ref.~\cite{Aoyama:2020ynm}.}
  \label{fig:SMEW2b}
\end{figure}

\begin{figure}[t]
\centerline{  \includegraphics[scale=.45]{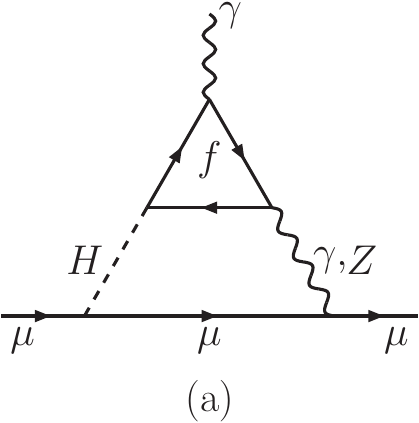}
\quad  \includegraphics[scale=.45]{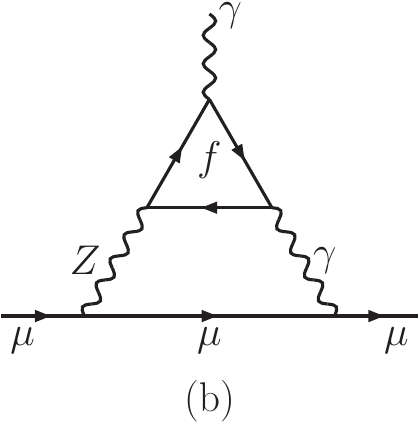}
\quad  \includegraphics[scale=.45]{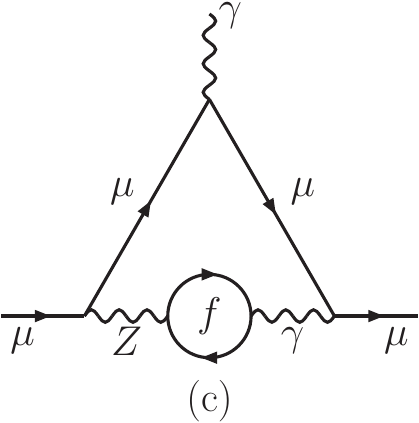}}
  \caption{Sample fermionic two-loop Feynman diagrams contributing to $\amuEW$. Figures taken from Ref.~\cite{Aoyama:2020ynm}.}
  \label{fig:SMEW2f}
\end{figure}

By definition, the EW SM contributions to $a_\mu$ comprise all SM contributions that are not
contained in the pure QED, HVP, or HLbL contributions.
They are given by Feynman diagrams
that contain at least one of the EW bosons $W$, $Z$, or the Higgs.
Figures~\ref{fig:SMEW1}--\ref{fig:SMEW2f} show sample
one-loop and two-loop diagrams.
Already in WP20~\cite{Aoyama:2020ynm}, the EW SM
contributions $\amuEW$ had a very small theoretical uncertainty of
$10^{-11}$, which had a negligible impact on the total SM theory uncertainty.
Here we summarize the current status of the EW SM contributions and
describe relevant recent updates. The updates are related to more accurate
measurements of input parameters entering Feynman diagrams, leading
to reduced parametric uncertainties, and to improved determinations of
hadronic EW contributions.
For further details, we refer to
Ref.~\cite{Aoyama:2020ynm}, the original papers described below, and
to the review Ref.~\cite{Jegerlehner:2009ry}.


The overall magnitude of the EW contributions can be obtained from the
essential factors entering the EW one-loop contribution from a
$W$-boson loop,
\begin{align}
  \label{eq:amuew1approx}
  \amuEWl \propto
  \frac{g_2^2}{16\pi^2  }
  \frac{m _\mu ^2}{M _W ^2} \sim 10^{-9}\,.
\end{align}
Here $g_2^2/16\pi^2$ corresponds to a loop factor involving the
weak SU(2) gauge coupling, and
$m_\mu ^2 / M _W ^2 \simeq 10 ^{-6}$ is a suppression factor arising
from the heavy $W$-boson. The appearance of two powers of
the ratio $m_\mu/M_W$ can be related to the need for a muon chirality
flip and the need for spontaneous electroweak symmetry breaking to
generate a nonvanishing muon mass and dipole moment, see
e.g., Ref.~\cite{Stockinger:2022ata}.

For a precise and systematic evaluation of the EW contributions, the
decomposition
\begin{align}
\label{EWsplitup}\amuEW &=
a_\mu^{\rm EW(1)} + \amub + \amuf + a_\mu^{\rm EW(\ge3)}
\end{align}
is useful.
Here $a_\mu^{\rm EW(1)}$ are the one-loop contributions, $\amub$ and
$\amuf$ are two-loop contributions, and $a_\mu^{\rm EW(\ge3)}$ are
three-loop corrections and higher.

The EW one-loop contributions arise from diagrams with $W$-,
$Z$-, or Higgs-boson exchange as shown in \cref{fig:SMEW1}. Depending on the
chosen gauge, diagrams with unphysical Goldstone bosons must also be
included. The full one-loop result can be written as
\begin{align}
  \label{eq:amuew1}
  \amuEWl &=
  \frac{G_{F}}{\sqrt{2}}
  \frac{m _\mu ^{2}}{8 \pi ^2}
  \Big[
    \frac{5}{3}
    +
    \frac{1}{3}(1-4 s_{\text{W}} ^2)^2
    \Big]=194.79(1)\times 10^{-11}\,,
\end{align}
where the SU(2) gauge coupling and the factor $1/M_W^2$ appearing in
\cref{eq:amuew1approx} have been replaced with the Fermi constant
$G_{F}$, which is more precisely measured. The on-shell weak mixing
angle  $s_{{W}}^2=1-M_W^2/M_Z^2$ is defined via the $W$- and
$Z$-boson pole masses. In the evaluation of $\amuEW$, the
$W$-boson mass is treated as an intermediate quantity that is itself
predicted within the SM as a function of $G_{F}$, $M_Z$, and further
inputs such as the Higgs-boson and top-quark masses. We use here the
current best-fit value $M_W=80.356(5)\GeV$~\cite{ParticleDataGroup:2024cfk}, 
which differs slightly from the value used in Ref.~\cite{Aoyama:2020ynm}. Though the
$W$-boson mass is the result of a calculation and thus unaffected by
discrepancies between different direct $M_W$ measurements~\cite{ParticleDataGroup:2024cfk}, it has an uncertainty
due to missing higher-order corrections and due to the parametric
dependence in particular on the top-quark mass. The numerical
impact of the modification on the number quoted above is below
$10^{-13}$. The quoted uncertainty covers the parametric uncertainty
from SM input parameters and neglected contributions that are
suppressed by additional powers of $m_\mu^2G_F$.

The bosonic two-loop contributions $\amub$ are defined by two-loop and
associated counterterm diagrams without closed fermion loop, as
in \cref{fig:SMEW2b}. These contributions have been computed in
Ref.~\cite{Czarnecki:1995sz} in the limit $M_H\gg M_W$, and with full
Higgs-boson mass dependence in Ref.~\cite{Heinemeyer:2004yq}. A
seminumerical result has been obtained in  Ref.~\cite{Gribouk:2005ee} and a fully numerical computation has been
carried out in
Ref.~\cite{Ishikawa:2018rlv}. Here we use the analytical result of
Ref.~\cite{Heinemeyer:2004yq},
updated in Ref.~\cite{Gnendiger:2013pva}, with  the PDG value  $M_H=125.20(11)\GeV$~\cite{ParticleDataGroup:2024cfk} and obtain
\begin{equation}
\label{Higgsbosonicres}
\amub=-19.962(3)\times 10^{-11}\,.
\end{equation}
The renormalization scheme used here reflects the chosen parameterization
of the one-loop result in terms of the Fermi constant $G_F$.
Thanks to the increased experimental accuracy of the SM input
parameters it is possible to provide here an additional digit
compared to Ref.~\cite{Aoyama:2020ynm}. The quoted uncertainties are
the parametric uncertainties obtained from varying the Higgs-boson
and $W$-boson masses by $1\sigma$ around their central values.

The fermionic two-loop contributions can be further subdivided into
\begin{equation}
  \label{EWfermsplit}
\amuf =
\amufrestH
+a_\mu^{\rm EW(2)}(e,u,d;\mu,c,s;\tau,t,b)
 + \amufrestnoH \, .
\end{equation}
The first term on the RHS of \cref{EWfermsplit}
denotes the Higgs-dependent fermion-loop diagrams
like in \cref{fig:SMEW2f}a.
The second term
denotes contributions from the
diagrams in  \cref{fig:SMEW2f}b with a
$\gamma^*$--$\gamma$--$Z^*$-subdiagram and fermions of the 1st, 2nd,
and 3rd generation in the inner loop. For these diagrams, quarks and
leptons must be combined because of gauge anomaly cancellation.
The third term collects all remaining
fermionic two-loop contributions, e.g., from the diagram in
\cref{fig:SMEW2f}c.

First we
focus on the Higgs-dependent fermion-loop corrections. They are given
by the so-called Barr--Zee diagrams of \cref{fig:SMEW2f}a with
Higgs--$\gamma$--$\gamma$ or Higgs--$\gamma$--$Z$ subdiagram. Various
limits of the result have been computed in
Ref.~\cite{Czarnecki:1995wq}; the exact result can be found, e.g., in
Ref.~\cite{Gnendiger:2013pva}. These Higgs-dependent fermion-loop
corrections are proportional to the Higgs--fermion Yukawa couplings and are
therefore largest for the top quark. For the evaluation we use the
quark masses of Ref.~\cite{ParticleDataGroup:2024cfk}, in particular
$m_t=172.57(29)\GeV$.  The full result for this class of
contributions is then
\begin{equation}
\amufrestH=-1.500(2)\times10^{-11}\,.
\label{Higgsfermionicres}
\end{equation}
Compared to Ref.~\cite{Aoyama:2020ynm} the result has a slightly
reduced magnitude due to the smaller top-quark mass.
The indicated
uncertainty is only the parametric uncertainty arising essentially from the
uncertainty of the input parameters $m_t$ and $M_H$. The theory
uncertainty from missing higher-order corrections to these
contributions could be estimated, e.g., by changing the renormalization
scheme of the quark mass definitions and would be significantly
larger. Since  it will be covered
by the general estimate of missing three-loop corrections below, it is not
included in
\cref{Higgsfermionicres}.

After the precise Higgs-boson mass measurement at the LHC, all the EW
contributions discussed so far have a tiny uncertainty. The
uncertainty of the remaining fermionic EW contributions appearing in
\cref{EWfermsplit} is larger because potentially large QCD
corrections affect the two-loop diagrams with quark loops. It has
become standard to generalize the perturbative loop counting and
define the quantities
$a_\mu^{\rm EW(2)}(e,u,d;\mu,c,s;\tau,t,b)
 + \amufrestnoH
$
more generally as the EW contributions given by the respective
two-loop diagrams plus appropriate higher-order, potentially
nonperturbative, QCD corrections. Early results of such an improved
treatment have been obtained in
Refs.~\cite{Peris:1995bb,Knecht:2002hr,Czarnecki:2002nt}, and the
results described and  quoted in Ref.~\cite{Aoyama:2020ynm} were the ones of
Ref.~\cite{Czarnecki:2002nt}.

More recently, Ref.~\cite{Hoferichter:2025yih} further improved the
evaluation of these hadronic EW contributions.
The largest of these remaining contributions are the ones involving
fermion loops inserted into a  $\gamma$--$\gamma$--$Z$
subdiagram, shown in \cref{fig:SMEW2f}b. Because
of the Furry theorem, here only the axial $Z$-boson coupling can
contribute. Therefore, the subdiagram constitutes a $VVA$ correlator
and gauge anomaly cancellation is important.
For 3rd-generation fermions, these diagrams can be evaluated perturbatively, and the
two-loop calculation and its uncertainty due to missing higher orders
have been discussed in Refs.~\cite{Czarnecki:2002nt,Gnendiger:2013pva}.
Reference~\cite{Hoferichter:2025yih} incorporated three-loop QCD corrections
of Ref.~\cite{Melnikov:2006qb} and obtained
\begin{equation}
\label{EWVVA3}
a_\mu^{\rm EW(2)}(\tau,t,b)=-8.12(1)\times10^{-11}\,.
\end{equation}
Compared to the pure two-loop result, the central value has shifted
slightly within the previous uncertainty interval, but the uncertainty
has been reduced by a factor 10.

For the first two generations of quarks it is appropriate to define
the $VVA$ Green functions  $\langle0|T
j^\mu(x)j^\nu(y)j_5^\rho(z)|0\rangle$, which can be expressed in terms
of two
scalar functions $w_{L,T}(Q^2)$ that only depend on the $Z^*$
momentum scale $Q^2$. The contribution to $a_\mu$ can then be obtained
from integrals over $w_{L,T}$.
References~\cite{Peris:1995bb,Knecht:2002hr,Czarnecki:2002nt} have investigated
constraints on these functions from nonrenormalization theorems and
from operator product expansions. For the actual calculation of $a_\mu$,
Ref.~\cite{Czarnecki:2002nt} used a model ansatz for the four
light-quark functions
$w_{L,T}^{[u,d],[s]}$ that is compatible with all
constraints, while  charm-quark and lepton loops are added
perturbatively. Reference~\cite{Hoferichter:2025yih} similarly treats the
light quarks nonperturbatively but uses a dispersive calculation
\cite{Ludtke:2024ase} for the $VVA$ Green function with $u,d$-quark loops.
The combined result for the 1st-generation quarks and leptons then reads
\begin{equation}
\label{EWVVA1}
a_\mu^{\rm EW(2)}(e,u,d)=-2.08(3)\times10^{-11}\,.
\end{equation}
Compared to Ref.~\cite{Czarnecki:2002nt} the uncertainty is reduced by
almost an order of magnitude.\footnote{This result is consistent with the Regge model evaluation discussed in \cref{sec:Regge}, which gives $a_\mu^{\rm EW(2)}(e,u,d)=-1.98(9)\times 10^{-11}$.}

For the 2nd-generation contribution to the $VVA$ Green function both
Ref.~\cite{Czarnecki:2002nt} and Ref.~\cite{Hoferichter:2025yih} use an
ansatz for the strange-quark contribution similar to the
1st-generation approach, combined with perturbation theory for the
charm-quark and the muon loop. Reference~\cite{Hoferichter:2025yih}
includes the perturbative three-loop correction of
Ref.~\cite{Melnikov:2006qb} to the charm-quark contribution.
In total, Ref.~\cite{Hoferichter:2025yih} finds the 2nd-generation
result
\begin{equation}
\label{EWVVA2}
a_\mu^{\rm EW(2)}(\mu,c,s)=-4.14(28)\times10^{-11}\,.
\end{equation}
Here the uncertainty is not significantly reduced compared to the
result of Ref.~\cite{Czarnecki:2002nt} quoted in
Ref.~\cite{Aoyama:2020ynm}. However,  particularly the 
pQCD three-loop corrections have shifted the  central value  by more
than the previously estimated uncertainty.

The non-Higgs-dependent contributions $\amufrestnoH$ have first been
computed in Ref.~\cite{Czarnecki:1995wq} neglecting the combination
$1-4s_{W}^2\to0$ appearing in the vectorial muon--$Z$
coupling. These leading contributions contain the enhancement factor
$m_t^2/M_W^2$. Subleading terms suppressed by $1-4s_{W}^2$  have been added in
Ref.~\cite{Czarnecki:2002nt}. They arise in part via corrections to
the $\rho$-parameter that are again enhanced by $m_t^2$, and in part
from the diagrams
with $\gamma$--$Z$ interaction shown in
\cref{fig:SMEW2f}c, which are logarithmically enhanced.
Based on Ref.~\cite{Czarnecki:2002nt}, Ref.~\cite{Gnendiger:2013pva} provides the
full analytic result. The required hadronic part of the $\gamma$--$Z$
interaction, $8\pi^2\bar{\Pi}^{\gamma Z}(-M_Z^2)$, is again subject to
nonperturbative QCD corrections. For the required result of this
quantity, Ref.~\cite{Aoyama:2020ynm} used the
value  $6.88(50)$ based on Ref.~\cite{Czarnecki:2002nt} and references
therein. More recent work shows that an improved data-based
evaluation yields a significantly lower value, e.g., Ref.~\cite{Jegerlehner:2017gek} quotes $5.87(4)$, and including lattice-QCD results~\cite{Ce:2022eix} to quantify SU(3)-breaking corrections Ref.~\cite{Hoferichter:2025yih} obtains $6.0(1)$.  Based on this last number,
the full result can
hence be written as
\begin{equation}
\amufrestnoH=\big[-4.09-0.23-0.26(10)\big]\times10^{-11} = -4.58(10)\times10^{-11}\,.
\label{fermrestnoHiggs}
\end{equation}
Here the three individual results correspond to the terms without
$1-4s_{W}^2$ suppression, the terms related to the
$\rho$-parameter, and the terms involving the  $\gamma$--$Z$ mixing
subdiagram, respectively. Compared to Ref.~\cite{Aoyama:2020ynm}, the
first result has changed because of the lower top-quark mass and the
last term has changed as a result of the improvement in
Ref.~\cite{Hoferichter:2025yih}.
The change is smaller than the quoted uncertainty, which
is unchanged and corresponds to an estimate of still neglected terms
which are suppressed by a factor $1-4s_{W}^2$ or $M_Z^2/m_t^2$.

Finally, $a_\mu^{\rm EW(\ge3)}$ collects   corrections beyond the
two-loop level which are not yet taken into account via the QCD
corrections within $a_\mu^{\rm EW(2)}(e,u,d;\mu,c,s;\tau,t,b)
 + \amufrestnoH
 $. These correspond to weak and QED corrections with and without fermion loops to $\amub$ and weak, QED, and QCD corrections to $\amufrestH$.
The leading-logarithmic corrections of the form $\simeq
G_F\alpha^2\log(M_Z/m_f)\log(M_Z/m_{f'})$, where $m_{f,f'}$ are light fermions, have been evaluated in
Refs.~\cite{Degrassi:1998es,Czarnecki:2002nt} using EFT and
RG methods. A surprising numerical cancellation
among the three-loop corrections
was observed in Ref.~\cite{Czarnecki:2002nt} if the
two-loop contributions are parameterized in terms of $G_F\,\alpha$ as done in
all the above results. In this
case the three-loop
logarithms are numerically negligible.  Hence,
\begin{equation}
a_\mu^{\rm EW(\ge3)}=0.00(20)\times10^{-11}\,.
\label{EWthreeloop}
\end{equation}
The uncertainty estimate from remaining unknown higher-order
contributions is from Ref.~\cite{Czarnecki:2002nt}. It is based on an analysis of  subleading weak logarithmic three-loop corrections, but is expected to also cover nonlogarithmic three-loop QCD contributions to $\amufrestH$.
 
In total, the full EW SM contribution to $a_\mu$ is obtained by summing
the one-loop contributions
\cref{eq:amuew1}, the bosonic two-loop contributions \cref{Higgsbosonicres},
the fermionic two-loop contributions
\cref{Higgsfermionicres,EWVVA3,EWVVA1,EWVVA2,fermrestnoHiggs}, and the leading three-loop
logarithms \cref{EWthreeloop}. The result is
\begin{equation}
\amuEW= \amuEWresult \times10^{-11}\,,
\label{amuEWNew}
\end{equation}
where we follow Ref.~\cite{Hoferichter:2025yih} and add the
individual uncertainties in quadrature.
After significant reduction of the hadronic EW uncertainties, the
remaining uncertainty is dominated about equally by 2nd-generation hadronic
contributions of the kind of \cref{fig:SMEW2f}b and by remaining
unknown higher-order contributions beyond the two-loop level.

\FloatBarrier

\clearpage

\section{Conclusions and outlook}
\label{sec:conclusionsWP}

\begin{table}[t]
\renewcommand{\arraystretch}{1.1}
\begin{centering}
\small
	\begin{tabular}{l  r r }
	\toprule
	   Contribution  & WP25 & WP20\\ \midrule
HVP LO (lattice) &  $\amuHVPLOresult$ & $7116(184)$\\
HVP LO ($e^+e^-,\tau$) &  \cref{tab:summary_ee} & $6931(40)^*$ \\
HVP NLO ($e^+e^-$) & $\amuHVPNLOresult$ & $-98.3(7)$\\
HVP NNLO ($e^+e^-$) & $\amuHVPNNLOresult$ & $12.4(1)$\\
HLbL (phenomenology) &  $\amuHLbLdataresult$ & $92(19)$\\
HLbL NLO (phenomenology) &   $\amuHLbLNLOdataresult$ & $2(1)$\\
HLbL (lattice) &  $\amuHLbLlatticeresult$ &  $82(35)$\\
HLbL (phenomenology + lattice) &  $\amuHLbLaverageresult$ & $90(17)$\\
\midrule
QED             &     $\amuQEDresult$ &  $116\,584\,718.931(104)$\\
EW     &    $\amuEWresult$  & $153.6(1.0)$\\
HVP (LO + NLO + NNLO)& $\amuHVPtotalresult$ & $6845(40)$ \\
HLbL (phenomenology + lattice + NLO) &   $\amuHLbLtotalresult$ & $92(18)$\\ 
      Total SM Value  &  $\amuSMresult$  & $ 116\, 591\, 810(43)$\\
        \bottomrule
        \renewcommand{\arraystretch}{1.0}
	\end{tabular}
	\caption{Comparison of the key results from this work (WP25), as given in \cref{tab:summary}, to the corresponding numbers from WP20~\cite{Aoyama:2020ynm} (in units of $10^{-11}$). Note that the ``HLbL (lattice)'' result from WP20 has been adapted to include the charm-loop contribution. The entry ``HVP (LO + NLO + NNLO)'' derives from  
    HVP LO (lattice) [WP25] and  HVP LO ($e^+e^-$) [WP20], respectively. The asterisk indicates that the LO HVP value from WP20 was based on $e^+e^-$ data only, while in \cref{tab:summary_ee} we also include the current status for $\tau$-based evaluations.}  
\label{tab:summary_comparison}
\end{centering}
\end{table}

In this second edition of the White Paper on the muon $g-2$, we have charted the progress that has been achieved since 2020 in evaluating the contributions from the electromagnetic (QED), electroweak (EW), and strong (QCD) interactions  to $a_\mu$.

Both the QED and EW contributions have changed very slightly since the first edition, as can be seen from \cref{tab:summary_comparison}. 
A discrepancy in the evaluation of a sub-class of the 10$^{\rm th}$-order QED contribution has been resolved~\cite{Volkov:2019phy,Volkov:2024yzc,Aoyama:2024aly}, leading to a tiny shift in the central value of $\amuQED$. At the same time, the current tension in the experimental determination of the fine-structure constant $\alpha$ is reflected in an increase of the error and a change in the last decimal.
The quoted uncertainty for the EW contribution has more than halved since WP20, thanks to a more precise determination of hadronic effects in the two-loop EW contributions~\cite{Hoferichter:2025fea}, while also the precision in input parameters such as the top-quark and Higgs-boson masses has increased.  
Contributions to $a_\mu$ from QCD are still by far the dominant sources of uncertainty. Here, much of the current debate is centered on hadronic vacuum polarization (HVP).
Significant developments in both data-driven, dispersive and lattice-QCD determinations of the HVP contribution to $a_\mu$ have fundamentally changed the picture since WP20.
This is reflected by the updated SM estimate, in which the result for the LO HVP contribution is now based on lattice-QCD calculations rather than the data-driven dispersive method.

For the data-driven approach, the CMD-3 measurements of the $e^+e^- \to \pi^+\pi^-$ cross section~\cite{CMD-3:2023alj,CMD-3:2023rfe} are higher than those of all other data sets. Despite substantial efforts, see the detailed discussions in \cref{sec:dataHVP}, the emerged discrepancies in this dominant channel are so far not understood and, unlike in WP20, cannot be accommodated any longer through any reasonable inflation of uncertainties in data combinations. The current situation is summarized in \cref{fig:summary_plot_ee}.
Resolving the puzzles will require new measurements, together with an improved understanding of higher-order radiative corrections and their implementation in MC generators. For the former, new data analyses, in particular for the two-pion channel, are expected from several experiments. For the latter, efforts are underway by several groups, coordinated by the {\it RadioMonteCarLow 2} initiative.
However, at this moment, the situation regarding the $\pi^+\pi^-$ data remains unresolved and prevents us from providing a common average for the data-driven dispersive estimate of $\amuHVPLO$.
In addition, we re-examined the role of isospin-breaking corrections to hadronic $\tau$ spectral-function data, providing a detailed assessment of our current understanding thereof, of promising work in progress and avenues for future improvements. 
We are confident that the latter will allow the use of hadronic $\tau$ decays as additional input for the critical two-pion channel in the future.
At the time of WP20 the precision of lattice-QCD calculations was not yet sufficient to impact the SM prediction, and the value quoted for HVP in the SM prediction of WP20 is entirely based on data-driven analyses of hadronic $e^+ e^-$ cross-section data.
In the meantime lattice-QCD calculations have matured significantly, allowing for a precise and robust first-principles calculation of the HVP contribution.
Two aspects are particularly important to achieve this. 
First, the introduction of window observables has proved instrumental for cross-checking and benchmarking lattice calculations of sub-contributions to HVP with a high level of precision. The individual windows isolate and separate the different technical challenges for lattice calculations and allow for tailored approaches for each window. A diverse set of methods with complementary systematic advantages and disadvantages employed by the different lattice-QCD collaborations has led to the consolidation of the individual window contributions one by one.
While this process highlights the consistency of the lattice approaches, significant tensions are observed between lattice and data-driven estimates for the intermediate and long-distance window observables. These tensions appear to originate from the dominant $\pi^+\pi^-$ channel, and would disappear if only CMD-3 data were used.
A second very important development in the lattice community is the broad adoption of blinding procedures to avoid confirmation bias. This is instrumental in establishing the reliability of the observed consolidation when comparing independent lattice-QCD results. The review of lattice-QCD results in WP25 is based on seventeen different papers from eight independent lattice-QCD collaborations \cite{ExtendedTwistedMass:2024nyi, MILC:2024ryz, Spiegel:2024dec, Boccaletti:2024guq, Kuberski:2024bcj, ExtendedTwistedMass:2022jpw, Wang:2022lkq, RBC:2023pvn, Borsanyi:2020mff, Ce:2022kxy, Aubin:2022hgm, Lehner:2020crt, RBC:2018dos, Djukanovic:2024cmq, FermilabLatticeHPQCD:2024ppc, RBC:2024fic, Giusti:2019xct}, including three almost complete lattice calculations of the entire LO HVP contribution \cite{Borsanyi:2020mff,RBC:2024fic,Djukanovic:2024cmq}.
All available results are combined in various ways, yielding consistent averages for $\amuHVPLO$---as our final SM prediction of the latter we take the average that includes the maximum number of independent lattice results from Refs.~\cite{ExtendedTwistedMass:2024nyi, MILC:2024ryz, Spiegel:2024dec, Boccaletti:2024guq, Kuberski:2024bcj, ExtendedTwistedMass:2022jpw, Wang:2022lkq, RBC:2023pvn, Borsanyi:2020mff, Ce:2022kxy, Aubin:2022hgm, Lehner:2020crt, RBC:2018dos, Djukanovic:2024cmq, FermilabLatticeHPQCD:2024ppc, RBC:2024fic, Giusti:2019xct}.
In summary, our consolidated average of lattice-QCD results provides a reliable determination of the LO HVP contribution to the SM prediction of $a_\mu$.

\begin{figure}[t]
\centering
\includegraphics[width=0.8\linewidth]{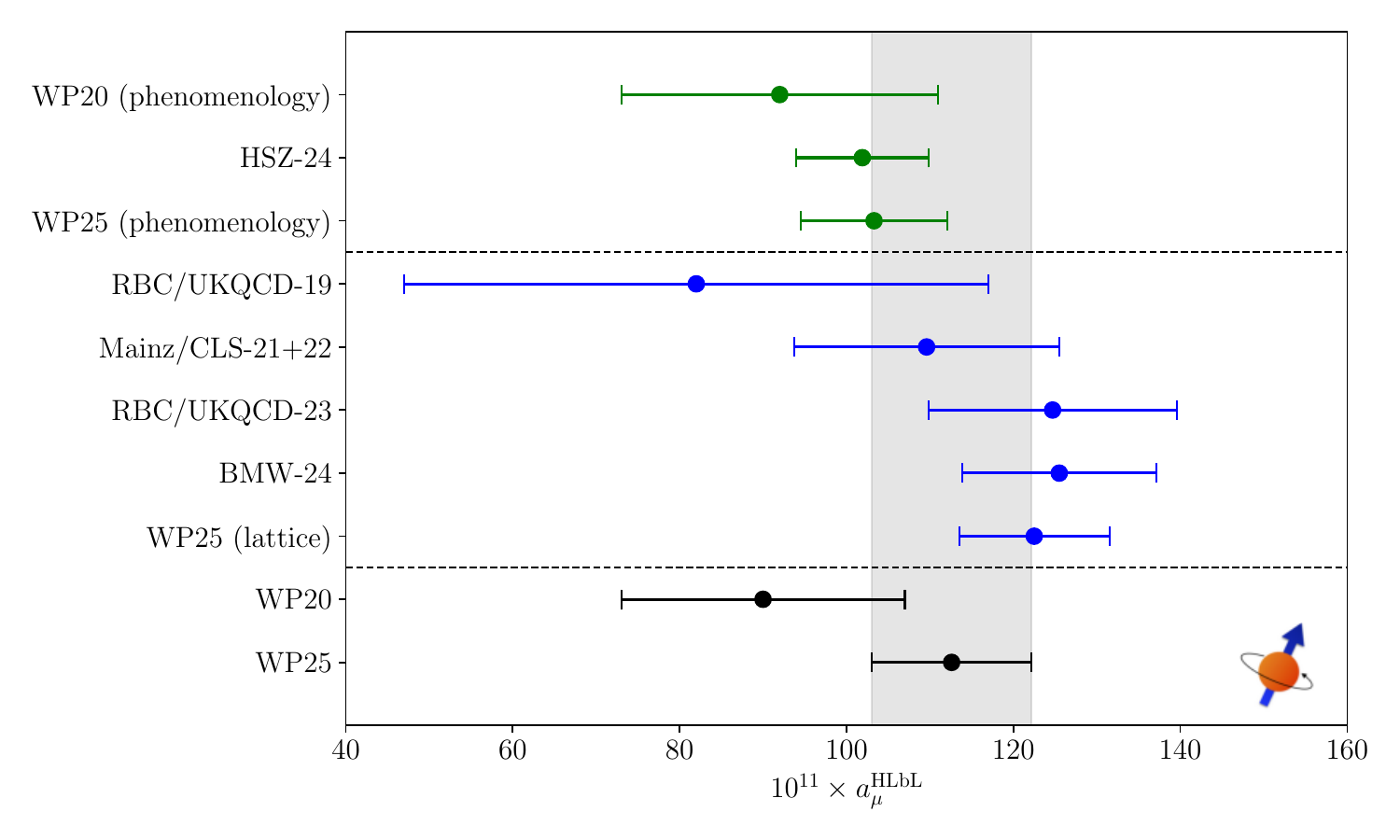}
\caption{Summary of HLbL evaluations, from data-driven methods (green), lattice QCD (blue), and combinations (black). The averages are from WP20~\cite{Aoyama:2020ynm} and WP25, respectively, the other points refer to HSZ-24~\cite{Hoferichter:2024vbu,Hoferichter:2024bae}, RBC/UKQCD-19~\cite{Blum:2019ugy}, Mainz/CLS-21+22~\cite{Chao:2021tvp,Chao:2022xzg}, RBC/UKQCD-23~\cite{Blum:2023vlm}, and BMW-24~\cite{Fodor:2024jyn}.}
\label{fig:summary_plot_hlbl}
\end{figure}

The hadronic light-by-light (HLbL) contribution was already provided as an average of data-driven and lattice QCD results in WP20. Since then both data-driven and lattice evaluations have been developed further such that in this White Paper an update with reduced uncertainty can be provided.
At the current level of precision the different lattice results as well as the lattice and data-driven average are consistent with each other (the latter two at the level of $1.5\sigma$), see \cref{fig:summary_plot_hlbl}.

\begin{figure}[t!]
\centering
\includegraphics[width=0.8\linewidth]{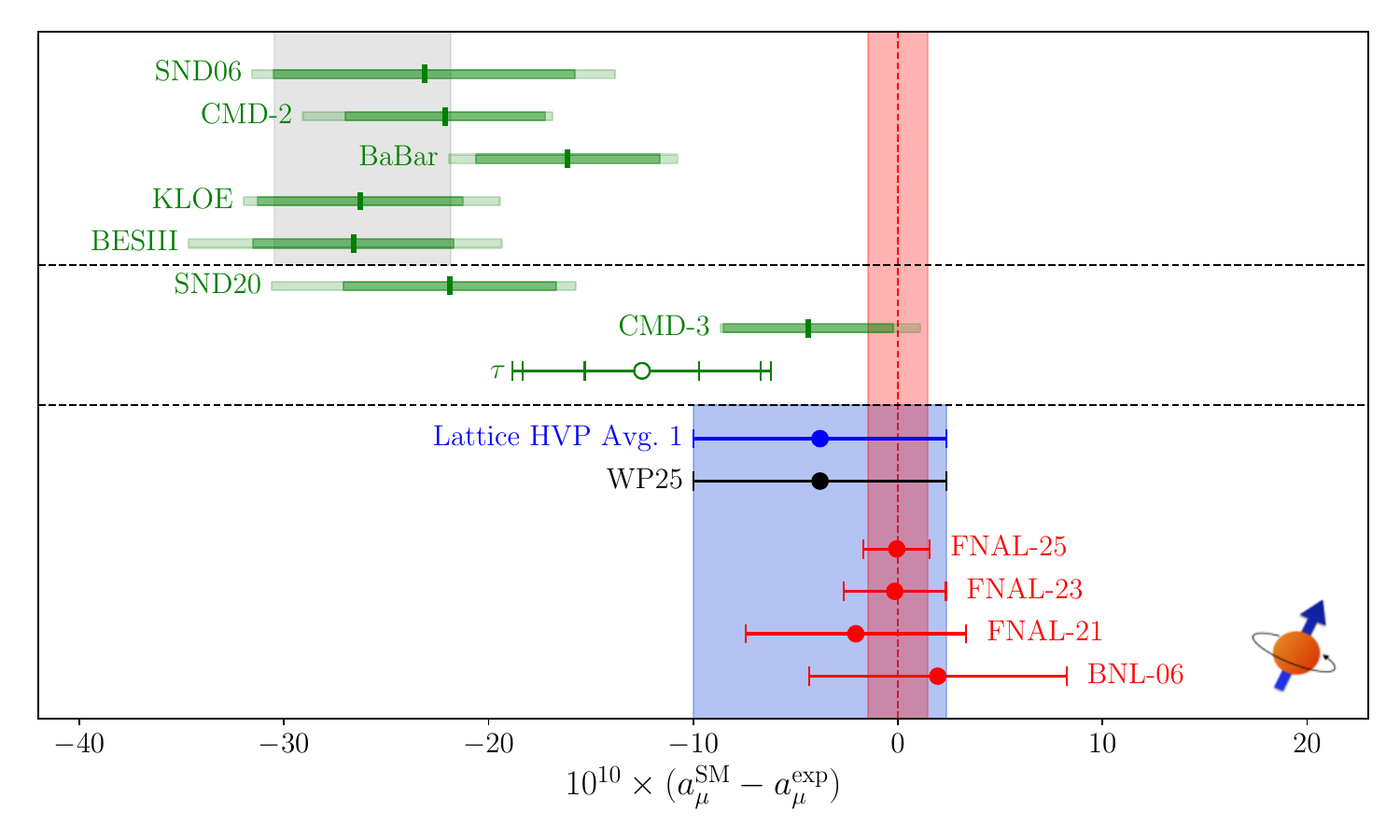}
\caption{Summary of the current SM prediction for $a_\mu$ in comparison to experiment (red band and data points). The final WP25 prediction is denoted in black and via the blue band, it derives from the LO HVP result defined by the lattice-QCD ``Avg.~1'' shown in blue, see \cref{eq:amuLOHVWP25}. The gray band indicates the WP20 result, based on the $e^+e^-$ experiments above the first dashed line. These experimental ranges, as well as the ones for SND20 and CMD-3 that appeared after WP20, are produced as in \cref{fig:summary_plot_data_hvp}; they are meant to illustrate the current situation, but cannot be interpreted as uncertainties with a proper statistical meaning. The $\tau$ point refers to \cref{LOHVP_tau}, the numerical results are collected in \cref{tab:summary_ee}. In all cases except for the gray WP20 band the LO HVP results are combined with WP25 values for the remaining contributions, as summarized in \cref{tab:summary}. The figure has been updated after the announcement of the final results from the Fermilab experiment, including the corrections to the previous experimental points as detailed in Ref.~\cite{Muong-2:2025xyk}.}
\label{fig:summary_plot_final}
\end{figure}

Adding the LO HVP average from lattice QCD, given in \cref{eq:amuLOHVWP25}, to the NLO and NNLO iterations from $e^+e^-$ data, given in \cref{HVPNLO,HVPNNLO}, we obtain for the total HVP contribution 
\begin{equation}
\label{HVPtotal}
\amuHVP=\amuHVPtotalresult \times 10^{-11}\,.
\end{equation}
Averaging the data-driven and lattice-QCD evaluations of the HLbL contribution, given in \cref{eq:amuHLbLdata,eq:amuHLbL_aver}, we obtain 
\begin{equation}
\label{eq:HLbL_comb}
\amuHLbL=\amuHLbLaverageresult\times 10^{-11}\qquad \text{(phenomenology + lattice)}\,,
\end{equation}
where the uncertainty includes a scale factor $S=1.5$. With this average, the NLO contribution in \cref{eq:amuHLbLNLOdata} slightly changes to $\amuHLbLNLO \text{(phenomenology + lattice)}=2.8(6)\times 10^{-11}$, and the total HLbL contribution becomes 
\begin{equation}
\label{eq:amuHLbLtotal}
\amuHLbL+\amuHLbLNLO =\amuHLbLtotalresult\times 10^{-11}\qquad \text{(phenomenology + lattice)}\,.
\end{equation}
Combining \cref{HVPtotal,eq:amuHLbLtotal} with the QED and EW contributions from \cref{amuQEDfinal,amuEWNew}, we obtain for the final SM prediction
\begin{equation}
\label{eq:amuSM}
\amuSM = \amuSMresult\times 10^{-11}\,,
\end{equation}
which can be compared to the current experimental average~\cite{\expref}\footnote{This paper was posted on arXiv on May 28, 2025.  \Cref{sec:dataHVP,sec:latticeHVP,sec:comparison,sec:dataHLbL,sec:latticeHLbL,sec:QED,sec:EW} and all numbers pertaining to the SM prediction have remained unchanged, but the experimental world average has been updated according to the E989 announcement on June 3, 2025~\cite{Muong-2:2025xyk}, and the description in \cref{sec:execsumm,sec:introWP,sec:conclusionsWP} has been adapted accordingly. In particular, the experimental results in abstract, \cref{tab:summary}, \cref{fig:summary_plot_data_hvp,fig:finalHVPsummary,fig:summary_plot_final}, and \cref{sec:conclusionsWP} have been updated.} 
\begin{equation}
\label{eq:amuexpresult}
    \amuexp=\amuexpresult\times 10^{-11}\,.
\end{equation}
At the current level of precision there is no tension between the SM
prediction and the experimental world average:
\begin{equation}
\label{amudiff}
 \Delta a_\mu\equiv\amuexp - \amuSM =\amudiffresult\times 10^{-11}\,.
\end{equation}
This marks a significant shift from the conclusions of WP20,
which is driven by the developments relating to the HVP LO contribution, as can be seen in \cref{tab:summary_comparison} and \cref{fig:summary_plot_final}.

By comparing the uncertainties of \cref{eq:amuexpresult,eq:amuSM} it is apparent that the precision of the SM prediction must be improved by about a factor four to match the precision of the current experimental average after the final result from the E989 experiment at Fermilab.
We expect progress on both data-driven and lattice methods applied to the hadronic contributions in the next few years. Resolving the tensions in the data-driven estimations of the HVP contribution is particularly important, and additional experimental results combined with further scrutiny of theory input such as from event generators should provide a path towards this goal.
Further progress in the calculation of isospin-breaking corrections, from both data-driven and lattice-QCD methods, 
should enable a robust SM prediction from $\tau$ data as well. 
For lattice-QCD calculations of HVP 
continuing efforts by the world-wide lattice community are expected to yield further significant improvements in precision and, hopefully, even better consolidation thanks to a diversity of methods. 
The future focus will be, in particular, on more precise evaluations of isospin-breaking effects and the noisy contributions at long distances.

The role of $a_\mu$ as a sensitive probe of the SM continues to evolve.
We stress that, even though a consistent picture has emerged regarding lattice calculations of HVP, the case for a continued assessment of the situation  remains very strong in view of the observed tensions among data-driven evaluations.
New and existing data on $e^+e^-$ hadronic cross sections from the main collaborations in the field, as well as new measurements of hadronic $\tau$ decays that will be performed at Belle~II, will be crucial not only for resolving the situation but also for pushing the precision of the SM prediction for $a_\mu$ to that of the direct measurement.
This must be complemented by new experimental efforts with completely different systematics, such as the MUonE experiment, aimed at measuring the LO HVP contribution, as well as an independent direct measurement of $a_\mu$, which is the goal of the E34 experiment at J-PARC.
The interplay of all these approaches, various experimental techniques and theoretical methods, may yield profound insights in the future, both regarding improved precision in the SM prediction and the potential role of physics beyond the SM.
Finally, the subtleties in the evaluation of the SM prediction for $a_\mu$ will also become relevant for the anomalous magnetic moment of the electron, once the experimental tensions in the determination of the fine-structure constant are resolved.

In this second-edition White Paper, WP25, the Muon $g-2$ Theory Initiative is presenting their new SM prediction for $a_\mu$, which constitutes a major change from WP20. For the future, the Theory Initiative remains dedicated to continuing to support the wide-ranging efforts and to compile updated predictions.

\clearpage

\section*{Acknowledgements}

We thank The University of Edinburgh for hosting the fifth plenary workshop~\cite{PlenaryWS-2022}. This workshop was supported by the Higgs Centre for Theoretical Physics, the Edinburgh Parallel Computing Centre, the Institute for Particle Physics Phenomenology (UKRI grant ST/T001011/1), UKLFT (UKRI grant ST/T000813/1), and STRONG-2020. This workshop was also sponsored by Eviden (formerly ATOS), NVIDIA, and StorJ. STRONG-2020 is a project part of the European Union's Horizon 2020 research and innovation program with grant agreement no.\ 824093.

We thank the University of Bern for hosting the sixth plenary workshop~\cite{PlenaryWS-2023}. This workshop was supported in part by the Albert Einstein Center for Fundamental Physics and the Swiss National Science Foundation (project nos.\ 200020\_200553, 200021\_200866, PCEFP2\_181117, and PCEFP2\_194272).

We thank KEK for hosting the seventh plenary workshop~\cite{PlenaryWS-2024}. This workshop was supported in part by KEK Theory Center, KEK Institute of Particle and Nuclear Studies, the Kobayashi--Maskawa Institute and Flavor Physics International research center in Nagoya University, and the U.S.--Japan Science and Technology Cooperation Program in High Energy Physics. We also thank the Wilhelm and Else Heraeus Foundation, which generously supported the participation of early career scientists at the Simon Eidelman School on Muon Dipole Moments and Hadronic Effects. The school was organized at the Kobayashi--Maskawa Institute in Nagoya during the week preceding the seventh plenary workshop. 

We thank M.~Petschlies for discussions. 

\begin{sloppypar}
The work in this paper was supported in part  
by the Agence nationale pour la recherche (French National Research Agency, ANR) under contract ANR-22-CE31-0011 ``HVP4NewPhys,''
by the Austrian Science Fund (FWF) through grant DOI 10.55776/PAT2089624, 10.55776/PAT7221623, 10.55776/I3845,
and the Doctoral Program Particles and Interactions, grant DOI 10.55776/W1252,
by Beatriz-Galindo support, University of Huelva, Huelva, Spain,
by CNPq (Brazilian Agency) grant no.\ 308979/2021-4,
by CNRS, 
by Conacyt,
by ICSC -- Centro Nazionale di Ricerca in High Performance Computing, Big Data and Quantum Computing, funded by European Union -- NextGenerationEU,
by Coordinaci\'on de la Investigaci\'on Cient\'ifica, Universidad Michoacana de San Nicol\'as de Hidalgo, Morelia, Mexico, grant  4.10, SECIHTI (Mexico), project CBF2023-2024-3544,
by the Deutsche Forschungsgemeinschaft (German Research Foundation, DFG) through the Cluster of Excellence ``Precision Physics, Fundamental Interactions and Structure of Matter'' (PRISMA$^+$ EXC 2118/1) funded within the German Excellence strategy (project no.\ 39083149),
through the Collaborative Research Center~1660 ``Hadrons and Nuclei as Discovery Tools,''
through the Research Unit FOR5327 ``Photon--photon interactions in the Standard Model and beyond: Exploiting the discovery potential from MESA to the LHC'' (grant 458854507),
through the Emmy Noether program (grant 449369623),
through project HI~2048/1-3 (project no.\ 399400745),
by the ERC grant MUON-101054515,
by the European Union's Horizon Europe Research and Innovation program under 
the Marie Sk\l{}odowska-Curie grant agreement no.\ 101106243, grant 
no.\ 824093 (H2020-INFRAIA-2018-1), and grant agreement no.\ 858199 (INTENSE),
through the EuroHPC Joint Undertaking computing time allocated under grant EHPC-REG-2022R03-166, 
by the Excellence Initiative of Aix-Marseille University -- A*MIDEX, a French ``Investissements d'Avenir'' program, through the Chaire d'Excellence program (AMX-22-RE-AB-052) and 
via grant AMX-22-RE-AB-052 ``AMUalphaNP,''
by FAPESP (S\~ao Paulo Research Foundation) grant nos.\ 2021/06756-6, 2020/15532-1, and 2022/02328-2,
by  FEDER UE through grants PID2023-147112NB-C21, through the ``Unit of Excellence Mar\'ia de Maeztu 2020-2023'' award to the Institute of Cosmos Sciences, grant CEX2019-000918-M,
by Generalitat de Catalunya (AGAUR) through grants 2021SGR01095 and  2021SGR00649, 
by the Serra H{\'u}nter program, through the ``PROMETEO'' program grant CIPROM/2022/66,
by Generalitat Valenciana (Spain) through the plan GenT program (CIDEIG/2023/12), through the ``PROMETEO'' program grant  
PROMETEO/2021/07,
by the Istituto Nazionale di Fisica Nucleare (INFN), through research projects ENP (Exploring New Physics) and LQCD123,
by the Japan Society for the Promotion of Science under grant nos.\ 
KAKENHI-16K05338, 20H05646, 20K03646, 20K03926, 20K03960, 20H05625, L23530, 22K21350,
by Junta de Andaluc\'\i a grant nos.\ POSTDOC\_21\_00136 and P18-FR-5057,
by the Leverhulme Trust, LIP-2021-014 (UK),
by the Italian Ministry of University and Research (MUR) and the European Union (EU) -- Next Generation EU, Mission 4, Component 1, PRIN 2022, CUP F53D23001480006, research project 2022TJFCYB -- Nonperturbative aspects of fundamental interactions, in the Standard Model and beyond,
by the Italian Ministero dell'Universit\`a e Ricerca (MUR) and European Union -- Next Generation EU through the research grant 2022ENJMRS under the program PRIN 2022,
by the Italian Ministero dell'Universit\`a e Ricerca (MUR) and European Union -- Next Generation EU through the research grant no.\ 20225X52RA ``MUS4GM2: Muon Scattering for $g-2$'' under the program Prin~2022,
by  MICIU/AEI/10.13039/501100011033,
by Spanish ``Agencia Estatal de Investigaci\'on,'' MCIN/AEI/10.13039/501100011033, through grant nos.\ PID2023-151418NB-I00, PID2020-114473GB-I00, PID2023-146220NB-I00, PID2020-114767GB-I00, PID2023-147072NB-I00, and CEX2023-001292-S,
by Spanish Ministry of Science and Innovation (MICINN) grants  PID2022-140440NB-C22,  PID2023-146142NB-I00,
by the Spanish MICIU (Ramon y Cajal program RYC2019-027605-I and project PID2022-136510NB-C31),
by the National Natural Science Foundation of China (NSFC) under grant no.\ 12125501,
by the Natural Sciences and Engineering Research Council of Canada,
by the ``Rita Levi Montalcini'' program for young researchers of the Italian Ministry of University and Research (MUR), The Royal Society (URF/R1/231503), 
by the Swiss National Science Foundation under grant nos.\ 
200020\_200553, 200020\_208222, 200021\_175761, PCEFP2\_181117, PCEFP2\_194272, TMCG-2\_213690, 200020\_207386,
Ambizione program (grant PZ00P2\_193383),
by SECIHTI (Mexico), project CBF2023-2024-3226,
by the UK Research and Innovation, Engineering and Physical Sciences Research Council, grant no.\ EP/X021971/1,
by the UK Science and Technology Facilities Council (STFC) under grant nos.\ ST/S000925/, ST/T000600/1, ST/T000988/1, ST/X000494/1, ST/X000699/1, and ST/Y509759/1,
by the U.S.\ Department of Energy, Office of Science, Office of High Energy Physics under award nos.\ DE-SC0010005, DE-SC0010120, DE-SC0010339, DE-SC0015655, DE-SC0013682, DE-SC0021147, and the ``High Energy Physics Computing Traineeship for Lattice Gauge Theory'' DE-SC0024053,
by the U.S.\ Department of Energy, Office of Science, Office of Nuclear Physics under award no.\ DE-FG02-00ER41132,
and by the U.S.\ National Science Foundation under grant nos.\ PHY20-13064, PHY23-10571, PHY-2309135 (to the Kavli Institute for Theoretical Physics, KITP), OAC-2311430.
This document was prepared in part using the resources of the Fermi National Accelerator
Laboratory (Fermilab), a U.S.\ Department of Energy, Office of Science, Office of High Energy
Physics HEP User Facility. Fermilab is managed by Fermi Forward Discovery Group, LLC, acting
under United States Department of Energy contract no.\ 89243024CSC000002.  
\end{sloppypar}

\clearpage

\bibliographystyle{apsrev4-1_mod}
\bibliography{SM}

\end{document}